\documentclass[12pt]{article}

\usepackage{varwidth}
\usepackage{lipsum}
\usepackage{threeparttable}
\usepackage{makecell}
\usepackage{amsfonts}
\usepackage{amsmath}
\usepackage{mathrsfs}
\usepackage{fancyhdr}
\usepackage{epsfig,color,morefloats}
\usepackage{xcolor}
\usepackage[authoryear]{natbib}
\usepackage{booktabs,multirow,multicol,longtable}
\usepackage{amsthm}
\usepackage{amssymb}
\usepackage{color}
\usepackage{lscape}
\usepackage{rotating}
\usepackage{caption}
\usepackage{subfigure}
\usepackage{graphicx}
\usepackage[breaklinks,colorlinks,linkcolor=blue,citecolor=blue,urlcolor=blue]{hyperref}
\usepackage{lineno}
\usepackage{cancel}
\usepackage{multirow}

\usepackage{setspace}

\usepackage{amsmath, amssymb, amsthm, latexsym,color, natbib}
\usepackage{fancyhdr}
\usepackage{algpseudocode}
\usepackage{algorithm}
\usepackage{algpseudocode}
\usepackage{enumitem}
\usepackage{geometry}
\usepackage{pdflscape}

\usepackage[resetlabels]{multibib}
\allowdisplaybreaks
\newcites{S}{References} 

\algnewcommand\INPUT{\item[\textbf{Input:}]}%
\algnewcommand\OUTPUT{\item[\textbf{Output:}]}%

\def\T{{\mathrm{\scriptscriptstyle \top} }}

\theoremstyle{definition}

\newtheorem{remark}{Remark}

\usepackage{pdflscape}

\makeatletter
\newcommand{\fdsy@scale}{1.0}
\newcommand\fdsy@mweight@normal{Book}
\newcommand\fdsy@mweight@small{Book}
\newcommand\fdsy@bweight@normal{Medium}
\newcommand\fdsy@bweight@small{Medium}

\DeclareFontFamily{U}{FdSymbolC}{}

\DeclareFontShape{U}{FdSymbolC}{m}{n}{
	<-7.1> s * [\fdsy@scale] FdSymbolC-\fdsy@mweight@small
	<7.1-> s * [\fdsy@scale] FdSymbolC-\fdsy@mweight@normal
}{}
\DeclareFontShape{U}{FdSymbolC}{b}{n}{
	<-7.1> s * [\fdsy@scale] FdSymbolC-\fdsy@bweight@small
	<7.1-> s * [\fdsy@scale] FdSymbolC-\fdsy@bweight@normal
}{}

\DeclareSymbolFont{arrows}{U}{FdSymbolC}{m}{n}
\SetSymbolFont{arrows}{bold}{U}{FdSymbolC}{b}{n}

\DeclareMathSymbol{\upvDash}{\mathrel}{arrows}{233}
\let\Vbar\upvDash
\DeclareMathSymbol{\upmodels}{\mathrel}{arrows}{237}
\makeatother

\DeclareMathSymbol{\upvDash}{\mathrel}{arrows}{233}
\let\Vbar\upvDash
\DeclareMathSymbol{\upmodels}{\mathrel}{arrows}{237}
\makeatother

\def\T{{\mathrm{\scriptscriptstyle \top} }}

\newcommand{\bA}{{\mathbf A}}
\newcommand{\bB}{{\mathbf B}}
\newcommand{\bX}{{\mathbf X}}
\newcommand{\bY}{{\mathbf Y}}
\newcommand{\bZ}{{\mathbf Z}}

\newcommand{\bU}{{\mathbf U}}
\newcommand{\bV}{{\mathbf V}}
\newcommand{\bW}{{\mathbf W}}
\newcommand{\bS}{{\mathbf S}}

\newcommand{\bT} {\boldsymbol{T}}
\newcommand{\balpha} {\boldsymbol{\alpha}}
\newcommand{\bbeta}  {\boldsymbol{\beta}}

\newcommand{\bdelta} {\boldsymbol{\delta}}

\newcommand{\bOmega}{\boldsymbol{\Omega}}

\newcommand{\bSigma}{\boldsymbol{\Sigma}}

\newcommand{\bgamma}{\boldsymbol{\gamma}}

\newcommand{\bTheta} {\boldsymbol{\Theta}}

\newcommand{\bxi} {\boldsymbol{\xi}}

\newcommand{\bzeta} {\boldsymbol{\zeta}}

\newcommand{\bzero}{{\mathbf 0}}

\newcommand{\bet}{\boldsymbol{\eta}}
\newcommand{\ve}{{\varepsilon}}
\newcommand{\cov}{{\rm Cov}}
\newcommand{\var}{\mbox{Var}}

\theoremstyle{plain}
\newtheorem{inequality}{Inequality}
\newtheorem{lemma}{Lemma}
\newtheorem{theorem}{Theorem}
\newtheorem{proposition}{Proposition}
\newtheorem{cd}{Condition}

\theoremstyle{definition}
\newtheorem{definition}{Definition}

\usepackage{authblk}

\allowdisplaybreaks

\makeatletter

\DeclareFontFamily{U}{FdSymbolC}{}

\DeclareFontShape{U}{FdSymbolC}{m}{n}{
	<-7.1> s * [\fdsy@scale] FdSymbolC-\fdsy@mweight@small
	<7.1-> s * [\fdsy@scale] FdSymbolC-\fdsy@mweight@normal
}{}
\DeclareFontShape{U}{FdSymbolC}{b}{n}{
	<-7.1> s * [\fdsy@scale] FdSymbolC-\fdsy@bweight@small
	<7.1-> s * [\fdsy@scale] FdSymbolC-\fdsy@bweight@normal
}{}

\DeclareSymbolFont{arrows}{U}{FdSymbolC}{m}{n}
\SetSymbolFont{arrows}{bold}{U}{FdSymbolC}{b}{n}

\DeclareMathSymbol{\upvDash}{\mathrel}{arrows}{233}
\let\Vbar\upvDash
\DeclareMathSymbol{\upmodels}{\mathrel}{arrows}{237}
\makeatother

\def\spacingset#1{\renewcommand{\baselinestretch}%
	{#1}\small\normalsize} \spacingset{1}

\newcommand{\blind}{1}

\def\singlespace{\def\baselinestretch{1}\@normalsize}

\begin{document}

\if1\blind
{
\spacingset{1.25}
  \title{\bf \Large Testing Independence and Conditional Independence in High Dimensions via  Coordinatewise Gaussianization}
\author[1,2]{Jinyuan Chang}
\author[1]{Yue Du}
\author[3]{Jing He}
\author[4]{Qiwei Yao}

\affil[1]{\it \small Joint Laboratory of Data Science and Business
Intelligence, Southwestern University of Finance and Economics, Chengdu, China}
\affil[2]{\it \small State Key Laboratory of Mathematical Sciences, Academy of Mathematics and Systems Science, Chinese Academy of Sciences, Beijing, China}
\affil[3]{\it \small Institute of Statistical Interdisciplinary Research, Southwestern University of Finance and Economics, Chengdu, China}
\affil[4]{\it \small Department of Statistics, The London School of Economics and Political Science, London, U.K.}

		\setcounter{Maxaffil}{0}
		
		\renewcommand\Affilfont{\itshape\small}
		\date{\vspace{-5ex}}
		\maketitle
	} \fi
	\if0\blind
	{
		\bigskip
		\bigskip
		\bigskip
		\begin{center}
			{
			\Large \bf  Testing Independence and Conditional Independence in High Dimensions via  Coordinatewise Gaussianization
			}
		\end{center}
		\medskip
	} \fi
 
\spacingset{1.3}
\begin{abstract}
 We propose new statistical tests, in high-dimensional settings, for testing the
		independence of two random vectors and their conditional independence
		given a third random vector. The key idea is simple, i.e., we first transform each component variable to the standard normal via its marginal empirical distribution, and we then test for independence and conditional independence of the transformed random vectors using appropriate $L_\infty$-type test statistics. While we are testing some necessary conditions 
		of the independence or the conditional independence, the new tests 
		outperform the 13 frequently used testing methods in a large scale simulation comparison. The advantage of the new tests can be summarized as follows: (i) they do not require any moment conditions, (ii) they allow arbitrary dependence structures of the components among the random vectors, and (iii) they allow
		the dimensions of random vectors to diverge at the exponential rates of the sample size. 
		The critical values of the proposed tests are determined
		by a computationally efficient multiplier bootstrap procedure. Theoretical analysis shows that the sizes of the proposed tests can be well controlled by the nominal significance level, and the proposed tests are also consistent under certain local alternatives. The finite sample performance of the new tests is illustrated via extensive simulation studies and a real data application.
\end{abstract}


\spacingset{1.69}
\setlength{\abovedisplayskip}{0.2\baselineskip}
\setlength{\belowdisplayskip}{0.2\baselineskip}
\setlength{\abovedisplayshortskip}{0.2\baselineskip}
\setlength{\belowdisplayshortskip}{0.2\baselineskip}

\section{Introduction}

Let $\bX \in \mathbb{R}^p$, $\bY \in \mathbb{R}^q$ and $\bZ\in\mathbb{R}^m$ be three random vectors. Given samples $\{(\bX_i,\bY_i,\bZ_i)\}_{i=1}^n$ with $(\bX_i,\bY_i,\bZ_i) \overset{\rm i.i.d.}\sim (\bX,\bY,\bZ)$, 
we are interested in the following two hypothesis testing problems:
\begin{itemize}
	\item (Hypothesis testing for independence)
	\begin{align}\label{eq:test}
		\mathbb{H}_0:\bX \Vbar   \bY~~~\mbox{versus}~~~\mathbb{H}_1:\bX  \not\Vbar   \bY	 \,.
        \end{align}
	\item (Hypothesis testing for conditional independence)
	\begin{align}\label{eq:testc}
		\mathbb{H}_0:\bX  \Vbar \bY \,|\, \bZ ~~~\mbox{ versus} ~~~ \mathbb{H}_1:  \bX  \not\Vbar \bY \,|\, \bZ\,.
	\end{align}
\end{itemize}
These two testing problems are of direct application in, among others, building statistical models including feature selection and simplification, causal inference, and understanding complex relationships in machine learning and data analysis for various practical problems. Due to their immense importance, a large number of the testing methods have been developed. In spite of this, we argue that there is still a justification for the proposed tests in this paper. Indeed the existing methods have demonstrated the successes under various settings and conditions, but none of them is predominately better than the others. Though it is prohibitively difficult, if not impossible, to construct a universally optimal test, we propose a new test, for each of (\ref{eq:test}) and (\ref{eq:testc}) respectively, under some mild conditions in high-dimensional settings, and  they uniformly outperform the 13 frequently used tests in
the extensive simulation studies.

Our new tests are based on coordinatewise Gaussianization and Gaussian approximation \citep{ {Chernozhukov2017},{chang2024central}} in the high-dimensional settings. Assuming all the marginal distributions of $\bX, \bY$ and $\bZ$ are continuous, we transform each component of $\bX, \bY$ and $\bZ$ to a standard normal random variable by its distribution function.  
Let $\bU, \bV$ and $\bW$ be the transformed
vectors of, respectively, $\bX, \bY$ and $\bZ$. We adopt the maximum
absolute pairwise sample covariance between the components of $\bU$ and those of $\bV$ as the statistic for testing independence hypothesis (\ref{eq:test}). Under the null hypothesis $\mathbb{H}_0$ in (\ref{eq:test}), all the covariances  between the components of $\bU$ and those of $\bV$ are 0. But the converse is not necessarily true.  For testing conditional independence hypothesis (\ref{eq:testc}),  we first fit regression models of $\bU$ and $\bV$ on $\bW$, and then adopt the maximum
absolute pairwise sample covariance between the components of the residuals for $\bU$ and those for $\bV$ as the statistic. Again we are testing  a necessary  
condition under the null hypothesis $\mathbb{H}_0$ in (\ref{eq:testc}). Nevertheless, the extensive simulation studies in Section \ref{sc:simulations} show that the proposed tests uniformly outperform the 13 frequently used tests. 


The null-distributions of the test statistics are evaluated in terms of the Gaussian approximation technique, which is implemented by a computationally efficient multiplier bootstrap scheme for computing the critical values of the tests. Our theoretical analysis shows that the sizes of the new tests can be correctly controlled by the prescribed nominal significance level, and they are also consistent under certain local alternatives.

The advantage exhibited by the proposed  tests can be summarized as follows: (a) They require no moment conditions on $\bX$, $\bY$ and $\bZ$, and, hence,  can be applied to heavy-tailed distributions. (b) They allow arbitrary dependence structures among the components of $\bX$, $\bY$ and $\bZ$. (c) The dimensions of $\bX$ and $\bY$ can diverge at the exponential rates of the sample size, and the dimension of $\bZ$ can diverge at a polynomial rate of the sample size.

The coordinatewise Gaussianization  is a widely used technique in statistical analysis, especially in high-dimensional settings. 
See, for example,  \cite{Liu2009}, \cite{Xue2012} and \cite{Mai2015b} for applications of coordinatewise Gaussianization in high-dimensional Gaussian graphical models and sufficient dimension reduction, and
\cite{Mai2022} for the theoretical guarantee of the coordinatewise Gaussianization methods. 

The literature on the tests of independence and conditional independence is large. 
The independence test has been well studied in the low-dimensional scenario. For example, \cite{Spearman1904},  \cite{Pearson1920}, \cite{Kendall1938}, \cite{Blum1961}, and \cite{Reshef2011} propose  various dependence measures when $p=q=1$. \cite{Wilks1935}, \cite{Hotelling1936}, \cite{Puri1971}, \cite{Hettmansperger1994}, \cite{Gieser1997}, \cite{Taskinen2003} and \cite{Taskinen2005} investigate the tests under the Gaussian or elliptically symmetric distributions with fixed $(p,q)$.
\cite{Gretton2008} consider a test based on the Hilbert-Schmidt independence criterion (HSIC). \cite{Bergsma2014} propose a consistent test based on a sign covariance. \cite{Lyons2013} and \cite{Jakobsen2017} deal with the tests in more general metric spaces. In the high-dimensional scenario with $p, q\gg n$, the distance correlations for characterizing the dependence between $\bX$ and $\bY$ have been proposed and the associated testing procedures for \eqref{eq:test} have been studied.  See \cite{Szekely2007}, \cite{Szekely2013},  \cite{Zhu2020}  and \cite{Gao2020}.
All the tests aforementioned require certain moment conditions on $\bX$ and $\bY$. To alleviate the moment restrictions, a projection correlation based test is considered by \cite{Zhu2017}, and some rank-based tests are presented by \cite{Heller2013},
\cite{Shi2020} and \cite{Deb2021}. 

Testing independence   \eqref{eq:test} is a special case of testing whether $\bX^{(1)}, \ldots, \bX^{(\ell)}$ are mutually independent with $\ell=2$,
where $\bX^{(1)}\in \mathbb{R}^{p_1}, \ldots, \bX^{(\ell)}\in \mathbb{R}^{p_{\ell}}$ are $\ell$ random vectors. Many existing works in the literature focus on this more general setting. 
 When $p_1=\cdots= p_{\ell} = 1$, \cite{Pfister2018} extend the HSIC test \citep{Gretton2008} to $\ell$-variate HSIC for $\ell >2$. See also \cite{Matteson2017}. \cite{Han2017}, \cite{Leung2018} and \cite{Yao2018} propose mutual independence tests for $\bX^{(1)}, \ldots, \bX^{(\ell)}$ when $\ell\gg n$.
When $p_1,\ldots, p_{\ell} > 1$, \cite{Jin2018} propose a test based on generalized distance covariance. \cite{Chakraborty2018} use joint distance covariance to quantify and to test the joint independence among $\ell$ random vectors. See \cite{Chang2023} for a more general form of this testing problem. 
Testing conditional independence \eqref{eq:testc} is more challenging, 
which relies on the properties of the controlling variables $\bZ$. There is also abundant literature on the conditional independence tests for fixed $(p,q,m)$. For the simplest case with $p=q=1$ and fixed $m$, partial correlation \citep{Lawrance1976} is the most commonly used measure for the conditional dependence between two normal variables with the effects of controlling variables being removed. However, in the non-Gaussian case, zero partial correlation coefficient is not necessarily equivalent to conditional independence. 
Various nonparametric tests have been developed in the literature, including  \cite{Kendall1942}, \cite{Goodman1959}, \cite{Veraverbeke2011}, \cite{Azadkia2021} and
\cite{Otneim2021}.
When $p, q, m > 1$, \cite{Su2008},  \cite{Corradi2012} and \cite{Huang2016}  construct the tests by comparing the conditional distributions under the null and the alternative hypotheses. \cite{Su2007} and \cite{Wang2018} introduce tests based on the conditional characteristic functions.  \cite{Fukumizu2007}, \cite{Zhang2012}, \cite{Doran2014} and \cite{Strobl2019} explore extensively various kernel based methods. 
\cite{Runge2018} propose a test based on conditional mutual information. 
 When $p$, $q$ or $m$ is potentially large, 
\cite{Berrett2020} introduce a conditional permutation test,  \cite{Szekely2014}, \cite{Wang2015} and \cite{Fan2020} construct the tests based on the extended conditional distance correlations.  
Furthermore, \cite{Shah2020} propose the generalized covariance measure based on the sample covariance between the residuals of the regressions $\bX$ and $\bY$ on $\bZ$.  Although both the test statistics of \cite{Shah2020}  and ours for conditional independence hypothesis \eqref{eq:testc} are based on the residuals of some regressions,  there are several fundamental differences worth noting. First, since \cite{Shah2020} consider the regressions $\bX$ and $\bY$ on $\bZ$ directly, it is necessary to impose certain moment conditions on the elements of $\bX$, $\bY$ and $\bZ$, whereas our proposed method considers the regressions $\bU$ and $\bV$ on $\bW$ which essentially eliminates the moment conditions on the elements of $\bX$, $\bY$ and $\bZ$ through coordinatewise Gaussianization, and can offer an advantage in dealing with heavy-tailed data.  Second,  the methods used to solve the regression problems are different, where \cite{Shah2020}  use kernel ridge regression, and we employ the feedforward neural network.   Third, the theoretical guarantee of \cite{Shah2020} requires that $m$ is fixed, while our proposed method allows $m$ to diverge with the sample size. 
Extensive simulation studies in Section \ref{sc:condindtestsimu} show that their method may not work for heavy-tailed data, but the proposed method performs exceptionally well with both high-dimensional and heavy-tailed data. 
For dependent data,
\cite{Zhou2022} propose a conditional independence test 
based on a projection approach. 

The rest of the paper is organized as follows. Section \ref{sec:CG} introduces the coordinatewise Gaussianization technique.  Section \ref{sec:indtest-m} introduces the proposed independence test. Section \ref{sc:CondIndTest} introduces the proposed conditional independence tests based on nonparametric regressions and linear regressions, respectively.
Section \ref{sc:ComputeFast} provides a computationally efficient multiplier bootstrap scheme for computing the critical values of the proposed tests. Section \ref{sec:theoretical} investigates the associated theoretical properties of the proposed tests. Section  \ref{sc:simulations} evaluates the finite-sample performance of the proposed tests via extensive simulation studies.  All technical proofs and a real data  example are relegated to the supplementary material. 
The used real data and the code for implementing our proposed tests are available at the GitHub repository: {\url{https://github.com/JinyuanChang-Lab/CoordinatewiseGaussianizationTest}}.

{\it Notation.} The notation $I(\cdot)$ denotes the indicator function. For any positive integer $k$, write $[k]=\{1, \ldots, k\}$, and denote by $\mathbf{I}_k$ the $k\times k$ identity matrix. For any $a,b \in \mathbb{R}$, let $\lceil a \rceil$ and $\lfloor a \rfloor$  denote, respectively, the smallest integer greater than or equal to $a$, and the largest integer less than or equal to $a$, and let $a \vee b$ and $a \wedge b$ denote, respectively, the larger and smaller number between $a$ and $b$.  
For a vector ${\bf a} = (a_1,\ldots, a_k)^{\T} \in \mathbb{R}^k$, let $|{\bf a}|_{0} = \sum_{i=1}^k I(a_i \neq 0)$, $|{\bf a}|_{1}=\sum_{i=1}^k|a_i|$,  $|{\bf a}|_2=(\sum_{i=1}^{k}a_{i}^2)^{1/2}$, and $|{\bf a}|_{\infty} = \max_{i \in [k]}|a_i|$ be its $L_0$-norm, $L_1$-norm, $L_2$-norm and $L_{\infty}$-norm, respectively. For a matrix $\bA=(A_{i,j})_{k_1\times k_2}$, we write $|\bA|_{\infty}=\max_{i\in[k_1],\,j\in[k_2]}|A_{i,j}|$. Denote by $\otimes$ the Kronecker product operator between matrices.  For any set $\mathcal{S}$, let $|\mathcal{S}|$ denote its cardinality.  Let $\mathcal{N}(\boldsymbol{\mu},\bB)$, $U(a,b)$ and  $t(c)$ denote, respectively, multi-dimensional normal distribution with mean vector $\boldsymbol{\mu}$ and covariance matrix $\bB$, the uniform distribution on $[a,b]$,  and the $t$-distribution with $c$ degrees of freedom. Let  $\Phi(\cdot)$ be the cumulative distribution function of the standard normal distribution $\mathcal{N}(0,1)$. For any two sequences of positive numbers $\{a_k\}$ and $\{b_k\}$, we write $a_k \lesssim b_k$ or $b_k\gtrsim a_k$ if $\limsup_{k\to \infty}a_k/b_k < \infty$, 
and write $a_{k}\ll b_{k}$ or $b_{k}\gg a_{k}$  if  $\limsup_{k\to \infty}a_k/b_k =0$.   Moreover, $a_{k} \asymp b_{k}$ means that $a_k\lesssim b_k$ and $b_k\lesssim a_k$  hold simultaneously. The sets of natural numbers, natural numbers including 0 and real numbers are denoted by $\mathbb{N}$, $\mathbb{N}_{0}$ and $\mathbb{R}$, respectively.  


\section{Coordinatewise Gaussianization}\label{sec:CG}
Let $\bX=(X_1,\ldots,X_p)^{\T} \sim F_{\bX}$, $\bY=(Y_1,\ldots,Y_q)^{\T} \sim F_{\bY}$ and $\bZ=(Z_1,\ldots,Z_m)^{\T} \sim F_{\bZ}$ be three generic random vectors. For each $j \in [p]$, $k \in [q]$ and $l\in[m]$, denote by $F_{\bX,j}(\cdot)$, $F_{\bY,k}(\cdot)$ and $F_{\bZ,l}(\cdot)$, respectively, the  distribution functions of $X_j$, $Y_k$ and $Z_l$. Assume all $F_{\bX,j}(\cdot)$, $F_{\bY,k}(\cdot)$ and $F_{\bZ,l}(\cdot)$ are continuous. Then $U_{j}\equiv\Phi^{-1}\{F_{\bX,j}(X_{j})\} $, $V_{k}\equiv\Phi^{-1}\{F_{\bY,k}(Y_{k})\}$ and $W_l\equiv\Phi^{-1}\{F_{\bZ,l}(Z_l)\}$ are  the standard normal random variables.  Put 
$
	\bU=(U_{1},\ldots,U_{p})^{\T}$, $\bV=(V_{1},\ldots,V_{q})^{\T}$ and $\bW=(W_1,\ldots,W_m)^{\T}$.
Since $\Phi^{-1}\{F_{\bX,j}(\cdot)\}$, $\Phi^{-1}\{F_{\bY,k}(\cdot)\}$ and $\Phi^{-1}\{F_{\bZ,l}(\cdot)\}$ are strictly monotone mappings, the hypotheses \eqref{eq:test} and \eqref{eq:testc} are equivalent to, respectively,
\begin{align}\label{eq:equind}
	\mathbb{H}_0:\bU \Vbar \bV  ~~~\mbox{versus} ~~~ \mathbb{H}_1: \bU \not\Vbar \bV\,, 
\end{align}
and
\begin{align}\label{eq:equiconind}
	\mathbb{H}_0:\bU \Vbar \bV \,|\, \bW  ~~~ \mbox{versus} ~~~ \mathbb{H}_1: \bU \not\Vbar \bV \,|\, \bW\,.
\end{align}


For each $i\in[n]$, write $\bX_i=(X_{i,1},\ldots,X_{i,p})^{\T}$, $\bY_i=(Y_{i,1},\ldots,Y_{i,q})^{\T}$ and $\bZ_i=(Z_{i,1},\ldots,Z_{i,m})^\T $, and define $\bU_i=(U_{i,1},\ldots,U_{i,p})^{\T}$, $\bV_i=(V_{i,1},\ldots,V_{i,q})^{\T}$ and $\bW_i=(W_{i,1},\ldots,W_{i,m})^{\T}$
with $U_{i,j}=\Phi^{-1}\{F_{\bX,j}(X_{i,j})\} $, $V_{i,k}=\Phi^{-1}\{F_{\bY,k}(Y_{i,k})\}$ and $W_{i,l}=\Phi^{-1}\{F_{\bZ,l}(Z_{i,l})\}$. Write $\mathcal{X}_{n}=\{\bX_1,\ldots,\bX_n\}$, $\mathcal{Y}_{n}=\{\bY_1,\ldots,\bY_n\}$ and $\mathcal{Z}_{n}=\{\bZ_1,\ldots,\bZ_n\}$. 
Given $(\mathcal{X}_n,\mathcal{Y}_n,\mathcal{Z}_n)$, we can approximate $\bU_{i}$, $\bV_{i}$ and $\bW_{i}$, respectively, by
	$\hat{\bU}_i=(\hat{U}_{i,1},\ldots,\hat{U}_{i,p})^{\T}$, $\hat{\bV}_i=(\hat{V}_{i,1},\ldots,\hat{V}_{i,q})^{\T}$ and $\hat{\bW}_i=(\hat{W}_{i,1},\ldots,\hat{W}_{i,m})^{\T}$
with 
\begin{align}\label{eq:hatU}
		\hat{U}_{i,j}=\Phi^{-1} \bigg\{\frac{n\hat{F}_{\bX,j}(X_{i,j})}{n+1}\bigg\}\,,  ~\hat{V}_{i,k}=\Phi^{-1} \bigg\{\frac{n\hat{F}_{\bY,k}(Y_{i,k})}{n+1}\bigg\}\,, ~\hat{W}_{i,l}=\Phi^{-1} \bigg\{\frac{n\hat{F}_{\bZ,l}(Z_{i,l})}{n+1}\bigg\} \,,
\end{align}
where 
	$\hat{F}_{\bX,j}(\cdot)=n^{-1}\sum_{s=1}^{n}I(X_{s,j}\le \cdot)$, $\hat{F}_{\bY,k}(\cdot)=n^{-1}\sum_{s=1}^{n}I(Y_{s,k}\le \cdot)$ and $\hat{F}_{\bZ,l}(\cdot)=n^{-1}\sum_{s=1}^{n}$ $I(Z_{s,l}\le \cdot)$. 
Multiplying them by $n(n+1)^{-1}$ in \eqref{eq:hatU} is to guarantee $|\hat{U}_{i,j}| < +\infty$, $|\hat{V}_{i,k}| < +\infty$ and $|\hat{W}_{i,l}| < +\infty$. 
In Sections \ref{sec:indtest-m} and \ref{sc:CondIndTest}, we will propose testing procedures for \eqref{eq:test} and \eqref{eq:testc} 
based on coordinatewise Gaussianization.

\section{Testing for Independence}\label{sec:indtest-m}
Note  that $\bgamma_i\equiv\bU_i\otimes\bV_i$ is a $d$-dimensional random vector with $d=pq$, and $\mathbb{E}(\bgamma_i) = \bzero$ under the null hypothesis $\mathbb{H}_0$ in \eqref{eq:equind}.
For given $\{(\bU_i,\bV_i)\}_{i=1}^n$, several studies have considered testing whether $\mathbb{E}(\bgamma_i)=\bzero$ holds; see, e.g., \cite{Chang2017}, \cite{Yang2024}, and the references therein. In practice, however, $\{(\bU_i, \bV_i)\}_{i=1}^{n}$ are unobservable, so we need to construct feasible statistics based on $\{(\hat{\bU}_i, \hat{\bV}_i)\}_{i=1}^{n}$. 
Let
$
    \hat{\bS}_n=n^{-1}\sum_{i=1}^{n}\hat{\bgamma}_i$ with $\hat{\bgamma}_i = \hat{\bU}_i \otimes \hat{\bV}_i$, where the components of $\hat{\bU}_i$ and $\hat{\bV}_i$ are specified in \eqref{eq:hatU}.  
The components of $\hat{\bS}_n$ can be viewed as all
the pairwise sample covariances between the components of
$\bU$ and those of $\bV$. 
We consider the test statistic 
\begin{align*}
	H_{n}=\sqrt{n}|\hat{\bS}_n|_{\infty} 
\end{align*}
for \eqref{eq:equind}, 
and reject $\mathbb{H}_0$ at the significance level $\alpha \in (0,1)$ if 
$
	H_{n} > {\rm cv}_{{\rm ind},\alpha}$, 
where $ {\rm cv}_{{\rm ind},\alpha}$ is the critical value satisfying $\mathbb{P} (H_{n} >{\rm cv}_{{\rm ind},\alpha}) = \alpha$ under $\mathbb{H}_0$.

Let $ \bSigma=\cov(\bgamma_i)$, which can be estimated by
$
    \hat{\bSigma}=n^{-1}\sum_{i=1}^{n}\hat{\bgamma}_i\hat{\bgamma}_i^{\T}-\bar{\hat{\bgamma}}\bar{\hat{\bgamma}}^{\T}
$
with $\bar{\hat{\bgamma}} = n^{-1}\sum_{i=1}^{n}\hat{\bgamma}_i$. For any $\alpha\in(0,1)$,  Proposition 1 in Appendix A of the supplementary material
indicates that ${\rm cv}_{{\rm ind},\alpha}$ can be approximated by 
\begin{equation}\label{eq:cv-ind-hat}
	\hat{{\rm cv}}_{{\rm ind},\alpha}=\inf\big\{t\ge 0:\mathbb{P}(|\hat{\bxi}|_{\infty} \leq t\,|\,\mathcal{X}_{n},\mathcal{Y}_n)\ge 1-\alpha\big\}
\end{equation}
for $\hat{\bxi}\,|\,\mathcal{X}_n,\mathcal{Y}_n\sim  \mathcal{N}(\boldsymbol{0},\hat{\bSigma})$. Section \ref{sc:ComputeFast} will introduce a multiplier bootstrap procedure to determine the critical value for the test, which  is computationally efficient in practice.  Notice that coordinatewise Gaussianization eliminates the need for moment conditions on $\bX$ and $\bY$, which enables the proposed independence test to handle heavy-tailed distributions.  

\begin{remark}
     The $L_{\infty}$-type test statistic $H_n$ can be extended to a more general test statistic:  
\begin{align}\label{eq:sta-general1}
    \tilde{H}_n = \max_{1\le j_1 < \cdots < j_T \le d}\sum_{t=1}^{T} \sqrt{n}| \hat{S}_{n,j_t}|\,,
\end{align}
where $\hat{\bS}_n  = (\hat{S}_{n,1}, \ldots, \hat{S}_{n,d})^{\T} =n^{-1}\sum_{i=1}^{n}\hat{\bgamma}_i $. Based on the Gaussian approximation technique \citep{Chernozhukov2017,chang2024central}, the associated critical value can be selected as $\inf\{t\ge 0:\mathbb{P}(\max_{1\le j_1 < \cdots < j_T \le d}\sum_{t=1}^{T} |\hat{\xi}_{j_t}| \leq t\,|\,\mathcal{X}_{n},\mathcal{Y}_n)\ge 1-\alpha\}$ with $\hat{\bxi} =(\hat{\xi}_1, \ldots, \hat{\xi}_{d})^{\T}$.  
When $T=1$, this new test statistic is identical to $H_n$.  Using the test statistic with the form \eqref{eq:sta-general1} has some technical advantages. More specifically, if $T$ is fixed or diverges slower than some certain rate, the associated test procedure is valid even if $p$ and $q$ diverge exponentially fast with  $n$, which does not require any structural assumptions imposed on the covariance matrices of $\bU$ and $\bV$. These advantages are quite important in practice. 
\end{remark}


\begin{remark}\label{rek:sufficient}
     Notice that $U_j$ and $V_k$ are Gaussian random variables for each $j\in[p]$ and $k\in[q]$. 
Our testing procedure actually selects $|\mathbb{E}(\bU\bV^{\T})|_{\infty}$  as a measure for $\bU \Vbar \bV$. If $\bU$ and $\bV$ are jointly Gaussian, then $|\mathbb{E}(\bU\bV^{\T}) |_{\infty}= 0$ if and only if  $\bU \Vbar \bV$. If $\bU$ and $\bV$ are not jointly Gaussian, our procedure essentially tests a necessary condition of the independence.  
However, regardless of whether $\bU$ and $\bV$ are jointly Gaussian or not, Theorem \ref{thm:1} in Section \ref{sec:indtest} shows that our proposed test can always control the size at the significance level. When $\bU \not\Vbar \bV$, Theorem \ref{thm:2} in Section \ref{sec:indtest} shows that the power of our proposed test will depend on the magnitude of $|\mathbb{E}(\bU\bV^{\T})|_{\infty}$. 
We admit that $|\mathbb{E}(\bU\bV^{\T})|_{\infty}$ may be equal to zero when $\bU \not\Vbar \bV$. In this case, we may consider to choose a different measure such that $\bU \Vbar \bV$ if and only if this measure between $\bU$ and $\bV$ equals to zero. See Section R.1 in the supplementary material for details. 
\end{remark}


\section{Testing for Conditional Independence} \label{sc:CondIndTest}
Given $(\bU_i,\bV_i,\bW_i)$, 
we consider two regression models: 
\begin{equation} \label{eq: generalReg}
    \bU_i={\bf f}(\bW_i) +\boldsymbol{\varepsilon}_i\,,~~~~~\bV_i={\bf g}( \bW_i)+\bdelta_i\,,
\end{equation}
where ${\bf f}(\bW_i) = \mathbb{E}(\bU_i\,|\,\bW_i)$, and ${\bf g}( \bW_i)
= \mathbb{E}(\bV_i\,|\,\bW_i)$. The null hypothesis $\mathbb{H}_0$ in \eqref{eq:equiconind} holds if and
only if $\boldsymbol{\varepsilon}_i \Vbar \bdelta_i\,|\,\bW_i$. In general, we can estimate ${\bf f}(\cdot)$ and ${\bf g}(\cdot)$ in \eqref{eq: generalReg} using feedforward neural networks, which will be introduced in Section \ref{sec:condTest-FNN}. It is well known that estimating nonparametric regression models using feedforward neural networks requires a substantially large number of observations, especially in high-dimensional scenarios. 
Alternatively, when the sample size $n$ is small, we can further consider to fit the following linear models: 
\begin{align}\label{eq:regressionUV}	\bU_i=\bA\bW_i+\boldsymbol{\varepsilon}_i\,,~~~~~\bV_i=\bB \bW_i+\bdelta_i \,,
\end{align}
with $\mathbb{E}(\boldsymbol{\varepsilon}_i\,|\,\bW_i)=\bzero$ and $\mathbb{E}(\bdelta_i\,|\,\bW_i)=\bzero$. If $(\bU_i, \bV_i,\bW_i)$ is jointly normal, (\ref{eq: generalReg}) reduces to the linear equations in \eqref{eq:regressionUV},  
and the null hypothesis $\mathbb{H}_0$ in \eqref{eq:equiconind}  holds if and only if $\cov(\boldsymbol{\varepsilon}_i,\bdelta_i)=\bzero$.  
We will proceed with the linear representation (\ref{eq:regressionUV}) in Section \ref{sec:cindtest-proc-linear}.  The simulation results in Section \ref{sc:condindtestsimu} indicate that the proposed conditional independence test based on the linear regressions performs well in most scenarios, and outperforms the proposed conditional independence test based on the nonparametric regressions in most cases when the sample size $n$ is small. 

\subsection{Conditional Independence Test based on Nonparametric Regressions} \label{sec:condTest-FNN}


Write $\boldsymbol{\ve}_{i}=(\ve_{i,1},\ldots, \ve_{i,p})^{\T}$ and $\boldsymbol{\delta}_{i}=(\delta_{i,1},\ldots,\delta_{i,q})^{\T}$. The component-wise forms of  \eqref{eq: generalReg} are as follows:
\begin{align}\label{eq:nonparametriic-model}
    U_{i,j}=f_{j}(\bW_i) + \varepsilon_{i,j}\,,\qquad V_{i,k}=g_{k}(\bW_i) + \delta_{i,k} \,,
\end{align}
where $f_j(\bW_i) =\mathbb{E}(U_{i,j}\,|\,\bW_i)$ and $g_k(\bW_i) =\mathbb{E}(V_{i,k}\,|\,\bW_i)$. 
Recall $\bU_i=(U_{i,1},\ldots,U_{i,p})^{\T}$, $\bV_i=(V_{i,1},\ldots,V_{i,q})^{\T}$ and $\bW_i=(W_{i,1},\ldots,W_{i,m})^{\T}$
with $U_{i,j}=\Phi^{-1}\{F_{\bX,j}(X_{i,j})\} $, $V_{i,k}=\Phi^{-1}\{F_{\bY,k}(Y_{i,k})\}$ and $W_{i,l}=\Phi^{-1}\{F_{\bZ,l}(Z_{i,l})\}$. 
Let $\mathcal{D}_1$, $\mathcal{D}_2$ and $\mathcal{D}_3$ be three disjoint subsets of $[n]$ with  $|\mathcal{D}_1|=n_1\asymp n$, $|\mathcal{D}_2|=n_2\asymp n$ and $|\mathcal{D}_3|=n_3\asymp n^{\kappa}$ for some constant $\kappa\in(0,1)$. Write $\mathcal{W}_{\mathcal{D}_j}=\{(\bX_i,\bY_i,\bZ_i): i\in \mathcal{D}_j\}$. Our testing procedure includes three steps:  Step 1 estimates $F_{\bX,j}(\cdot)$, $F_{\bY,k}(\cdot)$ and $F_{\bZ,l}(\cdot)$ based on $\mathcal{W}_{\mathcal{D}_1}$, Step 2 estimates $f_{j}$ and $g_{k}$ based on $\mathcal{W}_{\mathcal{D}_2}$, and  Step 3 calculates the test statistic and critical value based on $\mathcal{W}_{\mathcal{D}_3}$. See Section \ref{sec:step-1} for details. Section \ref{sec:selectn} will propose a data-driven procedure to select $(n_1,n_2,n_3)$ in practice.

\subsubsection{Testing Procedure}\label{sec:step-1}

Given the subsamples $\mathcal{W}_{\mathcal{D}_1}$, the empirical distribution functions 
     $\hat{F}_{\bX,j}(\cdot)= n_1^{-1}\sum_{s\in\mathcal{D}_1} I(X_{s,j} \le \cdot)$, $\hat{F}_{\bY,k}(\cdot)= n_1^{-1}\sum_{s\in\mathcal{D}_1} I(Y_{s,k} \le \cdot)$ and $\hat{F}_{\bZ,l}(\cdot)= n_1^{-1}\sum_{s\in\mathcal{D}_1} I(Z_{s,l} \le \cdot)$
provide the natural estimates for $F_{\bX,j}(\cdot)$, $F_{\bY,k}(\cdot)$ and $F_{\bZ,l}(\cdot)$. Since $\hat{F}_{\bX,j}(X_{i,j})$ may be equal to $0$ or $1$ for $i\in\mathcal{D}_2\cup\mathcal{D}_3$, we consider its truncated version as follows:
\begin{align}\label{eq:tructed-ecdf}
    \hat{F}_{\bX,j}^{(w)} (\cdot) =&~ \frac{1}{n_1}I\bigg\{\hat{F}_{\bX,j}  (\cdot) \le \frac{1}{n_1}\bigg\} +  \hat{F}_{\bX,j}  (\cdot)I\bigg\{\frac{1}{n_1} < \hat{F}_{\bX,j}  (\cdot) \le  \frac{n_1-1}{n_1}\bigg\} \notag\\
    &+\frac{n_1-1}{n_1} I\bigg\{\hat{F}_{\bX,j}  (\cdot) > \frac{n_1 -1}{n_1}\bigg\}  \,.
\end{align}
Analogously, we can define $\hat{F}_{\bY,k}^{(w)}(\cdot)$ and $\hat{F}_{\bZ,l}^{(w)}(\cdot)$ in the same manner. Then, for each $i\in[n]$, we can approximate $\bU_i$, $\bV_i$ and $\bW_i$, respectively, by $\hat{\bU}_{i}^{(w)}=(\hat{U}_{i,1}^{(w)}, \ldots, \hat{U}_{i,p}^{(w)})^{\T}$, $\hat{\bV}_{i}^{(w)}=(\hat{V}_{i,1}^{(w)}, \ldots, \hat{V}_{i,q}^{(w)})^{\T}$ and $
\hat{\bW}_{i}^{(w)}=(\hat{W}_{i,1}^{(w)}, \ldots, \hat{W}_{i,m}^{(w)})^{\T}$ with  $\hat{U}_{i,j}^{(w)}=\Phi^{-1}\{\hat{F}_{\bX,j}^{(w)} (X_{i,j})\}$, $\hat{V}_{i,k}^{(w)}=\Phi^{-1}\{\hat{F}_{\bY,k}^{(w)} (Y_{i,k})\}$ and $\hat{W}_{i,l}^{(w)}=\Phi^{-1}\{\hat{F}_{\bZ,l}^{(w)} (Z_{i,l})\}$, which guarantee $|\hat{U}_{i,j}^{(w)}| < +\infty$, $|\hat{V}_{i,k}^{(w)}| < +\infty$ and $|\hat{W}_{i,l}^{(w)}| < +\infty$.



Given an integer $\ell\geq0$, let $\mathcal{H}^{(\ell)}$ be the hierarchical neural networks proposed by \cite{Bauer2019}. See \eqref{eq:nnspace-hl} in Section \ref{subsec:condind_theory-nn} for its definition. Write  $ T_{\tilde{\beta}_n} \mathcal{H}^{(\ell)}=\{T_{\tilde{\beta}_{n}}h: h\in \mathcal{H}^{(\ell)}\}$,  where $(T_{\tilde{\beta}_n} h)(\boldsymbol{x}) =   \{|h(\boldsymbol{x})| \wedge \tilde{\beta}_n \} {\rm sign}\{h(\boldsymbol{x})\}$  with $\tilde{\beta}_n = (\log n)\log^{1/2}(\tilde{d}n)$ and $\tilde{d}=p\vee q\vee m$.  Given  $\{(\hat{\bU}_{i}^{(w)}, \hat{\bV}_{i}^{(w)}, \hat{\bW}_{i}^{(w)})\}_{i\in\mathcal{D}_2}$,  we can estimate $f_j$ and $g_k$ as
\begin{equation}\label{eq:fhj-ghk}
    \begin{split}
        \hat{f}_{j}(\cdot) =&~\arg\min_{h\in T_{\tilde{\beta}_n} \mathcal{H}^{(\ell)}} \frac{1}{n_2}\sum_{i\in \mathcal{D}_2} |\hat{U}_{i,j}^{(w)} - h(\hat{\bW}_i^{(w)})|^2\,, \\
        \hat{g}_{k}(\cdot) =&~\arg\min_{h\in T_{\tilde{\beta}_n} \mathcal{H}^{(\ell)}} \frac{1}{n_2}\sum_{i\in \mathcal{D}_2} |\hat{V}_{i,k}^{(w)} - h(\hat{\bW}_i^{(w)})|^2\,. 
    \end{split}
\end{equation}


Given  $\{(\hat{\bU}_{i}^{(w)}, \hat{\bV}_{i}^{(w)}, \hat{\bW}_{i}^{(w)})\}_{i\in\mathcal{D}_3}$,  let $\tilde{\bOmega}_{n} = n_{3}^{-1}\sum_{i\in\mathcal{D}_3} \tilde{\bet}_{i}$ with $\tilde{\bet}_{i}= \tilde{\boldsymbol{\ve}}_{i} \otimes \tilde{\bdelta}_{i}$, where 
$\tilde{\boldsymbol{\varepsilon}}_i=(\tilde{\ve}_{i,1},\ldots,\tilde{\ve}_{i,p})^{\T}$ and $\tilde{\bdelta}_i=(\tilde{\delta}_{i,1},\ldots,\tilde{\delta}_{i,q})^{\T}$ with $ \tilde{\varepsilon}_{i,j} = \hat{U}_{i,j}^{(w)} - \hat{f}_{j}(\hat{\bW}_{i}^{(w)})$ and $\tilde{\delta}_{i,k} = \hat{V}_{i,k}^{(w)} - \hat{g}_{k}(\hat{\bW}_{i}^{(w)})$ for $\hat{f}_j(\cdot)$ and $\hat{g}_{k}(\cdot)$ specified in \eqref{eq:fhj-ghk}. We consider the test statistic  
\begin{align*}
	\tilde{G}_{n}= \sqrt{n_{3}} |\tilde{\bOmega}_{n}|_{\infty}
\end{align*}
for \eqref{eq:equiconind}, and reject $\mathbb{H}_0$ at the significance level $\alpha \in(0,1)$ if $\tilde{G}_n> {\rm cv}_{{\rm cind},\alpha} $, where ${\rm cv}_{{\rm cind},\alpha} $ is the  critical value   satisfying $\mathbb{P}(\tilde{G}_{n} > {\rm cv}_{{\rm cind},\alpha} ) = \alpha$ under $\mathbb{H}_0$. 

Let $\bTheta =\cov(\bet_i)$ for $\bet_{i}= \boldsymbol{\ve}_{i} \otimes \bdelta_{i}$, which can be estimated by $\tilde{\bTheta}=n_3^{-1}\sum_{i\in\mathcal{D}_3}\tilde{\bet}_i\tilde{\bet}_{i}^{\T}-\bar{\tilde{\bet}} \bar{\tilde{\bet}}^{\T}$ with $\bar{\tilde{\bet}} = n_{3}^{-1}\sum_{i\in\mathcal{D}_3}\tilde{\bet}_i$. For any $\alpha\in(0,1)$,   Proposition 2 in Appendix B of the supplementary material  indicates that ${\rm cv}_{{\rm cind},\alpha} $ can be approximated by
\begin{equation}\label{eq:cv-cind-hat-nn}
	\hat{{\rm cv}}_{{\rm cind},\alpha}  =\inf\big\{t\ge 0:\mathbb{P}(|\tilde{\bzeta} |_{\infty} \leq t\,|\,\mathcal{X}_{n},\mathcal{Y}_n,\mathcal{Z}_n)\ge 1-\alpha\big\}
\end{equation}
for $\tilde{\bzeta} \,|\, \mathcal{X}_{n}, \mathcal{Y}_{n},\mathcal{Z}_{n} \sim  \mathcal{N}(\boldsymbol{0},\tilde{\bTheta})$.  
Section \ref{sc:ComputeFast} will introduce a multiplier bootstrap procedure to determine the critical value for the test, which is more computationally efficient.  Our theoretical analysis in Section \ref{subsec:condind_theory-nn} shows that the proposed conditional independence test based on nonparametric regressions has three advantages: (a) no moment conditions on $\bX$, $\bY$ and $\bZ$ are required,  (b) it allows arbitrary dependence structures among the components of $\bX$, $\bY$ and $\bZ$,  and (c) it allows the dimensions of $\bX$ and $\bY$  to grow exponentially with the sample size $n$, while allowing the dimension of $\bZ$ to grow polynomially with the sample size $n$.

\begin{remark}
    Notice that the key requirement for the validity of our proposed method is
\begin{equation}\label{eq:error-non1}
 \bigg|\frac{1}{\sqrt{n_3}}\sum_{i\in \mathcal{D}_3}\tilde{\boldsymbol{\varepsilon}}_i\tilde{\bdelta}_i^{\T}-\frac{1}{\sqrt{n_3}}\sum_{i\in \mathcal{D}_3}\boldsymbol{\varepsilon}_i\bdelta_i^{\T}\bigg|_\infty=o_{\rm p}(1)\,.
 \end{equation}
    As shown in Section \ref{subsec:condind_theory-nn}, the estimated functions $\hat{f}_j$ and $\hat{g}_k$ from  the hierarchical neural networks $\mathcal{H}^{(\ell)}$ can  satisfy \eqref{eq:error-non1} if $f_j$ and $g_k$ are $(\vartheta, C)$-smooth functions \citep{Bauer2019}. However,   $\mathcal{H}^{(\ell)}$ is not necessary for our proposed method.  More generally, any alternative function class can be used in place of $\mathcal{H}^{(\ell)}$, as long as the resulting estimates $\hat{f}_j$ and $\hat{g}_k$ satisfy \eqref{eq:error-non1}. 
\end{remark}

\subsubsection{A Data-driven Procedure for Selecting $(n_1,n_2,n_3)$}\label{sec:selectn}

To implement the testing procedure for conditional independence proposed in Section \ref{sec:step-1}, we need to determine $(n_1,n_2,n_3)$ in practice. Our theory requires $n_1\asymp n$, $n_2\asymp n$ and $n_3\asymp n^{\kappa}$ for some constant $\kappa\in(0,1)$. Since the test statistic $\tilde{G}_n$ is constructed based on $n_3$ samples, the selection of $n_3$ will play a key role in the size control of the proposed test. On the other hand, due to $n_1, n_2\gg n_3$, the approximation errors caused by $(\hat{\bU}_i^{(w)},\hat{\bV}_i^{(w)},\hat{\bW}_i^{(w)})$ to $(\bU_i,\bV_i,\bW_i)$ in Step 1 and $(\hat{f}_j,\hat{g}_k)$ to $(f_j,g_k)$ in Step 2 will be negligible in constructing the theoretical properties of $\tilde{G}_n$. Hence, we mainly focus on the selection of $n_3$. In practice, we always set $\mathcal{W}_{\mathcal{D}_1}=\{(\bX_i,\bY_i,\bZ_i)\}_{i=1}^{n_1}$ and $\mathcal{W}_{\mathcal{D}_2}=\{(\bX_i,\bY_i,\bZ_i)\}_{i=n_1+1}^{n_1+n_2}$ with $n_1=\lfloor n/3\rfloor$ and $n_2=\lfloor n/2\rfloor$, and target on selecting some samples from $\{(\bX_i,\bY_i,\bZ_i)\}_{i=n_1+n_2+1}^n$ to form $\mathcal{W}_{\mathcal{D}_3}$. More specifically, given $\mathcal{W}_{\mathcal{D}_1} \cup \mathcal{W}_{\mathcal{D}_2}$, we can obtain the estimate  $(\hat{f}_j, \hat{g}_k)$. Then, for each $i\in[n] $, we have $\tilde{\boldsymbol{\varepsilon}}_i=(\tilde{\ve}_{i,1},\ldots,\tilde{\ve}_{i,p})^{\T}$ and $\tilde{\bdelta}_i=(\tilde{\delta}_{i,1},\ldots,\tilde{\delta}_{i,q})^{\T}$ with $\tilde{\varepsilon}_{i,j} = \hat{U}_{i,j}^{(w)} - \hat{f}_{j}(\hat{\bW}_{i}^{(w)})$ and $\tilde{\delta}_{i,k} = \hat{V}_{i,k}^{(w)} - \hat{g}_{k}(\hat{\bW}_{i}^{(w)})$. Based on the idea of bootstrap, we present in Algorithm \ref{alg:select_tilden} a data-driven procedure for selecting $n_3$ in practice.

\begin{algorithm}[!ht]
	\setstretch{0.9}
	\caption{Selection of optimal $n_3$}\label{alg:select_tilden}
	\label{alg3}
	{\footnotesize
		\hspace*{0.02in} {\bf Input:}
  		(i) the number of repetitions $B$; and (ii) the significance level $\alpha$, (iii) the estimated functions $\{\hat{f}_j\}_{j=1}^{p}$ and $\{\hat{g}_k\}_{k=1}^{q}$. 
		\begin{algorithmic}[1]
			\For {$b\in[B]$}
			\State
			 Generate $\{\varsigma_{1,i,k}^{(b)}\}_{i,k=1}^n$, $\{\varsigma_{2,i,k}^{(b)}\}_{i,k=1}^n$ and $ \{\varsigma_{3,i,k}^{(b)}\}_{i,k=1}^n$ independently from  $\mathcal{N}(0,1)$. Compute 
			$
				\bW_{i}^{(b)}   = n^{-1/2}\sum_{k=1}^{n}\varsigma_{1,i,k}^{(b)} \hat{\bW}_k^{(w)}$, $  \boldsymbol{\ve}_{i}^{(b)}  =n^{-1/2}\sum_{k=1}^{n}\varsigma_{2,i,k}^{(b)} \tilde{\boldsymbol{\ve}}_{k}$ and $
				\bdelta_{i}^{(b)} =n^{-1/2}\sum_{k=1}^{n}\varsigma_{3,i,k}^{(b)} \tilde{\bdelta}_{k}$
			for each $i \in [n]$. Write   $\boldsymbol{\ve}_{i}^{(b)} =(\ve_{i,1}^{(b)} ,\ldots, \ve_{i,p}^{(b)})^{\T}$ and $ \bdelta_{i}^{(b)} =(\delta_{i,1}^{(b)} ,\ldots, \delta_{i,q}^{(b)} )^{\T} $.
			\State Calculate  $\bU_{i}^{(b)} =(U_{i,1}^{(b)} ,\ldots, U_{i,p}^{(b)} )^{\T} $ and $\bV_{i}^{(b)} =(V_{i,1}^{(b)} ,\ldots, V_{i,q}^{(b)} )^{\T}$ with $U_{i,j}^{(b)}  =\hat{f}_j( \bW_{i}^{(b)} ) + \ve_{i,j}^{(b)} $ and $ V_{i,k}^{(b)}  =\hat{g}_k( \bW_{i}^{(b)} ) + \delta_{i,k}^{(b)} $ for each $i \in [n]$. 
            \State Construct $\hat{\bU}_{i}^{(b)} =(\hat{U}_{i,1}^{(b)}, \ldots, \hat{U}_{i,p}^{(b)})^{\T}$ with $\hat{U}_{i,j}^{(b)} = U_{i,j}^{(b)}I(|U_{i,j}^{(b)}|\le M_1) + M_1 \cdot{\rm sign}(U_{i,j}^{(b)})I(|U_{i,j}^{(b)}|>M_1)$ and $M_1 =\Phi^{-1}(1-n_1^{-1})$. Analogously, construct  $\hat{\bV}_{i}^{(b)} =(\hat{V}_{i,1}^{(b)}, \ldots, \hat{V}_{i,q}^{(b)})^{\T} $ and $\hat{\bW}_{i}^{(b)} =(\hat{W}_{i,1}^{(b)}, \ldots, \hat{W}_{i,m}^{(b)})^{\T}$ in the same manner as $\hat{\bU}_{i}^{(b)}$ but with replacing $ \bU_{i}^{(b)}$ by $ \bV_{i}^{(b)}$ and $ \bW_{i}^{(b)}$, respectively. 
			\State For each $j\in[p]$ and $k\in[q]$, calculate $\hat{f}_{j}^{(b)} $ and $\hat{g}_{k}^{(b)} $ in the same manner as $\hat{f}_j$ and $\hat{g}_{k}$ specified in \eqref{eq:fhj-ghk} but with replacing $\{(\hat{\bU}_i^{(w)} ,  \hat{\bV}_i^{(w)} ,  \hat{\bW}_i^{(w)})\}_{i \in \mathcal{D}_2}$ by $\{(\hat{\bU}_{i}^{(b)}  ,  \hat{\bV}_{i}^{(b)} ,  \hat{\bW}_{i}^{(b)})\}_{i \in \mathcal{D}_2}$. 
			\State  For each $i\in[n]\backslash[n_1+n_2]$, calculate $ \tilde{\bet}_{i}^{(b)}  = \tilde{\boldsymbol{\ve}}_{i}^{(b)}  \otimes \tilde{\bdelta}_{i}^{(b)} $ with $\tilde{\boldsymbol{\ve}}_{i}^{(b)} =(\tilde{\ve}_{i,1}^{(b)} ,\ldots,\tilde{\ve}_{i,p}^{(b)} )^{\T} $ and $\tilde{\bdelta}_{i}^{(b)} =(\tilde{\delta}_{i,1}^{(b)} ,\ldots,\tilde{\delta}_{i,q}^{(b)} )^{\T}$, where
			$\tilde{\ve}_{i,j}^{(b)} = \hat{U}_{i,j}^{(b)} - \hat{f}_{j}^{(b)} (\hat{\bW}_{i}^{(b)} )$ and $ \tilde{\delta}_{i,k}^{(b)} = \hat{V}_{i,k}^{(b)} - \hat{g}_{k}^{(b)} (\hat{\bW}_{i}^{(b)} )$. 
		\For {$\tilde{\ell} \in [n-n_1-n_2]$}
		\State  Calculate the test statistic $\tilde{G}_{\tilde{\ell}}^{(b)}  = \sqrt{\tilde{\ell}}|\tilde{\bOmega}_{\tilde{\ell}}^{(b)} |_{\infty}$ with
		$
			\tilde{\bOmega}_{\tilde{\ell}}^{(b)}  = \tilde{\ell}^{-1} \sum_{i=n_1+n_2+ 1}^{n_1+n_2+\tilde{\ell}} \tilde{\bet}_{i}^{(b)} $. 
		\State  Calculate the critical value $\hat{{\rm cv}}_{{\rm cind},\alpha}^{(b)}$   in the same manner as $\hat{\rm cv}_{{\rm cind},\alpha}$ defined in \eqref{eq:cv-cind-hat-nn} but with replacing $\{\tilde{\bet}_i\}_{i\in \mathcal{D}_3}$ by $\{\tilde{\bet}_{i}^{(b)} \}_{i=n_1+n_2+ 1}^{n_1+n_2+\tilde{\ell}}$.
		\State  Calculate $a_{b}(\tilde{\ell})=I\{\tilde{G}_{\tilde{\ell}}^{(b)}  > \hat{{\rm cv}}_{{\rm cind},\alpha}^{(b)}  \}$.
		\EndFor
		\EndFor
		\State
		For each $\tilde{\ell}\in[n-n_1-n_2]$, calculate  $\bar{a}(\tilde{\ell})= B^{-1}\sum_{b=1}^{B}a_{b}(\tilde{\ell})$.
		\end{algorithmic}
		\hspace*{0.02in} {\bf Output:} $n_{3}^{\textrm{opt}} = \arg\min_{\tilde{\ell}\in[n-n_1-n_2]}|\bar{a}(\tilde{\ell}) -\alpha|$.
	}
\end{algorithm}

\begin{remark}
   The proposed conditional independence test based on nonparametric regressions employs a single sample-splitting.   As shown in Section R.4 in the supplementary material, the performance of the proposed method under null hypothesis is empirically insensitive to how to split all the samples into three disjoint parts as long as the size conditions of the three parts are satisfied. The
sample-splitting is introduced primarily for theoretical convenience in establishing the asymptotic properties of the proposed method as $n\rightarrow\infty$. Algorithm \ref{alg:select_tilden} is intended to ensure that the proposed method can achieve good size
control in finite samples. It requires  training additional $Bpq$ neural networks, which is computationally  intensive.   When $n$ is small, sample-splitting may also lead to power loss. In this case, we can consider in practice using the full sample at each step of the testing procedure. Numerical studies in Section R.4 of the supplementary material show that our proposed test with full sample works well when $n$ is small  ($n \le 100$). When $n$ is large, in order to improve the computational efficiency, we may consider setting $n_3=n-n_1-n_2$ directly in practice.

\end{remark}

\subsection{Conditional Independence Test based on Linear Regressions}\label{sec:cindtest-proc-linear} 
Recall $\hat{\bU}_i=(\hat{U}_{i,1},\ldots,\hat{U}_{i,p})^{\T}$, $\hat{\bV}_i=(\hat{V}_{i,1},\ldots,\hat{V}_{i,q})^{\T}$ and $\hat{\bW}_i=(\hat{W}_{i,1},\ldots,\hat{W}_{i,m})^{\T}$ with $\hat{U}_{i,j}$, $\hat{V}_{i,k}$ and $\hat{W}_{i,l}$ specified in \eqref{eq:hatU}. 
For $(\bA,\bB)$ specified in \eqref{eq:regressionUV}, we write $\bA=(\balpha_{1} , \ldots, \balpha_{p} )^{\T}$ and $\bB =(\bbeta_{1}, \ldots, \bbeta_{q})^{\T}$. We can estimate $\balpha_j$ and $\bbeta_k$ by the following Lasso estimators:
\begin{align}\label{eq:lasso}
\begin{split}	 
  \hat{\balpha}_{j}=&\,\arg\min_{\balpha\in\mathbb{R}^m}\bigg\{\frac{1}{n}\sum_{i=1}^{n}(\hat{U}_{i,j}-\balpha^{\T}\hat{\bW}_{i})^2 + 2\lambda_{\balpha,j}|\balpha|_{1}\bigg\}\,,  \\ \hat{\bbeta}_{k}=&\,\arg\min_{\bbeta\in\mathbb{R}^m}\bigg\{\frac{1}{n}\sum_{i=1}^{n}(\hat{V}_{i,k}-\bbeta^{\T}\hat{\bW}_{i})^2 + 2\lambda_{\bbeta,k}|\bbeta|_{1}\bigg\}\,,
\end{split}
\end{align}
where $\lambda_{\balpha,j}$ and $\lambda_{\bbeta,k}$ are the regularization parameters.   Let $ \hat{\bOmega}_{ n}= n^{-1}\sum_{i=1}^n\hat{\bet}_i $ 
with $\hat{\bet}_i=\hat{\boldsymbol{\ve}}_{i} \otimes \hat{\bdelta}_{i}$, where  $\hat{\boldsymbol{\varepsilon}}_i=(\hat{\ve}_{i,1},\ldots,\hat{\ve}_{i,p})^{\T}$ and   $\hat{\bdelta}_i=(\hat{\delta}_{i,1},\ldots,\hat{\delta}_{i,q})^{\T}$ with $\hat{\ve}_{i,j}= \hat{U}_{i,j}- \hat{\balpha}_{j}^{\T}\hat{\bW}_{i}$ and $\hat{\delta}_{i,k}= \hat{V}_{i,k}- \hat{\bbeta}_{k}^{\T}\hat{\bW}_{i}$.  We consider the test statistic 
\begin{align*}
\hat{G}_{n}=\sqrt{n}|\hat{\bOmega}_{n}|_{\infty} 
\end{align*}
for \eqref{eq:equiconind}, and  reject  $\mathbb{H}_0$  at the significance level $\alpha\in(0,1)$ if  $\hat{G}_{n} > {\rm cv}_{{\rm cind},\alpha}^{*}$, where ${\rm cv}_{{\rm cind},\alpha}^{*}$ is the critical value satisfying $\mathbb{P}(\hat{G}_{n} > {\rm cv}_{{\rm cind},\alpha}^{*}) = \alpha$  under  $\mathbb{H}_0$.

Recall $\bTheta = \cov (\bet_i)$, which can be estimated by $\hat{\bTheta}= n^{-1}\sum_{i=1}^{n}\hat{\bet}_i\hat{\bet}_i^{\T}-\bar{\hat{\bet}}\bar{\hat{\bet}}^{\T} $ with $\bar{\hat{\bet}} = n^{-1}\sum_{i=1}^{n}\hat{\bet}_i$. For any $\alpha\in(0,1)$, Proposition 3 in Appendix C of the supplementary material indicates that ${\rm cv}_{{\rm cind},\alpha}^{*}$ can be approximated by 
\begin{equation}\label{eq:cv-cind-hat}
	   \hat{{\rm cv}}_{{\rm cind},\alpha}^{*}   =\inf\big\{t\ge 0:\mathbb{P}(|\hat{\bzeta}|_{\infty} \leq t\,|\,\mathcal{X}_{n},\mathcal{Y}_n,\mathcal{Z}_n)\ge 1-\alpha\big\}
\end{equation}
for  $\hat{\bzeta} \,|\, \mathcal{X}_{n}, \mathcal{Y}_{n},\mathcal{Z}_{n} \sim \mathcal{N}(\boldsymbol{0},\hat{\bTheta})$. Section \ref{sc:ComputeFast} will introduce a multiplier bootstrap procedure to determine the critical value for the test, which is more computationally efficient.   The proposed conditional independence test based on linear regressions shares similar advantages with its nonparametric counterpart discussed in Section \ref{sec:step-1}. The main difference is that the current setting allows the dimensions of $\bX$, $\bY$ and $\bZ$ to grow exponentially with the sample size $n$. See our theoretical analysis in Section \ref{sec:theory-cindtest-linear} for details.


\section{Multiplier Bootstrap Procedure} \label{sc:ComputeFast}

To implement the proposed 
tests, we need to generate bootstrap samples of three $d$-dimensional normal random vectors $\hat{\bxi}\,|\,\mathcal{X}_n,\mathcal{Y}_n\sim  \mathcal{N}(\boldsymbol{0},\hat{\bSigma})$, $\tilde{\bzeta} \,|\, \mathcal{X}_{n}, \mathcal{Y}_{n},\mathcal{Z}_{n} \sim  \mathcal{N}(\boldsymbol{0},\tilde{\bTheta} )$, and $\hat{\bzeta} \,|\, \mathcal{X}_{n}, \mathcal{Y}_{n},\mathcal{Z}_{n} \sim  \mathcal{N}(\boldsymbol{0},\hat{\bTheta} )$. Let $\varpi_1,\ldots, \varpi_n \overset{\rm i.i.d.} \sim \mathcal{N}(0,1)$,
which are independent of $(\mathcal{X}_n,\mathcal{Y}_n,\mathcal{Z}_n)$. Then
\begin{align}\label{eq:Radem}
	\hat{\bxi}^{\dagger} = \frac{1}{\sqrt{n}}\sum_{i=1}^{n}\varpi_i(\hat{\bgamma}_i - \bar{\hat{\bgamma}})\,, ~~\tilde{\bzeta}^{\dagger} = \frac{1}{\sqrt{n_3}}\sum_{i\in \mathcal{D}_3}\varpi_i(\tilde{\bet}_i - \bar{\tilde{\bet}})  
	 ~~{\textrm{and}}~~ \hat{\bzeta}^{\dagger} = \frac{1}{\sqrt{n}}\sum_{i=1}^{n}\varpi_i(\hat{\bet}_i - \bar{\hat{\bet}}) 
\end{align}
satisfy $\hat{\bxi}^{\dagger}\,|\,\mathcal{X}_n,\mathcal{Y}_n\sim  \mathcal{N}(\boldsymbol{0},\hat{\bSigma})$, $\tilde{\bzeta}^{\dagger} \,|\, \mathcal{X}_{n}, \mathcal{Y}_{n},\mathcal{Z}_{n} \sim  \mathcal{N}(\boldsymbol{0},\tilde{\bTheta} )$, and $\hat{\bzeta}^{\dagger} \,|\, \mathcal{X}_{n}, \mathcal{Y}_{n},\mathcal{Z}_{n} \sim  \mathcal{N}(\boldsymbol{0},\hat{\bTheta} )$. For any $\alpha \in (0,1)$, the critical values  defined in \eqref{eq:cv-ind-hat},  \eqref{eq:cv-cind-hat-nn} and \eqref{eq:cv-cind-hat} 
are equal to, respectively, 
\begin{align}\label{eq:cv-tilde}
	\begin{split}
		&~~ \tilde{\rm cv}_{{\rm ind},\alpha} =\inf\big\{t\ge 0:\mathbb{P}(|\hat{\bxi}^{\dagger}|_{\infty} \leq t\,|\,\mathcal{X}_{n},\mathcal{Y}_n) \ge 1-\alpha\big\}\,,\\
		&\tilde{\rm cv}_{{\rm cind},\alpha} =\inf\big\{t\ge 0:\mathbb{P}(|\tilde{\bzeta}^{\dagger}|_{\infty} \leq t\,|\,\mathcal{X}_{n},\mathcal{Y}_n,\mathcal{Z}_n) \ge 1-\alpha\big\}\,,\\
		& \tilde{\rm cv}_{{\rm cind},\alpha}^{*} =\inf\big\{t\ge 0:\mathbb{P}(|\hat{\bzeta}^{\dagger}|_{\infty} \leq t\,|\,\mathcal{X}_{n},\mathcal{Y}_n,\mathcal{Z}_n) \ge 1-\alpha\big\}\,.
	\end{split}
\end{align}   
Empirically, $\tilde{\rm cv}_{{\rm ind},\alpha}$ can be selected as
the $\lfloor N\alpha \rfloor$-th largest value among $|\hat{\bxi}_{1}^{\dagger}|_{\infty},\ldots, |\hat{\bxi}_{N}^{\dagger}|_{\infty}$, where $N$ is a sufficiently large integer, and 
$\hat{\bxi}_1^{\dagger},\ldots,\hat{\bxi}_{N}^{\dagger}$ are generated independently by \eqref{eq:Radem}.  Analogously,  $\tilde{\rm cv}_{{\rm cind},\alpha}$ and $\tilde{\rm cv}_{{\rm cind},\alpha}^{*}$ can be determined  in the same manner.

Recall $d=pq$. When the dimension $d$ is large and the sample size $n$ is small, the Gaussian
approximation specified above may lead to  size distortions. See the numerical results in Section \ref{sc:simulations}. 
To improve the finite sample performance, we may consider two other types of
multipliers $\{\varpi_i\}_{i=1}^n$ in \eqref{eq:Radem} advocated by \cite{Deng2020}:
\begin{itemize}
	\item Mammen's multiplier \citep{Mammen1993}:   $\mathbb{P}\{\varpi_i=(1\pm \sqrt{5})/2\} =(\sqrt{5}\mp 1)/(2\sqrt{5})$.
	\item Rademacher multiplier:  $\mathbb{P}(\varpi_i =\pm 1)=1/2$. 
\end{itemize}

Theorem \ref{pro:12} in Section \ref{sec:theory-bootstrap}
 shows that $\tilde{\rm cv}_{{\rm ind},\alpha}$, $\tilde{\rm cv}_{{\rm cind},\alpha}$ and $\tilde{\rm cv}_{{\rm cind},\alpha}^{*}$ defined in \eqref{eq:cv-tilde} with either Mammen's multiplier or Rademacher multiplier are also asymptotically valid critical values.
Our extensive  simulation studies in Section \ref{sc:simulations} indicate that the 
Rademacher multiplier provides more accurate approximations in finite samples. Hence we recommend using Rademacher multiplier $\varpi_i$ in \eqref{eq:Radem}.

\section{Theoretical Analysis}\label{sec:theoretical}

In this section, we provide the theoretical analysis for the proposed independence test and conditional independence tests. 
\subsection{Independence Test}\label{sec:indtest}

\begin{theorem}\label{thm:1}
Let $p \lesssim n^{\varkappa_1}$ and $q\lesssim n^{\varkappa_2}$ for any given constants $\varkappa_1\geq0$ and $\varkappa_2\geq0$. Under the null hypothesis $\mathbb{H}_0$ in \eqref{eq:equind},  then  $\mathbb{P}(H_{n}> \hat{{\rm cv}}_{{\rm ind},\alpha}) \to \alpha$ as $n \to \infty$.
\end{theorem}

Theorem \ref{thm:1}  shows that the size of the proposed independence test can be correctly controlled by the significance level $\alpha\in(0,1)$. Recall $d=pq$.   Proposition 1 in Appendix A of the supplementary material indicates that Theorem \ref{thm:1} actually holds provided that $\log d \ll n^{\tilde{c}_1}$ for some constant $\tilde{c}_1\in(0,1)$. Assuming $p \lesssim n^{\varkappa_1}$ and $q\lesssim n^{\varkappa_2}$ is just to simplify the presentation. 
Write $\bSigma=\cov(\bgamma_i):=(\Sigma_{i,j})_{d\times d}$. Theorem \ref{thm:2} shows that the proposed independence test is consistent under certain local alternatives imposed on the magnitude of $ |\mathbb{E}(\bU_{i}\bV_{i}^{\T})|_{\infty}$.

\begin{theorem}\label{thm:2}
  Let $p \lesssim n^{\varkappa_1}$ and  $q\lesssim n^{\varkappa_2}$ for any given constants $\varkappa_1\geq0$ and $\varkappa_2\geq0$.  
Under the alternative hypothesis $\mathbb{H}_1$ in \eqref{eq:equind}, if $\min_{j\in[d]} \Sigma_{j,j} \ge c_1$ for some universal constant $c_1>0$, and $$|\mathbb{E}(\bU_{i}\bV_{i}^{\T})|_{\infty} \ge  4\sqrt{6} (1+\nu_n)n^{-1/2}(\log d)^{1/2}(\log n)/\sqrt{5}$$   with $\nu_n \geq c_2$, where $c_2>0$ is an arbitrarily prescribed universal constant, then 
 $\mathbb{P}(H_{n}> \hat{{\rm cv}}_{{\rm ind},\alpha}) \ge 1- 2d^{-\nu_n/2-\nu_n^2/16} - o(1)  $. 
 
\end{theorem}

If either $p$ or $q$ diverges with the sample size $n$, as long as $|\mathbb{E}(\bU_{i}\bV_{i}^{\T})|_{\infty} \geq Cn^{-1/2}(\log d)^{1/2}\log n$  under the alternative hypothesis $\mathbb{H}_1$ in \eqref{eq:equind} for some universal constant $C>4\sqrt{6/5}$,  Theorem \ref{thm:2} implies that the proposed independence test is a consistent test in the sense that its power approaches 1.  
If $d$ is a fixed constant, as long as $|\mathbb{E}(\bU_{i}\bV_{i}^{\T})|_{\infty} \gg n^{-1/2}\log n$ under the alternative hypothesis $\mathbb{H}_1$ in \eqref{eq:equind}, the proposed independence test is also a consistent test. As shown in Section A.3 of the supplementary material, Theorem \ref{thm:2} actually holds provided that $\log d \ll n^{\tilde{c}_2}$ for some constant $\tilde{c}_2\in(0,1)$.    Together with  Theorem \ref{thm:1}, we know that, even if the dimensions of $\bX$ and $\bY$ diverge exponentially with the sample size $n$, the proposed independence test can still correctly control the Type I error at the significance level $\alpha\in(0,1)$ and also have power approaching 1 under certain local alternatives.

\subsection{ Conditional Independence Test based on Nonparametric Regressions}\label{subsec:condind_theory-nn}

To establish the theoretical guarantee of the proposed conditional independence test based on nonparametric regressions, we assume that the regression functions $f_j$ and $g_k$ in \eqref{eq:nonparametriic-model} satisfy the $(\vartheta,C)$-smooth generalized hierarchical interaction model, which was introduced in \cite{Bauer2019}. This function class covers a wide variety of models frequently used in nonparametric regression, such as additive models, single-index models, and interaction models, which is enough for capturing the nonlinear dependence between $(\bU,\bV)$ and $\bW$ in practice. 
\cite{Bauer2019} establish the convergence rate of the regression estimates by using feedforward neural network under the $(\vartheta,C)$-smooth generalized hierarchical interaction model assumption, which provides the foundation of our theoretical results. See \cite{Bauer2019} for more discussions.
For the sake of completeness, we first introduce the definition of $(\vartheta,C)$-smooth generalized hierarchical interaction model.  
\begin{definition}[$(\vartheta,C)$-smooth function]\label{def:pc-smooth}
	 Let $\vartheta=\tilde{\vartheta} + s$ for some $\tilde{\vartheta} \in \mathbb{N}_{0}$ and $s\in(0,1]$. A function $h: \mathbb{R}^{m} \to \mathbb{R}$ is called $(\vartheta,C)$-smooth, if for every $\boldsymbol{r}=(r_1, \ldots, r_{m})^{\T} \in \mathbb{N}_{0}^{m}$ with $\sum_{j=1}^{m} r_{j}=\tilde{\vartheta}$, the partial derivative $\frac{\partial^{\tilde{\vartheta}} h}{\partial^{r_1} x_{1}\cdots\partial^{r_m} x_{m} }$ exists and satisfies
	\begin{align*}
		\bigg|\frac{\partial^{\tilde{\vartheta}} h}{\partial^{r_1} x_{1}\cdots\partial^{r_m} x_{m}}(\boldsymbol{x}) - \frac{\partial^{\tilde{\vartheta}} h}{\partial^{r_1} x_{1}\cdots\partial^{r_m} x_{m}}(\boldsymbol{z})\bigg| \le C |\boldsymbol{x}- \boldsymbol{z}|_2^{s}
	\end{align*}
	for all $\boldsymbol{x}=(x_1, \ldots, x_{m})^{\T}\in \mathbb{R}^{m}$ and $ \boldsymbol{z}=(z_1, \ldots, z_{m})^{\T} \in \mathbb{R}^{m}$.
\end{definition}

\begin{definition}[$(\vartheta,C)$-smooth generalized hierarchical interaction model]\label{def:hierarchical}
	 Let $m\in \mathbb{N}$, $m_{*} \in[m]$ and $f:\mathbb{R}^{m} \to \mathbb{R}$.
	\begin{itemize}
		\item[(i)]  We say that $f$ satisfies a generalized hierarchical interaction model of order $m_{*}$ and level 0, if there exist $h_1: \mathbb{R}^{m_{*}} \to \mathbb{R}$ and $\boldsymbol{\phi}_{1}, \ldots, \boldsymbol{\phi}_{m_{*}} \in \mathbb{R}^{m}$ such that  $f(\boldsymbol{x}) = h_1(\boldsymbol{\phi}_{1}^{\T}\boldsymbol{x},\ldots, \boldsymbol{\phi}_{m_{*}}^{\T}\boldsymbol{x})$ for all  $\boldsymbol{x} \in \mathbb{R}^{m}$.
		
	    \item[(ii)] We say that $f $ satisfies a generalized hierarchical interaction model of order $m_{*}$ and level $l+1$, if there exist $K\in \mathbb{N}$, $h_{k} : \mathbb{R}^{m_{*}} \to \mathbb{R}$ $(k\in[K])$ and $\tilde{h}_{1,k} ,\ldots,\tilde{h}_{m_{*},k}  : \mathbb{R}^{m} \to \mathbb{R}$ $(k\in[K])$ such that $\tilde{h}_{1,k} ,\ldots,\tilde{h}_{m_{*},k} $ $(k\in[K])$ satisfy a generalized hierarchical interaction model of order $m_{*}$ and level $l$, and $f(\boldsymbol{x}) =\sum_{k=1}^{K} h_{k}\{\tilde{h}_{1,k}(\boldsymbol{x}),\ldots, \tilde{h}_{m_{*},k}(\boldsymbol{x})\}$ for all $\boldsymbol{x} \in \mathbb{R}^{m}$.
        \item[(iii)] We say that the generalized hierarchical interaction model defined above is $(\vartheta,C)$-smooth, if all functions occurring in its definition are $(\vartheta,C)$-smooth according to Definition \ref{def:pc-smooth}. 
	\end{itemize}
\end{definition}

\begin{definition}\label{def:function-f}
Let $\mathcal{F}(m,m_{*}, l , K ,\vartheta,L, C,\tilde{C})$ be the set of functions $f:\mathbb{R}^{m} \to \mathbb{R}$, which satisfy the following conditions: $f$ satisfies a $(\vartheta, C)$-smooth generalized hierarchical interaction model of order $m_{*}$ and level $l$ as in Definition \ref{def:hierarchical} with $K\in \mathbb{N}$, $\vartheta= \tilde{\vartheta} + s$ for some $\tilde{\vartheta} \in \mathbb{N}_{0}$ and $s\in(0,1]$. All partial  derivatives of order less than or equal to $\tilde{\vartheta}$ of the functions $h_{k}$, $\tilde{h}_{j,k}$ given in Definition \ref{def:hierarchical}(ii) are bounded, that is,  each such function $h$ satisfies
 \begin{align*} 
     &\max_{\substack{j_1, \ldots, j_{m} \in\{0\}\cup[\tilde{\vartheta}],\\ j_1+\cdots+ j_m\le \tilde{\vartheta}}} \bigg| \frac{\partial^{j_1+\cdots+j_m} h}{\partial^{j_1}x_{1}\cdots \partial^{j_m}x_{m}}  \bigg|_{\infty} \le \tilde{C}    
	\end{align*}
for some constant $\tilde{C}>0$. And let all functions $h_{k}$ be Lipschitz continuous with Lipschitz constant $L>0$.

\end{definition}

\cite{Bauer2019} recommend using the hierarchical neural networks to estimate the $(\vartheta,C)$-smooth generalized hierarchical interaction model. Write $\boldsymbol{x}=(x_{1}, \ldots, x_{m})^{\T} \in \mathbb{R}^{m}$. For $M_{*} \in \mathbb{N}$, 
$m_{*} \in[m]$ and $\tilde{\alpha}_{n}>0$,  we denote by $\mathcal{F}^{\rm NN}_{M_{*},m_{*},m,\tilde{\alpha}_{n}}$ the set of all functions $h :\mathbb{R}^{m} \to \mathbb{R}$  that satisfy
\begin{align*}
	h(\boldsymbol{x}) =\sum_{i=1}^{M_{*}}\mu_{i}  \sigma\bigg\{\sum_{j=1}^{4m_{*}} \lambda_{i,j}  \sigma\bigg(\sum_{v=1}^{m}\theta_{i,j,v}x_{v} + \theta_{i,j,0}\bigg) + \lambda_{i,0} \bigg\} + \mu_{0}
\end{align*}
for some  $\mu_{i},\lambda_{i,j},\theta_{i,j,v} \in \mathbb{R}$, where $\sigma(x)=(1+e^{-x})^{-1}$ for any $x\in \mathbb{R}$, $ |\mu_{i}| \le \tilde{\alpha}_{n} $, $|\lambda_{i,j}| \le \tilde{\alpha}_{n}$ and $ |\theta_{i,j,v}| \le \tilde{\alpha}_{n}$ for each $i\in[M_{*}] \cup\{0\}$, $j\in[4m_{*}]\cup\{0\}$ and $v\in[m]\cup \{0\}$. For $l =0$, the space of hierarchical neural networks is defined by  $\mathcal{H}^{(0)} = \mathcal{F}^{\rm NN}_{M_{*}, m_{*},m,\tilde{\alpha}_{n}}$. For $l\ge 1$, define recursively
\begin{align}\label{eq:nnspace-hl}
	&\mathcal{H}^{(l)} =\bigg\{f:\mathbb{R}^{m} \to \mathbb{R}: f(\boldsymbol{x}) =\sum_{k=1}^{K} h_{k}\{\tilde{h}_{1,k}(\boldsymbol{x}),\ldots, \tilde{h}_{m_{*},k}(\boldsymbol{x})\} \notag\\
	&~~~~~~~~~~~~~~~~\textrm{for some}~ h_{k} \in \mathcal{F}^{\rm NN}_{M_{*}, m_{*},m_{*},\tilde{\alpha}_{n}} ~ \textrm{and}~ \tilde{h}_{j,k} \in \mathcal{H}^{(l-1)} \bigg\} 
\end{align} 
with $K\in \mathbb{N}$. 
Then, we impose the following condition on the regression models \eqref{eq:nonparametriic-model}.

\begin{cd}\label{cd:function-condition}
    For each $j\in[p]$ and $k\in[q]$, the functions $f_j,\, g_{k} \in \mathcal{F}(m,m_{*}, \ell, K,\vartheta, L,C,\tilde{C})$  with finite positive integers $m_{*}$,  $\ell$ and $K$, and some positive constants $\vartheta $, $L$, $C$ and $\tilde{C}$.  
\end{cd}

Condition \ref{cd:function-condition} is commonly assumed in the existing works of nonparametric regressions using deep neural networks, where they usually assume that the distribution of the predictor is supported on a bounded set. In our setting, although the predictor $\bW_{i}$ has unbounded support, as shown in Equation  (K.7) in the supplementary material, we have $\hat{\bW}_i^{(w)} \in [-\sqrt{2\log n_1},\sqrt{2\log n_1}]^{m}$ for sufficiently large $n_1$.

Recall $\bTheta = \cov (\bet_i)$ with $\bTheta =  (\Theta_{i,j})_{d\times d}$.   Write  $\varrho =    \vartheta+2m_{*}\tilde{\vartheta}+3m_{*} $, and  $(\tilde{\alpha}_{n},M_*)$ specified in \eqref{eq:nnspace-hl} as 
    $$
    \tilde{\alpha}_{n} =n^{c_{3} }~~{\rm and}~~M_{*}= {c_{4}} \lceil n^{m_{*}/ (4\vartheta + m_{*})} (m^2\log n)^{m_{*}(2\tilde{\vartheta}+3)/(2\vartheta)}\rceil$$ for some sufficiently large constants $c_{3}>0$ and $c_{4}>0$. Recall $n_3 \asymp n^{\kappa}$ for some constant $\kappa \in (0,1)$. 
    Theorem \ref{thm:nn-sta-h0}  shows that the size of the proposed conditional independence test based on nonparametric regressions can be correctly controlled by the significance level $\alpha\in(0,1)$.

\begin{theorem}\label{thm:nn-sta-h0}
	Let Condition {\rm\ref{cd:function-condition}} hold with $p \lesssim n^{\varkappa_1}$, $q\lesssim n^{\varkappa_2}$ and $m\lesssim n^{\varkappa_3}$ for any given constants 
\begin{align}\label{eq:k1-3-res}
    \varkappa_1\geq0, ~~\varkappa_2 \geq 0~~\textrm{and}~~ 0\le  \varkappa_3 < \min\bigg\{\frac{\vartheta}{\varrho}\bigg(\frac{4\vartheta}{4\vartheta+m_{*}}-\kappa\bigg), \frac{1-\kappa}{2} ,\frac{\kappa}{4} \bigg\}\,.
\end{align}   
Under the null hypothesis $\mathbb{H}_0$ in \eqref{eq:equiconind}, if $\min_{j\in[d]} \Theta_{j,j} \ge c_5$ for some universal constant $c_5>0$, then  $\mathbb{P}(\tilde{G}_{ n}> \hat{{\rm cv}}_{{\rm cind},\alpha} ) \to \alpha$
 as $n \to \infty$.
\end{theorem}

Recall $d=pq$. To obtain Theorem \ref{thm:nn-sta-h0},  Proposition 2 in Appendix B of the supplementary material indicates that $d$ needs to satisfy $\log d \ll n^{\tilde{c}_3}$ for some constant $\tilde{c}_3\in(0,1)$.
Assuming $p \lesssim n^{\varkappa_1}$ and $q\lesssim n^{\varkappa_2}$ is just to simplify the presentation. 
Write $\lambda(d,\alpha)=(2\log d)^{1/2}+ \{2\log(1/\alpha)\}^{1/2}$. Theorem \ref{thm:nn-sta-h1} shows that the proposed conditional independence test based on nonparametric regressions is consistent under certain local alternatives imposed on the magnitude of $ |\mathbb{E}(\boldsymbol{\varepsilon}_{i}\boldsymbol{\delta}_{i}^{\T})|_{\infty}$.

\begin{theorem}\label{thm:nn-sta-h1}
Let $n_3 \ge  n^{\kappa}$ and Condition {\rm\ref{cd:function-condition}} hold with $p \lesssim n^{\varkappa_1}$, $q\lesssim n^{\varkappa_2}$ and $m\lesssim n^{\varkappa_3}$ for any given constants $\varkappa_1$, $\varkappa_2$ and $\varkappa_3$ satisfying  \eqref{eq:k1-3-res}. Under the alternative hypothesis $\mathbb{H}_1$  in \eqref{eq:equiconind},  if $\min_{j\in[d]} \Theta_{j,j} \ge c_5$ for some universal constant $c_5>0$, and  
$$ |\mathbb{E}(\boldsymbol{\varepsilon}_{i}\boldsymbol{\delta}_{i}^{\T})|_{\infty} \ge  (1+\tilde{\epsilon}_{n})n^{-\kappa/2}\lambda(d,\alpha)\max_{j\in[d]}\Theta_{j,j}^{1/2}$$ with $\tilde{\epsilon}_{n}>0$  satisfying  $\tilde{\epsilon}_{n}^2\log d \to \infty$ as $n\to \infty$, then  $\mathbb{P} (\tilde{G}_{n}>  \hat{{\rm cv}}_{{\rm cind},\alpha} ) \to 1  $ as $n\to \infty$.
\end{theorem}

As long as $|\mathbb{E}(\boldsymbol{\varepsilon}_{i}\boldsymbol{\delta}_{i}^{\T})|_{\infty} \geq Cn^{-\kappa/2}(\log d)^{1/2}$ under the alternative hypothesis $\mathbb{H}_1$ in \eqref{eq:equiconind} for some universal constant $C>1$, Theorem \ref{thm:nn-sta-h1} implies that the proposed conditional independence test based on nonparametric regressions is a consistent test in the sense that its power approaches 1. 
As shown in Section B.3 of the supplementary material, to obtain Theorem \ref{thm:nn-sta-h1}, $d$ needs to satisfy $\log d \ll n^{\tilde{c}_4}$ for some constant $\tilde{c}_4\in(0,1)$. 
 Together with    Theorem \ref{thm:nn-sta-h0}, we know that, even if the dimensions of $\bX$ and $\bY$ diverge exponentially with the sample size $n$, and the dimension of  $\bZ$ diverges polynomially with the sample size $n$, the proposed conditional independence test based on nonparametric regressions can still correctly control the Type I error at the significance level $\alpha\in(0,1)$ and also have power approaching 1 under certain local alternatives.

 \subsection{Conditional Independence Test based on Linear Regressions}\label{sec:theory-cindtest-linear}

 Let $\bSigma_{W}={\rm Cov}(\bW)$. To establish the theoretical guarantee of the proposed conditional independence test based on linear regressions, we impose the following condition on the regression models \eqref{eq:regressionUV} and the regularization parameters $\lambda_{\balpha,j}$ and $\lambda_{\bbeta,k}$ involved in \eqref{eq:lasso}. Let $s=\max_{j\in[p]}|\balpha_j |_{0} \vee \max_{k\in[q]}|\bbeta_k |_{0}$.

\begin{cd}\label{cn:subgaussian}
{\rm(i)} There exist universal constants $c_6>0$ and $c_7>0$ such that  $\mathbb{P}(|\balpha_j^{\T}\bW_i|>x) \le c_6 e^{-c_7x^2}$ and $\mathbb{P}(|\bbeta_k^{\T}\bW_i|>x) \le c_6 e^{-c_7x^2}$ for any $x>0$, $i \in [n]$, $j\in[p]$ and $k\in[q]$. {\rm(ii)} 
The smallest eigenvalue of $\bSigma_W$ is uniformly bounded away from zero.  
{\rm (iii)} 
There exist two sufficiently large constants $c_8>0$ and $c_9>0$ such that $c_8n^{-1/2}\log^{1/2} (pm)\le \lambda_{\balpha,j}  \le c_9n^{-1/2}\log^{1/2} (pm)$ and  $c_8n^{-1/2}\log^{1/2} (qm)\le  \lambda_{\bbeta,k}  \le c_9n^{-1/2}\log^{1/2} (qm)$ for any $j\in[p]$ and $k\in[q]$.
\end{cd}

Write $\bTheta =  (\Theta_{i,j})_{d\times d}$.
Theorem \ref{thm:3}  shows that the size of the proposed conditional independence test based on linear regressions can be correctly controlled by the significance level $\alpha\in(0,1)$. 


\begin{theorem}\label{thm:3}
	Let Condition {\rm \ref{cn:subgaussian}} hold with $p \lesssim n^{\varkappa_1}$, $q\lesssim n^{\varkappa_2}$ and $m\lesssim n^{\varkappa_3}$ for any given constants $\varkappa_1\geq0$, $\varkappa_2\geq0$ and $\varkappa_3\geq0$. 
Under \eqref{eq:regressionUV} and  the null hypothesis $\mathbb{H}_0$ in \eqref{eq:equiconind}, if $s\ll n^{1/5}(\log n)^{-3}$ and  $\min_{j\in[d]} \Theta_{j,j} \ge c_5$ for some universal constant $c_5>0$,  then 
 $\mathbb{P} (\hat{G}_{n}> \hat{{\rm cv}}_{{\rm cind},\alpha}^{*} ) \to \alpha$
 as $n \to \infty$. 
\end{theorem}
 
Recall $\tilde{d}=p\vee q\vee m$.    
Proposition 3 in Appendix C of the supplementary material indicates that Theorem \ref{thm:3} actually holds provided that $\log \tilde{d} \ll  n^{\tilde{c}_5}$ for some constant $\tilde{c}_5\in(0,1)$. Assuming  $p \lesssim n^{\varkappa_1}$, $q\lesssim n^{\varkappa_2}$ and $m\lesssim n^{\varkappa_3}$ is just to simplify the presentation. 
 Theorem \ref{thm:4} shows that the proposed conditional independence test based on linear regressions is consistent under certain local alternatives imposed on the magnitude of $|\mathbb{E}(\boldsymbol{\varepsilon}_{i}\boldsymbol{\delta}_{i}^{\T})|_{\infty}$.

\begin{theorem}\label{thm:4}
   Let Condition {\rm \ref{cn:subgaussian}} hold with $p \lesssim n^{\varkappa_1}$, $q\lesssim n^{\varkappa_2}$ and  $m\lesssim n^{\varkappa_3}$ for any given constants $\varkappa_1\geq0$, $\varkappa_2\geq0$ and $\varkappa_3\geq0$.     
    Under \eqref{eq:regressionUV}	and the alternative hypothesis $\mathbb{H}_1$ in \eqref{eq:equiconind}, if $s\ll n^{1/5}(\log n)^{-1/2}$, $\min_{j\in [d]}\Theta_{j,j} \ge c_5 $ for some universal constant $c_5>0$,  and
$$|\mathbb{E}(\boldsymbol{\varepsilon}_{i}\boldsymbol{\delta}_{i}^{\T})|_{\infty} \ge  12\sqrt{3 \tilde{c}^{-1}} (\sqrt{2}+u_n)n^{-1/2}(\log \tilde{d})^{1/2}(\log n)/5$$
with $\tilde{c}=(1\wedge c_{7})/4$ and $u_n \ge c_{10}$, where $c_{10} >0$ is an arbitrarily prescribed  universal constant, then  
    $\mathbb{P} (\hat{G}_n> \hat{{\rm cv}}_{{\rm cind},\alpha}^{*} ) \ge 1-2\tilde{d}^{-\sqrt{2}u_n/2-u_n^2/16} - o(1) $. 

\end{theorem}

If either $p$,  $q$ or $m$ diverges with the sample size $n$, as long as $|\mathbb{E}(\boldsymbol{\varepsilon}_{i}\boldsymbol{\delta}_{i}^{\T})|_{\infty} \geq Cn^{-1/2}(\log \tilde{d})^{1/2}\log n$   under \eqref{eq:regressionUV} and the alternative hypothesis $\mathbb{H}_1$ in \eqref{eq:equiconind} for some universal constant $C>12\sqrt{6}/(5\sqrt{\tilde{c}})$,  
Theorem \ref{thm:4} implies that the proposed conditional independence test based on linear regressions is a consistent test in the sense that its power approaches 1.  
If $\tilde{d}$ is a fixed constant, as long as $ |\mathbb{E}(\boldsymbol{\varepsilon}_{i}\boldsymbol{\delta}_{i}^{\T})|_{\infty}\gg n^{-1/2}\log n$ under the alternative hypothesis $\mathbb{H}_1$ in \eqref{eq:equiconind}, the proposed conditional independence test based on linear regressions is also consistent.   As shown in Section C.3 of the supplementary material,  Theorem \ref{thm:4} actually holds provided that $\log \tilde{d} \ll n^{\tilde{c}_6}$ for some constant $\tilde{c}_6\in(0,1)$.  Together with  Theorem \ref{thm:3}, we know that, even if the dimensions of $\bX$, $\bY$ and $\bZ$ diverge exponentially with the sample size $n$, the proposed conditional independence test based on linear regressions can still correctly control the Type I error at the significance level $\alpha\in(0,1)$ and also have power approaching 1 under certain local alternatives.

\subsection{Multiplier Bootstrap Procedure}\label{sec:theory-bootstrap}

Theorem \ref{pro:12} shows that the null-distributions of the test statistics $H_n$, $\tilde{G}_n$ and $\hat{G}_n$ can be approximated, respectively, by the distributions of $\hat{\bxi}^{\dagger}$,  $\tilde{\bzeta}^{\dagger}$ and  $\hat{\bzeta}^{\dagger}$  defined in \eqref{eq:Radem} with either Mammen's multiplier or Rademacher multiplier. 

\begin{theorem}\label{pro:12}
Let $\hat{\bxi}^{\dagger}$,  $\tilde{\bzeta}^{\dagger}$ and  $\hat{\bzeta}^{\dagger}$ be defined in \eqref{eq:Radem}, with either Mammen's multiplier or Rademacher multiplier. 
Then the following three assertions hold.

{\rm (i)} Let the conditions of Theorem {\rm \ref{thm:1}} hold. Under the null hypothesis $\mathbb{H}_0$ in \eqref{eq:equind},  then
$ \sup_{z >0 } |\mathbb{P}(H_{n}>z )-\mathbb{P}(|\hat{\bxi}^{\dagger}|_{\infty}>z\,|\,\mathcal{X}_{n},\mathcal{Y}_{n}) |=o_{{\rm p}}(1)  $ as $n\to \infty$. 

{\rm (ii)} Let the conditions of Theorem {\rm \ref{thm:nn-sta-h0}} hold.   Under the null hypothesis $\mathbb{H}_0$ in \eqref{eq:equiconind}, then  
$ \sup_{z>0} |\mathbb{P}(\tilde{G}_{n} >z) - \mathbb{P}(| \tilde{\bzeta}^{\dagger}  |_{\infty} > z\,|\,\mathcal{X}_{n}, \mathcal{Y}_{n},\mathcal{Z}_{n})| =o_{{\rm p}}(1) $ as $n\to \infty$. 

{\rm (iii)} Let the conditions of Theorem {\rm \ref{thm:3}} hold.  Under \eqref{eq:regressionUV} and  the null hypothesis $\mathbb{H}_0$ in \eqref{eq:equiconind}, if $s\ll n^{1/6}(\log n)^{-13/6}$,   then  
$ \sup_{z>0}|\mathbb{P}(\hat{G}_n >z) - \mathbb{P}(|\hat{\bzeta}^{\dagger}|_{\infty} > z\,|\,\mathcal{X}_{n}, \mathcal{Y}_{n},\mathcal{Z}_{n})| =o_{{\rm p}}(1) $   
as $n\to \infty$. 
\end{theorem}

\section{Simulations}\label{sc:simulations}

In this section, we conduct numerical studies to evaluate the finite-sample performance of the proposed independence test and conditional independence tests. To implement the proposed tests, we always use the multiplier bootstrap procedure introduced in Section \ref{sc:ComputeFast} to calculate the associated critical values with $N=5000$. We  compare the performance of the three multipliers, i.e., Gaussian multiplier, Rademacher multiplier and Mammen’s multiplier. All simulation results are based on 2000 replications and  at the nominal significance level $\alpha= 0.05$.

\subsection{Independence Test}\label{sc:indtestsimu}

In this subsection, we evaluate the performance of the proposed independence test via five simulated examples which characterize different  types of dependence between the two random vectors  $\bX =(X_1, \ldots, X_p)^{\T} \in \mathbb{R}^p$ and $\bY =(Y_1, \ldots, Y_q)^{\T} \in \mathbb{R}^q$. We always set $p=q$ in Examples 1--5.

\begin{description} 
	\item[{\bf Example 1.}] 
    Draw $X_1, \ldots, X_{p}, \tilde{Y}_1, \ldots, \tilde{Y}_{q} \overset{\rm i.i.d.}\sim t(1)$. For $l\in[q]$, let $Y_{l} = \exp(X_l)I(l\in[K]) + \tilde{Y}_{l-K}I(l\in [q] \backslash [K])$. We set $K \in \{0, p/20,  p/10\}$. When $K=0$, $\bX \Vbar \bY$. Otherwise, $\bX \not \Vbar\bY$. 
  
	
	\item[{\bf Example 2.}] Let $\boldsymbol{\varphi}=(\varphi_1,\ldots,\varphi_p)^{\T}$ and $\tilde{\boldsymbol{\varphi}}=(\tilde{\varphi}_1,\ldots, \tilde{\varphi}_q)^{\T}$ with $\varphi_1,\ldots,\varphi_p,\tilde{\varphi}_1,\ldots, \tilde{\varphi}_q \overset{\rm i.i.d.}\sim t(1)$. Generate $\tau\sim \mathcal{N}(0,1)$ independently of $\boldsymbol{\varphi}$ and $\tilde{\boldsymbol{\varphi}}$. For  $j \in [p]$  and $l\in[q]$, let $X_j=0.2 \varphi_j+\tau I(j \in [K])$ and $Y_l=0.2 \tilde{\varphi}_l+\tau I(l \in [K])$. We set $K \in \{0, p/20, p/10\}$. When $K=0$, $\bX \Vbar \bY$. Otherwise, $\bX \not\Vbar \bY$. 
	
	\item[{\bf Example 3.}]   Draw $\tilde{X}_1, \ldots, \tilde{X}_p, \tilde{Y}_1, \ldots, \tilde{Y}_q, \tau_1, \ldots, \tau_K\overset{\rm i.i.d.} \sim U(0,2\pi)$. For  $j\in[p]$ and $l\in[q]$, let $X_j= \sin^2(\tau_j)I(j\in[K]) + \tilde{X}_{j}I(j \in[p]\backslash[K]) $ and $Y_l= \cos^2(\tau_l)I(l\in[K]) + \tilde{Y}_{l}I(l\in[q]\backslash[K])$. We set $K \in \{0, p/20, p/10\}$. When $K=0$, $\bX \Vbar \bY$. Otherwise, $\bX \not\Vbar \bY$. 
    
\end{description}

\begin{description} 
	\item[{\bf Example 4.}] Under the null hypothesis $\mathbb{H}_0$ in \eqref{eq:test}, generate $\boldsymbol{\varphi} = (\varphi_1, \ldots, \varphi_{p+q})^{\T} \sim \mathcal{N}(\mathbf{0},{\bf I}_{p+q})$. For $j \in [p]$ and $l \in [q]$, let $X_j =\varphi_j$ and $Y_l=\varphi_{p+l}$. Under the alternative hypothesis $\mathbb{H}_1$ in \eqref{eq:test}, generate $\boldsymbol{\varphi} \sim \mathcal{N}(\mathbf{0},\mathbf{R}^{*})$, where $\mathbf{R}^{*}$ is generated as follows. Let
	\begin{align*}
		\boldsymbol{\Delta} =\left(
		\begin{matrix}
			\mathbf{0} & \boldsymbol{\Delta}_{12} \\
			\boldsymbol{\Delta}_{12}^{\T} & \mathbf{0}
		\end{matrix}
		\right)  \in \mathbb{R}^{(p+q)\times (p+q)}
	\end{align*}
	be a random matrix, where $\boldsymbol{\Delta}_{12}\in \mathbb{R}^{p\times q}$ has only four nonzero entries. We set the locations of the four nonzero entries randomly in $\boldsymbol{\Delta}_{12}$, each with a magnitude randomly drawn from $U(0,1)$. To ensure positivity, let $\mathbf{R}^{*} = (1+\upsilon)\mathbf{I}_{p+q} + \boldsymbol{\Delta}$ with $\upsilon=\{-\lambda_{\min}(\mathbf{I}_{p+q}+\boldsymbol{\Delta})+0.05\}I\{\lambda_{\min}(\mathbf{I}_{p+q}+\boldsymbol{\Delta})\le 0\}$.  Then, for $j \in [p]$ and $l \in [q]$, let $X_j =\varphi_j$ and $Y_l=\varphi_{p+l}$.

	\item[{\bf Example 5.}] Write $\boldsymbol{\vartheta} = (\vartheta_1, \ldots, \vartheta_{p+q})^{\T}$. For $j \in [p]$ and $l \in [q]$, let $X_j =\vartheta_{j}^{1/3}$ and $Y_l=\vartheta_{p+l}^{1/3}$.
	Under the null hypothesis $\mathbb{H}_0$ in \eqref{eq:test}, generate $\boldsymbol{\vartheta}\sim \mathcal{N}(\mathbf{0},\mathbf{I}_{p+q})$. Under the alternative hypothesis $\mathbb{H}_1$ in \eqref{eq:test}, generate $\boldsymbol{\vartheta} \sim \mathcal{N}(\mathbf{0},\mathbf{R}^{*})$ with $\mathbf{R}^{*}$ specified in Example 4.
\end{description}

Example 1 is used in \cite{Zhu2017} for the monotone and nonlinear dependence  between $\bX$ and $\bY$. Example 2 is similar to the setting (V1) in the supplementary material of \cite{Deb2021}, which characterizes the monotone and linear dependence between $\bX$ and $\bY$. Their setting only considers the case with $K=p$, while our Example 2 is more general that can cover the cases with $K \neq p$.  In Examples 1 and 2, the distributions of $\bX$ and $\bY$ are heavy-tailed. Example 3 is  similar to Example A.4(iii) in the supplementary material of \cite{Zhu2020}, which characterizes the nonlinear and non-monotone dependence  between $\bX$ and $\bY$. In comparison to \cite{Zhu2020} that only consider the case with $K=p$, our Example 3 is more general which can cover the cases with $K\neq p$. Examples 4 and 5 extend the simulation settings in \cite{Han2017}, respectively, for data generated from the Gaussian distribution and the light-tailed Gaussian copula to the two-sample problem with $\boldsymbol{\Delta}_{12}$ being the cross covariance matrix between $\bX$ and $\bY$. 
These two examples can, respectively, characterize the linear and nonlinear dependence  between $\bX$ and $\bY$ under the sparse alternative.

We also compare the proposed independence test with eight other existing methods: (i) the test based on projection correlation (Pcor) in \cite{Zhu2017}, (ii) the test based on ranks of distances (rdCov) in \cite{Heller2013}, (iii) the test based on distance correlation (dCor) in \cite{Szekely2013}, (iv) the $k$-variate HSIC based test (dHSIC) in \cite{Pfister2018}, (v) the test based on the rank-based dependence matrix (JdCov\_R) in \cite{Chakraborty2018}, (vi) the generalized distance covariance based test (GdCov) in \cite{Jin2018}, (vii) the center-outward ranks and signs based test (Hallin) in \cite{Shi2020}, and (viii) the multivariate rank-based test (mrdCov) in \cite{Deb2021}. All simulations are implemented in \textsf{R}. The \textsf{R} codes for implementing the Pcor test are provided by the authors of \cite{Zhu2017}. The rdCov, dCor and dHSIC tests are implemented by calling the \textsf{R}-functions {\tt hhg.test}, {\tt dcorT.test} and {\tt dhsic.test} in the {\tt HHG}, {\tt energy} and {\tt dHSIC} packages, respectively. The JdCov\_R test is implemented by using the \textsf{R} codes provided in the supplementary material of \cite{Chakraborty2018}. The GdCov test is implemented by calling the \textsf{R}-function {\tt mdm\_test} in the \textsf{R}-package {\tt EDMeasure}. The \textsf{R} codes of the Hallin and mrdCov tests are, respectively, available in the supplementary materials of \cite{Shi2020} and \cite{Deb2021}.

We set $p=q \in\{100, 400, 1600 \}$ and $n\in\{50, 100\}$ in the simulations. Table \ref{tab:ind}  reports the empirical sizes and powers of the proposed independence test and the competing methods. In Example 1 with $K\in\{p/20,p/10\}$, since the dCor, dHSIC, GdCov, Hallin and mrdCov tests return invalid results for more than 20\% in the 2000 repetitions due to the heavy tails of the data, the associated results are reported by NA. Such a phenomenon indicates that these five tests may not work for the heavy-tailed data. 
For the proposed independence test, Rademacher multiplier has the best performance among the three choices of multipliers which can always control the sizes around the nominal significance level $0.05$ and also has the highest powers.  While Gaussian and Mammen's multipliers are under-sized in some scenarios, they still have quite good power performance in all the settings. For the competing methods, they can always control the sizes around the nominal level $0.05$ in all the settings. However, the competing methods (except the JdCov\_R test) have no powers in all the settings. 
The JdCov\_R test only has good power performance in Example 2, but it still underperforms the proposed method.  




\subsection{Conditional Independence Test}\label{sc:condindtestsimu}

In this subsection, we evaluate the performance of the proposed conditional independence tests based on nonparametric regressions (denoted by CI-FNN) and linear regressions (denoted by CI-Lasso), respectively, via five simulated examples which characterize different types of the conditional dependence between $\bX =(X_1, \ldots, X_p)^{\T} \in \mathbb{R}^p$ and $\bY =(Y_1, \ldots, Y_q)^{\T} \in \mathbb{R}^q$ given $\bZ = (Z_1, \ldots, Z_m)^{\T} \in \mathbb{R}^m$. We always set $p=q$ and $m\le p$ in  Examples 6--10.  
\begin{description}
    \item [{\bf Example 6.}]
     Let $\mathcal{C} = \{(s_{i,1}, s_{i,2}): 1\le s_{i,1} < s_{i,2} \le m, i\in[\tilde{s}]\}$ with $\tilde{s}=\min\{m(m-1)/2,p,q\}$. 
     Draw $Z_1,\ldots, Z_m, \tilde{X}_1,\ldots,  \tilde{X}_{p-\tilde{s}}, \tilde{Y}_1,\ldots, \tilde{Y}_{q-\tilde{s}} \overset{{\rm i.i.d.}}\sim t(2)$. Generate $\tau_{1}, \ldots, \tau_{K} \overset{{\rm i.i.d.}}\sim t(1)$ independently of $\{Z_l\}_{l=1}^{m}$, $\{\tilde{X}_j\}_{j=1}^{p-\tilde{s}}$ and $\{\tilde{Y}_k\}_{k=1}^{q-\tilde{s}}$. For $j\in[K]$, let $w_j =\tau_j + 3\tau_j^3$. For $j\in[p]$ and $k\in[q]$, let $X_j =(Z_{s_{j,1}} Z_{s_{j,2}})I(j\in [\tilde{s}]) + \tilde{X}_{j-\tilde{s}}I(j\in [p]\backslash[\tilde{s}]) + w_j I(j\in[K])$ and $Y_k =(Z_{s_{k,1}}+ Z_{s_{k,2}})I(k\in [\tilde{s}]) + \tilde{Y}_{k-\tilde{s}}I(k\in [q]\backslash[\tilde{s}]) + w_k I(k\in[K])$. We set $K \in \{0, p/10, p/5\}$. When $K=0$, $\bX \Vbar \bY\,|\,\bZ$. Otherwise,  $\bX \not\Vbar \bY\,|\,\bZ$. 
    \item [{\bf Example 7.}] Draw $Z_1,\ldots,Z_m\overset{{\rm i.i.d.}} \sim U(-1,1)$ and $\tilde{X}_1, \ldots, \tilde{X}_{p-m},  u_1,\ldots,u_q, \tau_{1}, \ldots, \tau_{q-m}\overset{{\rm i.i.d.}} \sim \mathcal{N}(0,1)$. Let $\nu_1,\ldots,\nu_m$ be independent random variables that are computed as the sum of 48 i.i.d. random variables from $U(-0.25,0.25)$. Assume $\{Z_l\}_{l=1}^{m}$, $\{\tilde{X}_j\}_{j=1}^{p-m}$, $\{u_k\}_{k=1}^{q}$, $\{\tau_s\}_{s=1}^{q-m}$, and   $\{\nu_l\}_{l=1}^{m}$ are mutually independent. For $j\in[p]$ and $k \in [q]$,  let $X_j=(Z_j+0.25Z_j^2+\nu_j)I(j\in[m]) + \tilde{X}_{j-m}I(j\in[p]\backslash[m])$ and $Y_k=(\beta X_k + Z_k + u_k)I(k\in[m]) + (\tau_{k-m} + \beta X_k + u_{k})I(k\in[q]\backslash[m])$ with $\beta=5\rho/(2\sqrt{1-\rho^2})$.  We set $\rho \in \{0, 0.7, 0.8\}$. When $\rho=0$, $\bX \Vbar \bY\,|\,\bZ$. Otherwise,  $\bX \not\Vbar \bY\,|\,\bZ$.
    \item [{\bf Example 8.}] Generate $Z_1,\ldots, Z_m, \tilde{X}_1,\ldots, \tilde{X}_{p-m}, \tilde{Y}_1,\ldots, \tilde{Y}_{q-m}, \nu_{1},\ldots,\nu_{m}, u_{1},\ldots, u_{m} \overset{\rm i.i.d.} \sim \mathcal{N}(0,1)$. 
     Draw $\tau_1,\ldots,\tau_K \overset{\rm i.i.d}\sim t(1)$ independently of $\{Z_l\}_{l=1}^{m}$, $\{\tilde{X}_j\}_{j=1}^{p-m}$, $\{\tilde{Y}_k\}_{k=1}^{q-m}$, $\{\nu_{l}\}_{l=1}^{m}$ and $\{u_{l}\}_{l=1}^{m}$. For $j \in [p]$ and $k \in [q]$, let 
	$X_j= \{\varphi_j +  \varphi_j^3/3 +  \tanh (\varphi_j/3)/2\}I(j\in[m]) + \tilde{X}_{j-m}I(j\in[p]\backslash[m]) + 3\tau_j I(j\in  [K])$ and $Y_k= \{\tilde{\varphi}_k + \tanh ( \tilde{\varphi}_k/3)\}^3I(k\in[m])   + \tilde{Y}_{k-m}I(k\in[q]\backslash[m]) + 3\tau_k I( k \in [K])$
	with $\varphi_j= \{0.7(Z_j^3/5+Z_j/2)+\tanh(\nu_{j})\}I(j\in[m])$ and $\tilde{\varphi}_k=\{(Z_k^3/4+Z_k)/3+u_{k}\}I(k\in[m])$. We set $K \in \{0, p/10, p/5\}$. When $K=0$, $\bX \Vbar \bY\,|\,\bZ$. Otherwise,  $\bX \not\Vbar \bY\,|\,\bZ$.
    \item [{\bf Example 9.}] Generate $\{Z_l\}_{l=1}^{m}$, $\{\tilde{X}_j\}_{j=1}^{p-m}$, $\{\tilde{Y}_k\}_{k=1}^{q-m}$, $\{\nu_{l}\}_{l=1}^{m}$ and $\{u_{l}\}_{l=1}^{m}$  in the same manner as Example 8. Draw $\tau_1,\ldots,\tau_K \overset{\rm i.i.d}\sim \mathcal{N}(0,1)$ independently of $\{Z_l\}_{l=1}^{m}$, $\{\tilde{X}_j\}_{j=1}^{p-m}$, $\{\tilde{Y}_k\}_{k=1}^{q-m}$, $\{\nu_{l}\}_{l=1}^{m}$ and $\{u_{l}\}_{l=1}^{m}$. 
    For $j \in  [p]$ and $k\in[q]$, let $\breve{X}_j=(\varphi_j +  \varphi_j^3/3)I(j \in [m]) + \tilde{X}_{j-m}I(j\in [p]\backslash[m]) $ and $\breve{Y}_k=\{\tilde{\varphi}_k + \tanh (\tilde{\varphi}_k /3) \} I(k \in [m]) + \tilde{Y}_{k-m}I(k\in [q]\backslash[m])$
	with $\varphi_j=\{0.5(Z_j^3/7+Z_j/2) + \tanh(\nu_j)\}I(j\in[m])$ and  $\tilde{\varphi}_k=\{(Z_k^3/2+Z_k)/3 +u_k\}I(k\in[m])$.  Then, let $X_j=  \{0.5\breve{X}_j +3 \cosh(\tau_j)\} I(j \in[K]) + \breve{X}_{j}I(j\in[p]\backslash[K])$ and  $Y_k= \{0.5\breve{Y}_k +3 \cosh(\tau_k^2)\} I(k \in[K]) + \breve{Y}_{k}I(k\in[q]\backslash[K])$ for $j \in [p]$ and $k \in [q]$.  We set $K \in \{0, p/10, p/5\}$. When $K=0$,  $\bX \Vbar \bY\,|\,\bZ$. Otherwise, $\bX \not\Vbar \bY\,|\,\bZ$.	
    \item [{\bf Example 10.}]   
    Draw  $Z_1,\ldots,Z_m, \tilde{X}_1,\ldots,\tilde{X}_{p-L}, \tilde{Y}_1,\ldots,\tilde{Y}_{q-L},
 \nu_1,\ldots,\nu_{L}, u_1,\ldots,u_{L}$, $ \tau_1,\ldots,\tau_K \overset{\rm i.i.d.}\sim \mathcal{N}(0,1)$ with  $L=p/4$ and $K\le L$. Let $\tilde{Z}= m^{-1}\sum_{i=1}^{m}Z_i $.  For $j \in [p]$ and $k \in [q]$, let $X_{j} =  \tanh\{\tilde{Z} + \nu_j + 3\tau_{j}I(j \in[K])\}I(j\in[L])  + \tilde{X}_{j-L} I(j\in [p]\backslash[L])$ and $Y_{k} = \{\tilde{Z} + u_k + 3\tau_{k}I(k \in[K])\}^3I(k \in[L])  + \tilde{Y}_{k-L} I(k\in [p]\backslash [L])$. We set $K \in \{0, p/10, p/5\}$.   When $K=0$,  $\bX \Vbar \bY\,|\,\bZ$. Otherwise, $\bX \not\Vbar \bY\,|\,\bZ$.  
    
\end{description}
 
Example 6 is similar to Example 10 in \cite{Wang2015}, where the latter only considers the fixed-dimensional scenario. Example 7 is similar to DGP1 in \cite{Su2012},  where the components of $\bX$ and $\bY$ are generated by the polynomial regression models on $\bZ$. Their setting only considers the case with $p=q=1$, and our Example 7 is more general which allows $p,q \ge 1$. Given $\bZ$, the random vectors $\bX$ and $\bY$ are linearly conditional correlated in Examples 6 and 7.  Example 8 is similar to the simulation setting provided in the \textsf{Matlab} codes of \cite{Zhang2012}, which characterizes the linear conditional dependence  between $\bX$ and $\bY$ given $\bZ$, under the nonlinear regression model settings of $\bX$ and $\bY$ on $\bZ$. Their 
setting only considers the case with $p=q=1$, while our Example 8 can cover more general cases with $p,q \ge 1$. Example 9 extends Example 7 in \cite{Wang2015} that only considers the case with $p = q = 1$ to more general cases with $p,q \ge 1$, which characterizes the nonlinear conditional dependence between $\bX$ and $\bY$ given $\bZ$  under the nonlinear regression model settings of $\bX$ and $\bY$ on $\bZ$. Example 10  extends the simulation setting in \cite{Runge2018} which only considers the case with $K=p=1$ to more general cases with $K\ne p$ and $p\ge 1$.

We also compare the finite-sample performance of the proposed conditional independence tests with five other existing methods: (i) the test based on the generalized covariance measure (GCM) in \cite{Shah2020}, (ii) the test based on the projective approach (PCD) in \cite{Zhou2022}, (iii) the randomized conditional independence test (RCIT) in \cite{Strobl2019}, (iv) the randomized conditional correlation test (RCoT) in \cite{Strobl2019}, and (v) the test based on conditional distance correlation (cdCov) in \cite{Wang2015}.  All simulations are implemented in \textsf{R}, except that the CI-FNN test is implemented in \textsf{Python}. In the CI-FNN test, $\hat{f}_j$ and $\hat{g}_k$ are  estimated by \eqref{eq:fhj-ghk} with the parameters $(\ell,K,m_{*},M_{*}) =(0,1,1,32)$. We set $n_1=\lfloor n/3 \rfloor$, $n_2=\lfloor n/2 \rfloor$ and $n_3 = n_3^{\rm opt}$, where $n_3^{\rm opt}$ is selected by  Algorithm \ref{alg:select_tilden} with $B=500$. In the  CI-Lasso test, the Lasso estimators $\hat{\balpha}_j$ and $\hat{\bbeta}_k$ are obtained by calling the \textsf{R}-functions {\tt glmnet} and {\tt cv.glmnet} in the {\tt glmnet} with the tuning parameters  $\lambda_{\balpha,j}$ and $\lambda_{\bbeta,k}$  being chosen by the default 10-fold cross validation method.  The GCM test is implemented by calling the \textsf{R}-function {\tt gcm.test} in the {\tt GeneralisedCovarianceMeasure} package. The codes of the PCD test are available in the supplementary material of \cite{Zhou2022}. The RCIT and RCoT tests are implemented by calling the \textsf{R}-functions {\tt RCIT} and {\tt RCoT} in the {\tt RCIT} package. The cdCov test is implemented by calling the \textsf{R}-function {\tt cdcov.test} in the {\tt cdcsis} package.

We set $p=q  \in\{100, 400, 1600 \}$, $m=5$ and $n\in\{  100,200 \}$ in the simulations. Table \ref{tab:cind} reports the empirical sizes and powers of the proposed conditional independence tests and the competing methods. Since the PCD test would return Inf/NaN values for the test statistics due to the curse of dimensionality issue for the kernel-based methods, the associated results are reported as NA.  When the sample size increases from $n=100$ to $n=200$, the proposed CI-FNN tests with three multipliers show significant improvements in both size control and power performance. This is consistent with the discussion in Section \ref{sc:CondIndTest}, where it is noted that fitting the feedforward neural network requires a substantial number of samples. Among the three choices of multipliers, same as the discussion in Section \ref{sc:indtestsimu} for the proposed independence test, the proposed CI-FNN test with Rademacher multiplier still has the best performance in all the settings with well-controlled sizes and the highest powers. The CI-FNN tests with Gaussian and  Mammen's multipliers are under-sized in most scenarios and exhibit reduced power when the sample size $n$ is small $(n=100)$. However, when $n$ increases to 200, they still have quite good power performance in all the settings. On the other hand, as discussed in Section \ref{sc:CondIndTest}, when the joint distribution of $(\bU_i,\bV_i,\bW_i)$ is close to normal, the CI-Lasso test can also be applied. It can be observed from Table \ref{tab:cind} that the CI-Lasso test with Rademacher multiplier has higher powers in most cases than the CI-FNN test with Rademacher multiplier, particularly when $n=100$. While the CI-Lasso tests with Gaussian and Mammen's multipliers are under-sized in most scenarios, their power performance is still quite good. Note that there are many more parameters to be estimated when fitting feedforward neural networks in the CI-FNN test than estimating Lasso estimators in the CI-Lasso test. For example, when $m=5$, to estimate a $f_{j}$, fitting a feedforward neural network needs to estimate 4929 parameters, while fitting a linear regression model only needs to estimate 5 parameters. This might also reduce the power performance of the CI-FNN test when $n$ is small. When $n$ increases to $200$, Table \ref{tab:cind} shows that the power performance of the CI-FNN and CI-Lasso tests becomes comparably good. 

For the competing methods, the RCIT, RCoT and cdCov tests fail to control the sizes around the nominal level in all the settings, since good approximation for the null distributions of the RCIT, RCoT and cdCov tests requires considerable sample size \citep{Runge2018, Strobl2019, Wang2015}.  The GCM test has good size control in the simulation settings except Example 6. For
Examples 6 and 8, the GCM test has no powers. The power performance of the GCM test in Example 9 is inferior to that of the CI-FNN and CI-Lasso tests with Rademacher multiplier. For Examples 7 and 10, the power performance of the GCM test is quite good and comparable to that of the CI-Lasso test. 

\section*{Supplementary Material}

Appendices A--P collect the technical proofs of all main theoretical results. Appendix Q presents the real-data analysis. Appendix R reports additional simulation studies on an extension of the proposed independence test; the effects of coordinatewise Gaussianization; comparisons with covariance-based methods under Gaussian data; and the effect of sample splitting in the proposed nonparametric conditional independence test.


\section*{Acknowledgments}
We are grateful to the editor, the associate editor, and three referees for their insightful comments and suggestions, which have led to significant improvement of our article.

\section*{Disclosure Statement}
The authors report there are no competing interest to declare. 

\section*{Funding}
J. Chang and Y. Du were supported in part by the National Natural Science
Foundation of China (Grant nos. 72125008 and  72495122).	
J. He were supported in part by the National Natural Science
Foundation of China (Grant nos. 72473114 and 72495121).
Q. Yao was supported in part by the U.K. Engineering and Physical Sciences Research
Council (Grant nos. EP/V007556/1 and EP/X002195/1).

\begin{landscape}
	\begin{table}[htbp]
		\scriptsize
		\caption{Empirical sizes (the rows with $K=0$ in Examples 1--3 and `{\rm null}' in Examples 4--5) and powers (the rows with $K=p/20$ and $p/10$ in Examples 1--3 and `{\rm alternative}' in Examples 4--5) of the  proposed independence test  and the comparing methods in Examples 1--5. All numbers reported below are multiplied by 100. The results reported by `NA' indicate that the associated tests return invalid results. 
        }
		\renewcommand\tabcolsep{4.0pt}
		\label{tab:ind}
		\begin{spacing}{1.3}
			\resizebox{23cm}{!}{
				\begin{threeparttable}
					\begin{tabular}{ccc|ccccccccccc|ccccccccccc}
						\hline
						\hline
						&     &  &\multicolumn{11}{c|}{$n=50$} & \multicolumn{11}{c}{$n=100$}   \\
						\hline 
						& \multirow{2}{*}{$p$}   & \multirow{2}{*} {Setting} & \multicolumn{3}{c}{Proposed Method} &  \multirow{2}{*}{Pcor} &  \multirow{2}{*}{rdCov} &  \multirow{2}{*}{dCor}  &  \multirow{2}{*}{dHSIC} &  \multirow{2}{*}{JdCov\_R} &  \multirow{2}{*}{GdCov} &  \multirow{2}{*}{Hallin} &  \multirow{2}{*}{mrdCov}& \multicolumn{3}{c}{Proposed Method} &  \multirow{2}{*}{Pcor} &  \multirow{2}{*}{rdCov} &  \multirow{2}{*}{dCor}  &  \multirow{2}{*}{dHSIC} &  \multirow{2}{*}{JdCov\_R} &  \multirow{2}{*}{GdCov} &  \multirow{2}{*}{Hallin} &  \multirow{2}{*}{mrdCov} \\
						
						&     &  & Gaussian & Mammen & Rademacher &   &   &    &  &   &    &  & & Gaussian & Mammen & Rademacher &   &   &    &  &   &    &  &  \\
						\hline 
						\multirow{9}{*}{Example 1} & \multirow{3}{*}{100}   & $K=0$ & 0.4  &  2.3 & 6.3 & 5.3   & 5.2   & 3.3   & 5.1   & 5.4   & 5.0   & 5.4   & 5.4  & 0.5&  2.6&  5.7 & 4.8   & 4.8   & 3.0   & 4.7   & 4.9   & 5.5   & 5.2   & 5.1\\
						&       & $K=p/20$ & 100.0 &  100.0  & 100.0  & 14.3  & 6.1   & NA    & NA    & 9.5 & NA    & NA    & NA &100.0 &  100.0& 100.0 & 35.7  & 8.2   & NA    & NA    & 9.4   & NA    & NA    & NA \\
						&       & $K=p/10$ & 100.0  &  100.0  & 100.0  & 27.0  & 13.2  & NA    & NA    & 15.4 & NA    & NA    & NA & 100.0 & 100.0 & 100.0 & 60.0  & 23.2  & NA    & NA    & 22.0  & NA    & NA    & NA \\
						\cline{2-25}          & \multirow{3}{*}{400}   & $K=0$ & 0.0 &  1.1 & 6.1 & 5.8   & 5.3   & 4.4   & 5.4   & 5.6   & 5.8   & 5.7   & 4.0  &0.0 & 1.3 &  5.3 & 5.8   & 5.1   & 4.0   & 5.4   & 4.9   & 6.3   & 4.8   & 5.1\\
						&       & $K=p/20$ & 100.0  & 100.0   & 100.0  & 6.1   & 5.4   & NA    & NA    & 7.4   & NA    & NA    & NA & 100.0 & 100.0 &  100.0 & 7.4   & 8.2   & NA    & NA    & 7.6   & NA    & NA    & NA \\
						&       & $K=p/10$ & 100.0  & 100.0   & 100.0  & 6.2   & 9.0   & NA    & NA    & 11.4  & NA    & NA    & NA & 100.0&  100.0& 100.0 & 10.7  & 17.7  & NA    & NA    & 13.1  & NA    & NA    & NA \\
						\cline{2-25}          & \multirow{3}{*}{1600}  & $K=0$ & 0.0 &  0.0 & 6.2 & 5.3   & 5.1   & 4.8   & 5.5   & 6.1   & 5.5   & 4.4   & 5.1 & 0.0 & 1.0  &  6.3 & 5.0   & 5.1   & 4.2   & 4.8   & 5.5   & 5.2   & 5.0   & 5.8 \\
						&       & $K=p/20$ & 100.0  & 100.0   & 100.0  & 4.5   & 4.8   & NA    & NA    & 7.5   & NA    & NA    & NA & 100.0 &100.0   &  100.0 & 4.9   & 6.6   & NA    & NA    & 9.2   & NA    & NA    & NA \\
						&       & $K=p/10$ & 100.0  & 100.0   & 100.0  & 3.4   & 7.4   & NA    & NA    & 11.8   & NA    & NA    & NA & 100.0 &100.0   &  100.0 & 3.7   & 12.5  & NA    & NA    & 12.2   & NA    & NA    & NA \\
						\hline
						\multirow{9}{*}{Example 2} & \multirow{3}{*}{100}   & $K=0$ &  0.4 &  1.4 & 4.5 & 5.1   & 4.2   & 3.6   & 4.4   & 5.5   & 5.8   & 5.2   & 5.5  &1.1 &  3.1&  5.7 & 4.6   & 4.8   & 3.5   & 5.6   & 5.5   & 5.2   & 4.7   & 5.3  \\
						&       & $K=p/20$ & 94.9 &  99.4 & 100.0 & 5.9   & 5.9   & 3.9   & 5.5   & 30.6  & 5.4   & 5.2   & 5.0  & 100.0 & 100.0 & 100.0 & 6.2   & 5.0   & 3.7   & 5.1   & 58.0  & 5.2   & 5.6   & 5.1  \\
						&       & $K=p/10$ & 99.8 &100.0   & 100.0 & 6.6   & 5.2   & 4.0   & 5.7   & 100.0  & 5.7   & 5.9   & 5.2  &100.0 &  100.0& 100.0 & 9.3   & 4.7   & 3.4   & 5.3   & 100.0  & 4.7   & 5.4   & 5.5  \\
						\cline{2-25}           & \multirow{3}{*}{400}   & $K=0$ & 0.0 &  1.1 &  4.4 & 4.7   & 4.3   & 4.2   & 4.5   & 5.6   & 5.1   & 4.9   & 5.1 &0.2 & 1.4 & 5.4 & 5.7   & 5.7   & 3.9   & 5.9   & 4.3   & 5.9   & 4.6   & 4.8  \\
						&       & $K=p/20$ & 98.9 &   100.0 & 100.0 & 5.4   & 5.0   & 4.2   & 5.7   & 94.8  & 5.5   & 5.2   & 8.1  &100.0 &  100.0& 100.0 & 6.1   & 5.0   & 4.3   & 5.2   & 99.9  & 5.7   & 4.5   & 5.4  \\
						&       & $K=p/10$ & 100.0 &  100.0 & 100.0 & 5.1   & 4.9   & 4.4   & 4.7   & 100.0  & 5.8   & 5.5   & 29.9 & 100.0&  100.0& 100.0 & 5.8   & 4.7   & 4.1   & 5.0   & 100.0  & 5.2   & 5.5   & 39.5  \\
						\cline{2-25}          & \multirow{3}{*}{1600} & $K=0$ & 0.0 & 0.4  &  5.9 & 4.9   & 5.5   & 3.6   & 5.2   & 5.3   & 5.8   & 4.7   & 5.1  & 0.0& 0.9 & 5.0  & 4.4   & 4.4   & 3.3   & 5.0   & 6.0   & 4.9   & 5.2   & 5.1\\
						&       & $K=p/20$ & 96.2 & 100.0  & 100.0 & 5.7   & 5.5   & 4.8   & 5.0   & 100.0  & 5.0   & 6.3   & 13.8 & 100.0&  100.0&100.0  & 5.1   & 4.0   & 3.7   & 5.3   & 100.0   & 5.3   & 5.3   & 22.6  \\
						&       & $K=p/10$ & 99.8 &  100.0 & 100.0& 5.8   & 4.6   & 4.3   & 5.6   & 100.0  & 5.2   & 4.9   & 40.2 & 100.0& 100.0& 100.0   & 5.6   & 4.4   & 3.9   & 6.0   & 100.0   & 5.8   & 4.8   & 66.0   \\
						\hline
						\multirow{9}{*}{Example 3} & \multirow{3}{*}{100}   & $K=0$ & 0.1 &  1.7 & 5.7  & 6.1   & 5.3   & 6.1   & 5.8   & 5.7   & 5.6   & 5.2   & 5.3 & 1.1&  2.3& 5.9 & 4.8   & 5.7   & 4.8   & 4.7   & 4.8   & 5.6   & 5.1   & 5.2   \\
						&       & $K=p/20$ & 100.0 & 100.0 &  100.0& 4.7   & 5.8   & 4.7   & 4.4   & 7.8   & 5.3   & 5.6   & 5.3  &100.0 &  100.0&100.0 & 5.1   & 4.3   & 5.4   & 5.3   & 8.4   & 5.2   & 5.6   & 4.8 \\
						&       & $K=p/10$ & 100.0 & 100.0  & 100.0 & 5.4   & 4.4   & 5.5   & 5.3   & 11.2   & 5.0   & 4.7   & 6.1 &100.0 &  100.0& 100.0 & 5.6   & 5.1   & 5.3   & 5.6   & 22.6   & 5.1   & 4.7   & 5.8  \\
						\cline{2-25}           & \multirow{3}{*}{400}   & $K=0$ & 0.1  & 0.8  & 5.4 & 5.5   & 5.4   & 5.5   & 5.5   & 6.7   & 5.3   & 4.9   & 5.2  & 0.3& 1.5 & 5.9  & 4.8   & 4.8   & 4.9   & 4.5   & 5.4   & 5.9   & 5.5   & 4.7\\
						&       & $K=p/20$ & 100.0 & 100.0 &100.0  & 4.9   & 5.2   & 4.9   & 4.7   & 6.8   & 4.9   & 4.4   & 4.7  &100.0 &  100.0& 100.0 & 5.5   & 4.6   & 5.4   & 5.2   & 8.6   & 5.1   & 4.8   & 5.9  \\
						&       & $K=p/10$ & 100.0 & 100.0  & 100.0 & 5.0   & 4.6   & 4.8   & 4.8   & 12.2  & 4.8   & 5.4   & 5.5 & 100.0&  100.0& 100.0 & 5.4   & 4.4   & 4.9   & 4.9   & 11.4   & 5.4   & 5.0   & 6.2  \\
						\cline{2-25}           & \multirow{3}{*}{1600}  & $K=0$ & 0.0  &  0.3 & 6.1 & 5.1   & 5.0   & 5.2   & 5.1   & 6.2   & 5.3   & 4.9   & 5.0 &0.0 &1.2  &  6.2 & 6.1   & 5.1   & 5.9   & 5.7   & 5.1   & 4.9   & 5.0   & 5.4 \\
						&       & $K=p/20$ & 100.0  & 100.0   & 100.0  & 5.2   & 5.3   & 4.8   & 4.7   & 6.0   & 6.2   & 6.0   & 4.8 &100.0 & 100.0 & 100.0 & 4.5   & 4.7   & 4.8   & 4.8   & 9.8   & 4.8   & 5.3   & 5.3  \\
						&       & $K=p/10$ & 100.0  &  100.0  & 100.0   & 5.4   & 4.5   & 5.1   & 5.2   & 9.6   & 5.0   & 5.1   & 4.6 &100.0 & 100.0 & 100.0 & 5.8   & 5.1   & 5.7   & 5.9   & 12.5   & 5.2   & 5.1   & 5.2   \\
						\hline   
						\multirow{6}{*}{Example 4} & \multirow{2}{*}{100}   & {\rm null} & 0.1 & 1.3  & 7.3  & 4.9   & 5.0   & 4.9   & 5.0   & 5.4   & 6.0   & 5.7   & 4.9 &0.4 & 1.6 &  5.6 & 5.2   & 4.9   & 4.9   & 4.9   & 4.4   & 5.5   & 5.9   & 5.2   \\
						&       & {\rm alternative}& 76.6 &  84.2 & 89.4 & 12.8  & 5.8   & 12.5  & 12.9  & 5.1   & 7.2   & 4.7   & 5.8 &95.2 & 96.4 & 97.1 & 26.5  & 6.3   & 26.8  & 25.6  & 5.6   & 8.9   & 5.8   & 6.0  \\
						\cline{2-25}           & \multirow{2}{*}{400}   & {\rm null} &  0.0 &  0.4 & 7.5 & 3.9   & 4.9   & 3.6   & 3.7   & 5.8   & 5.3   & 4.8   & 6.6 &0.0 & 1.7 & 5.5 & 5.1   & 4.9   & 4.9   & 4.7   & 5.1   & 5.5   & 4.7   & 6.0 \\
						&       & {\rm alternative} & 62.8 & 75.4  & 83.4 & 7.0   & 4.5   & 7.0   & 6.8   & 6.1   & 5.6   & 4.5   & 5.2  & 91.1&  93.0& 94.7 & 7.4   & 4.4   & 7.5   & 7.3   & 4.5   & 5.2   & 5.7   & 4.6  \\
						\cline{2-25}           & \multirow{2}{*}{1600}  & {\rm null} & 0.0 &  0.0 &  7.0 & 5.3   & 5.3   & 5.3   & 5.5   & 5.1   & 5.1   & 5.1   & 5.0 & 0.0& 0.9 &  6.4 & 4.5   & 3.3   & 4.5   & 4.9   & 5.6   & 5.3   & 5.0   & 5.4 \\
						&       & {\rm alternative} & 42.7 &  65.6 & 78.8 & 4.6   & 4.3   & 4.5   & 4.2   & 4.2   & 5.4   & 5.4   & 5.1  &88.2 & 91.1 & 92.4 & 6.0   & 4.3   & 5.8   & 5.9   & 4.6   & 5.7   & 4.8   & 4.4 \\
						\hline
						\multirow{6}{*}{Example 5} & \multirow{2}{*}{100}   & {\rm null} & 0.0  &   1.2  & 6.8  & 5.0   & 4.5   & 5.0   & 5.0   & 5.3   & 5.6   & 4.9   & 5.7 &0.6 & 1.9 &5.9  & 5.2   & 4.9   & 4.8   & 4.5   & 6.0   & 5.1   & 5.5   & 5.5  \\
						&       & {\rm alternative} & 77.5 &  83.9 & 88.7 & 11.7  & 5.5   & 11.5  & 11.3  & 5.6   & 6.4   & 6.0   & 4.8 & 95.1& 96.0 & 96.9 & 22.0  & 7.0   & 21.5  & 21.3  & 6.4   & 9.2   & 5.6   & 5.7  \\
						\cline{2-25}           & \multirow{2}{*}{400}   & {\rm null} & 0.0  & 0.5 & 7.0 & 6.1   & 4.8   & 6.0   & 5.9   & 4.7   & 5.0   & 5.6   & 6.7 & 0.1& 1.5 &  5.9 & 5.6   & 5.3   & 5.1   & 5.2   & 5.4   & 6.0   & 5.3   & 5.7 \\
						&       & {\rm alternative} & 62.5 & 74.5 & 83.3 & 6.8   & 5.3   & 6.7   & 6.8   & 5.6   & 5.9   & 5.1   & 5.3 &90.8 &  93.3& 94.7 & 8.0   & 5.9   & 7.8   & 7.5   & 6.0   & 5.0   & 5.2   & 4.7  \\
						\cline{2-25}         & \multirow{2}{*}{1600}  & {\rm null} &0.0   & 0.2 & 6.8 & 5.6   & 5.4   & 5.2   & 5.0   & 4.8   & 5.4   & 4.7   & 5.5 & 0.0& 0.6 & 6.2  & 5.9   & 5.3   & 6.0   & 6.0   & 5.6   & 5.1   & 5.5   & 4.3  \\
						&       & {\rm alternative} & 44.5 & 65.9 &  77.9 & 5.7   & 5.3   & 5.5   & 5.7   & 6.0   & 5.0   & 5.3   & 4.8 &87.7 & 90.9 & 92.6 & 5.0   & 4.9   & 5.0   & 5.0   & 4.6   & 5.8   & 5.5   & 5.2   \\
						\hline
						\hline
					\end{tabular}%
				\end{threeparttable}
			}
		\end{spacing}
	\end{table}
\end{landscape}

\begin{landscape}   
	\begin{table}[htbp]
		\scriptsize
		\caption{ Empirical sizes (the rows with $K=0$ in Examples 6 and 8--10, and $\rho=0$ in Example 7) and powers (the rows with $K=p/10$ and $p/5$ in Examples 6 and 8--10, $\rho=0.7$ and 0.8 in Example 7) of the proposed conditional independence tests and the comparing methods in Examples 6--10. All numbers reported below are multiplied by 100. The results reported by `NA' indicate that the associated tests return invalid results.  } 
		\label{tab:cind}%
		\begin{spacing}{1.1}
			\resizebox{23cm}{!}{
				\begin{threeparttable}
					\begin{tabular}{ccc|ccccccccccc|ccccccccccc}
						\hline
						\hline	   
						&     &  &\multicolumn{11}{c|}{$n=100$} & \multicolumn{11}{c}{$n=200$}   \\
						\hline 
						& \multirow{3}{*}{$p$}   & \multirow{3}{*} {Setting} & \multicolumn{6}{c}{Proposed Methods} &  \multirow{3}{*}{GCM}   &  \multirow{3}{*}{PCD}   &  \multirow{3}{*}{RCIT}  &  \multirow{3}{*}{RCoT}  &  \multirow{3}{*}{cdCov} & \multicolumn{6}{c}{Proposed Methods} &  \multirow{3}{*}{GCM}   &  \multirow{3}{*}{PCD}   &  \multirow{3}{*}{RCIT}  &  \multirow{3}{*}{RCoT}  &  \multirow{3}{*}{cdCov} \\
						&  &   & \multicolumn{2}{c}{Gaussian} & \multicolumn{2}{c}{Mammen} & \multicolumn{2}{c}{Rademacher} &     &     &   &    &  & \multicolumn{2}{c}{Gaussian} & \multicolumn{2}{c}{Mammen} & \multicolumn{2}{c}{Rademacher} &     &     &   &    & \\  
						&  &  & CI-FNN   & CI-Lasso & CI-FNN  & CI-Lasso & CI-FNN & CI-Lasso &   & & & & & CI-FNN   & CI-Lasso & CI-FNN  & CI-Lasso & CI-FNN & CI-Lasso &   & & & & \\
						\hline 
						\multirow{9}{*}{Example 6} & \multirow{3}{*}{100}   & $K=0$ & 0.3 & 0.8 & 3.2 & 2.5 & 7.3 & 7.1 & 0.2 & 1.1 & 99.6 & 99.6 & 35.1 & 0.9 & 2.3 & 2.1 & 4.1  & 7.0 & 6.0  & 0.2 & 0.4 & 19.2 & 23.4  &59.0\\
						&       & $K=p/10$ & 25.2 & 100.0 & 21.7 & 100.0 & 57.9 & 100.0 & 0.2 & NA & 99.6 & 99.2 & 18.9 & 90.2 &100.0  &90.5 & 100.0 & 92.6 & 100.0&0.1 &NA & 22.4 &21.1 & 40.6 \\
						&       & $K=p/5$ & 45.5 & 100.0& 34.6 & 100.0 & 79.9 & 100.0 & 0.2 & NA & 99.4 & 99.4 & 16.2 &96.7 & 100.0 & 94.5 & 100.0 & 97.6 & 100.0 & 0.0&NA & 27.2 & 27.1 & 39.5 \\
						\cline{2-25}
						& \multirow{3}{*}{400}    & $K=0$ & 0.0&0.1 &2.2 & 2.0&5.4 &6.6 &0.1 &5.4 &99.4 &99.1 &42.9 &0.0 &0.6 &1.2 & 2.1 &7.6 & 6.3& 0.0& 3.2&24.1 & 20.8 &66.3\\
						&       & $K=p/10$ & 23.4 & 100.0&27.6 & 100.0 & 70.8 & 100.0&0.2 &NA &99.6 &99.8 &17.1 &100.0 &100.0 & 100.0 &99.7 &100.0 &100.0 &0.1 & NA&24.3 &22.1 &40.6 \\
						&       & $K=p/5$ &69.6 & 100.0 & 95.6 & 100.0&99.7 & 100.0 &0.3 & NA& 99.5& 99.5& 18.7&100.0 & 100.0&100.0 & 100.0& 100.0& 100.0&0.0 & NA&24.9 &21.9 &35.5\\
						\cline{2-25}
						& \multirow{3}{*}{1600}  & $K=0$ &0.0 &0.0 &1.2 & 0.1 & 9.0& 8.9  &0.0 &NA & 99.4& 99.2&45.4 &0.5 &0.3 &1.5 &1.9 & 8.5 & 6.7 & 0.0& 3.2& 25.0&25.0&73.8 \\
						&       & $K=p/10$ & 42.5 & 100.0& 82.0 & 100.0& 97.0 & 100.0& 0.2& NA& 99.2& 99.2& 15.6& 100.0& 100.0& 100.0&100.0 &100.0  & 100.0&0.0 & NA&21.2 &20.2&40.4 \\
						&       & $K=p/5$ & 70.6 & 100.0& 97.2&100.0 & 99.4& 100.0& 0.0&NA & 98.8&98.6 &19.6 &100.0 &100.0 &100.0 &100.0 & 100.0& 100.0& 0.2& NA& 25.0&26.2 &42.8 \\
						\hline
						\multirow{9}{*}{Example 7} & \multirow{3}{*}{100}   & $\rho=0$ & 0.3 & 1.1 &  1.9 & 3.4 & 7.2 & 7.0 & 5.4 & 3.7 & 100.0 & 100.0 & 99.9 & 0.1 &  1.4  &1.0 & 3.2 & 5.1 & 6.2 & 4.0 & 1.8 & 55.9 & 54.6 &100.0 \\
						&       & $\rho=0.7$ & 73.4 & 100.0 & 47.9 & 100.0 & 97.9 & 100.0 & 100.0 & 0.0 & 100.0 &100.0 & 100.0 &99.0 & 100.0  &99.2 & 100.0  & 99.9 & 100.0& 100.0 & 0.0& 81.3& 83.7 &100.0 \\
						&       & $\rho=0.8$ & 87.0 & 100.0 &  60.4 & 100.0 & 99.3& 100.0 & 100.0 & 0.1 & 100.0 &  100.0  & 99.9 & 99.3 & 100.0 & 100.0 & 100.0& 100.0 &100.0 & 100.0 & 0.0& 83.1& 85.6& 100.0 \\
						\cline{2-25}
						& \multirow{3}{*}{400}   & $\rho=0$ & 0.2&0.0 &2.2 & 1.5&6.8 &6.5 &4.9 & 6.9&99.9 &100.0 &100.0 & 0.2 &0.3 & 1.2 &1.5 &7.0 &4.9  &4.4 & 4.7& 55.9&55.6 & 100.0\\
						&       & $\rho=0.7$ & 56.6 & 100.0&47.0 &100.0 & 97.6&100.0 &100.0 &54.6 & 99.9& 100.0&100.0 &100.0 &100.0 & 100.0& 100.0& 100.0&100.0 & 100.0& 1.8&65.0 &66.8 &100.0 \\
						&       & $\rho=0.8$ & 99.0 & 100.0& 100.0&100.0 &100.0 &100.0 &100.0 &98.4 & 100.0& 100.0&100.0 &100.0 & 100.0&100.0 & 100.0& 100.0&100.0 & 100.0& 20.5&63.7 &66.1 &100.0 \\
						\cline{2-25}
						& \multirow{3}{*}{1600}  & $\rho=0$ & 0.0 & 0.0& 1.2& 0.8& 8.4& 7.6& 6.4& 8.2& 100.0&100.0 & 100.0 & 0.3 & 0.4 &1.5 & 1.9 & 4.7 &6.2 &3.8 & 7.4& 55.6&57.4 &100.0 \\
						&       & $\rho=0.7$ & 74.5 & 100.0& 100.0& 100.0&100.0 &100.0 & 100.0&NA &100.0 &100.0 & 100.0& 100.0&100.0 & 100.0& 100.0&100.0 &100.0 &100.0 &94.2 &58.0 &57.2 &100.0 \\
						&       & $\rho=0.8$ & 96.6 & 100.0& 100.0 &100.0 &100.0 &100.0 & 100.0& NA&100.0 &100.0 &100.0 & 100.0&100.0 & 100.0&100.0 &100.0 &100.0 & 100.0& NA& 59.2& 58.2&100.0 \\
						\hline 
						\multirow{9}{*}{Example 8} & \multirow{3}{*}{100}   & $K=0$   & 0.0 & 0.5 & 2.7 & 2.2 & 7.8 & 5.5 &  4.6 & 0.5 & 99.9 & 99.9  & 95.7 & 0.4 &1.6 & 1.7 & 3.1 & 5.0 &5.4 &4.2 &0.3 & 38.4 & 40.9 & 99.8 \\
						&       & $K=p/10$ & 19.9 & 100.0 & 17.2 & 100.0 & 52.2& 100.0& 3.7 & 5.5 & 100.0 & 100.0 & 67.5 & 84.0 & 100.0  & 83.6 & 100.0 &89.0  & 100.0 & 4.1 & 0.3 &39.1 & 42.0 & 89.2\\
						&       & $K=p/5$ & 45.1 & 100.0 & 31.3 & 100.0 & 80.1& 100.0 & 4.3 &  15.4 & 99.9 & 99.9 & 67.8 & 95.5 & 100.0  &95.6 & 100.0 & 97.2 & 100.0  & 7.1& 7.2& 43.8&45.9 &89.5 \\
						\cline{2-25}
						& \multirow{3}{*}{400}  & $K=0$   &0.0 & 0.0&2.0 &1.0 & 7.5&5.1 &4.7 & 1.6&99.9 &100.0 &99.3 &0.2 &1.0 &1.4 &2.3 &6.4 &5.3 &4.4 & 1.7& 37.3&40.3 & 100.0\\
						&       & $K=p/10$ & 28.4 & 100.0 &27.4 &100.0 & 66.4 & 100.0&5.0 & NA&99.9 &100.0 &68.0 &100.0 & 100.0& 100.0&100.0 & 100.0 & 100.0&3.3 & NA&41.2 &39.9 &89.9 \\
						&       & $K=p/5$ &76.0 &100.0 & 97.2&100.0 &99.5 &100.0 &6.8 & NA& 100.0& 100.0&69.6 &100.0 &100.0 & 100.0 &100.0 &100.0 & 100.0&6.0 & NA& 42.3& 42.1&89.6 \\
						\cline{2-25}
						& \multirow{3}{*}{1600}  & $K=0$   &0.2 &0.0 &0.8 &0.4 &6.4 &6.4 &8.6 &3.0 &100.0 & 100.0&99.6 &0.5 &0.1 & 1.5 & 1.4 & 7.4 & 5.7&4.6 & 2.8&39.0 &42.8 &100.0 \\
						&       & $K=p/10$ &47.5 & 100.0 & 84.5& 100.0&98.5 & 100.0&6.2 & NA&100.0 & 100.0& 71.6&100.0 &100.0 &100.0 &100.0 &100.0 &100.0 &3.4 &NA & 40.8& 40.0&89.2 \\
						&       & $K=p/5$ & 76.4&100.0 &98.2 & 100.0& 100.0& 100.0&7.6 &NA &100.0 &100.0 &73.0   &100.0 &100.0 &100.0 & 100.0&100.0 & 100.0&4.6 & NA&35.6 & 42.2&86.9 \\
						\hline
						\multirow{9}{*}{Example 9} & \multirow{3}{*}{100}   & $K=0$   & 0.1 & 0.6 & 2.9&2.3 & 7.1 & 5.3 & 3.8& 4.0 & 99.9& 100.0 & 98.7 & 0.7& 1.7& 1.7 & 2.9 & 5.1& 6.2& 4.0& 3.6& 40.4& 37.2&100.0 \\
						&       & $K=p/10$ & 12.3 & 100.0 & 12.8& 100.0 &44.7 & 100.0 & 9.0&0.0 &  99.9 & 100.0 & 37.6  &85.9 & 100.0 &85.6 &100.0 & 89.6 & 100.0 &5.3 &0.0 & 41.4& 42.3&53.1 \\
						&       & $K=p/5$ & 23.1 & 100.0 &20.3 & 100.0 & 65.3& 100.0 & 18.4 & 0.0 &100.0 &100.0 & 38.5 &94.5  & 100.0   & 92.1 &100.0  &95.6 & 100.0 &9.7 &0.0 & 42.1& 40.8&49.3\\
						\cline{2-25}
						& \multirow{3}{*}{400}   & $K=0$   &0.0 & 0.0&1.8 & 1.0&5.6 &5.5 & 5.4& 4.8& 100.0&100.0 &99.8 &0.2 &0.9 &1.0 & 2.1 &6.4 &5.3 &2.3 &0.4 &39.8 &37.1 &100.0 \\
						&       & $K=p/10$ &19.2 &100.0 & 49.6&100.0 &74.8 &100.0 &18.8 &0.1 &100.0 &100.0 & 37.5 & 99.3 & 100.0& 100.0& 100.0&100.0 & 100.0 &10.1 &0.0& 39.6&39.4 &47.5 \\
						&       & $K=p/5$ & 37.1 &100.0 &77.0 &100.0 &94.5 &100.0 &48.1 &0.0 &100.0 & 100.0&34.3 &100.0 &100.0 &100.0 &100.0 &100.0 & 100.0&20.4 & 0.0&40.0 & 39.1& 48.6\\
						\cline{2-25}
						& \multirow{3}{*}{1600}  & $K=0$   &0.2 &0.0 &1.0 &0.3 & 7.0& 6.6& 6.2& 4.8&100.0 &100.0 &100.0 & 0.0 & 0.1 & 1.3 & 1.6 & 4.8 &6.0 &4.2 &3.8 & 38.0& 41.0&100.0 \\
						&       & $K=p/10$ & 14.0 & 100.0&51.2 & 100.0& 82.0& 100.0&48.4 &0.0 & 100.0&99.8 &31.2 & 99.6 &100.0 & 100.0&100.0 & 100.0&100.0 &20.2 &0.2 &39.6 &38.0 & 42.4\\
						&       & $K=p/5$ &26.4 & 100.0 &74.4 &100.0 & 95.6& 100.0 & 85.0&0.0 & 99.8& 100.0&30.4 & 100.0&100.0 & 100.0& 100.0&100.0 & 100.0& 55.4&0.0 & 41.6&40.0 &39.0 \\
						\hline 
						\multirow{9}{*}{Example 10}& \multirow{3}{*}{100}   & $K=0$   &0.3 & 0.7& 2.8& 2.6& 7.8 & 5.2 & 5.4&3.9 & 100.0&99.8 &97.7 &0.2 &1.7 & 2.1 & 3.1 & 5.9 &5.3 &4.0 &4.0 &39.1 & 41.6&100.0 \\
						&       & $K=p/10$ & 15.5 & 100.0 &15.4 &100.0 & 48.0& 100.0 & 100.0&4.9 & 100.0& 99.9& 93.7 & 83.9 & 100.0 &85.0 & 100.0 & 88.9 &100.0 &100.0 & 4.3&40.3 & 39.0&99.5 \\
						&       & $K=p/5$ & 24.6 & 100.0 & 26.8&100.0 &67.3 &100.0 &100.0 & 4.8&99.9 &100.0 &97.7 & 93.3 & 100.0 &92.2 & 100.0  &96.2 &100.0 &100.0 & 5.4&41.1 &43.4 &99.7 \\
						\cline{2-25}
						& \multirow{3}{*}{400}   & $K=0$   & 0.2 & 0.1& 1.0& 1.3&6.8 &6.1 &7.3 &3.2 &100.0 &100.0 &99.7 &0.0 &1.2 &1.4  &2.8 &6.2 &6.7 &3.9 &2.6 &38.6 & 42.0& 100.0\\
						&       & $K=p/10$ &28.3 &100.0 &58.5 &100.0 & 78.7&100.0 &100.0 &3.8 &100.0 &100.0 &99.5 & 99.7 &100.0 & 100.0&100.0 & 100.0& 100.0&100.0 &4.0 &41.9 &40.1 &100.0 \\
						&       & $K=p/5$ & 44.3 & 100.0& 81.4&100.0 &95.4 &100.0 &100.0 &5.4 &100.0 & 100.0& 99.9& 100.0&100.0 & 100.0& 100.0&100.0 & 100.0&100.0 &3.9 &40.0 & 39.6& 100.0\\
						\cline{2-25}
						& \multirow{3}{*}{1600}  & $K=0$   &0.0 &0.0 &2.0 &6.0 & 8.0& 6.8&9.8 &1.8 &100.0 &100.0 & 100.0 &0.0 & 0.1 &0.8  &1.6 & 6.0& 6.0& 4.2&1.0 &41.2 &39.4 &100.0 \\
						&       & $K=p/10$ & 16.5 & 100.0& 55.5& 100.0&84.5 &100.0 &100.0 & 2.2&100.0 & 100.0&100.0 & 100.0&100.0 &100.0 &100.0 &100.0 &100.0 &100.0 &2.4 &37.6 & 40.0&100.0 \\
						&       & $K=p/5$ &28.8 & 100.0 &79.4 &100.0 &96.6 & 100.0 & 100.0 &4.2 &100.0 &99.8 & 100.0&100.0 & 100.0& 100.0& 100.0&100.0 & 100.0& 100.0&5.2 &40.4 & 40.6& 100.0\\
						\hline
						\hline
					\end{tabular}%
				\end{threeparttable}
			}
		\end{spacing}
	\end{table}%
\end{landscape}

\clearpage
\onehalfspacing
\normalsize
\numberwithin{equation}{section}

\begin{center}
	{\bf  \Large
		Supplementary material for ``Testing Independence and Conditional Independence
		in High Dimensions via Coordinatewise
		Gaussianization" 
         by Jinyuan Chang, Yue Du, Jing He and Qiwei Yao
        }  \\
\end{center}

\setcounter{page}{1}
\renewcommand{\thepage}{S\arabic{page}}

\bigskip



\setcounter{section}{0}
\renewcommand{\thesection}{\Alph{section}}      
\makeatletter
\renewcommand{\theHsection}{\Alph{section}}     
\renewcommand{\theHsubsection}{\Alph{section}.\arabic{subsection}}
\renewcommand{\theHequation}{\Alph{section}.\arabic{equation}}
\renewcommand{\theHfigure}{\Alph{section}.\arabic{figure}}
\renewcommand{\theHtable}{\Alph{section}.\arabic{table}}
\newtheorem{lem}{Lemma} 
\numberwithin{lem}{section} 
\setcounter{lem}{0}
\renewcommand{\thelem}{\thesection\arabic{lem}} 
\makeatother

We first introduce some notation which will be used throughout the supplementary material. We use $C$, $C_1, \ldots$ to denote some generic positive constants that do not depend on $(n, p, q,m)$ which may be different in different uses.  For $f:\mathbb{R}^{m}\to \mathbb{R}$, the supremum norm of $f$ on a set $D\subset \mathbb{R}^{m}$ is denoted by $|f|_{\infty, D} =\sup_{\boldsymbol{x}\in D} |f(\boldsymbol{x})|$.   Given the natural numbers $k_1$ and $k_2$, denote by ${\rm C}_{k_1}^{k_2}$ the combination number, i.e., the number of ways to select $k_2$ distinct elements from a set of $k_1$ elements without regard to the order in which the elements are chosen. 
 For any $i\in[n]$, define
\begin{equation*} 
	\begin{split}
		&\hat{F}^{(i)}_{\bX,j}(X_{i,j})=\frac{1}{n-1}\sum_{s:\,s\ne i}I(X_{s,j}\le X_{i,j})\,, \ \ \hat{F}^{(i)}_{\bY,k}(Y_{i,k})=\frac{1}{n-1}\sum_{s:\,s\ne i}I(Y_{s,k}\le Y_{i,k})\,,\\
		&~~~~~~~~~~~~~~~~~~~~~~~~~~~\hat{F}^{(i)}_{\bZ,l}(Z_{i,l})=\frac{1}{n-1}\sum_{s:\,s\ne i}I(Z_{s,l}\le Z_{i,l})\,.
	\end{split}
\end{equation*}

\section{Proofs of Theorems \ref{thm:1} and \ref{thm:2}}\label{sec:pro1}

Recall  $\hat{\bSigma}= n^{-1}\sum_{i=1}^{n}\hat{\bgamma}_i\hat{\bgamma}_i^{\T}-\bar{\hat{\bgamma}}\bar{\hat{\bgamma}}^{\T}$
with $\bar{\hat{\bgamma}} = n^{-1}\sum_{i=1}^{n}\hat{\bgamma}_i$. To prove Theorem \ref{thm:1}, we need Proposition \ref{pro:1}, whose proof is given in Section \ref{sc:A1}.

\begin{proposition}\label{pro:1}
Let 
$\hat{\bxi}\,|  \,\mathcal{X}_n,\mathcal{Y}_n\sim  \mathcal{N}(\boldsymbol{0},\hat{\bSigma} )$. Under the null hypothesis $\mathbb{H}_0$ in  \eqref{eq:equind}, it holds that  
$$\sup_{z >0}\big|\mathbb{P}(H_{n}>z)-\mathbb{P}(|\hat{\bxi}|_{\infty}>z\,|\,\mathcal{X}_{n},\mathcal{Y}_n)\big|=o_{{\rm p}}(1)$$ as $n\rightarrow\infty$, provided that $\log d \ll  n^{1/8}(\log n)^{-1/4}$.
\end{proposition}

\subsection{Proof of Proposition \ref{pro:1}} \label{sc:A1}

The following Lemmas \ref{lem:uhv}--\ref{lem:covh} are needed in the proof of Proposition \ref{pro:1}, with their proofs given in Appendices \ref{sec:sub-huv}--\ref{sec:sub-sigmma-g-h0}, respectively.  Select $M_1=\sqrt{\kappa_1\log n}$ for some constant $  \kappa_1 \in (1,2)$, and define
\begin{align*}
\begin{split}
U_{i,j}^{*} =&~ U_{i,j}I(|U_{i,j}|\le M_1) + M_1 \cdot{\rm sign}(U_{i,j})I(|U_{i,j}|>M_1)\,, \\
V_{i,k}^{*} =&~ V_{i,k}I(|V_{i,k}|\le M_1) +   M_1 \cdot{\rm sign}(V_{i,k}) I(|V_{i,k}|>M_1)\,, \\
\end{split}
\end{align*}
where $U_{i,j}=\Phi^{-1}\{F_{\bX,j}(X_{i,j})\}$ and $V_{i,k}=\Phi^{-1}\{F_{\bY,k}(Y_{i,k})\}$ 
for $i\in[n]$, $j\in[p]$ and $k\in[q]$. 

\begin{lemma}\label{lem:uhv}
Under the null hypothesis $\mathbb{H}_0$ in \eqref{eq:equind},  it holds that
\begin{align*}
\max_{j\in[p],\,k\in[q]}\bigg|\frac{1}{\sqrt{n}}\sum_{i=1}^{n}(\hat{U}_{i,j}-U_{i,j}^{*})V_{i,k}^{*}\bigg| 
= &~ O_{\rm p}\{n^{-(\kappa_1-1)/2} (\log n)^{1/2}\} +   O_{\rm p} \{n^{-3/14} (\log n)^{1/2}\log  (dn)\}  \\
&+ O_{\rm p} \{n^{-1/7} (\log n)^{-1/4}\log^{1/2} (dn)\} \\
= &~ \max_{j\in[p],\,k\in[q]}\bigg|\frac{1}{\sqrt{n}}\sum_{i=1}^{n}(\hat{V}_{i,k}-V_{i,k}^{*})U_{i,j}^{*}\bigg| 
\end{align*}
provided that $ \log d \ll \min\{n^{1-\kappa_1/2} (\log n)^{-1/2}, n^{3/7}(\log n)^{-1} \}$.
\end{lemma}

\begin{lemma}\label{lem:huhv}
If $\kappa_1\in(1,8/5)$, then 
\begin{align*}
&\max_{j\in[p],\, k\in[q]}\bigg|\frac{1}{\sqrt{n}}\sum_{i=1}^{n}(\hat{U}_{i,j}-U_{i,j}^{*})(\hat{V}_{i,k}-V_{i,k}^{*}) \bigg| = O_{\rm p}\{n^{-(\kappa_1-1)/2} (\log n)^{1/2}\}
\end{align*}
provided that $ \log d \lesssim   n^{1-5\kappa_1/8}  \log n $.
\end{lemma}

\begin{lemma}\label{lem:usvs-uv}
It holds that 
\begin{align*}
\max_{j\in[p],\, k\in[q]}\bigg|\frac{1}{\sqrt{n}}\sum_{i=1}^{n}(U_{i,j}^{*}V_{i,k}^{*} - U_{i,j}V_{i,k})\bigg| =  O_{\rm p} \{n^{-(\kappa_1-1)/2}(\log n)^{1/2}\} + O_{\rm p}\{n^{-1/2}(\log d) \log (dn)\}  
\end{align*}
provided that $\log d \lesssim n^{1-\kappa_1/2}(\log n)^{-1/2}$.
\end{lemma}

\begin{lemma}\label{lem:covh}
It holds that \begin{align*}
|\hat{\bSigma} - \bSigma |_{\infty}  =   O_{\rm p}\{n^{-1/2} (\log n) (\log d)^{1/2}\log^{3/2}(dn)\}
\end{align*}
provided that   $\log d \lesssim  n^{1/3}$.
\end{lemma}

Recall $H_{n}=\sqrt{n}|\hat{\bS}_n|_{\infty}$ and  $\hat{\bS}_n= n^{-1}\sum_{i=1}^{n}\hat{\bgamma}_i$  with $\hat{\bgamma}_i = \hat{\bU}_i \otimes \hat{\bV}_i$. Define $\bS_{n}=n^{-1}\sum_{i=1}^{n}\bgamma_i$ with $\bgamma_i = \bU_i \otimes \bV_i$, and let $\bxi \sim  \mathcal{N}(\boldsymbol{0},\bSigma )$ with $\bSigma =\cov(\bgamma_i)$.
For any $x>0$ and $\upsilon>0$, it holds that
\begin{align*}
\mathbb{P} (\sqrt{n} |\hat{\bS}_n |_{\infty}>x ) =&~  \mathbb{P} (\sqrt{n} |\hat{\bS}_n |_{\infty}>x , \sqrt{n} |\hat{\bS}_n-\bS_n |_{\infty} > \upsilon  ) + \mathbb{P} (\sqrt{n} |\hat{\bS}_n |_{\infty}>x , \sqrt{n} |\hat{\bS}_n-\bS_n |_{\infty} \le \upsilon  )\\
\le &~ \mathbb{P} (\sqrt{n} |\hat{\bS}_n-\bS_n |_{\infty} > \upsilon  ) + \mathbb{P} (\sqrt{n} |\bS_n |_{\infty}>x-\upsilon )\,. 
\end{align*}
Thus, we have
\begin{align*}
\mathbb{P} (\sqrt{n} |\hat{\bS}_n |_{\infty}>x )-\mathbb{P} ( |\bxi |_{\infty}>x ) \le&~ \mathbb{P} (\sqrt{n} |\hat{\bS}_n-\bS_n |_{\infty} > \upsilon ) + \mathbb{P} (x-\upsilon < |\bxi |_{\infty}\le x ) \\
&+\mathbb{P} (\sqrt{n} |\bS_n |_{\infty}>x-\upsilon )- \mathbb{P} ( |\bxi |_{\infty}>x-\upsilon )\,. \notag
\end{align*}
On the other hand, for any $x>0$ and $\upsilon>0$, since
\begin{align*}
\mathbb{P} (\sqrt{n} |\hat{\bS}_n |_{\infty}>x ) \ge&~  \mathbb{P} (\sqrt{n} |\hat{\bS}_n |_{\infty}>x, 	\sqrt{n} |\bS_n |_{\infty}>x+\upsilon )\\
=&~\mathbb{P} (\sqrt{n} |\bS_n |_{\infty}>x+\upsilon )-\mathbb{P} (\sqrt{n} |\hat{\bS}_n |_{\infty}\le x , \sqrt{n} |\bS_n |_{\infty}>x+\upsilon )\\
\ge&~ \mathbb{P} (\sqrt{n} |\bS_n |_{\infty}>x+\upsilon ) - \mathbb{P} (\sqrt{n} |\hat{\bS}_n-\bS_n |_{\infty} > \upsilon )\,,
\end{align*}
we have 
\begin{align*}
\mathbb{P} (\sqrt{n} |\hat{\bS}_n |_{\infty}>x )-\mathbb{P} ( |\bxi |_{\infty}>x ) \ge& -\mathbb{P} (\sqrt{n} |\hat{\bS}_n-\bS_n |_{\infty} > \upsilon  ) - \mathbb{P} (x < |\bxi |_{\infty}\le x+\upsilon )\\
&+\mathbb{P} (\sqrt{n} |\bS_n |_{\infty}>x+\upsilon )- \mathbb{P} (  |\bxi |_{\infty}>x+\upsilon )\,.
\end{align*}
Therefore, due to $H_{n}=\sqrt{n}|\hat{\bS}_n|_{\infty}$, 
\begin{align*}
\sup_{x >0} \big|\mathbb{P} (H_{n}>x )-\mathbb{P} ( |\bxi |_{\infty}>x ) \big| \le&~ \sup_{x>0} \big| \mathbb{P} (\sqrt{n} |\bS_n |_{\infty}>x)- \mathbb{P} (  |\bxi |_{\infty}>x )\big|\notag\\
&+ \sup_{x>0}\mathbb{P} (x-\upsilon  < |\bxi |_{\infty}\le x) + \mathbb{P} (\sqrt{n} |\hat{\bS}_n-\bS_n |_{\infty} > \upsilon )\,.
\end{align*}
Recall $d=pq$. By Nazarov's inequality \citep[Lemma A.1]{Chernozhukov2017}, it holds that 
\begin{equation*}
\begin{aligned}
\sup_{x>0}\mathbb{P} (x-\upsilon < |\bxi |_{\infty}\le x ) \lesssim   \upsilon (\log d)^{1/2}\,.
\end{aligned}
\end{equation*}
Hence,
\begin{align}\label{eq:GAbound_a}
\sup_{x >0} \big|\mathbb{P} (H_{n}>x )-\mathbb{P} ( |\bxi |_{\infty}>x ) \big| \lesssim &~\sup_{x>0} \big| \mathbb{P} (\sqrt{n} |\bS_n |_{\infty}>x)- \mathbb{P} (  |\bxi |_{\infty}>x )\big|\notag\\
&+ \mathbb{P} (\sqrt{n} |\hat{\bS}_n-\bS_n |_{\infty} > \upsilon ) + \upsilon (\log d)^{1/2} \,.
\end{align}
Due to  
\begin{align*}
\frac{1}{n}\sum_{i=1}^{n} (\hat{U}_{i,j}\hat{V}_{i,k} - U_{i,j}V_{i,k} ) = &~\frac{1}{n}\sum_{i=1}^{n} (\hat{U}_{i,j} - U_{i,j}^{*} ) V_{i,k}^{*} + \frac{1}{n}\sum_{i=1}^{n}(\hat{V}_{i,k} - V_{i,k}^{*} ) U_{i,j}^{*} \notag\\
&+ \frac{1}{n}\sum_{i=1}^{n} (\hat{U}_{i,j} - U_{i,j}^{*} )(\hat{V}_{i,k} - V_{i,k}^{*} ) + \frac{1}{n}\sum_{i=1}^{n} (U_{i,j}^{*}V_{i,k}^{*} - U_{i,j}V_{i,k} ) 
\end{align*}
for any $j \in [p]$ and $k \in[q]$, by Lemmas \ref{lem:uhv}--\ref{lem:usvs-uv}, under the null hypothesis $\mathbb{H}_0$ in \eqref{eq:equind}, we have
\begin{align*}
&\sqrt{n} |\hat{\bS}_n-\bS_n |_{\infty} \\
&~~~~~~\le \max_{j \in [p],\,k \in[q]}\bigg | \frac{1}{\sqrt{n}}\sum_{i=1}^{n}(\hat{U}_{i,j} - U_{i,j}^{*}) V_{i,k}^{*}  \bigg|+ \max_{j \in [p],\,k \in[q]}\bigg | \frac{1}{\sqrt{n}}\sum_{i=1}^{n}(\hat{V}_{i,k} - V_{i,k}^{*}) U_{i,j}^{*} \bigg| \\
&~~~~~~~~~~+  \max_{j \in [p],\,k \in[q]}\bigg |\frac{1}{\sqrt{n}}\sum_{i=1}^{n}(\hat{U}_{i,j}-U_{i,j}^{*})(\hat{V}_{i,k}-V_{i,k}^{*})\bigg| +  \max_{j \in [p],\,k \in[q]}\bigg |\frac{1}{\sqrt{n}}\sum_{i=1}^{n}(U_{i,j}^{*}V_{i,k}^{*} - U_{i,j}V_{i,k} )\bigg| \\
&~~~~~~ =  O_{\rm p}\{n^{-(\kappa_1-1)/2} (\log n)^{1/2}\} +   O_{\rm p} \{n^{-3/14} (\log n)^{1/2}\log  (dn)\}  \\
&~~~~~~~~~~+ O_{\rm p} \{n^{-1/7} (\log n)^{-1/4}\log^{1/2} (dn)\} + O_{\rm p}\{n^{-1/2}(\log d)\log (dn)\}
\end{align*}
provided that $ \log d \lesssim  n^{1-5\kappa_1/8} \log n$ with $\kappa_1\in(1,8/5)$.
To make  $\mathbb{P}(\sqrt{n}|\hat{\bS}_n-\bS_n |_{\infty} >\upsilon)=o(1)$, it suffices to require $\upsilon \gg n^{-(\kappa_1-1)/2} (\log n)^{1/2}$, $\upsilon \gg n^{-3/14} (\log n)^{1/2}\log (dn)$, $\upsilon \gg n^{-1/7} (\log n)^{-1/4}\log^{1/2} (dn)$ and $\upsilon \gg n^{-1/2} (\log d) \log (dn)$. On the other hand, by \eqref{eq:GAbound_a}, to make $\sup_{x >0} |\mathbb{P} (H_{n}>x )-\mathbb{P} ( |\bxi |_{\infty}>x ) | =o(1)$ under the null hypothesis $\mathbb{H}_0$ in \eqref{eq:equind}, we need to require $\upsilon \ll (\log d)^{-1/2}$. Therefore, $(d,n)$ should satisfy 
\begin{align*}
\left\{
\begin{aligned}
& n^{-(\kappa_1-1)/2} (\log n)^{1/2}  \ll (\log d)^{-1/2}\,,\\
&    n^{-3/14} (\log n)^{1/2}\log (dn)  \ll  (\log d)^{-1/2}\,,\\
&  n^{-1/7} (\log n)^{-1/4}\log^{1/2} (dn)  \ll  (\log d)^{-1/2}\,,\\
&n^{-1/2} (\log d) \log (dn)  \ll (\log d)^{-1/2}\,,\\
&\log d \lesssim  n^{1-5\kappa_1/8}  \log n  \,,
\end{aligned}
\right.
\end{align*}
which implies
\begin{align}\label{eq:ind-nd}
\log d \ll \min\big\{n^{\kappa_1-1}(\log n)^{-1} ,\, n^{1/7}(\log n)^{-1/3}, \,n^{1-5\kappa_1/8} \log n \big\}\,.
\end{align}
Recall $\kappa_1\in(1,8/5)$. To allow $d$ to diverge with $n$ as fast as possible, we select $\kappa_1=48/35$. Hence, \eqref{eq:ind-nd} becomes $\log d \ll n^{1/7}(\log n)^{-1/3}$. By \eqref{eq:GAbound_a}, under the null hypothesis $\mathbb{H}_0$ in \eqref{eq:equind}, we have
\begin{align*}
\sup_{x >0} \big|\mathbb{P} (H_{n}>x )-\mathbb{P} ( |\bxi |_{\infty}>x ) \big| \lesssim \sup_{x>0} \big| \mathbb{P} (\sqrt{n} |\bS_n |_{\infty}>x)- \mathbb{P} (  |\bxi |_{\infty}>x )\big|+o(1)
\end{align*}
provided that $\log d \ll n^{1/7}(\log n)^{-1/3}$. Since $U_{i,j}, V_{i,k}\sim \mathcal{N}(0,1)$ are independent under the null hypothesis $\mathbb{H}_0$ in \eqref{eq:equind}, we know $\mathbb{E}(\bgamma_i)=\boldsymbol{0}$ for any $i\in[n]$ under the null hypothesis $\mathbb{H}_0$ in \eqref{eq:equind}. By Proposition 2.1 of \cite{Chernozhukov2017}, it holds that
\begin{align*}
\sup_{x >0}\big| \mathbb{P} (\sqrt{n} |\bS_n |_{\infty}>x )-\mathbb{P} ( |\bxi |_{\infty}>x ) \big| \lesssim  n^{-1/6}\log^{7/6}(dn) 
\end{align*}
under the null hypothesis $\mathbb{H}_0$ in \eqref{eq:equind}. Hence, 
\begin{align}\label{eq:ind-h0-1}
\sup_{x >0}\big|\mathbb{P}(H_{n}>x)-\mathbb{P}(|\bxi|_{\infty} >x)\big|=o(1)
\end{align}
provided that $\log d\ll n^{1/7}(\log n)^{-1/3}$. 

By triangle inequality, under the null hypothesis $\mathbb{H}_0$ in \eqref{eq:equind}, we have
\begin{align}\label{eq:hn-xih}
&\sup_{x >0}\big|\mathbb{P}(H_{n}>x)- \mathbb{P}(|\hat{\bxi}|_{\infty} >x \,|\,\mathcal{X}_n, \mathcal{Y}_n)\big|\notag\\
&~~~~~~~~~~\le\sup_{x >0}\big|\mathbb{P}(H_{n}>x)-\mathbb{P}(|\bxi|_{\infty} >x)\big| + \sup_{x >0}\big|\mathbb{P}(|\bxi|_{\infty} >x) - \mathbb{P}(|\hat{\bxi}|_{\infty} >x \,|\,\mathcal{X}_n, \mathcal{Y}_n)\big|\notag\\
&~~~~~~~~~~\le  \sup_{x >0}\big|\mathbb{P}(|\bxi|_{\infty} >x) - \mathbb{P}(|\hat{\bxi}|_{\infty} >x \,|\,\mathcal{X}_n, \mathcal{Y}_n)\big| + o(1) 
\end{align}
provided that $\log d\ll n^{1/7}(\log n)^{-1/3}$.  Let $\bxi^{{\rm ext}}=(\bxi^{\T},-\bxi^{\T})^{\T} = (\xi_1^{\rm ext},\ldots, \xi_{2d}^{\rm ext})^{\T}$ and $\hat{\bxi}^{{\rm ext}}=(\hat{\bxi}^{\T},-\hat{\bxi}^{\T})^{\T} = (\hat{\xi}^{{\rm ext}}_{1},\ldots,\hat{\xi}^{{\rm ext}}_{2d})^{\T}$ with $\bxi\sim \mathcal{N}({\bf 0},\bSigma)$ and $\hat{\bxi}\,|\, \mathcal{X}_{n}, \mathcal{Y}_n \sim \mathcal{N}({\bf 0},\hat{\bSigma})$. Write $\Delta_{n1}=|\hat{\bSigma}-\bSigma|_{\infty}$. By Lemma \ref{lem:covh}, $\Delta_{n1} =  O_{\rm p}\{n^{-1/2} (\log n) (\log d)^{1/2}\log^{3/2}(dn)\}$ provided that $\log d \lesssim n^{1/3} $. Then, by Lemma 3.1 of \cite{Chernozhukov2013}, it holds that
\begin{align}\label{eq:Prop2bound}
&\sup_{x>0}\big|\mathbb{P}(|\bxi|_{\infty} > x)-\mathbb{P}(|\hat{\bxi}|_{\infty} > x \,|\,\mathcal{X}_{n}, \mathcal{Y}_n)\big| \notag \\
&~~~~~~~~~~~~~=  \sup_{x>0}\bigg|\mathbb{P}\bigg(\max_{j \in [2d]}\xi^{{\rm ext}}_{j} >x\bigg) -\mathbb{P}\bigg(\max_{j \in [2d]}\hat{\xi}^{{\rm ext}}_{j} >x\,\bigg|\,\mathcal{X}_n, \mathcal{Y}_n\bigg)\bigg| \notag \\
&~~~~~~~~~~~~~\lesssim \Delta_{n1}^{1/3} \{1\vee \log(2d\Delta_{n1}^{-1})\}^{2/3} =o_{{\rm p}}(1) 
\end{align}
provided that $\log d \ll  n^{1/8}(\log n)^{-1/4}$. Together with \eqref{eq:hn-xih}, under the null hypothesis $\mathbb{H}_0$ in \eqref{eq:equind}, we have
\begin{align*}
\sup_{x >0}\big|\mathbb{P}(H_{n}>x)- \mathbb{P}(|\hat{\bxi}|_{\infty} >x \,|\,\mathcal{X}_n, \mathcal{Y}_n)\big|=o_{\rm p}(1)
\end{align*}
provided that $\log d \ll  n^{1/8}(\log n)^{-1/4}$.
Hence, we complete the proof of Proposition \ref{pro:1}. $\hfill\Box$

\subsection{Proof of Theorem \ref{thm:1}}\label{sec:sub-thm1}   
Given $\epsilon_0 >0$, let ${\rm cv}_{{\rm ind},\alpha}^{(\epsilon_0)}$ and  ${\rm cv}_{{\rm ind},\alpha}^{(-\epsilon_0)}$ be two positive constants such that $\mathbb{P}\{|\bxi|_{\infty} > {\rm cv}_{{\rm ind},\alpha}^{(\epsilon_0)}\} = \alpha+\epsilon_0$ and $\mathbb{P}\{|\bxi|_{\infty} > {\rm cv}_{{\rm ind},\alpha}^{(-\epsilon_0)}\} = \alpha-\epsilon_0$, respectively. Notice that $\hat{\rm cv}_{{\rm ind},\alpha}  =\inf\{t\in \mathbb{R}: \mathbb{P}(|\hat{\bxi}|_{\infty} > t\,|\,\mathcal{X}_n, \mathcal{Y}_n) \le \alpha\}$. Without loss of generality, we assume that $\mathbb{P}(|\hat{\bxi}|_{\infty} > \hat{\rm cv}_{{\rm ind},\alpha}\,|\,\mathcal{X}_n, \mathcal{Y}_n) =\alpha$.  Consider an event 
\begin{align*}
\mathcal{E}_{\epsilon_0} = \big\{{\rm cv}_{{\rm ind},\alpha}^{(\epsilon_0)}< \hat{\rm cv}_{{\rm ind},\alpha} < {\rm cv}_{{\rm ind},\alpha}^{(-\epsilon_0)}\big\}\,.
\end{align*}
We will next show $\mathbb{P}(\mathcal{E}_{\epsilon_0}) \to 1$ as $n \to \infty$.    Recall $d=pq$ with $p \lesssim n^{\varkappa_1}$ and $q\lesssim n^{\varkappa_2}$. For any given $\varkappa_1 \ge 0$ and $\varkappa_2 \ge 0$,  
if $ \hat{\rm cv}_{{\rm ind},\alpha} \le {\rm cv}_{{\rm ind},\alpha}^{(\epsilon_0)}$, by Proposition \ref{pro:1},   we have
\begin{align*}
\alpha = \mathbb{P}(|\hat{\bxi}|_{\infty} > \hat{\rm cv}_{{\rm ind},\alpha}\,|\,\mathcal{X}_n, \mathcal{Y}_n) \ge &~\mathbb{P}\big\{|\hat{\bxi}|_{\infty} > {\rm cv}_{{\rm ind},\alpha}^{(\epsilon_0)} \,|\,\mathcal{X}_n, \mathcal{Y}_n\big\}\\
= &~ \mathbb{P}\big\{|\bxi|_{\infty} > {\rm cv}_{{\rm ind},\alpha}^{(\epsilon_0)}\big\} + o_{\rm p}(1) = \alpha + \epsilon_0 + o_{\rm p}(1)\,,
\end{align*}
which is a contradictory with probability approaching one as $n\to \infty$. Analogously, for any given $\varkappa_1 \ge 0$ and $\varkappa_2 \ge 0$, if $ \hat{\rm cv}_{{\rm ind},\alpha} \ge {\rm cv}_{{\rm ind},\alpha}^{(-\epsilon_0)}$,
by Proposition \ref{pro:1} again,
\begin{align*}
\alpha = \mathbb{P}(|\hat{\bxi}|_{\infty} > \hat{\rm cv}_{{\rm ind},\alpha}\,|\,\mathcal{X}_n, \mathcal{Y}_n) \le &~\mathbb{P}\big\{|\hat{\bxi}|_{\infty} > {\rm cv}_{{\rm ind},\alpha}^{(-\epsilon_0)} \,|\,\mathcal{X}_n, \mathcal{Y}_n\big\}\\
= &~ \mathbb{P}\big\{|\bxi|_{\infty} > {\rm cv}_{{\rm ind},\alpha}^{(-\epsilon_0)}\big\} + o_{\rm p}(1) = \alpha - \epsilon_0 + o_{\rm p}(1)\,,
\end{align*}
which is also a contradictory with probability approaching one as $n\to \infty$. Hence, we have $\mathbb{P}(\mathcal{E}_{\epsilon_0}) \to 1$ as $n \to \infty$.  Then, under the null hypothesis $\mathbb{H}_0$ in \eqref{eq:equind}, for any given constants $\varkappa_1 \ge 0$ and $\varkappa_2 \ge 0$, together with \eqref{eq:ind-h0-1},  it holds that
\begin{align*}
\mathbb{P} (H_{n}> \hat{{\rm cv}}_{{\rm ind},\alpha}) \le &~\mathbb{P} (H_{n}> \hat{{\rm cv}}_{{\rm ind},\alpha}, \mathcal{E}_{\epsilon_0}) +  \mathbb{P}(\mathcal{E}_{\epsilon_0}^{\rm c}) \le \mathbb{P} \big\{H_{n}> {\rm cv}_{{\rm ind},\alpha}^{(\epsilon_0)}\big\} + o(1)\\
=&~ \mathbb{P}\big\{|\bxi|_{\infty} >  {\rm cv}_{{\rm ind},\alpha}^{(\epsilon_0)}\big\} +o(1) =\alpha+\epsilon_0 +o(1)\,,
\end{align*}
which implies
$\varlimsup_{n\to \infty}\mathbb{P} (H_{n}> \hat{{\rm cv}}_{{\rm ind},\alpha}) \le \alpha + \epsilon_0$ under the null hypothesis $\mathbb{H}_0$ in \eqref{eq:equind}. On the other hand, under the null hypothesis $\mathbb{H}_0$ in \eqref{eq:equind}, for any given $\varkappa_1 \ge 0$ and $\varkappa_2 \ge 0$, by \eqref{eq:ind-h0-1} again,   
\begin{align*}
\mathbb{P} (H_{n}> \hat{{\rm cv}}_{{\rm ind},\alpha}) \ge &~\mathbb{P} (H_{n}> \hat{{\rm cv}}_{{\rm ind},\alpha}, \mathcal{E}_{\epsilon_0})   \ge \mathbb{P} \big\{H_{n}> {\rm cv}_{{\rm ind},\alpha}^{(-\epsilon_0)}\big\}-\mathbb{P}(\mathcal{E}_{\epsilon_0}^{\rm c}) \\
=&~ \mathbb{P}\big\{|\bxi|_{\infty} >  {\rm cv}_{{\rm ind},\alpha}^{(-\epsilon_0)}\big\}-o(1) =\alpha -\epsilon_0 -o(1)\,,
\end{align*}
which implies
$\varliminf_{n\to \infty}\mathbb{P} (H_{n}> \hat{{\rm cv}}_{{\rm ind},\alpha}) \ge \alpha - \epsilon_0$ under the null hypothesis $\mathbb{H}_0$ in \eqref{eq:equind}. Hence, 
\[
\alpha-\epsilon_0\leq \varliminf_{n\to \infty}\mathbb{P} (H_{n}> \hat{{\rm cv}}_{{\rm ind},\alpha})\leq \varlimsup_{n\to \infty}\mathbb{P} (H_{n}> \hat{{\rm cv}}_{{\rm ind},\alpha}) \le \alpha + \epsilon_0 
\]
under the null hypothesis $\mathbb{H}_0$ in \eqref{eq:equind}.  Since $\varliminf_{n\to \infty}\mathbb{P} (H_{n}> \hat{{\rm cv}}_{{\rm ind},\alpha})$ and $\varlimsup_{n\to \infty}\mathbb{P} (H_{n}> \hat{{\rm cv}}_{{\rm ind},\alpha})$ do not depend on $\epsilon_0$, by letting $\epsilon_0 \to 0$, we have $\lim_{n\to \infty}\mathbb{P} (H_{n}> \hat{{\rm cv}}_{{\rm ind},\alpha})=\alpha$ under the null hypothesis $\mathbb{H}_0$ in \eqref{eq:equind}.  We complete the proof of Theorem \ref{thm:1}.  $\hfill\Box$

\subsection{Proof of Theorem \ref{thm:2}}\label{sec:thm:2} 

To prove Theorem \ref{thm:2}, we need Lemma  \ref{lem:h1f}  whose proof is given in Appendix \ref{sec:sub-uivi-h1}.  

\begin{lemma}\label{lem:h1f}
It holds that
\begin{align*}
\max_{j\in[p],\,k\in[q]} \bigg|\frac{1}{n}\sum_{i=1}^{n}(\hat{U}_{i,j}\hat{V}_{i,k} -U_{i,j} V_{i,k} ) \bigg|  \le  &~\max_{j\in[p],\, k\in[q]} \bigg|\frac{\sqrt{2\pi}}{n}\sum_{s=1}^{n}\big\{ \tilde{\delta}_{1,k}(U_{s,j}) + \tilde{\delta}_{2,j}(V_{s,k})\big\}\bigg| \\
& +O_{\rm p} \{n^{-5/8}(\log n)^{-1/4}\log^{1/2} (dn)\} +O_{\rm p}\{n^{-3/5}(\log n)^{1/2}\}  
\end{align*}
provided that $ \log d \lesssim n^{1/4} (\log n)^{-3/2} $, where
\begin{align*}
\tilde{\delta}_{1,k}(U_{s,j})=&~\mathbb{E} \big[e^{U_{i,j}^2/2} \big\{I(U_{s,j}\le U_{i,j})-\Phi(U_{i,j})\big\}V_{i,k}^{*}I\{|U_{i,j}|\le \sqrt{(\log n)/2}\} \,\big|\,U_{s,j} \big]\,, \\
\tilde{\delta}_{2,j}(V_{s,k})=&~\mathbb{E} \big[e^{V_{i,k}^2/2} \big\{I(V_{s,k}\le V_{i,k})-\Phi(V_{i,k})\big\}U_{i,j}^{*}I{\{|V_{i,k}|\le  \sqrt{(\log n)/2}\}} \,\big|\,V_{s,k} \big] 
\end{align*}
with $i\neq s$, and   
\begin{align*}
\begin{split}
	U_{i,j}^{*} =&~ U_{i,j}I\{|U_{i,j}|\le  \sqrt{6(\log n)/5}\} + \sqrt{6(\log n)/5} \cdot {\rm sign}(U_{i,j})I\{|U_{i,j}|>\sqrt{6(\log n)/5}\}\,, \\
	V_{i,k}^{*} =&~ V_{i,k}I\{|V_{i,k}|\le \sqrt{6(\log n)/5}\} +   \sqrt{6(\log n)/5} \cdot {\rm sign}(V_{i,k}) I\{|V_{i,k}|>\sqrt{6(\log n)/5}\}\,. \\
\end{split}
\end{align*}
\end{lemma}

Recall that $\hat{\bxi}\,|\, \mathcal{X}_{n}, \mathcal{Y}_n \sim  \mathcal{N}(\boldsymbol{0},\hat{\bSigma} )$. Write  $\hat{\bSigma}=(\hat{\Sigma}_{i,j})_{d\times d}$. As shown in \cite{Borell1975}, for any $u >0$, it holds that
\begin{equation}\label{eq:Epa}
\begin{aligned}
\mathbb{P}\big\{|\hat{\bxi}|_{\infty} > \mathbb{E}(|\hat{\bxi}|_{\infty}\,|\,\mathcal{X}_{n}, \mathcal{Y}_n)+u \,|\,\mathcal{X}_{n}, \mathcal{Y}_n \big\} \le \exp\bigg(-\frac{u^2}{2\max_{j \in [d]}\hat{\Sigma}_{j,j} }\bigg)\,.
\end{aligned}
\end{equation} 
For any $v_1 >0$, consider the event
\begin{equation*}
\begin{aligned}
\mathcal{E}_1(v_1)=\left\{\max_{j \in [d]}\frac{|\hat{\Sigma}_{j,j}^{1/2} - \Sigma_{j,j}^{1/2}|}{ \Sigma_{j,j}^{1/2}} \le v_1\right\}\,.
\end{aligned}
\end{equation*}
Since $\min_{j \in [d]}\Sigma_{j,j} \ge c_1$, by Lemma \ref{lem:covh}, we have 
\begin{align*}
\max_{j \in [d]}\frac{|\hat{\Sigma}_{j,j}^{1/2} - \Sigma_{j,j}^{1/2}|}{ \Sigma_{j,j}^{1/2}} = O_{\rm p}\{n^{-1/2} (\log n) (\log d)^{1/2}\log^{3/2}(dn)\}
\end{align*}
provided that  $\log d \lesssim n^{1/3} $.  Notice that 
\begin{equation}\label{eq:Ea}
\begin{aligned}
\mathbb{E}(|\hat{\bxi}|_{\infty}\,|\,\mathcal{X}_{n}, \mathcal{Y}_n) 
\le &~ \{1+(2\log d)^{-1}\}(2\log d)^{1/2} \max_{j \in [d]} \hat{\Sigma}_{j,j}^{1/2} \,.
\end{aligned}
\end{equation} 
Due to  $\hat{\rm cv}_{{\rm ind},\alpha}  =\inf\{t\in \mathbb{R}: \mathbb{P}(|\hat{\bxi}|_{\infty} > t\,|\,\mathcal{X}_n, \mathcal{Y}_n) \le \alpha\}$, by \eqref{eq:Epa} and \eqref{eq:Ea},
we have 
\begin{align}\label{eq:cvhat}
\hat{{\rm cv}}_{{\rm ind},\alpha} \le&~ \mathbb{E}(|\hat{\bxi}|_{\infty}\,|\,\mathcal{X}_{n}, \mathcal{Y}_n)+\{2\log(1/\alpha)\}^{1/2}\max_{j \in [d]}\hat{\Sigma}_{j,j}^{1/2}  \notag\\
\le&~ (1+v_1)\big[\{1+(2\log d)^{-1}\}(2\log d)^{1/2}+ \{2\log(1/\alpha)\}^{1/2}\big]\max_{j \in [d]}\Sigma_{j,j}^{1/2}
\end{align}  
restricted on $\mathcal{E}_1(v_1)$.
With selecting $v_1 = (1+2\log d)^{-1}$,  by Lemma \ref{lem:covh},  we have $\mathbb{P}\{\mathcal{E}_1^{\rm c}(v_1)\} \to 0$ provided that $\log d \ll n^{1/6}(\log n)^{-1/3}$,  and  $\hat{{\rm cv}}_{{\rm ind},\alpha} \le \{1+(\log d)^{-1}\}\lambda(d, \alpha) \max_{j \in [d]}\Sigma_{j,j}^{1/2}$ restricted on $\mathcal{E}_1(v_1)$, where $\lambda(d, \alpha)= (2\log d)^{1/2}+ \{2\log(1/\alpha)\}^{1/2}$.

Write $\boldsymbol{\mu}=\mathbb{E}(\bU_{i} \otimes \bV_{i})=(\mu_1,\ldots,\mu_d)^{\T}$. We sort $\{|\mu_l|\}_{l=1}^d$ in the decreasing order as $|\mu_{l_{1}^{*}}| \ge \cdots \ge |\mu_{l_{d}^{*}}|$.  Without loss of generality, we assume $\mu_{l_{1}^{*}}>0$.
Let $g$ be a bijective mapping from $\{(j,k): j\in[p],k \in[q]\}$ to $[d]$, such that $g(j,k)=l$.  There exist $j^{*}\in [p]$ and ${k}^{*} \in [q]$ such that $g(j^{*}, k^{*})=l_1^{*}$.
For any $v_2 >0$, consider the event
\begin{align*}	 
&~~~\mathcal{E}_2(v_2)=\bigg\{\max_{j \in [p],\,k\in[q]}\bigg|\frac{1}{\sqrt{n}}\sum_{i=1}^{n} (\hat{U}_{i,j}\hat{V}_{i,k} -U_{i,j} V_{i,k} )\bigg|\le v_2\bigg\}\,.
\end{align*} 
Recall $H_{n}=\sqrt{n}|\hat{\bS}_n|_{\infty}$ and  $\hat{\bS}_n= n^{-1}\sum_{i=1}^{n}\hat{\bgamma}_i$  with $\hat{\bgamma}_i = \hat{\bU}_i \otimes \hat{\bV}_i$. Write $\hat{\bgamma}_i=(\hat{\gamma}_{i,1},\ldots, \hat{\gamma}_{i,d})^{\T}$. 
Therefore, under the alternative hypothesis $\mathbb{H}_1$ in \eqref{eq:equind}, with selecting $v_1 = (1+2\log d)^{-1}$, it holds that 
\begin{align}\label{eq:hnd-1}
&\mathbb{P} (H_{n}> \hat{{\rm cv}}_{{\rm ind},\alpha}) \notag\\
&~~\ge \mathbb{P} \bigg(\frac{1}{\sqrt{n}}\sum_{i=1}^{n}\hat{\gamma}_{i,l_{1}^{*}} > \hat{{\rm cv}}_{{\rm ind},\alpha}\bigg)\notag\\
&~~\ge \mathbb{P} \bigg[\frac{1}{\sqrt{n}}\sum_{i=1}^{n}\{U_{i,j^{*}} V_{i,k^{*}} - \mathbb{E}(U_{i,j^{*}} V_{i,k^{*}}) + (\hat{U}_{i,j^{*}}\hat{V}_{i,k^{*}} -U_{i,j^{*}} V_{i,k^{*}} )\} + \sqrt{n}\mu_{l_1^{*}} > \hat{{\rm cv}}_{{\rm ind},\alpha},\, \mathcal{E}_2(v_2)\bigg]\notag\\
&~~\ge \mathbb{P} \bigg[\frac{1}{\sqrt{n}}\sum_{i=1}^{n}\{U_{i,j^{*}} V_{i,k^{*}} - \mathbb{E}(U_{i,j^{*}} V_{i,k^{*}})\}   > - \sqrt{n}\mu_{l_1^{*}} + \hat{{\rm cv}}_{{\rm ind},\alpha} + v_2,\, \mathcal{E}_2(v_2)\bigg] \notag \\
&~~\ge\mathbb{P} \bigg[ \frac{1}{\sqrt{n}}\sum_{i=1}^{n}\{U_{i,j^{*}} V_{i,k^{*}} - \mathbb{E}(U_{i,j^{*}} V_{i,k^{*}})\} \\
&~~~~~~~~~~~~~~~~~~~~ > - \sqrt{n}\mu_{l_1^{*}} + \{1+(\log d)^{-1}\}\lambda(d, \alpha) \max_{j \in [d]}\Sigma_{j,j}^{1/2} +v_2 ,\,\mathcal{E}_1(v_1),\mathcal{E}_2(v_2)\bigg]\notag\\
&~~\ge 1- \mathbb{P} \bigg[\frac{1}{\sqrt{n}}\sum_{i=1}^{n}\{U_{i,j^{*}} V_{i,k^{*}} - \mathbb{E}(U_{i,j^{*}} V_{i,k^{*}})\} \notag\\
&~~~~~~~~~~~~~~~~~~~~~~~~~ \le - \sqrt{n}\mu_{l_{1}^{*}}  + \{1+(\log d)^{-1}\}\lambda(d, \alpha) \max_{j \in [d]}\Sigma_{j,j}^{1/2} +v_2 \bigg]  -o(1)- \mathbb{P}\{\mathcal{E}_2^{\rm c}(v_2)\} \notag
\end{align}
provided that $\log d \ll n^{1/6}(\log n)^{-1/3}$. 
Recall $U_{i,j}, U_{s,j} \sim \mathcal{N}(0,1)$ are independent for any $s\ne i$. For $\tilde{\delta}_{1,k}(U_{s,j})$ and $\tilde{\delta}_{2,j}(V_{s,k})$ defined in Lemma \ref{lem:h1f}, it holds that $\mathbb{E}\{\tilde{\delta}_{1,k}(U_{s,j})\}=0 $, $\mathbb{E}\{\tilde{\delta}_{2,j}(V_{s,k})\}=0 $, $|\tilde{\delta}_{1,k}(U_{s,j})| \le \sqrt{6/(5\pi)}\log n$ and $|\tilde{\delta}_{2,j}(V_{s,k})| \le  \sqrt{6/(5\pi)}\log n$.
By Bonferroni inequality and  Hoeffding's inequality,  it holds that
\begin{align}\label{eq:d1-d2-tail}
\mathbb{P}\bigg[\max_{j\in[p],\,k\in[q]}\bigg|\frac{\sqrt{2\pi}}{n}\sum_{s=1}^{n} \big\{\tilde{\delta}_{1,k}(U_{s,j}) + \tilde{\delta}_{2,j}(V_{s,k}) \big\}\bigg| > x\bigg]  \le 2d\exp\bigg\{-\frac{5nx^2}{96(\log n)^2}\bigg\}
\end{align} 
for any $x>0$. By Lemma \ref{lem:h1f},  we have
\begin{align*}
\max_{j\in[p],\,k\in[q]} \bigg|\frac{1}{\sqrt{n}}\sum_{i=1}^{n} (\hat{U}_{i,j}\hat{V}_{i,k} -U_{i,j} V_{i,k}) \bigg| 
\le &\, \max_{j\in[p],\, k\in[q]} \bigg|\frac{\sqrt{2\pi }}{\sqrt{n}}\sum_{s=1}^{n}\big\{ \tilde{\delta}_{1,k}(U_{s,j}) + \tilde{\delta}_{2,j}(V_{s,k})\big\}\bigg|\\
&+O_{\rm p} \{n^{-1/8}(\log n)^{-1/4}\log^{1/2} (dn)\} +O_{\rm p}\{n^{-1/10}(\log n)^{1/2}\}
\end{align*}
provided that $ \log d \lesssim n^{1/4} (\log n)^{-3/2} $. Recall $\nu_n \ge c_2$ for some universal constant $c_2>0$. Selecting $v_2 = 4\sqrt{6}  (1+\nu_n/2)(\log d)^{1/2}(\log n) /\sqrt{5}$, by  \eqref{eq:d1-d2-tail}, we have 
\begin{align}\label{eq:hnd-2}
\mathbb{P}\{\mathcal{E}_2^{\rm c}(v_2)\}  \le&~  \mathbb{P}\bigg[\max_{j\in[p],\,k\in[q]}\bigg|\frac{\sqrt{2\pi}}{\sqrt{n}}\sum_{i=1}^{n} \big\{\tilde{\delta}_{1,k}(U_{i,j}) + \tilde{\delta}_{2,j}(V_{i,k}) \big\}\bigg| > \frac{4\sqrt{6}}{\sqrt{5}}\bigg(1+\frac{\nu_n}{4}\bigg)(\log d)^{1/2}\log n\bigg]\notag \\
&+ \mathbb{P}\bigg[  O_{{\rm p}}\{n^{-1/8}(\log n)^{-1/4}\log^{1/2}  (dn)\}    >  \frac{\sqrt{6}\nu_n}{2\sqrt{5}}  (\log d)^{1/2} \log n \bigg] \notag \\
&+ \mathbb{P}\bigg[  O_{{\rm p}}\{n^{-1/10}(\log n)^{1/2}\}     >  \frac{\sqrt{6}\nu_n}{2\sqrt{5}}  (\log d)^{1/2} \log n \bigg] \notag \\
\le&~  2d^{-\nu_n/2-\nu_n^2/16} + o(1)  
\end{align}
provided that $ \log d \lesssim n^{1/4} (\log n)^{-3/2} $. Notice that $\lambda(d, \alpha) \ll (\log d )^{1/2} \log n$. Due to  $\mu_{l_{1}^{*}}  \ge  4\sqrt{6} (1+\nu_n)n^{-1/2}(\log d)^{1/2}(\log n)/\sqrt{5}$ under the alternative hypothesis $\mathbb{H}_1$ in \eqref{eq:equind}, we have
\begin{align*}
\sqrt{n}\mu_{l_{1}^{*}}  - \{1+(\log d)^{-1}\}\lambda(d, \alpha) \max_{j \in [d]}\Sigma_{j,j}^{1/2} -v_2 \ge&~ 2\sqrt{6} \nu_n  (\log d)^{1/2} (\log n)/\sqrt{5} - C\lambda(d, \alpha)\\
\ge&~  \nu_n (\log d)^{1/2} (\log n) 
\end{align*}  
for sufficiently large $n$. Since $\min_{j \in [d]}\Sigma_{j,j} \ge c_1$, we have $c_1 \le \var(U_{i,j^{*}}V_{i,k^{*}}) \le 3$. It follows from the Central Limit Theorem that  $n^{-1/2} \sum_{i=1}^{n}\{U_{i,j^{*}} V_{i,k^{*}} - \mathbb{E}(U_{i,j^{*}} V_{i,k^{*}})\}\{\var(U_{i,j^{*}}V_{i,k^{*}})\}^{-1/2}\to \mathcal{N}(0,1)$ in distribution. Then, due to $\nu_n (\log d)^{1/2}\log n \to \infty$,  under the alternative hypothesis $\mathbb{H}_1$ in \eqref{eq:equind}, for any sufficiently large $n$,
\begin{align*}
&\mathbb{P} \bigg[\frac{1}{\sqrt{n}}\sum_{i=1}^{n}\{U_{i,j^{*}} V_{i,k^{*}} - \mathbb{E}(U_{i,j^{*}} V_{i,k^{*}})\} \le - \sqrt{n}\mu_{l_{1}^{*}}  + \{1+(\log d)^{-1}\}\lambda(d, \alpha) \max_{j \in [d]}\Sigma_{j,j}^{1/2} +v_2 \bigg]\\
&~~~~~~~~\le \mathbb{P}\bigg[\frac{1}{\sqrt{n}}\sum_{i=1}^{n}\{U_{i,j^{*}} V_{i,k^{*}} - \mathbb{E}(U_{i,j^{*}} V_{i,k^{*}})\} \le   -  \nu_n (\log d)^{1/2}\log n  \bigg] \\
&~~~~~~~~= \mathbb{P}\bigg\{\frac{1}{\sqrt{n}}\sum_{i=1}^{n}\frac{U_{i,j^{*}} V_{i,k^{*}} - \mathbb{E}(U_{i,j^{*}} V_{i,k^{*}})} {\sqrt{\var(U_{i,j^{*}}V_{i,k^{*}})}} \le   -  \frac{ \nu_n (\log d)^{1/2}\log n}{\sqrt{\var(U_{i,j^{*}}V_{i,k^{*}})} } \bigg\} \to 0\,.
\end{align*} 
Together with \eqref{eq:hnd-1} and \eqref{eq:hnd-2}, under the alternative hypothesis $\mathbb{H}_1$ in \eqref{eq:equind}, it holds that
\begin{align*}
\mathbb{P} (H_{n}> \hat{{\rm cv}}_{{\rm ind},\alpha}) \ge 1- 2d^{-\nu_n/2-\nu_n^2/16}- o(1) 
\end{align*}
provided that  $\log d \ll n^{1/6}(\log n)^{-1/3}$.  Since $d=pq$ with $p \lesssim n^{\varkappa_1}$ and $q\lesssim n^{\varkappa_2}$, the restriction $\log d \ll n^{1/6}(\log n)^{-1/3}$ holds automatically.   We complete the proof of Theorem \ref{thm:2}.
$\hfill\Box$

\section{Proofs of  Theorems \ref{thm:nn-sta-h0} and \ref{thm:nn-sta-h1}}\label{sec:pro:nn-sta-h0}   
Recall $\tilde{d}=p\vee q\vee m$, $\bTheta =  (\Theta_{i,j})_{d\times d}$ and  $\tilde{\bTheta}=n_3^{-1}\sum_{i\in\mathcal{D}_3}\tilde{\bet}_i\tilde{\bet}_{i}^{\T}-\bar{\tilde{\bet}} \bar{\tilde{\bet}}^{\T} $ with $\bar{\tilde{\bet}} = n_{3}^{-1}\sum_{i\in\mathcal{D}_3}\tilde{\bet}_i$. 
To prove Theorem \ref{thm:nn-sta-h0}, we need Proposition \ref{pro:nn-sta-h0}  with its proof given in Section \ref{sec:sub-nn-sta-h0}.

\begin{proposition}\label{pro:nn-sta-h0}
Let  $\tilde{\bzeta}  \,|\, \mathcal{X}_{n}, \mathcal{Y}_{n},\mathcal{Z}_{n} \sim \mathcal{N}(\bzero, \tilde{\bTheta})$.  Select $(\tilde{\alpha}_{n}, M_{*})$ specified in \eqref{eq:nnspace-hl} as $\tilde{\alpha}_{n} =n^{c_{3}}$ and  $M_{*}= c_{4} \lceil n^{m_{*}/ (4\vartheta + m_{*})} (m^2\log n)^{m_{*}(2\tilde{\vartheta}+3)/(2\vartheta)}\rceil$ for some sufficiently large constants $c_{3}>0$ and $c_{4}>0$. Under Condition  {\rm\ref{cd:function-condition}} and the null hypothesis $\mathbb{H}_{0}$ in \eqref{eq:equiconind}, if $\min_{j\in[d]} \Theta_{j,j} \ge c_5$ for some universal constant $c_5>0$, and 
\begin{align}\label{eq:pro-32-cd-n}
& \log \tilde{d} \ll \min \big\{ n^{\vartheta/(4\vartheta+m_{*})-\kappa/4} (\log n)^{-1-\varrho/(8\vartheta)} ,\,  n^{2\kappa/15}(\log n)^{-14/15}\,, \notag\\
&~~~~~~~~~~~~~~~~~~n^{1/16}(\log n)^{-7/8},\, n^{4\vartheta/(68\vartheta+17m_{*})}(\log n)^{-16/17-\varrho/(34\vartheta)}  \big\} \,, \notag\\
&m \ll \min\big[n^{\vartheta\{4\vartheta/(4\vartheta+m_{*})-\kappa\}/\varrho } (\log n )^{-4\vartheta/\varrho -1/2} \{\log(\tilde{d}n)\}^{-4\vartheta/\varrho} \,, \notag\\
&~~~~~~~~~~~~~~~n^{(1-\kappa)/2}(\log n)^{-1} \{\log(\tilde{d}n)\}^{-3/2}\,,\, n^{\kappa/4}(\log n)^{-1} \{\log(\tilde{d}n)\}^{-5/4} \,,\\
&~~~~~~~~~~~~~~~n^{4\vartheta^2/\{\varrho(4\vartheta+m_{*})\}} (\log n)^{-16\vartheta/\varrho-1/2}\{\log (\tilde{d}n)\}^{-17\vartheta/\varrho}\,,\notag \\
&~~~~~~~~~~~~~~~ n^{\kappa/2}(\log n)^{-7/2}\{\log (\tilde{d}n )\}^{-15/4},\, n^{1/2}(\log n)^{-7}\{\log (\tilde{d}n)\}^{-8}\big]   \notag 
\end{align} 
with $\varrho =    \vartheta+2m_{*}\tilde{\vartheta}+3m_{*} $, then
\begin{align*} 
\sup_{z>0}\big|\mathbb{P} (\tilde{G}_{n}>z) - \mathbb{P}(|\tilde{\bzeta}|_{\infty} > z\,|\,\mathcal{X}_{n},\mathcal{Y}_n,\mathcal{Z}_n)\big| =o_{\rm p}(1) 
\end{align*}
as $n\rightarrow \infty$. 
\end{proposition}



\subsection{Proof of Proposition \ref{pro:nn-sta-h0}}\label{sec:sub-nn-sta-h0}
Recall $\tilde{d}=p\vee q\vee m$. 
To prove Proposition  \ref{pro:nn-sta-h0}, we need Lemmas \ref{lem:nn-eh-e-delta}--\ref{lem:nn-cov-theta} with their proofs given in Appendices \ref{sec:nn-eh-e-delta}--\ref{sec:nn-cov-theta}, respectively.

\begin{lemma}\label{lem:nn-eh-e-delta}
Let $\hat{f}_{j} $ and $\hat{g}_{k} $ be the estimates specified in \eqref{eq:fhj-ghk}  with $(m_{*},K)$ as in the definitions of $f_j$ and $g_{k}$,   $\tilde{\alpha}_{n} =n^{c_{3}} $ and $M_{*}= c_{4} \lceil n^{m_{*}/ (4\vartheta + m_{*})}(m^2 \log n)^{m_{*}(2\tilde{\vartheta}+3)/(2\vartheta)}\rceil$ for some sufficiently large constants $c_{3}>0$ and $c_{4}>0$.   Under Condition  {\rm\ref{cd:function-condition}},
it  holds that
\begin{align*}
&\max_{j\in[p],\,k\in[q]}\bigg|\frac{1}{n_{3}} \sum_{t\in \mathcal{D}_3} ( \tilde{\varepsilon}_{t,j} - \varepsilon_{t,j})\delta_{t,k}\bigg| \\
&~~~~~=   O_{\rm p}\{n^{-\kappa/2-\vartheta/(4\vartheta+m_{*})}  (m^2\log n)^{(\vartheta+2m_{*}\tilde{\vartheta}+3m_{*} )/(8\vartheta)} (\log n)\log^{7/4}(\tilde{d}n) \}+ O_{\rm p}\{ n^{-1/2} m \log(\tilde{d}n)\}\\ 
&~~~~~~~~ +O_{\rm p}\{n^{-\kappa/2-1/4} m^{1/2}  (\log n)^{1/2}\log^{3/2}  (\tilde{d}n) \} + O_{\rm p}\{n^{-\kappa}m^2 (\log n)\log^2(\tilde{d}n) \} \\
&~~~~~= \max_{j\in[p],\,k\in[q]}\bigg|\frac{1}{n_{3}} \sum_{t\in \mathcal{D}_3} ( \tilde{\delta}_{t,k} - \delta_{t,k})\varepsilon_{t,j}\bigg|
\end{align*}
provided that  $ \log (\tilde{d}n) \ll  n^{1-\kappa} (\log n)^{-1/2}$ and $m \lesssim n$.
\end{lemma}

\begin{lemma}\label{lem:nn-epsdts-epdt}
Under the conditions of Lemma {\rm\ref{lem:nn-eh-e-delta}},  it holds that
\begin{align*}
&\max_{j\in[p],\, k\in[q]}\bigg|\frac{1}{n_{3}} \sum_{t\in \mathcal{D}_3} ( \tilde{\varepsilon}_{t,j} - \varepsilon_{t,j})( \tilde{\delta}_{t,k}-\delta_{t,k})\bigg| \\
&~~~~~~ = O_{\rm p}\{n^{-2\vartheta/(4\vartheta+m_{*})} (m^2\log n)^{(\vartheta+2m_{*}\tilde{\vartheta}+3m_{*} )/(4\vartheta)} (\log n)^{2}\log^{3/2} (\tilde{d}n) \} \notag\\
&~~~~~~~~~ + O_{\rm p}\{n^{-\kappa/2-\vartheta/(4\vartheta+m_{*})} (m^2\log n)^{(\vartheta+2m_{*}\tilde{\vartheta}+3m_{*} )/(8\vartheta)} (\log n)^{2}\log^{7/4} (\tilde{d}n) \} \notag\\
&~~~~~~~~~ +O_{\rm p}\{n^{-1/2}m (\log n) \log  (\tilde{d}n)\}  +  O_{\rm p}\{n^{-\kappa} m^2   (\log n)^2\log^2 (\tilde{d}n) \} \notag\\
&~~~~~~~~~ + O_{\rm p}\{n^{-\kappa/2-1/4}m^{1/2} (\log n)^{3/2} \log^{3/2} (\tilde{d}n)\}
\end{align*}
provided that $\log(\tilde{d}n) \ll n^{1-\kappa}(\log n)^{-1/2} $ and $m \lesssim n$. 

\end{lemma}

\begin{lemma}\label{lem:nn-cov-theta}
Under the conditions of Lemma {\rm \ref{lem:nn-eh-e-delta}},  it holds that
\begin{align*} 
|\tilde{\bTheta}-\bTheta|_{\infty} =&~ O_{\rm p}\{n^{-\vartheta/(4\vartheta+m_{*})} (m^2\log n)^{(\vartheta+2m_{*}\tilde{\vartheta}+3m_{*})/(8\vartheta)} (\log n)^4\log^{9/4}(\tilde{d}n) \} \notag \\
& +  O_{\rm p}\{n^{-1/4}  m^{1/2}  (\log n)^{7/2}\log^{2} (\tilde{d}n) \}  + O_{\rm p}\{n^{-\kappa/2}m (\log n)^{7/2}\log^{7/4}  (\tilde{d}n) \}
\end{align*}
provided that  $m \ll \min [n^{4\vartheta^2/\{\varrho(4\vartheta+m_{*})\}} (\log n)^{-4\vartheta/\varrho-1/2}\{\log (\tilde{d}n)\}^{-3\vartheta/\varrho}, n^{\kappa/2}(\log n)^{-1}\{\log (\tilde{d}n )\}^{-1} ]$  and $\log(\tilde{d}n) \ll \min\{n^{1-\kappa} (\log n)^{-1/2},n^{\kappa/3}, n^{4\vartheta/(12\vartheta+3m_{*})}(\log n)^{-4/3-\varrho/(6\vartheta)} \}$ with $\varrho =    \vartheta+2m_{*}\tilde{\vartheta}+3m_{*} $. 
\end{lemma}

Let $\breve{\bOmega}_{n} = n_{3}^{-1}\sum_{i\in\mathcal{D}_3}\bet_i$ with $\bet_i =\boldsymbol{\varepsilon}_i \otimes \bxi_i$. Recall  $\bzeta \sim \mathcal{N}(\boldsymbol{0}, \bTheta)$ with $\bTheta=\cov(\bet_i)$,  $\tilde{\bOmega}_{n} = n_{3}^{-1}\sum_{i\in\mathcal{D}_3} \tilde{\bet}_{i}$ with $\tilde{\bet}_{i} = \tilde{\boldsymbol{\ve}}_{i} \otimes \tilde{\bdelta}_{i}$, and $\tilde{G}_n = \sqrt{n_3}|\tilde{\bOmega}_{n}|_{\infty}$. Recall $d=pq$. Using the similar arguments for the derivation of \eqref{eq:GAbound_a}, it holds that 
\begin{align*} 
\sup_{x>0}\big|\mathbb{P}(\tilde{G}_{n}>x) - \mathbb{P}(|\bzeta|_{\infty} > x)\big| \lesssim &~  \sup_{x>0}\big|\mathbb{P}(\sqrt{n_{3}}|\breve{\bOmega}_{n}|_{\infty}>x)-\mathbb{P}(|\bzeta |_{\infty} > x) \big| \notag\\
&   + \mathbb{P}(\sqrt{n_{3}}|\tilde{\bOmega}_{n} - \breve{\bOmega}_{n}|_{\infty} > u) + u(\log d)^{1/2} 
\end{align*}
for any $u>0$. Notice that $\tilde{\varepsilon}_{t,j}\tilde{\delta}_{t,k} - \varepsilon_{t,j}\delta_{t,k}= (\tilde{\varepsilon}_{t,j} - \varepsilon_{t,j})\delta_{t,k}  +  (\tilde{\delta}_{t,k} - \delta_{t,k})\varepsilon_{t,j} +(\tilde{\varepsilon}_{t,j} - \varepsilon_{t,j})(\tilde{\delta}_{t,k}-\delta_{t,k})$. Recall $n_3 \asymp n^{\kappa}$ for some constant $0<\kappa<1$. By Lemmas \ref{lem:nn-eh-e-delta} and \ref{lem:nn-epsdts-epdt}, we have
\begin{align*}
\sqrt{n_{3}}|\tilde{\bOmega}_{n} - \breve{\bOmega}_{n}|_{\infty} =&~ O_{\rm p}\{n^{\kappa/2-2\vartheta/(4\vartheta+m_{*})} (m^2\log n)^{(\vartheta+2m_{*}\tilde{\vartheta}+3m_{*} )/(4\vartheta)} (\log n)^{2}\log^{3/2} (\tilde{d}n) \} \notag\\
& + O_{\rm p}\{n^{-\vartheta/(4\vartheta+m_{*})} (m^2\log n)^{(\vartheta+2m_{*}\tilde{\vartheta}+3m_{*} )/(8\vartheta)} (\log n)^{2}\log^{7/4} (\tilde{d}n) \} \notag\\
& +O_{\rm p}\{n^{\kappa/2-1/2}m (\log n) \log  (\tilde{d}n)\}  +  O_{\rm p}\{n^{-\kappa/2} m^2   (\log n)^2\log^2 (\tilde{d}n) \} \notag\\
& + O_{\rm p}\{n^{-1/4}m^{1/2} (\log n)^{3/2} \log^{3/2} (\tilde{d}n)\}
\end{align*} 
provided that $\log(\tilde{d}n) \ll n^{1-\kappa}(\log n)^{-1/2} $ and $m \lesssim n$. Recall $|f_j|_{\infty} \le \tilde{C}$, it holds that
\begin{align*}
\mathbb{P}(|\ve_{i,j}|>x) = &~\mathbb{P}\{|U_{i,j} - f_j(\bW_i)|>x\}  \leq \mathbb{P}\bigg(|U_{i,j}| >\frac{x}{2}\bigg) + \mathbb{P}\bigg\{|f_j(\bW_i)|>\frac{x}{2}\bigg\} \notag\\
\leq&~ 2e^{-x^2/4} + C_1e^{-x^2/4} \le C_2 e^{- x^2/4}
\end{align*}
for any $x>0$, $i \in [n]$ and $j \in [p]$. Analogously, we also have $\mathbb{P}(|\delta_{i,k}|>x) \le C_2 e^{- x^2/4}$ for any $x>0$, $i \in [n]$ and $k \in [q]$. Recall $d=pq$ and $\tilde{d} =p\vee q\vee m$.  Parallel to the proof of Proposition \ref{pro:1}, to ensure  $\sup_{x>0}|\mathbb{P}(\tilde{G}_{n}>x) - \mathbb{P}(|\bzeta|_{\infty} > x)|=o(1)$ under the null hypothesis  $\mathbb{H}_0$ in \eqref{eq:equiconind}, we know $(\tilde{d},m,n)$ should satisfy 
\begin{align*}
\left\{
\begin{aligned}
&n^{\kappa/2-2\vartheta/(4\vartheta+m_{*})} (m^2\log n)^{(\vartheta+2m_{*}\tilde{\vartheta}+3m_{*} )/(4\vartheta)} (\log n)^{2}\log^{3/2} (\tilde{d}n)  \ll (\log \tilde{d})^{-1/2}\,,\\
&n^{-\vartheta/(4\vartheta+m_{*})} (m^2\log n)^{(\vartheta+2m_{*}\tilde{\vartheta}+3m_{*} )/(8\vartheta)} (\log n)^{2}\log^{7/4} (\tilde{d}n) \ll (\log \tilde{d})^{-1/2}\,,\\
&n^{\kappa/2-1/2}m (\log n) \log  (\tilde{d}n) \ll (\log \tilde{d})^{-1/2} \,,\\
& n^{-\kappa/2} m^2   (\log n)^2\log^2 (\tilde{d}n) \ll (\log \tilde{d})^{-1/2} \,,\\
&n^{-1/4}m^{1/2} (\log n)^{3/2} \log^{3/2} (\tilde{d}n) \ll (\log \tilde{d})^{-1/2} \,,\\
&\log(\tilde{d}n) \ll n^{1-\kappa}(\log n)^{-1/2} \,\\
&\log (\tilde{d}n) \ll n^{\kappa/7} \,,\\
& m \lesssim n \,,
\end{aligned}
\right.
\end{align*}
which implies  
\begin{align}\label{eq:m-res}
&~~~~~~~~~~~~~ \log \tilde{d} \ll \min\{  n^{\kappa/7} ,\,  n^{\vartheta/(4\vartheta+m_{*}) -\kappa/4} (\log n)^{-1-\varrho/(8\vartheta)}\}\,, \notag\\
&m \ll \min\big[n^{\vartheta\{4\vartheta/(4\vartheta+m_{*})-\kappa\}/\varrho } (\log n )^{-4\vartheta/\varrho -1/2} \{\log(\tilde{d}n)\}^{-4\vartheta/\varrho} \,, \notag\\
&~~~~~~~~~~~~~~~n^{(1-\kappa)/2}(\log n)^{-1} \{\log(\tilde{d}n)\}^{-3/2}\,,\, n^{\kappa/4}(\log n)^{-1} \{\log(\tilde{d}n)\}^{-5/4}  \big] 
\end{align}
with $\varrho =    \vartheta+2m_{*}\tilde{\vartheta}+3m_{*} $.

Parallel to the arguments for the proof of Proposition \ref{pro:1},  under the null hypothesis  $\mathbb{H}_0$ in \eqref{eq:equiconind},
\begin{align}\label{eq:tilde-g-tilde-zeta}
&\sup_{x >0}\big|\mathbb{P}(\tilde{G}_{n}>x)- \mathbb{P}(|\tilde{\bzeta}|_{\infty} >x \,|\,\mathcal{X}_n, \mathcal{Y}_n,\mathcal{Z}_n)\big|\notag \notag \\
&~~~~~~~~~~\le  \sup_{x >0}\big|\mathbb{P}(|\bzeta|_{\infty} >x) - \mathbb{P}(|\tilde{\bzeta}|_{\infty} >x \,|\,\mathcal{X}_n, \mathcal{Y}_n,\mathcal{Z}_n)\big| + o(1) 
\end{align}
provided that \eqref{eq:m-res} holds. Write $\tilde{\Delta}_{n2} = |\tilde{\bTheta} -\bTheta |_{\infty}$. Recall $d=pq$, $\tilde{d} =p \vee q \vee m$ and $\varrho =    \vartheta+2m_{*}\tilde{\vartheta}+3m_{*} $.  Using the similar arguments for derivation of \eqref{eq:Prop2bound}, by Lemma \ref{lem:nn-cov-theta}, we have 
\begin{align*}
\sup_{x>0}\big|\mathbb{P}(|\bzeta|_{\infty} > x)-\mathbb{P}(|\tilde{\bzeta}|_{\infty} > x \,|\,\mathcal{X}_{n}, \mathcal{Y}_n,\mathcal{Z}_{n})\big|  
\lesssim \tilde{\Delta}_{n2}^{1/3} \{1\vee \log(2d\tilde{\Delta}_{n2}^{-1})\}^{2/3} =o_{{\rm p}}(1) 
\end{align*}
provided that   
\begin{align*}
& \log \tilde{d} \ll \min \big\{ n^{1-\kappa} (\log n)^{-1/2} ,\, n^{1/16}(\log n)^{-7/8},\, n^{2\kappa/15}(\log n)^{-14/15}\,,\\
&~~~~~~~~~~~~~~~~~~~n^{4\vartheta/(68\vartheta+17m_{*})}(\log n)^{-16/17-\varrho/(34\vartheta)}  \big\}\,,  \\
& m\ll  \min \big[n^{4\vartheta^2/\{\varrho(4\vartheta+m_{*})\}} (\log n)^{-16\vartheta/\varrho-1/2}\{\log (\tilde{d}n)\}^{-17\vartheta/\varrho}\,, \\
&~~~~~~~~~~~~~~~ n^{\kappa/2}(\log n)^{-7/2}\{\log (\tilde{d}n )\}^{-15/4},\, n^{1/2}(\log n)^{-7}\{\log (\tilde{d}n)\}^{-8}\big] \,.   \notag
\end{align*}     
Together with \eqref{eq:tilde-g-tilde-zeta},  under the null hypothesis  $\mathbb{H}_0$ in \eqref{eq:equiconind},  we have
\begin{align*}
\sup_{x >0}\big|\mathbb{P}(\tilde{G}_{n}>x)- \mathbb{P}(|\tilde{\bzeta}|_{\infty} >x \,|\,\mathcal{X}_n, \mathcal{Y}_n,\mathcal{Z}_n)\big|\notag= o_{\rm p}(1) 
\end{align*}
provided that 
\begin{align}\label{eq:pro-32-cd}
& \log \tilde{d} \ll \min \big\{ n^{\vartheta/(4\vartheta+m_{*})-\kappa/4} (\log n)^{-1-\varrho/(8\vartheta)} ,\,  n^{2\kappa/15}(\log n)^{-14/15}\,, \notag\\
&~~~~~~~~~~~~~~~~~~~n^{1/16}(\log n)^{-7/8},\, n^{4\vartheta/(68\vartheta+17m_{*})}(\log n)^{-16/17-\varrho/(34\vartheta)}  \big\} \,, \notag\\
&m \ll \min\big[n^{\vartheta\{4\vartheta/(4\vartheta+m_{*})-\kappa\}/\varrho } (\log n )^{-4\vartheta/\varrho -1/2} \{\log(\tilde{d}n)\}^{-4\vartheta/\varrho} \,, \notag\\
&~~~~~~~~~~~~~~~n^{(1-\kappa)/2}(\log n)^{-1} \{\log(\tilde{d}n)\}^{-3/2}\,,\, n^{\kappa/4}(\log n)^{-1} \{\log(\tilde{d}n)\}^{-5/4} \,,\\
&~~~~~~~~~~~~~~~n^{4\vartheta^2/\{\varrho(4\vartheta+m_{*})\}} (\log n)^{-16\vartheta/\varrho-1/2}\{\log (\tilde{d}n)\}^{-17\vartheta/\varrho}\,,\notag \\
&~~~~~~~~~~~~~~~ n^{\kappa/2}(\log n)^{-7/2}\{\log (\tilde{d}n )\}^{-15/4},\, n^{1/2}(\log n)^{-7}\{\log (\tilde{d}n)\}^{-8}\big] \,.  \notag 
\end{align}   
Hence, we complete the proof of Proposition \ref{pro:nn-sta-h0}.
$\hfill\Box$

\subsection{Proof of Theorem \ref{thm:nn-sta-h0}}

The proof of Theorem \ref{thm:nn-sta-h0} is almost identical to that of Theorem \ref{thm:1} given in  Section \ref{sec:sub-thm1}. Hence, we omit it here. $\hfill\Box$

\subsection{Proof of Theorem \ref{thm:nn-sta-h1}}\label{sec:thm:nn-sta-ha}
For any $v_3 >0$, consider the event
\begin{equation*}
\begin{aligned}
\mathcal{E}_3(v_3)=\left\{\max_{j \in [d]}\frac{|\tilde{\Theta}_{j,j}^{1/2} - \Theta_{j,j}^{1/2}|}{ \Theta_{j,j}^{1/2}} \le v_3 \right\}\,.
\end{aligned}
\end{equation*}
Due to  $\hat{\rm cv}_{{\rm cind},\alpha}  =\inf\{t\in \mathbb{R}: \mathbb{P}(|\tilde{\bzeta}|_{\infty} > t\,|\,\mathcal{X}_n, \mathcal{Y}_n, \mathcal{Z}_{n}) \le \alpha\}$, parallel to \eqref{eq:cvhat}, 
\begin{align*}
\hat{{\rm cv}}_{{\rm cind},\alpha} 
\le 
(1+v_3)\big[\{1+(2\log d)^{-1}\}(2\log d)^{1/2}+ \{2\log(1/\alpha)\}^{1/2}\big]\max_{j \in [d]}\Theta_{j,j}^{1/2}
\end{align*}
restricted on $\mathcal{E}_3(v_3)$.  Recall $d=pq$,   $\tilde{d}=p\vee q\vee m$ and  $\varrho =    \vartheta+2m_{*}\tilde{\vartheta}+3m_{*} $. With selecting $v_3 = (1+2\log d)^{-1}$,  by Lemma \ref{lem:nn-cov-theta},  we have $\mathbb{P}\{\mathcal{E}_3^{\rm c}(v_3)\} \to 0$ provided that 
\begin{align}\label{eq:e5-cd} 
&  \log \tilde{d} \ll \min \big\{ n^{1-\kappa} (\log n)^{-1/2} ,\, n^{1/12}(\log n)^{-7/6},\, n^{2\kappa/11}(\log n)^{-14/11} \,, \notag\\
&~~~~~~~~~~~~~~~~~~~  n^{4\vartheta/(52\vartheta+13m_{*})}(\log n)^{-16/13-\varrho/(26\vartheta)}  \big\}\,,  \\
& m\ll  \min \big[n^{4\vartheta^2/\{\varrho(4\vartheta+m_{*})\}} (\log n)^{-16\vartheta/\varrho-1/2}\{\log (\tilde{d}n)\}^{-13\vartheta/\varrho}\,, \notag\\
&~~~~~~~~~~~~~~~~ n^{1/2}(\log n)^{-7} \{\log (\tilde{d}n )\}^{-6},\, n^{\kappa/2}(\log n)^{-7/2}\{\log (\tilde{d}n )\}^{-11/4}\big] \,,  \notag
\end{align} 
and  $\hat{{\rm cv}}_{{\rm cind},\alpha}  \le \{1+(\log d)^{-1}\}\lambda(d, \alpha) \max_{j \in [d]}\Theta_{j,j}^{1/2}$ restricted on $\mathcal{E}_3(v_3)$, where $\lambda(d, \alpha)= (2\log d)^{1/2}+ \{2\log(1/\alpha)\}^{1/2}$.  Recall $\bOmega = \mathbb{E}(\boldsymbol{\ve}_{i} \otimes \bdelta_{i} ) =(\Omega_1,\ldots, \Omega_d)^{\T}$. We sort $\{|\Omega_{l}|\}_{l=1}^{d}$ in the decreasing order as $|\Omega_{l_{1}^{*}}| \ge \cdots \ge |\Omega_{l_{d}^{*}}|$.
Without loss of generality, we assume $\Omega_{l_{1}^{*}}>0$. Let $g$ be a bijective mapping from $\{(j,k): j\in[p],k \in[q]\}$ to $[d]$, such that $g(j,k)=l$. Then there exist $j^{*}\in [p]$ and ${k}^{*} \in [q]$ such that $g(j^{*}, k^{*})=l_1^{*}$.  For any $v_4>0$, consider the event 
\begin{align*}
\mathcal{E}_4(v_4) =\bigg\{ \max_{j\in[p],\,k\in[q]} \bigg|\frac{1}{\sqrt{n_{3}}}\sum_{i\in \mathcal{D}_3} (\tilde{\varepsilon}_{i,j} \tilde{\delta}_{i,k}   -\varepsilon_{i,j} \delta_{i,k}) \bigg| \le v_4\bigg\}\,.
\end{align*}
Recall $\tilde{G}_{n} = \sqrt{n_3} |\tilde{\bOmega}_{n} |_{\infty}$ and $\tilde{\bOmega}_{n}  = n_{3}^{-1}\sum_{i\in \mathcal{D}_3} \tilde{\bet}_{i}$ with $\tilde{\bet}_{i} = \tilde{\boldsymbol{\ve}}_{i}  \otimes \tilde{\bdelta}_{i} $. 
Parallel to \eqref{eq:hnd-1}, under the alternative hypothesis $\mathbb{H}_{1}$ in \eqref{eq:equiconind}, with selecting $v_3=(1+2\log d)^{-1}$, we have 
\begin{align*}
\mathbb{P} (\tilde{G}_{n}> \hat{{\rm cv}}_{{\rm cind},\alpha} ) \ge&~ 1-\mathbb{P} \bigg[\frac{1}{\sqrt{n_{3}}}\sum_{i \in \mathcal{D}_3}\{\varepsilon_{i,j^{*}} \delta_{i,k^{*}} - \mathbb{E}(\varepsilon_{i,j^{*}} \delta_{i,k^{*}})\} \notag\\ 
&~~~~~~~~~~~~~~~~~~~~~~~ \le -\sqrt{n_{3}}\Omega_{l_{1}^{*}} +\{1+(\log d)^{-1}\}\lambda(d,\alpha)\max_{j \in [d]}\Theta_{j,j}^{1/2}+v_4 \bigg]\notag \\
&~-o(1)
- \mathbb{P}\{\mathcal{E}_4^{\rm c}(v_4)\}  
\end{align*}
provided that \eqref{eq:e5-cd} holds.  By Lemmas \ref{lem:nn-eh-e-delta} and \ref{lem:nn-epsdts-epdt}, for some constant $v_4>0$, we have 
\begin{align*}
\mathbb{P}\{\mathcal{E}_4^{\rm c}(v_4)\}  \to 0
\end{align*}
provided that 
\begin{align*} 
&~  \log \tilde{d} \ll \min\{  n^{\kappa/4}(\log n)^{-1} ,\,  n^{4\vartheta/(12\vartheta+3m_{*}) -\kappa/3} (\log n)^{-4/3-\varrho/(6\vartheta)}\}\,, \notag\\
&m \ll \min\big[n^{\vartheta\{4\vartheta/(4\vartheta+m_{*})-\kappa\}/\varrho } (\log n )^{-4\vartheta/\varrho -1/2} \{\log(\tilde{d}n)\}^{-3\vartheta/\varrho} \,, \notag\\
&~~~~~~~~~~~~~~ n^{(1-\kappa)/2}(\log n)^{-1} \{\log(\tilde{d}n)\}^{-1}\,,\, n^{\kappa/4}(\log n)^{-1} \{\log(\tilde{d}n)\}^{-1}  \big] \,. 
\end{align*}
Recall $n_3\asymp n^{\kappa}$ for some constant $0<\kappa<1$. Since $\Omega_{l_{1}^{*}} \ge (1+\tilde{\epsilon}_{n})n^{-\kappa/2}\lambda(d,\alpha)\max_{j \in [d]}\Theta_{j,j}^{1/2}$ and $\tilde{\epsilon}_n^2\log d \to \infty$ as $n\to \infty$  under the alternative hypothesis  $\mathbb{H}_1$ in \eqref{eq:equiconind}, we have
\begin{align*}
\sqrt{n_{3}}\Omega_{l_{1}^{*}} -\{1+(\log d)^{-1}\}\lambda(d,\alpha)\max_{j \in [d]}\Theta_{j,j}^{1/2} \ge \{\tilde{\epsilon}_n-(\log d)^{-1}\} \lambda(d,\alpha)\max_{j \in [d]}\Theta_{j,j}^{1/2} \to \infty \,.
\end{align*}
Under the alternative hypothesis $\mathbb{H}_{1}$ in \eqref{eq:equiconind}, it holds that
\begin{align*} 
&\mathbb{P} \bigg[\frac{1}{\sqrt{n_{3}}}\sum_{i\in \mathcal{D}_3}\{\varepsilon_{i,j^{*}} \delta_{i,k^{*}} - \mathbb{E}(\varepsilon_{i,j^{*}} \delta_{i,k^{*}}) \} \\
&~~~~~~~~~~~~~~\le -\sqrt{n_{3}}\Omega_{l_{1}^{*}} +\{1+(\log d)^{-1}\}\lambda(d,\alpha)\max_{j \in [d]}\Theta_{j,j}^{1/2}+v_4 \bigg] \to 0\,. \notag
\end{align*}
Hence, under the alternative hypothesis $\mathbb{H}_{1}$ in \eqref{eq:equiconind}, we have 
\begin{align*}
\mathbb{P} (\tilde{G}_{n}> \hat{{\rm cv}}_{{\rm cind},\alpha} ) \to 1  
\end{align*}
provided that
\begin{align}\label{eq:nn-h1-cd} 
&~~~~~~~~~~~   \log \tilde{d} \ll \min\{  n^{4\vartheta/(12\vartheta+3m_{*}) -\kappa/3} (\log n)^{-4/3-\varrho/(6\vartheta)} ,\,n^{1/12}(\log n)^{-7/6}  \,, \notag\\
&~~~~~~~~~~~~~~~~~~~~~~~~~~~~~   n^{2\kappa/11}(\log n)^{-14/11} ,\, n^{4\vartheta/(52\vartheta+13m_{*})}(\log n)^{-16/13-\varrho/(26\vartheta)}  \}\,, \notag\\
&m \ll \min\big[n^{\vartheta\{4\vartheta/(4\vartheta+m_{*})-\kappa\}/\varrho } (\log n )^{-4\vartheta/\varrho -1/2} \{\log(\tilde{d}n)\}^{-3\vartheta/\varrho} ,\, n^{(1-\kappa)/2}(\log n)^{-1} \{\log(\tilde{d}n)\}^{-1}\,, \notag\\
&~~~~~~~~~~~~~~ n^{\kappa/4}(\log n)^{-1} \{\log(\tilde{d}n)\}^{-1} ,\, n^{\kappa/2}(\log n)^{-7/2}\{\log (\tilde{d}n)\}^{-11/4} \,,\\
&~~~~~~~~~~~~~~ n^{4\vartheta^2/\{\varrho(4\vartheta+m_{*})\}} (\log n)^{-16\vartheta/\varrho-1/2}\{\log (\tilde{d}n)\}^{-13\vartheta/\varrho},\,  n^{1/2}(\log n)^{-7} \{\log (\tilde{d}n )\}^{-6}\big] \,. \notag 
\end{align}
Recall $\tilde{d} =p \vee  q\vee m$ with $p \lesssim n^{\varkappa_1}$, $q\lesssim n^{\varkappa_2}$ and $m\lesssim n^{\varkappa_3}$. For any given constants $\varkappa_1>0$, $\varkappa_2>0$ and
$0 \le \varkappa_3 < \min[\vartheta\{4\vartheta/(4\vartheta+m_{*})-\kappa\}/\varrho, (1-\kappa)/2 ,\kappa/4 ]$, the restrictions  \eqref{eq:nn-h1-cd} hold automatically.  
 We complete the proof of Theorem \ref{thm:nn-sta-h1}. 
$\hfill\Box$

\section{Proofs of  Theorems \ref{thm:3} and \ref{thm:4}}\label{subsec:p3-proof}  
Recall $\tilde{d} =p\vee q\vee m$, $\bTheta =  (\Theta_{i,j})_{d\times d}$ and  $\hat{\bTheta}= n^{-1}\sum_{i=1}^{n}\hat{\bet}_i\hat{\bet}_i^{\T}-\bar{\hat{\bet}}\bar{\hat{\bet}}^{\T}$ 
with $\bar{\hat{\bet}} = n^{-1}\sum_{i=1}^{n}\hat{\bet}_i$.  To prove Theorem \ref{thm:3}, we need Proposition \ref{pro:2} with its proof given in Section \ref{subsec:p21-proof}.  

\begin{proposition}\label{pro:2}
Let  $\hat{\bzeta}\,|\, \mathcal{X}_{n}, \mathcal{Y}_{n},\mathcal{Z}_{n} \sim \mathcal{N}(\bzero, \hat{\bTheta})$. Under  Condition {\rm \ref{cn:subgaussian}}, \eqref{eq:regressionUV}	  and the null hypothesis $\mathbb{H}_0$ in \eqref{eq:equiconind}, if $\min_{j\in[d]} \Theta_{j,j} \ge c_5$ for some universal constant $c_5>0$, $s\lesssim n^{1/5}(\log n)^{-2}$ and $\log \tilde{d} \ll \min\{n^{1/10}(s\log n)^{-1/2}, n^{1/8}(s^2\log n)^{-1/4}\}$, 
then it holds that $$\sup_{z>0}\big|\mathbb{P} (\hat{G}_n > z)-\mathbb{P}(|\hat{\bzeta}|_{\infty} > z \,|\,\mathcal{X}_{n}, \mathcal{Y}_n,\mathcal{Z}_{n})\big|  =o_{{\rm p}}(1)$$ 
as $n\rightarrow \infty$. 
\end{proposition}

\subsection{Proof of Proposition \ref{pro:2}}\label{subsec:p21-proof}  
To prove Proposition \ref{pro:2}, we need Lemmas \ref{lem:epsdeth-epsdet}--\ref{lem:cov} with their proofs given in Appendices \ref{sec:sub-eps-epsh}--\ref{subsec:G-thetah}, respectively.

\begin{lemma}\label{lem:epsdeth-epsdet}
Assume \eqref{eq:regressionUV} and Condition {\rm \ref{cn:subgaussian}} hold. Then
\begin{align*}
\frac{1}{n}\sum_{i=1}^{n} (\hat{\ve}_{i,j}\hat{\delta}_{i,k} -  \ve_{i,j}\delta_{i,k}  ) 
=   \frac{\sqrt{2\pi}(n-1)}{n(n+1)}\sum_{s=1}^{n} \big\{\tilde{\delta}_{4,k}(U_{s,j}) + \tilde{\delta}_{5,j}(V_{s,k})\big\}  + {\rm Rem}_1(j,k)
\end{align*} 
with
\begin{align*}
\max_{j\in[p],\, k\in[q]}|{\rm Rem}_1(j,k)| =  O_{\rm p} \{s  n^{-7/10} \log^{3/2}(\tilde{d}n)\}  +  O_{\rm p} \{s^{1/2} n^{-13/20}(\log n)^{-3/4} \log(\tilde{d}n)\}
\end{align*}
provided that  $s \lesssim n^{3/10}(\log \tilde{d})^{1/2}$ and $ \log  \tilde{d} \ll  n^{1/10}(\log n)^{-1/2}$, where
\begin{align*}
\tilde{\delta}_{4,k}(U_{s,j})=&~\mathbb{E} \big[e^{U_{i,j}^2/2}  \big\{I(U_{s,j}\le U_{i,j})-\Phi(U_{i,j})\big\}\delta_{i,k}  I\{|U_{i,j}|\le \sqrt{3(\log n)/5}\}\,\big|\,U_{s,j} \big]\,,\\
\tilde{\delta}_{5,j}(V_{s,k})=&~\mathbb{E} \big[e^{V_{i,k}^2/2}  \big\{I(V_{s,k}\le V_{i,k})-\Phi(V_{i,k})\big\}\varepsilon_{i,j}I\{|V_{i,k}|\le \sqrt{3(\log n)/5}\} \,\big|\,V_{s,k} \big]
\end{align*}
with  $i\ne s$.
\end{lemma}

\begin{lemma}\label{lem:delta-t-4}	
Assume \eqref{eq:regressionUV} and Condition {\rm \ref{cn:subgaussian}(i)} hold. Then
\begin{align*}
\frac{1}{n}\sum_{s=1}^{n}\big\{\tilde{\delta}_{4,k}(U_{s,j}) + \tilde{\delta}_{5,j}(V_{s,k})\big\} =  \frac{1}{n}\sum_{s=1}^{n}\big\{\tilde{\delta}_{44,k}(U_{s,j}) + \tilde{\delta}_{54,j}(V_{s,k})\big\} + {\rm Rem}_{2}(j,k)
\end{align*}
with 
\begin{align*}
\max_{j\in[p],\,k\in[q]}|{\rm Rem}_{2}(j,k)| =O_{\rm p} \{n^{-4/5}(\log n)^{1/4} (\log \tilde{d})^{1/2} \} 
\end{align*}  provided that $\log \tilde{d} \lesssim n$, where	\begin{align*}
\tilde{\delta}_{44,k}(U_{s,j}) =&~ \mathbb{E} \big[e^{U_{i,j}^2/2} \big\{I(U_{s,j}\le U_{i,j})-\Phi(U_{i,j})\big\} \delta_{i,k} 
I\{|U_{i,j}|\le \sqrt{3(\log n)/5}\}I(|\delta_{i,k}|\le \tilde{M}) \,\big|\,U_{s,j} \big]\,,\\
\tilde{\delta}_{54,k}(V_{s,k}) =&~ \mathbb{E} \big[e^{V_{i,k}^2/2} \big\{I(V_{s,k}\le V_{i,k})-\Phi(V_{i,k})\big\} \varepsilon_{i,j}I\{|V_{i,k}|\le \sqrt{3(\log n)/5}\} I(|\varepsilon_{i,j}|\le \tilde{M}) \,\big|\,V_{s,k} \big]
\end{align*}
with  $i \ne s$ and $\tilde{M}=\sqrt{9(\log n)/(10\tilde{c})} $ for $\tilde{c}=(1 \wedge c_7)/4$.
\end{lemma}

\begin{lemma}\label{lem:cov}
Assume \eqref{eq:regressionUV} and Condition {\rm \ref{cn:subgaussian}} hold. Then
\begin{align*}
|\hat{\bTheta}-\bTheta|_{\infty} =   O_{\rm p}\{s^{2}n^{-1/2} (\log n) (\log \tilde{d})^{1/2}\log^{3/2}(\tilde{d}n)\}  
\end{align*} 
provided that  $s \lesssim n^{3/10}(\log \tilde{d})^{1/2}$ and $ \log  \tilde{d}  \ll n^{1/10}(\log n)^{-1/2}$.
\end{lemma}

Recall $\hat{G}_n = \sqrt{n} |\hat{\bOmega}_{n}|_{\infty}$ and $\hat{\bOmega}_{n} = n^{-1}\sum_{i=1}^{n} \hat{\bet}_{i}$ with $\hat{\bet}_{i}= \hat{\boldsymbol{\ve}}_{i} \otimes \hat{\bdelta}_{i}$. Define $\bOmega_{n} = n^{-1}\sum_{i=1}^{n}\bet_i$ with $\bet_i =\boldsymbol{\varepsilon}_i \otimes \bxi_i$,   and let $\bzeta \sim \mathcal{N}(\boldsymbol{0}, \bTheta)$ with $\bTheta=\cov(\bet_i)$. Recall $d=pq$. Parallel to 
\eqref{eq:GAbound_a}, for any $u>0$, we have 
\begin{align}\label{eq:omega-h}
\sup_{x>0}\big|\mathbb{P}(\hat{G}_n>x) - \mathbb{P}(|\bzeta|_{\infty} > x)\big| \lesssim &~  \sup_{x>0}\big|\mathbb{P}(\sqrt{n}|\bOmega_{n}|_{\infty}>x)-\mathbb{P}(|\bzeta |_{\infty} > x) \big| \notag\\
&   + \mathbb{P}(\sqrt{n}|\hat{\bOmega}_{n} - \bOmega_{n}|_{\infty} > u) + u(\log d)^{1/2}\,.
\end{align}
Since $U_{i,j}=\balpha_j^{\T} \bW_i + \ve_{i,j}$ and $V_{i,k}=\bbeta_{k}^{\T} \bW_i + \delta_{i,k}$ with $\mathbb{E}(\ve_{i,j}\,|\, \bW_i)= 0 = \mathbb{E}(\delta_{i,k}\,|\, \bW_i)$, under the null hypothesis  $\mathbb{H}_0$ in \eqref{eq:equiconind}, we know the following two assertions hold: (i) $U_{i,j}$ and $\delta_{i,k}$ are conditionally independent given $\bW_i$, and  (ii) $V_{i,k}$ and $\ve_{i,j}$ are conditionally independent given $\bW_i$. 
Hence, for any $s\ne i $ and $a\in \mathbb{R}$,
\begin{align*}
&\mathbb{E} \big[e^{U_{i,j}^2/2}  \big\{I(U_{s,j}\le U_{i,j})-\Phi(U_{i,j})\big\} \delta_{i,k} I\{|U_{i,j}|\le \sqrt{3(\log n)/5}\} \,\big|\,U_{s,j} =a\big]\\
&~~~~~~~~~~= \mathbb{E} \big[e^{U_{i,j}^2/2} \big\{I(a\le U_{i,j})-\Phi(U_{i,j})\big\} \delta_{i,k} I\{|U_{i,j}|\le \sqrt{3(\log n)/5}\}  \big]\\
&~~~~~~~~~~=   \mathbb{E} \big\{\mathbb{E} \big[e^{U_{i,j}^2/2} \big\{I(a\le U_{i,j})-\Phi(U_{i,j})\big\}I\{|U_{i,j}|\le \sqrt{3(\log n)/5}\}  \,|\, \bW_i \big]\mathbb{E}(\delta_{i,k} \,|\, \bW_i)\big\}=0\,,
\end{align*}
which implies $\tilde{\delta}_{4,k}(U_{s,j})=0$  under the null hypothesis $\mathbb{H}_0$ in \eqref{eq:equiconind}. Analogously, we also have $\tilde{\delta}_{5,j}(V_{s,k})=0$ under the null hypothesis $\mathbb{H}_0$ in \eqref{eq:equiconind}. By Lemma \ref{lem:epsdeth-epsdet},    we have 
\begin{align*}
\sqrt{n}|\hat{\bOmega}_{n} - \bOmega_{n}|_{\infty} = &~  O_{\rm p} \{s  n^{-1/5} \log^{3/2}(\tilde{d}n)\}  +  O_{\rm p} \{s^{1/2} n^{-3/20}(\log n)^{-3/4} \log(\tilde{d}n)\}
\end{align*}   
provided that   $s \lesssim n^{3/10}(\log \tilde{d})^{1/2}$ and $ \log \tilde{d} \ll n^{1/10}(\log n)^{-1/2}$. To make 
$\mathbb{P}(\sqrt{n}|\hat{\bOmega}_{n} - \bOmega_{n }|_{\infty}>u)=o(1)$, it suffices to require  $u\gg \max\{s  n^{-1/5} \log^{3/2}(\tilde{d}n), s^{1/2} n^{-3/20}(\log n)^{-3/4} \log(\tilde{d}n)\}$. On the other hand, since $\tilde{d}=p\vee q \vee m$, by \eqref{eq:omega-h}, to make $\sup_{x>0}|\mathbb{P}( \hat{G}_n > x) - \mathbb{P}(|\bzeta|_{\infty} > x)|=o(1)$ under the null hypothesis  $\mathbb{H}_0$ in \eqref{eq:equiconind}, we need to require $u \ll (\log \tilde{d})^{-1/2}$. Therefore, $(\tilde{d},n)$ should satisfy
\begin{align*}
\left\{
\begin{aligned}
&s n^{-1/5} \log^{3/2}(\tilde{d}n) \ll (\log \tilde{d})^{-1/2}\,,\\
&s^{1/2} n^{-3/20}(\log n)^{-3/4} \log(\tilde{d}n) \ll (\log \tilde{d})^{-1/2}\,,\\
&\log \tilde{d} \ll n^{1/10}(\log n)^{-1/2}
\end{aligned}
\right.
\end{align*}
with  $s \lesssim n^{3/10}(\log \tilde{d})^{1/2}$.  By \eqref{eq:omega-h}, under the null hypothesis  $\mathbb{H}_0$ in \eqref{eq:equiconind},  it holds that
\begin{align*}
\sup_{x>0}\big|\mathbb{P}(\hat{G}_n >x) - \mathbb{P}(|\bzeta|_{\infty} > x)\big| \lesssim \sup_{x>0}\big|\mathbb{P}(\sqrt{n}|\bOmega_{n}|_{\infty}>x)-\mathbb{P}(|\bzeta |_{\infty} > x) \big| + o(1) 
\end{align*}
provided that 
$\log \tilde{d} \ll n^{1/10}(s\log n)^{-1/2}$ and $s \lesssim n^{1/5} (\log n)^{-2}$. Recall $U_{i,j}\sim \mathcal{N}(0,1)$. By Condition \ref{cn:subgaussian}(i), we have
\begin{align}\label{eq:ve_tail}
\mathbb{P}(|\ve_{i,j}|>x) = &~\mathbb{P}(|U_{i,j} - \balpha_j^{\T}\bW_i|>x)  \leq \mathbb{P}\bigg(|U_{i,j}| >\frac{x}{2}\bigg) + \mathbb{P}\bigg(|\balpha_j^{\T}\bW_i|>\frac{x}{2}\bigg) \notag\\
\leq&~ 2e^{-x^2/4} + c_6e^{-c_7x^2/4} \le C_1 e^{-\tilde{c} x^2}
\end{align}
for any $x>0$, $i \in [n]$ and $j \in [p]$, where $\tilde{c}=(1\wedge c_7)/4$.
Identically, we also have $\mathbb{P}(|\delta_{i,k}|>x)\le C_1 e^{-\tilde{c} x^2} $ for any $x>0$, $i \in [n]$ and $k \in [q]$.
By  Lemma 2 of \cite{Chang2013}, it holds that 
\begin{align}\label{eq:eps-delta-tail}
\mathbb{P} ( |\ve_{i,j}\delta_{i,k}| > x ) \le 2C_1e^{-\tilde{c} x}
\end{align}
for any $x>0$. Recall $\min_{j\in [d]}\Theta_{j,j} \ge c_5$. 
By Proposition 2.1 of \cite{Chernozhukov2017}, it holds that
\begin{align*}
\sup_{x>0}\big|\mathbb{P}(\sqrt{n}|\bOmega_{n}|_{\infty}>x)-\mathbb{P}(|\bzeta |_{\infty} > x) \big| \lesssim  n^{-1/6}\log^{7/6}(\tilde{d}n)\,. 
\end{align*}
Then, under the null hypothesis  $\mathbb{H}_0$ in \eqref{eq:equiconind}, we have 
\begin{align*}
\sup_{x>0}\big|\mathbb{P}(\hat{G}_n >x) - \mathbb{P}(|\bzeta|_{\infty} > x)\big|=o(1)
\end{align*}
provided that $\log \tilde{d} \ll  n^{1/10}(s\log n)^{-1/2} $ and $s \lesssim n^{1/5} (\log n)^{-2}$. 

Parallel to \eqref{eq:hn-xih},   under the null hypothesis $\mathbb{H}_0$ in \eqref{eq:equiconind}, 
\begin{align}\label{eq:gn-zetah}
&\sup_{x >0}\big|\mathbb{P}(\hat{G}_n>x)- \mathbb{P}(|\hat{\bzeta}|_{\infty} >x \,|\,\mathcal{X}_n, \mathcal{Y}_n,\mathcal{Z}_n)\big|\notag\\
&~~~~~~~~~~\le  \sup_{x >0}\big|\mathbb{P}(|\bzeta|_{\infty} >x) - \mathbb{P}(|\hat{\bzeta}|_{\infty} >x \,|\,\mathcal{X}_n, \mathcal{Y}_n,\mathcal{Z}_n)\big| + o(1) 
\end{align}
provided that $\log \tilde{d} \ll  n^{1/10}(s\log n)^{-1/2} $ and $s \lesssim n^{1/5} (\log n)^{-2}$. 
Write $\Delta_{n2} = |\hat{\bTheta} -\bTheta |_{\infty}$.
By Lemma \ref{lem:cov}, $\Delta_{n2} =  O_{\rm p}\{s^{2}n^{-1/2} (\log n) (\log \tilde{d})^{1/2}\log^{3/2}(\tilde{d}n)\}$ provided that $\log \tilde{d} \ll n^{1/10}(\log n)^{-1/2}$ and $s\lesssim n^{3/10}(\log \tilde{d})^{1/2}$. Recall $d=pq$ and $\tilde{d}=p\vee q\vee m$. Parallel to \eqref{eq:Prop2bound}, it holds that
\begin{align*}
\sup_{x>0}\big|\mathbb{P}(|\bzeta|_{\infty} > x)-\mathbb{P}(|\hat{\bzeta}|_{\infty} > x \,|\,\mathcal{X}_{n}, \mathcal{Y}_n,\mathcal{Z}_{n})\big|  
\lesssim \Delta_{n2}^{1/3} \{1\vee \log(2d\Delta_{n2}^{-1})\}^{2/3} =o_{{\rm p}}(1) 
\end{align*}
provided that $\log \tilde{d} \ll \min\{n^{1/10}(\log n)^{-1/2}, n^{1/8}(s^2\log n)^{-1/4}\}$ and $s\lesssim n^{1/4}(\log n)^{-5/2}$. Together with \eqref{eq:gn-zetah},  under the null hypothesis  $\mathbb{H}_0$ in \eqref{eq:equiconind}, we have 
\begin{align*}
\sup_{x >0}\big|\mathbb{P}(\hat{G}_n>x)- \mathbb{P}(|\hat{\bzeta}|_{\infty} >x \,|\,\mathcal{X}_n, \mathcal{Y}_n,\mathcal{Z}_n)\big|=o_{\rm p}(1)
\end{align*}
provided that $\log \tilde{d} \ll \min\{n^{1/10}(s\log n)^{-1/2}, n^{1/8}(s^2\log n)^{-1/4}\}$ and $s\lesssim n^{1/5}(\log n)^{-2}$. Hence, we complete the proof of Proposition \ref{pro:2}.  $\hfill\Box$

\subsection{Proof of Theorem \ref{thm:3}}

The proof of Theorem \ref{thm:3} is almost identical to that of Theorem \ref{thm:1} given in  Section \ref{sec:sub-thm1}. Hence, we omit it here. $\hfill\Box$   

\subsection{Proof of Theorem \ref{thm:4}}\label{subsec:th4-proof}

For any $v_5 >0$, consider the event
\begin{equation*}
\begin{aligned}
\mathcal{E}_5(v_5)=\left\{\max_{j \in [d]}\frac{|\hat{\Theta}_{j,j}^{1/2} - \Theta_{j,j}^{1/2}|}{ \Theta_{j,j}^{1/2}} \le v_5 \right\}\,.
\end{aligned}
\end{equation*}
Due to  $\hat{\rm cv}_{{\rm cind},\alpha}^{*}  =\inf\{t\in \mathbb{R}: \mathbb{P}(|\hat{\bzeta}|_{\infty} > t\,|\,\mathcal{X}_n, \mathcal{Y}_n, \mathcal{Z}_{n}) \le \alpha\}$, parallel to \eqref{eq:cvhat}, 
\begin{align*}
\hat{{\rm cv}}_{{\rm cind},\alpha}^{*} 
\le 
(1+v_5)\big[\{1+(2\log d)^{-1}\}(2\log d)^{1/2}+ \{2\log(1/\alpha)\}^{1/2}\big]\max_{j \in [d]}\Theta_{j,j}^{1/2}
\end{align*}
restricted on $\mathcal{E}_5(v_5)$. Recall  $\tilde{d}=p\vee q\vee m$. With selecting $v_5 = (1+2\log d)^{-1}$,  by Lemma \ref{lem:cov},  we have $\mathbb{P}\{\mathcal{E}_5^{\rm c}(v_5)\} \to 0$ provided that $\log \tilde{d} \ll \min\{n^{1/6}(s^{2}\log n)^{-1/3}, n^{1/10}(\log n)^{-1/2}\}$ and $s\lesssim n^{1/4}(\log n)^{-2}$, and  $\hat{{\rm cv}}_{{\rm cind},\alpha}^{*} \le \{1+(\log d)^{-1}\}\lambda(d, \alpha) \max_{j \in [d]}\Theta_{j,j}^{1/2}$ restricted on $\mathcal{E}_5(v_5)$, where $\lambda(d, \alpha)= (2\log d)^{1/2}+ \{2\log(1/\alpha)\}^{1/2}$. 

Write $\bOmega = \mathbb{E}(\boldsymbol{\ve}_{i} \otimes \bdelta_{i} ) =(\Omega_1,\ldots, \Omega_d)^{\T}$. We sort $\{|\Omega_{l}|\}_{l=1}^{d}$ in the decreasing order as $|\Omega_{l_{1}^{*}}| \ge \cdots \ge |\Omega_{l_{d}^{*}}|$.
Without loss of generality, we assume $\Omega_{l_{1}^{*}}>0$. Let $g$ be a bijective mapping from $\{(j,k): j\in[p],k \in[q]\}$ to $[d]$, such that $g(j,k)=l$. Then there exist $j^{*}\in [p]$ and ${k}^{*} \in [q]$ such that $g(j^{*}, k^{*})=l_1^{*}$. For any $v_6>0$, consider the event 
\begin{align*}
\mathcal{E}_6(v_6) =\bigg\{ \max_{j\in[p],\,k\in[q]} \bigg|\frac{1}{\sqrt{n}}\sum_{i=1}^{n} (\hat{\varepsilon}_{i,j}\hat{\delta}_{i,k} -\varepsilon_{i,j} \delta_{i,k}) \bigg| \le v_6\bigg\}\,.
\end{align*}
Recall $\hat{G}_n = \sqrt{n} |\hat{\bOmega}_{n}|_{\infty}$ and $\hat{\bOmega}_{n} = n^{-1}\sum_{i=1}^{n} \hat{\bet}_{i}$ with $\hat{\bet}_{i}= \hat{\boldsymbol{\ve}}_{i} \otimes \hat{\bdelta}_{i}$. 
Parallel to \eqref{eq:hnd-1}, under the alternative hypothesis $\mathbb{H}_{1}$ in \eqref{eq:equiconind}, with selecting $v_5=(1+2\log d)^{-1}$, it holds that 
\begin{align}\label{eq:gnc}
\mathbb{P} (\hat{G}_n> \hat{{\rm cv}}_{{\rm cind},\alpha}^{*}) \ge&~ 1-\mathbb{P} \bigg[\frac{1}{\sqrt{n}}\sum_{i=1}^{n}\{\varepsilon_{i,j^{*}} \delta_{i,k^{*}} - \mathbb{E}(\varepsilon_{i,j^{*}} \delta_{i,k^{*}})\} \notag\\ 
&~~~~~~~~~~~~~~~~~~~~~~~ \le -\sqrt{n}\Omega_{l_{1}^{*}} +\{1+(\log d)^{-1}\}\lambda(d,\alpha)\max_{j \in [d]}\Theta_{j,j}^{1/2}+v_6 \bigg]\notag \\
&-o(1)
- \mathbb{P}\{\mathcal{E}_6^{\rm c}(v_6)\}  
\end{align}
provided that $\log \tilde{d} \ll \min\{n^{1/6}(s^{2}\log n)^{-1/3}, n^{1/10}(\log n)^{-1/2}\}$ and $s\lesssim n^{1/4}(\log n)^{-2}$.  By Lemmas \ref{lem:epsdeth-epsdet} and \ref{lem:delta-t-4},  we have
\begin{align*}
\max_{j\in[p],\,k\in[q]} \bigg|\frac{1}{\sqrt{n}}\sum_{i=1}^{n} (\hat{\varepsilon}_{i,j}\hat{\delta}_{i,k} -\varepsilon_{i,j} \delta_{i,k}) \bigg| \le &~ \max_{j\in[p],\, k\in[q]} \bigg|\frac{\sqrt{2\pi }}{\sqrt{n}}\sum_{s=1}^{n}\big\{ \tilde{\delta}_{44,k}(U_{s,j}) + \tilde{\delta}_{54,j}(V_{s,k})\big\}\bigg|  \\
& + O_{\rm p} \{s  n^{-1/5} \log^{3/2}(\tilde{d}n)\} + O_{\rm p} \{s^{1/2} n^{-3/20}(\log n)^{-3/4} \log(\tilde{d}n)\}
\end{align*}
provided that $\log \tilde{d} \ll n^{1/10}(\log n)^{-1/2}$ and  $s \lesssim n^{3/10}(\log \tilde{d})^{1/2}$. 
Recall $\tilde{d}=p\vee q\vee m $ and $\tilde{M}=\sqrt{9(\log n)/(10\tilde{c})} $ with $\tilde{c} =(1\wedge c_7)/4$. 
Analogous to the derivation of \eqref{eq:tail-delta-4}  in Section \ref{sec:sub-s4} for the proof of Lemma \ref{lem:cov}, it holds that
\begin{align}\label{eq:delta4445-tail}
\mathbb{P}\bigg[\max_{j\in[p],\,k\in[q]}\bigg|\frac{\sqrt{2\pi}}{n}\sum_{s=1}^{n} \big\{\tilde{\delta}_{44,k}(U_{s,j}) + \tilde{\delta}_{54,j}(V_{s,k})\big\}\bigg| > x\bigg]  \le 2\tilde{d}^2\exp\bigg\{-\frac{25\tilde{c}nx^2}{432(\log n)^2}\bigg\}
\end{align} 
for any $x>0$. Recall $u_n \ge c_{10}$ for some universal constant $c_{10}>0$.
Selecting $v_6 = 12\sqrt{3 \tilde{c}^{-1}}  (\sqrt{2}+u_n/2)(\log \tilde{d})^{1/2}(\log n)/5 $, by \eqref{eq:delta4445-tail}, we have
\begin{align}\label{eq:e4c}
\mathbb{P}\{\mathcal{E}_6^{\rm c}(v_6)\}  \le&~  \mathbb{P}\bigg[\max_{j\in[p],\,k\in[q]}\bigg|\frac{\sqrt{2\pi}}{\sqrt{n}}\sum_{i=1}^{n} \big\{\tilde{\delta}_{44,k}(U_{s,j}) + \tilde{\delta}_{54,j}(V_{s,k}) \big\}\bigg| > \frac{12\sqrt{3}}{5\sqrt{\tilde{c}}} \bigg(\sqrt{2}+\frac{u_n}{4}\bigg)(\log \tilde{d})^{1/2}\log n\bigg]\notag \\
&+ \mathbb{P}\bigg[ O_{\rm p} \{s  n^{-1/5} \log^{3/2}(\tilde{d}n)\} >  \frac{3\sqrt{3}u_n}{10\sqrt{\tilde{c}}}  (\log \tilde{d})^{1/2}\log n \bigg] \notag\\
&+ \mathbb{P}\bigg[ O_{\rm p} \{s^{1/2} n^{-3/20}(\log n)^{-3/4} \log(\tilde{d}n)\} >  \frac{3\sqrt{3}u_n}{10\sqrt{\tilde{c}}}  (\log \tilde{d})^{1/2}\log n \bigg] \notag\\
\le&~  2\tilde{d}^{-\sqrt{2}u_n/2-u_n^2/16} + o(1) 
\end{align}
provided that $\log \tilde{d} \ll \min\{n^{1/10}(\log n)^{-1/2}, s^{-1}n^{1/5}\log n\}$ and $s \ll n^{1/5}(\log \tilde{d})^{1/2}(\log n)^{-1/2}$.
Due to  $\Omega_{l_{1}^{*}} \ge  12 \sqrt{3 \tilde{c}^{-1}} (\sqrt{2}+u_n) n^{-1/2}(\log \tilde{d})^{1/2}(\log n)/5$ under the alternative hypothesis $\mathbb{H}_{1}$ in \eqref{eq:equiconind}, we have
\begin{align*}
\sqrt{n}\tilde{\Omega}_{l_{1}^{*}}  - \{1+(\log d)^{-1}\}\lambda(d, \alpha) \max_{j \in [d]}\Theta_{j,j}^{1/2} -v_6 \ge&~ 6\sqrt{3 \tilde{c}^{-1}}  u_n  (\log \tilde{d})^{1/2} (\log n)/5 - C\lambda(d, \alpha)\\
\ge&~ \sqrt{3 \tilde{c}^{-1}}   u_n (\log \tilde{d})^{1/2} \log n 
\end{align*}  
for sufficiently large $n$. By \eqref{eq:eps-delta-tail}, it holds that $c_5< \var(\ve_{i,j^{*}}\delta_{i,k^{*}}) \le C_2$ for some positive constant $C_2>c_5$. It follows from the Central Limit Theorem that $n^{-1/2} \sum_{i=1}^{n}\{\varepsilon_{i,j^{*}} \delta_{i,k^{*}} - \mathbb{E}(\varepsilon_{i,j^{*}} \delta_{i,k^{*}})\}\{\var(\ve_{i,j^{*}}\delta_{i,k^{*}})\}^{-1/2}\to \mathcal{N}(0,1)$ in distribution.  Then, due to $u_n (\log \tilde{d})^{1/2}\log n \to \infty$,  under the alternative hypothesis $\mathbb{H}_{1}$ in \eqref{eq:equiconind}, for any sufficiently large $n$,
\begin{align}\label{eq:h1-gaussian-0}
&\mathbb{P} \bigg[\frac{1}{\sqrt{n}}\sum_{i=1}^{n}\{\varepsilon_{i,j^{*}} \delta_{i,k^{*}} - \mathbb{E}(\varepsilon_{i,j^{*}} \delta_{i,k^{*}}) \} \le -\sqrt{n}\Omega_{l_{1}^{*}} +\{1+(\log d)^{-1}\}\lambda(d,\alpha)\max_{j \in [d]}\Theta_{j,j}^{1/2}+v_6 \bigg] \notag\\
&~~~~~\le \mathbb{P}\bigg[\frac{1}{\sqrt{n}}\sum_{i=1}^{n}\{\varepsilon_{i,j^{*}} \delta_{i,k^{*}} - \mathbb{E}(\varepsilon_{i,j^{*}} \delta_{i,k^{*}}) \} \le   -  \sqrt{\frac{3} {\tilde{c}} }  u_n (\log \tilde{d})^{1/2}\log n\bigg]\\
&~~~~~= \mathbb{P}\bigg\{\frac{1}{\sqrt{n}}\sum_{i=1}^{n}\frac{\varepsilon_{i,j^{*}} \delta_{i,k^{*}} - \mathbb{E}(\varepsilon_{i,j^{*}} \delta_{i,k^{*}}) }{\sqrt{\var(\ve_{i,j^{*}}\delta_{i,k^{*}})} } \le   -  \frac{\sqrt{3} u_n (\log \tilde{d})^{1/2}\log n} {\sqrt{ \tilde{c}\var(\ve_{i,j^{*}}\delta_{i,k^{*}})}}\bigg\}   \to 0\,. \notag
\end{align}
Together with \eqref{eq:gnc} and \eqref{eq:e4c}, under the alternative hypothesis $\mathbb{H}_{1}$ in \eqref{eq:equiconind}, it holds that
\begin{align*}
\mathbb{P} (\hat{G}_n> \hat{{\rm cv}}_{{\rm cind},\alpha}^{*}) \ge 1- 2\tilde{d}^{-\sqrt{2}u_n/2-u_n^2/16}- o(1) 
\end{align*}
provided that  
\begin{align}\label{eq:d-s-cd}
    &\log \tilde{d} \ll \min\{n^{1/6}(s^2\log n)^{-1/3},\, n^{1/10}(\log n)^{-1/2},\,s^{-1}n^{1/5}\log n\}\,, \notag\\
    &~~~~~~~~~s\ll  \min\{n^{1/4}(\log n)^{-2},\, n^{1/5}(\log \tilde{d})^{1/2}(\log n)^{-1/2}\} \,.
\end{align}
Recall $\tilde{d} =p \vee  q\vee m$ with $p \lesssim n^{\varkappa_1}$, $q\lesssim n^{\varkappa_2}$ and $m\lesssim n^{\varkappa_3}$. If $s\ll n^{1/5}(\log n)^{-1/2} $, the restrictions \eqref{eq:d-s-cd} hold for any constants $\varkappa_1 \ge 0$, $\varkappa_2\ge 0$ and $\varkappa_3 \ge 0$. We then have Theorem \ref{thm:4}. $\hfill\Box$

\section{Proof of Theorem \ref{pro:12} }\label{sec:pro:12}
\subsection{Proof of Theorem \ref{pro:12}(i)}
 By triangle inequality and Proposition \ref{pro:1}, under the null hypothesis $\mathbb{H}_0$ in \eqref{eq:equind},   we have
\begin{align}\label{eq:mut-xi}
&\sup_{x >0}\big|\mathbb{P}(H_{n}>x)- \mathbb{P}(|\hat{\bxi}^{\dagger}|_{\infty} >x \,|\,\mathcal{X}_n, \mathcal{Y}_n)\big| \notag\\
&~~~~~\le\sup_{x >0}\big|\mathbb{P}(H_{n}>x) - \mathbb{P}(|\hat{\bxi}|_{\infty} >x \,|\,\mathcal{X}_n, \mathcal{Y}_n)\big| \notag\\
&~~~~~~~~~ + \sup_{x>0}\big| \mathbb{P}(|\hat{\bxi}|_{\infty} > x \,|\,\mathcal{X}_{n}, \mathcal{Y}_n) - \mathbb{P}(|\hat{\bxi}^{\dagger}|_{\infty} > x\,|\,\mathcal{X}_{n}, \mathcal{Y}_n)\big| \notag\\
&~~~~~ \le \sup_{x>0}\big| \mathbb{P}(|\hat{\bxi}|_{\infty} > x \,|\,\mathcal{X}_{n}, \mathcal{Y}_n) - \mathbb{P}(|\hat{\bxi}^{\dagger}|_{\infty} > x\,|\,\mathcal{X}_{n}, \mathcal{Y}_n)\big| + o_{\rm p}(1)  
\end{align}
provided that $\log d \ll  n^{1/8}(\log n)^{-1/4}$. Recall $U_{i,j}, V_{i,k} \sim \mathcal{N}(0,1)$. Under the null hypothesis $\mathbb{H}_0$ in \eqref{eq:equind}, 
$\Sigma_{j,j} =1 $ for any $j\in[d]$.  By Lemma \ref{lem:covh}, we have  $\min_{j \in [d]}\hat{\Sigma}_{j,j} \ge 1/2 $ with probability approaching  one provided that $\log d \ll  n^{1/4}(\log n)^{-1/2}$. By \eqref{eq:uh-bound} in Section \ref{sec:sub-I3} for the proof of Lemma \ref{lem:uhv}, we have $\max_{i\in[n],\,j\in[p]}|\hat{U}_{i,j}| \le  \sqrt{2\log (n+1)}$. Analogously, we also have $\max_{i\in[n],\,k\in[q]}|\hat{V}_{i,k}| \le  \sqrt{2\log (n+1)}$. Recall $\hat{\bgamma}_i = \hat{\bU}_i \otimes \hat{\bV}_i$ and $\bar{\hat{\bgamma}}=n^{-1}\sum_{i=1}^n \hat{\bgamma}_i$. Hence, it holds that $\max_{i \in [n]}|\hat{\bgamma}_i-\bar{\hat{\bgamma}}|_{\infty} \lesssim \log n$. 
For either Mammen's or Rademacher multiplier $ \varpi_i $, we have 
$\max_{i \in [n]}| \varpi_i |\le C$, which implies 
\begin{align}\label{eq:gamma-bound}
\max_{i \in [n]}| \varpi_i (\hat{\bgamma}_{i} - \bar{\hat{\bgamma}})|_{\infty} \lesssim \log n\,.
\end{align}
Applying Proposition 2.1 of \cite{Chernozhukov2017} with $B_n= \tilde{C} (\log n)^3$ for some universal constant $\tilde{C}>0$, it holds that 
\begin{align*}
\sup_{x>0}\big|\mathbb{P}(|\hat{\bxi}|_{\infty} > x \,|\,\mathcal{X}_{n},\mathcal{Y}_{n}) -\mathbb{P}(|\hat{\bxi}^{\dagger}|_{\infty} > x\,|\,\mathcal{X}_{n},\mathcal{Y}_{n})\big| = O_{\rm p}\{ n^{-1/6}\log^{7/6}(dn) \log n\}
\end{align*}
provided that $\log d \ll  n^{1/4}(\log n)^{-1/2}$, which implies
\begin{align*}
\sup_{x>0}\big|\mathbb{P}(|\hat{\bxi}|_{\infty} > x \,|\,\mathcal{X}_{n},\mathcal{Y}_{n})-\mathbb{P}(|\hat{\bxi}^{\dagger}|_{\infty} > x\,|\,\mathcal{X}_{n},\mathcal{Y}_{n})\big| =o_{\rm p}(1)
\end{align*}
provided that $\log d \ll  n^{1/7}(\log n)^{-6/7}$. By \eqref{eq:mut-xi}, if $\log d \ll  n^{1/8}(\log n)^{-1/4}$, under the null hypothesis $\mathbb{H}_0$ in \eqref{eq:equind}, we have 
\begin{align*}
\sup_{x >0}\big|\mathbb{P}(H_{n}>x)- \mathbb{P}(|\hat{\bxi}^{\dagger}|_{\infty} >x \,|\,\mathcal{X}_n, \mathcal{Y}_n)\big|=o_{\rm p}(1)\,.
\end{align*}
 Since $d=pq$ with $p \lesssim n^{\varkappa_1}$ and $q \lesssim n^{\varkappa_2}$, the restriction $\log d \ll  n^{1/8}(\log n)^{-1/4}$ holds automatically for any constants $\varkappa_1 \ge 0$ and  $\varkappa_2\ge 0$. Hence, we complete the proof of Theorem \ref{pro:12}(i).
$\hfill\Box$

\subsection{Proof of Theorem \ref{pro:12}(ii)}   
 Parallel to \eqref{eq:mut-xi}, by Proposition  \ref{pro:nn-sta-h0}, under the null hypothesis  $\mathbb{H}_0$ in \eqref{eq:equiconind},  we have
\begin{align}\label{eq:mut-zeta-nn}
&\sup_{x >0}\big|\mathbb{P}(\tilde{G}_{n}>x)- \mathbb{P}(|\tilde{\bzeta}^{\dagger}|_{\infty} >x \,|\,\mathcal{X}_n, \mathcal{Y}_n,\mathcal{Z}_n)\big| \notag\\
&~~~~~~~~~~\le  \sup_{x >0}\big|\mathbb{P}(|\tilde{\bzeta}|_{\infty} > x\,|\,\mathcal{X}_n, \mathcal{Y}_n,\mathcal{Z}_n) - \mathbb{P}(|\tilde{\bzeta}^{\dagger}|_{\infty} > x\,|\,\mathcal{X}_n, \mathcal{Y}_n,\mathcal{Z}_n)\big| + o_{\rm p}(1)  
\end{align}
provided that \eqref{eq:pro-32-cd-n} holds. Recall     $\tilde{d}=p\vee q\vee m$ and  $\varrho =    \vartheta+2m_{*}\tilde{\vartheta}+3m_{*} $. Since $\min_{j \in [d]}\Theta_{j,j} \geq c_5$, by Lemma \ref{lem:nn-cov-theta}, if  
\begin{align}\label{eq:theta-rate} 
& \log \tilde{d} \ll \min \big\{ n^{1-\kappa} (\log n)^{-1/2} ,\, n^{1/8}(\log n)^{-7/4} ,\,n^{2\kappa/7}(\log n)^{-2}   \,,\notag \\
&~~~~~~~~~~~~~~~~~~~~  n^{4\vartheta/(36\vartheta+9m_{*})}(\log n)^{-16/9-\varrho/(18\vartheta)}  \big\}\,,  \notag\\
& m\ll  \min \big[ n^{1/2}(\log n)^{-7}\{\log (\tilde{d}n )\}^{-4},\,  n^{\kappa/2}(\log n)^{-7/2}\{\log (\tilde{d}n )\}^{-7/4}\,,\\
&~~~~~~~~~~~~~~~~  n^{4\vartheta^2/\{\varrho(4\vartheta+m_{*})\}} (\log n)^{-16\vartheta/\varrho-1/2}\{\log (\tilde{d}n)\}^{-9\vartheta/\varrho} \big] \,,  \notag
\end{align} 
it holds that $\min_{j \in [d]}\tilde{\Theta}_{j,j} \geq c_5/2$  with probability approaching  one. Notice that $\tilde{\ve}_{i,j}= \hat{U}_{i,j}^{(w)}- \hat{f}_j(\hat{\bW}_{i}^{(w)})$ and $\tilde{\delta}_{i,k}= \hat{V}_{i,k}^{(w)}- \hat{g}_{k}(\hat{\bW}_{i}^{(w)})$. Recall  $\max_{i\in\mathcal{D}_3,\,j\in[p]}|\hat{U}_{i,j}^{(w)}| \le  \sqrt{2\log n_1}$, $\max_{i\in\mathcal{D}_3,\,k\in[q]}|\hat{V}_{i,k}^{(w)}| \le  \sqrt{2\log n_1}$, $\max_{i\in\mathcal{D}_3,\,j\in[p]}|\hat{f}_j(\hat{\bW}_{i}^{(w)})| \le \tilde{\beta}_n$ and $\max_{i\in\mathcal{D}_3,\,k\in[q]}|\hat{g}_{k}(\hat{\bW}_{i}^{(w)})| \le  \tilde{\beta}_n$ with $n_1\asymp n$ and $\tilde{\beta}_n =(\log n) \log^{1/2}(\tilde{d}n)$. We have $\max_{i \in \mathcal{D}_3, \,j\in[p]}|\tilde{\varepsilon}_{i,j}| \lesssim  (\log n) \log^{1/2}(\tilde{d}n)$ and $\max_{i \in \mathcal{D}_3, \,k\in[q]}|\tilde{\delta}_{i,k}| \lesssim (\log n) \log^{1/2}(\tilde{d}n)$. Recall $\tilde{\bet}_{i} = \tilde{\boldsymbol{\ve}}_{i}  \otimes \tilde{\bdelta}_{i} $ and $\bar{\tilde{\bet}} = n_{3}^{-1}\sum_{i\in\mathcal{D}_3}\tilde{\bet}_i$.  Parallel to \eqref{eq:gamma-bound}, for either Mammen's or Rademacher multiplier $ \varpi_i $, we can also show
$\max_{i \in \mathcal{D}_3}| \varpi_i (\tilde{\bet}_{i} - \bar{\tilde{\bet}})|_{\infty} \lesssim (\log n)^2 \log(\tilde{d}n)$.   Recall 
$n_3 =n^{\kappa}$ for some constant $0<1 <\kappa$. Applying Proposition 2.1 of \cite{Chernozhukov2017} with 
$B_n= \tilde{C}(\log n)^{6} \log^{3}(\tilde{d}n)$ for some universal constant $\tilde{C}>0$, we have 
\begin{align*}
\sup_{x >0}\big|\mathbb{P}(|\tilde{\bzeta}|_{\infty} > x\,|\,\mathcal{X}_n, \mathcal{Y}_n,\mathcal{Z}_n)- \mathbb{P}(|\tilde{\bzeta}^{\dagger}|_{\infty} > x\,|\,\mathcal{X}_n, \mathcal{Y}_n,\mathcal{Z}_n)\big| =O_{\rm p}\{ n^{-\kappa/6}\log^{13/6}(\tilde{d}n)(\log n)^2 \}
\end{align*}
provided that \eqref{eq:theta-rate} holds. 
Therefore, 
\begin{align*}
\sup_{x >0}\big|\mathbb{P}(|\tilde{\bzeta}|_{\infty} > x\,|\,\mathcal{X}_n, \mathcal{Y}_n,\mathcal{Z}_n) -\mathbb{P}(|\tilde{\bzeta}^{\dagger}|_{\infty} > x\,|\,\mathcal{X}_n, \mathcal{Y}_n,\mathcal{Z}_n)\big| =o_{{\rm p}}(1)
\end{align*}
provided that $\log \tilde{d} \ll \  n^{\kappa/13} (\log n)^{-12/13} $ and  \eqref{eq:theta-rate} holds.  By \eqref{eq:mut-zeta-nn}, under the null hypothesis  $\mathbb{H}_0$ in \eqref{eq:equiconind},  it holds that
\begin{align*}
\sup_{x >0}\big|\mathbb{P}(\tilde{G}_{n}>x)- \mathbb{P}(|\tilde{\bzeta}^{\dagger}|_{\infty} >x \,|\,\mathcal{X}_n, \mathcal{Y}_n,\mathcal{Z}_n)\big| =o_{{\rm p}}(1)
\end{align*} 
provided that 
\begin{align}\label{eq:dn-3-boot-cd}
& \log \tilde{d} \ll \min \big\{ n^{\vartheta/(4\vartheta+m_{*})-\kappa/4} (\log n)^{-1-\varrho/(8\vartheta)} ,\,  n^{\kappa/13}(\log n)^{-12/13}\,, \notag\\
&~~~~~~~~~~~~~~~~~~~n^{1/16}(\log n)^{-7/8},\, n^{4\vartheta/(68\vartheta+17m_{*})}(\log n)^{-16/17-\varrho/(34\vartheta)}  \big\} \,, \notag\\
&m \ll \min\big[n^{\vartheta\{4\vartheta/(4\vartheta+m_{*})-\kappa\}/\varrho } (\log n )^{-4\vartheta/\varrho -1/2} \{\log(\tilde{d}n)\}^{-4\vartheta/\varrho} \,, \notag\\
&~~~~~~~~~~~~~~~n^{(1-\kappa)/2}(\log n)^{-1} \{\log(\tilde{d}n)\}^{-3/2}\,,\, n^{\kappa/4}(\log n)^{-1} \{\log(\tilde{d}n)\}^{-5/4} \,,\\
&~~~~~~~~~~~~~~~n^{4\vartheta^2/\{\varrho(4\vartheta+m_{*})\}} (\log n)^{-16\vartheta/\varrho-1/2}\{\log (\tilde{d}n)\}^{-17\vartheta/\varrho}\,,\notag \\
&~~~~~~~~~~~~~~~ n^{\kappa/2}(\log n)^{-7/2}\{\log (\tilde{d}n )\}^{-15/4},\, n^{1/2}(\log n)^{-7}\{\log (\tilde{d}n)\}^{-8}\big]    \notag 
\end{align} 
with $\varrho =    \vartheta+2m_{*}\tilde{\vartheta}+3m_{*} $. Recall $\tilde{d} =p \vee  q\vee m$ with $p \lesssim n^{\varkappa_1}$, $q\lesssim n^{\varkappa_2}$ and $m\lesssim n^{\varkappa_3}$. For any given constants  $\varkappa_1 \ge 0$, $\varkappa_2 \ge 0$ and $ 0 \le \varkappa_3 < \min[\vartheta\{4\vartheta/(4\vartheta+m_{*})-\kappa\}/\varrho, (1-\kappa)/2 ,\kappa/4 ]$, the restrictions \eqref{eq:dn-3-boot-cd} hold automatically. 
Hence, we complete the proof of Theorem \ref{pro:12}(ii).
$\hfill\Box$   

\subsection{Proof of Theorem \ref{pro:12}(iii)}
 Parallel to \eqref{eq:mut-xi},  by Proposition  \ref{pro:2}, under the null hypothesis  $\mathbb{H}_0$ in \eqref{eq:equiconind},  we have
\begin{align}\label{eq:mut-zeta}
&\sup_{x >0}\big|\mathbb{P}(\hat{G}_n>x)- \mathbb{P}(|\hat{\bzeta}^{\dagger}|_{\infty} >x \,|\,\mathcal{X}_n, \mathcal{Y}_n,\mathcal{Z}_n)\big| \notag\\
&~~~~~~~~~~\le  \sup_{x >0}\big|\mathbb{P}(|\hat{\bzeta}|_{\infty} > x\,|\,\mathcal{X}_n, \mathcal{Y}_n,\mathcal{Z}_n) - \mathbb{P}(|\hat{\bzeta}^{\dagger}|_{\infty} > x\,|\,\mathcal{X}_n, \mathcal{Y}_n,\mathcal{Z}_n)\big| + o_{\rm p}(1) 
\end{align}
provided that  $\log \tilde{d} \ll \min\{n^{1/10}(s\log n)^{-1/2}, n^{1/8}(s^2\log n)^{-1/4}\}$ and $s\lesssim n^{1/5}(\log n)^{-2}$. 
Since $\min_{j \in [d]}\Theta_{j,j} \geq c_5$, by Lemma \ref{lem:cov}, if $\log \tilde{d} \ll \min\{n^{1/10}(\log n)^{-1/2}, n^{1/4}(s^2\log n)^{-1/2}\}$ and $s\lesssim n^{1/4}(\log n)^{-3/2}$, it holds that $\min_{j \in [d]}\hat{\Theta}_{j,j} \geq c_5/2$  with probability approaching  one.
Notice that $\hat{\ve}_{i,j}= \hat{U}_{i,j}- \hat{\balpha}_{j}^{\T}\hat{\bW}_{i}$ and $\hat{\delta}_{i,k}= \hat{V}_{i,k}- \hat{\bbeta}_{k}^{\T}\hat{\bW}_{i}$. Recall $\max_{i\in[n],\,j\in[p]}|\hat{U}_{i,j}| \le  \sqrt{2\log (n+1)}$ and  $\max_{i\in[n],\,k\in[q]}|\hat{V}_{i,k}| \le  \sqrt{2\log (n+1)}$. Analogously, it holds that $\max_{i\in[n],\,l\in[m]}|\hat{W}_{i,l}| \le  \sqrt{2\log (n+1)}$. 
By Lemma \ref{lem:coeff}, we have $\max_{i \in [n], \,j\in[p]}|\hat{\varepsilon}_{i,j}| \lesssim \sqrt{s\log (n+1)}$ and $\max_{i \in [n], \,k\in[q]}|\hat{\delta}_{i,k}| \lesssim \sqrt{s\log (n+1)}$ provided that $s\ll n^{1/2}(\log n)^{-1}\{\log(\tilde{d}n)\}^{-1}$ and $\log \tilde{d} \ll n^{1/10}(\log n)^{-1/2}$. Recall $\hat{\bet}_{i} = \hat{\boldsymbol{\ve}}_{i}  \otimes \hat{\bdelta}_{i} $ and $\bar{\hat{\bet}} = n^{-1}\sum_{i=1}^{n}\hat{\bet}_i$. Parallel to \eqref{eq:gamma-bound}, for either Mammen's or Rademacher multiplier $ \varpi_i $, we can also show
$\max_{i \in [n]}| \varpi_i (\hat{\bet}_{i} - \bar{\hat{\bet}})|_{\infty} \lesssim s\log n$.  Applying Proposition 2.1 of \cite{Chernozhukov2017} with 
$B_n=\tilde{C}(s \log n)^{3}$ for some universal constant $\tilde{C}>0$, it holds that 
\begin{align*}
\sup_{x >0}\big|\mathbb{P}(|\hat{\bzeta}|_{\infty} > x\,|\,\mathcal{X}_n, \mathcal{Y}_n,\mathcal{Z}_n)- \mathbb{P}(|\hat{\bzeta}^{\dagger}|_{\infty} > x\,|\,\mathcal{X}_n, \mathcal{Y}_n,\mathcal{Z}_n)\big| =O_{\rm p}\{ sn^{-1/6}\log^{7/6}(dn)\log n\}
\end{align*}
provided that $\log \tilde{d} \ll \min\{n^{1/10}(\log n)^{-1/2}, n^{1/4}(s^2\log n)^{-1/2}\}$ and $s\lesssim n^{1/4}(\log n)^{-3/2}$. Recall $\tilde{d}=p\vee q\vee m$.
Therefore, 
\begin{align*}
\sup_{x >0}\big|\mathbb{P}(|\hat{\bzeta}|_{\infty} > x\,|\,\mathcal{X}_n, \mathcal{Y}_n,\mathcal{Z}_n) -\mathbb{P}(|\hat{\bzeta}^{\dagger}|_{\infty} > x\,|\,\mathcal{X}_n, \mathcal{Y}_n,\mathcal{Z}_n)\big| =o_{{\rm p}}(1)
\end{align*}
provided that $\log \tilde{d} \ll \min\{n^{1/10}(\log n)^{-1/2}, n^{1/7}(s\log n)^{-6/7}\}$ and $s\ll n^{1/6}(\log n)^{-13/6}$. 
By \eqref{eq:mut-zeta}, under the null hypothesis  $\mathbb{H}_0$ in \eqref{eq:equiconind}, it holds that
\begin{align*}
\sup_{x >0}\big|\mathbb{P}(\hat{G}_n>x)- \mathbb{P}(|\hat{\bzeta}^{\dagger}|_{\infty} >x \,|\,\mathcal{X}_n, \mathcal{Y}_n,\mathcal{Z}_n)\big| =o_{{\rm p}}(1)
\end{align*} 
provided that 
\begin{align}\label{eq:ds-cd-boot}
    &\log \tilde{d} \ll \min\{n^{1/10}(s\log n)^{-1/2}, n^{1/8}(s^2\log n)^{-1/4},  n^{1/7}(s\log n)^{-6/7}\}\,, \notag\\
    &~~~~~~~~~~~~~~~~~~~~~~~~~~~~~ s\ll n^{1/6}(\log n)^{-13/6} \,.
\end{align}
Recall $\tilde{d} =p \vee  q\vee m$ with $p \lesssim n^{\varkappa_1}$, $q\lesssim n^{\varkappa_2}$ and $m\lesssim n^{\varkappa_3}$. If $s\ll n^{1/6}(\log n)^{-13/6}$, the restrictions \eqref{eq:ds-cd-boot} hold automatically for any constants $\varkappa_1 \ge 0$, $\varkappa_2 \ge 0$ and $\varkappa_3 \ge 0$.   Hence, we complete the proof of Theorem \ref{pro:12}(iii).
$\hfill\Box$

\section{Some useful inequalities for the proofs of auxiliary lemmas}\label{sec:lem}

To prove the auxiliary lemmas, we first introduce some  inequalities. 

\begin{inequality}[Dvoretzky–Kiefer–Wolfowitz inequality \citep{Massart1990}]\label{ieq:D-ecdf}
Let $\{\varphi_i\}_{i=1}^n$ be independent and identically distributed random variables with the distribution function $F_{\varphi}$. Write $\hat{F}_{\varphi}(x)=n^{-1}\sum_{i=1}^{n}I(\varphi_{i}\le x)$. For any $z>0$, it holds that 
\begin{align*}
\mathbb{P}\bigg\{\sup_{x\in \mathbb{R}} |\hat{F}_{\varphi}(x)- F_{\varphi}(x)| > z\bigg\} \le 2\exp(-2nz^2)\,.
\end{align*}
\end{inequality}

Let $\{{\psi}_i\}$ be a sequence of independent random variables on a measurable space $(S,\mathscr{S})$ and let $\{\psi_i^{(j)}\}$, $j\in[k]$, be $k$ independent copies of $\{\psi_i\}$. Let $f_{i_1,\ldots ,i_k}$ be families of functions of $k$ variables taking $S\times \cdots \times S$ into a Banach space $(\mathcal{B},\|\cdot\|)$. For any real valued measurable function $h$ on $S\times \cdots \times S$ and any random variables $\tilde{\psi}_{1},\ldots,\tilde{\psi}_{k}$ on the measurable space $(S,\mathscr{S})$, let 
$\mathbb{E}\{h(\tilde{\psi}_{1},\ldots,\tilde{\psi}_{k})\}$ be the expected value with respect to all the random variables $\tilde{\psi}_{1},\ldots,\tilde{\psi}_{k}$, and denote by  $\mathbb{E}_J\{h(\tilde{\psi}_{1},\ldots,\tilde{\psi}_{k})\}$  the expected value with respect to the random variables $\tilde{\psi}_j$'s with $j\in J\subset[k]$. We have the following inequalities. The proofs of Inequalities \ref{ieq:decoupling-u} and \ref{ieq:u-sta-2} are given, respectively, in \cite{de la Pena1995} and \cite{Gine2000}. 

\begin{inequality}[Decoupling inequality, Theorem 1 of \cite{de la Pena1995}]\label{ieq:decoupling-u}
For all $n\ge k \ge 2$ and $t>0$, there exists a numerical constant $C_k^*\in(0,\infty)$ depending on $k$ only so that
\begin{align*}
&\mathbb{P}\bigg\{\bigg\|\sum_{1\le i_1 \ne  \cdots \ne i_k \le n}f_{i_1,\ldots , i_k} \big(\psi^{(1)}_{i_1},\ldots, \psi^{(1)}_{i_k} \big)\bigg\| \ge t \bigg\}\\
&~~~~~~\le C_k^*\mathbb{P}\bigg\{C_k^*\bigg\|\sum_{1\le i_1 \ne  \cdots \ne i_k \le n}f_{i_1,\ldots , i_k} \big(\psi^{(1)}_{i_1},\ldots,  \psi^{(k)}_{i_k}\big) \bigg\|\ge t\bigg \}\,.
\end{align*}
\end{inequality}

\begin{inequality}[Theorem 3.3 of \cite{Gine2000}]\label{ieq:u-sta-2}
Let $h_{i_1,i_2}$ be real valued measurable functions on $S\times S$. There exists a universal constant $L_1\in(0,\infty)$ such that, if $h_{i_1,i_2}$ are bounded canonical kernels, then 
\begin{align*}
\mathbb{P}\bigg\{\bigg|\sum_{i_1,i_2=1 }^{n}h_{i_1,i_2}\big(\psi_{i_1}^{(1)}, \psi_{i_2}^{(2)}\big)\bigg|\geq t\bigg\} \le L_1 \exp \bigg\{ -\frac{1}{L_1} \min\bigg( \frac{t^2}{E^2}, \frac{t}{D}, \frac{t^{2/3}}{B^{2/3}}, \frac{t^{1/2}}{A^{1/2}} \bigg) \bigg\}
\end{align*}
for any $t>0$, where 
\begin{align*}
&~~~~~~~~~~~~A=\max_{i_1,i_2\in[n]}\sup_{x,y \in S  }\big| h_{i_1,i_2}(x,y)\big|\,, \quad	E^2=\sum_{i_1,i_2=1}^{n}\mathbb{E}\big\{h_{i_1,i_2}^2\big(\psi_{i_1}^{(1)},\psi_{i_2}^{(2)}\big)\big\}\,,\\	&~B^2=\bigg[\max_{i_2\in[n]}\sup_{y \in S  } \sum_{i_1=1}^{n}\mathbb{E}_{\{1\}}\big\{h_{i_1,i_2}^2\big(\psi_{i_1}^{(1)},y\big)\big\} \bigg] \vee    
\bigg[ \max_{i_1\in[n]} \sup_{x \in S }\sum_{i_2=1}^{n}\mathbb{E}_{\{2\}}\big\{h_{i_1,i_2}^2\big(x,\psi_{i_2}^{(2)}\big)\big\} \bigg]\,,\\
&D=\sup\bigg[ \mathbb{E}\bigg\{\sum_{i_1,i_2=1}^{n}h_{i_1,i_2}\big(\psi_{i_1}^{(1)},\psi_{i_2}^{(2)}\big)\tilde{\varphi}_{i_1}\big(\psi_{i_1}^{(1)}\big)\varphi_{i_2}\big(\psi_{i_2}^{(2)}\big)\bigg\} \\
&~~~~~~~~~~~~~~~~~~~~~~~~~~~~~~~~~~~~~~~~: \mathbb{E}\bigg\{\sum_{i_1=1}^{n}\tilde{\varphi}_{i_1}^2\big(\psi_{i_1}^{(1)}\big)\bigg\}\le 1,~
\mathbb{E}\bigg\{\sum_{i_2=1}^{n}\varphi_{i_2}^2\big(\psi_{i_2}^{(2)}\big)\bigg\}\le 1	\bigg]\,.
\end{align*}
\end{inequality}

\section{Proof of Lemma \ref{lem:uhv}}\label{sec:sub-huv}
To prove Lemma \ref{lem:uhv}, we need Lemma \ref{lem:ecdf} with its proof given in Section \ref{sec:sub-ecdf}.

\newtheorem{lemm}{Lemma}
\setcounter{lemm}{0}
\renewcommand{\thelemm}{F\arabic{lemm}}
\begin{lemm}\label{lem:ecdf}
There exist universal constants $K_1>0$ and $K_2>0$ such that, for any $x>0$,  
\begin{align*}
\max_{i\in[n],\, j\in[p]} \mathbb{P}\big\{| \hat{F}^{(i)}_{\bX,j}(X_{i,j})-F_{\bX,j}(X_{i,j}) | > x \big\} \leq&\,  K_1\exp(-K_2nx^2)\,,\\
\max_{i\in[n], \,k\in[q]}\mathbb{P}\big\{| \hat{F}^{(i)}_{\bY,k}(Y_{i,k})-F_{\bY,k}(Y_{i,k}) | > x \big\} \leq&\, K_1\exp(-K_2nx^2)\,.
\end{align*}
\end{lemm}

Recall $\hat{U}_{i,j}=\Phi^{-1} \{n(n+1)^{-1}\hat{F}_{\bX,j}(X_{i,j})\}$,  $U_{i,j}=\Phi^{-1}\{F_{\bX,j}(X_{i,j})\}$  and $U_{i,j}^{*} =  U_{i,j}I(|U_{i,j}|\le M_1) + M_1 \cdot{\rm sign}(U_{i,j})I(|U_{i,j}|>M_1)$ with $M_1=\sqrt{\kappa_1\log n}$ for some constant $\kappa_1 \in(1, 2)$. Given  $ M_2 =\sqrt{\kappa_2\log n}$ for some constant $\kappa_2 \in(0,1)$, we have
\begin{align}\label{eq:id}
&\frac{1}{n}\sum_{i=1}^{n} (\hat{U}_{i,j} - U_{i,j}^{*}) V_{i,k}^{*} \notag\\
&~~~~~~= \frac{1}{n}\sum_{i=1}^{n} (\hat{U}_{i,j} - U_{i,j}^{*}) V_{i,k}^{*} I(|U_{i,j}|\le M_1) + \frac{1}{n}\sum_{i=1}^{n} (\hat{U}_{i,j} - U_{i,j}^{*}) V_{i,k}^{*} I(|U_{i,j}| > M_1) \notag\\
&~~~~~~=  \underbrace{\frac{1}{n}\sum_{i=1}^{n} \bigg[\Phi^{-1} \bigg\{\frac{n}{n+1}\hat{F}_{\bX,j}(X_{i,j})\bigg\} - \Phi^{-1}\{F_{\bX,j}(X_{i,j})\} \bigg] V_{i,k}^{*} I(|U_{i,j}|\le M_2)}_{\textrm{I}_{1}(j,k)}\notag  \\
&~~~~~~~~~~~+ \underbrace{\frac{1}{n}\sum_{i=1}^{n} \bigg[\Phi^{-1} \bigg\{\frac{n}{n+1}\hat{F}_{\bX,j}(X_{i,j})\bigg\} - \Phi^{-1}\{F_{\bX,j}(X_{i,j})\} \bigg] V_{i,k}^{*} I(M_2 < |U_{i,j}|\le M_1)}_{\textrm{I}_{2}(j,k)}\notag  \\
&~~~~~~~~~~~ + \underbrace{\frac{1}{n}\sum_{i=1}^{n} (\hat{U}_{i,j} - U_{i,j}^{*}) V_{i,k}^{*} I(|U_{i,j}| > M_1) }_{\textrm{I}_{3}(j,k)}  \,.  
\end{align}
As we will show in Sections \ref{sec:sub-I1}--\ref{sec:sub-I3}, 
\begin{align}\label{eq:bdI1}
\max_{j\in[p],\,k\in[q]}|{\rm I}_{1}(j,k)|=  O_{\rm p} \{n^{-1+\kappa_2/2} (\log n)^{1/2}\log (dn)\} 
\end{align}
under the null hypothesis  $\mathbb{H}_0$ in \eqref{eq:equind} provided that $ \log d \ll n^{1-\kappa_2}(\log n)^{-1} $, 
\begin{align}\label{eq:i2}
\max_{j\in[p],\,k\in[q]}|{\rm I}_{2}(j,k)| =O_{\rm p} \{n^{-1/2-\kappa_2/4} (\log n)^{-1/4}\log^{1/2} (dn)\}  
\end{align}
provided that $ \log d \ll n^{1-\kappa_1/2} (\log n)^{-1/2}$, 
and
\begin{align}\label{eq:bdI2}
\max_{j\in[p],\,k\in[q]}|{\rm I}_{3}(j,k)|= O_{\rm p}\{n^{-\kappa_1/2} (\log n)^{1/2}\}
\end{align}
provided that  $ \log d \lesssim n^{1-\kappa_1/2}(\log n)^{-1/2}$.
Together with \eqref{eq:bdI1}--\eqref{eq:bdI2}, \eqref{eq:id} implies
\begin{align*}
\max_{j\in[p],\,k\in[q]}\bigg|\frac{1}{\sqrt{n}}\sum_{i=1}^{n}(\hat{U}_{i,j}-U_{i,j}^{*})V_{i,k}^{*}\bigg| =&~  O_{\rm p} \{n^{-(1 -\kappa_2)/2} (\log n)^{1/2}\log (dn)\} \\
&+  O_{\rm p} \{n^{-\kappa_2/4} (\log n)^{-1/4}\log^{1/2} (dn)\}  + O_{\rm p}\{n^{-(\kappa_1-1)/2} (\log n)^{1/2}\}
\end{align*}
under the null hypothesis  $\mathbb{H}_0$ in \eqref{eq:equind} provided that $ \log d \ll \min\{n^{1-\kappa_1/2} (\log n)^{-1/2}, n^{1-\kappa_2}(\log n)^{-1} \}$.
In our Gaussian approximation theory used in the proof of Proposition \ref{pro:1}, we need to require the selected $\kappa_2\in(0,1)$ to satisfy the conditions: 
$$
n^{-(1 -\kappa_2)/2} (\log n)^{1/2}\log (dn) \ll (\log d)^{-1/2}~~\mbox{and}~~n^{-\kappa_2/4} (\log n)^{-1/4}\log^{1/2} (dn) \ll (\log d)^{-1/2}\,,$$ which are equivalent to 
\begin{align*}
\log d \ll \min\{n^{(1-\kappa_2)/3}(\log n)^{-1/3}, n^{\kappa_2/4}(\log n)^{1/4}\}\,.
\end{align*}
To allow $d$ to diverge with $n$ as fast as possible, we select $\kappa_2 = 4/7$. With such selected $\kappa_2$, under the null hypothesis  $\mathbb{H}_0$ in \eqref{eq:equind},  we have
\begin{align}\label{eq:lem-2-reta}
\max_{j\in[p],\,k\in[q]}\bigg|\frac{1}{\sqrt{n}}\sum_{i=1}^{n}(\hat{U}_{i,j}-U_{i,j}^{*})V_{i,k}^{*}\bigg| =&~ O_{\rm p}\{n^{-(\kappa_1-1)/2} (\log n)^{1/2}\} +   O_{\rm p} \{n^{-3/14} (\log n)^{1/2}\log  (dn)\}  \notag\\
&+ O_{\rm p} \{n^{-1/7} (\log n)^{-1/4}\log^{1/2} (dn)\} 
\end{align}
under the null hypothesis  $\mathbb{H}_0$ in \eqref{eq:equind}, provided that $ \log d \ll \min\{n^{1-\kappa_1/2} (\log n)^{-1/2}, n^{3/7}(\log n)^{-1} \}$.
Identically, we can also show such convergence rate holds for $\max_{j\in[p],\,k\in[q]}|n^{-1/2}\sum_{i=1}^{n}(\hat{V}_{i,k}-V_{i,k}^{*})U_{i,j}^{*}|$. We complete the proof of Lemma \ref{lem:uhv}. 
$\hfill\Box$

\subsection{Proof of Lemma \ref{lem:ecdf}}\label{sec:sub-ecdf}

Recall
\begin{align*}
\hat{F}^{(i)}_{\bX,j}(X_{i,j})=\frac{1}{n-1}\sum_{s:\,s\ne i}I(X_{s,j}\le X_{i,j}) 
= \frac{n}{n-1}\hat{F}_{\bX,j}(X_{i,j}) -\frac{1}{n-1}
\end{align*}
with $\hat{F}_{\bX,j}(x)=n^{-1}\sum_{s=1}^nI(X_{s,j}\leq x)$.
For any $x>2(n-1)^{-1}$, we then have
\begin{align*}
&\mathbb{P}\big\{| \hat{F}^{(i)}_{\bX,j}(X_{i,j})-F_{\bX,j}(X_{i,j}) | > x \big\} \\
&~~~~~~~= 	\mathbb{P}\big\{ | \hat{F}_{\bX,j}(X_{i,j})-F_{\bX,j}(X_{i,j})-(n-1)^{-1} +(n-1)^{-1}\hat{F}_{\bX,j}(X_{i,j}) | > x \big\} \\
&~~~~~~~\le \mathbb{P}\big\{ | \hat{F}_{\bX,j}(X_{i,j})-F_{\bX,j}(X_{i,j}) | > x-2(n-1)^{-1} \big\}\\ 
&~~~~~~~\le  2\exp\big[-2n\{x-2(n-1)^{-1}\}^2\big]\\
&~~~~~~~\le 2 \exp\bigg\{ -(2-C)nx^2 + \frac{8nx}{n-1}\bigg\} \exp(-Cnx^2)\,,	
\end{align*}
where the second inequality follows by Inequality \ref{ieq:D-ecdf}.  
Restricting $C\in(0,2)$, we have 
\begin{align*}
\exp\bigg\{ -(2-C)nx^2 + \frac{8nx}{n-1}\bigg\} \le \exp\bigg\{\frac{16n}{(2-C)(n-1)^2}\bigg\}\le \frac{\bar{C}}{2}
\end{align*}
for any $n\geq2$, which implies that there exist universal constants $C>0$ and $\bar{C}>0$ such that 
\begin{align*}
\mathbb{P}\big\{| \hat{F}^{(i)}_{\bX,j}(X_{i,j})-F_{\bX,j}(X_{i,j}) | > x \big\} \leq \bar{C}\exp(-Cnx^2)
\end{align*}
for any $n\geq 2$ and $x>2(n-1)^{-1}$. For above specified $C>0$, there exists a universal constant $\breve{C}>0$ such that 
\begin{align*}
\breve{C} \le \exp\bigg\{-\frac{4Cn}{(n-1)^2}\bigg\}
\end{align*}
for any $n\geq 2$. Select a universal constant $\tilde{C}>\breve{C}^{-1}$. Then, for any $n\geq2$ and $0<x\leq 2(n-1)^{-1}$,
\begin{align*}
\mathbb{P}\big\{| \hat{F}^{(i)}_{\bX,j}(X_{i,j})-F_{\bX,j}(X_{i,j}) | > x \big\} \leq 1 \le \tilde{C} \exp\{-4Cn/(n-1)^2\} \le \tilde{C}\exp(-Cnx^2)\,.
\end{align*}
Hence,  for any $n\geq 2$ and $x>0$, it holds that 
\begin{align*}
\mathbb{P}\big\{| \hat{F}^{(i)}_{\bX,j}(X_{i,j})-F_{\bX,j}(X_{i,j}) | > x \big\} \leq (\bar{C}\vee\tilde{C})\exp(-Cnx^2)\,.
\end{align*}
Analogously, we can also establish the same upper bound for $\mathbb{P}\{| \hat{F}^{(i)}_{\bY,k}(Y_{i,k})-F_{\bY,k}(Y_{i,k}) | > x \}$. 
We complete the proof of Lemma \ref{lem:ecdf}.
$\hfill\Box$
\subsection{Convergence rate of ${\max_{j\in[p],\,k\in[q]}|{\rm I_{1}}(j,k)|}$}\label{sec:sub-I1}
For any $l \in \mathbb{Z}_{+}$, let $f^{(l)}(x)$ be the $l$-th derivative of $f(x)$. When there is no confusion, we also denote the first and second derivatives of $f(x)$ by $f'(x)$ and $f''(x)$, respectively. Notice that $\Phi^{-1}(x)$ is infinitely differentiable at any $x\in (0,1)$. By direct calculation, we have
\begin{align}\label{eq:Derivative12}
&(\Phi^{-1})'(x) = \sqrt{2\pi}\exp\bigg[ \frac{1}{2}\big\{\Phi^{-1}(x)\big\}^2\bigg]
\end{align}
for any $x\in (0,1)$.  Let $P_{l}(x)$ be a polynomial in $x$ of degree $l$ satisfying
$P_{0}(x)=1$ and  $P_{l}(x)=P_{l-1}'(x)+lxP_{l-1}(x)$ for any $l\in \mathbb{Z}_{+}$. By mathematical induction, we can show $(\Phi^{-1})^{(l)}(x) = P_{l-1}\{\Phi^{-1}(x)\} \{(\Phi^{-1})'(x)\}^{l}$  for any $l \in \mathbb{Z}_{+}$ and $x\in (0,1)$, and there exists a universal constant $\bar{C}>0$ such that 
\begin{align}\label{eq:derivative_n}
\big|(\Phi^{-1})^{(l)}(x)\big| \leq  &~ \bar{C}^{l} l! |\Phi^{-1}(x)|^{l-1} \exp\bigg[\frac{l}{2}\big\{\Phi^{-1}(x)\big\}^2\bigg] \,.
\end{align}
Notice that
\begin{align*}
&{\rm {I}_{1}}(j,k)\\
&~~~=  \underbrace{\frac{1}{n}\sum_{i=1}^{n}(\Phi^{-1})^{'}\{F_{\bX,j}(X_{i,j})\} \bigg\{\frac{n}{n+1}\hat{F}_{\bX,j}(X_{i,j})- F_{\bX,j}(X_{i,j})  \bigg\}V_{i,k}^{*}I{(|U_{i,j}| \le M_2)}}_{\textrm{I}_{11}(j,k)} \notag\\ 
&~~~~~~+ \underbrace{\sum_{l=2}^{\infty}\frac{1}{n\cdot l!}\sum_{i=1}^{n}(\Phi^{-1})^{(l)}\{F_{\bX,j}(X_{i,j})\}\bigg\{\frac{n}{n+1}\hat{F}_{\bX,j}(X_{i,j})-  F_{\bX,j}(X_{i,j}) \bigg\}^{l}V_{i,k}^{*}I{(|U_{i,j}| \le M_2)} }_{\textrm{I}_{12}(j,k)} \,.
\end{align*}
As we will show in Sections \ref{sec:sub-i11} and \ref{sec:sub-i12}, 
\begin{align}\label{eq:i11}
\max_{j\in[p],\,k\in[q]}|{\rm I}_{11}(j,k)|= O_{\rm p}(n^{-1} M_1e^{M_2^2/2} \log d)
\end{align}
under the null hypothesis  $\mathbb{H}_0$ in \eqref{eq:equind} provided that $\log d \lesssim ne^{-M_2^2/2} M_2$, and 
\begin{align}\label{eq:i12}
\max_{j\in[p],\, k\in[q]}|{\rm I}_{12}(j,k)| = O_{\rm p} \{n^{-1}M_1 e^{M_2^2/2}\log (dn)\}
\end{align}
provided that $ \log (dn) \ll ne^{-M_2^2} M_2^{-2}$. Recall   $ M_1 =\sqrt{\kappa_1\log n}$ and $ M_2 =\sqrt{\kappa_2\log n}$ for some constants $\kappa_1 \in(1,2)$ and $\kappa_2\in(0,1)$. Combining \eqref{eq:i11} and \eqref{eq:i12}, we have \eqref{eq:bdI1} holds. $\hfill\Box$

\subsubsection{Proof of \eqref{eq:i11}}\label{sec:sub-i11}
Recall
\begin{align*}
\hat{F}^{(i)}_{\bX,j}(X_{i,j})=\frac{1}{n-1}\sum_{s:\,s\ne i}I(X_{s,j}\le X_{i,j})\,.
\end{align*}
Then, for any $i\in [n]$ and $j \in [p]$, we have 
\begin{align}\label{eq:n+1f-f}
\frac{n}{n+1} \hat{F}_{\bX,j}(X_{i,j}) -  F_{\bX,j}(X_{i,j})=&~  \frac{1}{n+1}\sum_{s=1}^{n}I(X_{s,j}\le X_{i,j}) - F_{\bX,j}(X_{i,j})\\
=&~ \frac{n-1}{n+1} \big\{\hat{F}^{(i)}_{\bX,j}(X_{i,j}) -F_{\bX,j}(X_{i,j})\big\} - \frac{2}{n+1}F_{\bX,j}(X_{i,j}) + \frac{1}{n+1} \,.\notag
\end{align}
By \eqref{eq:Derivative12} and $U_{i,j}=\Phi^{-1}\{F_{\bX,j}(X_{i,j})\}$, it then holds that  
\begin{align}\label{eq:i11-dec}
{\rm I_{11}}(j,k)=&~ \frac{n-1}{n(n+1)}\sum_{i=1}^{n}(\Phi^{-1})^{'}\{F_{\bX,j}(X_{i,j})\} \big\{\hat{F}^{(i)}_{\bX,j}(X_{i,j})- F_{\bX,j}(X_{i,j})  \big\}V_{i,k}^{*}I{(|U_{i,j}| \le M_2)} \notag\\
&+ \frac{1}{n(n+1)}\sum_{i=1}^{n}(\Phi^{-1})^{'}\{F_{\bX,j}(X_{i,j})\} \big\{1-2 F_{\bX,j}(X_{i,j}) \big\}V_{i,k}^{*}I{(|U_{i,j}| \le M_2)} \notag\\
=&~\frac{1}{n(n+1)}\sum_{1\leq i_1 \ne i_2 \leq n} \bigg[(\Phi^{-1})^{'} \{F_{\bX,j}(X_{i_1,j})\} \big\{I(X_{i_2,j} \le X_{i_1,j}) -F_{\bX,j}(X_{i_1,j}) \big\}V_{i_1,k}^{*}\notag\\
&~~~~~~~~~~~~~~~~~~~~~~~~~~~~~~~\times I{(|U_{i_1,j}| \le M_2)}	\bigg] \notag \\
&+ \frac{1}{n(n+1)}\sum_{i=1}^{n}(\Phi^{-1})^{'}\{F_{\bX,j}(X_{i,j})\} \big\{1-2 F_{\bX,j}(X_{i,j}) \big\}V_{i,k}^{*}I{(|U_{i,j}| \le M_2)}  \notag\\
=&~ \underbrace{\frac{\sqrt{2\pi}}{n(n+1)}\sum_{1\leq i_1 \ne i_2 \leq n}e^{U_{i_1,j}^2/{2}}\big\{I(U_{i_2,j} \le U_{i_1,j}) -\Phi(U_{i_1,j})\big\}  V_{i_1,k}^{*}I{(|U_{i_1,j}| \le M_2)}}_{\textrm{I}_{111}(j,k)} \notag\\
&+ \underbrace{\frac{\sqrt{2\pi}}{n(n+1)}\sum_{i=1}^{n}e^{U_{i,j}^2/{2}} \big\{1-2\Phi(U_{i,j})\big\} V_{i,k}^{*}I{(|U_{i,j}| \le M_2)} }_{\textrm{I}_{112}(j,k)} \,.
\end{align}
Given $(j,k)$, write $\bT_{i}=(U_{i,j},V_{i,k})$  for any $i\in[n]$. Define 
\begin{align*}
\bar{\varpi}_1(\bT_{i_1},\bT_{i_2}) = &~ e^{U_{i_1,j}^2/2} \big\{I(U_{i_2,j} \le U_{i_1,j}) -\Phi(U_{i_1,j})\big\} V_{i_1,k}^{*}I{(|U_{i_1,j}| \le M_2)}
\end{align*}
for any $i_1 \neq i_2$. Recall $V_{i,k}^{*} =  V_{i,k}I(|V_{i,k}|\le M_1) +   M_1 \cdot{\rm sign}(V_{i,k}) I(|V_{i,k}|>M_1) $. Such defined $\bar{\varpi}_1(\cdot,\cdot)$ is a bounded kernel. Let $\{\bT_{i}^{(1)}\}$ and $\{\bT_{i}^{(2)}\}$ be two independent copies of $\{\bT_{i}\}$ with $\bT_{i}^{(1)}=\{U_{i,j}^{(1)},V_{i,k}^{(1)}\}$ and $\bT_{i}^{(2)}=\{U_{i,j}^{(2)},V_{i,k}^{(2)}\}$. We define $V_{i,k}^{(1),*}$ in the same manner as $V_{i,k}^{*}$ but with replacing $V_{i,k}$ by $V_{i,k}^{(1)}$. Then we also have $\mathbb{E}\{V_{i,k}^{(1),*}\}=0$.
Since $U_{i,j},V_{i,k} \sim \mathcal{N}(0,1)$ are independent under the null hypothesis  $\mathbb{H}_0$ in \eqref{eq:equind}, we have
\begin{align*}
&\mathbb{E}_{\{1\}}\big\{\bar{\varpi}_1\big(\bT_{i_1}^{(1)},\bT_{i_2}^{(2)}\big)\big\}\\
&~~~= 	\mathbb{E}\big(e^{\{U_{i_1,j}^{(1)}\}^2/2} \big[I\big\{U_{i_2,j}^{(2)} \le U_{i_1,j}^{(1)}\big\} -\Phi\big(U_{i_1,j}^{(1)}\big)\big]I{\big\{|U_{i_1,j}^{(1)}| \le M_2\big\}} \,\big|\,U_{i_2,j}^{(2)}\big)   \mathbb{E}\big\{V_{i_1,k}^{(1),*} \big\} =0 \,,\\	
&\mathbb{E}_{\{2\}}\big\{\bar{\varpi}_1\big(\bT_{i_1}^{(1)},\bT_{i_2}^{(2)}\big)\big\}\\
&~~~= e^{\{U_{i_1,j}^{(1)}\}^2/2}  V_{i_1,k}^{(1),*}I{\big\{|U_{i_1,j}^{(1)}| \le M_2\big\}} \mathbb{E}\big[I\big\{U_{i_2,j}^{(2)} \le U_{i_1,j}^{(1)}\big\} -\Phi\big(U_{i_1,j}^{(1)}\big)  \,|\,U_{i_1,j}^{(1)}\big]=0\,,
\end{align*}
which implies $\bar{\varpi}_1(\cdot,\cdot)$ is a bounded  canonical kernel. Due to 
\begin{align*}
{\rm I}_{111}(j,k)=\frac{\sqrt{2\pi}}{n(n+1)}\sum_{1\le i_1 \ne i_2 \le n}  \bar{\varpi}_1(\bT_{i_1},\bT_{i_2})\,,
\end{align*} 
by  Inequalities \ref{ieq:decoupling-u} and \ref{ieq:u-sta-2}, we have
\begin{align*}
\mathbb{P}\{|{\rm I}_{111}(j,k)| \ge x\} \le &~ C_1 \mathbb{P}\bigg\{C_1\bigg|\sum_{1\le i_1\ne i_2\le n} \bar{\varpi}_1\big(\bT_{i_1}^{(1)},\bT_{i_2}^{(2)}\big)\bigg| \ge \frac{n(n+1)x}{\sqrt{2\pi}}\bigg\}\\
\le&~ C_2\exp\bigg\{-\frac{1}{C_2}\min \bigg(\frac{n^2M_2x^2}{e^{M_2^2/2}}, \frac{nx}{M_1e^{M_2^2/2}},\frac{nx^{2/3}}{M_1^{2/3}e^{M_2^2/3}},
\frac{nx^{1/2}}{M_1^{1/2}e^{M_2^2/4}} \bigg)\bigg\}
\end{align*}
for any $x>0$ under the null hypothesis  $\mathbb{H}_0$ in \eqref{eq:equind}.
Recall $d=pq$. Notice that above inequality holds for any $j\in[p]$ and $k\in[q]$.  Hence, it holds that
\begin{align}\label{eq:i111}
\max_{j\in[p],\,k\in[q]}|{\rm I}_{111}(j,k)| =  O_{\rm p}(n^{-1} M_1e^{M_2^2/2} \log d)
\end{align}
provided that $\log d \lesssim n$.

Recall $U_{i,j},V_{i,k} \sim \mathcal{N}(0,1)$ and $V_{i,k}^{*} =  V_{i,k}I(|V_{i,k}|\le M_1) +   M_1 \cdot{\rm sign}(V_{i,k}) I(|V_{i,k}|>M_1) $. Let 
\begin{align*}
\mu_1(i,j,k) = \mathbb{E} \big[e^{U_{i,j}^2/2}  \big\{1-2\Phi(U_{i,j})\big\} V_{i,k}^{*}I{(|U_{i,j}| \le M_2)}\big]\,.
\end{align*} 
We have
\begin{align*}
&\max_{i\in[n],\,j\in[p],\, k\in[q]} |\mu_1(i,j,k)| \le M_1  \max_{i\in[n],\, j\in[p]}\mathbb{E} \big\{e^{U_{i,j}^2/2}   I{(|U_{i,j}| \le M_2)}\big\} \le M_1 M_2 \,,\\
&\max_{i\in[n],\,i\in[n],\, k\in[q]}\var \big[e^{U_{i,j}^2/2}   \big\{1-2\Phi(U_{i,j})\big\}  V_{i,k}^{*}I{(|U_{i,j}| \le M_2)}\big] \\
&~~~~~~~~~~~~~~~~~~~~~~~~~~\le M_1^2 \max_{i\in[n],\,j\in[p]}\mathbb{E} \big\{e^{U_{i,j}^2}   I{(|U_{i,j}| \le M_2)}\big\}   \lesssim M_1^2M_2^{-1}e^{M_2^2/2}\,.
\end{align*}
Recall $d=pq$. By Bonferroni inequality and Bernstein inequality, for any $x>0$, it holds that
\begin{align}\label{eq:i112-tail}
&\mathbb{P}\bigg(\max_{j\in[p],\, k\in[q]}\bigg|\frac{1}{n}\sum_{i=1}^{n} \big[e^{U_{i,j}^2/2} \big\{1-2\Phi(U_{i,j})\big\} V_{i,k}^{*}I{(|U_{i,j}| \le M_2)} -  \mu_1(i,j,k)\big]\bigg| > x\big)\notag\\
&~~~~~~~~~~  \le 2d \exp\bigg(-\frac{nx^2}{C_3M_1^{2}M_2^{-1} e^{M_2^2/2} + C_4M_1e^{M_2^2/2}x }\bigg)\,,
\end{align}
which implies
\begin{align*}
&\max_{j\in[p],\, k\in[q]}\bigg|\frac{1}{n}\sum_{i=1}^{n} \big[e^{U_{i,j}^2/2}  \big\{1-2\Phi(U_{i,j})\big\} V_{i,k}^{*}I{(|U_{i,j}| \le M_2)} -  \mu_1(i,j,k)\big]\bigg| \\
&~~~~~~~~~=O_{\rm p} \{n^{-1/2}M_1 M_2^{-1/2}e^{M_2^2/4}(\log d)^{1/2} \} + O_{\rm p} (n^{-1}M_1e^{M_2^2/2} \log d)\,.
\end{align*}
Then we have
\begin{align*}
\max_{j\in[p],\,k\in[q]}|{\rm I}_{112}(j,k)| =  O_{\rm p}(n^{-1} M_1 M_2) 
\end{align*}
provided that $\log d \lesssim ne^{-M_2^2/2} M_2$. 
Together with \eqref{eq:i111}, by \eqref{eq:i11-dec}, we complete the proof of \eqref{eq:i11}. $\hfill\Box$

\subsubsection{Proof of \eqref{eq:i12}}\label{sec:sub-i12}
Define the event
\begin{align*}
\mathcal{H}_1 = \bigg\{ \max_{i\in[n],\,j\in[p]}|\hat{F}^{(i)}_{\bX,j}(X_{i,j})-  F_{\bX,j}(X_{i,j})| \le C_5 n^{-1/2}\log^{1/2}(pn) \bigg\}
\end{align*}
with $C_5 =2K_2^{-1/2}$, where $K_2$ is specified in Lemma \ref{lem:ecdf}. Restricted on $\mathcal{H}_1$, by  \eqref{eq:derivative_n} and \eqref{eq:n+1f-f}, it holds that
\begin{align}\label{eq:i12-bound}
|{\rm I}_{12}(j,k)| \le&~ \sum_{l=2}^{\infty}  M_1 C_6^{l} \bigg\{\frac{\log(pn)}{n}\bigg\}^{l/2}  \bigg\{ \frac{1}{n }\sum_{i=1}^{n}|U_{i,j}|^{l-1}   e^{lU_{i,j}^2/2}  I{(|U_{i,j}| \le M_2)} \bigg\} \notag\\
\le &~   \sum_{l=2}^{\infty} \bigg\{\frac{C_7 M_2  e^{M_2^2/2}\log^{1/2}(pn)}{n^{1/2}}   \bigg\}^{l-2}  \times \frac{M_1M_2 \log(pn)}{n}  \times   \frac{1}{n }\sum_{i=1}^{n}  e^{U_{i,j}^2} I{(|U_{i,j}| \le M_2)} \notag\\
\le &~ \frac{C_8 M_1M_2 \log(pn)}{n}  \times  \frac{1}{n }\sum_{i=1}^{n}   e^{U_{i,j}^2} I{(|U_{i,j}| \le M_2)} 
\end{align}
provided that $ \log (pn) \ll ne^{-M_2^2} M_2^{-2}$, which implies
\begin{align}\label{eq:i12-tail-h1}
&\mathbb{P} \bigg\{\max_{j\in[p],\, k\in[q]}|{\rm I}_{12}(j,k)| > \frac{C_{\epsilon}M_1 e^{M_2^2/2}\log (pn)}{n}, \,\mathcal{H}_1\bigg\} \notag\\
&~~~~~~~~\le\mathbb{P} \bigg\{\max_{j\in[p]} \frac{1}{n }\sum_{i=1}^{n}   e^{U_{i,j}^2 } I{(|U_{i,j}| \le M_2)} > \frac{C_{\epsilon}e^{M_2^2/2}}{C_8M_2}\bigg\}\,.
\end{align}
Recall $U_{i,j} \sim \mathcal{N}(0,1)$. We then have 
\begin{align*}
\max_{i\in[n],\,j\in[p] }\mathbb{E}\big\{ e^{U_{i,j}^2}I(|U_{i,j}| \le M_2)\big\} \lesssim &~ M_2^{-1} e^{ M_2^2/2}\,,\\ 
\max_{i\in[n],\,j\in[p] }\var\big\{e^{U_{i,j}^2}I(|U_{i,j}| \le M_2)\big\} \lesssim&~ M_2^{-1} e^{3M_2^2/2}\,.
\end{align*}
By Bonferroni inequality and Bernstein inequality, it holds that
\begin{align}\label{eq:i12-tail}
&\mathbb{P} \bigg( \max_{j\in[p]}\bigg|\frac{1}{n}\sum_{i=1}^{n} \big[ e^{U_{i,j}^2} I(|U_{i,j}| \le M_2) - \mathbb{E}\{e^{U_{i,j}^2} I(|U_{i,j}| \le M_2)\} \big] \bigg| > x\bigg)\notag\\
&~~~~~~~~~\le 2p\exp\bigg(-\frac{nx^2}{C_{9}M_2^{-1} e^{3M_2^2/2} + C_{10}e^{M_2^2}x }\bigg)
\end{align}
for any $x>0$, which implies
\begin{align*}
&\max_{j\in[p]}\bigg|\frac{1}{n}\sum_{i=1}^{n} \big[ e^{U_{i,j}^2} I(|U_{i,j}| \le M_2) - \mathbb{E}\{ e^{U_{i,j}^2} I(|U_{i,j}| \le M_2)\} \big] \bigg| \\
&~~~~~~~~~~~~~~~~= O_{\rm p} \big\{n^{-1/2}M_2^{-1/2}e^{3M_2^2/4}(\log p)^{1/2} \big\} +  O_{\rm p} \big(n^{-1}e^{M_2^2} \log p \big) \,.
\end{align*}
We then have
\begin{align}\label{eq:bern-bound}
\max_{j\in[p]} \frac{1}{n}\sum_{i=1}^{n}  e^{U_{i,j}^2} I(|U_{i,j}| \le M_2)   = O_{\rm p}(M_2^{-1}e^{M_2^2/2})  
\end{align}
provided that $ \log p \lesssim ne^{-M_2^2/2} M_2^{-1}$. Hence, for any $\epsilon >0$, there exists $C_{\epsilon} >0$ such that
\begin{align*}
\mathbb{P} \bigg\{\max_{j\in[p]} \frac{1}{n }\sum_{i=1}^{n}   e^{ U_{i,j}^2 } I{(|U_{i,j}| \le M_2)} > \frac{C_{\epsilon}e^{M_2^2/2}}{C_8M_2}\bigg\} \le \epsilon\,,
\end{align*}
which implies, by \eqref{eq:i12-tail-h1},
\begin{align*}
\mathbb{P} \bigg\{\max_{j\in[p],\, k\in[q]}|{\rm I}_{12}(j,k)| > \frac{C_{\epsilon}M_1 e^{M_2^2/2}\log (pn)}{n}, \,\mathcal{H}_1\bigg\} \le \epsilon 
\end{align*}
provided that $ \log (pn) \ll ne^{-M_2^2} M_2^{-2}$.
Recall $C_5 =2K_2^{-1/2}$. By Lemma \ref{lem:ecdf}, we have 
\begin{align}\label{eq:h1-c}
\mathbb{P}(\mathcal{H}_{1}^{\rm c}) \le&~ np  \max_{i\in[n],\, j\in[p]} \mathbb{P}\big\{| \hat{F}^{(i)}_{\bX,j}(X_{i,j})-F_{\bX,j}(X_{i,j}) | > C_5 n^{-1/2}\log^{1/2}(pn) \big\} \notag\\
\le&~ np   K_1\exp\{-K_2C_5^2 \log(pn)\} \le K_1 (pn)^{-3} \,.
\end{align}
Therefore, if $ \log (pn) \ll ne^{-M_2^2} M_2^{-2}$, it then holds that
\begin{align*}
& \mathbb{P} \bigg\{\max_{j\in[p],\, k\in[q]}|{\rm I}_{12}(j,k)| > \frac{C_{\epsilon}M_1 e^{M_2^2/2}\log (pn)}{n}\bigg\} \\
&~~~~~~~~~~~~~~ \leq \mathbb{P} \bigg\{\max_{j\in[p],\, k\in[q]}|{\rm I}_{12}(j,k)| >\frac{C_{\epsilon}M_1 e^{M_2^2/2}\log (pn)}{n}, \,\mathcal{H}_1\bigg\} + \mathbb{P}(\mathcal{H}_{1}^{\rm c})\le \epsilon + K_1 (pn)^{-3}
\end{align*}
for any $\epsilon >0$, which implies
\begin{align*}
\max_{j\in[p],\, k\in[q]}|{\rm I}_{12}(j,k)| = O_{\rm p} \{n^{-1}M_1 e^{M_2^2/2}\log (pn)\}
\end{align*}
provided that $ \log (pn) \ll ne^{-M_2^2} M_2^{-2}$. Recall $d=pq$. We complete the proof of \eqref{eq:i12}.
$\hfill\Box$

\subsection{Convergence rate of $\max_{j\in[p],\,k\in[q]}|{\rm I}_{2}(j,k)|$}\label{sec:sub-I2}
Notice  that $\Phi^{-1}(x)$ is infinitely differentiable at any $x\in (0,1)$. We have
\begin{align*}
{\rm I}_{2}(j,k) =&~\sum_{l=1}^{\infty}\frac{1}{n\cdot l!}\sum_{i=1}^{n}\bigg[(\Phi^{-1})^{(l)}\{F_{\bX,j}(X_{i,j})\}\bigg\{\frac{n}{n+1}\hat{F}_{\bX,j}(X_{i,j})-  F_{\bX,j}(X_{i,j}) \bigg\}^{l}\\
&~~~~~~~~~~~~~~~~~~\times V_{i,k}^{*}I{( M_2<|U_{i,j}| \le M_1)} \bigg]\,.
\end{align*}
Let $K(U_{i,j}, p,n) = 4 n^{-1/2}  [\Phi(U_{i,j}) \{1-\Phi(U_{i,j})\}]^{1/2} \log^{1/2}(pn)  + 7n^{-1} \log(pn)$. Define the event 
\begin{align*}
\mathcal{H}_2 =  \bigcap_{i\in[n],\,j\in[p]}\big\{|\hat{F}_{\bX,j}^{(i)}(X_{i,j})-  F_{\bX,j}(X_{i,j})| \le K(U_{i,j}, p,n)\big\} \,.
\end{align*}
Notice that 
\begin{align*}
\hat{F}^{(i)}_{\bX,j}(X_{i,j})-  F_{\bX,j}(X_{i,j}) = &~ \frac{1}{n-1}\sum_{s:\,s\ne i} \big\{I(X_{s,j}\le X_{i,j})  - F_{\bX,j}(X_{i,j})\big\}  \\
=&~  \frac{1}{n-1}\sum_{s:\,s\ne i} \big\{I(U_{s,j}\le U_{i,j})  - \Phi(U_{i,j})\big\} \,.
\end{align*}
By  Bernstein inequality, it holds that
\begin{align*}
\mathbb{P} \bigg[  \bigg|\frac{1}{n-1}\sum_{s:\,s\ne i} \big\{I(U_{s,j}\le U_{i,j})  - \Phi(U_{i,j})\big\}\bigg| > x   \,\bigg|\, U_{i,j}\bigg]  \le 2  \exp\bigg[-\frac{(n-1)x^2}{2\Phi(U_{i,j}) \{1-\Phi(U_{i,j})\} + x }\bigg] 
\end{align*}
for any $x>0$.   For sufficiently large $n$, we have 
\begin{align}\label{eq:h2-c}
\mathbb{P}(\mathcal{H}_2^{\rm c}) = &~\mathbb{P}\bigg[ \bigcup_{i\in[n], \,j\in[p]}\big\{|\hat{F}^{(i)}_{\bX,j}(X_{i,j})-  F_{\bX,j}(X_{i,j})| > K(U_{i,j}, p,n)\big\}\bigg] \notag\\
\le &~ \sum_{i=1}^{n}\sum_{j=1}^{p} \mathbb{E} \bigg( \mathbb{P} \bigg[ \bigg|\frac{1}{n-1}\sum_{s:\,s\ne i} \big\{I(U_{s,j}\le U_{i,j})  - \Phi(U_{i,j})\big\}\bigg| > K(U_{i,j}, p,n)   \,\bigg|\, U_{i,j}\bigg]\bigg) \notag\\
\le &~ 2 np \max_{i\in[n],\,j\in[p]}\mathbb{E} \bigg( \exp\bigg[-\frac{(n-1)K^2(U_{i,j}, p,n)}{4\Phi(U_{i,j}) \{1-\Phi(U_{i,j})\}  }\bigg] +  \exp\bigg\{-\frac{(n-1)K(U_{i,j}, p,n)}{2}\bigg\} \bigg) \notag\\
\le &~  4(np)^{-2}\,.
\end{align}
Restricted on $\mathcal{H}_2$, by \eqref{eq:n+1f-f}, it holds that
\begin{align}\label{eq:n+1fh-f-l}
&\bigg|\frac{n}{n+1}\hat{F}_{\bX,j}(X_{i,j})-  F_{\bX,j}(X_{i,j})\bigg|^{l} \notag\\
&~~~~~~~~\le 3^{l}\bigg\{|\hat{F}^{(i)}_{\bX,j}(X_{i,j})-  F_{\bX,j}(X_{i,j})|^{l} +   \bigg|\frac{2F_{\bX,j}(X_{i,j}) }{n+1}\bigg|^{l} +  \bigg|\frac{1 }{n+1}\bigg|^{l} \bigg\} \notag\\
&~~~~~~~~\le C_{11}^{l}\bigg| \frac{\Phi(U_{i,j}) \{1-\Phi(U_{i,j})\}\log (pn)}{n}\bigg|^{l/2} + C_{12}^{l} \bigg|\frac{\log(pn)}{n} \bigg|^{l}\,.
\end{align}
By \eqref{eq:derivative_n}, we have
\begin{align}\label{eq:i2-dec}
&~ |{\rm I}_{2} (j,k)|\notag \\
\le&~  \sum_{l=1}^{\infty}   \frac{M_1(C_{11}\bar{C})^{l}}{n }\sum_{i=1}^{n} |U_{i,j}|^{l-1} e^{lU_{i,j}^2/2} \bigg| \frac{\Phi(U_{i,j}) \{1-\Phi(U_{i,j})\}\log (pn)}{n}\bigg|^{l/2} I{(M_2 <|U_{i,j}| \le M_1)}\notag\\
&+\sum_{l=1}^{\infty}  \frac{M_1(C_{12}\bar{C})^{l}}{n }\sum_{i=1}^{n} |U_{i,j}|^{l-1} e^{lU_{i,j}^2/{2}}\bigg|\frac{\log(pn)}{n} \bigg|^{l} I{(M_2 <|U_{i,j}| \le M_1)} \notag \\
\le&~ \sum_{l=1}^{\infty}  \frac{M_1C_{13}^{l}}{n }\sum_{i=1}^{n} |U_{i,j}|^{l/2-1} e^{lU_{i,j}^2/{4} }\bigg|\frac{\log(pn)}{n} \bigg|^{l/2}I{(M_2 < |U_{i,j}| \le M_1)} \notag\\
&+\sum_{l=1}^{\infty}  \frac{M_1C_{14}^{l}}{n }\sum_{i=1}^{n} |U_{i,j}|^{l-1} e^{lU_{i,j}^2/{2}}\bigg|\frac{\log(pn)}{n} \bigg|^{l} I{(M_2 <|U_{i,j}| \le M_1)} \notag\\
\le&~  \sum_{l=1}^{\infty}\bigg\{  \frac{C_{15}M_1^{1/2} e^{M_1^2/4}\log^{1/2}(pn)}{n^{1/2}}  \bigg\}^{l-1} \times\frac{M_1\log^{1/2}(pn)}{n^{1/2}M_2^{1/2}}\times
\frac{1}{n}\sum_{i=1}^{n} e^{U_{i,j}^2/4}I{(M_2 < |U_{i,j}| \le M_1)} \notag\\
&+  \sum_{l=1}^{\infty}\bigg\{ \frac{C_{16}M_1 e^{M_1^2/2} \log(pn)}{n} \bigg\}^{l-1}  \times \frac{M_1\log (pn)}{n} \times\frac{1}{n} \sum_{i=1}^{n} e^{U_{i,j}^2/{2}}I{(M_2 < |U_{i,j}| \le M_1)}  \notag\\
\leq&~ \bigg\{ \frac{C_{17}M_1\log^{1/2}(pn)}{n^{1/2}M_2^{1/2}} + \frac{C_{18}e^{M_1^2/4}M_1\log (pn)}{n} \bigg\}\times
\frac{1}{n}\sum_{i=1}^{n} e^{U_{i,j}^2/{4}}I{(M_2 < |U_{i,j}| \le M_1)}  \notag\\
\leq&~  \frac{C_{19}M_1\log^{1/2}(pn)}{n^{1/2}M_2^{1/2}}\times
\frac{1}{n}\sum_{i=1}^{n} e^{U_{i,j}^2/{4}}I{(M_2 < |U_{i,j}| \le M_1)} 
\end{align}
provided that $\log (pn) \ll n e^{-M_1^2/2}M_1^{-1}$, where the second step is  due to $\Phi(x)=1-\Phi(-x)$ for any $x\in\mathbb{R}$  and  the inequality $1-\Phi(x) \le x^{-1}\phi(x)$ for any $x>0$.
Recall $U_{i,j} \sim \mathcal{N}(0,1)$. Then
\begin{align*}
&\max_{i\in[n],\,j\in[p]} \mathbb{E}\big\{ e^{U_{i,j}^2/4} I(M_2< |U_{i,j}| \le M_1)\big\}  \lesssim  M_2^{-1} e^{- M_2^2/4}\,,\\ 
&~~~~~\max_{i\in[n],\,j\in[p]} \var\big\{e^{U_{i,j}^2/4}  I(M_2< |U_{i,j}| \le M_1)\big\}   \lesssim  M_1 \,.
\end{align*}
Using the similar arguments for the derivation of \eqref{eq:bern-bound}, it holds that
\begin{align*}
&\max_{j\in[p]} \frac{1}{n}\sum_{i=1}^{n}  e^{U_{i,j}^2/4} I(M_2< |U_{i,j}| \le M_1)    = O_{\rm p} (M_2^{-1} e^{- M_2^2/4}) 
\end{align*}
provided that $\log p \lesssim ne^{-M_1^2/4}e^{-M_2^2/4}M_2^{-1}$.
As shown in \eqref{eq:h2-c}, $\mathbb{P}(\mathcal{H}_2^{\rm c}) \to 0$ as $n \to \infty$.  Hence, applying the similar arguments in Section \ref{sec:sub-i12} for deriving the convergence rate of $\max_{j\in[p],\,k\in[q]}|{\rm I}_{12}(j,k)|$, we have
\begin{align*} 
\max_{j\in[p],\, k\in[q]}|{\rm I}_{2}(j,k)| = O_{\rm p} \{n^{-1/2}M_1M_2^{-3/2} e^{-M_2^2/4}\log^{1/2} (pn)\}
\end{align*}
provided that $ \log (pn) \ll n e^{-M_1^2/2} M_1^{-1}$. Recall $d=pq$. Then, we complete the proof of \eqref{eq:i2}.
$\hfill\Box$

\subsection{Convergence rate of ${\max_{j\in[p],\,k\in[q]}|{\rm I}_{3}(j,k)|}$}\label{sec:sub-I3}
Recall $\hat{U}_{i,j}=\Phi^{-1} \{n(n+1)^{-1}\hat{F}_{\bX,j}(X_{i,j})\}$ and $n(n+1)^{-1} \hat{F}_{\bX,j}(X_{i,j})$ takes $n$ values $\{k(n+1)^{-1}:k\in[n]\}$. Due to  $-\sqrt{2\log (n+1)}\le \Phi^{-1}\{(n+1)^{-1}\} < \Phi^{-1}\{1-(n+1)^{-1}\} \le \sqrt{2\log (n+1)}$ for sufficiently large $n$, we have 
\begin{align}\label{eq:uh-bound}
\max_{i\in[n],\,j\in[p]}|\hat{U}_{i,j}| \le  \sqrt{2\log (n+1)}\,.
\end{align}
Recall $U_{i,j}^{*} =  U_{i,j}I(|U_{i,j}|\le M_1) +   M_1 \cdot{\rm sign}(U_{i,j}) I(|U_{i,j}|>M_1)$ with $M_1 =\sqrt{\kappa_1\log n}$  for some constant $\kappa_1\in(1,2)$.  Then  $\max_{i\in[n],\,j\in[p]}|U_{i,j}^{*}|\le M_1 < \sqrt{2\log n} < \sqrt{2\log (n+1)}$. Therefore,
\begin{align}\label{eq:Uhatstar}
\max_{i\in[n],\,j\in[p]}|\hat{U}_{i,j} - U_{i,j}^{*}| \le 2 \sqrt{2\log (n+1)}\,.
\end{align}
Analogously, we also have $\max_{i\in[n],\,k\in[q]}|V_{i,k}^{*}|\le \sqrt{2\log (n+1)}$. By \eqref{eq:id}, we have
\begin{align*}
|{\rm I}_{3}(j,k)| \le 4 \log (n+1) \times \frac{1}{n}\sum_{i=1}^{n} I(|U_{i,j}| > M_1)\,.
\end{align*}
Due to $U_{i,j} \sim \mathcal{N}(0,1)$, then
\begin{align}\label{eq:i(u>m)}
\max_{i\in[n],\,j\in[p]}\mathbb{E}\{I(|U_{i,j}| > M_1)\} \lesssim&~ M_1^{-1} e^{-M_1^2/2}\,, \notag\\  \max_{i\in[n],\,j\in[p]}\var\{I(|U_{i,j}| > M_1)\} \lesssim &~ M_1^{-1} e^{-M_1^2/2}\,.
\end{align}
Identical to the derivation of \eqref{eq:bern-bound}, we have
\begin{align}\label{eq:iu-M1}
\max_{j\in[p]}\bigg|\frac{1}{n}\sum_{i=1}^{n} I(|U_{i,j}| > M_1)\bigg| = O_{\rm p}(M_1^{-1} e^{-M_1^2/2})
\end{align}
provided that $ \log p \lesssim n e^{-M_1^2/2} M_1^{-1}$. Hence, it holds that 
\begin{align*}
\max_{j\in[p],\, k\in[q]}|{\rm I}_{3}(j,k)| =  O_{\rm p}(M_1^{-1} e^{-M_1^2/2} \log n)
\end{align*}
provided that $ \log p \lesssim n e^{-M_1^2/2} M_1^{-1}$. Recall $M_1=\sqrt{\kappa_1\log n}$ for some constant $\kappa_1\in(1,2)$ and $d=pq$. We complete the proof of \eqref{eq:bdI2}.
$\hfill\Box$

\section{Proof of Lemma \ref{lem:huhv}}\label{sec:sub-huhv}
Recall $\hat{U}_{i,j}=\Phi^{-1} \{n(n+1)^{-1}\hat{F}_{\bX,j}(X_{i,j})\}$, $\hat{V}_{i,k}=\Phi^{-1} \{n(n+1)^{-1}\hat{F}_{\bY,k}(Y_{i,k})\}$, $U_{i,j}^{*} =  U_{i,j}I(|U_{i,j}|\le M_1) + M_1 \cdot{\rm sign}(U_{i,j})I(|U_{i,j}|>M_1)$, $V_{i,k}^{*} =  V_{i,k}I(|V_{i,k}|\le M_1) + M_1 \cdot{\rm sign}(V_{i,k})I(|V_{i,k}|>M_1)$,  $U_{i,j}=\Phi^{-1}\{F_{\bX,j}(X_{i,j})\}$ and $V_{i,k}=\Phi^{-1}\{F_{\bY,k}(Y_{i,k})\}$, where $M_1=\sqrt{\kappa_1\log n}$ for some constant $\kappa_1  \in(1,2)$. We have
\begin{align*}
&\frac{1}{n}\sum_{i=1}^{n}(\hat{U}_{i,j}-U_{i,j}^{*})(\hat{V}_{i,k}-V_{i,k}^{*})\\
&~~~~~~= \frac{1}{n}\sum_{i=1}^{n}(\hat{U}_{i,j}-U_{i,j})(\hat{V}_{i,k}-V_{i,k})I(|U_{i,j}|\le M_1) I(|V_{i,k}|\le M_1)  \\
&~~~~~~~~~+  \frac{1}{n}\sum_{i=1}^{n}(\hat{U}_{i,j}-U_{i,j}^{*})(\hat{V}_{i,k}-V_{i,k}^{*}) \big\{I(|U_{i,j}|\le M_1) I(|V_{i,k}|> M_1) +  I(|U_{i,j}|> M_1)\big\}  \\
&~~~~~~=\frac{1}{n}\sum_{i=1}^{n} \bigg[\Phi^{-1} \bigg\{\frac{n}{n+1}\hat{F}_{\bX,j}(X_{i,j})\bigg\} - \Phi^{-1}\{F_{\bX,j}(X_{i,j})\} \bigg]\\
&~~~~~~~~~~\underbrace{~~~~~~~~\times \bigg[\Phi^{-1} \bigg\{\frac{n}{n+1}\hat{F}_{\bY,k}(Y_{i,k})\bigg\} - \Phi^{-1}\{F_{\bY,k}(Y_{i,k})\} \bigg] I(|U_{i,j}|\le M_1) I(|V_{i,k}|\le M_1) }_{\textrm{J}_{1}(j,k)}\\
&~~~~~~~~~+  \underbrace{\frac{1}{n}\sum_{i=1}^{n}(\hat{U}_{i,j}-U_{i,j}^{*})(\hat{V}_{i,k}-V_{i,k}^{*}) \big\{I(|U_{i,j}|\le M_1) I(|V_{i,k}|> M_1) +  I(|U_{i,j}|> M_1)\big\} }_{\textrm{J}_{2}(j,k)}
\end{align*}
As we will show in Sections \ref{sec:sub-J1} and \ref{sec:sub-J2}, 
\begin{align}\label{eq:j1}
\max_{j\in[p],\, k\in[q]}|{\rm J}_{1}(j,k)| =  O_{\rm p} \{n^{-(1-\kappa_1/8)} (\log n)^{-1/2}\log (dn)\} 
\end{align}
provided that $ \log d \ll n^{1-\kappa_1/2} (\log n)^{-1/2} $,  and
\begin{align}\label{eq:j2}
\max_{j\in[p],\, k\in[q]}|{\rm J}_{2}(j,k)| =  O_{\rm p}\{n^{-\kappa_1/2} (\log n)^{1/2}\}
\end{align}
provided that $ \log d \lesssim n^{1-\kappa_1/2} (\log n)^{-1/2}$. 
Hence, we have 
\begin{align*}
&\max_{j\in[p],\, k\in[q]}\bigg|\frac{1}{\sqrt{n}}\sum_{i=1}^{n}(\hat{U}_{i,j}-U_{i,j}^{*})(\hat{V}_{i,k}-V_{i,k}^{*}) \bigg| = O_{\rm p}\{n^{-(\kappa_1-1)/2} (\log n)^{1/2}\}
\end{align*}
provided that $ \log d \lesssim   n^{1-5\kappa_1/8}  \log n $ with $\kappa_1<8/5$.
We complete the proof of Lemma \ref{lem:huhv}.
$\hfill\Box$

\subsection{Convergence rate of ${\max_{j\in[p],\,k\in[q]}|{\rm J}_{1}(j,k)|}$}\label{sec:sub-J1}
Notice  that $\Phi^{-1}(x)$ is infinitely differentiable at any $x\in (0,1)$. Given $M_2=\sqrt{\kappa_2\log n}$ for some constant $\kappa_2\in(0,1)$, we have
\begin{align*}
&~{\rm J}_{1}(j,k) \\
=&~ \frac{1}{n}\sum_{i=1}^{n}\bigg[ \sum_{l=1}^{\infty}\frac{1}{ l!}(\Phi^{-1})^{(l)}\{F_{\bX,j}(X_{i,j})\}\bigg\{\frac{n}{n+1}\hat{F}_{\bX,j}(X_{i,j})-  F_{\bX,j}(X_{i,j}) \bigg\}^{l} I{(|U_{i,j}| \le M_1)} \bigg]\\
&~~~~~~~~\times \bigg[ \sum_{s=1}^{\infty}\frac{1}{ s!}(\Phi^{-1})^{(s)}\{F_{\bY,k}(Y_{i,k})\}\bigg\{\frac{n}{n+1}\hat{F}_{\bY,k}(Y_{i,k})-  F_{\bY,k}(Y_{i,k}) \bigg\}^{s} I{(|V_{i,k}| \le M_1)} \bigg] \\
=&~ \sum_{l=1}^{\infty} \sum_{s=1}^{\infty}   \frac{1}{n}\sum_{i=1}^{n} \bigg[ \frac{1}{  l!}(\Phi^{-1})^{(l)}\{F_{\bX,j}(X_{i,j})\}\bigg\{\frac{n}{n+1}\hat{F}_{\bX,j}(X_{i,j})-  F_{\bX,j}(X_{i,j}) \bigg\}^{l} I{(|U_{i,j}| \le M_2)}   \\
&~~\underbrace{~~~~~~~~~~~~~~~~~\times  \frac{1}{  s!}(\Phi^{-1})^{(s)}\{F_{\bY,k}(Y_{i,k})\}\bigg\{\frac{n}{n+1}\hat{F}_{\bY,k}(Y_{i,k})-  F_{\bY,k}(Y_{i,k}) \bigg\}^{s} I{(|V_{i,k}| \le M_2)}  \bigg] }_{\textrm{J}_{11}(j,k)}\\
&+\sum_{l=1}^{\infty} \sum_{s=1}^{\infty}   \frac{1}{n}\sum_{i=1}^{n} \bigg[ \frac{1}{  l!}(\Phi^{-1})^{(l)}\{F_{\bX,j}(X_{i,j})\}\bigg\{\frac{n}{n+1}\hat{F}_{\bX,j}(X_{i,j})-  F_{\bX,j}(X_{i,j}) \bigg\}^{l} I{( |U_{i,j}| \le M_2)}   \\
&~~~\underbrace{~~~~~~~~~~~~~~~\times  \frac{1}{  s!}(\Phi^{-1})^{(s)}\{F_{\bY,k}(Y_{i,k})\}\bigg\{\frac{n}{n+1}\hat{F}_{\bY,k}(Y_{i,k})-  F_{\bY,k}(Y_{i,k}) \bigg\}^{s} I{(M_2 <|V_{i,k}| \le M_1)}  \bigg] }_{\textrm{J}_{12}(j,k)}\\
&+\sum_{l=1}^{\infty} \sum_{s=1}^{\infty}   \frac{1}{n}\sum_{i=1}^{n} \bigg[ \frac{1}{  l!}(\Phi^{-1})^{(l)}\{F_{\bX,j}(X_{i,j})\}\bigg\{\frac{n}{n+1}\hat{F}_{\bX,j}(X_{i,j})-  F_{\bX,j}(X_{i,j}) \bigg\}^{l} I{(M_2<|U_{i,j}| \le M_1)}   \\
&~~~\underbrace{~~~~~~~~~~~~~~~~~\times  \frac{1}{  s!}(\Phi^{-1})^{(s)}\{F_{\bY,k}(Y_{i,k})\}\bigg\{\frac{n}{n+1}\hat{F}_{\bY,k}(Y_{i,k})-  F_{\bY,k}(Y_{i,k}) \bigg\}^{s} I{( |V_{i,k}| \le M_2)}  \bigg] }_{\textrm{J}_{13}(j,k)}\\
&+\sum_{l=1}^{\infty} \sum_{s=1}^{\infty}   \frac{1}{n}\sum_{i=1}^{n} \bigg[ \frac{1}{  l!}(\Phi^{-1})^{(l)}\{F_{\bX,j}(X_{i,j})\}\bigg\{\frac{n}{n+1}\hat{F}_{\bX,j}(X_{i,j})-  F_{\bX,j}(X_{i,j}) \bigg\}^{l} I{(M_2<|U_{i,j}| \le M_1)}   \\
&~~~\underbrace{~~~~~~~~~~~~~~\times  \frac{1}{  s!}(\Phi^{-1})^{(s)}\{F_{\bY,k}(Y_{i,k})\}\bigg\{\frac{n}{n+1}\hat{F}_{\bY,k}(Y_{i,k})-  F_{\bY,k}(Y_{i,k}) \bigg\}^{s} I{(M_2 <|V_{i,k}| \le M_1)}  \bigg]}_{\textrm{J}_{14}(j,k)}\,.
\end{align*}
As we will show in Sections \ref{sec:sub-j11}--\ref{sec:sub-j14},
\begin{align}\label{eq:h0-j11}
\max_{j\in[p],\,k\in[q]}|{\rm J}_{11}(j,k)| =O_{\rm p} \{n^{-1}M_2^{-1} e^{M_2^2/2}\log (dn)\}
\end{align}
provided that $ \log (dn) \ll ne^{-M_2^2} M_2^{-2}$, 
\begin{align}\label{eq:h0-j12}
\max_{j\in[p],\, k\in[q]}|{\rm J}_{12}(j,k)|= O_{\rm p} \{n^{-1}M_1^{1/2}M_2^{-1} e^{M_2^2/4}\log (dn)\} = \max_{j\in[p],\, k\in[q]}|{\rm J}_{13}(j,k)|
\end{align}
provided that $ \log (dn) \ll \min\{ne^{-M_1^2/2} M_1^{-1}, ne^{-M_2^2} M_2^{-2} \}$, and
\begin{align}\label{eq:h0-j14}
\max_{j\in[p],\, k\in[q]}|{\rm J}_{14 }(j,k)|=   O_{\rm p}\{n^{-1}  M_{1}M_{2}^{-1}\log(dn)\} 
\end{align}
provided that $ \log (dn) \ll ne^{-M_1^2/2} M_1^{-1}$.  Recall  $ M_1 =\sqrt{\kappa_1\log n}$ and  $ M_2 =\sqrt{\kappa_2\log n}$ for some constants $\kappa_1 \in(1,2)$ and $\kappa_2 \in (0,1)$. Together with \eqref{eq:h0-j11}--\eqref{eq:h0-j14}, we have 
\begin{align*} 
\max_{j\in[p],\, k\in[q]}|{\rm J}_{1}(j,k)| =  O_{\rm p} \{n^{-(1-\kappa_2/2)} (\log n)^{-1/2}\log (dn)\} 
\end{align*}
provided that $ \log d \ll \min\{n^{1-\kappa_1/2} (\log n)^{-1/2}, n^{1-\kappa_2}(\log n)^{-1} \}$. We complete the proof of \eqref{eq:j1} with selecting $\kappa_2 =\kappa_1/4$. $\hfill\Box$

\subsubsection{Proof of \eqref{eq:h0-j11}}\label{sec:sub-j11}
Recall $\mathcal{H}_1$   defined in Section \ref{sec:sub-i12} for the proof of Lemma \ref{lem:uhv}.  Analogously, define the event 
\begin{align*}
\mathcal{H}_3 = \bigg\{ \max_{i\in[n],\,k\in[q]}|\hat{F}^{(i)}_{\bY,k}(Y_{i,k})-  F_{\bY,k}(Y_{i,k})| \le \tilde{C} n^{-1/2}\log^{1/2}(qn) \bigg\}
\end{align*} 
with $\tilde{C}=2K_2^{-1/2}$,  where $K_2$ is specified in Lemma \ref{lem:ecdf}. 
Recall $d=pq$. Restricted on $\mathcal{H}_1 \bigcap \mathcal{H}_{3}$,  by \eqref{eq:derivative_n}, it holds that
\begin{align}\label{eq:I11b}
|{\rm J}_{11}(j,k)| \le&~  
\sum_{l=1}^{\infty}\sum_{s=1}^{\infty} C_1^{l+s} \bigg\{\frac{\log(dn)}{n}\bigg\}^{(l+s)/2}  \frac{1}{n } \sum_{i=1}^{n}\bigg\{|U_{i,j}|^{l-1}|V_{i,k}|^{s-1}    e^{lU_{i,j}^2/2}\ e^{sV_{i,k}^2/2} \notag\\
&~~~~~~~~~~~~~~~~~~~~~~~~~~~~~~~~~~~~~~~~~~~~~~~~~ \times   I{(|U_{i,j}| \le M_2)} I{(|V_{i,k}| \le M_2)}\bigg\}  \notag\\
\le&~    \sum_{l=1}^{\infty} \bigg\{\frac{C_2 M_2e^{M_2^2/2}  \log^{1/2}(dn) }{n^{1/2}}\bigg\}^{l-1}  \times \sum_{s=1}^{\infty}\bigg\{\frac{C_2 M_2e^{M_2^2/2} \log^{1/2}(dn) }{n^{1/2}}\bigg\}^{s-1}  \times \frac{\log(dn)}{n}   \notag\\
&~~~~\times  
\frac{1}{n }\sum_{i=1}^{n}   e^{U_{i,j}^2/2} e^{V_{i,k}^2/2} I{(|U_{i,j}| \le M_2)} I{(|V_{i,k}| \le M_2)} \notag\\
\le&~ \frac{C_3 \log(dn)}{n}\times   \frac{1}{n }\sum_{i=1}^{n}   e^{U_{i,j}^2/2} e^{V_{i,k}^2/2} I{(|U_{i,j}| \le M_2)} I{(|V_{i,k}| \le M_2)}
\end{align}
provided that $ \log (dn) \ll ne^{-M_2^2} M_2^{-2}$.  Recall $U_{i,j}, V_{i,k} \sim \mathcal{N}(0,1)$. By  Cauchy-Schwarz inequality, we then have 
\begin{align*}
&\max_{i\in[n],\,j\in[p],\, k\in[q]}\mathbb{E}\big\{ e^{U_{i,j}^2/2} e^{V_{i,k}^2/2} I{(|U_{i,j}| \le M_2)} I{(|V_{i,k}| \le M_2)} \big\} \\
&~~~~~ \le \max_{i\in[n],\,j\in[p] } \big[\mathbb{E}\big\{ e^{U_{i,j}^2} I{(|U_{i,j}| \le M_2)} \big\}\big]^{1/2}   \max_{i\in[n] ,\,k\in[q] }\big[\mathbb{E}\big\{  e^{V_{i,k}^2}I{(|V_{i,k}| \le M_2)} \big\}\big]^{1/2}  \lesssim  M_2^{-1} e^{M_2^2/2}\,,\\ 
&\max_{i\in[n],\,j\in[p],\, k\in[q]}\var\big\{ e^{U_{i,j}^2/2} e^{V_{i,k}^2/2} I{(|U_{i,j}| \le M_2)} I{(|V_{i,k}| \le M_2)} \bigg\}  \\
&~~~~~ \le \max_{i\in[n],\,j\in[p] }\big[\mathbb{E}\big\{ e^{2U_{i,j}^2} I{(|U_{i,j}| \le M_2)} \big\}\big]^{1/2}  \max_{i\in[n],\,k\in[q]}\big[\mathbb{E}\big\{  e^{2V_{i,k}^2}I{(|V_{i,k}| \le M_2)} \big\}\big]^{1/2} \lesssim  M_2^{-1} e^{3M_2^2/2}\,. 
\end{align*}
Analogous to the derivation of \eqref{eq:bern-bound}, it holds that
\begin{align}\label{eq:u2v2}
\max_{j\in[p],\, k\in[q]}  \frac{1}{n }\sum_{i=1}^{n}   e^{U_{i,j}^2/2}  e^{V_{i,k}^2/2} I{(|U_{i,j}| \le M_2)} I{(|V_{i,k}| \le M_2)} = O_{\rm p} ( M_2^{-1} e^{M_2^2/2})
\end{align}
provided that $\log d \lesssim ne^{-M_2^2/2}M_2^{-1}$. 
Recall $\mathbb{P}(\mathcal{H}_{1}^{\rm c}) \le K_1 (pn)^{-3}$ by \eqref{eq:h1-c}. Similarly, we also have $\mathbb{P}(\mathcal{H}_{3}^{\rm c}) \le K_1 (qn)^{-3}$.
Hence, applying the similar arguments in Section \ref{sec:sub-i12} for deriving the convergence rate of $\max_{j\in[p],\,k\in[q]}|{\rm I}_{12}(j,k)|$,
we can show
\begin{align*}
\max_{j\in[p],\, k\in[q]}|{\rm J}_{11}(j,k)| = O_{\rm p} \{n^{-1}M_2^{-1} e^{M_2^2/2}\log (dn)\}
\end{align*}
provided that $ \log (dn) \ll ne^{-M_2^2} M_2^{-2}$. We complete the proof of \eqref{eq:h0-j11}. $\hfill\Box$

\subsubsection{Proof of \eqref{eq:h0-j12}}\label{sec:sub-j12}
Let $K(V_{i,k}, q,n) = 4 n^{-1/2}  [\Phi(V_{i,k}) \{1-\Phi(V_{i,k})\}]^{1/2} \log^{1/2}(qn)  + 7n^{-1} \log(qn)$. Define the event 
\begin{align}\label{eq:h4-def}
\mathcal{H}_4 =  \bigcap_{i\in[n],\,k\in[q]}\big\{|\hat{F}^{(i)}_{\bY,k}(Y_{i,k})-  F_{\bY,k}(Y _{i,k})| \le K(V_{i,k}, q,n)\big\} \,.
\end{align}
Similar to \eqref{eq:n+1fh-f-l}, restricted on $\mathcal{H}_{4}$, we have
\begin{align}\label{eq:n+1fh-f-s}
\bigg|\frac{n}{n+1}\hat{F}_{\bY,k}(Y_{i,k})-  F_{\bY,k}(Y_{i,k})\bigg|^{s} \le  C_4^{s}\bigg| \frac{\Phi(V_{i,k}) \{1-\Phi(V_{i,k})\}\log (qn)}{n}\bigg|^{s/2} + C_5^{s} \bigg|\frac{\log(qn)}{n} \bigg|^{s}\,.
\end{align}
Recall $d=pq$, and $\mathcal{H}_1$ defined in Section \ref{sec:sub-i12} for the proof of Lemma \ref{lem:uhv}. Restricted on $\mathcal{H}_{1} \bigcap\mathcal{H}_{4}$,  by \eqref{eq:derivative_n}, it holds that
\begin{align*}
&~|{\rm J}_{12}(j,k)| \\
\le &~ \sum_{l=1}^{\infty}\sum_{s=1}^{\infty} C_6^{l+s} \bigg\{\frac{\log(dn)}{n}\bigg\}^{(l+s)/2}  \frac{1}{n } \sum_{i=1}^{n}\big\{|U_{i,j}|^{l-1}|V_{i,k}|^{s-1}    e^{lU_{i,j}^2/2} e^{sV_{i,k}^2/2}  \\
&~\underbrace{~~~~~~~~~~~~~~~~~~~~~~~~~~~~\times  \big[\Phi(V_{i,k}) \{1-\Phi(V_{i,k})\}\big]^{s/2} I{(|U_{i,j}| \le M_2)} I{(M_2 <|V_{i,k}| \le M_1)} \big\} }_{\textrm{J}_{121}(j,k)} \\
&+ \sum_{l=1}^{\infty}\sum_{s=1}^{\infty} C_7^{l+s} \bigg\{\frac{\log(dn)}{n}\bigg\}^{(l+2s)/2}  \frac{1}{n } \sum_{i=1}^{n}\big\{|U_{i,j}|^{l-1}|V_{i,k}|^{s-1}    e^{lU_{i,j}^2/2} e^{sV_{i,k}^2/2} \\
&~~~\underbrace{~~~~~~~~~~~~~~~~~~~~~~~~~~~~~~~~~~~~~~~~~~~~~~~~~~~~~~~\times I{(|U_{i,j}| \le M_2)} I{(M_2 <|V_{i,k}| \le M_1)}\big\}}_{\textrm{J}_{122}(j,k)}\,.
\end{align*}
Due to $1-\Phi(x) \le x^{-1}\phi(x)$ for any $x>0$, we have 
\begin{align}\label{eq:J121}
|\textrm{J}_{121}(j,k)|  \le&~  \sum_{l=1}^{\infty}\sum_{s=1}^{\infty} C_8^{l+s} \bigg\{\frac{\log(dn)}{n}\bigg\}^{(l+s)/2}  \frac{1}{n } \sum_{i=1}^{n}\big\{|U_{i,j}|^{l-1}|V_{i,k}|^{s/2-1}    e^{lU_{i,j}^2/2}e^{sV_{i,k}^2/4}  \notag\\
&~~~~~~~~~~~~~~~~~~~~~~~~~~~~~~~~~~~~~~~~~~~~~~~\times   I{(|U_{i,j}| \le M_2)} I{(M_2 <|V_{i,k}| \le M_1)} \big\}  \notag\\
\le&~ \frac{\log(dn)}{nM_2^{1/2}} \times  \sum_{l=1}^{\infty} \bigg\{\frac{C_9 \log^{1/2}(dn) M_2 e^{M_2^2/2}}{n^{1/2}}\bigg\}^{l-1}  \times   \sum_{s=1}^{\infty} \bigg\{\frac{C_9 \log^{1/2}(dn) M_1^{1/2}e^{M_1^2/4}}{n^{1/2}}\bigg\}^{s-1}   \notag \\
&\times  
\frac{1}{n }\sum_{i=1}^{n}   e^{U_{i,j}^2/2}  e^{V_{i,k}^2/4} I{(|U_{i,j}| \le M_2)} I{(M_2<|V_{i,k}| \le M_1)} \\
\le &~ \frac{C_{10}\log(dn)}{nM_2^{1/2}}
\times  \frac{1}{n }\sum_{i=1}^{n}   e^{U_{i,j}^2/2}e^{V_{i,k}^2/4} I{(|U_{i,j}| \le M_2)} I{(M_2<|V_{i,k}| \le M_1)} \notag
\end{align}
provided that $ \log (dn) \ll \min\{ne^{-M_1^2/2} M_1^{-1}, ne^{-M_2^2} M_2^{-2} \}$. Recall $U_{i,j}, V_{i,k} \sim \mathcal{N}(0,1)$. By  Cauchy-Schwarz inequality, it holds that 
\begin{align*}
&\max_{i\in[n],\,j\in[p],\, k\in[q]}\mathbb{E}\big\{ e^{U_{i,j}^2/2}  e^{V_{i,k}^2/4} I{(|U_{i,j}| \le M_2)} I{(M_2<|V_{i,k}| \le M_1)} \big\} \\
&~~~~~~~ \le \max_{i\in[n],\,j\in[p] }\big[\mathbb{E}\big\{ e^{U_{i,j}^2} I{(|U_{i,j}| \le M_2)} \big\}\big]^{1/2}  \max_{i\in[n],\,k\in[q]}\big[\mathbb{E}\big\{  e^{V_{i,k}^2/2}I{(M_2 < |V_{i,k}| \le M_1)} \big\}\big]^{1/2} \\
&~~~~~~~ \lesssim  M_1^{1/2} M_2^{-1/2} e^{M_2^2/4}\,,\\ 
&\max_{i\in[n],\,j\in[p],\, k\in[q]}\var\big\{ e^{U_{i,j}^2/2} e^{V_{i,k}^2/4} I{(|U_{i,j}| \le M_2)} I{(M_2 <|V_{i,k}| \le M_1)} \big\} \\
&~~~~~~~ \le \max_{i\in[n],\,j\in[p] }\big[\mathbb{E}\big\{ e^{2U_{i,j}^2} I{(|U_{i,j}| \le M_2)} \big\}\big]^{1/2}  \max_{i\in[n],\,k\in[q]}\big[\mathbb{E}\big\{  e^{V_{i,k}^2}I{(M_2 <|V_{i,k}| \le M_1)} \big\}\big]^{1/2} \\
&~~~~~~~ \lesssim  M_{1}^{-1/2} M_2^{-1/2} e^{3M_2^2/4}  e^{M_1^2/4}\,. 
\end{align*}
Analogous to the derivation of \eqref{eq:bern-bound}, we can show
\begin{align}\label{eq:u2v4}
&\max_{j\in[p],\, k\in[q]} \frac{1}{n}\sum_{i=1}^{n}   e^{U_{i,j}^2/2}  e^{V_{i,k}^2/4}  I{(|U_{i,j}| \le M_2)} I{(M_2<|V_{i,k}| \le M_1)} \notag \\
&~~~~~~~~~~= O_{\rm p} (M_1^{1/2} M_2^{-1/2} e^{M_2^2/4})
\end{align}
provided that $\log d \lesssim nM_1^{1/2}M_2^{-1/2}e^{-M_1^2/4}e^{-M_2^2/4}$. By \eqref{eq:J121}, it holds that
\begin{align*}
\max_{j\in[p],\, k\in[q]}|{\rm J}_{121}(j,k)| = O_{\rm p} \{n^{-1}M_1^{1/2}M_2^{-1} e^{M_2^2/4}\log (dn)\}
\end{align*}
provided that $ \log (dn) \ll \min\{ne^{-M_1^2/2} M_1^{-1}, ne^{-M_2^2} M_2^{-2} \}$.  
Analogously,  it holds that
\begin{align*}
|\textrm{J}_{122}(j,k)| 
\le&~  \frac{\log^{3/2}(dn)}{n^{3/2}} \sum_{l=1}^{\infty} \bigg\{\frac{C_{11} \log^{1/2}(dn) M_2 e^{M_2^2/2}}{n^{1/2}}\bigg\}^{l-1}  \times   \sum_{s=1}^{\infty} \bigg\{\frac{C_{11}  \log(dn) M_1e^{M_1^2/2}}{n}\bigg\}^{s-1}     \notag \\
&~~~~\times  
\frac{1}{n }\sum_{i=1}^{n}   e^{U_{i,j}^2/2}e^{V_{i,k}^2/2} I{(|U_{i,j}| \le M_2)} I{(M_2<|V_{i,k}| \le M_1)}   \\
\le&~   \frac{C_{12}e^{M_1^2/4}\log^{3/2}(dn)}{n^{3/2}}\times \frac{1}{n}
\sum_{i=1}^{n}   e^{U_{i,j}^2/2}e^{V_{i,k}^2/4} I{(|U_{i,j}| \le M_2)} I{(M_2<|V_{i,k}| \le M_1)}  
\end{align*}
provided that $ \log (dn) \ll \min\{ne^{-M_1^2/2} M_1^{-1}, ne^{-M_2^2} M_2^{-2} \}$.   By \eqref{eq:u2v4} again, we also have
\begin{align*}
\max_{j\in[p],\, k\in[q]}|{\rm J}_{122}(j,k)| = O_{\rm p} \{n^{-3/2} M_{1}^{1/2}M_2^{-1/2}e^{M_1^2/4} e^{M_2^2/4}\log^{3/2} (dn)\}
\end{align*}
provided that $ \log (dn) \ll \min\{ne^{-M_1^2/2} M_1^{-1}, ne^{-M_2^2} M_2^{-2} \}$.
Notice that, restricted on $\mathcal{H}_1 \bigcap \mathcal{H}_{4}$, 
\begin{align*}
\max_{j\in[p],\, k\in[q]}|{\rm J}_{12}(j,k)| \le  \max_{j\in[p],\, k\in[q]}|{\rm J}_{121}(j,k)| + \max_{j\in[p],\, k\in[q]}|{\rm J}_{122}(j,k)| \,.
\end{align*}
Recall $\mathbb{P}(\mathcal{H}_{1}^{\rm c}) \le K_1 (pn)^{-3}$ by \eqref{eq:h1-c}. Identical to \eqref{eq:h2-c}, we have $\mathbb{P}(\mathcal{H}_{4}^{\rm c}) \le 4(qn)^{-2}$. Hence, applying the same  arguments in Section \ref{sec:sub-i12} for deriving the convergence rate of $\max_{j\in[p],\,k\in[q]}|{\rm I}_{12}(j,k)|$,
we can show
\begin{align*}
\max_{j\in[p],\, k\in[q]}|{\rm J}_{12}(j,k)| = O_{\rm p} \{n^{-1}M_{1}^{1/2}M_2^{-1} e^{M_2^2/4}\log (dn)\} 
\end{align*}
provided that $ \log (dn) \ll \min\{ne^{-M_1^2/2} M_1^{-1}, ne^{-M_2^2} M_2^{-2} \}$. Using the similar arguments, we can also show such convergence rate holds for $\max_{j\in[p],k\in[q]}|{\rm J}_{13}(j,k)|$. Then \eqref{eq:h0-j12} holds. $\hfill\Box$

\subsubsection{Proof of \eqref{eq:h0-j14}}\label{sec:sub-j14}
Recall $\mathcal{H}_{2}$ defined in Section \ref{sec:sub-I2}  for the proof of Lemma \ref{lem:uhv} and $\mathcal{H}_4$ given in \eqref{eq:h4-def}. Restricted on $\mathcal{H}_{2} \bigcap\mathcal{H}_{4}$,  by \eqref{eq:derivative_n}, \eqref{eq:n+1fh-f-l} and \eqref{eq:n+1fh-f-s}, we have
\begin{align}\label{eq:j14-dec} 
&~|{\rm J}_{14}(j,k)| \notag\\
\le &~ \sum_{l=1}^{\infty}\sum_{s=1}^{\infty} C_{13}^{l+s} \frac{1}{n}\sum_{i=1}^{n}\bigg( |U_{i,j}|^{l-1} e^{lU_{i,j}^2/2} \bigg[ \bigg| \frac{\Phi(U_{i,j}) \{1-\Phi(U_{i,j})\}\log (pn)}{n}\bigg|^{l/2} +  \bigg|\frac{\log(pn)}{n} \bigg|^{l} \bigg] \notag\\
&~~~~~~~~~~~~~~~~~~~~~~~~~~\times  |V_{i,k}|^{s-1} e^{sV_{i,k}^2/2} \bigg[ \bigg| \frac{\Phi(V_{i,k}) \{1-\Phi(V_{i,k})\}\log (qn)}{n}\bigg|^{s/2} +  \bigg|\frac{\log(qn)}{n} \bigg|^{s} \bigg]\notag \\
&~~~~~~~~~~~~~~~~~~~~~~~~~~\times I{(M_2 <|U_{i,j}|,\,|V_{i,k}| \le M_1)}  \bigg)\notag \\
= &~  \sum_{l=1}^{\infty}\sum_{s=1}^{\infty} C_{13}^{l+s} \frac{1}{n}\sum_{i=1}^{n}\bigg[ |U_{i,j}|^{l-1} |V_{i,k}|^{s-1}  e^{lU_{i,j}^2/2}e^{sV_{i,k}^2/2} \bigg| \frac{\Phi(U_{i,j}) \{1-\Phi(U_{i,j})\}\log (pn)}{n}\bigg|^{l/2}  \notag \\
&~~\underbrace{~~~~~~~~~~~~~~~~~~~~~~~~\times  \bigg| \frac{\Phi(V_{i,k}) \{1-\Phi(V_{i,k})\}\log (qn)}{n}\bigg|^{s/2}  I{(M_2 <|U_{i,j}|,\,|V_{i,k}| \le M_1)}  \bigg]}_{\textrm{J}_{141}(j,k)} \notag\\
&+\sum_{l=1}^{\infty}\sum_{s=1}^{\infty} C_{13}^{l+s}  \frac{1}{n}\sum_{i=1}^{n}\bigg[ |U_{i,j}|^{l-1}|V_{i,k}|^{s-1} e^{lU_{i,j}^2/2}  e^{sV_{i,k}^2/2}  \bigg| \frac{\Phi(U_{i,j}) \{1-\Phi(U_{i,j})\}\log (pn)}{n}\bigg|^{l/2}  \notag \\
&~~~\underbrace{~~~~~~~~~~~~~~~~~~~~~~~~\times  \bigg|\frac{\log (qn)}{n}\bigg|^{s}  I{(M_2 <|U_{i,j}|,\,|V_{i,k}| \le M_1)}  \bigg]~~~~~~~~~~~~~~~~~~~~~~~~~~}_{\textrm{J}_{142}(j,k)} \notag \\
&+\sum_{l=1}^{\infty}\sum_{s=1}^{\infty} C_{13}^{l+s} \frac{1}{n}\sum_{i=1}^{n}\bigg[ |U_{i,j}|^{l-1}|V_{i,k}|^{s-1} e^{lU_{i,j}^2/2}  e^{sV_{i,k}^2/2} \bigg| \frac{\Phi(V_{i,k}) \{1-\Phi(V_{i,k})\}\log (qn)}{n}\bigg|^{s/2} \notag \\
&~~~\underbrace{~~~~~~~~~~~~~~~~~~~~~~~~~\times \bigg| \frac{\log (pn)}{n}\bigg|^{l}   I{(M_2 <|U_{i,j}| ,\,|V_{i,k}|\le M_1)}  \bigg]~~~~~~~~~~~~~~~~~~~~~~~~~~}_{\textrm{J}_{143}(j,k)} \notag\\
&+\sum_{l=1}^{\infty}\sum_{s=1}^{\infty} C_{13}^{l+s} \frac{1}{n}\sum_{i=1}^{n}\bigg\{ |U_{i,j}|^{l-1}|V_{i,k}|^{s-1} e^{lU_{i,j}^2/2}e^{sV_{i,k}^2/2} \bigg| \frac{\log (pn)}{n}\bigg|^{l}  \bigg| \frac{\log (qn)}{n}\bigg|^{s} \notag\\
&~~~\underbrace{~~~~~~~~~~~~~~~~~~~~~~~~~~\times   I{(M_2 <|U_{i,j}|,\,|V_{i,k}| \le M_1)}  \bigg\}~~~~~~~~~~~~~~~~~~~~~}_{\textrm{J}_{144}(j,k)}    \,.
\end{align}
Recall $d=pq$. Due to $1-\Phi(x) \le x^{-1}\phi(x)$ for any $x>0$, we have
\begin{align}\label{eq:J141}
|\textrm{J}_{141}(j,k)|  \le  &~ \sum_{l=1}^{\infty}\sum_{s=1}^{\infty} C_{14}^{l+s} \bigg\{\frac{\log(dn)}{n}\bigg\}^{(l+s)/2} \frac{1}{n}\sum_{i=1}^{n}\bigg\{|U_{i,j}|^{l/2-1} |V_{i,k}|^{s/2-1}e^{lU_{i,j}^2/4}e^{sV_{i,k}^2/4}   \notag\\
&~~~~~~~~~~~~~~~~~~~~~~~~~~~~~~~~~~~~~~~~~~~~~~~~\times I{(M_2 <|U_{i,j}| ,\,|V_{i,k}|\le M_1)}  \bigg\} \notag\\
\le &~  \frac{\log (dn)}{nM_2}    \sum_{l=1}^{\infty} \bigg\{ \frac{C_{15}M_1^{1/2}e^{M_1^2/4}\log^{1/2}(dn)}{n^{1/2}}\bigg\}^{l-1} \times  \sum_{s=1}^{\infty} \bigg\{ \frac{C_{15}M_1^{1/2}e^{M_1^2/4}\log^{1/2}(dn)}{n^{1/2}}\bigg\}^{s-1} \notag \\
&~~~~~\times \frac{1}{n}\sum_{i=1}^{n} e^{U_{i,j}^2/4}e^{V_{i,k}^2/4} I{(M_2 <|U_{i,j}|,\,|V_{i,k}| \le M_1)}   \\
\le&~  \frac{C_{16}\log (dn)}{nM_{2}}\times \frac{1}{n}\sum_{i=1}^{n}e^{U_{i,j}^2/4}e^{V_{i,k}^2/4} I{(M_2 <|U_{i,j}|,\,|V_{i,k}| \le M_1)}   \notag
\end{align}
provided that $\log(dn) \ll nM_1^{-1}e^{-M_1^2/2}$. Due to $U_{i,j}, V_{i,k} \sim \mathcal{N}(0,1)$, by  Cauchy-Schwarz inequality, it holds that
\begin{align*}
&\max_{i\in[n],\,j\in[p],\, k\in[q]}\mathbb{E}\big\{ e^{U_{i,j}^2/4}e^{V_{i,k}^2/4} I{(M_2<|U_{i,j}|,\,|V_{i,k}|\le M_1)}   \big\} \\
&~~~~~\le \max_{i\in[n],\,j\in[p]}\big[\mathbb{E}\big\{ e^{U_{i,j}^2/2} I{(|U_{i,j}| \le M_1)} \big\}\big]^{1/2}   \max_{i\in[n],\,k\in[q] }\big[\mathbb{E}\big\{ e^{V_{i,k}^2/2} I{(|V_{i,k}| \le M_1)} \big\}\big]^{1/2}  \lesssim    M_1 \,,\\ 
&\max_{i\in[n],\,j\in[p],\, k\in[q]}\var\big\{ e^{U_{i,j}^2/4}e^{V_{i,k}^2/4} I{(M_2 <|U_{i,j}|,\,|V_{i,k}| \le M_1)}  \big\}  \\
&~~~~~\le  \max_{i\in[n],\,j\in[p] }\big[\mathbb{E}\big\{ e^{U_{i,j}^2} I{( |U_{i,j}| \le M_1)} \big\}\big]^{1/2} \max_{i\in[n],\,k\in[q]}\big[\mathbb{E}\big\{  e^{V_{i,k}^2}I{( |V_{i,k}| \le M_1)} \big\}\big]^{1/2}  \lesssim   M_1^{-1} e^{M_1^2/2} \,.
\end{align*}
Using the similar arguments for the derivation of \eqref{eq:bern-bound}, we have
\begin{align}\label{eq:u4v4} 
&\max_{j\in[p],\, k\in[q]}  \frac{1}{n }\sum_{i=1}^{n}   e^{U_{i,j}^2/4}e^{V_{i,k}^2/4}  I{(M_2<|U_{i,j}| ,\,|V_{i,k}|\le M_1)}   = O_{\rm p} ( M_1)
\end{align}
provided that $\log d \lesssim ne^{-M_1^2/2}M_1$. By \eqref{eq:J141}, it holds that
\begin{align}\label{eq:j141}
\max_{j\in[p], \,k\in[q]}|{\rm J}_{141}(j,k)| =O_{\rm p}\{n^{-1}M_{1}M_{2}^{-1}\log(dn)\} 
\end{align}
provided that $\log(dn)\ll ne^{-M_1^2/2}M_1^{-1}$. Analogously,  we have
\begin{align*}
&~|\textrm{J}_{142}(j,k)| \notag\\
\le&~ \sum_{l=1}^{\infty}\sum_{s=1}^{\infty} C_{17}^{l+s} \bigg\{\frac{\log(dn)}{n}\bigg\}^{(l+2s)/2}   \frac{M_{1}^{s-1}}{n}\sum_{i=1}^{n}\big\{|U_{i,j}|^{l/2-1}e^{lU_{i,j}^2/4}e^{sV_{i,k}^2/2} I{(M_2 <|U_{i,j}|,\,|V_{i,k}| \le M_1)}\\
\le&~ \frac{\log^{3/2} (dn)}{n^{3/2}M_{2}^{1/2}}  \times \sum_{l=1}^{\infty} \bigg\{ \frac{C_{18}M_1^{1/2}e^{M_1^2/4}\log^{1/2}(dn)}{n^{1/2}}\bigg\}^{l-1}  \times  \sum_{s=1}^{\infty} \bigg\{ \frac{C_{18}M_1e^{M_1^2/2}\log(dn)}{n}\bigg\}^{s-1}  \notag \\
&~~~~~\times \frac{1}{n}\sum_{i=1}^{n}e^{U_{i,j}^2/4}e^{V_{i,k}^2/2} I{(M_2 <|U_{i,j}|,\,|V_{i,k}| \le M_1)}   \\
\le &~ \frac{C_{19}e^{M_1^2/4}\log^{3/2} (dn)}{n^{3/2}M_{2}^{1/2}} \times \frac{1}{n} \sum_{i=1}^{n} e^{U_{i,j}^2/4}e^{V_{i,k}^2/4} I{(M_2 <|U_{i,j}|,\,|V_{i,k}| \le M_1)}   
\end{align*}
provided that $\log(dn)\ll ne^{-M_1^2/2}M_1^{-1}$.  By \eqref{eq:u4v4}, 
\begin{align}\label{eq:j142} 
\max_{j\in[p], \,k\in[q]}|{\rm J}_{142}(j,k)| =O_{\rm p}\{n^{-3/2}M_{1}M_{2}^{-1/2}e^{M_1^2/4}\log^{3/2}(dn)\} 
\end{align}
provided that $\log(dn)\ll ne^{-M_1^2/2}M_1^{-1}$. Analogously, we can also show such convergence rate holds for $	\max_{j\in[p], \,k\in[q]}|{\rm J}_{143}(j,k)|$. If $\log(dn)\ll ne^{-M_1^2/2}M_1^{-1}$, it holds that
\begin{align*} 
|\textrm{J}_{144}(j,k)| 
\le&~  \sum_{l=1}^{\infty}\sum_{s=1}^{\infty} C_{20}^{l+s} \bigg\{\frac{\log(dn)}{n}\bigg\}^{l+s}   \frac{M_{1}^{l+s-2}}{n}\sum_{i=1}^{n}e^{lU_{i,j}^2/2}e^{sV_{i,k}^2/2} I{(M_2 <|U_{i,j}| ,\,|V_{i,k}|\le M_1)}    \notag \\
\le&~  \frac{\log^{2} (dn)}{n^{2}} \times  \sum_{l=1}^{\infty}   \bigg\{ \frac{C_{21}M_1e^{M_1^2/2}\log(dn)}{n}\bigg\}^{l-1}  \times \sum_{s=1}^{\infty}  \bigg\{\frac{C_{21}M_1e^{M_1^2/2}\log(dn)}{n}\bigg\}^{s-1}  \\
&~~~~\times  \frac{1}{n}\sum_{i=1}^{n} e^{U_{i,j}^2/2}e^{V_{i,k}^2/2} I{(M_2 <|U_{i,j}|,\,|V_{i,k}| \le M_1)}  \notag\\
\le &~ \frac{C_{22}e^{M_1^2/2}\log^{2} (dn)}{n^{2}} \times \frac{1}{n} \sum_{i=1}^{n} e^{U_{i,j}^2/4}e^{V_{i,k}^2/4} I{(M_2 <|U_{i,j}| ,\,|V_{i,k}|\le M_1)}  \,.
\end{align*}
By \eqref{eq:u4v4} again, 
\begin{align}\label{eq:j144}
\max_{j\in[p], \,k\in[q]}|{\rm J}_{144}(j,k)| =O_{\rm p}\{n^{-2}M_1 e^{M_1^2/2}\log^2(dn) \} 
\end{align}
provided that $\log(dn)\ll ne^{-M_1^2/2}M_1^{-1}$. Notice that, restricted on $\mathcal{H}_{2} \bigcap\mathcal{H}_{4}$, by \eqref{eq:j14-dec},
\begin{align*}
\max_{j\in[p], \,k\in[q]}|{\rm J}_{14}(j,k)|  \le&~ \max_{j\in[p], \,k\in[q]}|{\rm J}_{141}(j,k)|+ \max_{j\in[p], \,k\in[q]}|{\rm J}_{142}(j,k)|\\
&+ \max_{j\in[p], \,k\in[q]}|{\rm J}_{143}(j,k)|+\max_{j\in[p], \,k\in[q]}|{\rm J}_{144}(j,k)| \,.
\end{align*}
Since $\mathbb{P}(\mathcal{H}_{2}^{\rm c}) \le 4 (pn)^{-2}$ and  $\mathbb{P}(\mathcal{H}_{4}^{\rm c}) \le 4 (qn)^{-2}$, applying the similar arguments in Section \ref{sec:sub-i12} for deriving the convergence rate of $\max_{j\in[p],\,k\in[q]}|{\rm I}_{12}(j,k)|$, 
together with \eqref{eq:j141}--\eqref{eq:j144}, we have
\begin{align*}
& \max_{j\in[p],\, k\in[q]}|{\rm J}_{14}(j,k)| = O_{\rm p}\{n^{-1}M_1M_2^{-1}\log(dn)\}  
\end{align*}
provided that $ \log (dn) \ll ne^{-M_1^2/2} M_1^{-1}$. We complete the proof of \eqref{eq:h0-j14}.
$\hfill\Box$

\subsection{Convergence rate of ${\max_{j\in[p],\,k\in[q]}|{\rm J}_{2}(j,k)|}$}\label{sec:sub-J2}
As shown in \eqref{eq:Uhatstar}, it holds that $\max_{i\in[n],\,j\in[p]}|\hat{U}_{i,j} - U_{i,j}^{*}| \le 2 \sqrt{2\log (n+1)}$. Analogously, we also have $\max_{i\in[n],\,k\in[q]}|\hat{V}_{i,k} -V_{i,k}^{*}| \le 2 \sqrt{2\log (n+1)}$. Then
\begin{align*}
|{\rm J}_{2}(j,k)| \le 8\log(n+1) \bigg\{\frac{1}{n}\sum_{i=1}^{n} I(|V_{i,k}|> M_1) +  \frac{1}{n}\sum_{i=1}^{n} I(|U_{i,j}|> M_1)\bigg\}\,.
\end{align*}
Recall $V_{i,k}\sim N(0,1)$. Identical to \eqref{eq:iu-M1}, it holds that
\begin{align*} 
\max_{k\in[q]}\bigg|\frac{1}{n}\sum_{i=1}^{n} I(|V_{i,k}| > M_1)\bigg| = O_{\rm p}(M_1^{-1} e^{-M_1^2/2})
\end{align*}
provided that $ \log q \lesssim n e^{-M_1^2/2} M_1^{-1}$. Recall $d=pq$ and $M_1=\sqrt{\kappa_1\log n}$ for some constant $  \kappa_1 \in(1, 2)$. Together with \eqref{eq:iu-M1}, we complete the proof of  \eqref{eq:j2}.
$\hfill\Box$

\section{Proof of Lemma \ref{lem:usvs-uv}}\label{sec:sub-usvs-uv} 
Recall $U_{i,j}^{*} =  U_{i,j}I(|U_{i,j}|\le M_1) + M_1 \cdot{\rm sign}(U_{i,j})I(|U_{i,j}|>M_1)$ and $V_{i,k}^{*} =  V_{i,k}I(|V_{i,k}|\le M_1) + M_1 \cdot{\rm sign}(V_{i,k})I(|V_{i,k}|>M_1)$, where $M_1=\sqrt{\kappa_1\log n}$ for some constant $\kappa_1 \in(1, 2)$. Define $\acute{U}_{i,j} =  U_{i,j}-M_1 \cdot{\rm sign}(U_{i,j})$ and $\acute{V}_{i,k}  =  V_{i,k}-M_1 \cdot{\rm sign}(V_{i,k})$. We have  $U_{i,j}^{*} - U_{i,j} =  -\acute{U}_{i,j}I(|U_{i,j}|>M_1)$ and $V_{i,k}^{*} -V_{i,k} = -\acute{V}_{i,k}I(|V_{i,k}|>M_1)$.  Hence, it holds that
\begin{align}\label{eq:usvs-dec}
&\frac{1}{n}\sum_{i=1}^{n}(U_{i,j}^{*}V_{i,k}^{*} - U_{i,j}V_{i,k}) \notag \\
&~~~~~~~ =  \frac{1}{n}\sum_{i=1}^{n}(U_{i,j}^{*} - U_{i,j}) V_{i,k}^{*} +  \frac{1}{n}\sum_{i=1}^{n}(V_{i,k}^{*} -V_{i,k})U_{i,j}^{*}  - \frac{1}{n}\sum_{i=1}^{n}(U_{i,j}^{*} - U_{i,j})(V_{i,k}^{*} -V_{i,k}) \notag\\
&~~~~~~~=  - \underbrace{\frac{1}{n}\sum_{i=1}^{n}  \acute{U}_{i,j} I(|U_{i,j}|>M_1) V_{i,k}^{*}   }_{\textrm{K}_{1}(j,k)} -   \underbrace{\frac{1}{n}\sum_{i=1}^{n}  \acute{V}_{i,k} I(|V_{i,k}|>M_1) U_{i,j}^{*}   }_{\textrm{K}_{2}(j,k)} \notag\\
&~~~~~~~~~~~~- \underbrace{\frac{1}{n}\sum_{i=1}^{n}  \acute{U}_{i,j}\acute{V}_{i,k} I(|U_{i,j}|,\,|V_{i,k}|>M_1) }_{\textrm{K}_{3}(j,k)} \,.
\end{align}
Given $Q>M_1$, it holds that
\begin{align}\label{eq:k1-dec}
{\rm K}_{1}(j,k) =&~ ~\frac{1}{n}\sum_{i=1}^{n} \underbrace{\big[\acute{U}_{i,j} I(M_1<|U_{i,j}|\le Q) V_{i,k}^{*} -\mathbb{E}\{\acute{U}_{i,j} I(M_1<|U_{i,j}|\le Q) V_{i,k}^{*}\}\big] }_{\textrm{K}_{11}(i,j,k)}\notag\\
&+\frac{1}{n}\sum_{i=1}^{n} \underbrace{\acute{U}_{i,j} I(|U_{i,j}|>Q) V_{i,k}^{*}  }_{\textrm{K}_{12}(i,j,k)} +\underbrace{\mathbb{E}\{\acute{U}_{i,j} I(M_1<|U_{i,j}|\le Q) V_{i,k}^{*}\} }_{\textrm{K}_{13}(i,j,k)}\,.
\end{align}
Recall $U_{i,j} \sim \mathcal{N}(0,1)$, $\acute{U}_{i,j} =  U_{i,j}-M_1 \cdot{\rm sign}(U_{i,j})$ and $V_{i,k}^{*} =  V_{i,k}I(|V_{i,k}|\le M_1) + M_1 \cdot{\rm sign}(V_{i,k})I(|V_{i,k}|>M_1)$. Notice that
\begin{align*}
&\max_{i\in[n],\,j\in[p],\, k\in[q]}\var\{\acute{U}_{i,j} I(M_1< |U_{i,j}|\le Q) V_{i,k}^{*}\} \\
&~~~~~~~~~~~\le M_1^2 \max_{i\in[n],\,j\in[p] }\mathbb{E}\{\acute{U}_{i,j}^2 I(|U_{i,j}|>M_1)\} \lesssim M_1^3e^{-M_1^2/2}\,.
\end{align*}
Recall $d=pq$. By
Bonferroni inequality and Bernstein inequality, it holds that
\begin{align*}
\mathbb{P}\bigg\{\max_{j\in[p],\, k\in[q]}\bigg|\frac{1}{n}\sum_{i=1}^{n} {\rm K}_{11}(i,j,k)\bigg|> x\bigg\} \le 2d \exp\bigg(-\frac{nx^2}{C_1M_{1}^{3}e^{-M_1^2/2} + C_2 QM_1x}\bigg)
\end{align*}
for any  $x>0$, which implies
\begin{align}\label{eq:k11}
\max_{j\in[p], \,k\in[q]}\bigg|\frac{1}{n}\sum_{i=1}^{n} {\rm K}_{11}(i,j,k)\bigg|=O_{\rm p} \{n^{-1/2}M_1^{3/2}e^{-M_1^2/4}(\log d)^{1/2}\} + O_{\rm p} (n^{-1}QM_1 \log d)\,.
\end{align}
Due to $U_{i,j}\sim \mathcal{N}(0,1)$, for any $x>0$,  by the Bonferroni inequality, we have
\begin{align*} 
\mathbb{P}\bigg\{\max_{j\in[p], \,k\in[q]}\bigg|\frac{1}{n}\sum_{i=1}^{n} {\rm K}_{12}(i,j,k)\bigg|> x\bigg\} \le&~   \mathbb{P}\bigg(\max_{i\in[n],\, j\in[p]}|U_{i,j}| > Q\bigg)\\
\le&~  np \max_{i\in[n],\, j\in[p]}\mathbb{P}(|U_{i,j}| > Q)\lesssim nd 
e^{-CQ^2}\,,
\end{align*}
which implies 
\begin{align}\label{eq:k12}
\max_{j\in[p], \,k\in[q]}\bigg|\frac{1}{n}\sum_{i=1}^{n} {\rm K}_{12}(i,j,k)\bigg|=o_{\rm p}(n^{-1})
\end{align}
provided that $\log(dn) \lesssim Q^2$. Furthermore, 
\begin{align*}
\max_{i\in[n],\,j\in[p],\, k\in[q]}|{\rm K}_{13}(i,j,k)| \le  M_1 \max_{i\in[n],\,j\in[p]}\mathbb{E}\{|\acute{U}_{i,j}| I(M_1 <|U_{i,j}| \le Q)\} \lesssim   M_1 e^{-M_1^2/2} \,.
\end{align*}
By selecting $Q =C_* \log^{1/2} (dn)$ for some sufficiently large constant $C_*> \kappa_1$, together with \eqref{eq:k11} and \eqref{eq:k12}, by \eqref{eq:k1-dec}, we then have 
\begin{align}\label{eq:k1}
\max_{j\in[p],\, k\in[q]}|{\rm K}_{1}(j,k)| =  O_{\rm p} \{n^{-1}M_1 (\log d) \log^{1/2} (dn) \} + O_{\rm p}(M_1e^{-M_1^2/2}) 
\end{align}
provided that $\log d \lesssim n M_1^{-1}e^{-M_1^2/2}$. Using the similar arguments, we can also show such convergence rate holds for $\max_{j\in[p],\, k\in[q]}|{\rm K}_{2}(j,k)|$. 

Analogously, given $Q>M_1$, it holds that
\begin{align}\label{eq:k3-dec}
{\rm K}_{3}(j,k) =&~ \underbrace{\frac{1}{n}\sum_{i=1}^{n} \big[ \acute{U}_{i,j}\acute{V}_{i,k} I(M_1<|U_{i,j}|,\,|V_{i,k}|\le Q) -\mathbb{E}\{\acute{U}_{i,j}\acute{V}_{i,k} I(M_1 < |U_{i,j}|,\,|V_{i,k}|\le Q) \}\big] }_{\textrm{K}_{31}(j,k)}\notag\\
&+\underbrace{\frac{1}{n}\sum_{i=1}^{n} \acute{U}_{i,j}\acute{V}_{i,k} I(M_1<|U_{i,j}|\le Q) I(|V_{i,k}|>Q) }_{\textrm{K}_{32}(j,k)} \\ &+\underbrace{\frac{1}{n}\sum_{i=1}^{n} \acute{U}_{i,j}\acute{V}_{i,k} I(|U_{i,j}|>Q) I(|V_{i,k}|>M_1) }_{\textrm{K}_{33}(j,k)}+\underbrace{\mathbb{E}\{\acute{U}_{i,j}\acute{V}_{i,k} I(M_1 < |U_{i,j}|,\,|V_{i,k}|\le Q)  \}}_{\textrm{K}_{34}(j,k)} \notag \,. 
\end{align}
Recall $U_{i,j}, V_{i,k} \sim \mathcal{N}(0,1)$, $\acute{U}_{i,j} =  U_{i,j}-M_1 \cdot{\rm sign}(U_{i,j})$ and $\acute{V}_{i,k}  =  V_{i,k}-M_1 \cdot{\rm sign}(V_{i,k})$. By  Cauchy-Schwarz inequality, we have 
\begin{align*}
&\max_{i\in[n],\, j\in[p], \,k\in[q]}\var\{\acute{U}_{i,j}\acute{V}_{i,k} I(M_1<|U_{i,j}|,\,|V_{i,k}|\le Q)  \}\\
&~~~~~~~~\le \max_{i\in[n], \,j\in[p]}\big(\mathbb{E}[\{M_1 \cdot{\rm sign}(U_{i,j}) -U_{i,j}\}^4 I( M_1 <|U_{i,j}| \le Q) ] \big)^{1/2}\\
&~~~~~~~~~~~~\times \max_{i\in[n], \,k\in[q]} \big(\mathbb{E}[ \{M_1 \cdot{\rm sign}(V_{i,k}) -V_{i,k}\}^4 I(M_1 < |V_{i,k}|\le Q)]\big) ^{1/2}\\
&~~~~~~~~\le   C_3\max_{i\in[n], \,j\in[p]}\big[M_1^4 \mathbb{P}(|U_{i,j}|> M_1) + \mathbb{E}\{U_{i,j}^4 I( |U_{i,j}|>M_1 )\}\big]^{1/2}\\
&~~~~~~~~~~~~\times \max_{i\in[n],\, k\in[q]} \big[M_1^4 \mathbb{P}(|V_{i,k}|> M_1) + \mathbb{E}\{V_{i,k}^4 I(  |V_{i,k}| >M_1 )\}\big]^{1/2}\\
&~~~~~~~~\lesssim M_{1}^3 e^{-M_1^2/2}\,.
\end{align*}
Recall $d=pq$. Analogous to the derivation of \eqref{eq:k11}, it holds that
\begin{align*}
\max_{j\in[p], \,k\in[q]}| {\rm K}_{31}(j,k)|=O_{\rm p} \{n^{-1/2}M_1^{3/2}e^{-M_1^2/4}(\log d)^{1/2}\} + O_{\rm p} (n^{-1}Q^2 \log d)\,.
\end{align*}
Using the similar arguments for the derivation of \eqref{eq:k12}, we also have $\max_{j\in[p], \,k\in[q]} | {\rm K}_{32}(j,k) |=o_{\rm p}(n^{-1}) =\max_{j\in[p], \,k\in[q]} | {\rm K}_{33}(j,k) |$ provided that $\log(dn) \lesssim Q^2$. Furthermore, by  Cauchy-Schwarz inequality,
\begin{align*}
\max_{j\in[p],\, k\in[q]}|{\rm K}_{34}(j,k)| \le&~  \max_{i\in[n],\,j\in[p]} [\mathbb{E}\{\acute{U}_{i,j}^2 I(|U_{i,j}|>M_1)\} ]^{1/2} \max_{i\in[n],\,k\in[q]} [\mathbb{E}\{\acute{V}_{i,k}^2I( |V_{i,k}|>M_1)\} ]^{1/2} \\
\lesssim &~   M_1 e^{-M_1^2/2}   \,.
\end{align*}
With selecting $Q =C_{*} \log^{1/2} (dn)$ for some sufficiently large constant $C_{*}> \kappa_1$, by \eqref{eq:k3-dec}, we then have 
\begin{align*}
\max_{j\in[p],\, k\in[q]}|{\rm K}_{3}(j,k)| =  O_{\rm p} \{n^{-1} (\log d) \log (dn) \} + O_{\rm p}(M_{1}e^{-M_1^2/2}) 
\end{align*}
provided that $\log d \lesssim n M_1^{-1}e^{-M_1^2/2}$. Recall $M_1=\sqrt{\kappa_1\log n}$ for some constant $\kappa_1 \in(1, 2)$. Together with \eqref{eq:k1}, by \eqref{eq:usvs-dec}, it holds that
\begin{align*}
&\max_{j\in[p],\, k\in[q]}\bigg|\frac{1}{\sqrt{n}}\sum_{i=1}^{n}(U_{i,j}^{*}V_{i,k}^{*} - U_{i,j}V_{i,k})\bigg| \\
&~~~~~~~~~\le\sqrt{n}\bigg\{\max_{j\in[p],\, k\in[q]} |{\rm K}_{1}(j,k)|+ \max_{j\in[p],\, k\in[q]}|{\rm K}_{2}(j,k)| + \max_{j\in[p],\, k\in[q]}|{\rm K}_{3}(j,k)|\bigg\}\\
&~~~~~~~~~=  O_{\rm p} \{n^{-(\kappa_1-1)/2}(\log n)^{1/2}\} + O_{\rm p}\{n^{-1/2}(\log d) \log (dn)\}  
\end{align*}
provided that $\log d \lesssim n^{1-\kappa_1/2}(\log n)^{-1/2}$. We complete the proof of Lemma \ref{lem:usvs-uv}.
$\hfill\Box$

\section{Proof of Lemma \ref{lem:covh}}\label{sec:sub-sigmma-g-h0}  
Recall $\bSigma= \mathbb{E}(\bgamma_i\bgamma_i^{\T}) -  \mathbb{E}(\bgamma_i)\mathbb{E}(\bgamma_i^{\T})$ and $\hat{\bSigma}=n^{-1}\sum_{i=1}^{n}\hat{\bgamma}_i\hat{\bgamma}_i^{\T}-(n^{-1}\sum_{i=1}^{n}\hat{\bgamma}_i)(n^{-1}\sum_{i=1}^{n}\hat{\bgamma}_i)^{\T}$ with $\bgamma_i=\bU_i \otimes \bV_i$ and   $\hat{\bgamma}_i=\hat{\bU}_i \otimes \hat{\bV}_i$. Then
\begin{align}\label{eq:SigmaGamma}
|\hat{\bSigma} - \bSigma|_{\infty}\le&~ \underbrace{\max_{j,l\in[p],\,k,t\in[q]}\bigg|\frac{1}{n}\sum_{i=1}^{n}\hat{U}_{i,j}\hat{U}_{i,l}\hat{V}_{i,k}\hat{V}_{i,t}-\frac{1}{n}\sum_{i=1}^{n}U_{i,j}U_{i,l}V_{i,k}V_{i,t}\bigg|}_{\textrm{R}_{1}}\notag\\ 
&+ \underbrace{\max_{j,l\in[p],\,k,t\in[q]}\bigg|\frac{1}{n}\sum_{i=1}^{n}U_{i,j}U_{i,l}V_{i,k}V_{i,t}-\mathbb{E}(U_{i,j}U_{i,l}V_{i,k}V_{i,t})\bigg|}_{\textrm{R}_{2}}\\
&+ \underbrace{\max_{j,l \in[p],\, k, t\in [q]}\bigg| \bigg(\frac{1}{n}\sum_{i=1}^{n}U_{i,j}V_{i,k}\bigg)\bigg(\frac{1}{n}\sum_{i=1}^{n}U_{i,l}V_{i,t}\bigg) - \mathbb{E}(U_{i,j}V_{i,k})\mathbb{E}(U_{i,l}V_{i,t})\bigg|}_{\textrm{R}_{3}}\notag\\
&+\underbrace{\max_{j,l \in[p], \,k, t\in [q]}\bigg| \bigg(\frac{1}{n}\sum_{i=1}^{n}\hat{U}_{i,j}\hat{V}_{i,k}\bigg)\bigg(\frac{1}{n}\sum_{i=1}^{n}\hat{U}_{i,l}\hat{V}_{i,t}\bigg)- \bigg(\frac{1}{n}\sum_{i=1}^{n}U_{i,j}V_{i,k}\bigg)\bigg(\frac{1}{n}\sum_{i=1}^{n}U_{i,l}V_{i,t}\bigg)\bigg|}_{\textrm{R}_{4}} \notag\,.
\end{align}
As we will show in Sections \ref{sub:sec-h0-R1}--\ref{sub:sec-h0-R4}, 
\begin{align}\label{eq:h0-R1}
{\rm R}_{1}=   O_{\rm p}\{n^{-1/2} (\log n) (\log d)^{1/2}\log^{3/2}(dn)\}
\end{align}
provided that $\log d \lesssim n^{5/12}(\log n)^{-1/2} $, 
\begin{align}
&{\rm R}_{2} = O_{\rm p}\{n^{-1/2}(\log d)^{1/2}\} + O_{\rm p}\{n^{-1}\log^2 (dn) \log d\}\,,\label{eq:h0-R2}\\
&~~~~~~~~~~~~~~~~{\rm R}_{3}=O_{{\rm p}}\{n^{-1/2}(\log d)^{1/2}\}\label{eq:r3}
\end{align}
provided that $\log d \lesssim  n^{1/3}$, and
\begin{align}\label{eq:h0-R4}
{\rm R}_{4}=   O_{\rm p} \{n^{-1/2}(\log n) (\log d)^{1/2} \log^{1/2}(dn)\}
\end{align}
provided that  $\log d \lesssim  n^{1/3}$.
Together with \eqref{eq:h0-R1}--\eqref{eq:h0-R4}, it follows from \eqref{eq:SigmaGamma} that
\begin{align*}
|\hat{\bSigma} - \bSigma |_{\infty}  =  O_{\rm p}\{n^{-1/2} (\log n) (\log d)^{1/2}\log^{3/2}(dn)\}
\end{align*}
provided that $\log d \lesssim  n^{1/3}$.
We complete the proof of Lemma \ref{lem:covh}.
$\hfill\Box$

\subsection{Convergence rate of ${\rm R}_{1}$}\label{sub:sec-h0-R1}
Notice that
\begin{align}\label{eq:4expend}
&~\frac{1}{n}\sum_{i=1}^{n}\big(\hat{U}_{i,j}\hat{U}_{i,l}\hat{V}_{i,k}\hat{V}_{i,t}-U_{i,j}U_{i,l}V_{i,k}V_{i,t}\big)\notag \\
=&~\frac{1}{n}\sum_{i=1}^{n}
(\hat{U}_{i,j}-U_{i,j})(\hat{U}_{i,l}-U_{i,l})(\hat{V}_{i,k}-V_{i,k})(\hat{V}_{i,t}-V_{i,t})  \notag\\
&+\frac{1}{n}\sum_{i=1}^{n}(\hat{U}_{i,j}-U_{i,j})(\hat{U}_{i,l}-U_{i,l})(\hat{V}_{i,k}-V_{i,k})V_{i,t}  + \frac{1}{n}\sum_{i=1}^{n}
(\hat{U}_{i,j}-U_{i,j})(\hat{V}_{i,k}-V_{i,k})(\hat{V}_{i,t}-V_{i,t})U_{i,l}\notag\\
&+ \frac{1}{n}\sum_{i=1}^{n}
(\hat{U}_{i,l}-U_{i,l})(\hat{V}_{i,k}-V_{i,k})(\hat{V}_{i,t}-V_{i,t})U_{i,j} +\frac{1}{n}\sum_{i=1}^{n}(\hat{U}_{i,j}-U_{i,j})(\hat{U}_{i,l}-U_{i,l})(\hat{V}_{i,t}-V_{i,t})V_{i,k}  \notag\\
&+ \frac{1}{n}\sum_{i=1}^{n}(\hat{U}_{i,j}-U_{i,j})(\hat{U}_{i,l}-U_{i,l})V_{i,k}V_{i,t} + \frac{1}{n}\sum_{i=1}^{n}
(\hat{U}_{i,j}-U_{i,j})(\hat{V}_{i,t}-V_{it})U_{i,l}V_{i,k}\\
&+\frac{1}{n}\sum_{i=1}^{n}
(\hat{U}_{i,l}-U_{i,l})(\hat{V}_{i,k}-V_{i,k})U_{i,j}V_{i,t} + \frac{1}{n}\sum_{i=1}^{n}
(\hat{U}_{i,l}-U_{i,l})(\hat{V}_{i,t}-V_{i,t})U_{i,j}V_{i,k}\notag\\
&+\frac{1}{n}\sum_{i=1}^{n}
(\hat{U}_{i,j}-U_{i,j})(\hat{V}_{i,k}-V_{i,k})U_{i,l}V_{i,t} +\frac{1}{n}\sum_{i=1}^{n}
(\hat{V}_{i,k}-V_{i,k})(\hat{V}_{i,t}-V_{i,t})U_{i,j}U_{i,l} \notag \\
&+ \frac{1}{n}\sum_{i=1}^{n}(\hat{U}_{i,j}-U_{i,j})U_{i,l}V_{i,k}V_{i,t} + \frac{1}{n}\sum_{i=1}^{n}(\hat{U}_{i,l}-U_{i,l})U_{i,j}V_{i,k}V_{i,t}\notag\\
& +\frac{1}{n}\sum_{i=1}^{n}(\hat{V}_{i,k}-V_{i,k})U_{i,j}U_{i,l}V_{i,t} + \frac{1}{n}\sum_{i=1}^{n}(\hat{V}_{i,t}-V_{i,t})U_{i,j}U_{i,l}V_{i,k} \notag\,.
\end{align}
To derive the convergence rate of  ${\rm R}_{1}$, by the symmetry, we only consider the convergence rates of  the following terms:
\begin{align}\label{eq:R1terms}
\textrm{R}_{11} & = \max_{j,l\in[p],\,k,t\in[q]}\bigg|\frac{1}{n}\sum_{i=1}^{n}(\hat{U}_{i,j}-U_{i,j})U_{i,l}V_{i,k}V_{i,t}\bigg| \,,\notag\\
\textrm{R}_{12} & = \max_{j,l\in[p],\,k,t\in[q]}\bigg|\frac{1}{n}\sum_{i=1}^{n}(\hat{U}_{i,j}-U_{i,j})(\hat{U}_{i,l}-U_{i,l})V_{i,k}V_{i,t}\bigg|\,, \notag \\
\textrm{R}_{13} & = \max_{j,l\in[p],\,k,t\in[q]}\bigg|\frac{1}{n}\sum_{i=1}^{n}(\hat{U}_{i,j}-U_{i,j})(\hat{V}_{i,k}-V_{i,k})U_{i,l}V_{i,t}\bigg|\,, \\
\textrm{R}_{14} & = \max_{j,l\in[p],\,k,t\in[q]}\bigg|\frac{1}{n}\sum_{i=1}^{n}(\hat{U}_{i,j}-U_{i,j})(\hat{U}_{i,l}-U_{i,l})(\hat{V}_{i,k}-V_{i,k})V_{i,t}\bigg|\,, \notag\\
\textrm{R}_{15} & = \max_{j,l\in[p],\,k,t\in[q]}\bigg|\frac{1}{n}\sum_{i=1}^{n}(\hat{U}_{i,j}-U_{i,j})(\hat{U}_{i,l}-U_{i,l})(\hat{V}_{i,k}-V_{i,k})(\hat{V}_{i,t}-V_{i,t})\bigg| \notag\,.
\end{align}
As we will show in Sections \ref{sec:sub-r11}--\ref{sec:sub-r15},
\begin{align}
&~~~~{\rm R}_{11} = 
O_{\rm p}\{n^{-1/2} (\log n) (\log d)^{1/2}\log^{3/2}(dn)\}\,,\label{eq:r11}\\
&{\rm R}_{12} =  
O_{\rm p}\{n^{-1/2} (\log n) (\log d)^{1/2}\log^{3/2}(dn)\}={\rm R}_{13}  \,,\label{eq:r12}\\
&{\rm R}_{14} =  
O_{\rm p}\{n^{-1/2} (\log n) (\log d)^{1/2}\log^{3/2}(dn)\} = {\rm R}_{15}\label{eq:r14}  
\end{align}  
provided that $\log d \lesssim n^{5/12}(\log n)^{-1/2}$. Combining with \eqref{eq:r11}--\eqref{eq:r14}, by \eqref{eq:4expend} and \eqref{eq:R1terms}, we have \eqref{eq:h0-R1} holds. $\hfill\Box$

\subsubsection{ Convergence rate of ${\rm R}_{11}$}\label{sec:sub-r11}
Recall $\hat{U}_{i,j}=\Phi^{-1} \{n(n+1)^{-1}\hat{F}_{\bX,j}(X_{i,j})\}$ and $\hat{V}_{i,k}=\Phi^{-1} \{n(n+1)^{-1}\hat{F}_{\bY,k}(Y_{i,k})\}$. Given $M_1=\sqrt{\kappa_1\log n}$ for some constant $ \kappa_1 \in(1, 2)$, define $U_{i,j}^{*} =  U_{i,j}I(|U_{i,j}|\le M_1) + M_1 \cdot{\rm sign}(U_{i,j})I(|U_{i,j}|>M_1)$  and $V_{i,k}^{*} =  V_{i,k}I(|V_{i,k}|\le M_1) + M_1 \cdot{\rm sign}(V_{i,k})I(|V_{i,k}|>M_1)$. Let
\begin{align*}
\hat{U}_{i,j}^{*}  = \hat{U}_{i,j} -U_{i,j}^{*} \,, ~
\hat{V}_{i,k}^{*} =  \hat{V}_{i,k}- V_{i,k}^{*}\,,  ~ \tilde{U}_{i,j} = U_{i,j} - U_{i,j}^{*}\,, ~ \tilde{V}_{i,k} = V_{i,k} - V_{i,k}^{*}\,.
\end{align*} 
Then, we have $\hat{U}_{i,j}-U_{i,j} = \hat{U}_{i,j}^{*}-\tilde{U}_{i,j}$ and $\hat{V}_{i,k}-V_{i,k}=\hat{V}_{i,k}^{*}-\tilde{V}_{i,k}$. Hence, it holds that
\begin{align}\label{eq:R25terms}
{\rm R}_{11} =&~\max_{j,l\in[p],\,k,t\in[q]}\bigg|\frac{1}{n}\sum_{i=1}^{n}(\hat{U}_{i,j}^{*}-\tilde{U}_{i,j})U_{i,l}V_{i,k}V_{i,t}\bigg|\notag\\
\le&~ \underbrace{\max_{j,l\in[p],\,k,t\in[q]}\bigg|\frac{1}{n}\sum_{i=1}^{n}\tilde{U}_{i,j}U_{i,l}V_{i,k}V_{i,t}\bigg|}_{\textrm{R}_{111}} 
+ \underbrace{\max_{j,l\in[p],\,k,t\in[q]}\bigg|\frac{1}{n}\sum_{i=1}^{n}\hat{U}_{i,j}^{*}U_{i,l}V_{i,k}V_{i,t}\bigg|}_{\textrm{R}_{112}}\,.
\end{align}
Recall $\acute{U}_{i,j} = U_{i,j} -  M_1 \cdot{\rm sign}(U_{i,j})$. Since $U_{i,j}^{*} =  U_{i,j}I(|U_{i,j}|\le M_1) + M_1 \cdot{\rm sign}(U_{i,j})I(|U_{i,j}|>M_1)$, we have $\tilde{U}_{i,j} = U_{i,j} - U_{i,j}^{*} = \acute{U}_{i,j}I(|U_{i,j}|> M_1)$.  Given $Q>M_1$,  it holds that
\begin{align*}
{\rm R}_{111}\le &~ \underbrace{\max_{j,l\in[p],\, k,t \in[q]}\frac{1}{n}\sum_{i=1}^{n}|\acute{U}_{i,j}U_{i,l}V_{i,k}V_{i,t}| I(M_1<|U_{i,j}|\le Q)  I(|U_{i,l}|,\,|V_{i,k}|,\,|V_{i,t}|\le Q)  }_{\textrm{R}_{1111}} \\
&+\underbrace{\max_{j,l\in[p],\, k,t \in[q]}\frac{1}{n}\sum_{i=1}^{n}|\acute{U}_{i,j}U_{i,l}V_{i,k}V_{i,t}| I(M_1<|U_{i,j}|\le Q)  I(|U_{i,l}|,\,|V_{i,k}|\le Q)   I(|V_{i,t}|> Q) }_{\textrm{R}_{1112}}\\
&+ \underbrace{\max_{j,l\in[p],\, k,t \in[q]}\frac{1}{n}\sum_{i=1}^{n}|\acute{U}_{i,j}U_{i,l}V_{i,k}V_{i,t}| I( M_1 < |U_{i,j}| \le Q) I( |U_{i,l}| \le Q)  I(|V_{i,k}|>Q)}_{\textrm{R}_{1113}}\\
& + \underbrace{\max_{j,l\in[p],\, k,t \in[q]}\frac{1}{n}\sum_{i=1}^{n}|\acute{U}_{i,j}U_{i,l} V_{i,k}V_{i,t}| I( M_1 < |U_{i,j}| \le Q) I(|U_{i,l}|>Q)}_{\textrm{R}_{1114}} \\
&+\underbrace{\max_{j,l\in[p],\, k,t \in[q]}\frac{1}{n}\sum_{i=1}^{n}|\acute{U}_{i,j}U_{i,l}V_{i,k}V_{i,t}| I(|U_{i,j}|> Q)}_{\textrm{R}_{1115}}\,.
\end{align*}
Due to  $U_{i,j}\sim \mathcal{N}(0,1)$, we have
\begin{align*}
&\max_{i\in[n],\,j,l\in[p],\, k,t \in[q]}\mathbb{E}\big\{|\acute{U}_{i,j}U_{i,l} V_{i,k} V_{i,t}| I( M_1 <|U_{i,j}| \le Q) I(|U_{i,l}|,\,|V_{i,k}|,\,|V_{i,t}| \le Q)   \big\}\\
&~~~~~~~~~~~~ \le Q^3 \max_{i\in[n],\,j \in[p] }\mathbb{E} \big\{ |\acute{U}_{i,j}|I( M_1 <|U_{i,j}| \le Q)  \big\} \lesssim Q^3 e^{-M_1^2/2}\,,\\
&\max_{i\in[n],\,j,l\in[p],\, k,t \in[q]}\var\big\{|\acute{U}_{i,j}U_{i,l} V_{i,k} V_{i,t}| I( M_1 <|U_{i,j}| \le Q) I(|U_{i,l}|,\,|V_{i,k}|,\,|V_{i,t}| \le Q) \big\}\\
&~~~~~~~~~~~~  \le  Q^6\max_{i\in[n],\,j \in[p] }\mathbb{E} \big\{|\acute{U}_{i,j}|^2  I( M_1 <|U_{i,l}| \le Q)  \big\}  \lesssim Q^6 M_1e^{-M_1^2/2}\,.
\end{align*}
Recall $d=pq$.
Using the similar arguments for the derivation of \eqref{eq:bern-bound}, it holds that
\begin{align*}
{\rm R}_{1111} = O_{\rm p}(Q^3 e^{-M_1^2/2}) + O_{\rm p} (n^{-1} Q^4 \log d)
\end{align*}
provided that $\log d \lesssim nM_1^{-1}e^{-M_1^2/2}$.
Recall  $ V_{i,k}\sim \mathcal{N}(0,1)$. Analogous to the derivation of \eqref{eq:k12}, it holds that ${\rm R}_{1112}=o_{\rm p}(n^{-1}) = {\rm R}_{1113}$ and ${\rm R}_{1114}=o_{\rm p}(n^{-1}) = {\rm R}_{1115}$ provided that $\log (dn) \lesssim Q^2$. With selecting $Q=\tilde{C}\log^{1/2}(dn)$ for some sufficiently large constant $\tilde{C} >\kappa_1$, we have
\begin{align}\label{eq:r111}
{\rm R}_{111} = O_{\rm p}\{ e^{-M_1^2/2} \log^{3/2} (dn)\} + O_{\rm p} \{n^{-1} (\log d) \log^2(dn)\}  
\end{align}
provided that $\log d \lesssim nM_1^{-1}e^{-M_1^2/2}$.  

Given $Q> M_1$, it holds that
\begin{align*}
{\rm R}_{112} \le &~ \underbrace{\max_{j,l\in[p],\,k,t\in[q]}\bigg|\frac{1}{n}\sum_{i=1}^{n}\hat{U}_{i,j}^{*}U_{i,l}V_{i,k}V_{i,t}I(|U_{i,l}|,\,|V_{i,k}|,\,|V_{i,t}| \le Q) \bigg|}_{\textrm{R}_{1121}}\\
&+\underbrace{\max_{j,l\in[p],\,k,t\in[q]}\bigg|\frac{1}{n}\sum_{i=1}^{n}\hat{U}_{i,j}^{*}U_{i,l}V_{i,k}V_{i,t}I(|U_{i,l}|,\,|V_{i,k}| \le Q)I(|V_{i,t}|> Q)\bigg|}_{\textrm{R}_{1122}}\\
&+ \underbrace{\max_{j,l\in[p],\,k,t\in[q]}\bigg|\frac{1}{n}\sum_{i=1}^{n}\hat{U}_{i,j}^{*}U_{i,l}V_{i,k}V_{i,t}I(|U_{i,l}| \le Q)I(|V_{i,k}| > Q)\bigg|}_{\textrm{R}_{1123}}\\
&+ \underbrace{\max_{j,l\in[p],\,k,t\in[q]}\bigg|\frac{1}{n}\sum_{i=1}^{n}\hat{U}_{i,j}^{*}U_{i,l}V_{i,k}V_{i,t}I(|U_{i,l}|> Q) \bigg|}_{\textrm{R}_{1124}}\,.
\end{align*}
Recall $\hat{U}_{i,j}^{*}  = \hat{U}_{i,j} -U_{i,j}^{*}$ with $\hat{U}_{i,j}=\Phi^{-1} \{n(n+1)^{-1}\hat{F}_{\bX,j}(X_{i,j})\}$ and $U_{i,j}^{*} =  U_{i,j}I(|U_{i,j}|\le M_1) + M_1 \cdot{\rm sign}(U_{i,j})I(|U_{i,j}|>M_1)$. We have
\begin{align*}
{\rm R}_{1121} \le Q^3 \max_{j\in[p]} \frac{1}{n}\sum_{i=1}^{n} |\hat{U}_{i,j}^{*}| = Q^3 \max_{j\in[p]}\frac{1}{n}\sum_{i=1}^{n} | \hat{U}_{i,j} -U_{i,j}^{*}|\,.
\end{align*}
Recall $M_1=\sqrt{\kappa_1 \log n}$ for some constant $ \kappa_1 \in(1,2)$. Repeating the proofs for Lemmas 6 and 7 of \cite{Mai2022}, we can also show 
\begin{align*}
&\mathbb{P}\bigg\{\max_{j\in[p]} \bigg|\frac{1}{n}\sum_{i=1}^{n} | \hat{U}_{i,j} -U_{i,j}^{*}| - \mathbb{E}\bigg(\frac{1}{n}\sum_{i=1}^{n} | \hat{U}_{i,j} -U_{i,j}^{*}|\bigg) \bigg| > x\bigg\} \le C_1 p \exp\bigg(-\frac{C_2nx^2}{\log n}\bigg)
\end{align*}
for any $x>0$, and
\begin{align*}
\max_{j\in[p]}\mathbb{E}\bigg(\frac{1}{n}\sum_{i=1}^{n} | \hat{U}_{i,j} -U_{i,j}^{*}|\bigg) \lesssim \frac{\log n}{\sqrt{n}}\,,
\end{align*}
which implies
\begin{align}\label{eq:uh-us}
\max_{j\in[p]}\frac{1}{n}\sum_{i=1}^{n} | \hat{U}_{i,j} -U_{i,j}^{*}| = O_{\rm p}\{n^{-1/2} (\log n) (\log p)^{1/2}\}\,.
\end{align}
Recall $d=pq$. We then have
\begin{align*}
{\rm R}_{1121} = O_{\rm p}\{Q^3n^{-1/2} (\log n) (\log d)^{1/2}\}\,.
\end{align*} 
Recall  $U_{i,j}, V_{i,k}\sim \mathcal{N}(0,1)$. Analogous to the derivation of \eqref{eq:k12}, we also have ${\rm R}_{1122} =o_{\rm p}(n^{-1})$, ${\rm R}_{1123} =o_{\rm p}(n^{-1})$  and ${\rm R}_{1124} =o_{\rm p}(n^{-1})$ provided that $\log(dn) \lesssim Q^2$.  With selecting $Q = \tilde{C} \log^{1/2}(dn)$ for some sufficiently large constant $\tilde{C} > \kappa_1$, we have
\begin{align*}
{\rm R}_{112} = O_{\rm p}\{n^{-1/2} (\log n) (\log d)^{1/2}\log^{3/2}(dn)\} \,.
\end{align*} 
Recall  $M_1=\sqrt{\kappa_1 \log n}$ for some constant $\kappa_1 \in(1,2)$. Together with \eqref{eq:r111}, with selecting $\kappa_1=7/6$, by \eqref{eq:R25terms}, we have \eqref{eq:r11} holds.
$\hfill\Box$

\subsubsection{ Convergence rates of ${\rm R}_{12}$ and ${\rm R}_{13}$}\label{sec:sub-r12}
Due to $\hat{U}_{i,j}-U_{i,j} = \hat{U}_{i,j}^{*}-\tilde{U}_{i,j}$, we have 
\begin{align}\label{eq:R23terms} 
{\rm R}_{12} = &~\max_{j,l\in[p],\,k,t\in[q]}\bigg|\frac{1}{n}\sum_{i=1}^{n}(\hat{U}_{i,j}^{*}-\tilde{U}_{i,j})(\hat{U}_{i,l}^{*}-\tilde{U}_{i,l})V_{i,k}V_{i,t}\bigg| \notag\\
\le&~ \underbrace{\max_{j,l\in[p],\,k,t\in[q]}\bigg|\frac{1}{n}\sum_{i=1}^{n}\tilde{U}_{i,j}\tilde{U}_{i,l}V_{i,k}V_{i,t}\bigg|}_{\textrm{R}_{121}} + 2 \underbrace{\max_{j,l\in[p],\,k,t\in[q]}\bigg|\frac{1}{n}\sum_{i=1}^{n}\hat{U}_{i,j}^{*}\tilde{U}_{i,l}V_{i,k}V_{i,t}\bigg|}_{\textrm{R}_{122}}   \\
& + \underbrace{\max_{j,l\in[p],\,k,t\in[q]}\bigg|\frac{1}{n}\sum_{i=1}^{n}\hat{U}_{i,j}^{*}\hat{U}_{i,l}^{*}V_{i,k}V_{i,t}\bigg|}_{\textrm{R}_{123}}  \notag\,.
\end{align} 
Recall $\tilde{U}_{i,j} = U_{i,j} - U_{i,j}^{*} = \{U_{i,j} -  M_1 \cdot{\rm sign}(U_{i,j})\}I(|U_{i,j}|> M_1)$. Using the similar arguments for deriving the convergence rate of ${\rm R}_{111}$ in Section \ref{sec:sub-r11}, we can also show 
\begin{align*}
{\rm R}_{121} = O_{\rm p}\{ e^{-M_1^2/2} \log^{3/2} (dn)\} + O_{\rm p} \{n^{-1} (\log d) \log^2(dn)\}  
\end{align*}
provided that $\log d \lesssim nM_1^{-1}e^{-M_1^2/2}$. Analogous to the derivation of  ${\rm R}_{112}$ in Section \ref{sec:sub-r11}, we have 
\begin{align*}
{\rm R}_{122}=O_{\rm p}\{n^{-1/2} (\log n) (\log d)^{1/2}\log^{3/2}(dn)\}\,.
\end{align*}
Recall  $M_1=\sqrt{\kappa_1 \log n}$ for some constant $\kappa_1 \in(1,2)$ and  $\hat{U}_{i,j}^{*}  = \hat{U}_{i,j} -U_{i,j}^{*}$. Given $Q>M_1$, by  \eqref{eq:Uhatstar}, it holds that
\begin{align} \label{eq:Uijstar}
\max_{i\in[n],\,j\in[p] }|\hat{U}_{i,j}^{*}| \le 2\sqrt{2\log(n+1)}\le \bar{C}M_1 < \bar{C}Q
\end{align}
for some universal constant $\bar{C}>0$, and
\begin{align*}
{\rm R}_{123} \le \bar{C} Q \max_{j\in[p],\,k,t\in[q]}\frac{1}{n}\sum_{i=1}^{n} |\hat{U}_{i,j}^{*}V_{i,k}V_{i,t}|\,.
\end{align*}
Applying the similar arguments for deriving the convergence rate of ${\rm R}_{112}$ in Section \ref{sec:sub-r11}, we have
\begin{align*}
\max_{j\in[p],\,k,t\in[q]}\frac{1}{n}\sum_{i=1}^{n} |\hat{U}_{i,j}^{*}V_{i,k}V_{i,t}| = O_{\rm p}\{Q^2n^{-1/2}(\log n)(\log d)^{1/2}\}
\end{align*}
provided that $\log(dn) \lesssim Q^2$.  With selecting $Q = \tilde{C} \log^{1/2}(dn)$ for some sufficiently large constant $\tilde{C} > \kappa_1$, it holds that
\begin{align*}
{\rm R}_{123}=O_{\rm p}\{n^{-1/2} (\log n) (\log d)^{1/2}\log^{3/2}(dn)\} \,.
\end{align*}
Hence, by \eqref{eq:R23terms}, it holds that
\begin{align*}
{\rm R}_{12} = 
O_{\rm p}\{n^{-1/2} (\log n) (\log d)^{1/2}\log^{3/2}(dn)\} 
\end{align*} 
provided that $\log d \lesssim  n^{1-\kappa_1/2}(\log n)^{-1/2} $.  Using the similar arguments, we can also show such convergence rate holds for ${\rm R}_{13}$. With selecting $\kappa_1=7/6$, we have \eqref{eq:r12} holds. 
$\hfill\Box$

\subsubsection{ Convergence rate of ${\rm R}_{14}$}\label{sec:sub-r14}
Due to  $\hat{U}_{i,j}-U_{i,j} = \hat{U}_{i,j}^{*}-\tilde{U}_{i,j}$ and $\hat{V}_{i,k}-V_{i,k}=\hat{V}_{i,k}^{*}-\tilde{V}_{i,k}$,  we have
\begin{align*}
{\rm R_{14}} =&~\max_{j,l\in[p],\,k,t\in[q]}\bigg|\frac{1}{n}\sum_{i=1}^{n}(\hat{U}_{i,j}^{*}-\tilde{U}_{i,j})(\hat{U}_{i,l}^{*}-\tilde{U}_{i,l})(\hat{V}_{i,k}^{*}-\tilde{V}_{i,k})V_{i,t}\bigg|\,.
\end{align*}
Notice that
\begin{align*}
&~\frac{1}{n}\sum_{i=1}^{n}(\hat{U}_{i,j}^{*}-\tilde{U}_{i,j})(\hat{U}_{i,l}^{*}-\tilde{U}_{i,l})(\hat{V}_{i,k}^{*}-\tilde{V}_{i,k})V_{i,t} \\
=&~ \frac{1}{n}\sum_{i=1}^{n}\hat{U}_{i,j}^{*}\hat{U}_{i,l}^{*}\hat{V}_{i,k}^{*}V_{i,t}  -  \frac{1}{n}\sum_{i=1}^{n}\hat{U}_{i,j}^{*}\hat{U}_{i,l}^{*}\tilde{V}_{i,k}V_{i,t}  -  \frac{1}{n}\sum_{i=1}^{n}\hat{U}_{i,j}^{*}\tilde{U}_{i,l}\hat{V}_{i,k}^{*}V_{i,t}  + \frac{1}{n}\sum_{i=1}^{n}\hat{U}_{i,j}^{*}\tilde{U}_{i,l}\tilde{V}_{i,k}V_{i,t}\notag\\
& -   \frac{1}{n}\sum_{i=1}^{n}\tilde{U}_{i,j}\hat{U}_{i,l}^{*}\hat{V}_{i,k}^{*}V_{i,t}  +  \frac{1}{n}\sum_{i=1}^{n}\tilde{U}_{i,j}\hat{U}_{i,l}^{*}\tilde{V}_{i,k}V_{i,t} +  \frac{1}{n}\sum_{i=1}^{n}\tilde{U}_{i,j}\tilde{U}_{i,l}\hat{V}_{i,k}^{*}V_{i,t}  -  \frac{1}{n}\sum_{i=1}^{n}\tilde{U}_{i,j}\tilde{U}_{i,l}\tilde{V}_{i,k}V_{i,t}\,. \notag
\end{align*}
In order to  derive the convergence rate of ${\rm R}_{14}$, by the symmetry, we only need to consider the convergence rates of the following terms:
\begin{align}\label{eq:R22terms}
{\rm{R}_{141}} = &\max_{j,l\in[p],\,k,t\in[q]}\bigg|\frac{1}{n}\sum_{i=1}^{n}\tilde{U}_{i,j}\tilde{U}_{i,l}\tilde{V}_{i,k}V_{i,t}\bigg| \,, \quad
{\rm{R}_{142}} =\max_{j,l\in[p],\,k,t\in[q]}\bigg|\frac{1}{n}\sum_{i=1}^{n}\hat{U}_{i,j}^{*}\hat{U}_{i,l}^{*}\hat{V}_{i,k}^{*}V_{i,t}\bigg|\,,  \notag\\
{\rm{R}_{143}}=&\max_{j,l\in[p],\,k,t\in[q]}\bigg|\frac{1}{n}\sum_{i=1}^{n}\hat{U}_{i,j}^{*}\hat{U}_{i,l}^{*}\tilde{V}_{i,t}V_{i,k}\bigg|\,, \quad {\rm{R}_{144}}=\max_{j,l\in[p],\,k,t\in[q]}\bigg|\frac{1}{n}\sum_{i=1}^{n}\hat{U}_{i,j}^{*}\tilde{U}_{i,l}\hat{V}_{i,t}^{*}V_{i,k}\bigg|\,,\\
{\rm{R}_{145}}=&\max_{j,l\in[p],\,k,t\in[q]}\bigg|\frac{1}{n}\sum_{i=1}^{n}\hat{U}_{i,j}^{*}\tilde{U}_{i,l}\tilde{V}_{i,k}V_{i,t}\bigg|\,, \quad {\rm{R}_{146}}=\max_{j,l\in[p],\,k,t\in[q]}\bigg|\frac{1}{n}\sum_{i=1}^{n}\tilde{U}_{i,j}\tilde{U}_{i,l}\hat{V}_{i,k}^{*}V_{i,t}\bigg|\notag \,.
\end{align}
Recall $\tilde{U}_{i,j} = U_{i,j} - U_{i,j}^{*} = \{U_{i,j} -  M_1 \cdot{\rm sign}(U_{i,j})\}I(|U_{i,j}|> M_1)$ and $\tilde{V}_{i,k} = V_{i,k} - V_{i,k}^{*} = \{V_{i,k} -  M_1 \cdot{\rm sign}(V_{i,k})\}I(|V_{i,k}|> M_1)$. Using the similar arguments for deriving the convergence rate of ${\rm R}_{111}$ in Section \ref{sec:sub-r11}, we also have 
\begin{align*}
{\rm R}_{141} = O_{\rm p}\{ e^{-M_1^2/2} \log^{3/2} (dn)\} + O_{\rm p} \{n^{-1} (\log d) \log^2(dn)\}  
\end{align*}
provided that $\log d \lesssim nM_1^{-1}e^{-M_1^2/2}$. Identical to \eqref{eq:Uijstar}, we also have $|\hat{V}_{i,k}^{*}| \le \bar{C}Q$ for some universal constant $\bar{C}>0$. Applying the similar arguments for the derivation of  ${\rm R}_{123}$ in Section \ref{sec:sub-r12},  it holds that 
\begin{align*}
&{\rm R}_{142}=O_{\rm p}\{n^{-1/2} (\log n) (\log d)^{1/2}\log^{3/2}(dn)\} ={\rm R}_{143}\,,\\
&~~~~~{\rm R}_{144}=O_{\rm p}\{n^{-1/2} (\log n) (\log d)^{1/2}\log^{3/2}(dn)\} \,.
\end{align*}   Analogous to the derivation of  ${\rm R}_{122}$ in Section \ref{sec:sub-r12}, we have 
\begin{align*}
{\rm R}_{145}=O_{\rm p}\{n^{-1/2} (\log n) (\log d)^{1/2}\log^{3/2}(dn)\} ={\rm R}_{146}\,.
\end{align*}
Recall  $M_1=\sqrt{\kappa_1 \log n}$ for some constant $ \kappa_1 \in(1, 2)$. Hence, by \eqref{eq:R22terms},  with selecting $\kappa_1=7/6$, we know the first equation in  \eqref{eq:r14} holds.
$\hfill\Box$

\subsubsection{Convergence rate of ${\rm R}_{15}$}\label{sec:sub-r15}
Due to $\hat{U}_{i,j}-U_{i,j} = \hat{U}_{i,j}^{*}-\tilde{U}_{i,j}$ and $\hat{V}_{i,k}-V_{i,k}=\hat{V}_{i,k}^{*}-\tilde{V}_{i,k}$, we have
\begin{align*}
&~\frac{1}{n}\sum_{i=1}^{n}(\hat{U}_{i,j}-U_{i,j})(\hat{U}_{i,l}-U_{i,l})(\hat{V}_{i,k}-V_{i,k})(\hat{V}_{i,t}-V_{i,t})\\
=&~\frac{1}{n}\sum_{i=1}^{n}(\hat{U}_{i,j}^{*}-\tilde{U}_{i,j})(\hat{U}_{i,l}^{*}-\tilde{U}_{i,l})(\hat{V}_{i,k}^{*}-\tilde{V}_{i,k})(\hat{V}_{i,t}^{*}-\tilde{V}_{i,t})\\
=&~\frac{1}{n}\sum_{i=1}^{n}  \hat{U}_{i,j}^{*}\hat{U}_{i,l}^{*}\hat{V}_{i,k}^{*}\hat{V}_{i,t}^{*}  -  \frac{1}{n}\sum_{i=1}^{n}  \hat{U}_{i,j}^{*}\hat{U}_{i,l}^{*}\hat{V}_{i,k}^{*}\tilde{V}_{i,t}  - \frac{1}{n}\sum_{i=1}^{n}  \hat{U}_{i,j}^{*}\hat{U}_{i,l}^{*}\tilde{V}_{i,k}\hat{V}_{i,t}^{*}  + \frac{1}{n}\sum_{i=1}^{n}  \hat{U}_{i,j}^{*}\hat{U}_{i,l}^{*}\tilde{V}_{i,k}\tilde{V}_{i,t} \\
&-\frac{1}{n}\sum_{i=1}^{n}  \hat{U}_{i,j}^{*}\tilde{U}_{i,l}\hat{V}_{i,k}^{*}\hat{V}_{i,t}^{*}  +  \frac{1}{n}\sum_{i=1}^{n}  \hat{U}_{i,j}^{*}\tilde{U}_{i,l}\hat{V}_{i,k}^{*}\tilde{V}_{i,t}  + \frac{1}{n}\sum_{i=1}^{n}  \hat{U}_{i,j}^{*}\tilde{U}_{i,l}\tilde{V}_{i,k}\hat{V}_{i,t}^{*} - \frac{1}{n}\sum_{i=1}^{n}  \hat{U}_{i,j}^{*}\tilde{U}_{i,l}\tilde{V}_{i,k}\tilde{V}_{i,t}\\
&-\frac{1}{n}\sum_{i=1}^{n}  \tilde{U}_{i,j} \hat{U}_{i,l}^{*}\hat{V}_{i,k}^{*}\hat{V}_{i,t}^{*}  +  \frac{1}{n}\sum_{i=1}^{n}  \tilde{U}_{i,j} \hat{U}_{i,l}^{*}\hat{V}_{i,k}^{*}\tilde{V}_{i,t}  + \frac{1}{n}\sum_{i=1}^{n}  \tilde{U}_{i,j} \hat{U}_{i,l}^{*}\tilde{V}_{i,k}\hat{V}_{i,t}^{*}  - \frac{1}{n}\sum_{i=1}^{n}  \tilde{U}_{i,j} \hat{U}_{i,l}^{*}\tilde{V}_{i,k}\tilde{V}_{i,t} \\
&+\frac{1}{n}\sum_{i=1}^{n}  \tilde{U}_{i,j} \tilde{U}_{i,l}\hat{V}_{i,k}^{*}\hat{V}_{i,t}^{*}  -  \frac{1}{n}\sum_{i=1}^{n}  \tilde{U}_{i,j} \tilde{U}_{i,l}\hat{V}_{i,k}^{*}\tilde{V}_{i,t}  - \frac{1}{n}\sum_{i=1}^{n}  \tilde{U}_{i,j} \tilde{U}_{i,l}\tilde{V}_{i,k}\hat{V}_{i,t}^{*} + \frac{1}{n}\sum_{i=1}^{n}  \tilde{U}_{i,j} \tilde{U}_{i,l}\tilde{V}_{i,k}\tilde{V}_{i,t}\,.
\end{align*}
To derive the convergence rate  of ${\rm R}_{15}$, by the symmetry, we only need to consider the convergence rates of the following terms:
\begin{align}\label{eq:R11terms}
&\textrm{R}_{151} = \max_{j,l\in[p],\,k,t\in[q]}\bigg|\frac{1}{n}\sum_{i=1}^{n}\tilde{U}_{i,j}\tilde{U}_{i,l}\tilde{V}_{i,k}\tilde{V}_{i,t}\bigg|\,, \quad \textrm{R}_{152}  = \max_{j,l\in[p],\,k,t\in[q]}\bigg|\frac{1}{n}\sum_{i=1}^{n}\hat{U}_{i,j}^{*}\tilde{U}_{i,l}\tilde{V}_{i,k}\tilde{V}_{i,t}\bigg| \notag\\
&\textrm{R}_{153}  = \max_{j,l\in[p],\,k,t\in[q]}\bigg|\frac{1}{n}\sum_{i=1}^{n}\hat{U}_{i,j}^{*}\hat{U}_{i,l}^{*}\tilde{V}_{i,t}\tilde{V}_{i,k}\bigg|\,,  \quad
\textrm{R}_{154}  = \max_{j,l\in[p],\,k,t\in[q]}\bigg|\frac{1}{n}\sum_{i=1}^{n}\hat{U}_{i,j}^{*}\tilde{U}_{i,l}\hat{V}_{i,k}^{*}\tilde{V}_{i,t}\bigg|\,,  \\
&  
\textrm{R}_{155}  = \max_{j,l\in[p],\,k,t\in[q]}\bigg|\frac{1}{n}\sum_{i=1}^{n}\hat{U}_{i,j}^{*}\hat{U}_{i,l}^{*}\hat{V}_{i,k}^{*}\tilde{V}_{i,t}\bigg|\,, \quad \textrm{R}_{156}  = \max_{j,l\in[p],\,k,t\in[q]}\bigg|\frac{1}{n}\sum_{i=1}^{n}\hat{U}_{i,j}^{*}\hat{U}_{i,l}^{*}\hat{V}_{i,k}^{*}\hat{V}_{i,t}^{*}\bigg|\notag\,.
\end{align}
Recall $\tilde{U}_{i,j} = U_{i,j} - U_{i,j}^{*} = \{U_{i,j} -  M_1 \cdot{\rm sign}(U_{i,j})\}I(|U_{i,j}|> M_1)$ and $\tilde{V}_{i,k} = V_{i,k} - V_{i,k}^{*} = \{V_{i,k} -  M_1 \cdot{\rm sign}(V_{i,k})\}I(|V_{i,k}|> M_1)$. Using the similar arguments for deriving the convergence rate of ${\rm R}_{111}$ in Section \ref{sec:sub-r11}, we also have 
\begin{align*}
{\rm R}_{151} = O_{\rm p}\{ e^{-M_1^2/2} \log^{3/2} (dn)\} + O_{\rm p} \{n^{-1} (\log d) \log^2(dn)\}  
\end{align*}
provided that $\log d \lesssim nM_1^{-1}e^{-M_1^2/2}$. 
Analogous to the derivation of ${\rm R}_{122}$ in Section \ref{sec:sub-r12}, we have
\begin{align*}
{\rm R}_{152} = O_{\rm p}\{n^{-1/2} (\log n) (\log d)^{1/2} \log^{3/2} (dn)\}\,.
\end{align*}
Recall $|\hat{U}_{i,j}^{*}|\le \bar{C}Q$ and $|\hat{V}_{i,k}^{*}|\le \bar{C}Q$ for some universal constant $\bar{C}>0$. Using the similar arguments for deriving the convergence rate of ${\rm R}_{123}$ in Section \ref{sec:sub-r12}, it holds that  
\begin{align*}
&{\rm R}_{153} = O_{\rm p}\{n^{-1/2} (\log n) (\log d)^{1/2} \log^{3/2} (dn)\} ={\rm R}_{154}\,,\\
&{\rm R}_{155} = O_{\rm p}\{n^{-1/2} (\log n) (\log d)^{1/2} \log^{3/2} (dn)\} ={\rm R}_{156}\,.
\end{align*}
Recall  $M_1=\sqrt{\kappa_1 \log n}$ for some constant $ \kappa_1 \in(1, 2)$. By \eqref{eq:R11terms}, with selecting $\kappa_1=7/6$, we know the second equation of \eqref{eq:r14} holds.
$\hfill\Box$

\subsection{Convergence rate of ${\rm R}_{2}$}\label{sub:sec-h0-R2}
For any $Q>0$, it holds that
\begin{align*}
&~\frac{1}{n}\sum_{i=1}^{n}\big\{U_{i,j}U_{i,l}V_{i,k}V_{i,t} -\mathbb{E}(U_{i,j}U_{i,l}V_{i,k}V_{i,t})\big\} \\
=&~\frac{1}{n}\sum_{i=1}^{n}\big[U_{i,j} U_{i,l}  V_{i,k}  V_{i,t} I(|U_{i,j}|,\,|U_{i,l}|,\,|V_{i,k}|,\,|V_{i,t}|\le Q)\\
&\underbrace{~~~~~~~~~~~~~~-\mathbb{E}\big\{U_{i,j} U_{i,l}  V_{i,k}  V_{i,t} I(|U_{i,j}|,\,|U_{i,l}|,\,|V_{i,k}|,\,|V_{i,t}|\le Q)\big\}\big]}_{\textrm{R}_{21}(j,l,k,t)} \\
&+\underbrace{\frac{1}{n}\sum_{i=1}^{n}U_{i,j}  U_{i,l}  V_{i,k}  V_{i,t} I(|U_{i,j}|,\,|U_{i,l}|,\,|V_{i,k}|\le Q)I(|V_{i,t}|> Q) }_{\textrm{R}_{22}(j,l,k,t)} \\
& + \underbrace{\frac{1}{n}\sum_{i=1}^{n}U_{i,j} U_{i,l}  V_{i,k} I(|U_{i,j}|,\,|U_{i,l}|\le Q) I(|V_{i,k}|> Q) V_{i,t}}_{\textrm{R}_{23}(j,l,k,t)} \\
&+ \underbrace{\frac{1}{n}\sum_{i=1}^{n}U_{i,j}I(|U_{i,j}|\le Q) U_{i,l} I(|U_{i,l}|> Q) V_{i,k} V_{i,t}}_{\textrm{R}_{24}(j,l,k,t)}+\underbrace{\frac{1}{n}\sum_{i=1}^{n}U_{i,j}I(|U_{i,j}|> Q) U_{i,l}   V_{i,k} V_{i,t}}_{\textrm{R}_{25}(j,l,k,t)}\\
&- \underbrace{\big[\mathbb{E}(U_{i,j}U_{i,l}V_{i,k}V_{i,t}) - \mathbb{E}\big\{U_{i,j} U_{i,l}  V_{i,k}  V_{i,t} I(|U_{i,j}|,\,|U_{i,l}|,\,|V_{i,k}|,\,|V_{i,t}|\le Q)\big\}\big]}_{\textrm{R}_{26}(j,l,k,t)}\,.
\end{align*}
Recall $U_{i,j}, V_{i,k}\sim \mathcal{N}(0,1)$ and  $d=pq$. Since $\var\{U_{i,j} U_{i,l}  V_{i,k}  V_{i,t} I(|U_{i,j}|,\,|U_{i,l}|,\,|V_{i,k}|,\,|V_{i,t}|\le Q)\} \le C_1$,  by Bernstein inequality, it holds that
\begin{align}\label{eq:R21}
\max_{j,l\in [p],\,k,t \in[q]}|{\rm R}_{21}(j,l,k,t)|=O_{{\rm p}}\{n^{-1/2}(\log d)^{1/2}\} + O_{\rm p} (n^{-1}Q^4\log d)\,.
\end{align}
Analogous to the derivation of \eqref{eq:k12}, if $\log (dn) \lesssim Q^2$, then 
\begin{align*}
&\max_{j,l\in [p],\,k,t \in[q]}|{\rm R}_{22}(j,l,k,t)| =o_{\rm p}(n^{-1}) =\max_{j,l\in [p],\,k,t \in[q]}|{\rm R}_{23}(j,l,k,t)|\,,\\
&\max_{j,l\in [p],\,k,t \in[q]}|{\rm R}_{24}(j,l,k,t)|=o_{\rm p}(n^{-1})=\max_{j,l\in [p],\,k,t \in[q]}|{\rm R}_{25}(j,l,k,t)|\,.\notag
\end{align*}
Furthermore, 
\begin{align*}
\max_{j,l\in [p],\,k,t \in[q]}|{\rm R}_{26}(j,l,k,t)| \lesssim &~  \max_{i\in[n],\,j\in[p]}\big[\mathbb{E}\{I(|U_{i,j}|> Q)\}\big]^{1/2} + \max_{i\in[n],\,k\in[q]}\big[\mathbb{E}\{I(|V_{i,k}|> Q)\}\big]^{1/2}\\  \lesssim &~ Q^{-1/2}e^{-Q^2/4} \,.
\end{align*}
Together with \eqref{eq:R21}, by selecting $Q=\tilde{C} \log^{1/2} (dn)$ for some sufficiently large constant $\tilde{C}>0$,  we then have \eqref{eq:h0-R2} holds.
$\hfill\Box$

\subsection{Convergence rate of ${\rm R}_3$}\label{sub:sec-h0-R3}
Notice that
\begin{align}\label{eq:R3-dep}
{\rm R}_3 \le &~ 2\underbrace{\max_{j,l\in[p],\, k,t\in[q]} \bigg|\frac{1}{n}\sum_{i=1}^{n}\big\{U_{i,j}V_{i,k} -\mathbb{E}(U_{i,j}V_{i,k})\big\} \mathbb{E}(U_{i,l}V_{i,t})\bigg|}_{\textrm{R}_{31}} \notag\\
&+\underbrace{\max_{j\in[p],\, k\in[q]} \bigg|\frac{1}{n}\sum_{i=1}^{n}\big\{U_{i,j}V_{i,k} -\mathbb{E}(U_{i,j}V_{i,k})\big\}\bigg|^2}_{\textrm{R}_{32}}\,. 
\end{align}
Notice that
\begin{align*}
{\rm R}'_3(j,k):=&~\frac{1}{n}\sum_{i=1}^{n}\big\{U_{i,j}V_{i,k} -\mathbb{E}(U_{i,j}V_{i,k})\big\} \\
=&~\underbrace{\frac{1}{n}\sum_{i=1}^{n}\big[U_{i,j}V_{i,k} I(|U_{i,j}|,\,|V_{i,k}|\le Q) -\mathbb{E}\{U_{i,j}V_{i,k} I(|U_{i,j}|,\,|V_{i,k}|\le Q)\}\big]}_{\textrm{R}'_{31}(j,k)} \\
&+\underbrace{\frac{1}{n}\sum_{i=1}^{n} U_{i,j}V_{i,k}I(|U_{i,j}|\le Q)I(|V_{i,k}|> Q)}_{\textrm{R}'_{32}(j,k)} + \underbrace{\frac{1}{n}\sum_{i=1}^{n} U_{i,j}V_{i,k}I(|U_{i,j}|> Q)}_{\textrm{R}'_{33}(j,k)} \\
&-\underbrace{\big[\mathbb{E}(U_{i,j}V_{i,k}) - \mathbb{E}\{U_{i,j}V_{i,k} I(|U_{i,j}|,\,|V_{i,k}|\le Q)\}\big]}_{\textrm{R}'_{34}(j,k)}
\end{align*}
Recall $d=pq$. Due to $U_{i,j}, V_{i,k} \sim \mathcal{N}(0,1)$, then $|\mathbb{E}(U_{i,j} V_{i,k})| \le 1$ and $  \var(U_{i,j}V_{i,k}) \le 3 $,   by Bonferroni inequality and Bernstein inequality,  it holds that
\begin{align}\label{eq:r31'-tail}
\mathbb{P}\bigg\{\max_{j\in[p], \,k\in [q]} | {\rm R}'_{31}(j,k) | > x\bigg\}\le
2d\exp\bigg(-\frac{nx^2}{C_{1}+C_2Q^2x} \bigg)  
\end{align}
for any $x>0$, which implies
\begin{align*}
\max_{j\in[p], \,k\in [q]} | {\rm R}'_{31}(j,k) |=O_{{\rm p}}\{n^{-1/2}(\log d)^{1/2}\} + O_{{\rm p}}(n^{-1}Q^2\log  d)\,.
\end{align*}
Using the similar arguments for the derivation of \eqref{eq:k12}, we have 
$\max_{j\in[p], \,k\in [q]} | {\rm R}'_{32}(j,k) |=o_{\rm p}(n^{-1})=\max_{j\in[p], \,k\in [q]} | {\rm R}'_{33}(j,k) |$ provided that $\log (dn) \lesssim Q^2$. Furthermore,
\begin{align*}
\max_{j\in [p],\,k \in[q]}|{\rm R}'_{34}(j ,k)| \lesssim &~  \max_{i\in[n],\,j\in[p]}\big[\mathbb{E}\{I(|U_{i,j}|> Q)\}\big]^{1/2} + \max_{i\in[n],\,k\in[q]}\big[\mathbb{E}\{I(|V_{i,k}|> Q)\}\big]^{1/2}\\  \lesssim &~ Q^{-1/2}e^{-Q^2/4} \,.
\end{align*}
By selecting $Q =\tilde{C} \log^{1/2} (dn)$ for some  sufficiently large constant $\tilde{C}>0$, it holds that
\begin{align}\label{eq:uv-bound-2}
\max_{j\in[p], \,k\in [q]} | {\rm R}'_{3}(j,k) |=O_{{\rm p}}\{n^{-1/2}(\log d)^{1/2}\}  
\end{align}
provided that $\log d\lesssim n^{1/3}$. Then ${\rm R}_{31}=O_{{\rm p}}\{n^{-1/2}(\log d)^{1/2}\}$ and ${\rm R}_{32}=O_{{\rm p}}(n^{-1}\log d)$ provided that $\log d \lesssim n^{1/3}$. Then, by \eqref{eq:R3-dep}, we have \eqref{eq:r3} holds.  
$\hfill\Box$

\subsection{Convergence rate of ${\rm R}_4$}\label{sub:sec-h0-R4}
Notice that 
\begin{align}\label{eq:r4-dep}
{\rm R}_{4}
\le &~  2\underbrace{\max_{j,l\in[p],\,k,t\in[q]}\bigg|\bigg(\frac{1}{n}\sum_{i=1}^{n}\hat{U}_{i,j}\hat{V}_{i,k} - \frac{1}{n}\sum_{i=1}^{n}U_{i,j}V_{i,k} \bigg)\bigg(\frac{1}{n}\sum_{i=1}^{n}U_{i,l}V_{i,t}\bigg)\bigg| }_{\textrm{R}_{41}}\notag\\
&~+ \underbrace{\max_{j\in[p],\,k\in[q]}\bigg|\frac{1}{n}\sum_{i=1}^{n}\hat{U}_{i,j}\hat{V}_{i,k} - \frac{1}{n}\sum_{i=1}^{n}U_{i,j}V_{i,k} \bigg|^2 }_{\textrm{R}_{42}}\,,
\end{align}
and $ \hat{U}_{i,j}\hat{V}_{i,k}-  U_{i,j}V_{i,k} = (\hat{U}_{i,j}-U_{i,j})V_{i,k} +  (\hat{V}_{i,k}-V_{i,k})U_{i,j}  +   (\hat{U}_{i,j}-U_{i,j})(\hat{V}_{i,k}-V_{i,k})$.
Due to $\hat{U}_{i,j}-U_{i,j} = \hat{U}_{i,j}^{*}-\tilde{U}_{i,j}$ and $\hat{V}_{i,k}-V_{i,k}=\hat{V}_{i,k}^{*}-\tilde{V}_{i,k}$, 
we have
\begin{align}\label{eq:r4'-dep}
{\rm R}_{4}' :=&~ \max_{j \in[p],\,k \in[q]}\bigg|\frac{1}{n}\sum_{i=1}^{n}(\hat{U}_{i,j}\hat{V}_{i,k}-  U_{i,j}V_{i,k})\bigg|\notag\\
\le&~ \max_{j \in[p],\,k \in[q]}\bigg|\frac{1}{n}\sum_{i=1}^{n}(\hat{U}_{i,j}^{*}-\tilde{U}_{i,j})V_{i,k}\bigg|+ \max_{j \in[p],\,k \in[q]}\bigg|\frac{1}{n}\sum_{i=1}^{n}(\hat{V}_{i,k}^{*}-\tilde{V}_{i,k})U_{i,k}\bigg| \notag \\
&+ \max_{j \in[p],\,k \in[q]}\bigg|\frac{1}{n}\sum_{i=1}^{n} (\hat{U}_{i,j}^{*}-\tilde{U}_{i,j})(\hat{V}_{i,k}^{*}-\tilde{V}_{i,k})\bigg|\,.
\end{align}
To derive the convergence rate of ${\rm R}_{4}'$, by the symmetry, we only consider the convergence rates of the following terms:
\begin{align*}
&{\rm R}'_{41}=  \max_{j \in[p],\,k \in[q]}\bigg|\frac{1}{n}\sum_{i=1}^{n} \tilde{U}_{i,j}V_{i,k}\bigg| \,, \qquad {\rm R}'_{42}=  \max_{j \in[p],\,k \in[q]}\bigg|\frac{1}{n}\sum_{i=1}^{n} \tilde{U}_{i,j}\tilde{V}_{i,k}\bigg| \,, \\
&{\rm R}'_{43}=\max_{j \in[p],\,k \in[q]}\bigg|\frac{1}{n}\sum_{i=1}^{n}\hat{U}_{i,j}^{*} V_{i,k}\bigg|\,, \qquad {\rm R}'_{44}=\max_{j \in[p],\,k \in[q]}\bigg|\frac{1}{n}\sum_{i=1}^{n}\hat{U}_{i,j}^{*} \tilde{V}_{i,k}\bigg|\,,  \\
& {\rm R}'_{45}=\max_{j \in[p],\,k \in[q]}\bigg|\frac{1}{n}\sum_{i=1}^{n}\hat{U}_{i,j}^{*} \hat{V}_{i,k}^{*}\bigg|\,.
\end{align*}
As we will show in Sections \ref{sec:sub-r41'} and \ref{sec:sub-r43'},
\begin{align}
{\rm R}'_{41} =  O_{\rm p}\{ e^{-M_1^2/2}\log^{1/2}(dn)\} + O_{\rm p} \{n^{-1} (\log d) \log(dn)\} ={\rm R}'_{42}\,, \label{eq:r41'}
\end{align}
provided that $\log d \lesssim nM_1^{-1}e^{-M_1^2/2}$, and 
\begin{align}
&{\rm R}'_{43} = O_{\rm p} \{n^{-1/2}(\log n) (\log d)^{1/2} \log^{1/2}(dn)\} ={\rm R}'_{44} \,, \label{eq:r43'}\\
&~~~~~~~~~{\rm R}'_{45} = O_{\rm p} \{n^{-1/2}(\log n)^{3/2} (\log d)^{1/2} \}  \label{eq:r45'}\,. 
\end{align}
Recall  $M_1=\sqrt{\kappa_1 \log n}$ for some constant $ \kappa_1 \in(1, 2)$. Hence, by \eqref{eq:r4'-dep}, it holds that
\begin{align*}
{\rm R}'_{4} =  O_{\rm p} \{n^{-1/2}(\log n) (\log d)^{1/2} \log^{1/2}(dn)\}
\end{align*}
provided that  $\log d \lesssim  n^{1-\kappa_1/2}(\log n)^{-1/2}$. 
By \eqref{eq:uv-bound-2}, due to $|\mathbb{E}(U_{i,j} V_{i,k})| \le 1$, we then have $\max_{j\in[p],\, k\in[q]}|n^{-1}\sum_{i=1}^{n}U_{i,j}V_{i,k}|=O_{\rm p}(1)$ provided that $\log d\lesssim n^{1/3}$. By \eqref{eq:r4-dep}, with selecting $\kappa_1=7/6$, we have
\begin{align*}
{\rm R}_{4} =  O_{\rm p} \{n^{-1/2}(\log n) (\log d)^{1/2} \log^{1/2}(dn)\}
\end{align*}
provided that $\log d \lesssim   n^{1/3}$.  Then \eqref{eq:h0-R4} holds.

\subsubsection{Convergence rates of ${\rm R}'_{41}$ and ${\rm R}'_{42}$}\label{sec:sub-r41'}
Recall $\tilde{U}_{i,j} = \acute{U}_{i,j}I(|U_{i,j}| > M_1)$ with $\acute{U}_{i,j} = U_{i,j} -  M_1 \cdot{\rm sign}(U_{i,j})$.  Given $Q> M_1$, we have
\begin{align*}
{\rm R}'_{41} \le &~ \underbrace{\max_{j \in[p],\,k \in[q]}\bigg|\frac{1}{n}\sum_{i=1}^{n} \acute{U}_{i,j}I(M_1<|U_{i,j}|\le Q) V_{i,k}I(|V_{i,k}|\le Q) \bigg|}_{{\textrm R}_{411}'}\\
&+ \underbrace{\max_{j \in[p],\,k \in[q]}\bigg|\frac{1}{n}\sum_{i=1}^{n} \acute{U}_{i,j}I(M_1<|U_{i,j}|\le Q) V_{i,k}I(|V_{i,k}|> Q)\bigg|}_{{\textrm R}_{412}'} \\
&+ \underbrace{\max_{j \in[p],\,k \in[q]}\bigg| \frac{1}{n}\sum_{i=1}^{n} \acute{U}_{i,j}I(|U_{i,j}|> Q) V_{i,k}\bigg|}_{{\textrm R}_{413}'}\,.
\end{align*}
Due to  $U_{i,j}, V_{i,k} \sim \mathcal{N}(0,1)$, it then holds that
\begin{align*}
&\max_{i\in[n],\,j \in[p],\,k \in[q]}|\mathbb{E}\{\acute{U}_{i,j}I(M_1<|U_{i,j}|\le Q) V_{i,k}I(|V_{i,k}|\le Q)\}|  \\
&~~~~~~~~~~~~~\le  Q \max_{i\in[n],\,j \in[p] }\mathbb{E}\{|\acute{U}_{i,j}|I(M_1<|U_{i,j}|\le Q) \}   
\lesssim Q  e^{-M_1^2/2}\,,\\
&\max_{i\in[n],\,j \in[p],\,k \in[q]}\var\{\acute{U}_{i,j}I(M_1<|U_{i,j}|\le Q) V_{i,k}I(|V_{i,k}|\le Q)\} \\
&~~~~~~~~~~~~~\le Q^2 \max_{i\in[n],\,j \in[p] }\mathbb{E}\{|\acute{U}_{i,j}|^2I(M_1<|U_{i,j}|\le Q) \} 
\lesssim Q^2 M_1e^{-M_1^2/2}\,.
\end{align*}
Recall $d=pq$. Using the similar arguments for deriving the convergence rate of 
${\rm R}_{1111}$ in Section \ref{sec:sub-r11}, we have
\begin{align*}
{\rm R}'_{411} = O_{\rm p}(Q e^{-M_1^2/2}) + O_{\rm p} (n^{-1} Q^2 \log d)
\end{align*}
provided that $\log d \lesssim n M_1^{-1}e^{-M_1^2/2}$.
Analogous to the derivation of \eqref{eq:k12}, we also have ${\rm R}'_{412} =o_{\rm p}(n^{-1}) ={\rm R}'_{413}$ provided that $\log(dn) \lesssim Q^2$.  With selecting $Q = \tilde{C} \log^{1/2}(dn)$ for some sufficiently large constant $\tilde{C}> \kappa_1$, then 
\begin{align*}
{\rm R}'_{41} =  O_{\rm p}\{ e^{-M_1^2/2}\log^{1/2}(dn)\} + O_{\rm p} \{n^{-1} (\log d) \log(dn)\}
\end{align*}
provided that $\log d \lesssim nM_1^{-1}e^{-M_1^2/2}$. 
Using the similar arguments, we can also show such convergence rate holds for ${\rm R}'_{42}$. Hence,
\eqref{eq:r41'} holds. 
$\hfill\Box$

\subsubsection{Convergence rates of ${\rm R}'_{43}$,  ${\rm R}'_{44}$ and ${\rm R}'_{45}$ }\label{sec:sub-r43'}
Given $Q > M_1$, we have
\begin{align*}
{\rm R}'_{43} \le  \underbrace{\max_{j \in[p],\,k \in[q]}\bigg|\frac{1}{n}\sum_{i=1}^{n}\hat{U}_{i,j}^{*} V_{i,k}I(|V_{i,k}| \le Q)\bigg|}_{\textrm{R}'_{431}} + \underbrace{ \max_{j \in[p],\,k \in[q]}\bigg|\frac{1}{n}\sum_{i=1}^{n}\hat{U}_{i,j}^{*} V_{i,k}I(|V_{i,k}| > Q)\bigg| }_{\textrm{R}'_{432}} \,.
\end{align*}
Recall $\hat{U}_{i,j}^{*} = \hat{U}_{i,j}- U_{i,j}^{*}$. By \eqref{eq:uh-us}, it holds that
\begin{align*}
{\rm R}'_{431} \le Q \max_{j\in[p]}\frac{1}{n}\sum_{i=1}^{n}|\hat{U}_{i,j}- U_{i,j}^{*}| = O_{\rm p} \{n^{-1/2}Q(\log n)(\log p)^{1/2} \}\,.
\end{align*}
Analogous to the derivation of \eqref{eq:k12}, we also have ${\rm R}'_{432} =o_{\rm p}(n^{-1})$ provided that $\log(pn) \lesssim Q^2$.  Recall $d=pq$. With selecting $Q = \tilde{C} \log^{1/2}(dn)$ for some sufficiently large constant $\tilde{C}> \kappa_1$,  it holds that
\begin{align*}
{\rm R}'_{43} = O_{\rm p} \{n^{-1/2}(\log n) (\log d)^{1/2} \log^{1/2}(dn)\}\,.  
\end{align*}
Using the similar arguments, we can also show such convergence rate holds for ${\rm R}'_{44}$. Then  \eqref{eq:r43'} holds. 
Due to $|\hat{V}_{i,k}^{*}|\le 2\sqrt{2\log(n+1)}$, by \eqref{eq:uh-us}, 
we have
\begin{align*}
{\rm R}'_{45} \lesssim \sqrt{\log n} \max_{j\in[p]}\frac{1}{n}\sum_{i=1}^{n}|\hat{U}_{i,j}^{*}|  =\sqrt{\log n} \max_{j\in[p]}\frac{1}{n}\sum_{i=1}^{n}|\hat{U}_{i,j}- U_{i,j}^{*}| = O_{\rm p} \{n^{-1/2}(\log n)^{3/2}(\log d)^{1/2} \}\,.
\end{align*}
Then  \eqref{eq:r45'} holds.
$\hfill\Box$

\section{Proof of Lemma \ref{lem:h1f}}\label{sec:sub-uivi-h1} 
Let $M_1=\sqrt{\kappa_1 \log n}$ and $M_2=\sqrt{\kappa_2 \log n}$ with $\kappa_1=6/5$ and $\kappa_2=1/2$.
Then  $U_{i,j}^{*} =  U_{i,j}I(|U_{i,j}|\le M_1) + M_1 \cdot{\rm sign}(U_{i,j})I(|U_{i,j}|>M_1)$, $V_{i,k}^{*} =  V_{i,k}I(|V_{i,k}|\le M_1) + M_1 \cdot{\rm sign}(V_{i,k})I(|V_{i,k}|>M_1)$, and
\begin{align*}
\tilde{\delta}_{1,k}(U_{s,j})=&~\mathbb{E} \big[e^{U_{i,j}^2/2} \big\{I(U_{s,j}\le U_{i,j})-\Phi(U_{i,j})\big\}V_{i,k}^{*} I(|U_{i,j}|\le M_2)\,\big|\,U_{s,j} \big]\,, \\
\tilde{\delta}_{2,j}(V_{s,k})=&~\mathbb{E} \big[e^{V_{i,k}^2/2} \big\{I(V_{s,k}\le V_{i,k})-\Phi(V_{i,k})\big\}U_{i,j}^{*} I(|V_{i,k}|\le M_2)\,\big|\,V_{s,k} \big]
\end{align*}
with $i\ne s$. It holds that
\begin{align}\label{eq:Pi-dec}
\frac{1}{n} \sum_{i=1}^{n}  (\hat{U}_{i,j}\hat{V}_{i,k} - U_{i,j}V_{i,k}) =&~ \underbrace{\frac{1}{n} \sum_{i=1}^{n}\bigg\{(\hat{U}_{i,j}-U_{i,j}^{*})V_{i,k}^{*} - \frac{1}{n+1}\sum_{s:\,s\ne i}\sqrt{2\pi}\tilde{\delta}_{1,k}(U_{s,j}) \bigg\}}_{\textrm{K}'_{1}(j,k)}  \notag\\
&+ \underbrace{\frac{1}{n} \sum_{i=1}^{n}\bigg\{(\hat{V}_{i,k}-V_{i,k}^{*})U_{i,j}^{*} - \frac{1}{n+1}\sum_{s:\,s\ne i}\sqrt{2\pi}\tilde{\delta}_{2,j}(V_{s,k})\bigg\} }_{\textrm{K}'_{2}(j,k)} \notag\\
&+  \underbrace{\frac{1}{n} \sum_{i=1}^{n} (\hat{U}_{i,j}-U_{i,j}^{*})(\hat{V}_{i,k}-V_{i,k}^{*}) }_{\textrm{K}'_{3}(j,k)}  +\underbrace{\frac{1}{n} \sum_{i=1}^{n}  (U_{i,j}^{*}V_{i,k}^{*} - U_{i,j}V_{i,k})   }_{\textrm{K}'_{4}(j,k)} \notag\\
&+ \frac{\sqrt{2\pi}(n-1)}{n(n+1)} \sum_{s=1}^{n}\big\{ \tilde{\delta}_{1,k}(U_{s,j}) + \tilde{\delta}_{2,j}(V_{s,k})\big\}\,.
\end{align}
By Lemma \ref{lem:huhv}, we have
\begin{align*}
\max_{j\in[p],\,k\in[q]}|{\rm K}'_{3}(j,k)|=  O_{\rm p}\{n^{-3/5} (\log n)^{1/2}\}
\end{align*}
provided that $ \log d \lesssim n^{1/4} \log n$.
By Lemma \ref{lem:usvs-uv}, it holds that
\begin{align*}
\max_{j\in[p],\, k\in[q]}|{\rm K}'_{4}(j,k)| = O \{n^{-3/5}(\log n)^{1/2}\} +  O_{\rm p} \{n^{-1} (\log d) \log (dn) \}  
\end{align*}
provided that $\log d \lesssim n^{2/5}(\log n)^{-1/2}$.
As we will show in Section  \ref{sec:sub-k1'}, 
\begin{align}\label{eq:k1'}
\max_{j\in[p],\,k\in[q]}|{\rm K}'_{1}(j,k)| =&~ ~O_{\rm p} \{n^{-5/8} (\log n)^{-1/4}\log^{1/2} (dn)\}  + O_{\rm p}\{n^{-3/5} (\log n)^{1/2}\}  \notag\\
=&~\max_{j\in[p],\,k\in[q]}|{\rm K}'_{2}(j,k)|
\end{align}
provided that $ \log d \lesssim n^{1/4} (\log n)^{-3/2}$. 
Hence, by \eqref{eq:Pi-dec},  we have
\begin{align*}
\max_{j\in[p],\, k\in[q]} \bigg|\frac{1}{n} \sum_{i=1}^{n}  (\hat{U}_{i,j}\hat{V}_{i,k} - U_{i,j}V_{i,k}) \bigg| \le &~ \max_{j\in[p],\, k\in[q]} \bigg|\frac{\sqrt{2\pi}}{n} \sum_{s=1}^{n}\big\{ \tilde{\delta}_{1,k}(U_{s,j}) + \tilde{\delta}_{2,j}(V_{s,k})\big\} \bigg| \\
&+   O_{\rm p} \{n^{-5/8}(\log n)^{-1/4}\log^{1/2} (dn)\} +O_{\rm p}\{n^{-3/5}(\log n)^{1/2}\}
\end{align*}
provided that $\log d\lesssim n^{1/4} (\log n)^{-3/2} $.
We complete the proof of Lemma \ref{lem:h1f}.
$\hfill\Box$

\subsection{Convergence rates of $\max_{j\in[p],\,k\in[q]}|{\rm K}'_{1}(j,k)|$ and $\max_{j\in[p],\,k\in[q]}|{\rm K}'_{2}(j,k)|$}\label{sec:sub-k1'}
Recall $\hat{U}_{i,j}=\Phi^{-1} \{n(n+1)^{-1}\hat{F}_{\bX,j}(X_{i,j})\}$,   $U_{i,j}=\Phi^{-1}\{F_{\bX,j}(X_{i,j})\}$ and $U_{i,j}^{*} =  U_{i,j}I(|U_{i,j}|\le M_1) + M_1 \cdot{\rm sign}(U_{i,j})I(|U_{i,j}|>M_1)$, where $M_1=\sqrt{\kappa_1\log n}$ with $ \kappa_1=6/5$. We have
\begin{align*}
{\rm K}'_1(j,k) =&~ \frac{1}{n}\sum_{i=1}^{n} \bigg\{(\hat{U}_{i,j} - U_{i,j}^{*}) V_{i,k}^{*} I(|U_{i,j}|\le M_1) +  (\hat{U}_{i,j} - U_{i,j}^{*}) V_{i,k}^{*} I(|U_{i,j}| > M_1) \\
&~~~~~~~~~~~~-\frac{1}{n+1}\sum_{s:\,s\ne i}\sqrt{2\pi}\tilde{\delta}_{1,k}(U_{s,j})\bigg\}\notag\\
=&~  \frac{1}{n}\sum_{i=1}^{n} \bigg(\bigg[\Phi^{-1} \bigg\{\frac{n}{n+1}\hat{F}_{\bX,j}(X_{i,j})\bigg\} - \Phi^{-1}\{F_{\bX,j}(X_{i,j})\} \bigg] V_{i,k}^{*} I(|U_{i,j}|\le M_2) \\
&\underbrace{~~~~~~~~~~~~- \frac{1}{n+1}\sum_{s:\,s\ne i}\sqrt{2\pi}\tilde{\delta}_{1,k}(U_{s,j})\bigg)~~~~~~~~~~~~~~~~~~~~~~~~~~~~~~~~~~~~~~~~~~~}_{\textrm{K}'_{11}(j,k)}\notag  \\
&+ \underbrace{\frac{1}{n}\sum_{i=1}^{n} \bigg[\Phi^{-1} \bigg\{\frac{n}{n+1}\hat{F}_{\bX,j}(X_{i,j})\bigg\} - \Phi^{-1}\{F_{\bX,j}(X_{i,j})\} \bigg] V_{i,k}^{*} I(M_2 < |U_{i,j}|\le M_1)}_{\textrm{K}'_{12}(j,k)}\notag  \\
& + \underbrace{\frac{1}{n}\sum_{i=1}^{n} (\hat{U}_{i,j} - U_{i,j}^{*}) V_{i,k}^{*} I(|U_{i,j}| > M_1) }_{\textrm{K}'_{13}(j,k)}  \,.  
\end{align*} 
Notice that ${\rm K}'_{12}(j,k)={\rm I}_{2}(j,k)$ and ${\rm K}'_{13}(j,k)={\rm I}_{3}(j,k)$ for ${\rm I}_{2}(j,k)$ and ${\rm I}_{3}(j,k)$ defined in \eqref{eq:id}  with $\kappa_1=6/5$ and $\kappa_2=1/2$. Since the convergence rates of $\max_{j\in[p], \, k\in[q]}|{\rm I}_{2}(j,k)|$ and $\max_{j\in[p], \, k\in[q]}|{\rm I}_{3}(j,k)|$ obtained in Sections \ref{sec:sub-I2} and \ref{sec:sub-I3} do not depend on
whether or not the null hypothesis $\mathbb{H}_0$ in \eqref{eq:equind} holds, we still have 
\begin{align*}
\max_{j\in[p],\,k\in[q]}|{\rm K}'_{12}(j,k)| =O_{\rm p} \{n^{-5/8} (\log n)^{-1/4}\log^{1/2} (dn)\}  
\end{align*}
provided that $ \log d \ll n^{2/5} (\log n)^{-1/2}$, 
and
\begin{align*}
\max_{j\in[p],\,k\in[q]}|{\rm K}'_{13}(j,k)|= O_{\rm p}\{n^{-3/5} (\log n)^{1/2}\}
\end{align*}
provided that $ \log d \lesssim n^{2/5}(\log n)^{-1/2}$.
As we will show in Section \ref{sec:sub-k11'},  
\begin{align}\label{eq:k11'}
\max_{j\in[p],\, k\in[q]}|{\rm K}'_{11}(j,k)| =O_{\rm p} \{n^{-3/4}(\log n)^{1/2}\log (dn)\}
\end{align}
provided that $ \log d \ll n^{1/2}(\log n)^{-1}$.
Hence, we have
\begin{align*} 
\max_{j\in[p],\,k\in[q]}|{\rm K}'_{1}(j,k)| =&~O_{\rm p} \{n^{-5/8} (\log n)^{-1/4}\log^{1/2} (dn)\}  + O_{\rm p}\{n^{-3/5} (\log n)^{1/2}\} 
\end{align*}
provided that $ \log  d \lesssim n^{1/4}(\log n)^{-3/2}$. Analogously, we can also show such convergence rate holds for $\max_{j\in[p],\,k\in[q]}|{\rm K}'_{2}(j,k)|$. 
We complete the proof of \eqref{eq:k1'}.
$\hfill\Box$

\subsubsection{Proof of \eqref{eq:k11'}} \label{sec:sub-k11'}   
By the Taylor's expression,  \eqref{eq:Derivative12} and  \eqref{eq:n+1f-f}, it holds that
\begin{align*}
&~{\rm K}'_{11}(j,k)\\
= &~ \frac{1}{n}\sum_{i=1}^{n}\bigg[(\Phi^{-1})^{'}\{F_{\bX,j}(X_{i,j})\} \bigg\{\frac{n}{n+1}\hat{F}_{\bX,j}(X_{i,j})- F_{\bX,j}(X_{i,j})  \bigg\}V_{i,k}^{*}I{(|U_{i,j}| \le M_2)} \\
&~~~~~~~~~~~- \frac{1}{n+1}\sum_{s:\,s\ne i}\sqrt{2\pi}\tilde{\delta}_{1,k}(U_{s,j}) \bigg]\\ 
&+ \sum_{l=2}^{\infty}\frac{1}{n\cdot l!}\sum_{i=1}^{n}(\Phi^{-1})^{(l)}\{F_{\bX,j}(X_{i,j})\}\bigg\{\frac{n}{n+1}\hat{F}_{\bX,j}(X_{i,j})-  F_{\bX,j}(X_{i,j}) \bigg\}^{l}V_{i,k}^{*}I{(|U_{i,j}| \le M_2)} \\
=&~ \underbrace{\frac{\sqrt{2\pi}}{n(n+1)}\sum_{1\leq i_1 \ne i_2 \leq n} \big\{e^{U_{i_1,j}^2/2} \big\{I(U_{i_2,j} \le U_{i_1,j}) -\Phi(U_{i_1,j})\big\}  V_{i_1,k}^{*}I{(|U_{i_1,j}| \le M_2)} -\tilde{\delta}_{1,k}(U_{i_2,j})\big\}}_{\textrm{K}'_{111}(j,k)} \\
&+ \underbrace{\frac{\sqrt{2\pi}}{n(n+1)}\sum_{i=1}^{n}e^{U_{i,j}^2/2}\big\{1-2\Phi(U_{i,j})\big\} V_{i,k}^{*}I{(|U_{i,j}| \le M_2)} }_{\textrm{K}'_{112}(j,k)} \\
&+ \underbrace{\sum_{l=2}^{\infty}\frac{1}{n\cdot l!}\sum_{i=1}^{n}(\Phi^{-1})^{(l)}\{F_{\bX,j}(X_{i,j})\}\bigg\{\frac{n}{n+1}\hat{F}_{\bX,j}(X_{i,j})-  F_{\bX,j}(X_{i,j}) \bigg\}^{l}V_{i,k}^{*}I{(|U_{i,j}| \le M_2)} }_{\textrm{K}'_{113}(j,k)} 
\end{align*} 
Given  $(j,k)$, write $\bT_{i}=(U_{i,j},V_{i,k})$ for $i\in[n]$,  and define 
\begin{align*}
\bar{\varpi}_2(\bT_{i_1},\bT_{i_2}) = \tilde{\varpi}_2(\bT_{i_1},\bT_{i_2})  - \tilde{\delta}_{1,k}(U_{i_2,j}) 
\end{align*}
with 
\begin{align*} 
&\tilde{\varpi}_2(\bT_{i_1},\bT_{i_2}) = e^{U_{i_1,j}^2/2}  \big\{I(U_{i_2,j} \le U_{i_1,j}) -\Phi(U_{i_1,j}) \big\} V^{*}_{i_1,k} I{(|U_{i_1,j}| \le M_2)} \,.
\end{align*}	
Then
\begin{align*}
{\rm K}'_{111}(j,k)=\frac{\sqrt{2\pi}}{n(n+1)}\sum_{1\le i_1\ne i_2 \le n} \bar{\varpi}_2(\bT_{i_1},\bT_{i_2})\,.
\end{align*}
Recall $V_{i,k}^{*} =  V_{i,k}I(|V_{i,k}|\le M_1) +   M_1 \cdot{\rm sign}(V_{i,k}) I(|V_{i,k}|>M_1) $. Such defined $\bar{\varpi}_2(\cdot,\cdot)$ is a bounded kernel. Let $\{\bT_{i}^{(1)}\}$ and $\{\bT_{i}^{(2)}\}$  be two independent copies of $\{\bT_{i}\}$ with $\bT_{i}^{(1)}=\{U_{i_1,j}^{(1)},V_{i_1,k}^{(1)}\}$ and  $\bT_{i}^{(2)}=\{U_{i_2,j}^{(2)},V_{i_2,k}^{(2)}\}$.  We define $V_{i,k}^{(1),*}$ in the same manner as $V_{i,k}^{*}$ but with replacing $V_{i,k}$ by $V_{i,k}^{(1)}$. Recall $U_{i,j} \sim \mathcal{N}(0,1)$. We have 
\begin{align*}
&\mathbb{E}_{\{2\}}\big\{\tilde{\varpi}_2\big(\bT_{i_1}^{(1)},\bT_{i_2}^{(2)}\big)\big\} \\
&~~~~ = \mathbb{E} \big[I\big\{U_{i_2,j}^{(2)}\le U_{i_1,j}^{(1)}\big\} - \Phi(U_{i_1,j}^{(1)}) \,\big|\, U_{i_1,j}^{(1)}\big]   e^{\{U_{i_1,j}^{(1)}\}^2/2}  V_{i_1,k}^{(1),*} 
I\big\{\big|U_{i_1,j}^{(1)}\big| \le M_2\big\} =0\,.
\end{align*}
Then 
\begin{align*}
\mathbb{E}_{\{1\}}\big\{\bar{\varpi}_2\big(\bT_{i_1}^{(1)},\bT_{i_2}^{(2)}\big) \big\} =&~ \mathbb{E}_{\{1\}}\big\{\tilde{\varpi}_2\big(\bT_{i_1}^{(1)},\bT_{i_2}^{(2)}\big)\big\}- \tilde{\delta}_{1,k}\big(U_{i_2,j}^{(2)}\big)= \tilde{\delta}_{1,k}\big(U_{i_2,j}^{(2)}\big)-\tilde{\delta}_{1,k}\big(U_{i_2,j}^{(2)}\big)=0\,,\\
\mathbb{E}_{\{2\}}\big\{\bar{\varpi}_2\big(\bT_{i_1}^{(1)},\bT_{i_2}^{(2)}\big) \big\} = &~\mathbb{E}_{\{2\}}\big\{\tilde{\varpi}_2\big(\bT_{i_1}^{(1)},\bT_{i_2}^{(2)}\big)\big\}- \mathbb{E}\big\{\tilde{\delta}_{1,k}\big(U_{i_2,j}^{(2)}\big)\big\}= -\mathbb{E}\big\{\tilde{\varpi}_2\big(\bT_{i_1}^{(1)},\bT_{i_2}^{(2)}\big) \big\}=0\,,
\end{align*}
which implies $\bar{\varpi}_2(\cdot,\cdot)$ is a bounded canonical kernel. By Inequalities \ref{ieq:decoupling-u} and \ref{ieq:u-sta-2}, we have  
\begin{align}\label{eq:k111-tail}
\mathbb{P}\{|{\rm K}'_{111}(j,k)| \ge x\} \le&~ C_1 \mathbb{P}\bigg\{C_1 \bigg|\sum_{1\le i_1\ne i_2 \le n} \bar{\varpi}_2\big(\bT_{i_1}^{(1)},\bT_{i_2}^{(2)}\big)\bigg| \ge  \frac{n(n+1)x}{\sqrt{2\pi}}\bigg\} \\
\le &~ C_2\exp \bigg\{-\frac{1}{C_2}\min\bigg(\frac{n^2M_2x^2}{M_1^2e^{M_2^2/2}}, \frac{nx}{ M_1e^{M_2^2/2}}, \frac{nx^{2/3}}{M_1^{2/3}e^{M_2^2/3}}, \frac{nx^{1/2}}{M_1^{1/2}e^{M_2^2/4}} \bigg)\bigg\}\notag
\end{align}
for any $x>0$.
Recall $d=pq$. Notice that above inequality holds for any $j\in[p]$ and $k\in[q]$. Hence, we have
\begin{align}\label{eq:k111'}
\max_{j\in[p],\,k\in[q]}|{\rm K}'_{111} (j,k)|= O_{{\rm p}}\big(n^{-1}M_1e^{M_2^2/2}\log d\big)
\end{align}
provided that $\log d \lesssim n$. Notice that  ${\rm K}'_{113}(j,k)={\rm I}_{12}(j,k)$ and ${\rm K}'_{112}(j,k)={\rm I}_{112}(j,k)$ for  ${\rm I}_{12}(j,k)$ and ${\rm I}_{112}(j,k)$ defined in Sections \ref{sec:sub-I1} and \ref{sec:sub-i11}, respectively, with $\kappa_1=6/5$ and $\kappa_2=1/2$. Since the convergence rates of $\max_{j\in[p], \, k\in[q]}|{\rm I}_{112}(j,k)|$ and $\max_{j\in[p], \, k\in[q]}|{\rm I}_{12}(j,k)|$ obtained in Sections  \ref{sec:sub-i11} and \ref{sec:sub-i12}  do not depend on
whether or not the null hypothesis  $\mathbb{H}_0$ in \eqref{eq:equind} holds, we still have 
\begin{align*}
\max_{j\in[p],\,k\in[q]}|{\rm K}'_{112}(j,k)| =  O_{\rm p}(n^{-1} M_1 M_2) 
\end{align*}
provided that $\log d \lesssim ne^{-M_2^2/2} M_2$, and 
\begin{align*}
\max_{j\in[p],\, k\in[q]}|{\rm K}'_{113}(j,k)| = O_{\rm p} \{n^{-1}M_1 e^{M_2^2/2}\log (dn)\}
\end{align*}
provided that $ \log (dn) \ll ne^{-M_2^2} M_2^{-2}$. Together with \eqref{eq:k111'}, it holds that
\begin{align*}
\max_{j\in[p],\, k\in[q]}|{\rm K}'_{11}(j,k)| \le &~ 	\max_{j\in[p],\,k\in[q]}|{\rm K}'_{111}(j,k)| + 	\max_{j\in[p],\,k\in[q]}|{\rm K}'_{112}(j,k)| +	\max_{j\in[p],\,k\in[q]}|{\rm K}'_{113}(j,k)| \\
=&~ O_{\rm p} \{n^{-1}M_1 e^{M_2^2/2}\log (dn)\}
\end{align*}
provided that $ \log (dn) \ll ne^{-M_2^2} M_2^{-2}$. Recall $M_1=\sqrt{6(\log n)/5}$ and $M_2=\sqrt{(\log n)/2}$. Then 
\begin{align*}
\max_{j\in[p],\, k\in[q]}|{\rm K}'_{11}(j,k)| = O_{\rm p} \{n^{-3/4}(\log n)^{1/2}  \log (dn)\}
\end{align*}
provided that $ \log d \ll n^{1/2} (\log n)^{-1}$. We have \eqref{eq:k11'} holds.
$\hfill\Box$

\section{Proof of  Lemma \ref{lem:nn-eh-e-delta}}\label{sec:nn-eh-e-delta}  
In order to prove Lemma \ref{lem:nn-eh-e-delta}, we need Lemmas \ref{lem:uh-u-delta}--\ref{lem:fhwh-fwh}, with their proofs given in Sections \ref{sec:sub-uh-u-delta}--\ref{sec:sub-fhwh-fwh}, respectively. Recall $\tilde{d}=p\vee q\vee m$.

\begin{lem}\label{lem:uh-u-delta}
	If $\log (\tilde{d}n) \ll n^{1-\kappa}(\log n)^{-1/2}$, then
	\begin{align*}
		\max_{j\in[p],\,k\in[q]}\bigg|\frac{1}{n_{3}} \sum_{t\in \mathcal{D}_3} (\hat{U}_{t,j}^{(w)}- U_{t,j})\delta_{t,k}\bigg| =&~ O_{\rm p}\{ n^{-\kappa}  \log ^2(\tilde{d}n) \} + O_{\rm p}\{ n^{-1/2} \log(\tilde{d}n)\} \\
		=&~\max_{j\in[p],\,k\in[q]}\bigg| \frac{1}{n_{3}} \sum_{t\in \mathcal{D}_3} (\hat{V}_{t,k}^{(w)}- V_{t,k})\varepsilon_{t,j} \bigg|\,.
	\end{align*} 
\end{lem}  

\begin{lem}\label{lem:fwh-fw-delta}
	Under Condition {\rm\ref{cd:function-condition}}, it holds that
	\begin{align*} 
		&\max_{j\in[p],\,k\in[q]}\bigg|\frac{1}{n_{3}} \sum_{t\in \mathcal{D}_3} \{f_{j} (\hat{\bW}_{t}^{(w)}) - f_{j} (\bW_{t})\}\delta_{t,k} \bigg|\\
		&~~~~~~~~~~~~~~ =  O_{\rm p}\{n^{-\kappa} m^{2} \log (\tilde{d}n)\}  + O_{\rm p} \{ n^{-1/2} m \log  (\tilde{d}n)\} \\
		&~~~~~~~~~~~~~~~~~~~~~~~~~~~~~ =\max_{j\in[p],\,k\in[q]}\bigg|\frac{1}{n_{3}} \sum_{t\in \mathcal{D}_3} \{g_{k} (\hat{\bW}_{t}^{(w)}) - g_{k} (\bW_{t})\}\varepsilon_{t,j} \bigg|
	\end{align*}
	provided that $\log (\tilde{d}n) \ll n^{1-\kappa}(\log n)^{-1/2}$. 
\end{lem}   

\begin{lem}\label{lem:fhwh-fwh}
	Let $\hat{f}_{j} $ and $\hat{g}_{k} $ be  the estimates specified in \eqref{eq:fhj-ghk}  with $(m_{*},K)$ as in the definitions of $f_j$ and $g_k$, $\tilde{\alpha}_{n} =n^{c_{3}} $ and $M_{*}= c_{4} \lceil n^{m_{*}/ (4\vartheta + m_{*})} (m^2 \log n)^{m_{*}(2\tilde{\vartheta}+3)/(2\vartheta)}\rceil$  for some sufficiently large constants $c_{3}>0$ and $c_{4}>0$.  Under Condition  {\rm\ref{cd:function-condition}},  it holds that   
	\begin{align*}
		&\max_{j\in[p],\,k\in[q]}\bigg|\frac{1}{n_{3}} \sum_{t\in \mathcal{D}_3} \{\hat{f}_{j} (\hat{\bW}_{t}^{(w)}) - f_{j} (\hat{\bW}_{t}^{(w)})\}\delta_{t,k}\bigg|\\ 
		&~~~~~~= O_{\rm p}\{n^{-\kappa/2-\vartheta/(4\vartheta+m_{*})}  (m^2\log n)^{(\vartheta+2m_{*}\tilde{\vartheta}+3m_{*} )/(8\vartheta)} (\log n)\log^{7/4}(\tilde{d}n) \}  \notag \\
		&~~~~~~~~~~~ +  O_{\rm p}\{n^{-\kappa/2-1/4} m^{1/2}  (\log n)^{1/2}\log^{3/2}  (\tilde{d}n) \}  + O_{\rm p}\{n^{-\kappa}m (\log n)\log^2(\tilde{d}n) \} \\
		&~~~~~~~~~~~~~~  = \max_{j\in[p],\,k\in[q]}\bigg|\frac{1}{n_{3}} \sum_{t\in \mathcal{D}_3} \{\hat{g}_{k} (\hat{\bW}_{t}^{(w)}) - g_{k} (\hat{\bW}_{t}^{(w)})\}\varepsilon_{t,j}\bigg|
	\end{align*}
	provided that  $ \log (\tilde{d}n) \ll  n^{1-\kappa} (\log n)^{-1/2}$ and $m \lesssim n$.     
	
\end{lem}

Recall $\varepsilon_{i,j} = U_{i,j}- f_{j}(\bW_i)$ and $ \tilde{\varepsilon}_{i,j} = \hat{U}_{i,j}^{(w)} - \hat{f}_{j}(\hat{\bW}_{i}^{(w)})$.  By Lemmas \ref{lem:uh-u-delta}--\ref{lem:fhwh-fwh}, we have
\begin{align*}
	&\max_{j\in[p],\,k\in[q]}\bigg|\frac{1}{n_{3}} \sum_{t\in \mathcal{D}_3} ( \tilde{\varepsilon}_{t,j} - \varepsilon_{t,j})\delta_{t,k}\bigg| \\
	&~~~~~~~= \max_{j\in[p],\,k\in[q]}\bigg|\frac{1}{n_{3}} \sum_{t\in \mathcal{D}_3} (\hat{U}_{t,j}^{(w)}- U_{t,j})\delta_{t,k} - \frac{1}{n_{3}} \sum_{t\in \mathcal{D}_3} \{\hat{f}_{j} (\hat{\bW}_{t}^{(w)}) - f_{j} (\bW_{t})\} \delta_{t,k}\bigg|\\
	&~~~~~~~\le \max_{j\in[p],\,k\in[q]}\bigg|\frac{1}{n_{3}} \sum_{t\in \mathcal{D}_3} (\hat{U}_{t,j}^{(w)}- U_{t,j})\delta_{t,k}\bigg| +\max_{j\in[p],\,k\in[q]}\bigg|  \frac{1}{n_{3}} \sum_{t\in \mathcal{D}_3} \{f_{j} (\hat{\bW}_{t}^{(w)}) - f_{j} (\bW_{t})\}\delta_{t,k}\bigg|\\
	&~~~~~~~~~~~+\max_{j\in[p],\,k\in[q]}\bigg|\frac{1}{n_{3}} \sum_{t\in \mathcal{D}_3} \{\hat{f}_{j} (\hat{\bW}_{t}^{(w)}) - f_{j} (\hat{\bW}_{t}^{(w)})\}\delta_{t,k}\bigg|\\
	&~~~~~~~=   O_{\rm p}\{n^{-\kappa/2-\vartheta/(4\vartheta+m_{*})}  (m^2\log n)^{(\vartheta+2m_{*}\tilde{\vartheta}+3m_{*} )/(8\vartheta)} (\log n)\log^{7/4}(\tilde{d}n) \}  \notag \\
	&~~~~~~~~~~~ +O_{\rm p}\{n^{-\kappa/2-1/4} m^{1/2}  (\log n)^{1/2}\log^{3/2}  (\tilde{d}n) \}\\
	&~~~~~~~~~~~ + O_{\rm p}\{n^{-\kappa}m^2 (\log n)\log^2(\tilde{d}n) \} + O_{\rm p}\{ n^{-1/2} m \log(\tilde{d}n)\}
\end{align*}
provided that   $ \log (\tilde{d}n) \ll  n^{1-\kappa} (\log n)^{-1/2}$ and $m \lesssim n$.  Analogously, we can show such convergence rate also holds for $	\max_{j\in[p],\,k\in[q]} | n_{3}^{-1} \sum_{t\in \mathcal{D}_3} ( \tilde{\delta}_{t,k} - \delta_{t,k})\varepsilon_{t,j}|$. Therefore, we complete the proof of Lemma \ref{lem:nn-eh-e-delta}.
$\hfill\Box$

\subsection{Proof of Lemma \ref{lem:uh-u-delta}}\label{sec:sub-uh-u-delta}
Define $U_{t,j}^{*} = U_{t,j}I(|U_{t,j}|\le M_1) + M_1 \cdot{\rm sign}(U_{t,j})I(|U_{t,j}|>M_1)$ with $M_1 = \sqrt{2\log n_{3}}$.  Given $Q>0$, we have
\begin{align*}
	\frac{1}{n_{3}} \sum_{t\in \mathcal{D}_3} (\hat{U}_{t,j}^{(w)}- U_{t,j})\delta_{t,k} = &~	\underbrace{\frac{1}{n_{3}}\sum_{t\in \mathcal{D}_3} \{\hat{U}_{t,j}^{(w)} - U_{t,j}^{*}\} \delta_{t,k}I(|U_{t,j}|\le M_1)I(|\delta_{t,k}|\le Q) }_{\tilde{\textrm{H}}_{1}(j,k)}\\
	&+\underbrace{	\frac{1}{n_{3}}\sum_{t\in \mathcal{D}_3} \{\hat{U}_{t,j}^{(w)} - U_{t,j}^{*}\} \delta_{t,k}I(|U_{t,j}|> M_1)I(|\delta_{t,k}|\le Q) }_{\tilde{\textrm{H}}_{2}(j,k)} \\
	&+ \underbrace{	\frac{1}{n_{3}}\sum_{t\in \mathcal{D}_3} (U_{t,j}^{*}-U_{t,j})  \delta_{t,k}I(|\delta_{t,k}|\le Q) }_{\tilde{\textrm{H}}_{3}(j,k)} \\
	&+ 	\underbrace{\frac{1}{n_{3}}\sum_{t\in \mathcal{D}_3} \{\hat{U}_{t,j}^{(w)} - U_{t,j}\} \delta_{t,k}I(|\delta_{t,k}|> Q) }_{\tilde{\textrm{H}}_{4}(j,k)} \,.
\end{align*}
Recall $d=pq$ and $n_3\asymp n^{\kappa}$ for some constant $0<\kappa<1$. Analogous to the derivation of \eqref{eq:k1} with $M_1 = \sqrt{2\log n_{3}}$, we have
\begin{align*}
	\max_{j\in[p],\,k\in[q]}|\tilde{{\rm H}}_{3}(j,k)| = &~ O_{\rm p} \{Qn_{3}^{-1}(\log n_{3})^{1/4} (\log d)^{1/2}\} + O_{\rm p} (Q^2n_{3}^{-1} \log d)\\
	= &~ O_{\rm p} \{Qn^{-\kappa}(\log n)^{1/4} (\log d)^{1/2}\} + O_{\rm p} (Q^2n^{-\kappa} \log d)  
\end{align*}
provided that $\log (d n) \lesssim Q^2$.    
Recall $\delta_{t,k} = V_{t,k} - g_k(\bW_t)$, $V_{i,k} \sim \mathcal{N}(0,1)$ and $|g_k|_{\infty} \le \tilde{C}$.  It holds that
\begin{align}\label{eq:delta-tail-nn}
    \mathbb{P}(|\delta_{t,k}|>x) = &~\mathbb{P}\{|V_{t,k} - g_k(\bW_t)|>x\}  \leq \mathbb{P}\bigg(|V_{t,k}| >\frac{x}{2}\bigg) + \mathbb{P}\bigg\{|g_k(\bW_t)|>\frac{x}{2}\bigg\} \notag\\
    \leq&~ 2e^{-x^2/4} + C_1e^{-x^2/4} \le C_2 e^{- x^2/4}
\end{align}  
for any $x>0$, $t\in[n]$ and $k\in[q]$.  Then, for any $x>0$, we have
\begin{align*}
    \mathbb{P}\bigg\{\max_{j\in[p],\, k\in[q]}|\tilde{\rm H}_{4}(j,k)|>x\bigg\} \le \max_{t\in[n],\,k\in[q]} nq \mathbb{P}(|\delta_{t,k}|> Q) \lesssim  nqe^{- Q^2/4} \,,
\end{align*}
which implies  
\begin{align*}
	\max_{j\in[p],\,k\in[q]}|\tilde{{\rm H}}_{4}(j,k)| = o_{\rm p}(n^{-1}) 
\end{align*}
provided that $\log (d n) \lesssim Q^2$. 
As we will show in Sections \ref{sec:sub-g11} and \ref{sec:sub-g12},  
\begin{align}\label{eq:g11}
	\max_{j\in[p],\,k \in [q]}|\tilde{{\rm H}}_{1}(j,k)| =   O_{\rm p}\{Q n^{-(1+\kappa)/2}\log^{3/2}(pn) \} + O_{\rm p}\{Qn^{-1/2}\log^{1/2}(pn)\}
\end{align}
provided that $ \log (pn) \ll n^{1-\kappa}(\log n)^{-1/2}$, and 
\begin{align}\label{eq:g12}
	\max_{j\in[p],\,k\in[q]}|\tilde{{\rm H}}_{2}(j,k)| =   O_{\rm p}\{ Qn^{-\kappa} (\log n)^{1/2}\log p \}\,.
\end{align}  
Recall $\tilde{d}=p\vee q\vee m$. By selecting  $Q =C \log^{1/2} (\tilde{d} n)$ for some sufficiently large constant $C > 0$, it holds that
\begin{align*}
	\max_{j\in[p],\,k \in [q]}\bigg|\frac{1}{n_{3}} \sum_{t\in \mathcal{D}_3} (\hat{U}_{t,j}^{(w)}- U_{t,j})\delta_{t,k}\bigg|  
	\le&~ \max_{j\in[p],\,k\in[q]}|\tilde{{\rm H}}_{1}(j,k)| + \max_{j\in[p],\,k\in[q]}|\tilde{{\rm H}}_{2}(j,k)|\\ &+\max_{j\in[p],\,k\in[q]}|\tilde{{\rm H}}_{3}(j,k)| + \max_{j\in[p],\,k\in[q]}|\tilde{{\rm H}}_{4}(j,k)|\\ 
	=&~ O_{\rm p}\{ n^{-\kappa}  \log ^2(\tilde{d}n)\}  + O_{\rm p}\{ n^{-1/2} \log(\tilde{d}n)\}
\end{align*}
provided that $\log (\tilde{d}n) \ll n^{1-\kappa}(\log n)^{-1/2}$.  Identically, we can also show such convergence rate holds for $\max_{j\in[p],\,k \in [q]} |n_{3}^{-1} \sum_{t\in \mathcal{D}_3} (\hat{V}_{t,k}^{(w)}- V_{t,k})\varepsilon_{t,j}|$.  
$\hfill\Box$

\subsubsection{ Proof of \eqref{eq:g11} }\label{sec:sub-g11}

Recall $\hat{U}_{i,j}^{(w)} = \Phi^{-1} \{\hat{F}_{\bX,j}^{(w)} (X_{i,j})\}$, $U_{i,j}=\Phi^{-1}\{F_{\bX,j}(X_{i,j})\}$ and  $U_{i,j}^{*} = U_{i,j}I(|U_{i,j}|\le M_1) + M_1 \cdot{\rm sign}(U_{i,j})I(|U_{i,j}|>M_1)$ with $\hat{F}_{\bX,j}^{(w)}(X_{i,j})$  defined in \eqref{eq:tructed-ecdf} and $M_1 = \sqrt{2\log n_{3}}$.  
Let 
$$K(U_{i,j}, p,n_{1}) = 4 n_{1}^{-1/2}  [\Phi(U_{i,j}) \{1-\Phi(U_{i,j})\}]^{1/2} \log^{1/2}(pn_{1})  + 7n_1^{-1} \log(pn_{1})\,.$$
Define the event 
\begin{align*}
	\mathcal{H}_{5} =  \bigcap_{i\in\mathcal{D}_3,\,j\in[p]}\big\{|\hat{F}_{\bX,j}(X_{i,j})-  F_{\bX,j}(X_{i,j})| \le K(U_{i,j}, p,n_{1})\big\} \,.
\end{align*}
Recall $\mathcal{D}_1, \mathcal{D}_2$ and $\mathcal{D}_3$ are three disjoint subsets of $[n]$ with $|\mathcal{D}_1|=n_1\asymp n$, $|\mathcal{D}_2|=n_2\asymp n$ and $|\mathcal{D}_3|=n_3\asymp n^{\kappa}$  for some constant $0<\kappa<1$ and $n_1+n_2+n_3=n$. Similar to \eqref{eq:h2-c}, we have 
\begin{align}\label{eq:h5-c}
	\mathbb{P}(\mathcal{H}_{5}^{\rm c}) = &~\mathbb{P}\bigg[ \bigcup_{i\in\mathcal{D}_3, \,j\in[p]}\big\{|\hat{F}_{\bX,j}(X_{i,j})-  F_{\bX,j}(X_{i,j})| > K(U_{i,j}, p,n_{1})\big\}\bigg] \notag\\
	\le &~ \sum_{i\in\mathcal{D}_3}\sum_{j=1}^{p} \mathbb{E} \bigg( \mathbb{P} \bigg[ \bigg|\frac{1}{n_{1}}\sum_{s\in \mathcal{D}_1} \big\{I(U_{s,j}\le U_{i,j})  - \Phi(U_{i,j})\big\}\bigg| > K(U_{i,j}, p,n_{1})   \,\bigg|\, U_{i,j}\bigg]\bigg) \notag\\
	\le &~ 2 n_{3}p \max_{i\in \mathcal{D}_3,\,j\in[p]}\mathbb{E} \bigg( \exp\bigg[-\frac{n_{1}K^2(U_{i,j}, p,n_{1})}{4\Phi(U_{i,j}) \{1-\Phi(U_{i,j})\}  }\bigg] +  \exp\bigg\{-\frac{n_{1}K(U_{i,j}, p,n_{1})}{2}\bigg\} \bigg) \notag\\
	\le &~  4(n_{1}p)^{-2}\,.
\end{align}
Restricted on $\mathcal{H}_{5}$, for any integer $l\ge 0$, it holds that
\begin{align}\label{eq:fwh-f-dec}
	|\hat{F}_{\bX,j}^{(w)}(X_{i,j})-  F_{\bX,j}(X_{i,j})|^{l} \le&~ 2^{l}|\hat{F}_{\bX,j}^{(w)}(X_{i,j})-  \hat{F}_{\bX,j} (X_{i,j})|^{l} + 2^{l}|\hat{F}_{\bX,j} (X_{i,j}) - F_{\bX,j}(X_{i,j})|^{l} \notag\\
	\le &~ C_{1}^{l}\bigg| \frac{\Phi(U_{i,j}) \{1-\Phi(U_{i,j})\}\log (pn_{1})}{n_{1}}\bigg|^{l/2} + C_{2}^{l} \bigg|\frac{\log(pn_{1})}{n_{1}} \bigg|^{l}\,.
\end{align}
Given some constant $M_2\in (0, M_1)$, restricted on $\mathcal{H}_{5}$, by \eqref{eq:derivative_n} and \eqref{eq:fwh-f-dec}, 
\begin{align}\label{eq:h1t-dec}
	|\tilde{{\rm H}}_{1}(j,k)|  \le &~  \frac{Q}{n_{3}}\sum_{i\in \mathcal{D}_3} |\hat{U}_{i,j}^{(w)} - U_{i,j}| I(|U_{i,j}|\le M_1) \notag\\
	\le &~ \sum_{l=1}^{\infty}\frac{Q}{n_{3}\cdot l!}\sum_{i\in \mathcal{D}_3}\big|(\Phi^{-1})^{(l)}\{F_{\bX,j}(X_{i,j})\}\big| \big|\hat{F}_{\bX,j}^{(w)}(X_{i,j})-  F_{\bX,j}(X_{i,j}) \big|^{l}  I(|U_{i,j}|\le M_1) \notag\\
	\le &~ \sum_{l=1}^{\infty}\frac{Q}{n_{3}  }\sum_{i\in \mathcal{D}_3} \bar{C}^{l} |U_{i,j}|^{l-1} e^{lU_{i,j}^2/2} \big|\hat{F}_{\bX,j}^{(w)}(X_{i,j})-  F_{\bX,j}(X_{i,j}) \big|^{l}  I(|U_{i,j}|\le M_1) \notag\\
	\le &~ \sum_{l=1}^{\infty}\frac{Q}{n_{3}  }\sum_{i\in \mathcal{D}_3} C_3^{l} |U_{i,j}|^{l-1} e^{lU_{i,j}^2/2} \bigg|\frac{\log (pn_{1})}{n_{1}} \bigg|^{l/2}  I(|U_{i,j}|\le M_2) \notag\\
	&+ \sum_{l=1}^{\infty}\frac{Q}{n_{3}  }\sum_{i\in \mathcal{D}_3} C_4^{l} |U_{i,j}|^{l-1} e^{lU_{i,j}^2/2} \bigg|\frac{\log (pn_{1})}{n_{1}} \bigg|^{l}  I(M_2 < |U_{i,j}|\le M_1) \notag\\
	&+ \sum_{l=1}^{\infty}\frac{Q}{n_{3}  }\sum_{i\in \mathcal{D}_3} C_5^{l} |U_{i,j}|^{l/2-1} e^{lU_{i,j}^2/4} \bigg|\frac{\log (pn_{1})}{n_{1}} \bigg|^{l/2}  I(M_2 < |U_{i,j}|\le M_1) \notag \\
	\le&~ \sum_{l=1}^{\infty} \bigg\{\frac{C_6 M_2  e^{M_2^2/2}\log^{1/2}(pn_{1})}{n_{1}^{1/2}}   \bigg\}^{l-1} \times \frac{Q\log^{1/2}(pn_{1})}{n_{1}^{1/2}} \times \frac{1}{n_{3} }\sum_{i\in \mathcal{D}_3}  e^{U_{i,j}^2/2} I (|U_{i,j}|\le M_2) \notag\\
	&+ \sum_{l=1}^{\infty} \bigg\{\frac{C_7 M_1  e^{M_1^2/2}\log (pn_{1})}{n_{1} }   \bigg\}^{l-1} \times \frac{Q\log(pn_{1})}{n_{1}} \times \frac{1}{n_{3} }\sum_{i\in \mathcal{D}_3}  e^{U_{i,j}^2/2} I (M_2 < |U_{i,j}|\le M_1) \notag\\
	&+ \sum_{l=1}^{\infty} \bigg\{\frac{C_8 M_1^{1/2}  e^{M_1^2/4}\log^{1/2} (pn_{1})}{n_{1}^{1/2} }   \bigg\}^{l-1}  \times \frac{Q\log^{1/2}(pn_{1})}{n_{1}^{1/2}M_2^{1/2}} \notag\\ &~~~~~\times \frac{1}{n_{3} }\sum_{i\in \mathcal{D}_3}  e^{U_{i,j}^2/4} I (M_2 <|U_{i,j}|\le M_1) \notag\\
	\le&~\frac{C_{9}Q\log^{1/2}(pn_{1})}{n_{1}^{1/2}} \times   \frac{1}{n_{3} }\sum_{i\in \mathcal{D}_3}  e^{U_{i,j}^2/4} I ( |U_{i,j}|\le M_1)  
\end{align}
provided that $\log (pn_1)\ll n_{1}M_{1}^{-1}e^{-M_1^2/2}$, where the fourth step is due to $1-\Phi(x) \le x^{-1} \phi(x)$ for $x>0$. Recall $U_{i,j} \sim \mathcal{N}(0,1)$. 
We then have $ \mathbb{E}\{ e^{U_{i,j}^2/4} I( |U_{i,j}| \le M_1)\}  \le  C_{10}$ and $  \var\{e^{U_{i,j}^2/4}  I( |U_{i,j}| \le M_1)\}\lesssim  M_1$. By Bonferroni inequality and Bernstein inequality, it holds that
\begin{align*} 
	\max_{j\in[p]}\bigg|   \frac{1}{n_{3} }\sum_{i\in \mathcal{D}_3}  e^{U_{i,j}^2/4} I ( |U_{i,j}|\le M_1)\bigg| =O_{\rm p} \{n_{3}^{-1/2}M_1^{1/2} (\log p)^{1/2} \} +  O_{\rm p}(n_{3}^{-1}e^{M_1^2/4} \log p ) + O(1) \,. 
\end{align*}
As shown in \eqref{eq:h5-c}, $\mathbb{P}(\mathcal{H}_{5}^{\rm c}) \to 0$ as $n_1\to \infty$.  Hence, applying the similar arguments in Section \ref{sec:sub-i12} for deriving the convergence rate of $\max_{j\in[p],\,k\in[q]}|{\rm I}_{12}(j,k)|$, by \eqref{eq:h1t-dec},
we can show
\begin{align*}
	\max_{j\in[p],\,k\in[q]}|\tilde{{\rm H}}_{1}(j,k)| =&~ O_{\rm p}  \{Qn_{3}^{-1/2} n_{1}^{-1/2}M_1^{1/2} \log (pn_{1}) \} +  O_{\rm p}\{Qn_{3}^{-1} n_{1}^{-1/2}e^{M_1^2/4} \log^{3/2}(pn_{1})  \} \\
	&+ O_{\rm p}\{Qn_{1}^{-1/2}\log^{1/2}(pn_{1})\}
\end{align*}
provided that $ \log (pn_{1}) \ll n_{1}e^{-M_1^2/2} M_1^{-1}$. Recall  $M_{1}= \sqrt{2\log n_{3}}$, $n_1\asymp n$ and $n_3\asymp n^{\kappa}$ for some constant $0<\kappa<1$.
Hence, we have \eqref{eq:g11} holds. 
$\hfill\Box$

\subsubsection{ Proof of \eqref{eq:g12} }\label{sec:sub-g12}
Recall $\hat{U}_{i,j}^{(w)} = \Phi^{-1} \{\hat{F}_{\bX,j}^{(w)} (X_{i,j})\}$ and $n_{1}^{-1} \le \hat{F}_{\bX,j}^{(w)} (X_{i,j}) \le (n_{1}-1)n_{1}^{-1}$. Due to  $-\sqrt{2\log n_{1}}\le \Phi^{-1}(n_{1}^{-1}) < \Phi^{-1}(1-n_{1}^{-1}) \le \sqrt{2\log n_{1}}$, we have  
\begin{align}\label{eq:uh-w}
	\max_{i\in \mathcal{D}_3,\,j\in[p]}|\hat{U}_{i,j}^{(w)}| \le \sqrt{2\log n_{1}} 
\end{align}
for sufficiently large $n_{1}$. Recall $U_{i,j}^{*}=U_{i,j}I(|U_{i,j}| \le M_1) + M_1\cdot {\rm sign}(U_{i,j})I(|U_{i,j}|>M_1)$ with $M_1=\sqrt{2\log n_3}$, $n_1\asymp n$ and $n_3\asymp n^{\kappa}$ for some constant $0<\kappa<1$. Then $\max_{i\in\mathcal{D}_3,\, j\in[p]}|U_{i,j}^{*}| \le \sqrt{2\log n_3} \le C_{11}\sqrt{\log n}$.  Hence, 
\begin{align*}
	\max_{j\in[p],\,k\in[q]}|\tilde{{\rm H}}_{2}(j,k)| \le C_{12}  Q\sqrt{\log n}   \times \max_{j\in[p]} \frac{1}{n_{3}}\sum_{i \in \mathcal{D}_3} I(|U_{i,j}| > M_1)\,.
\end{align*}  
By \eqref{eq:i(u>m)},  Bonferroni inequality and Bernstein inequality, it holds that
\begin{align*}
	\max_{j\in[p] }\frac{1}{n_{3}}\sum_{i\in\mathcal{D}_3} I(|U_{i,j}| > M_1) = &~O_{\rm p}\{n_{3}^{-1/2} M_1^{-1/2}e^{-M_1^2/4} (\log p)^{1/2}\}  +   O_{\rm p}(n_{3}^{-1} \log p ) \\
	&+ O_{\rm p}(M_1^{-1} e^{-M_1^2/2}) \\
	=&~ O_{\rm p } (n^{-\kappa}\log p)\,.
\end{align*}
Hence, we have \eqref{eq:g12} holds. 
$\hfill\Box$

\subsection{Proof of Lemma \ref{lem:fwh-fw-delta}}\label{sec:sub-lem:fwh-fw-delta}
Given $Q>0$, we have 
\begin{align*}
	\frac{1}{n_{3}} \sum_{t\in \mathcal{D}_3} \{f_{j} (\hat{\bW}_{t}^{(w)}) - f_{j} (\bW_{t})\}\delta_{t,k} = &~ \underbrace{\frac{1}{n_{3}} \sum_{t\in \mathcal{D}_3} \{f_{j} (\hat{\bW}_{t}^{(w)}) - f_{j} (\bW_{t})\}\delta_{t,k}I(|\delta_{t,k}|\le Q)}_{\textrm{H}_{1}(j,k)} \\
	&+\underbrace{ \frac{1}{n_{3}} \sum_{t\in \mathcal{D}_3} \{f_{j} (\hat{\bW}_{t}^{(w)}) - f_{j} (\bW_{t})\}\delta_{t,k}I(|\delta_{t,k}|> Q) }_{\textrm{H}_{2}(j,k)}\,.
\end{align*}
Recall $d=pq$. Using the similar arguments for deriving the convergence rate of  $\max_{j\in[p],\, k\in[q]}|\tilde{{\rm H}}_{4}(j,k)|$  in Section \ref{sec:sub-uh-u-delta} for the proof of Lemma \ref{lem:uh-u-delta}, it holds that
\begin{align*}
	\max_{j\in[p],\,k\in[q]}|{\rm H}_{2}(j,k)| = o_{\rm p}(n^{-1}) 
\end{align*}
provided that $\log (d n) \lesssim Q^2$. As we will show in Section  \ref{sec:sub-g21}, 
\begin{align}\label{eq:g21}
	\max_{j\in[p],\,k\in[q]}|{\rm H}_{1}(j,k)| = O_{\rm p}\{n^{-\kappa}Qm^{2} \log^{1/2} (mn)\}  + O_{\rm p} \{ n^{-1/2} Qm\log^{1/2} (mn)\}  
\end{align}
provided that $\log (mn) \ll n^{1-\kappa} (\log  n)^{-1/2}$. Recall $\tilde{d}=p\vee q \vee m$. By selecting $Q=\breve{C}\log^{1/2}(\tilde{d}n)$ for some sufficiently large constant $\breve{C}>0$, we have 
\begin{align*}
	\max_{j\in[p],\,k\in[q]}\bigg|\frac{1}{n_{3}} \sum_{t\in \mathcal{D}_3} \{f_{j} (\hat{\bW}_{t}^{(w)}) - f_{j} (\bW_{t})\}\delta_{t,k}\bigg| =  O_{\rm p}\{n^{-\kappa} m^{2} \log (\tilde{d}n)\}  + O_{\rm p} \{ n^{-1/2} m \log  (\tilde{d}n)\}  
\end{align*}
provided that $\log (\tilde{d}n) \ll n^{1-\kappa} (\log n)^{-1/2}$. Identically, we can also show such convergence rate holds for $\max_{j\in[p],\,k\in[q]}|n_{3}^{-1} \sum_{t\in \mathcal{D}_3} \{g_{k} (\hat{\bW}_{t}^{(w)}) - g_{k} (\bW_{t})\}\varepsilon_{t,j} |$. Hence, we complete the proof of Lemma \ref{lem:fwh-fw-delta}.
$\hfill\Box$

\subsubsection{Convergence rate of $\max_{j\in[p],\,k\in[q]}|{\rm H}_{1}(j,k)|$}\label{sec:sub-g21}
We first show that for any  $f_j:\mathbb{R}^{m} \to \mathbb{R}$ satisfies a $(\vartheta, C, \breve{C})$-smooth generalized hierarchical interaction model of finite order $m_{*}$ and finite level $\ell$ according to  Condition \ref{cd:function-condition}, it holds that
\begin{align}\label{eq:fwh-fw-bound}
	|f_{j}(\hat{\bW}_t^{(w)}) - f_{j}(\bW_t)| \le \tilde{C} |\hat{\bW}_{t}^{(w)} - \bW_{t}|_1
\end{align}
for any $t \in \mathcal{D}_3$ and $j\in[p]$, where $\tilde{C}>0$ is some universal constant that does not depend on the selection of $f_{j}$. 
If $\ell=0$, by Definition \ref{def:hierarchical}, $f_{j}(\boldsymbol{x})$ can be expressed by 
\begin{align*}
	f_{j}(\boldsymbol{x}) = h_1^{(j)}(\boldsymbol{\phi}_{1}^{(j),\T} \boldsymbol{x}, \ldots, \boldsymbol{\phi}_{m_{*}}^{(j),\T} \boldsymbol{x})\,,~~~~ \boldsymbol{x} \in \mathbb{R}^m\,,
\end{align*}
where $h_1^{(j)} $ is a $(\vartheta, C)$-smooth function and $\boldsymbol{\phi}_{1}^{(j)},\ldots, \boldsymbol{\phi}_{m_{*}}^{(j)}\in \mathbb{R}^{m}$  with $\max_{k\in [m_{*}]}|\boldsymbol{\phi}_{k}^{(j)}|_{\infty} \le \breve{C}$. By Condition \ref{cd:function-condition}, $h_{1}^{(j)} $ is Lipschitz continuous with Lipschitz constant $L>0$. We then have 
\begin{align*}
	|f_{j}(\hat{\bW}_t^{(w)}) - f_{j}(\bW_t)| \le&~ L  \sum_{k=1}^{m_{*}} \big|\boldsymbol{\phi}_{k}^{(j),\T}\hat{\bW}_{t}^{(w)} - \boldsymbol{\phi}_{k}^{(j),\T}\bW_{t} \big| \\
	\le &~ 
	Lm_{*} \cdot\max_{k\in [m_{*}]}|\boldsymbol{\phi}_{k}^{(j)}|_{\infty} \cdot |\hat{\bW}_{t}^{(w)} - \bW_{t}|_1  \equiv C_1|\hat{\bW}_{t}^{(w)} - \bW_{t}|_1
\end{align*}
for any $t \in \mathcal{D}_3$ and $j\in[p]$, which means  \eqref{eq:fwh-fw-bound} holds  when $\ell=0$. We assume \eqref{eq:fwh-fw-bound} holds for $\ell=l$. When $\ell=l+1$, by Definition \ref{def:hierarchical}, there exists a finite constant $K \in \mathbb{N}$ such that
\begin{align*}
	f_{j}(\boldsymbol{x}) = \sum_{k=1}^{K} h_{k}^{(j)}\big\{\tilde{h}_{1,k}^{(j)}(\boldsymbol{x}),\ldots, \tilde{h}_{m_{*},k}^{(j)}(\boldsymbol{x})\big\}\,, ~~~~ \boldsymbol{x} \in \mathbb{R}^{m} \,,
\end{align*}
where, for any $k\in[K]$, $h_{k}^{(j)}: \mathbb{R}^{m_{*}} \to \mathbb{R}$  and $\tilde{h}_{1,k}^{(j)}, \ldots, \tilde{h}_{m_{*},k}^{(j)}: \mathbb{R}^m \to \mathbb{R}$ are  $(\vartheta, C)$-smooth functions with $\tilde{h}_{1,k}^{(j)}, \ldots\, \tilde{h}_{m_{*},k}^{(j)}$  satisfying a generalized hierarchical interaction model of order $m_{*}$ and level $l$.  Since \eqref{eq:fwh-fw-bound} holds for $\ell=l$, we have
\begin{align*}
	\big| \tilde{h}_{i,k}^{(j)}(\hat{\bW}_{t}^{(w)}) -\tilde{h}_{i,k}^{(j)}(\bW_{t})\big| \le \tilde{C} |\hat{\bW}_{t}^{(w)} - \bW_{t}|_1
\end{align*}
for any $t \in \mathcal{D}_3$, $i \in [m_{*}]$ and $k \in [K]$. By Condition \ref{cd:function-condition}, for any  $k\in[K]$,  $h_{k}^{(j)} $ is  Lipschitz continuous with Lipschitz constant $L>0$. It then holds that 
\begin{align*}
	|f_{j}(\hat{\bW}_t^{(w)}) - f_{j}(\bW_t)|  \le &~  \sum_{k=1}^{K} L  \sum_{i=1}^{m_{*}} \big| \tilde{h}_{i,k}^{(j)}(\hat{\bW}_{t}^{(w)}) -\tilde{h}_{i,k}^{(j)}(\bW_{t})\big|  \\
	\le &~ L K m_{*} \cdot \tilde{C} |\hat{\bW}_{t}^{(w)} - \bW_{t}|_1 \equiv C_2   |\hat{\bW}_{t}^{(w)} - \bW_{t}|_1
\end{align*}
for any $t \in \mathcal{D}_3$ and $j\in[p]$. Hence, we have \eqref{eq:fwh-fw-bound} holds  when $\ell = l+1$. Based on the mathematical induction, we know \eqref{eq:fwh-fw-bound} holds for given $\ell$ specified in Condition \ref{cd:function-condition}. 

Define $W_{i,j}^{*} = W_{i,j}I(|W_{i,j}|\le M_1) + M_1 \cdot{\rm sign}(W_{i,j})I(|W_{i,j}|>M_1)$ with $M_1 = \sqrt{2\log n_{3}}$. Recall $\bW_{t}=(W_{t,1}, \ldots, W_{t,m})^{\T}$ and $\hat{\bW}_{t}^{(w)}=(\hat{W}_{t,1}^{(w)}, \ldots, \hat{W}_{t,m}^{(w)})^{\T}$. By \eqref{eq:fwh-fw-bound}, we have
\begin{align*}
	|{\rm H}_{1}(j,k)| \le&~    \frac{C_3Q}{n_{3}} \sum_{t\in \mathcal{D}_3}  |\hat{\bW}_{t}^{(w)} - \bW_{t}|_1  = \frac{C_3Q}{n_{3}} \sum_{t\in \mathcal{D}_3}    \sum_{s=1}^{m}|\hat{W}_{t,s}^{(w)} -W_{t,s}| \\
	\le &~  \frac{C_3Q}{n_{3}} \sum_{t\in \mathcal{D}_3} \bigg\{ \sum_{s=1}^{m}|\hat{W}_{t,s}^{(w)} -W_{t,s}^{*}| + \sum_{s=1}^{m}|W_{t,s}^{*} -W_{t,s}|\bigg\}\\
	=&~  \underbrace{\frac{C_3Q}{n_{3}} \sum_{t\in \mathcal{D}_3}  \sum_{s=1}^{m}|\hat{W}_{t,s}^{(w)} -W_{t,s}^{*}| I(|\bW_{t}|_{\infty} \le M_1)}_{\textrm{H}_{11}} \\
	&+ \underbrace{\frac{C_3Q}{n_{3}} \sum_{t\in \mathcal{D}_3}   \sum_{s=1}^{m}|\hat{W}_{t,s}^{(w)} -W_{t,s}^{*}| I(|\bW_{t}|_{\infty} > M_1) }_{\textrm{H}_{12}} +  \underbrace{\frac{C_3Q}{n_{3}} \sum_{t\in \mathcal{D}_3}   \sum_{s=1}^{m}|W_{t,s}^{*} -W_{t,s}| }_{\textrm{H}_{13}}\,.
\end{align*}
As we will show in Sections \ref{sec:sub-g211}--\ref{sec:sub-g213},
\begin{align}\label{eq:g211}
	|{\rm H}_{11}|  =&~  O_{\rm p}\{n_{1}^{-1/2}n_{3}^{-1/2}QmM_1^{1/2}\log^{1/2}(mn_1)\} \\
	& + O_{\rm p} \{n_1^{-1/2}n_{3}^{-1}Qme^{M_1^2/4}\log^{1/2}(mn_1)\}  + O_{\rm p}\{n_1^{-1/2}Qm\log^{1/2}(mn_1 )\} \notag
\end{align}
provided that $\log (mn_1) \ll n_1M_1^{-1}e^{-M_1^2/2}$, and
\begin{align}
	|{\rm H}_{12}| =&~ O_{\rm p}\{n_{3}^{-1/2}m^{3/2}QM_1^{-1/2}e^{-M_1^2/4}(\log n_{1})^{1/2}\}  + O_{\rm p}\{n_{3}^{-1}mQ(\log n_{1})^{1/2}\} \label{eq:g212} \\
	& + O_{\rm p}\{m^{2}Q M_1^{-1}e^{-M_1^2/2}(\log n_{1})^{1/2}\}\,,  \notag\\
	|{\rm H}_{13}| = &~O_{\rm p}(n_{3}^{-1/2}mQM_1^{1/2}e^{-M_1^2/4}) + O_{\rm p}\{n_{3}^{-1}mQ\log^{1/2}(n_{3}m)\} + O_{\rm p}(mQ e^{-M_1^2/2}) \label{eq:g213}\,.
\end{align}
Recall $M_1 = \sqrt{2\log n_{3}}$, $n_1\asymp n$ and $n_3\asymp n^{\kappa}$ for some constant $0<\kappa<1$. 
Combining with \eqref{eq:g211}--\eqref{eq:g213}, we have 
\begin{align*} 
	\max_{j\in[p],\,k\in[q]}|{\rm H}_{1}(j,k)| = O_{\rm p}\{n^{-\kappa}Qm^{2} \log^{1/2} (mn)\}  + O_{\rm p} \{ n^{-1/2} Qm\log^{1/2} (mn)\}  
\end{align*}  
provided that  $\log (mn) \ll n^{1-\kappa} (\log n)^{-1/2}$. Hence, \eqref{eq:g21} holds.
$\hfill\Box$

\subsubsection{Proof of \eqref{eq:g211}}\label{sec:sub-g211}
Recall $\hat{W}_{i,j}^{(w)} = \Phi^{-1} \{\hat{F}_{\bZ,j}^{(w)} (Z_{i,j})\}$,  $W_{i,j}=\Phi^{-1}\{F_{\bZ,j}(Z_{i,j})\}$ and $W_{i,j}^{*}=W_{i,j}I(|W_{i,j}|\le M_1) + M_1\cdot {\rm sign}(W_{i,j})I(|W_{i,j}|>M_1)$ with $M_1 = \sqrt{2\log n_{3}}$, where $\hat{F}_{\bZ,j}^{(w)}(Z_{i,j})$ is the truncated empirical distribution function defined in the same manner as  \eqref{eq:tructed-ecdf} based on the data in $\mathcal{W}_{\mathcal{D}_1}$.
Let $$K(W_{i,j}, m,n_{1}) = 4 n_{1}^{-1/2}  [\Phi(W_{i,j}) \{1-\Phi(W_{i,j})\}]^{1/2} \log^{1/2}(mn_{1})  + 7n_{1}^{-1} \log(mn_{1})\,.$$
Define the event 
\begin{align}\label{eq:h8-def}
	\mathcal{H}_{6} =  \bigcap_{i\in \mathcal{D}_3,\,j\in[m]}\big\{|\hat{F}_{\bZ,j}(Z_{i,j})-  F_{\bZ,j}(Z_{i,j})| \le K(W_{i,j}, m,n_{1})\big\} \,.
\end{align}
Restricted on $\mathcal{H}_{6}$,  given some constant $M_2 \in(0,M_1)$,  it holds that
\begin{align*}
	|{\rm H}_{11}| = &~ \frac{C_3Q}{n_{3}} \sum_{t\in \mathcal{D}_3}  \sum_{j=1}^{m}|\Phi^{-1} \{\hat{F}_{\bZ,j}^{(w)} (Z_{t,j})\} -\Phi^{-1}\{F_{\bZ,j}(Z_{t,j})\}|  I(|\bW_{t}|_{\infty} \le M_1)\\
	\le&~ \sum_{l=1}^{\infty} \frac{C_3Q}{n_{3}\cdot l!} \sum_{t\in \mathcal{D}_3}  \sum_{j=1}^{m} |(\Phi^{-1})^{(l)}\{F_{\bZ,j}(Z_{t,j})\}||\hat{F}_{\bZ,j}^{(w)} (Z_{t,j})- F_{\bZ,j}(Z_{t,j})|^{l}I(|\bW_{t}|_{\infty} \le M_1)\\
	\le&~ \frac{C_3Q}{n_{3}} \sum_{t\in \mathcal{D}_3}  \sum_{j=1}^{m} \sum_{l=1}^{\infty}C_{4}^{l} |W_{t,j}|^{l-1}e^{lW_{t,j}^2/2}\bigg[\bigg| \frac{\Phi(W_{t,j}) \{1-\Phi(W_{t,j})\}\log (mn_{1})}{n_{1}}\bigg|^{l/2} \\
	&~~~~~~~~~~~~~~~~~~~~~~~~~~~~~~~~~~~~~~~~~~~~~~~~~~~~~+   \bigg|\frac{\log(mn_{1})}{n_{1}} \bigg|^{l}\bigg]  I(|W_{t,j}| \le M_1)\\
	\le&~  \underbrace{\frac{C_5Q}{n_{3}} \sum_{t\in \mathcal{D}_3}  \sum_{j=1}^{m}\sum_{l=1}^{\infty}C_{6}^{l} |W_{t,j}|^{l-1}e^{lW_{t,j}^2/2}\bigg| \frac{\log (mn_{1})}{n_{1}}\bigg|^{l/2}  I(|W_{t,j}| \le M_2) }_{\textrm{H}_{111}}\\
	&+ \frac{C_5Q}{n_{3}} \sum_{t\in \mathcal{D}_3} \bigg[ \sum_{j=1}^{m}\sum_{l=1}^{\infty}C_{7}^{l} |W_{t,j}|^{l-1}e^{lW_{t,j}^2/2}\bigg| \frac{\Phi(W_{t,j}) \{1-\Phi(W_{t,j})\}\log (mn_{1})}{n_{1}}\bigg|^{l/2} \\
	&~~~\underbrace{~~~~~~~~~~~~~~~~~~~~~~~~\times I(M_2<|W_{t,j}| \le M_1) \bigg]~~~~~~~~~~~~~~~~~~~~~~~~~~~~~~~~~~}_{\textrm{H}_{112}}\\
	&+\underbrace{ \frac{C_5Q}{n_{3}} \sum_{t\in \mathcal{D}_3} \sum_{j=1}^{m}\sum_{l=1}^{\infty}C_{8}^{l} |W_{t,j}|^{l-1}e^{lW_{t,j}^2/2}\bigg| \frac{ \log (mn_{1})}{n_{1}}\bigg|^{l}  I(M_2<|W_{t,j}| \le M_1) }_{\textrm{H}_{113}}\,,
\end{align*}
where the third step holds using \eqref{eq:derivative_n} and the similar arguments for deriving  \eqref{eq:fwh-f-dec}, and  
the last step holds provided that $\log(mn_{1}) \lesssim n_{1}$. Notice that
\begin{align}\label{eq:h111-bound}
	|{\rm H}_{111}| 
	\le&~ \sum_{l=1}^{\infty} \bigg\{\frac{C_9M_{2}e^{M_2^2/2}\log^{1/2}(mn_{1})}{n_{1}^{1/2}}\bigg\}^{l-1}  \times \frac{C_5Q\log^{1/2} (mn_{1})}{n_{1}^{1/2} } \notag\\ &~~\times \frac{1}{n_{3}} \sum_{t\in \mathcal{D}_3}  \sum_{j=1}^{m} e^{W_{t,j}^2/2} I(|W_{t,j}| \le M_2) \notag\\
	\le&~ \frac{C_{10}Q  \log^{1/2}(mn_{1})}{n_{1}^{1/2} } \times \frac{1}{n_3}\sum_{t\in \mathcal{D}_3} \sum_{j=1}^{m} e^{W_{t,j}^2/4} I(|W_{t,j}| \le M_2)
\end{align}
provided that $\log (mn_{1}) \ll n_{1} $.  On the other hand, 
\begin{align}\label{eq:h112-bound}
	|{\rm H}_{112}| \le&~   \frac{C_{5}Q}{n_3}\sum_{t\in \mathcal{D}_3}  \sum_{j=1}^{m}\sum_{l=1}^{\infty}C_{11}^{l} |W_{t,j}|^{l/2-1}e^{lW_{t,j}^2/4}\bigg| \frac{\log (mn_{1})}{n_{1}}\bigg|^{l/2} I(M_2 < |W_{t,j}| \le M_1)\notag\\
	\le &~   \sum_{l=1}^{\infty}\bigg\{\frac{C_{12}M_{1}^{1/2}e^{M_1^2/4}\log^{1/2}(mn_{1})}{n_{1}^{1/2}}\bigg\}^{l-1} \times  \frac{C_{5}Q\log^{1/2} (mn_{1}) }{n_{1}^{1/2}M_{2}^{1/2} }\notag\\
	&~~~~~ \times\frac{1}{n_3}\sum_{t\in \mathcal{D}_3} \sum_{j=1}^{m} e^{W_{t,j}^2/4}I( M_2<|W_{t,j}| \le M_1) \notag\\
	\le&~  \frac{C_{13}Q\log^{1/2}(mn_1)}{n_1^{1/2}}\times \frac{1}{n_3}\sum_{t\in \mathcal{D}_3} \sum_{j=1}^{m}e^{W_{t,j}^2/4}I( |W_{t,j}| \le M_1)  
\end{align}
provided that $\log (mn_{1}) \ll n_{1}M_1^{-1}e^{-M_1^2/2}$,  where the first step is due to $1-\Phi(x) \le x^{-1} \phi(x)$ for $x>0$.
Furthermore, it holds that
\begin{align*}
	|{\rm H}_{113}| 
	\le&~  \sum_{l=1}^{\infty} \bigg\{\frac{C_{14}M_{1}e^{M_1^2/2}\log(mn_{1})}{n_{1}}\bigg\}^{l-1}  \times \frac{C_5Q\log (mn_{1}) }{n_{1}}  \times \frac{1}{n_{3}} \sum_{t\in \mathcal{D}_3}  \sum_{j=1}^{m} e^{W_{t,j}^2/2} I(|W_{t,j}| \le M_1)\\
	\le&~ \frac{C_{15}Q  \log^{1/2} (mn_{1})}{n_{1}^{1/2} } \times \frac{1}{n_3}\sum_{t\in \mathcal{D}_3}  \sum_{j=1}^{m} e^{W_{t,j}^2/4}  I( |W_{t,j}| \le M_1) 
\end{align*}
provided that $\log (mn_{1}) \ll n_{1}M_1^{-1}e^{-M_1^2/2}$. 
Together with \eqref{eq:h111-bound} and \eqref{eq:h112-bound}, restricted on $\mathcal{H}_{6}$, 
\begin{align}\label{eq:h11-bound}
	|{\rm H}_{11}| \le  |{\rm H}_{111}| + |{\rm H}_{112}| + |{\rm H}_{113}| \le \frac{C_{16}Q  \log^{1/2}(mn_{1})}{n_{1}^{1/2} } \times   \frac{1}{n_3}\sum_{t\in \mathcal{D}_3}\sum_{j=1}^{m}  e^{W_{t,j}^2/4} I( |W_{t,j}| \le M_1) 
\end{align}
provided that $\log (mn_{1}) \ll n_{1}M_1^{-1}e^{-M_1^2/2}$. 
Due to $W_{t,j}\sim \mathcal{N}(0,1)$, then  $\mathbb{E} \{ \sum_{j=1}^{m} e^{W_{t,j}^2/4} I( |W_{t,j}| \le M_1) \} \lesssim m$.
By Cauchy-Schwarz inequality, we have
\begin{align}\label{eq:sum-variance}
	\var\bigg\{ \sum_{j=1}^{m} e^{W_{t,j}^2/4}  I(|W_{t,j}| \le M_1)\bigg\} \le&~ \mathbb{E}\bigg[\bigg\{\sum_{j=1}^{m} e^{W_{t,j}^2/4} I(|W_{t,j}| \le M_1)\bigg\}^{2}\bigg]\notag\\
	=&~ \sum_{j=1}^{m}\mathbb{E}\big\{ e^{W_{t,j}^2/2} I(|W_{t,j}| \le M_1)\big\}\notag\\
	&+ \sum_{1\le j\ne k \le m} \mathbb{E}\big\{ e^{W_{t,j}^2/4}e^{W_{t,k}^2/4} I(|W_{t,j}|\le M_1) I(|W_{t,k}|\le M_1) \big\}\notag\\
	\lesssim &~m^2 M_1\,.
\end{align}
By Bonferroni inequality and Bernstein inequality, it holds that
\begin{align*}
	\bigg|\frac{1}{n_{3}}\sum_{t\in \mathcal{D}_3}  \sum_{j=1}^{m} e^{W_{t,j}^2/4} I(|W_{t,j}|\le M_1)\bigg| =  O_{\rm p}(n_{3}^{-1/2}mM_1^{1/2}) + O_{\rm p}(n_{3}^{-1}me^{M_1^2/4})   + O (m)\,.
\end{align*}
Analogous to \eqref{eq:h5-c}, we have $\mathbb{P}(\mathcal{H}_{6}^{\rm c}) \to 0$ as $n_1\to \infty$. Hence, applying the similar arguments in Section \ref{sec:sub-i12} for deriving the convergence rate of $\max_{j\in[p],\,k\in[q]}|{\rm I}_{12}(j,k)|$, by \eqref{eq:h11-bound}, we have
\begin{align*}
	|{\rm H}_{11}|  =&~  O_{\rm p}\{n_{1}^{-1/2}n_{3}^{-1/2}QmM_1^{1/2}\log^{1/2}(mn_{1})\} \\
	&+ O_{\rm p} \{n_{1}^{-1/2}n_{3}^{-1}Q me^{M_1^2/4}\log^{1/2}(mn_{1})\}  + O_{\rm p}\{n_{1}^{-1/2}Qm\log^{1/2}(mn_{1})\}
\end{align*}
provided that $\log (mn_{1}) \ll n_{1}M_1^{-1}e^{-M_1^2/2}$. Hence, \eqref{eq:g211} holds.
$\hfill\Box$

\subsubsection{Proof of \eqref{eq:g212} }\label{sec:sub-g212}
Analogous to the derivation of \eqref{eq:uh-w}, we can also show $\max_{i\in\mathcal{D}_3,\,j\in[m]}|\hat{W}_{i,j}^{(w)}| \le \sqrt{2\log n_{1}} $ for sufficiently large $n_1$. Recall $W_{i,j}^{*}=W_{i,j}I(|W_{i,j}|\le M_1) + M_1\cdot {\rm sign}(W_{i,j})I(|W_{i,j}|>M_1)$ with $M_1 = \sqrt{2\log n_{3}}$. Due to $n_1\asymp n$ and $n_3\asymp n^{\kappa}$ for some constant $0<\kappa<1$,  then
\begin{align}\label{eq:h12-dec}
	|{\rm H}_{12}| \le C_{17}Q m \sqrt{\log n_{1}}\times\frac{1}{n_{3}} \sum_{t\in \mathcal{D}_3}  I(|\bW_{t}|_{\infty} > M_1)\,.
\end{align}
Since $W_{i,j} \sim \mathcal{N}(0,1)$, then $\mathbb{E}\{I(|\bW_{t}|_{\infty} > M_1)\}   \le 2m M_{1}^{-1}e^{-M_1^2/2}$ and  $ \var\{I(|\bW_{t}|_{\infty} > M_1)\}   \le 2m M_{1}^{-1}e^{-M_1^2/2}$. By Bonferroni inequality and Bernstein inequality, it holds that
\begin{align*}
	\bigg|\frac{1}{n_{3}} \sum_{t\in \mathcal{D}_3}  I(|\bW_{t}|_{\infty} > M_1)\bigg| =  O_{\rm p}(n_{3}^{-1/2}m^{1/2}M_1^{-1/2}e^{-M_1^2/4}) + O_{\rm p}(n_{3}^{-1} )  + O_{\rm p}(m M_1^{-1}e^{-M_1^2/2})\,.
\end{align*}
Hence, by \eqref{eq:h12-dec}, we have \eqref{eq:g212} holds.
$\hfill\Box$

\subsubsection{ Proof of \eqref{eq:g213} }\label{sec:sub-g213}
Recall $W_{i,j}^{*}=W_{i,j}I(|W_{i,j}|\le M_1) + M_1\cdot {\rm sign}(W_{i,j})I(|W_{i,j}|>M_1)$ with $M_1 = \sqrt{2\log n_{3}}$. Given $Q_1>M_1$, we have
\begin{align*}
	{\rm H}_{13} =&~ \underbrace{\frac{C_3Q}{n_{3}} \sum_{t\in \mathcal{D}_3}  \sum_{j=1}^{m}|W_{t,j}^{*} -W_{t,j}| I(|\bW_{t}|_{\infty} \le Q_1)}_{\textrm{H}_{131} } \\
	&+\underbrace{\frac{C_3Q}{n_{3}} \sum_{t\in \mathcal{D}_3}  \sum_{j=1}^{m}|W_{t,j}^{*} -W_{t,j}| I(|\bW_{t}|_{\infty} > Q_1) }_{\textrm{H}_{132} } \,.
\end{align*}
Since $W_{i,j} \sim \mathcal{N}(0,1)$ and $|W_{i,j}^{*}-W_{i,j}|\le  |W_{i,j}|I(|W_{i,j}|> M_1)$, using the similar arguments for the derivation of \eqref{eq:sum-variance}, it holds that
\begin{align*}
	&~~~~~~\mathbb{E}\bigg\{ \sum_{j=1}^{m}|W_{t,j}^{*} -W_{t,j}| I(|\bW_{t}|_{\infty} \le Q_1)\bigg\} \le   \mathbb{E}\bigg(\sum_{j=1}^{m}|W_{t,j}^{*} -W_{t,j}|\bigg) \lesssim m e^{-M_1^2/2}\,,\\
	& \var\bigg\{ \sum_{j=1}^{m}|W_{t,j}^{*} -W_{t,j}| I(|\bW_{t}|_{\infty} \le Q_1) \bigg\} \le   \mathbb{E}\bigg\{\bigg(\sum_{j=1}^{m}|W_{t,j}^{*} -W_{t,j}|\bigg)^2\bigg\} \lesssim m^2M_1 e^{-M_1^2/2}\,.
\end{align*}
By Bonferroni inequality and Bernstein inequality, we have
\begin{align*}
	|{\rm H}_{131}| = O_{\rm p}(n_{3}^{-1/2}mQM_1^{1/2}e^{-M_1^2/4})  + O_{\rm p}(n_{3}^{-1}mQQ_1) + O_{\rm p}(mQ e^{-M_1^2/2})\,.
\end{align*}
Furthermore, it holds that  
\begin{align*}
	\mathbb{P}({\rm H}_{132}> x)  \le n_{3}\max_{t\in \mathcal{D}_3}\mathbb{P}(|\bW_{t}|_{\infty} > Q_1) \le n_{3}m \max_{t\in \mathcal{D}_3,\,j\in[m]} \mathbb{P}(|W_{t,j}|> Q_1) \le n_{3}m e^{-Q_1^2/2}  
\end{align*}
for any $x>0$. We have 
\begin{align*}
	|{\rm H}_{132}|= o_{\rm p}(n_{3}^{-1}) 
\end{align*}
provided that $\log(n_{3}m)\lesssim Q_1^2$. Selecting $Q_1=\tilde{C}\log^{1/2}(n_{3}m)$ for some  sufficiently large constant $\tilde{C}>0$, it holds that
\begin{align*}
	|{\rm H}_{13}| = &~O_{\rm p}(n_{3}^{-1/2}mQM_1^{1/2}e^{-M_1^2/4})  + O_{\rm p}\{n_{3}^{-1}mQ\log^{1/2}(n_{3}m)\} + O_{\rm p}(mQ e^{-M_1^2/2})\,.
\end{align*}
We then have \eqref{eq:g213} holds.
$\hfill\Box$

\subsection{Proof of Lemma \ref{lem:fhwh-fwh}}\label{sec:sub-fhwh-fwh}
To prove Lemma \ref{lem:fhwh-fwh}, we need Lemma \ref{lem:network-hiera}, the proof of which is given in Section \ref{sec:sub-network-hiera}.

\begin{lem}\label{lem:network-hiera}
	Let $\bX \in [-a_n,a_n]^{m}$ be a random vector, and  $f \in \mathcal{F}(m,m_{*}, \ell,K,\vartheta,  L,C, \tilde{C})$ with  the parameters $(m,m_{*}, \ell,K,\vartheta, L,C, \tilde{C})$   specified in Lemma {\rm\ref{lem:fhwh-fwh}}, where  $\vartheta= \tilde{\vartheta} + s$ for some  $\tilde{\vartheta} \in \mathbb{N}_{0}$ and $s\in(0,1]$. Select $N\in \mathbb{N}_{0}$ such that $ N \ge \tilde{\vartheta}$.  Let $M_{n} \in \mathbb{N}$ and $ a_{n} \in [1,M_{n}]$ be increasing such that $ m^{2N+3}a_{n}^{2N+3}  \ll M_{n}^{\vartheta} $.  For any $c>0$ and $\eta_{n} \in(0,1)$, let $\mathcal{H}^{(\ell)}$ be defined in \eqref{eq:nnspace-hl} with $(K,m,m_{*})$ as in  the definition of $f $, $M_{*}=(N+1)(M_{n}+1)^{m_{*}} \cdot {\rm C}_{m_{*}+N}^{m_{*}}$ and 
	$\tilde{\alpha}_{n} = \bar{C}(c\eta_{n})^{-1} m^{\tilde{\vartheta}} M_{n}^{m_{*}+2+\vartheta(2N+3)}$  for some sufficiently large constant $\bar{C}>0$. For all $n$ greater than a certain $n_0(c) \in \mathbb{N}$,  there exists a neural network $t \in \{t\in \mathcal{H}^{(\ell)}: |t|_{\infty,[-a_{n},a_{n}]^{m}\backslash \boldsymbol{H}} \le \tilde{\beta}_{n}\}$ such that     
	\begin{align*}
		|t(\boldsymbol{x})-f(\boldsymbol{x})| \le \breve{C}_{1} M_{n}^{-\vartheta} m^{2N+3}a_{n}^{2N+3} \,,  ~~~~\boldsymbol{x} \in  [-a_n,a_n]^{m} \backslash \boldsymbol{H}
	\end{align*}  
	holds with    $\tilde{\beta}_n = (\log n)\log^{1/2}(\tilde{d}n)$ and  $\mathbb{P}(\bX \in \boldsymbol{H}) \le c\eta_n$,	where $\boldsymbol{H} \subset [-a_n,a_n]^{m}$ and $\breve{C}_1>0$ is a universal constant only depending on $(m_{*}, N)$.  
\end{lem}

Given $Q>0$, it holds that
\begin{align*}
	\frac{1}{n_{3}} \sum_{t\in \mathcal{D}_3} \{\hat{f}_{j} (\hat{\bW}_{t}^{(w)}) - f_{j} (\hat{\bW}_{t}^{(w)})\}\delta_{t,k}  = &~  \underbrace{\frac{1}{n_{3}} \sum_{t\in \mathcal{D}_3} \{\hat{f}_{j} (\hat{\bW}_{t}^{(w)}) - f_{j} (\hat{\bW}_{t}^{(w)})\}\delta_{t,k} I(|\delta_{t,k}| \le Q) }_{\textrm{G}_{1}(j,k)} \\
	&+  \underbrace{\frac{1}{n_{3}} \sum_{t\in \mathcal{D}_3} \{\hat{f}_{j} (\hat{\bW}_{t}^{(w)}) - f_{j} (\hat{\bW}_{t}^{(w)})\}\delta_{t,k} I(|\delta_{t,k}| > Q)}_{\textrm{G}_{2}(j,k)}\,.
\end{align*}
Recall $\tilde{d}=p\vee q\vee m$. Analogous to the derivation of the convergence rate of $\max_{j\in[p],\,k\in[q]}|\tilde{{\rm H}}_{4}(j,k)|$ in Section \ref{sec:sub-uh-u-delta} for the proof of Lemma \ref{lem:uh-u-delta}, it holds that
\begin{align*}
	\max_{j\in[p],\,k\in[q]}|{\rm G}_{2}(j,k)|  = o_{\rm p}(n^{-1}) 
\end{align*}
provided that $\log (\tilde{d} n) \lesssim Q^2$. As we will show in Section \ref{sec:sub-g1},  
\begin{align}\label{eq:g1}
	\max_{j\in[p],\,k\in[q]}|{\rm G}_1(j,k)| = &~ O_{\rm p}\{n^{-\kappa/2-\vartheta/(4\vartheta+m_{*})} Q (m^2\log n)^{(\vartheta+2m_{*}\tilde{\vartheta}+3m_{*} )/(8\vartheta)} \tilde{\beta}_n\log^{3/4}(\tilde{d}n) \}  \notag \\
	& +  O_{\rm p}\{n^{-\kappa/2-1/4} Qm^{1/2}  \tilde{\beta}_n^{1/2}\log^{3/4}  (\tilde{d}n) \} \\
	&+ O_{\rm p}\{n^{-\kappa}m Q  \tilde{\beta}_n\log  (\tilde{d}n) \} +  O( \tilde{\beta}_nQ  e^{-\breve{c}Q^2} ) \notag
\end{align}  
provided that $ \log (\tilde{d}n) \ll n^{1-\kappa}(\log n)^{-1/2}$ and $m \lesssim n$, where $\breve{c}>0$ is a universal constant.  Recall $\tilde{\beta}_n = (\log n)\log^{1/2}(\tilde{d}n)$.  By selecting $Q=\bar{C}\log^{1/2}(\tilde{d}n)$ for some sufficiently large constant $\bar{C}> 0$, we have  
\begin{align*}
	&\max_{j\in[p],\,k\in[q]}\bigg|	\frac{1}{n_{3}} \sum_{t\in \mathcal{D}_3} \{\hat{f}_{j} (\hat{\bW}_{t}^{(w)}) - f_{j} (\hat{\bW}_{t}^{(w)})\}\delta_{t,k}\bigg|\\
	&~~~~~~= O_{\rm p}\{n^{-\kappa/2-\vartheta/(4\vartheta+m_{*})}  (m^2\log n)^{(\vartheta+2m_{*}\tilde{\vartheta}+3m_{*} )/(8\vartheta)} (\log n)\log^{7/4}(\tilde{d}n) \}  \notag \\
	&~~~~~~~~ +  O_{\rm p}\{n^{-\kappa/2-1/4} m^{1/2}  (\log n)^{1/2}\log^{3/2}  (\tilde{d}n) \}  + O_{\rm p}\{n^{-\kappa}m (\log n)\log^2(\tilde{d}n) \}
\end{align*}   
provided that  $ \log (\tilde{d}n) \ll n^{1-\kappa}(\log n)^{-1/2}$ and $m \lesssim n$.   Analogously, we can also show such convergence rate also holds for $\max_{j\in[p],\,k\in[q]}|n_{3}^{-1} \sum_{t\in \mathcal{D}_3} \{\hat{g}_{k} (\hat{\bW}_{t}^{(w)}) - g_{k} (\hat{\bW}_{t}^{(w)})\}\varepsilon_{t,j}|$.
Hence, we complete the proof of Lemma \ref{lem:fhwh-fwh}.  
$\hfill\Box$

\subsubsection{Convergence rate of $\max_{j\in[p],\, k\in[q]}|{\rm G}_{1}(j,k)|$}\label{sec:sub-g1}
Recall $\mathcal{W}_{\mathcal{D}_j}=\{(\bX_i,\bY_i,\bZ_i): i\in \mathcal{D}_j\}$ for $j\in[3]$, where $\mathcal{D}_1, \mathcal{D}_2$ and $\mathcal{D}_3$ are three disjoint subsets of $[n]$ with $|\mathcal{D}_1|=n_1\asymp n$, $|\mathcal{D}_2|=n_2\asymp n$ and $|\mathcal{D}_3|=n_3\asymp n^{\kappa}$  for some constant $0<\kappa<1$ and $n_1+n_2+n_3=n$. 
Notice that $\hat{\bW}_{t}^{(w)}=(\hat{W}_{t,1}^{(w)},\ldots, \hat{W}_{t,m}^{(w)})^{\T}$ with $\hat{W}_{t,j}^{(w)}=\Phi^{-1}\{\hat{F}_{\bZ,j}^{(w)}(Z_{t,j})\}$, where $\hat{F}_{\bZ,j}^{(w)}(Z_{t,j})$ is the truncated empirical distribution function defined in the same manner as \eqref{eq:tructed-ecdf} based on the data in $\mathcal{W}_{\mathcal{D}_1}$. 
For any $t\in \mathcal{D}_3$, define
\begin{align*}
	&~~~~~~\tilde{\mu}_{1,j} = \mathbb{E}\big[\{\hat{f}_{j} (\hat{\bW}_{t}^{(w)}) - f_{j} (\hat{\bW}_{t}^{(w)})\}\delta_{t,k}I(|\delta_{t,k}|\le Q) \,|\, \mathcal{W}_{\mathcal{D}_1}, \mathcal{W}_{\mathcal{D}_2} \big] \,, \\
	&\tilde{\sigma}^2_{1,j}  =  \mathbb{E}  \big\{ \big[\{\hat{f}_{j} (\hat{\bW}_{t}^{(w)}) - f_{j} (\hat{\bW}_{t}^{(w)})\}\delta_{t,k} I(|\delta_{t,k}| \le Q)   -\tilde{\mu}_{1,j}  \big]^2\,|\, \mathcal{W}_{\mathcal{D}_1}, \mathcal{W}_{\mathcal{D}_2} \big\}\,.
\end{align*}   
Due to  $W_{t,j}=\Phi^{-1}\{F_{\bZ,j}(Z_{t,j})\}$ and $\mathbb{E}(\delta_{t,k}\,|\, \bW_{t})=0$, then $\mathbb{E}(\delta_{t,k}\,|\, \bZ_{t})=0$.  Notice that $\hat{f}_{j}$ is specified in \eqref{eq:fhj-ghk} based on the data in $\mathcal{W}_{\mathcal{D}_1}\cup \mathcal{W}_{\mathcal{D}_2}$.
Since $\mathcal{W}_{\mathcal{D}_1}$, $\mathcal{W}_{\mathcal{D}_2}$ and $\mathcal{W}_{\mathcal{D}_3}$ are independent, for any $t\in\mathcal{D}_3$,  we have
\begin{align*}
	&  \mathbb{E}\big[\{\hat{f}_{j} (\hat{\bW}_{t}^{(w)}) - f_{j} (\hat{\bW}_{t}^{(w)})\}\delta_{t,k}\, |\, \mathcal{W}_{\mathcal{D}_1}, \mathcal{W}_{\mathcal{D}_2}\big] \\
	&~~~~~~~~~~=  \mathbb{E}\big[\{\hat{f}_{j} (\hat{\bW}_{t}^{(w)}) - f_{j} (\hat{\bW}_{t}^{(w)})\}\times \mathbb{E}(\delta_{t,k}\,|\, \mathcal{W}_{\mathcal{D}_1}, \mathcal{W}_{\mathcal{D}_2}, \bZ_{t})\, |\, \mathcal{W}_{\mathcal{D}_1}, \mathcal{W}_{\mathcal{D}_2}\big] \\
	&~~~~~~~~~~=   \mathbb{E}\big[\{\hat{f}_{j} (\hat{\bW}_{t}^{(w)}) - f_{j} (\hat{\bW}_{t}^{(w)})\} \times   \mathbb{E}(\delta_{t,k}\,|\, \bZ_{t})   \, |\, \mathcal{W}_{\mathcal{D}_1}, \mathcal{W}_{\mathcal{D}_2}\big] =0\,.
\end{align*}
Recall $\max_{t\in \mathcal{D}_3,\,j\in[m]}|\hat{W}_{t,j}^{(w)}|\le \sqrt{2\log n_1}$.
Since $\max_{t\in\mathcal{D}_3,\,j\in[p]}|\hat{f}_{j}(\hat{\bW}_{t}^{(w)})|  \le  \tilde{\beta}_{n}$ and $|f_j|_{\infty}  \le \tilde{C}$ with $\tilde{\beta}_{n} =(\log n)\log^{1/2}(\tilde{d}n)$ and $\tilde{C}$ specified in Condition \ref{cd:function-condition}, by \eqref{eq:delta-tail-nn}, for any $t\in \mathcal{D}_3$, we have
\begin{align}\label{eq:mu-z-bound}
	|\tilde{\mu}_{1,j} | =&~  \big|\mathbb{E}\big[\{\hat{f}_{j} (\hat{\bW}_{t}^{(w)}) - f_{j} (\hat{\bW}_{t}^{(w)})\}\delta_{t,k}I(|\delta_{t,k}|> Q) \,|\, \mathcal{W}_{\mathcal{D}_1}, \mathcal{W}_{\mathcal{D}_2}\big]\big| \notag\\
	\le&~   2\tilde{\beta}_{n} \mathbb{E}\{|\delta_{t,k} | I(|\delta_{t,k}|> Q) \} \notag\\
	\le &~ 2\tilde{\beta}_{n} \bigg\{Q\mathbb{P}(|\delta_{t,k}|>Q) +   \int_{Q}^{\infty}   \mathbb{P}(|\delta_{t,k}|>x) \,{\rm d}x \bigg\} \notag\\
	\le  &~  C_1 \tilde{\beta}_{n} Q  e^{-\breve{c}Q^2} 
\end{align}
for sufficiently large $n$, where $\breve{c}>0$ is a universal constant.  Furthermore,  
\begin{align}\label{eq:sigma-t1}
	\tilde{\sigma}^2_{1,j} \le&~  \mathbb{E} \big[\{\hat{f}_{j} (\hat{\bW}_{t}^{(w)}) - f_{j} (\hat{\bW}_{t}^{(w)})\}^2\delta_{t,k}^2 I(|\delta_{t,k}| \le Q)   \,|\, \mathcal{W}_{\mathcal{D}_1}, \mathcal{W}_{\mathcal{D}_2} \big]\notag\\
	\le&~   Q^2\mathbb{E} \big[\{\hat{f}_{j} (\hat{\bW}_{t}^{(w)}) - f_{j} (\hat{\bW}_{t}^{(w)})\}^2\,|\, \mathcal{W}_{\mathcal{D}_1}, \mathcal{W}_{\mathcal{D}_2} \big] \,.
\end{align}
Recall $\mathcal{H}^{(\ell)}$ defined in \eqref{eq:nnspace-hl} and  $\mathbb{E}(\varepsilon_{t,j}\,|\,\bW_{t})=0$ with $\varepsilon_{t,j}=U_{t,j}-f_{j}(\bW_{t})$.
Due to  $W_{t,j}=\Phi^{-1}\{F_{\bZ,j}(Z_{t,j})\}$, then $\mathbb{E}(\varepsilon_{t,j}\,|\, \bZ_{t})=0$. For any $t\in\mathcal{D}_3$, it holds that
\begin{align*}
	&\mathbb{E}  \big[\{\hat{f}_{j} (\hat{\bW}_{t}^{(w)}) - f_{j} (\hat{\bW}_{t}^{(w)})\}^2 \,|\, \mathcal{W}_{\mathcal{D}_1}, \mathcal{W}_{\mathcal{D}_2}\big]\\
	&~~~~~~~ =  \mathbb{E} \big[\{\hat{f}_{j} (\hat{\bW}_{t}^{(w)}) - \hat{U}_{t,j}^{(w)}\}^2 \,|\, \mathcal{W}_{\mathcal{D}_1}, \mathcal{W}_{\mathcal{D}_2}\big]  - \mathbb{E} \big[\{f_{j} (\hat{\bW}_{t}^{(w)}) - \hat{U}_{t,j}^{(w)}\}^2 \,|\, \mathcal{W}_{\mathcal{D}_1}, \mathcal{W}_{\mathcal{D}_2}\big]\\
	&~~~~~~~~~~- 2 \mathbb{E}  \big[\{\hat{f}_{j} (\hat{\bW}_{t}^{(w)}) - f_{j} (\hat{\bW}_{t}^{(w)})\} \{f_{j} (\hat{\bW}_{t}^{(w)}) - \hat{U}_{t,j}^{(w)}\} \,|\, \mathcal{W}_{\mathcal{D}_1}, \mathcal{W}_{\mathcal{D}_2}\big]\\
	&~~~~~~~=   \mathbb{E}  \big[\{\hat{f}_{j} (\hat{\bW}_{t}^{(w)}) - \hat{U}_{t,j}^{(w)}\}^2 \,|\, \mathcal{W}_{\mathcal{D}_1}, \mathcal{W}_{\mathcal{D}_2} \big] - \inf_{h\in T_{\tilde{\beta}_n} \mathcal{H}^{(\ell)}} \mathbb{E} \big[\{h (\hat{\bW}_{t}^{(w)}) - \hat{U}_{t,j}^{(w)}\}^2 \,|\, \mathcal{W}_{\mathcal{D}_1} \big]\\
	&~~~~~~~~~~+ \inf_{h\in T_{\tilde{\beta}_n} \mathcal{H}^{(\ell)}} \mathbb{E}  \big[\{h (\hat{\bW}_{t}^{(w)}) - \hat{U}_{t,j}^{(w)}\}^2 \,|\, \mathcal{W}_{\mathcal{D}_1} \big] - \mathbb{E}  \big[\{f_{j} (\hat{\bW}_{t}^{(w)}) - \hat{U}_{t,j}^{(w)}\}^2 \,|\, \mathcal{W}_{\mathcal{D}_1}, \mathcal{W}_{\mathcal{D}_2}\big]\\
	&~~~~~~~~~~- 2 \mathbb{E}  \big[\{\hat{f}_{j} (\hat{\bW}_{t}^{(w)}) - f_{j} (\hat{\bW}_{t}^{(w)})\} \{f_{j} (\hat{\bW}_{t}^{(w)}) - \hat{U}_{t,j}^{(w)}\} \,|\, \mathcal{W}_{\mathcal{D}_1}, \mathcal{W}_{\mathcal{D}_2}\big]\\
	&~~~~~~~= \mathbb{E} \big[\{\hat{f}_{j} (\hat{\bW}_{t}^{(w)}) - \hat{U}_{t,j}^{(w)}\}^2 \,|\, \mathcal{W}_{\mathcal{D}_1}, \mathcal{W}_{\mathcal{D}_2}\big] - \inf_{h\in T_{\tilde{\beta}_n} \mathcal{H}^{(\ell)}} \mathbb{E}  \big[\{h (\hat{\bW}_{t}^{(w)}) - \hat{U}_{t,j}^{(w)}\}^2 \,|\, \mathcal{W}_{\mathcal{D}_1} \big]\\
	&~~~~~~~~~~+  \inf_{h\in T_{\tilde{\beta}_n} \mathcal{H}^{(\ell)}} \bigg(\mathbb{E}  \big[\{h (\hat{\bW}_{t}^{(w)}) - f_{j} (\hat{\bW}_{t}^{(w)}) \}^2 \,|\, \mathcal{W}_{\mathcal{D}_1} \big]\\
	&~~~~~~~~~~~~~~~~~~~~~~~~~~~~~~  + 2\mathbb{E} \big[\{h (\hat{\bW}_{t}^{(w)}) - f_{j} (\hat{\bW}_{t}^{(w)})\} \{f_{j} (\hat{\bW}_{t}^{(w)}) - \hat{U}_{t,j}^{(w)}\} \,|\, \mathcal{W}_{\mathcal{D}_1}, \mathcal{W}_{\mathcal{D}_2} \big] \bigg)\\
	&~~~~~~~~~~- 2 \mathbb{E} \big[\{\hat{f}_{j} (\hat{\bW}_{t}^{(w)}) - f_{j} (\hat{\bW}_{t}^{(w)})\} \{f_{j} (\hat{\bW}_{t}^{(w)}) - \hat{U}_{t,j}^{(w)}\} \,|\, \mathcal{W}_{\mathcal{D}_1}, \mathcal{W}_{\mathcal{D}_2}\big]\\
	&~~~~~~~ \le \underbrace{\inf_{h\in T_{\tilde{\beta}_n} \mathcal{H}^{(\ell)}} \mathbb{E}  \big[\{h (\hat{\bW}_{t}^{(w)}) - f_{j} (\hat{\bW}_{t}^{(w)}) \}^2 \,|\, \mathcal{W}_{\mathcal{D}_1} \big] }_{\textrm{G}_{11}(j)}\\
	&~~~~~~~~~~+ \underbrace{\mathbb{E}  \big[\{\hat{f}_{j} (\hat{\bW}_{t}^{(w)}) - \hat{U}_{t,j}^{(w)}\}^2 \,|\, \mathcal{W}_{\mathcal{D}_1}, \mathcal{W}_{\mathcal{D}_2} \big] - \inf_{h\in T_{\tilde{\beta}_n} \mathcal{H}^{(\ell)}} \mathbb{E} \big[\{h (\hat{\bW}_{t}^{(w)}) - \hat{U}_{t,j}^{(w)}\}^2 \,|\, \mathcal{W}_{\mathcal{D}_1} \big]}_{ \textrm{G}_{12}(j) } \\
	&~~~~~~~~~~- 2\underbrace{ \mathbb{E}  \big[\{\hat{f}_{j} (\hat{\bW}_{t}^{(w)}) - f_{j} (\hat{\bW}_{t}^{(w)})\} \{f_{j} (\hat{\bW}_{t}^{(w)}) - f_{j} (\bW_{t}) + U_{t,j}- \hat{U}_{t,j}^{(w)}\} \,|\, \mathcal{W}_{\mathcal{D}_1}, \mathcal{W}_{\mathcal{D}_2}\big] }_{\textrm{G}_{13}(j)}\\
	&~~~~~~~~~~+ 2  \sup_{h\in T_{\tilde{\beta}_n} \mathcal{H}^{(\ell)} } \mathbb{E} \big|\big[\{h (\hat{\bW}_{t}^{(w)}) - f_{j} (\hat{\bW}_{t}^{(w)})\} \\
	&~~~~~~~~~~~~~~~~~\underbrace{ ~~~~~~~~~~~~~~~ \times  \{f_{j} (\hat{\bW}_{t}^{(w)}) - f_{j} (\bW_{t}^{(w)}) + U_{t,j} - \hat{U}_{t,j}^{(w)}\} \,|\, \mathcal{W}_{\mathcal{D}_1}, \mathcal{W}_{\mathcal{D}_2} \big] \big| }_{\textrm{G}_{14}(j)} \,,
\end{align*}
where the last step is due to 
\begin{align*}
	& \mathbb{E} \big[\{\hat{f}_{j} (\hat{\bW}_{t}^{(w)}) - f_{j} (\hat{\bW}_{t}^{(w)})\} \{f_{j} (\bW_{t}) - U_{t,j}\} \, |\,  \mathcal{W}_{\mathcal{D}_1}, \mathcal{W}_{\mathcal{D}_2}\big]\\
	&~~~~~~= \mathbb{E} \big[\{\hat{f}_{j} (\hat{\bW}_{t}^{(w)}) - f_{j} (\hat{\bW}_{t}^{(w)})\} \times\mathbb{E}\{f_{j} (\bW_{t}) - U_{t,j}\,\big|\,   \mathcal{W}_{\mathcal{D}_1}, \mathcal{W}_{\mathcal{D}_2} ,\bZ_{t}\} \, |\,  \mathcal{W}_{\mathcal{D}_1}, \mathcal{W}_{\mathcal{D}_2}\big]\\
	&~~~~~~= \mathbb{E} \big[\{\hat{f}_{j} (\hat{\bW}_{t}^{(w)}) - f_{j} (\hat{\bW}_{t}^{(w)})\} \times\mathbb{E} ( \varepsilon_{t,j} \,|\,\bZ_{t} ) \, |\, \mathcal{W}_{\mathcal{D}_1}, \mathcal{W}_{\mathcal{D}_2}\big] =0 
\end{align*}
for any $t\in \mathcal{W}_{\mathcal{D}_3}$, and $\mathbb{E} [\{h (\hat{\bW}_{t}^{(w)}) - f_{j} (\hat{\bW}_{t}^{(w)})\} \{ f_{j} (\bW_{t}) -U_{t,j} \} \,|\, \mathcal{W}_{\mathcal{D}_1}, \mathcal{W}_{\mathcal{D}_2}]=0$ for any $t\in \mathcal{W}_{\mathcal{D}_3}$ analogously.
As we will show in Sections \ref{sec:sub-g12-bound}--\ref{sec:sub-g13-tail}, for  some sufficiently large constants $\tilde{C}_1, \tilde{C_2}, \tilde{C}_3>0$,
\begin{align}
	&~~~~~~~~~~~~~~~~~~~~~~~\mathbb{P}\bigg\{\max_{j\in[p]} {\rm G}_{11}(j)  > \frac{\tilde{C}_1}{n^{2\vartheta/(4\vartheta+m_{*})}}   \bigg\} \lesssim n^{-1}  \label{eq:g12-bound} \,,\\  
	&\mathbb{P}\bigg\{ \max_{j\in[p]}{\rm G}_{12}(j) > \frac{ \tilde{C}_2(m^2\log n)^{(\vartheta+2m_{*}\tilde{\vartheta}+3m_{*} )/(4\vartheta)}  \tilde{\beta}_{n}^2\log^{1/2} (\tilde{d}n)}{n^{ 2\vartheta /(4\vartheta+m_{*})}}  \bigg\} \lesssim  (\tilde{d}n)^{-2} \label{eq:g11-tail}
\end{align}
provided that $\tilde{\beta}_n \ll n$ and $m\lesssim n$, and 
\begin{align}
	&\mathbb{P}\bigg\{\max_{j\in[p]}|{\rm G}_{13}(j)| > \frac{\tilde{C}_3m\tilde{\beta}_{n}\log^{1/2}(\tilde{d}n)}{n^{1/2}} + \frac{\tilde{C}_3m^2 \tilde{\beta}_{n}}{n^{\kappa}} \bigg\} \lesssim  (\tilde{d}n)^{-2}\,, \label{eq:g13-tail}\\
	&\mathbb{P}\bigg\{\max_{j\in[p]}|{\rm G}_{14}(j)| > \frac{\tilde{C}_3m\tilde{\beta}_{n}\log^{1/2}(\tilde{d}n)}{n^{1/2}} + \frac{\tilde{C}_3m^2 \tilde{\beta}_{n}}{n^{\kappa}} \bigg\} \lesssim  (\tilde{d}n)^{-2} \,, \label{eq:g14-tail}
\end{align}
provided that $ \log (\tilde{d}n) \ll n^{1-\kappa}(\log n)^{-1/2}$.  Let
\begin{align*}
	K(n, m ,\tilde{d}) = \bigg\{ \frac{ \tilde{C}_4(m^2\log n)^{(\vartheta+2m_{*}\tilde{\vartheta}+3m_{*} )/(4\vartheta)}  \tilde{\beta}_{n}^2\log^{1/2} (\tilde{d}n)}{n^{ 2\vartheta /(4\vartheta+m_{*})}}   + \frac{\tilde{C}_4m\tilde{\beta}_{n}\log^{1/2}(\tilde{d}n)}{n^{1/2}}+\frac{\tilde{C}_4m^2 \tilde{\beta}_n }{n^{\kappa}}\bigg\}^{1/2} 
\end{align*}  
for some sufficiently large constant $\tilde{C}_4> 0$.  
Recall $\tilde{\beta}_{n} = (\log n)\log^{1/2}(\tilde{d}n) $.  Due to  $\mathbb{E} [\{\hat{f}_{j} (\hat{\bW}_{t}^{(w)}) - f_{j} (\hat{\bW}_{t}^{(w)})\}^2\, |\, \mathcal{W}_{\mathcal{D}_1}, \mathcal{W}_{\mathcal{D}_2} ] \le {\rm G}_{11}(j) + {\rm G}_{12}(j) + 2|{\rm G}_{13}(j)| + 2|{\rm G}_{14}(j)|$, it holds that
\begin{align}\label{eq:sigma2-z}
	& \mathbb{P}\bigg(\max_{j\in[p]}\mathbb{E} \big[\{\hat{f}_{j} (\hat{\bW}_{t}^{(w)}) - f_{j} (\hat{\bW}_{t}^{(w)})\}^2\, |\, \mathcal{W}_{\mathcal{D}_1}, \mathcal{W}_{\mathcal{D}_2} \big] >  K^2(n, m ,\tilde{d}) \bigg) \notag\\
	&~~~~~~~~~\le  \mathbb{P}\bigg\{\max_{j\in[p]}  {\rm G}_{11}(j)  >  \frac{K^2(n, m ,\tilde{d})}{6} \bigg\}  + \mathbb{P}\bigg\{\max_{j\in[p]}{\rm G}_{12}(j) >  \frac{K^2(n, m ,\tilde{d})}{6}\bigg\} \notag\\
	&~~~~~~~~~~~~~~~+ \mathbb{P}\bigg\{\max_{j\in[p]}|{\rm G}_{13}(j)| >  \frac{K^2(n, m ,\tilde{d})}{6}\bigg\}  + \mathbb{P}\bigg\{\max_{j\in[p]}|{\rm G}_{14}(j)| >  \frac{K^2(n, m ,\tilde{d})}{6}\bigg\} \notag\\
	&~~~~~~~~~\lesssim  n^{-1 } 
\end{align} 
provided that  $ \log (\tilde{d}n) \ll n^{1-\kappa}(\log n)^{-1/2}$ and $m\lesssim n$. Consider the event
\begin{align*}
	\mathcal{G}_1=\bigg\{\max_{j\in[p]}\tilde{\sigma}_{1,j}^2 \le Q^2 K^2(n, m ,\tilde{d})\bigg\}\,.
\end{align*}   By  \eqref{eq:sigma-t1},  
\begin{align*}
	\mathbb{P}(\mathcal{G}_1^{\rm c}) =\mathbb{P}\bigg\{\max_{j\in[p]}\tilde{\sigma}_{1,j}^{2}  > Q^2K^2(n, m ,\tilde{d}) \bigg\} \lesssim   n^{-1 }
\end{align*}
provided that   $ \log (\tilde{d}n) \ll n^{1-\kappa}(\log n)^{-1/2}$ and $m\lesssim n$. Recall $\tilde{d}=p\vee q\vee m$, $n_3\asymp n^{\kappa}$ for some constant $0<\kappa<1$,    $\max_{t\in \mathcal{D}_3,\,j\in[p]}|\hat{f}_{j}(\hat{\bW}_{t}^{(w)})|  \le  \tilde{\beta}_{n}$ and $ | f_{j}|_{\infty} \le \tilde{C}$. 
By Bonferroni inequality and Bernstein inequality, for any $x>0$,  we have 
\begin{align*}
	&\mathbb{P}\bigg(\max_{j\in[p]}\bigg|\frac{1}{n_{3}} \sum_{t\in \mathcal{D}_3} \big[\{\hat{f}_{j} (\hat{\bW}_{t}^{(w)}) - f_{j} (\hat{\bW}_{t}^{(w)})\}\delta_{t,k} I(|\delta_{t,k}| \le Q) - \tilde{\mu}_{1,j} \big]\bigg| >x \bigg)\\
	&~~~~~=   \mathbb{P}\bigg[\max_{j\in[p]}\bigg|\frac{1}{  n_3}  \sum_{t\in \mathcal{D}_3}  \frac{\{\hat{f}_{j} (\hat{\bW}_{t}^{(w)}) - f_{j} (\hat{\bW}_{t}^{(w)})\}\delta_{t,k} I(|\delta_{t,k}| \le Q) -\tilde{\mu}_{1,j} }{QK(n, m ,\tilde{d})} \bigg|> \frac{x}{QK(n, m ,\tilde{d})} \bigg]\\
	&~~~~~\le  p\max_{j\in[p]}\mathbb{P}\bigg(\bigg|\frac{1}{ n_3}  \sum_{t\in \mathcal{D}_3} \bigg[ \frac{\{\hat{f}_{j} (\hat{\bW}_{t}^{(w)}) - f_{j} (\hat{\bW}_{t}^{(w)})\}\delta_{t,k} I(|\delta_{t,k}| \le Q)}{QK(n, m ,\tilde{d})}\\
	&~~~~~~~~~~~~~~~~~~~~~~~~~~~~~~~~~~~~~ - \frac{ \tilde{\mu}_{1,j} }{QK(n, m ,\tilde{d})} \bigg]\bigg| > \frac{x}{QK(n, m ,\tilde{d})}\,,~ \mathcal{G}_1 \bigg)  + \mathbb{P}(\mathcal{G}_1^{\rm c})\\
	&~~~~~\le p\max_{j\in[p]} \mathbb{E}\bigg( \mathbb{P}\bigg[\bigg|\frac{1}{ n_3}  \sum_{t\in \mathcal{D}_3}  \frac{\{\hat{f}_{j} (\hat{\bW}_{t}^{(w)}) - f_{j} (\hat{\bW}_{t}^{(w)})\}\delta_{t,k} I(|\delta_{t,k}| \le Q)-\tilde{\mu}_{1,j} }{QK(n, m ,\tilde{d})} \bigg| \\
	&~~~~~~~~~~~~~~~~~~~~~~~~~~~~~~~~~  > \frac{x}{QK(n, m ,\tilde{d})}\,, ~\mathcal{G}_1 \,\bigg|\, \mathcal{W}_{ \mathcal{D}_1},  \mathcal{W}_{ \mathcal{D}_2} \bigg] \bigg)   + \mathbb{P}(\mathcal{G}_1^{\rm c})\\
	&~~~~~\le  \tilde{C}_6\tilde{d}\exp\bigg\{-\frac{  n^{\kappa} x^2}{\tilde{C}_5Q^2K^2(n, m ,\tilde{d})+ \tilde{C}_5Q\tilde{\beta}_{n}x} \bigg\} + \tilde{C}_6 n^{-1 }
\end{align*}
provided that   $ \log (\tilde{d}n) \ll n^{1-\kappa}(\log n)^{-1/2}$ and $m \lesssim n$, which implies  
\begin{align}\label{eq:fhwh-fwh-cov}
	&\max_{j\in[p]}\bigg|\frac{1}{n_{3}} \sum_{t\in \mathcal{D}_3} \big[\{\hat{f}_{j} (\hat{\bW}_{t}^{(w)}) - f_{j} (\hat{\bW}_{t}^{(w)})\}\delta_{t,k} I(|\delta_{t,k}| \le Q) - \tilde{\mu}_{1,j} \big]\bigg| \notag \\
	&~~~~~~~~  = O_{\rm p}\{n^{-\kappa/2-\vartheta/(4\vartheta+m_{*})} Q (m^2\log n)^{(\vartheta+2m_{*}\tilde{\vartheta}+3m_{*} )/(8\vartheta)} \tilde{\beta}_n\log^{3/4}(\tilde{d}n) \}  \\
	&~~~~~~~~~~~~ +  O_{\rm p}\{n^{-\kappa/2-1/4} Qm^{1/2}  \tilde{\beta}_n^{1/2}\log^{3/4}  (\tilde{d}n) \}  + O_{\rm p}\{n^{-\kappa}m Q  \tilde{\beta}_n\log  (\tilde{d}n) \}   \notag
\end{align}
provided that   $ \log (\tilde{d}n) \ll n^{1-\kappa}(\log n)^{-1/2}$ and $m \lesssim n$. Hence, together with \eqref{eq:mu-z-bound}, we have \eqref{eq:g1} holds.
$\hfill\Box$

\subsubsection{Proof of \eqref{eq:g12-bound}}\label{sec:sub-g12-bound}
Recall $\max_{t\in \mathcal{D}_3,\,j\in[m]}|\hat{W}_{t,j}^{(w)}|\le \sqrt{2\log n_1}$.  Then   $\hat{\bW}_{t}^{(w)} \in [-\sqrt{2\log n_1}, \sqrt{2\log n_1}]^{m}$ with $n_1\asymp n$. With selecting $N=\tilde{\vartheta}$,  $M_{n}=\lceil n^{1/(4\vartheta+m_{*})} (m^2\log n)^{(2\tilde{\vartheta}+3)/(2\vartheta)} \rceil$ and  $\eta_{n}\asymp n^{-1}$,  let $\mathcal{H}^{(\ell)}$ be defined in \eqref{eq:nnspace-hl} with $(m_{*},K)$ as in  the definition of $f_j $, $\tilde{\alpha}_{n} =n^{c_{3}} $ and 
$$M_{*}= c_{4} \lceil n^{m_{*}/ (4\vartheta + m_{*})}  (m^2\log n)^{m_{*}(2\tilde{\vartheta}+3)/(2\vartheta)} \rceil$$  for some sufficiently large constants $c_{3}>0$ and $c_{4}>0$.  By Lemma \ref{lem:network-hiera},   there exists a neural network  
$h^{*} \in \{ t\in \mathcal{H}^{(\ell)}: |t|_{\infty,[-\sqrt{2\log n_1}, \sqrt{2\log n_1}]^{m}\backslash \breve{\boldsymbol{D}}} \le \tilde{\beta}_{n}\}$ with $\tilde{\beta}_n =(\log n)\log^{1/2}(\tilde{d}n)$ such that  
\begin{align}\label{eq:nn-hstar-f}
	&|h^{*}(\boldsymbol{x})-f_{j}(\boldsymbol{x})| \le   \frac{C_1 }{n^{ \vartheta/(4\vartheta+m_{*})}}  \,,  ~~~~~\boldsymbol{x} \in[-\sqrt{2\log n_1 }, \sqrt{2\log n_1}]^{m}\backslash\breve{\boldsymbol{D}}
\end{align}
holds with $\mathbb{P}(\hat{\bW}^{(w)}_{t} \in \breve{\boldsymbol{D}})\le C_{2}\eta_{n}$ for sufficiently large $n$, where $\breve{\boldsymbol{D}} \subset [-\sqrt{2\log n_1}, \sqrt{2\log n_1}]^{m}$.  
Write $\breve{\boldsymbol{D}}^{\rm c} =[-\sqrt{2\log n_1}, \sqrt{2\log n_1}]^{m}\backslash\breve{\boldsymbol{D}}$. By \eqref{eq:nn-hstar-f}, it holds that
\begin{align}\label{eq:hstar-bound}
	\mathbb{E}  \big[\{h^{*} (\hat{\bW}_{t}^{(w)}) - f_{j} (\hat{\bW}_{t}^{(w)}) \}^2  I(\hat{\bW}^{(w)}_{t} \in \breve{\boldsymbol{D}}^{\rm c}) \,|\, \mathcal{W}_{\mathcal{D}_1} \big]  \le  \frac{C_{3}}{n^{2\vartheta/(4\vartheta+m_{*})}}   
\end{align}
for any $j\in[p]$ and $t\in \mathcal{D}_3$. Let 
$$\breve{h}  =\arg\min_{h \in T_{\tilde{\beta}_n} \mathcal{H}^{(\ell)}} \mathbb{E}  \big[\{h (\hat{\bW}_{t}^{(w)}) - f_{j} (\hat{\bW}_{t}^{(w)}) \}^2  I(\hat{\bW}^{(w)}_{t} \in\breve{\boldsymbol{D}}^{\rm c}) \,|\, \mathcal{W}_{\mathcal{D}_1} \big] \,.$$  
Since  $T_{\tilde{\beta}_n} h^{*}  =h^{*} $ for any $\boldsymbol{x} \in \breve{\boldsymbol{D}}^{\rm c}$,  it holds that
\begin{align*}
	&\mathbb{E}  \big[\{\breve{h} (\hat{\bW}_{t}^{(w)}) - f_{j} (\hat{\bW}_{t}^{(w)}) \}^2  I(\hat{\bW}^{(w)}_{t} \in \breve{\boldsymbol{D}}^{\rm c}) \,|\, \mathcal{W}_{\mathcal{D}_1} \big] \\
	&~~~~~~~~~~ \le \mathbb{E}  \big[\{h^{*} (\hat{\bW}_{t}^{(w)}) - f_{j} (\hat{\bW}_{t}^{(w)}) \}^2  I(\hat{\bW}^{(w)}_{t} \in\breve{\boldsymbol{D}}^{\rm c}) \,|\, \mathcal{W}_{\mathcal{D}_1} \big] \,.
\end{align*}
Due to  
\begin{align*}
	{\rm G}_{11}(j) \le&~ \mathbb{E}  \big[\{\breve{h} (\hat{\bW}_{t}^{(w)}) - f_{j} (\hat{\bW}_{t}^{(w)}) \}^2  I(\hat{\bW}^{(w)}_{t} \in\breve{\boldsymbol{D}}^{\rm c}) \,|\, \mathcal{W}_{\mathcal{D}_1} \big] \\
	&+ \mathbb{E}  \big[\{\breve{h} (\hat{\bW}_{t}^{(w)}) - f_{j} (\hat{\bW}_{t}^{(w)}) \}^2  I(\hat{\bW}^{(w)}_{t} \in\breve{\boldsymbol{D}}) \,|\, \mathcal{W}_{\mathcal{D}_1} \big] \\
	\le &~ \mathbb{E}  \big[\{h^{*} (\hat{\bW}_{t}^{(w)}) - f_{j} (\hat{\bW}_{t}^{(w)}) \}^2  I(\hat{\bW}^{(w)}_{t} \in\breve{\boldsymbol{D}}^{\rm c}) \,|\, \mathcal{W}_{\mathcal{D}_1} \big] \\
	&+ \mathbb{E}  \big[\{\breve{h} (\hat{\bW}_{t}^{(w)}) - f_{j} (\hat{\bW}_{t}^{(w)}) \}^2  I(\hat{\bW}^{(w)}_{t} \in\breve{\boldsymbol{D}}) \,|\, \mathcal{W}_{\mathcal{D}_1}  \big]  \,,
\end{align*}
by \eqref{eq:hstar-bound}, we have
\begin{align*}
	&\mathbb{P}\bigg\{\max_{j\in[p]} {\rm G}_{11}(j)  > \frac{\tilde{C}_1 }{n^{2\vartheta/(4\vartheta+m_{*})}}   \bigg\}\\
	&~~~~~~~~\le \mathbb{P}\bigg(\max_{j\in[p]}\mathbb{E} \big[\{\breve{h} (\hat{\bW}_{t}^{(w)}) - f_{j} (\hat{\bW}_{t}^{(w)}) \}^2 I(\hat{\bW}^{(w)}_{t} \in\breve{\boldsymbol{D}}) \,|\, \mathcal{W}_{\mathcal{D}_1} \big] > 0\bigg)   \\
	&~~~~~~~~\le  \mathbb{P}(\hat{\bW}^{(w)}_{t} \in \breve{\boldsymbol{D}}) \le C_{4}n^{-1} \,,
\end{align*} 
where $\tilde{C}_1 >C_3$ is some sufficiently large constant.  Hence, we have \eqref{eq:g12-bound}.
$\hfill\Box$

\subsubsection{Proof of \eqref{eq:g11-tail}}\label{sec:sub-g11-tail}
For any $t\in\mathcal{D}_3$,  by  the definition of $\hat{f}_{j}$ given in \eqref{eq:fhj-ghk}, we have
\begin{align*}
	{\rm G}_{12}(j) \le  &~ \sup_{ h\in T_{\tilde{\beta}_n} \mathcal{H}^{(\ell)} }  \bigg(\mathbb{E}  \big[\{\hat{f}_{j} (\hat{\bW}_{t}^{(w)}) - \hat{U}_{t,j}^{(w)}\}^2 \,|\, \mathcal{W}_{\mathcal{D}_1}, \mathcal{W}_{\mathcal{D}_2}\big] - \frac{1}{n_2} \sum_{i\in \mathcal{D}_{2}} \{\hat{f}_{j} (\hat{\bW}_{i}^{(w)}) - \hat{U}_{i,j}^{(w)}\}^2 \notag\\
	&~~~~~~~~~~~~~~~~~~~~~~+ \frac{1}{n_2} \sum_{i\in \mathcal{D}_{2}} \{\hat{f}_{j} (\hat{\bW}_{i}^{(w)}) - \hat{U}_{i,j}^{(w)}\}^2  - \frac{1}{n_2} \sum_{i\in \mathcal{D}_{2}} \{h (\hat{\bW}_{i}^{(w)}) - \hat{U}_{i,j}^{(w)}\}^2 \notag\\
	&~~~~~~~~~~~~~~~~~~~~~~+  \frac{1}{n_2} \sum_{i\in \mathcal{D}_{2}} \{h (\hat{\bW}_{i}^{(w)}) - \hat{U}_{i,j}^{(w)}\}^2 - \mathbb{E}  \big[\{h (\hat{\bW}_{t}^{(w)}) - \hat{U}_{t,j}^{(w)}\}^2 \,|\, \mathcal{W}_{\mathcal{D}_1} \big] \bigg) \notag\\
	\le  &~ 
	\bigg|\mathbb{E}  \big[\{\hat{f}_{j} (\hat{\bW}_{t}^{(w)}) - \hat{U}_{t,j}^{(w)}\}^2 \,|\, \mathcal{W}_{\mathcal{D}_1}, \mathcal{W}_{\mathcal{D}_2}\big] - \frac{1}{n_2} \sum_{i\in \mathcal{D}_{2}} \{\hat{f}_{j} (\hat{\bW}_{i}^{(w)}) - \hat{U}_{i,j}^{(w)}\}^2 \bigg| \\
	&+ \sup_{ T_{\tilde{\beta}_n} \mathcal{H}^{(\ell)} } \bigg| \frac{1}{n_2} \sum_{i\in \mathcal{D}_{2}}  \{h (\hat{\bW}_{i}^{(w)}) - \hat{U}_{i,j}^{(w)}\}^2 - \mathbb{E} \big[\{h (\hat{\bW}_{t}^{(w)}) - \hat{U}_{t,j}^{(w)}\}^2 \,|\, \mathcal{W}_{\mathcal{D}_1}  \big]  \bigg|\\
	\le &~  2 \underbrace{\sup_{ h \in T_{\tilde{\beta}_{n}}\mathcal{H}^{(\ell)} } \bigg| \frac{1}{n_2} \sum_{i\in \mathcal{D}_{2}} \big(\{h (\hat{\bW}_{i}^{(w)}) - \hat{U}_{i,j}^{(w)}\}^2 - \mathbb{E}  \big[\{h (\hat{\bW}_{i}^{(w)}) - \hat{U}_{i,j}^{(w)}\}^2 \,|\, \mathcal{W}_{\mathcal{D}_1} \big]\big) \bigg|}_{\breve{\textrm{G}}_{12}(j)}\,.
\end{align*}

Let $\mathcal{G}$ be a set of functions $\mathbb{R}^{m} \to \mathbb{R}$ and $\Psi_{n}=\{\psi_{1},\ldots,\psi_{n}\}$ be given i.i.d. random variables. 
For given $\epsilon>0$, denote by $\mathcal{N}_1 (\epsilon, \mathcal{G},   \Psi_{n})$ the minimal $N \in \mathbb{N}$ such that there exist $\tilde{g}_1, \ldots, \tilde{g}_N \in \mathcal{G}$ with the property that for every $\tilde{g} \in \mathcal{G}$ there is a $j=j(\tilde{g}) \in[N]$ such that $n^{-1}\sum_{i=1}^n |\tilde{g} (\psi_i )- \tilde{g}_j (\psi_i ) |  <\epsilon$, and denote by $\mathcal{M}_1 (\epsilon, \mathcal{G}, \Psi_n )$ the maximal $N \in \mathbb{N}$ such that there exist 
$\tilde{g}_1, \ldots, \tilde{g}_N \in \mathcal{G}$ with $n^{-1} \sum_{i=1}^n |\tilde{g}_j (\psi_i )- \tilde{g}_k (\psi_i ) | \ge \epsilon$ for all $1 \leq j<k \leq N$. 
Furthermore, denote by $\mathcal{N} (\epsilon, \mathcal{G},   |\cdot|_{\infty, \boldsymbol{D}})$  the minimal $N \in \mathbb{N}$ such that there exist  $\tilde{g}_1, \ldots, \tilde{g}_N\in \mathcal{G}$ with the property that for every $\tilde{g} \in \mathcal{G}$ there is a $j=j(\tilde{g}) \in[N]$ such that $\sup_{\boldsymbol{x} \in \boldsymbol{D}} |\tilde{g} (\boldsymbol{x})- \tilde{g}_j (\boldsymbol{x}) |  <\epsilon $. Recall $\max_{i\in \mathcal{D}_2,\,j\in[p]}|\hat{U}_{i,j}^{(w)}|\le \sqrt{2\log n_1}$ and $\tilde{\beta}_n =(\log n)\log^{1/2}(\tilde{d}n)$. Define
\begin{align*}
	\mathcal{G}_{2} = \big\{g: \mathbb{R}^{m} \times    \mathbb{R} \to [0,4\tilde{\beta}_n^2]: \exists  h \in T_{\tilde{\beta}_{n}}\mathcal{H}^{(\ell)} ~\text{such that}~ g(\boldsymbol{x},y) =|h(\boldsymbol{x}) - y|^2 \big\}\,.
\end{align*}  
By Theorem 9.1 of \cite{Gyorfi2002}, it holds that
\begin{align}\label{eq:covering-number-tail}
	&\mathbb{P}\{\breve{{\rm G}}_{12}(j) > x\,|\,\mathcal{W}_{\mathcal{D}_1}\} \notag\\
	&~~~~~~=\mathbb{P}\bigg\{\sup_{ h \in T_{\tilde{\beta}_{n}}\mathcal{H}^{(\ell)} } \bigg| \frac{1}{n_2} \sum_{i\in \mathcal{D}_{2}} \big(\{h (\hat{\bW}_{i}^{(w)}) - \hat{U}_{i,j}^{(w)}\}^2 - \mathbb{E} \big[\{h (\hat{\bW}_{i}^{(w)}) - \hat{U}_{i,j}^{(w)}\}^2\,|\, \mathcal{W}_{\mathcal{D}_1}\big]\big) \bigg| > x \,\bigg|\, \mathcal{W}_{  \mathcal{D}_1} \bigg\} \notag \\
	&~~~~~~\le  \mathbb{P}\bigg(\sup_{ a \in  \mathcal{G}_{2} } \bigg| \frac{1}{n_2} \sum_{i\in \mathcal{D}_{2}} \big[a(\hat{\bW}_i^{(w)}, \hat{U}_{i,j}^{(w)}) - \mathbb{E}  \big\{a(\hat{\bW}_i^{(w)}, \hat{U}_{i,j}^{(w)}) \,|\, \mathcal{W}_{\mathcal{D}_1}\big\} \big] \bigg| > x \,\bigg|\, \mathcal{W}_{ \mathcal{D}_1} \bigg) \notag\\
	&~~~~~~\le 8 \mathbb{E}\bigg(\mathcal{N}_1\bigg[\frac{x}{8} , \mathcal{G}_2, \{(\hat{\bW}_{i}^{(w)}, \hat{U}_{i,j}^{(w)} )\}_{i\in \mathcal{D}_2} \bigg] \,\bigg|\, \mathcal{W}_{\mathcal{D}_1}\bigg) \times  \exp\bigg\{-\frac{n_2x^2}{128 (4\tilde{\beta}_{n}^2)^2}\bigg\}
\end{align}
for any $x>0$.  Recall $\hat{\bW}_{i}^{(w)} \in[-\sqrt{2\log n_{1}}, \sqrt{2\log n_{1} }]^{m}$. By Lemma 9.2 and  Equation (10.21) of \cite{Gyorfi2002}, we  have
\begin{align*}
	&\mathcal{N}_1\bigg[\frac{x}{8} , \mathcal{G}_2, \{(\hat{\bW}_{i}^{(w)}, \hat{U}_{i,j}^{(w)} )\}_{i\in \mathcal{D}_2} \bigg] \\
	&~~~~~~\le  \mathcal{M}_1\bigg[\frac{x}{8} , \mathcal{G}_2, \{(\hat{\bW}_{i}^{(w)}, \hat{U}_{i,j}^{(w)} )\}_{i\in \mathcal{D}_2} \bigg] 
	\le  \mathcal{M}_1\bigg[\frac{x}{32\tilde{\beta}_{n}} , T_{\tilde{\beta}_{n}}\mathcal{H}^{(\ell)}, \{\hat{\bW}_{i}^{(w)} \}_{i\in \mathcal{D}_2} \bigg]\\
	&~~~~~~\le   \mathcal{N}_1\bigg[\frac{x}{64\tilde{\beta}_{n}} , T_{\tilde{\beta}_{n}}\mathcal{H}^{(\ell)}, \{\hat{\bW}_{i}^{(w)} \}_{i\in \mathcal{D}_2} \bigg] 
	\le  \mathcal{N}\bigg\{\frac{x}{64\tilde{\beta}_{n}} , T_{\tilde{\beta}_{n}}\mathcal{H}^{(\ell)}, |\cdot|_{\infty,[-\sqrt{2\log n_{1}}, \sqrt{2\log n_{1} }]^{m}} \bigg\}\,.
\end{align*}
Recall $\tilde{d}=p\vee q\vee m$ and $n_2\asymp n$.  By  \eqref{eq:covering-number-tail}, for some sufficiently large constant $C_1>0$,   
\begin{align}\label{eq:g11-tail-cov}
	&\mathbb{P}\bigg( \breve{{\rm G}}_{12}(j) > \frac{C_1\tilde{\beta}_{n}^2}{\sqrt{n_2}}  \log^{1/2} \bigg[\tilde{d}n \cdot \mathcal{N}\bigg\{\frac{1}{64n_2\tilde{\beta}_{n}} , T_{\tilde{\beta}_{n}}\mathcal{H}^{(\ell)}, |\cdot|_{\infty,[-\sqrt{2\log n_{1}}, \sqrt{2\log n_{1} }]^{m}}\bigg\}\bigg] \, \bigg| \, \mathcal{W}_{ \mathcal{D}_1} \bigg) \notag\\
	&~~~~~~~~~\le C_2(\tilde{d}n)^{-3}\,.
\end{align}
By Equation (8) of  \cite{Bauer2019}, the  neural network $  \mathcal{H}^{(\ell)}$ has at most  
\begin{align*}
	\bigg\{\sum_{j=1}^{\ell}m_{*}^{j-1} K^{j} + (m_{*}K)^{\ell}\bigg\} \cdot [M_{*}\{4m_{*}(m+2)+2\}+1] 
\end{align*}
weights.  Parallel to the proof of Lemma 2 in  \cite{Bauer2019}, it holds that
\begin{align*}
	\log \mathcal{N} \bigg\{\frac{1}{64n_2\tilde{\beta}_{n}} , \mathcal{H}^{(\ell)}, |\cdot|_{\infty, [-\sqrt{2\log n_1}, \sqrt{2\log n_1}]^{m}} \bigg\} \le&~  C_3 M_{*}m \log n  
\end{align*} 
provided that $\tilde{\beta}_n \ll n$ and $m\lesssim n$. 
Recall $M_{*}= c_{4} \lceil n^{m_{*}/ (4\vartheta + m_{*})}  (m^2\log n)^{m_{*}(2\tilde{\vartheta}+3)/(2\vartheta)} \rceil$ for some sufficiently large constant $ c_{4} >0$,   $\tilde{d}=p\vee q\vee m$ and $n_2\asymp n$.   Then, it holds that
\begin{align*}
	&n_2^{-1/2}\tilde{\beta}_{n}^2 \log^{1/2} \bigg[\tilde{d}n \cdot \mathcal{N}\bigg\{\frac{1}{64n_2\tilde{\beta}_{n}} , T_{\tilde{\beta}_{n}}\mathcal{H}^{(\ell)}, |\cdot|_{\infty,[-\sqrt{2\log n_{1}}, \sqrt{2\log n_{1} }]^{m} }\bigg\}\bigg] \\
	&~~~~~~~~~~~~ \lesssim n^{-1/2}  \tilde{\beta}_{n}^2\big\{\log^{1/2}(\tilde{d}n) + (M_{*}m  \log n )^{1/2}\big\} \\
	&~~~~~~~~~~~~\lesssim  \frac{ (m^2\log n)^{(\vartheta+2m_{*}\tilde{\vartheta}+3m_{*} )/(4\vartheta)}  \tilde{\beta}_{n}^2\log^{1/2} (\tilde{d}n)}{n^{ 2\vartheta /(4\vartheta+m_{*})}}   \,.
\end{align*}
Together with \eqref{eq:g11-tail-cov}, it holds that
\begin{align*}
	&\mathbb{P}\bigg\{ \max_{j\in[p]}{\rm G}_{12}(j) >  \frac{\tilde{C}_2 (m^2\log n)^{(\vartheta+2m_{*}\tilde{\vartheta}+3m_{*} )/(4\vartheta)}  \tilde{\beta}_{n}^2\log^{1/2} (\tilde{d}n)}{n^{ 2\vartheta /(4\vartheta+m_{*})}} \bigg\}\\
	&~~~~~\le \mathbb{E}\bigg[\mathbb{P}\bigg\{ \max_{j\in[p]}\breve{{\rm G}}_{12}(j) >  \frac{\tilde{C_2} (m^2\log n)^{(\vartheta+2m_{*}\tilde{\vartheta}+3m_{*} )/(4\vartheta)}  \tilde{\beta}_{n}^2\log^{1/2} (\tilde{d}n)}{2n^{ 2\vartheta /(4\vartheta+m_{*})}} \,\bigg|\, \mathcal{W}_{\mathcal{D}_1}\bigg\} \bigg] \le C_4(\tilde{d}n)^{-2} 
\end{align*}
for some sufficiently large constant $\tilde{C}_2>0$, provided that $\tilde{\beta}_n \ll n$ and $m\lesssim n$.  
$\hfill\Box$

\subsubsection{Proofs of \eqref{eq:g13-tail} and \eqref{eq:g14-tail}}\label{sec:sub-g13-tail}
Notice that $\max_{t\in\mathcal{D}_3,\,j\in[p]}|\hat{f}_{j}(\hat{\bW}_{t}^{(w)})|  \le  \tilde{\beta}_{n}$ and $|f_{j}|_{\infty} \le \tilde{C}$. For any $t\in\mathcal{D}_3$, we have 
\begin{align}\label{eq:g13-dec}
	|{\rm G}_{13}(j)| 
	\le 2 \tilde{\beta}_{n}\underbrace{\mathbb{E}  \big(|\hat{U}_{t,j}^{(w)} - U_{t,j} | \,|\, \mathcal{W}_{\mathcal{D}_1}\big) }_{\textrm{G}_{131}(j)}  + 2\tilde{\beta}_{n} \underbrace{\mathbb{E}  \big\{| f_{j} (\hat{\bW}_{t}^{(w)}) - f_{j} (\bW_{t})| \,|\, \mathcal{W}_{\mathcal{D}_1}\big\} }_{\textrm{G}_{132}(j)} \,.
\end{align}
Recall  $U_{t,j} \sim \mathcal{N}(0,1)$, $\max_{t\in\mathcal{D}_3,\,j\in[p]}|\hat{U}_{t,j}^{(w)}|\le \sqrt{2\log n_1}$, $n_1\asymp n$ and  $n_3 \asymp n^{\kappa}$ for some constant $0<\kappa<1$. Given $M_1=\sqrt{2 \log n_3}$, it holds that
\begin{align}\label{eq:exp-hu-u}
	{\rm G}_{131}(j) =&~ \mathbb{E}  \big\{|\hat{U}_{t,j}^{(w)}- U_{t,j}| I(|U_{t,j}|\le M_1) \,|\, \mathcal{W}_{\mathcal{D}_1}\big\} + \mathbb{E}  \big\{|\hat{U}_{t,j}^{(w)} - U_{t,j} | I(|U_{t,j}| > M_1) \,|\, \mathcal{W}_{\mathcal{D}_1}\big\} \notag\\
	\le&~ \mathbb{E}  \big\{|\hat{U}_{t,j}^{(w)} -U_{t,j} | I(|U_{t,j}|\le M_1) \,|\, \mathcal{W}_{\mathcal{D}_1}\big\} + \mathbb{E} \big\{|U_{t,j}| I(|U_{t,j}| > M_1)\big\} \notag\\
	&+ (2\log n_1)^{1/2} \mathbb{E} \big\{ I(|U_{t,j}| > M_1)\big\}\\ 
	\le&~ \mathbb{E}  \big\{|\hat{U}_{t,j}^{(w)} -U_{t,j} | I(|U_{t,j}|\le M_1) \,|\, \mathcal{W}_{\mathcal{D}_1}\big\} + C_{1} n^{-\kappa} \,.\notag
\end{align}
Recall $\tilde{d}=p\vee q\vee m$. Let 
$$K(U_{t,j}, \tilde{d},n_{1}) = 4 n_{1}^{-1/2}  [\Phi(U_{t,j}) \{1-\Phi(U_{t,j})\}]^{1/2} \log^{1/2}(\tilde{d}n_{1})  + 7n_1^{-1} \log(\tilde{d}n_{1})\,.$$
Using the similar arguments for the derivation of the convergence rate of $\max_{j\in[p],\,k\in[q]}|\tilde{{\rm H}}_{1}(j,k)|$ in Section \ref{sec:sub-g11} for the proof of Lemma \ref{lem:uh-u-delta}, we have
\begin{align*}
	&\mathbb{E} \big[|\hat{U}_{t,j}^{(w)} -U_{t,j}|I(|U_{t,j}|\le M_1) I\{|\hat{F}_{\bX,j}(X_{t,j})-  F_{\bX,j}(X_{t,j})| \le K(U_{t,j}, \tilde{d},n_1)\} \,|\, \mathcal{W}_{\mathcal{D}_1} \big] \notag\\
	&~~~~  \le   \frac{C_{2}\log^{1/2}(\tilde{d}n_1)}{n_1^{1/2}} \times \mathbb{E}\big\{ e^{U_{t,j}^2/4}I( |U_{t,j}| \le M_1)\big\} \le    \frac{C_{3}\log^{1/2}(\tilde{d}n)}{n^{1/2}} 
\end{align*}
provided that $ \log (\tilde{d}n) \ll n^{1-\kappa} (\log n)^{-1/2}$. Since
\begin{align*}
	&\mathbb{E} \{|\hat{U}_{t,j}^{(w)} -U_{t,j}|I(|U_{t,j}|\le M_1) \,|\, \mathcal{W}_{\mathcal{D}_1}\}\\
	&~~~~~~=\mathbb{E} \big[|\hat{U}_{t,j}^{(w)} -U_{t,j}|I(|U_{t,j}|\le M_1) I\{|\hat{F}_{\bX,j}(X_{t,j})-  F_{\bX,j}(X_{t,j})| \le K(U_{t,j}, \tilde{d},n_1)\} \,|\, \mathcal{W}_{\mathcal{D}_1}\big]\\
	&~~~~~~~~~+\mathbb{E} \big[|\hat{U}_{t,j}^{(w)} -U_{t,j}|I(|U_{t,j}|\le M_1) I\{|\hat{F}_{\bX,j}(X_{t,j})-  F_{\bX,j}(X_{t,j})| > K(U_{t,j}, \tilde{d},n_1)\} \,|\, \mathcal{W}_{\mathcal{D}_1}\big]\,,
\end{align*}
analogous to \eqref{eq:h5-c}, it holds that
\begin{align}\label{eq:exp-hu-u-tail}
	&\mathbb{P}\bigg[\max_{j\in[p]}\mathbb{E} \{|\hat{U}_{t,j}^{(w)} -U_{t,j} |I(|U_{t,j}|\le M_1) \,|\, \mathcal{W}_{\mathcal{D}_1}\} > \frac{2C_{3}\log^{1/2}(\tilde{d}n)}{n^{1/2}} \bigg] \notag\\
	&~~~~~~~~~\le \mathbb{P}\bigg\{\max_{t\in\mathcal{D}_3,\,j\in[p]}|\hat{F}_{\bX,j}(X_{t,j})-  F_{\bX,j}(X_{t,j})| > K(U_{t,j}, \tilde{d},n_1)\bigg\} \le 4(\tilde{d}n)^{-2} 
\end{align}
provided that $ \log (\tilde{d}n) \ll n^{1-\kappa}(\log n)^{-1/2}$. By \eqref{eq:exp-hu-u}, we have
\begin{align}\label{g131-tail}
	\mathbb{P}\bigg\{\max_{j\in[p]}{\rm G}_{131}(j) >  \frac{2C_{3}\log^{1/2}(\tilde{d}n)}{n^{1/2}} + \frac{ C_1}{n^{\kappa}}  \bigg\}  \lesssim (\tilde{d}n)^{-2} 
\end{align}
provided that $ \log (\tilde{d}n) \ll n^{1-\kappa}(\log n)^{-1/2}$. Recall $\bW_{t}=(W_{t,1}, \ldots, W_{t,m})^{\T}$, $\hat{\bW}_{t}^{(w)}=(\hat{W}_{t,1}^{(w)}, \ldots, \hat{W}_{t,m}^{(w)})^{\T}$, $\max_{t\in\mathcal{D}_3,\,j\in[m]}|\hat{W}_{t,j}^{(w)}|\le \sqrt{2\log n_1}$, $n_1\asymp n$ and  $n_3 \asymp n^{\kappa}$ for some constant $0<\kappa<1$. Furthermore, by \eqref{eq:fwh-fw-bound} and $W_{t,j} \sim \mathcal{N}(0,1)$, given $M_1=\sqrt{2 \log n_3}$, for any $t\in \mathcal{D}_3$, 
\begin{align}\label{eq:exp-fhw-w}
	{\rm G}_{132}(j) \le&~ C_{4}\mathbb{E} \{|\hat{\bW}_{t}^{(w)} - \bW_{t}|_1 \,|\, \mathcal{W}_{\mathcal{D}_1} \} = C_{4} \mathbb{E} \bigg\{  \sum_{j=1}^{m}|\hat{W}_{t,j}^{(w)} -W_{t,j}|  \,\bigg|\, \mathcal{W}_{\mathcal{D}_1} \bigg\}\notag\\
	\le&~ C_{4} \mathbb{E} \bigg\{  \sum_{j=1}^{m}|\hat{W}_{t,j}^{(w)} -W_{t,j}|  I(|\bW_{t}|_{\infty} \le M_1) \,\bigg|\, \mathcal{W}_{\mathcal{D}_1}\bigg\}\notag\\
	&+  C_{4}\mathbb{E} \bigg[\bigg\{  \sum_{j=1}^{m}|\hat{W}_{t,j}^{(w)}|  +  \sum_{j=1}^{m}|W_{t,j}| \bigg\} I(|\bW_{t}|_{\infty} > M_1) \,\bigg|\, \mathcal{W}_{\mathcal{D}_1}\bigg]\\
	\le&~ C_{4} \mathbb{E} \bigg\{  \sum_{j=1}^{m}|\hat{W}_{t,j}^{(w)} -W_{t,j}| I(|\bW_{t}|_{\infty} \le M_1) \,\bigg|\, \mathcal{W}_{\mathcal{D}_1} \bigg\} + \frac{C_{5}m^2}{n^{\kappa} } \,.\notag
\end{align}
Let 
$$K(W_{t,j}, \tilde{d},n_{1}) = 4 n_{1}^{-1/2}  [\Phi(W_{t,j}) \{1-\Phi(W_{t,j})\}]^{1/2} \log^{1/2}(\tilde{d}n_{1})  + 7n_1^{-1} \log(\tilde{d}n_{1})\,.$$
Using the similar arguments for the derivation of the convergence rate of $ |{\rm H}_{11} |$ in Section \ref{sec:sub-g211} for the proof of Lemma \ref{lem:fwh-fw-delta},  it holds that
\begin{align*}
	&\mathbb{E} \bigg[  \sum_{j=1}^{m}|\hat{W}_{t,j}^{(w)} -W_{t,j}| I(|\bW_{t}|_{\infty} \le M_1) \prod_{k=1}^{m}I\{|\hat{F}_{\bZ,k}(Z_{t,k})-  F_{\bZ,k}(Z_{t,k})| \le K(W_{t,k}, \tilde{d},n_1)\} \,\bigg|\, \mathcal{W}_{\mathcal{D}_1}\bigg]\\
	&~~~~\le \frac{C_{6}\log^{1/2}(\tilde{d}n_1)}{n_1^{1/2}}\times  \mathbb{E}\bigg\{  \sum_{j=1}^{m}  e^{W_{t,j}^2/4} I(  |W_{t,j}| \le M_1) \bigg\}  \le \frac{C_{7}m\log^{1/2}(\tilde{d}n)}{n^{1/2}}
\end{align*}
provided that $ \log (\tilde{d}n) \ll n^{1-\kappa}(\log n)^{-1/2}$. Hence, similar to \eqref{eq:h5-c} and \eqref{eq:exp-hu-u-tail}, we have
\begin{align}\label{eq:exp-fhw-w-tail}
	&\mathbb{P}\bigg[\mathbb{E} \bigg\{  \sum_{j=1}^{m}|\hat{W}_{t,j}^{(w)} -W_{t,j}| I(|\bW_{t}|_{\infty} \le M_1) \,\bigg|\, \mathcal{W}_{\mathcal{D}_1}\bigg\} > \frac{2C_{7}m\log^{1/2}(\tilde{d}n)}{n^{1/2}} \bigg] \notag\\
	&~~~~~~~~~~~\le \mathbb{P}\bigg\{\max_{t\in\mathcal{D}_3,\,j\in[m]}|\hat{F}_{\bZ,j}(Z_{t,j})-  F_{\bZ,j}(Z_{t,j})| > K(W_{t,j}, \tilde{d},n_1)\bigg\} \le 4(\tilde{d}n)^{-2}\,.
\end{align}
Then, by \eqref{eq:exp-fhw-w}, it holds that
\begin{align}\label{g132-tail}
	\mathbb{P}\bigg\{\max_{j\in[p]}{\rm G}_{132}(j) >  \frac{2C_{7}C_4m\log^{1/2}(\tilde{d}n)}{n^{1/2}} +  \frac{C_{5}m^2}{n^{\kappa}} \bigg\}  \lesssim (\tilde{d}n)^{-2} 
\end{align}
provided that $ \log (\tilde{d}n) \ll n^{1-\kappa}(\log n)^{-1/2}$. Thus, combining \eqref{g131-tail} and \eqref{g132-tail}, by \eqref{eq:g13-dec}, it holds that 
\begin{align*}
	&\mathbb{P}\bigg\{\max_{j\in[p]}|{\rm G}_{13}(j)| > \frac{C_{8}m\tilde{\beta}_{n}\log^{1/2}(\tilde{d}n)}{n^{1/2}} + \frac{C_{8}m^2\tilde{\beta}_{n} }{n^{\kappa}} \bigg\}\\
	&~~~~~~~\le \mathbb{P}\bigg\{\max_{j\in[p]}\tilde{\beta}_{n}{\rm G}_{131}(j) >  \frac{C_{8}m\tilde{\beta}_{n}\log^{1/2}(\tilde{d}n)}{4n^{1/2}} + \frac{C_{8}m^2\tilde{\beta}_{n} }{4n^{\kappa}} \bigg\} \\
	&~~~~~~~~~~~~+ \mathbb{P}\bigg\{ \max_{j\in[p]}\tilde{\beta}_{n} {\rm G}_{132}(j) >  \frac{C_{8}m\tilde{\beta}_{n}\log^{1/2}(\tilde{d}n)}{4n^{1/2}} +  \frac{C_{8}m^2\tilde{\beta}_{n} }{4n^{\kappa}} \bigg\} \\
	&~~~~~~~\lesssim (\tilde{d}n)^{-2} 
\end{align*}
for some sufficiently large constant $C_{8}>0$, provided that $ \log (\tilde{d}n) \ll n^{1-\kappa} (\log n)^{-1/2}$, which implies \eqref{eq:g13-tail} holds. On the other hand, due to $|f_{j}|_{\infty} \le \tilde{C}$ and $|h|_{\infty} \le \tilde{\beta}_n$ for any $h\in T_{\tilde{\beta}_n} \mathcal{H}^{(\ell)}$,  we have 
\begin{align*} 
	|{\rm G}_{14}(j)| 
	\le 2 \tilde{\beta}_{n}\mathbb{E}  \big(|\hat{U}_{t,j}^{(w)} - U_{t,j} | \,|\, \mathcal{W}_{\mathcal{D}_1}\big)   + 2\tilde{\beta}_{n} \mathbb{E}  \big\{| f_{j} (\hat{\bW}_{t}^{(w)}) - f_{j} (\bW_{t})| \,|\, \mathcal{W}_{\mathcal{D}_1}\big\} 
\end{align*}
for any $t\in\mathcal{D}_3$. Hence, we also know  \eqref{eq:g14-tail} holds. 
$\hfill\Box$

\subsubsection{Proof of Lemma \ref{lem:network-hiera}}\label{sec:sub-network-hiera}
To prove Lemma \ref{lem:network-hiera}, we need Lemmas \ref{lem:polynomials}--\ref{lem:network-pc}, whose proofs are given in Sections \ref{sec:sub-polynomials}--\ref{sec:sub-lem:network-pc}, respectively.

\begin{lem}\label{lem:polynomials}
	Write $\boldsymbol{x}=(x_1,\ldots, x_{\tilde{m}})^{\T}$ for some general integer $\tilde{m}\geq1$. Let $\tilde{N}\in \mathbb{N}_{0}$, and $\mathcal{P}_{\tilde{N}}$ be the linear span of all monomials of the form $\prod_{k=1}^{\tilde{m}}x_{k}^{\tilde{r}_{k}}$ for some $\tilde{r}_{1}, \ldots, \tilde{r}_{\tilde{m}} \in \mathbb{N}_{0}$ and $\sum_{k=1}^{\tilde{m}}\tilde{r}_{k} \le \tilde{N}$. Let $f \in \mathcal{P}_{\tilde{N}}$, and $m_{1}, \ldots, m_{{\rm C}_{\tilde{m}+\tilde{N}}^{\tilde{m}}}$ denote all monomials in $\mathcal{P}_{\tilde{N}}$. Define $r_i \in \mathbb{R}$, $i\in [ {\rm C}_{\tilde{m}+\tilde{N}}^{\tilde{m}}]$, by
	\begin{align*}
		f(\boldsymbol{x}) = \sum_{i=1}^{{\rm C}_{\tilde{m}+\tilde{N}}^{\tilde{m}}} r_{i}  m_{i}(\boldsymbol{x}) \,,
	\end{align*}
	and set $\bar{r}(f)= \max_{i\in[{\rm C}_{\tilde{m}+N}^{\tilde{m}}]}|r_{i}|$.  For any $R>0$ and  $\tilde{a}_n \ge 1$, there exists a neural network of the type 
	\begin{align*}
		s(\boldsymbol{x}) =\sum_{l=1}^{(\tilde{N}+1){\rm C}_{\tilde{m}+\tilde{N}}^{\tilde{m}}} \tilde{b}_{l}  \sigma\bigg(\sum_{v=1}^{\tilde{m}} \tilde{a}_{l,v}x_{v} + \tilde{a}_{l,0} \bigg)
	\end{align*}
	with $\sigma(x)=(1+e^{-x})^{-1}$ for any $x\in \mathbb{R}$, such that
	\begin{align*}
		|f(\boldsymbol{x}) -s(\boldsymbol{x})| \le \tilde{C}_1 ({\rm C}_{\tilde{m}+\tilde{N}}^{\tilde{m}})^2 \cdot  \frac{\bar{r}(f)\cdot  \tilde{a}_n^{\tilde{N}+1}}{R}
	\end{align*}
	holds for all $\boldsymbol{x}\in [-\tilde{a}_n,\tilde{a}_n]^{\tilde{m}}$, and the coefficients of this neural network satisfy
	\begin{align*}
		|\tilde{b}_{l}| \le \tilde{C}_{2} {\rm C}_{\tilde{m}+\tilde{N}}^{\tilde{m}} R^{\tilde{N}} \bar{r}(f) ~~{\textrm{and}}~~|\tilde{a}_{l,v}| \le  \frac{\tilde{C}_{3}\tilde{a}_n}{R(\tilde{m}+1)}   
	\end{align*}
	for all $l\in [(\tilde{N}+1){\rm C}_{\tilde{m}+N}^{\tilde{m}}]$ and $v\in[\tilde{m}]\cup \{0\}$, where $\tilde{C}_1>0$, $\tilde{C}_2>0$ and $\tilde{C}_3>0$ are some universal constants only depending on $(\tilde{m},\tilde{N})$.   
\end{lem}

\begin{lem}\label{lem:network-cooef}
	Write $\boldsymbol{x}=(x_1, \ldots, x_{\tilde{m}})^{\T}$ for some general integer $\tilde{m}\geq1$. Let $K\subset \mathbb{R}^{\tilde{m}}$ be a polytope bounded by hyperplanes $\boldsymbol{v}_{j}^{\T} \boldsymbol{x} + w_{j} \le 0$ for any $j\in[H]$, where $\boldsymbol{v}_{1}, \ldots, \boldsymbol{v}_{H} \in \mathbb{R}^{\tilde{m}}$ and $w_{1}, \ldots, w_{H}\in \mathbb{R}$. 
	Let $\tilde{a}_n\ge 1$, and $\tilde{M}_n\in \mathbb{N}$ be sufficiently large (independent of the size of $\tilde{a}_n$, but $\tilde{a}_n\le \tilde{M}_n$ must hold).  For any $\delta>0$, define
	\begin{align*}
		K_{\delta}^{\rm o}:=& \,\big\{\boldsymbol{x} \in \mathbb{R}^{\tilde{m}}:\, \boldsymbol{v}_{j}^{\T}\boldsymbol{x} + w_{j} \le -\delta ~\textrm{for all}~ j\in[H]\big\}\,,\\
		K_{\delta}^{\rm c}:=& \,\big\{\boldsymbol{x} \in \mathbb{R}^{\tilde{m}}:\, \boldsymbol{v}_{j}^{\T}\boldsymbol{x} + w_{j} \ge \delta ~\textrm{for some}~ j\in[H]\big\}\,.
	\end{align*}
	Let $\tilde{N} \in \mathbb{N}_0$ and $\tilde{N} \ge \tilde{\vartheta}$, where  $\vartheta= \tilde{\vartheta} + s$ for $\tilde{\vartheta} \in \mathbb{N}_{0}$ and $s\in(0,1]$ with $\vartheta$ given in Lemma {\rm\ref{lem:network-hiera}}. 	Let $f:\mathbb{R}^{\tilde{m}} \to \mathbb{R}$ be a polynomial from $\mathcal{P}_{\tilde{N}}$ with $\bar{r}(f)$ defined as in Lemma {\rm\ref{lem:polynomials}}. Then, there exists a function 
	\begin{align*}
		t(\boldsymbol{x}) =\sum_{j=1}^{(\tilde{N}+1){\rm C}_{\tilde{m}+\tilde{N}}^{\tilde{m}}} \mu_{j}   \sigma\bigg\{ \sum_{l=1}^{2\tilde{m}+H} \lambda_{j,l}  \sigma\bigg( \sum_{v=1}^{\tilde{m}} \theta_{l,v} x_{v} + \theta_{l,0}\bigg) + \lambda_{j,0} \bigg\} 
	\end{align*}
	with $\sigma(x)=(1+e^{-x})^{-1}$ for any $x\in \mathbb{R}$,  such that 
	\begin{align}
		&|t(\boldsymbol{x}) -f(\boldsymbol{x})| \le   \frac{\tilde{C}_{4} H\tilde{a}_n^{\tilde{N}+3} ({\rm C}_{\tilde{m}+\tilde{N}}^{\tilde{m}})^2 \bar{r}(f)} {(\tilde{M}_n+1)^{\vartheta} }\,,~~~\boldsymbol{x} \in K_{\delta}^{\rm o} \cap [-\tilde{a}_n,\tilde{a}_n]^{\tilde{m}}\,,\label{eq:lem-t312-1}\\
		&~~~~~~~~~~~~~|t(\boldsymbol{x})| \le \frac{\tilde{C}_5 ({\rm C}_{\tilde{m}+\tilde{N}}^{\tilde{m}})^2 \bar{r}(f) }{(\tilde{M}_n+1)^{2\vartheta+\tilde{m}+1} }\,, ~~~ \boldsymbol{x} \in K_{\delta}^{\rm c} \cap [-\tilde{a}_n,\tilde{a}_n]^{\tilde{m}}\,, \label{eq:lem-t312-2}\\
		&~~~~~~~~~~~~~~\,|t(\boldsymbol{x})| \le \tilde{C}_{6}  ({\rm C}_{\tilde{m}+\tilde{N}}^{\tilde{m}})^2 \bar{r}(f) (\tilde{M}_n+1)^{\tilde{N}\vartheta} \,, ~~~  \boldsymbol{x} \in \mathbb{R}^{\tilde{m}}\,, \label{eq:lem-t312-3}
	\end{align} 
	where $\tilde{C}_4>0$, $\tilde{C}_5>0$ and $\tilde{C}_6>0$ are some universal constants only depending on $(\tilde{m},\tilde{N})$.  Here the coefficients satisfy
	\begin{align*}
		&~~~~~~~~~~~~|\mu_{j}|\le  \tilde{C}_{2} {\rm C}_{\tilde{m}+\tilde{N}}^{\tilde{m}} \bar{r}(f)(\tilde{M}_n+1)^{\tilde{N}\vartheta} \,, \qquad |\lambda_{j,l}| \le   \tilde{C}_{7} (\tilde{M}_n+1)^{\tilde{m}+1+\vartheta(\tilde{N}+2)}\,,\\
		&|\theta_{l,v}| \le  \max\bigg[ \frac{1}{(\tilde{M}_n+1)^{\vartheta(\tilde{N}+1)}}, \frac{(\tilde{M}_n+1)^{\tilde{m}+1+\vartheta(2\tilde{N}+3)}}{\delta} \cdot \max\{|\boldsymbol{v}_{1}|_{\infty}, \ldots,|\boldsymbol{v}_{H}|_{\infty}, |w_{1}|, \ldots, |w_{H}| \}\bigg] 
	\end{align*} 
	for all $j\in [(\tilde{N}+1){\rm C}_{\tilde{m}+\tilde{N}}^{\tilde{m}}]$, $l\in[2\tilde{m}+H] \cup\{0\}$ and $v\in [\tilde{m}] \cup\{0\}$, where $\tilde{C}_2>0$ is specified in Lemma {\rm\ref{lem:polynomials}},  and  $\tilde{C}_7>0$ is a universal constant only depending on $(\tilde{m},\tilde{N})$. 
\end{lem}

\begin{lem}\label{lem:taylor}
	Let $\tilde{m} \ge 1$ be a general integer, and $f :\mathbb{R}^{\tilde{m}} \to \mathbb{R}$ be a $(\vartheta, C)$-smooth function with $(\vartheta,C)$ given in Lemma {\rm\ref{lem:network-hiera}}, where $\vartheta= \tilde{\vartheta} + s$ for $\tilde{\vartheta} \in \mathbb{N}_{0}$ and $s\in(0,1]$.  Let $p_{\tilde{\vartheta}} $ be the Taylor   polynomial of the total degree $\tilde{\vartheta}$ around $\boldsymbol{x}_{0}$ with $\boldsymbol{x}_{0} =(x_{0,1},\ldots, x_{0,\tilde{m}})^{\T}\in \mathbb{R}^{\tilde{m}}$, i.e.,
	\begin{align*}
		p_{\tilde{\vartheta}}(\boldsymbol{x}) = \sum_{\substack{j_1, \ldots, j_{\tilde{m}} \in\{0\}\cup[\tilde{\vartheta}],\\j_1+\cdots+ j_{\tilde{m}}\le \tilde{\vartheta}}} \bigg\{\frac{1}{j_1! \cdots j_{\tilde{m}}!} \cdot \frac{\partial^{j_1+\cdots+j_{\tilde{m}}} f}{\partial^{j_1}x_{1}\cdots \partial^{j_{\tilde{m}}}x_{\tilde{m}}}(\boldsymbol{x}_{0}) \cdot (x_{1}-x_{0,1})^{j_1} \cdots (x_{\tilde{m}}- x_{0,\tilde{m}})^{j_{\tilde{m}}}\bigg\}\,,
	\end{align*} 
	where $\boldsymbol{x}=(x_{1},\ldots,x_{\tilde{m}})^{\T}$. For any $\boldsymbol{x} \in \mathbb{R}^{\tilde{m}}$, it holds that
	\begin{align*}
		|f(\boldsymbol{x}) - p_{\tilde{\vartheta}}(\boldsymbol{x})| \le C\tilde{C}_{8} \tilde{m}^{\tilde{\vartheta}} \cdot |\boldsymbol{x} -\boldsymbol{x}_{0}|_{2}^{\vartheta} \,,
	\end{align*}
	where $\tilde{C}_{8}>0$ is a universal constant only depending on $\tilde{\vartheta}$.
\end{lem}

\begin{lem}\label{lem:network-pc}
	Write $\boldsymbol{x}=(x_1, \ldots, x_{\tilde{m}})^{\T}$ for some general integer $\tilde{m}\geq1$. Let $\tilde{a}_n\ge 1$, and $\tilde{M}_n\in \mathbb{N}$ be sufficiently large (independent of the size of $\tilde{a}_n$, but $\tilde{a}_n\le \tilde{M}_n$ must hold). Let $f:\mathbb{R}^{\tilde{m}} \to \mathbb{R}$ be a $(\vartheta, C)$-smooth function with $(\vartheta,C)$ given in Lemma {\rm\ref{lem:network-hiera}}, which satisfies
	\begin{align}\label{eq:lemm-partial-a}
		\max_{\substack{j_1, \ldots, j_{\tilde{m}} \in\{0\}\cup[\tilde{\vartheta}],\\j_1+\cdots+ j_{\tilde{m}}\le \tilde{\vartheta}}} \bigg| \frac{\partial^{j_1+\cdots+j_{\tilde{m}}} f}{\partial^{j_1}x_{1}\cdots \partial^{j_{\tilde{m}}}x_{\tilde{m}}}  \bigg|_{\infty, [-2\tilde{a}_n,2\tilde{a}_n]^{\tilde{m}} }  \le  B 
	\end{align}
	with some universal constant $B > 0$. Let $\tilde{N} \in \mathbb{N}_0$ and $\tilde{N} \ge \tilde{\vartheta}$, where  $\vartheta= \tilde{\vartheta} + s$ for $\tilde{\vartheta} \in \mathbb{N}_{0}$ and $s\in(0,1]$, and  $\mu$ be an arbitrary measure on $(\mathbb{R}^{\tilde{m}},\mathcal{B}(\mathbb{R}^{\tilde{m}}))$ such that 
	$$
	\mu(\underbrace{\mathbb{R}\times\cdots\times\mathbb{R}}_{j-1}\times[-\tilde{a}_n,\tilde{a}_n]\times\underbrace{\mathbb{R}\times\cdots\times \mathbb{R}}_{\tilde{m}-j})\leq1\,,~~~j\in[\tilde{m}]\,,
	$$
	where $\mathcal{B}(\mathbb{R}^{\tilde{m}})$ is the Borel sets of $\mathbb{R}^{\tilde{m}}$. Then, for any $\tilde{\eta}_n  \in(0,1)$, there exist a measurable set $\boldsymbol{D}\subset[-\tilde{a}_n,\tilde{a}_n]^{\tilde{m}}$ and a neural network of the type
	\begin{align*}
		t(\boldsymbol{x}) =\sum_{j=1}^{(\tilde{N}+1)( \tilde{M}_n +1)^{\tilde{m}} {\rm C}_{\tilde{m}+\tilde{N}}^{\tilde{m}}} \mu_{j}   \sigma\bigg\{ \sum_{l=1}^{4\tilde{m}} \lambda_{j,l}  \sigma\bigg( \sum_{v=1}^{\tilde{m}} \theta_{j,l,v} x_{v} + \theta_{j,l,0}\bigg) + \lambda_{j,0} \bigg\} 
	\end{align*}
	with $\sigma(x)=(1+e^{-x})^{-1}$ for any $x\in \mathbb{R}$,	such that $\mu(\boldsymbol{D})\leq \tilde{\eta}_n$ and 
	\begin{align*}
		|t(\boldsymbol{x}) - f(\boldsymbol{x})| \le  \tilde{C}_{9}  \tilde{M}_n  ^{-\vartheta}\{({\rm C}_{\tilde{m}+\tilde{N}}^{\tilde{m}})^3 + \tilde{m}^{\tilde{\vartheta} + \vartheta/2}\}   \tilde{a}_n^{\tilde{N}+3 + \tilde{\vartheta}} 
	\end{align*}
	holds for $\boldsymbol{x}\in [-\tilde{a}_n,\tilde{a}_n]^{\tilde{m}}\setminus\boldsymbol{D}$, where $\tilde{C}_9>0$ is a universal constant only depending on $(\tilde{m},\tilde{N},B)$. The coefficients of $t(\boldsymbol{x})$ satisfy 
	\begin{align*}
		&|\mu_{j}|\le  \tilde{C}_{10} ({\rm C}_{\tilde{m}+\tilde{N}}^{\tilde{m}} )^2   \tilde{a}_n^{\tilde{\vartheta}} (\tilde{M}_n+1)^{\tilde{N}\vartheta}   \,, \qquad |\lambda_{j,l}| \le   \tilde{C}_{7} (\tilde{M}_n+1)^{\tilde{m}+1+\vartheta(\tilde{N}+2)}\,,  \\
		&~~~~~~~~~~~~~~~~~~~~~~ |\theta_{j,l,v}| \le    4\tilde{\eta}_n^{-1}\tilde{m} (\tilde{M}_n+1)^{\tilde{m}+2+\vartheta(2\tilde{N}+3)}
	\end{align*}  
	for all $j\in [(\tilde{N}+1)(\tilde{M}_n+1)^{\tilde{m}}{\rm C}_{\tilde{m}+\tilde{N}}^{\tilde{m}}]$,  $l\in[4\tilde{m}]\cup\{0\}$ and $v\in[\tilde{m}]\cup\{0\}$. Here $\tilde{C}_{10}>0$ is a universal constant only depending on $(\tilde{m},\tilde{N},B)$, and $\tilde{C}_7>0$ is specified in Lemma  {\rm\ref{lem:network-cooef}}.
\end{lem}

We will prove Lemma \ref{lem:network-hiera} by mathematical induction. If $\ell=0$, by Definition \ref{def:hierarchical}, $f(\boldsymbol{x})$ can be expressed by  $f(\boldsymbol{x}) = h_1(\boldsymbol{\phi}_{1}^{\T}\boldsymbol{x}, \ldots, \boldsymbol{\phi}_{m_{*}}^{\T}\boldsymbol{x})$, where $h_1$ is a $(\vartheta,C)$-smooth function and $\boldsymbol{\phi}_{1}, \ldots,\boldsymbol{\phi}_{m_{*}} \in \mathbb{R}^{m}$. Let $\bar{\boldsymbol{s}}(\boldsymbol{x})=(\boldsymbol{\phi}_{1}^{\T}\boldsymbol{x}, \ldots, \boldsymbol{\phi}_{m_{*}}^{\T}\boldsymbol{x})^{\T}$. Based on Definition \ref{def:function-f}, it holds that $\max_{k\in[m_{*}]}|\boldsymbol{\phi}_{k}|_{\infty} \le \tilde{C}$, which implies   $\bar{\boldsymbol{s}}(\boldsymbol{x}) \in [-\tilde{C}ma_{n}, \tilde{C}ma_{n}]^{m_{*}}$ for any $\boldsymbol{x}\in[-a_n,a_n]^{m}$. 
Applying Lemma \ref{lem:network-pc} with selecting    $(\tilde{m},\tilde{a}_n, \tilde{M}_n,   \tilde{N},B )=(m_{*}, \breve{C}ma_{n},M_{n}, N, \tilde{C})$  and $\mu(\cdot)=\mathbb{P}\{\bar{\boldsymbol{s}}(\bX)\in\cdot\}$, there exist a measurable set $\tilde{\boldsymbol{D}}_0 \subset [-\tilde{C}ma_{n}, \tilde{C}ma_{n}]^{m_{*}}$  and a neural network of the type
\begin{align}\label{eq:h1hat-def}
	\hat{h}_1(\tilde{\boldsymbol{x}}) =   \sum_{j=1}^{(N+1)(M_{n}+1)^{m_{*}} \cdot  {\rm C}_{m_{*}+N}^{m_{*}}} \mu_{j}  \sigma\bigg\{ \sum_{l=1}^{4m_{*}} \lambda_{j,l}   \sigma\bigg( \sum_{k=1}^{m_{*}} \theta_{j,l,k} \tilde{x}_{k} + \theta_{j,l,0}\bigg) + \lambda_{j,0} \bigg\} 
\end{align}
with $\sigma(x)=(1+e^{-x})^{-1}$ for any $x\in \mathbb{R}$,	 such that $\mathbb{P}\{\bar{\boldsymbol{s}}(\bX) \in \tilde{\boldsymbol{D}}_{0}\} \le c\eta_{n}$ and  
\begin{align}\label{eq:h1h-h1}
	|\hat{h}_1(\tilde{\boldsymbol{x}}) - h_1(\tilde{\boldsymbol{x}})| \le&~ \tilde{C}_{11} M_{n}^{-\vartheta} \{({\rm C}_{m_{*}+N}^{m_{*}})^3 + m_{*}^{\tilde{\vartheta} + \vartheta/2}\}  (\tilde{C}m  a_{n}) ^{N+3 + \tilde{\vartheta}}   \\
	\le&~  \tilde{C}_{12}M_{n}^{-\vartheta}  m^{2N+3 }a_{n}^{2N+3 }  \,,~~~ \tilde{\boldsymbol{x}}  \in [-\tilde{C}m a_{n}, \tilde{C}m a_{n}]^{m_{*}} \backslash \tilde{\boldsymbol{D}}_{0} \notag 
\end{align}
holds  with the coefficients bounded as therein, where  $\tilde{C}_{11}>0$ and $\tilde{C}_{12}>0$ are some universal constants only depending on $(m_{*}, N,\tilde{C})$.
Let $\tilde{t}(\boldsymbol{x}) = \hat{h}_1\{\bar{\boldsymbol{s}}({\boldsymbol{x}})\}$. By \eqref{eq:h1h-h1}, it holds that
\begin{align*}
	| \tilde{t}(\boldsymbol{x}) -f(\boldsymbol{x}) |\le   \tilde{C}_{12}M_{n}^{-\vartheta}  m^{2N+3 }a_{n}^{2N+3 }   \,,~~~~ \boldsymbol{x}\in [-a_{n}, a_{n}]^{m}  \backslash \boldsymbol{D}_{0}\,,
\end{align*}
where $\boldsymbol{D}_{0} =\{\boldsymbol{x}\in \mathbb{R}^{m}:\bar{\boldsymbol{s}}(\boldsymbol{x}) \in \tilde{\boldsymbol{D}}_0 \}$ with $\mathbb{P}(\bX \in \boldsymbol{D}_{0}) \le \mathbb{P}\{\bar{\boldsymbol{s}}(\bX) \in \tilde{\boldsymbol{D}}_0\} \le c \eta_{n}$. Write $\boldsymbol{D}_0^{\rm c} = [-a_{n}, a_{n}]^{m}  \backslash \boldsymbol{D}_{0}$. Let
\begin{align}\label{tx-def-nn}
	t(\boldsymbol{x}) =  \tilde{t}(\boldsymbol{x}) \bigg(\frac{|f |_{\infty,\boldsymbol{D}_0^{\rm c}}}{|\tilde{t} |_{\infty,\boldsymbol{D}_0^{\rm c}}} \wedge 1 \bigg) \,.
\end{align}
Due to  $|f|_{\infty} \le \tilde{C}$   and $\tilde{\beta}_n = (\log n)\log^{1/2}(\tilde{d}n)$, then $|t|_{\infty,\boldsymbol{D}_0^{\rm c}} \le |f|_{\infty} \le \tilde{\beta}_{n}$ when $n$ is sufficiently large.  Since
\begin{align}\label{eq:tx-fx-bound-tilde}
	|t -f |_{\infty, \boldsymbol{D}_0^{\rm c}} \le |t - \tilde{t} |_{\infty,\boldsymbol{D}_0^{\rm c}} + |\tilde{t}  - f |_{\infty,\boldsymbol{D}_0^{\rm c}}  
	\le 2 |\tilde{t} - f |_{\infty,\boldsymbol{D}_0^{\rm c}}\,,
\end{align}
we have
\begin{align*}
	|t(\boldsymbol{x})-f(\boldsymbol{x})|\le  2 \tilde{C}_{12}M_{n}^{-\vartheta}  m^{2N+3 }a_{n}^{2N+3 }   \,,~~~~ \boldsymbol{x}\in [-a_{n}, a_{n}]^{m}  \backslash \boldsymbol{D}_{0} \,.
\end{align*}
Write $\boldsymbol{\phi}_{k}=(\phi_{k,1}, \ldots, \phi_{k,m})^{\T}$   and 
$$\tilde{\mu}_{j}=\mu_{j}\bigg(\frac{|f |_{\infty,\boldsymbol{D}_0^{\rm c}}}{|\tilde{t} |_{\infty,\boldsymbol{D}_0^{\rm c}}} \wedge 1\bigg)\,.$$    
By \eqref{eq:h1hat-def} and \eqref{tx-def-nn}, we have
\begin{align*}
	t(\boldsymbol{x}) =&~  \sum_{j=1}^{(N+1)(M_{n}+1)^{m_{*}} \cdot {\rm C}_{m_{*}+N}^{m_{*}}} \tilde{\mu}_{j}   \sigma\bigg\{ \sum_{l=1}^{4m_{*}} \lambda_{j,l}  \sigma\bigg( \sum_{k=1}^{m_{*}} \theta_{j,l,k}   \boldsymbol{\phi}_{k}^{\T} \boldsymbol{x} + \theta_{j,l,0}\bigg) + \lambda_{j,0} \bigg\} \\
	=&~\sum_{j=1}^{(N+1)(M_{n}+1)^{m_{*}} \cdot {\rm C}_{m_{*}+N}^{m_{*}}} \tilde{\mu}_{j}   \sigma\bigg\{ \sum_{l=1}^{4m_{*}} \lambda_{j,l} \sigma\bigg(\sum_{v=1}^{m} \sum_{k=1}^{m_{*}}  \phi_{k,v}\theta_{j,l,k} x_{v} + \theta_{j,l,0}\bigg) + \lambda_{j,0} \bigg\} \\
	=&~\sum_{j=1}^{(N+1)(M_{n}+1)^{m_{*}} \cdot {\rm C}_{m_{*}+N}^{m_{*}}} \tilde{\mu}_{j}   \sigma\bigg\{ \sum_{l=1}^{4m_{*}} \lambda_{j,l}   \sigma\bigg(\sum_{v=1}^{m} \tilde{\theta}_{j,l,v} x_{v} + \tilde{\theta}_{j,l,0}\bigg) + \lambda_{j,0} \bigg\} \,,
\end{align*}
where $\tilde{\theta}_{j,l,v} = \sum_{k=1}^{m_{*}} \phi_{k,v}\theta_{j,l,k}$ and $\tilde{\theta}_{j,l,0}=\theta_{j,l,0}$. Recall $\tilde{\alpha}_{n} = \bar{C}(c\eta_{n})^{-1} m^{\tilde{\vartheta}} M_{n}^{m_{*}+2+\vartheta(2N+3)}$  for some  sufficiently large constant $\bar{C}>0$ . By Lemma \ref{lem:network-pc}, for sufficiently large $n$, it holds that
\begin{align*}
	&~|\tilde{\mu}_{j}| \le  \tilde{C}_{13} ({\rm C}_{m_{*}+N}^{m_{*}} )^2  (\tilde{C}ma_{n})^{\tilde{\vartheta}} (M_{n}+1)^{N\vartheta}   \le \tilde{\alpha}_{n}\,,\\
	&~~~~~~~ |\lambda_{j,l}| \le   \tilde{C}_{14} (M_{n}+1)^{m_{*}+1+\vartheta(N+2)} \le \tilde{\alpha}_{n}\,,\\
	&|\tilde{\theta}_{j,l,v}| \le    4 \tilde{C} (c\eta_{n})^{-1}m_{*}^2 (M_{n}+1)^{m_{*}+2+\vartheta(2N+3)}  \le \tilde{\alpha}_{n}
\end{align*}
for any $j\in [(N+1)(M_{n}+1)^{m_{*}}{\rm C}_{m_{*}+N}^{m_{*}}]$, $l\in[4m_{*}]\cup\{0\}$ and $v\in[m]\cup\{0\}$, where  $\tilde{C}_{13}>0$ and $\tilde{C}_{14}>0$ are  some universal constants only depending on $(m_{*}, N,\tilde{C})$.   Notice that all coefficients of $t(\boldsymbol{x})$ can be bounded by $\tilde{\alpha}_{n}$. Hence,  $t(\boldsymbol{x}) \in \mathcal{H}^{(0)}=\mathcal{F}^{\rm NN}_{M_{*},m_{*},m,\tilde{\alpha}_{n}} $ with $M_{*}= (N+1)(M_{n}+1)^{m_{*}}  \cdot {\rm C}_{m_{*}+N}^{m_{*}}$, which means that the assertion of Lemma \ref{lem:network-hiera} holds for $\ell=0$.

We assume the assertion of Lemma \ref{lem:network-hiera} holds for $\ell =\bar{l}-1$. When $\bar{l}\ge 1$, by Definition \ref{def:hierarchical}, $f(\boldsymbol{x})$ can be expressed by  $f(\boldsymbol{x})=\sum_{k=1}^{K}h_{k}\{\tilde{h}_{1,k}(\boldsymbol{x}), \ldots, \tilde{h}_{m_{*}, k}(\boldsymbol{x})\} $, where all the  $\tilde{h}_{j,k}$ satisfy  $(\vartheta,C)$-smooth generalized hierarchical interaction model of order $m_{*}$ and level $\bar{l}-1$.   It follows Definition \ref{def:function-f} that  $\tilde{h}_{j,k} \in \mathcal{F}(m,m_{*}, \bar{l}-1 ,K,\vartheta,  L,C, \tilde{C})$.  
Then there exists a neural network  $\hat{\tilde{h}}_{j,k}  \in \{t\in\mathcal{H}^{(\bar{l}-1)}: |t|_{\infty,[-a_{n}, a_{n}]^{m}  \backslash \boldsymbol{D}_{j,k}} \le \tilde{\beta}_{n} \}$ 
such that   
\begin{align}\label{eq:htilde-jk} 
	|\hat{\tilde{h}}_{j,k}(\boldsymbol{x}) - \tilde{h}_{j,k}(\boldsymbol{x})| \le  C_{*}  M_{n}^{-\vartheta}  m^{2N+3 }a_{n}^{2N+3 }   \,, ~~~~ \boldsymbol{x}\in [-a_{n}, a_{n}]^{m}  \backslash \boldsymbol{D}_{j,k}  
\end{align}   
holds with $\mathbb{P}(\bX \in  \boldsymbol{D}_{j,k}) \le c\eta_{n}(2Km_{*})^{-1}$, where $\boldsymbol{D}_{j,k} \subset [-a_{n}, a_{n}]^{m}$ and $C_{*}>0$ is a universal constant only depending on $(m_{*}, N)$.  Write  $\hat{\tilde{\boldsymbol{h}}}_{k}(\boldsymbol{x}) = \{\hat{\tilde{h}}_{1,k}(\boldsymbol{x}), \ldots, \hat{\tilde{h}}_{m_{*},k}(\boldsymbol{x})\}^{\T}$ and $\bar{h}_{k,\max}=\max_{j\in[m_{*}] }|\tilde{h}_{j,k}|_{\infty,[-a_{n},a_{n}]^{m}}$. Due to $f \in  \mathcal{F}(m,m_{*}, \bar{l} ,K,\vartheta,  L,C, \tilde{C})$, then $\bar{h}_{k, \max} \le \tilde{C}$.  Since $ m^{2N+3}a_{n}^{2N+3} \ll M_{n}^{\vartheta} $, by \eqref{eq:htilde-jk}, it holds that
\begin{align*}
	\hat{\tilde{\boldsymbol{h}}}_{k}(\boldsymbol{x}) \in [-\tilde{C}-\tilde{C}_{15}, \tilde{C}+\tilde{C}_{15}]^{m_{*}}\,,~~~  \boldsymbol{x} \in [-a_{n}, a_{n}]^{m} \setminus \bigg(\bigcup_{j\in[m_{*}]   } \boldsymbol{D}_{j,k}\bigg) \,,
\end{align*}
where $\tilde{C}_{15}>0$ is a universal constant only depending on $(m_{*}, N)$.   For each given  $k\in[K]$, applying Lemma \ref{lem:network-pc} with selecting $(\tilde{m},\tilde{a}_n, \tilde{M}_n, \tilde{N},B)=(m_{*}, \tilde{C}_{15}+\tilde{C} ,M_n,  N,\tilde{C})$ and $\mu(\cdot)=\mathbb{P}\{\hat{\tilde{\boldsymbol{h}}}_{k}(\bX)\in\cdot\}$, there exist a measurable set $\tilde{\boldsymbol{D}}_{k} \subset [-\tilde{C}_{15}-\tilde{C} , \tilde{C}_{15}+\tilde{C} ]^{m_{*}}$ and a neural network
\begin{align}\label{eq:hh-k}
	\hat{h}_{k} (\breve{\boldsymbol{x}}) =   \sum_{j=1}^{(N+1)(M_{n}+1)^{m_{*}} \cdot {\rm C}_{m_{*}+N}^{m_{*}}} \mu_{j}^{(k)}  \sigma\bigg\{ \sum_{l=1}^{4m_{*}} \lambda_{j,l}^{(k)}  \sigma\bigg( \sum_{s=1}^{m_{*}} \theta_{j,l,s}^{(k)} \breve{x}_{s} + \theta_{j,l,0}^{(k)}\bigg) + \lambda_{j,0}^{(k)} \bigg\}  
\end{align}
with $\sigma(x)=(1+e^{-x})^{-1}$ for any $x\in \mathbb{R}$, such that $\mathbb{P}\{\hat{\tilde{\boldsymbol{h}}}_{k}(\bX) \in\tilde{\boldsymbol{D}}_{k}\} \le c \eta_{n}(2K)^{-1}$ and  
\begin{align}\label{eq:hx-hxh}
	| \hat{h}_{k}(\breve{\boldsymbol{x}}) -h_{k}(\breve{\boldsymbol{x}}) |  \le &~  \tilde{C}_{16}  M_{n}^{ -\vartheta} \{({\rm C}_{m_{*}+N}^{m_{*}})^3 + (m_{*})^{\tilde{\vartheta} + \vartheta/2}\}  (\tilde{C}_{15}+\tilde{C} )^{N+3 + \tilde{\vartheta}}   \\
	\le &~   \tilde{C}_{17}  M_{n}^{-\vartheta}      \,, ~~~~  \breve{\boldsymbol{x}}   \in [-\tilde{C}_{15}-\tilde{C} , \tilde{C}_{15}+\tilde{C} ]^{m_{*}} \backslash \tilde{\boldsymbol{D}}_{k} \notag
\end{align} 
with  the coefficients satisfying 
\begin{align*}
	&|\mu_{j}^{(k)}| \le  \tilde{C}_{18} ({\rm C}_{m_{*}+N}^{m_{*}} )^2  (\tilde{C}_{15}+\tilde{C} )^{\tilde{\vartheta}} (M_{n}+1)^{N\vartheta} \le \tilde{\alpha}_{n}\,,\\
	&~~~~~~~~ |\lambda_{j,l}^{(k)}| \le   \tilde{C}_{19} (M_{n}+1)^{m_{*}+1+\vartheta(N+2)} \le \tilde{\alpha}_{n}\,,\\
	&|\theta_{j,l,v}^{(k)}| \le     8K (c\eta_{n})^{-1} m_{*} (M_{n}+1)^{m_{*}+2+\vartheta(2N+3)}  \le \tilde{\alpha}_{n}
\end{align*}
for any $j\in [(N+1)(M_{n}+1)^{m_{*}}{\rm C}_{m_{*}+N}^{m_{*}}]$, $l\in[4m_{*}]\cup\{0\}$ and $v\in[m^*]\cup\{0\}$.  Here $\tilde{C}_{16}>0$,  $\tilde{C}_{17}>0$, $\tilde{C}_{18}>0$ and $\tilde{C}_{19}>0$ are some universal constants only depending on $(m_{*}, N)$. Thus, we know $\hat{h}_{k} \in \mathcal{F}^{\rm NN}_{M_{*},m_{*},m_{*},\tilde{\alpha}_{n}}$ with $M_{*}=(N+1)(M_{n}+1)^{m_{*}}\cdot{\rm C}_{m_{*}+N}^{m_{*}}$.  By \eqref{eq:hx-hxh}, it holds that
\begin{align}\label{eq:hk-hat-hk}
	&| \hat{h}_{k}\{\hat{\tilde{\boldsymbol{h}}}_{k}({\boldsymbol{x}})\} - h_{k}\{\hat{\tilde{\boldsymbol{h}}}_{k}(\boldsymbol{x})\} | \le   \tilde{C}_{17} M_{n}^{-\vartheta}     \,,  ~~\boldsymbol{x}\in [-a_{n}, a_{n}]^{m}\setminus \bigg\{\bigg(\bigcup_{j\in[m_{*}] } \boldsymbol{D}_{j,k} \bigg)\cup \boldsymbol{D}_{k}\bigg\} \,,
\end{align}
where $\boldsymbol{D}_{k} =\{\boldsymbol{x}\in \mathbb{R}^{m}:\hat{\tilde{\boldsymbol{h}}}_{k}(\boldsymbol{x}) \in \tilde{\boldsymbol{D}}_k \}$ with $\mathbb{P}(\bX \in \boldsymbol{D}_{k}) \le \mathbb{P}\{\hat{\tilde{\boldsymbol{h}}}_{k}(\bX) \in \tilde{\boldsymbol{D}}_k\} \le c \eta_{n}(2K)^{-1}$. Write $\bar{\boldsymbol{D}}^{\rm c} = [-a_{n}, a_{n}]^{m}\backslash \{(\cup_{j\in[m_{*}],k\in[K]} \boldsymbol{D}_{j,k} )\cup (\cup_{k\in[K]}\boldsymbol{D}_{k})\}$. Let 
\begin{align*}
	t(\boldsymbol{x}) =  \tilde{t}(\boldsymbol{x}) \bigg(\frac{|f|_{\infty,\bar{\boldsymbol{D}}^{\rm c}}}{|\tilde{t}|_{\infty,\bar{\boldsymbol{D}}^{\rm c}}} \wedge 1 \bigg)~~{\rm with}~~ \tilde{t}(\boldsymbol{x}) = \sum_{k=1}^{K}\hat{h}_{k}\{\hat{\tilde{\boldsymbol{h}}}_{k}({\boldsymbol{x}})\}  \,.
\end{align*}
Then $|t|_{\infty,\bar{\boldsymbol{D}}^{\rm c}} \le |f|_{\infty } \le \tilde{\beta}_{n}$.  Recall  $\hat{\tilde{\boldsymbol{h}}}_{k}(\boldsymbol{x}) = \{\hat{\tilde{h}}_{1,k}(\boldsymbol{x}), \ldots, \hat{\tilde{h}}_{m_{*},k}(\boldsymbol{x})\}^{\T}$. By \eqref{eq:hh-k}, we have 
\begin{align*}
	t(\boldsymbol{x}) = \sum_{k=1}^{K}\sum_{j=1}^{(N+1)(M_{n}+1)^{m_{*}} \cdot {\rm C}_{m_{*}+N}^{m_{*}}}  \tilde{\mu}_{j}^{(k)}  \sigma\bigg\{ \sum_{l=1}^{4m_{*}} \lambda_{j,l}^{(k)}   \sigma\bigg( \sum_{s=1}^{m_{*}} \theta_{j,l,s}^{(k)} \hat{\tilde{h}}_{s,k}(\boldsymbol{x}) + \theta_{j,l,0}^{(k)}\bigg) + \lambda_{j,0}^{(k)} \bigg\} 
\end{align*}
with
$$\tilde{\mu}_{j}^{(k)}= \mu_{j}^{(k)}   \bigg(\frac{|f|_{\infty,\bar{\boldsymbol{D}}_{k}^{\rm c}}}{|\tilde{t}|_{\infty,\bar{\boldsymbol{D}}_{k}^{\rm c}} } \wedge 1 \bigg)\,.$$
Due to  $\hat{\tilde{h}}_{j,k}\in \mathcal{H}^{(\bar{l}-1)}$,  by  \eqref{eq:nnspace-hl}, we  have $t(\boldsymbol{x})\in \mathcal{H}^{(\bar{l})}$.   Notice that $h_{k}$ is Lipschitz continuous with Lipschitz constant $L>0$. By \eqref{eq:htilde-jk} and \eqref{eq:hk-hat-hk}, it holds that
\begin{align*}
	|\tilde{t}(\boldsymbol{x}) - f(\boldsymbol{x})   | 
	\le&~ \bigg| \sum_{k=1}^{K}\hat{h}_{k}\{\hat{\tilde{h}}_{1,k}(\boldsymbol{x}), \ldots, \hat{\tilde{h}}_{m_{*}, k}(\boldsymbol{x})\} -\sum_{k=1}^{K}h_{k}\{\hat{\tilde{h}}_{1,k}(\boldsymbol{x}), \ldots, \hat{\tilde{h}}_{m_{*}, k}(\boldsymbol{x})\} \bigg| \notag \\
	&+ \bigg| \sum_{k=1}^{K}h_{k}\{\hat{\tilde{h}}_{1,k}(\boldsymbol{x}), \ldots, \hat{\tilde{h}}_{m_{*}, k}(\boldsymbol{x})\} -\sum_{k=1}^{K} h_{k}\{\tilde{h}_{1,k}(\boldsymbol{x}), \ldots, \tilde{h}_{m_{*}, k}(\boldsymbol{x})\} \bigg| \notag\\
	\le&~ \sum_{k=1}^{K} | \hat{h}_{k}\{\hat{\tilde{\boldsymbol{h}}}_{k}({\boldsymbol{x}})\} - h_{k}\{\hat{\tilde{\boldsymbol{h}}}_{k}(\boldsymbol{x})\} | + \sum_{k=1}^{K} L\cdot  \sum_{j=1}^{m_{*}} |  \hat{\tilde{h}}_{j,k}(\boldsymbol{x}) - \tilde{h}_{j,k}(\boldsymbol{x})|   \\
	\le&~  K  \tilde{C}_{17}  M_{n}^{-\vartheta}  + KLm_{*} \cdot C_{*} M_{n}^{-\vartheta}  m^{2N+3}a_{n}^{2N+3}    \\
	\le &~  \tilde{C}_{20} M_{n}^{-\vartheta} m^{2N+3}a_{n}^{2N+3}  \,, ~~~~ \boldsymbol{x} \in \bar{\boldsymbol{D}}^{\rm c} \,, 
\end{align*}
where $\tilde{C}_{20}>0$ is a universal constant only depending on $(m_{*}, N)$. Using the similar arguments for deriving  \eqref{eq:tx-fx-bound-tilde}, we have  
\begin{align*}
	|t(\boldsymbol{x})-f(\boldsymbol{x})| \le  2\tilde{C}_{20} M_{n}^{-\vartheta} m^{2N+3}a_{n}^{2N+3}  \,,~~~~ \boldsymbol{x} \in \bar{\boldsymbol{D}}^{\rm c}\,.
\end{align*}   
Moreover, it holds that
\begin{align*}
	\mathbb{P} \bigg\{\bX \in \bigg(\bigcup_{j\in[m_{*}],k\in[K]} \boldsymbol{D}_{j,k}\bigg) \cup \bigg(\bigcup_{k\in[K]} \boldsymbol{D}_{k}\bigg)\bigg \} \le&~ \sum_{j\in[m_{*}], k\in[K]} \mathbb{P} (\bX \in \boldsymbol{D}_{j,k}) + \sum_{k\in[K]}\mathbb{P} (\bX \in \boldsymbol{D}_{k})\\
	\le &~  \sum_{j\in[m_{*}], k\in[K]} \frac{c\eta_{n}}{2Km_{*}} + \sum_{k\in[K]}\frac{c\eta_{n}}{2K} =c\eta_{n}\,.
\end{align*}
Hence, we have Lemma \ref{lem:network-hiera} holds for $\ell =\bar{l}$. Based on the mathematical induction, we know Lemma \ref{lem:network-hiera} holds for given $\ell$. We complete the proof of Lemma \ref{lem:network-hiera}.
$\hfill\Box$

\subsubsection{Proof of Lemma \ref{lem:polynomials}}\label{sec:sub-polynomials} 

The proof of Lemma \ref{lem:polynomials} follows in a straightforward way from the proof of Lemma 5 and Remark 2 in \cite{Bauer2019}. Hence, we
omit it here.
$\hfill\Box$

\subsubsection{Proof of Lemma \ref{lem:network-cooef}}\label{sec:sub-network-cooef}

Select $R = (\tilde{M}_n+1)^{\vartheta}$ for some sufficiently large $\tilde{M}_n \in \mathbb{N}$ and $\vartheta$ given in Lemma \ref{lem:network-hiera}. Consider
\begin{align*}
	s(\boldsymbol{x}) =\sum_{j=1}^{(\tilde{N}+1){\rm C}_{\tilde{m}+\tilde{N}}^{\tilde{m}}} \tilde{b}_{j}  \sigma\bigg(\sum_{l=1}^{\tilde{m}} \tilde{a}_{j,l}x_{l} + \tilde{a}_{j,0} \bigg)\,,
\end{align*}
with $|\tilde{b}_{j}| \le \tilde{C}_{2} {\rm C}_{\tilde{m}+\tilde{N}}^{\tilde{m}} R^{\tilde{N}} \bar{r}(f)$ and $|\tilde{a}_{j,l} |\le  \tilde{C}_{3}\tilde{a}_nR^{-1}(\tilde{m}+1)^{-1} $ for $j\in[(\tilde{N}+1){\rm C}_{\tilde{m}+\tilde{N}}^{\tilde{m}}]$ and $ l\in [\tilde{m}]\cup \{0\}$, where $\bar{r}(f)$,   $\tilde{C}_{2}>0$ and $\tilde{C}_3>0$ are specified in Lemma \ref{lem:polynomials}.  Write $\boldsymbol{v}_{j}=(v_{j,1},\ldots, v_{j,\tilde{m}})^{\T}$ for  $j \in [H]$, and $B= (\tilde{M}_n+1)^{\tilde{m}+1+\vartheta(\tilde{N}+2)}$. We define 
\begin{align}\label{eq:tx-1-rep}
	&t(\boldsymbol{x}) =  \sum_{j=1}^{(\tilde{N}+1){\rm C}_{\tilde{m}+\tilde{N}}^{\tilde{m}}} \tilde{b}_{j}   \sigma\bigg\{\sum_{l=1}^{\tilde{m}} \tilde{a}_{j,l} \sum_{k=1}^{2}\gamma_{k}  \sigma(\varrho_{k}x_{l} ) - B \sum_{l'=1}^{H}\sigma\bigg(\sum_{s=1}^{\tilde{m}} a _{l',s}x_{s} + a_{l',0}\bigg) + \tilde{a}_{j,0} \bigg\}\,,
\end{align}
where $ a_{l',s}= v_{l',s}(\varsigma\delta)^{-1}$ and $ a_{l',0} = w_{l'}(\varsigma\delta)^{-1}$ with $\varsigma=(\tilde{M}_n+1)^{-(\tilde{m}+1)-\vartheta(2\tilde{N}+3)}$ for $l'\in [H]$ and $ s\in[\tilde{m}]$. 

Recall $K_{\delta}^{\rm o} = \{\boldsymbol{x} \in \mathbb{R}^{\tilde{m}}:\, \boldsymbol{v}_{j}^{\T}\boldsymbol{x} + w_{j} \le -\delta ~\textrm{for all}~ j\in[H] \}$. For any $\boldsymbol{x} \in K_{\delta}^{\rm o} \cap [-\tilde{a}_n,\tilde{a}_n]^{\tilde{m}}$, 
\begin{align*}
	|t(\boldsymbol{x})-f(\boldsymbol{x})| \le&~ \underbrace{\bigg|t(\boldsymbol{x} ) - \sum_{j=1}^{(\tilde{N}+1){\rm C}_{\tilde{m}+\tilde{N}}^{\tilde{m}}} \tilde{b}_{j}   \sigma\bigg\{\sum_{l=1}^{\tilde{m}} \tilde{a}_{j,l} \sum_{k=1}^{2}\gamma_{k}  \sigma(\varrho_{k}x_{l}  )  + \tilde{a}_{j,0} \bigg\}\bigg| }_{\tilde{\textrm{T}}_{1}(\boldsymbol{x})} \\
	&+ \underbrace{\bigg| \sum_{j=1}^{(\tilde{N}+1){\rm C}_{\tilde{m}+\tilde{N}}^{\tilde{m}}} \tilde{b}_{j}   \sigma\bigg\{\sum_{l=1}^{\tilde{m}} \tilde{a}_{j,l} \sum_{k=1}^{2}\gamma_{k} \sigma(\varrho_{k}x_{l}  )  + \tilde{a}_{j,0} \bigg\} - s(\boldsymbol{x})\bigg|}_{\tilde{\textrm{T}}_{2}(\boldsymbol{x})}  + \underbrace{|s(\boldsymbol{x})-f(\boldsymbol{x})|}_{\tilde{\textrm{T}}_{3}(\boldsymbol{x})} \,.
\end{align*}
By Lemma \ref{lem:polynomials}, we have
\begin{align*}
	\tilde{{\rm T}}_3(\boldsymbol{x})  \le \frac{\tilde{C}_1 ({\rm C}_{\tilde{m}+\tilde{N}}^{\tilde{m}})^2   \tilde{a}_n^{\tilde{N}+1} \bar{r}(f) }{( \tilde{M}_n+1)^{\vartheta}}\,,~~~~  \boldsymbol{x} \in [-\tilde{a}_n,\tilde{a}_n]^{\tilde{m}}\,,
\end{align*}
where $\tilde{C}_1>0$ is specified in Lemma \ref{lem:polynomials}.  
Due to $ a_{l',s}= v_{l',s}(\varsigma\delta)^{-1}$ and $ a_{l',0} = w_{l'}(\varsigma\delta)^{-1}$  for $l'\in [H]$ and $ s\in[\tilde{m}]$, then $\sum_{s=1}^{\tilde{m}} a _{l',s}x_{s} + a_{l',0} \le -\varsigma^{-1}$
for  any $\boldsymbol{x} \in K_{\delta}^{\rm o}$ and $l'\in[H]$.  Since $|\sigma(x)| \le |x|^{-1}$ for any $x<0$, we have 
\begin{align*}
	\bigg|\sum_{l'=1}^{H} \sigma\bigg(\sum_{s=1}^{\tilde{m}} a _{l',s}x_{s} + a_{l',0}\bigg)\bigg| \le \sum_{l'=1}^{H} \bigg|\sigma\bigg(\sum_{s=1}^{\tilde{m}} a _{l',s}x_{s} + a_{l',0}\bigg)\bigg|  \le H\varsigma \,.
\end{align*}
Since  $\sigma $ is Lipschitz continuous with the Lipschitz constant $C_{*}$,  it holds that 
\begin{align*}
	\tilde{{\rm T}}_1(\boldsymbol{x}) \le &~C_{*} B \sum_{j=1}^{(\tilde{N}+1){\rm C}_{\tilde{m}+\tilde{N}}^{\tilde{m}}} |\tilde{b}_{j}|   \bigg|  \sum_{l'=1}^{H}\sigma\bigg(\sum_{s=1}^{\tilde{m}} a _{l',s}x_{s} + a_{l',0}\bigg)\bigg| \\
	\le&~ \bar{C}_{1}  ({\rm C}_{\tilde{m}+\tilde{N}}^{\tilde{m}})^2 \bar{r}(f) R^{\tilde{N}} BH \varsigma 
	=  \frac{\bar{C}_{1} H ({\rm C}_{\tilde{m}+\tilde{N}}^{\tilde{m}})^2 \bar{r}(f) }{(\tilde{M}_n+1)^{\vartheta} }\,,~~~~ \boldsymbol{x} \in K_{\delta}^{\rm o}\cap[-\tilde{a}_n,\tilde{a}_n]^{\tilde{m}} \,,
\end{align*}
where $\bar{C}_1>0$ is a universal constant only depending on $(\tilde{m},\tilde{N})$. Select $\tilde{R}=(\tilde{M}_n+1)^{\vartheta(\tilde{N}+1)}$. 
By Lemma 4 of \cite{Bauer2019}, there exist coefficients $(\gamma_1,\gamma_2,\varrho_1,\varrho_2)$ satisfying $|\gamma_{k}| \le C_1\tilde{R} $ and  $|\varrho_{k}| \le  \tilde{R}^{-1}$ for some universal constant $C_1>0$ independent of $(\tilde{m},\tilde{N})$,  such that 
\begin{align*}
	\bigg| \sum_{k=1}^{2}\gamma_{k}  \sigma(\varrho_{k}x  ) -x\bigg| \le \frac{C_{2}\tilde{a}_n^2}{\tilde{R}} \,, ~~~~ x \in [-\tilde{a}_n,\tilde{a}_n]\,,
\end{align*}
where $C_2>0$ is a universal constant  independent of $(\tilde{m},\tilde{N})$.  Since  $\sigma$ is Lipschitz continuous  with the Lipschitz constant $C_{*}$,  it then holds that
\begin{align*}
	\tilde{{\rm T}}_2(\boldsymbol{x}) \le&~ C_{*}\sum_{j=1}^{(\tilde{N}+1){\rm C}_{\tilde{m}+\tilde{N}}^{\tilde{m}}} |\tilde{b}_{j}|   \bigg| \sum_{l=1}^{\tilde{m}}\tilde{a}_{j,l} \bigg\{\sum_{k=1}^{2}\gamma_{k}  \sigma(\varrho_{k}x_{l}  ) -x_{l} \bigg\}\bigg|\\
	\le&~ \bar{C}_2  {\rm C}_{\tilde{m}+\tilde{N}}^{\tilde{m}} R^{\tilde{N}} \bar{r}(f) \cdot \sum_{j=1}^{(\tilde{N}+1){\rm C}_{\tilde{m}+\tilde{N}}^{\tilde{m}} }\sum_{l=1}^{\tilde{m}}  |\tilde{a}_{j,l} | \bigg|\sum_{k=1}^{2}\gamma_{k}\ \sigma(\varrho_{k}x_{l}  ) -x_{l} \bigg| \\
	\le&~ \bar{C}_3  ({\rm C}_{\tilde{m}+\tilde{N}}^{\tilde{m}})^2 R^{\tilde{N}} \bar{r}(f) \tilde{m} \cdot  \frac{ \tilde{a}_n}{R(\tilde{m}+1)}   \cdot \frac{ \tilde{a}_n^2}{\tilde{R}}
	\le \frac{\bar{C}_3  \tilde{a}_n^3 ({\rm C}_{\tilde{m}+\tilde{N}}^{\tilde{m}})^2 \bar{r}(f)}{(\tilde{M}_n+1)^{\vartheta} }\,,~~~~\boldsymbol{x} \in [-\tilde{a}_n,\tilde{a}_n]^{\tilde{m}}\,,
\end{align*}
where $\bar{C}_2>0$ and $\bar{C}_3>0$ are some universal constants only depending on $(\tilde{m},\tilde{N})$.  
Hence,
\begin{align*}
	|t(\boldsymbol{x})-f(\boldsymbol{x})| \le \tilde{{\rm T}}_1(\boldsymbol{x}) + \tilde{{\rm T}}_2(\boldsymbol{x})+ \tilde{{\rm T}}_3(\boldsymbol{x}) \le \frac{\bar{C}_4  H ({\rm C}_{\tilde{m}+\tilde{N}}^{\tilde{m}})^2   \tilde{a}_n^{\tilde{N}+3}\bar{r}(f)}  {(  \tilde{M}_n+1)^{\vartheta} }\,, ~~~~ \boldsymbol{x} \in K_{\delta}^{\rm o} \cap [-\tilde{a}_n,\tilde{a}_n]^{\tilde{m}}\,,
\end{align*}
where $\bar{C}_4>0$ is a universal constant only depending on $(\tilde{m},\tilde{N})$. Then, we have \eqref{eq:lem-t312-1}. 

Recall $K_{\delta}^{\rm c}=\{\boldsymbol{x} \in \mathbb{R}^{\tilde{m}}:\, \boldsymbol{v}_{j}^{\T}\boldsymbol{x} + w_{j} \ge \delta ~\textrm{for some} ~ j\in[H]\}$, $ a_{l',s}= v_{l',s}(\varsigma\delta)^{-1}$ and $ a_{l',0} = w_{l'}(\varsigma\delta)^{-1}$ for $l'\in [H]$ and $s\in[\tilde{m}]$. For any  $\boldsymbol{x} \in K_{\delta}^{\rm c}$, there exits  $j_{*}\in[H]$ such that $\sum_{s=1}^{\tilde{m}} a _{j_{*},s}x_{s} + a_{j_{*},0} \ge {\varsigma}^{-1}$. Since $|\sigma(x)-1| \le x^{-1}$ for any $x>0$, then $|\sigma(\sum_{s=1}^{\tilde{m}} a _{j_{*},s}x_{s} + a_{j_{*},0})-1| \le \varsigma$, 
which implies
\begin{align*}
	\sum_{l'=1}^{H}\sigma\bigg(\sum_{s=1}^{\tilde{m}} a _{l',s}x_{s} + a_{l',0}\bigg) \ge \sigma\bigg(\sum_{s=1}^{\tilde{m}} a _{j_{*},s}x_{s} + a_{j_{*},0}\bigg) \ge 1-\varsigma\,.
\end{align*}
For $t(\cdot)$ defined in \eqref{eq:tx-1-rep}, we restrict the coefficients $(\gamma_{1},\gamma_{2})$ satisfying $|\gamma_{k}| \le C_1\tilde{R} $  with $C_1>0$ specified above. Since $\sigma $ is nondecreasing and $\sigma \in (0,1)$,  for any $\boldsymbol{x} \in K_{\delta}^{\rm c} \cap [-\tilde{a}_n,\tilde{a}_n]^{\tilde{m}}$, we have  
\begin{align*}
	|t(\boldsymbol{x})| \le&~ \sum_{j=1}^{(\tilde{N}+1){\rm C}_{\tilde{m}+\tilde{N}}^{\tilde{m}}} |\tilde{b}_{j} |   \sigma\bigg\{\sum_{l=1}^{\tilde{m}} \tilde{a}_{j,l} \sum_{k=1}^{2}\gamma_{k}  \sigma(\varrho_{k}x_{l}  ) - B \sum_{l'=1}^{H}\sigma\bigg(\sum_{s=1}^{\tilde{m}} a _{l',s}x_{s} + a_{l',0}\bigg) + \tilde{a}_{j,0} \bigg\}\\
	\le &~ \sum_{j=1}^{(\tilde{N}+1){\rm C}_{\tilde{m}+\tilde{N}}^{\tilde{m}}} |\tilde{b}_{j} |   \sigma\bigg\{2\tilde{C}_{3}C_1\tilde{m}\cdot\frac{ \tilde{a}_n}{R(\tilde{m}+1)} \cdot \tilde{R} - B(1-\varsigma) +   \frac{\tilde{C}_{3} \tilde{a}_n}{R(\tilde{m}+1)}   \bigg\}\\
	\le &~ \sum_{j=1}^{(\tilde{N}+1){\rm C}_{\tilde{m}+\tilde{N}}^{\tilde{m}}} |\tilde{b}_{j} |   \sigma\bigg\{ \frac{ \bar{C}_{5} \tilde{m} \tilde{a}_n \tilde{R}}{R(\tilde{m}+1)}    - B(1-\varsigma) \bigg\} \le  \sum_{j=1}^{(\tilde{N}+1){\rm C}_{\tilde{m}+\tilde{N}}^{\tilde{m}}} |\tilde{b}_{j} |   \sigma\bigg( \frac{ \bar{C}_{5}  \tilde{a}_n \tilde{R}}{R}    - B(1-\varsigma) \bigg)\,, 
\end{align*}
where $\bar{C}_5>0$ is a universal constant only depending on $(\tilde{m},\tilde{N})$. Recall $R=(\tilde{M}_n+1)^{\vartheta}$, $\tilde{R}=(\tilde{M}_n+1)^{\vartheta(\tilde{N}+1)}$, $B=(\tilde{M}_n+1)^{\tilde{m}+1+\vartheta(\tilde{N}+2)}$ and $\varsigma=(\tilde{M}_n+1)^{-(\tilde{m}+1)-\vartheta(2\tilde{N}+3)}$. Since $\tilde{a}_n \le \tilde{M}_n$, for sufficiently large $M_n \in \mathbb{N}$, it holds that
\begin{align*}
	\frac{ \bar{C}_{5}\tilde{a}_n \tilde{R}}{R }    \le  \bar{C}_{5}  (\tilde{M}_n+1)^{1+\vartheta\tilde{N}}  
	=  \bar{C}_5( \tilde{M}_n+1)^{-\tilde{m}-2\vartheta} B \le  B(3/4- \varsigma) \,. 
\end{align*}	
Due to $|\sigma(x)| \le |x|^{-1}$ for any $x<0$, then
\begin{align*}
	|t(\boldsymbol{x})|  
	\le &~ \sum_{j=1}^{(\tilde{N}+1){\rm C}_{\tilde{m}+\tilde{N}}^{\tilde{m}}} |\tilde{b}_{j} |   \sigma \bigg(-\frac{B}{4}\bigg)\le 4 \tilde{C}_{2}(\tilde{N}+1) ({\rm C}_{\tilde{m}+\tilde{N}}^{\tilde{m}})^2 \bar{r}(f) R^{\tilde{N}} B^{-1} \\
	\le&~  \frac{\bar{C}_{6}({\rm C}_{\tilde{m}+\tilde{N}}^{\tilde{m}})^2 \bar{r}(f)}{ ( \tilde{M}_n+1)^{2\vartheta+\tilde{m}+1} } \,, ~~~~~~~~ \boldsymbol{x} \in K_{\delta}^{\rm c} \cap[-\tilde{a}_n,\tilde{a}_n]^{\tilde{m}}\,,
\end{align*}
where $\bar{C}_{6}>0$ is a universal constant only depending on $(\tilde{m},\tilde{N})$. Hence, we have \eqref{eq:lem-t312-2}. 
Furthermore, it also holds that
\begin{align*}
	|t(\boldsymbol{x})| \le&~  \sum_{j=1}^{(\tilde{N}+1){\rm C}_{\tilde{m}+\tilde{N}}^{\tilde{m}}} |\tilde{b}_{j} |  \le \tilde{C}_{2} (\tilde{N}+1) ({\rm C}_{\tilde{m}+\tilde{N}}^{\tilde{m}})^2 \bar{r}(f) R^{\tilde{N}} \\
	\le&~ \bar{C}_{7}  ({\rm C}_{\tilde{m}+\tilde{N}}^{\tilde{m}})^2 \bar{r}(f) (\tilde{M}_n+1)^{\tilde{N}\vartheta}\,, ~~~~~~~~ \boldsymbol{x} \in \mathbb{R}^{\tilde{m}}\,,
\end{align*}
where $\bar{C}_{7}>0$ is a universal constant only depending on $(\tilde{m},\tilde{N})$. Thus, \eqref{eq:lem-t312-3}  holds. 

For $t(\cdot)$ defined in \eqref{eq:tx-1-rep}, we also restrict the coefficients $(\varrho_1,\varrho_2)$ satisfying $|\varrho_{k}| \le  \tilde{R}^{-1}$. We  can reformulate it as
\begin{align*}
	t(\boldsymbol{x}) =\sum_{j=1}^{(\tilde{N}+1){\rm C}_{\tilde{m}+\tilde{N}}^{\tilde{m}}} \mu_{j}   \sigma\bigg\{ \sum_{l=1}^{2\tilde{m}+H} \lambda_{j,l}  \sigma\bigg( \sum_{v=1}^{\tilde{m}} \theta_{l,v} x_{v} + \theta_{l,0}\bigg) + \lambda_{j,0} \bigg\}
\end{align*}
with $\mu_{j}=\tilde{b}_{j}$ for $j\in[(\tilde{N}+1){\rm C}_{\tilde{m}+\tilde{N}}^{\tilde{m}}]$, and
\begin{align*}
	\lambda_{j,l}=&\,\left\{
	\begin{aligned}
		\tilde{a}_{j,0}\,, ~~~~~~~~~~~~~~~~~~~~~~~~~~~ &{\rm if}~j\in[(\tilde{N}+1){\rm C}_{\tilde{m}+\tilde{N}}^{\tilde{m}}]\,,\,l=0\,,\\
		\tilde{a}_{j,\lceil l/2 \rceil } \cdot \gamma_{2-l+2\lfloor l/2 \rfloor} \,, ~~~~~~~~~&{\rm if}~j\in[(\tilde{N}+1){\rm C}_{\tilde{m}+\tilde{N}}^{\tilde{m}}]\,,\,l \in[2\tilde{m}]\,,\\
		-B\,,~~~~~~~~~~~~~~~~~~~~~~~~~~~ &{\rm if}~j\in[(\tilde{N}+1){\rm C}_{\tilde{m}+\tilde{N}}^{\tilde{m}}]\,,\,l \in [2\tilde{m}+H]\backslash[2\tilde{m}]\,,
	\end{aligned}
	\right. \\
	\theta_{l,v}=&\,\left\{
	\begin{aligned}
		0\,, ~~~~~~~~~~~~~~~~~~~~~~~~~~~~~~ &{\rm if}~l\in[2\tilde{m}]\,, \,v=0\,,\\
		\varrho_{2-l+2\lfloor l/2 \rfloor} \cdot I(\lceil l/2 \rceil=v) \,, ~~&{\rm if}~l \in[2\tilde{m}]\,,\, v\in[\tilde{m}]\,,\\
		a_{l-2\tilde{m},v}\,,~~~~~~~~~~~~~~~~~~~~~~~ &{\rm if}~l \in [2\tilde{m}+H]\backslash[2\tilde{m}]\,,\,v\in[\tilde{m}]\cup \{0\}\,,
	\end{aligned}
	\right.
\end{align*}
where the coefficients satisfy
\begin{align*}
	&~~~~~~~~~~~~~~~~~~~~~~~~~~~~~~~~~~~~~~|\mu_{j}|\le \tilde{C}_{2} {\rm C}_{\tilde{m}+\tilde{N}}^{\tilde{m}} \bar{r}(f)( \tilde{M}_n+1)^{\tilde{N}\vartheta} \,,\\
	&~~~~~~~~~~~~~~~|\lambda_{j,l}| \le  \max \bigg\{ \frac{\tilde{C}_3\tilde{a}_n}{R(\tilde{m}+1)} ,  \frac{ \tilde{C}_3 C_1\tilde{R}\tilde{a}_n}{R(\tilde{m}+1)}   , B\bigg\} \le \bar{C}_{8}( \tilde{M}_n+1)^{\tilde{m}+1+\vartheta(\tilde{N}+2)}\,,\\
	&|\theta_{l,v}| \le  \max\bigg[   \frac{1}{(\tilde{M}_n+1)^{\vartheta(\tilde{N}+1)}}, \frac{( \tilde{M}_n+1)^{\tilde{m}+1+\vartheta(2\tilde{N}+3)}}{\delta} \cdot \max\{|\boldsymbol{v}_{1}|_{\infty}, \ldots,|\boldsymbol{v}_{H}|_{\infty}, |w_{1}|, \ldots, |w_{H}| \}\bigg]  
\end{align*}
for $j\in [(\tilde{N}+1){\rm C}_{\tilde{m}+\tilde{N}}^{\tilde{m}}]$, $l\in[2\tilde{m}+H]\cup\{0\}$ and $v\in [\tilde{m}] \cup\{0\}$. Here  $\bar{C}_8>0$ is a universal constant only depending on $(\tilde{m},\tilde{N})$. Hence, we complete the proof of Lemma \ref{lem:network-cooef}.
$\hfill\Box$

\subsubsection{Proof of Lemma \ref{lem:taylor}}\label{sec:sub-taylor}
Recall $\vartheta= \tilde{\vartheta} + s$ for $\tilde{\vartheta} \in \mathbb{N}_{0}$ and $s\in(0,1]$. If $\tilde{\vartheta} =0$, Lemma \ref{lem:taylor} holds by the definition of $(\vartheta, C)$-smooth function. If $\tilde{\vartheta} \ge 1$, following the proof of Lemma 1 in \cite{Kohler2014}, we have
\begin{align*}
	|f(\boldsymbol{x}) - p_{\tilde{\vartheta}}(\boldsymbol{x})| \le&~ \sum_{\substack{j_1, \ldots, j_{\tilde{m}} \in\{0\}\cup[\tilde{\vartheta}],\\j_1+\cdots+ j_{\tilde{m}} = \tilde{\vartheta}}} \bigg\{\frac{\tilde{\vartheta}}{j_1! \cdots j_{\tilde{m}}!} \cdot  |\boldsymbol{x} -\boldsymbol{x}_{0}|_{2}^{\tilde{\vartheta}} \cdot C \int_{0}^{1} (1-t)^{\tilde{\vartheta}-1} t^{s} |\boldsymbol{x} -\boldsymbol{x}_{0}|_{2}^{s} \,{\rm d}t \bigg\}\\
	\le &~  \sum_{\substack{j_1, \ldots, j_{\tilde{m}} \in\{0\}\cup[\tilde{\vartheta}],\\j_1+\cdots+ j_{\tilde{m}} = \tilde{\vartheta}}} \bigg\{\frac{\tilde{\vartheta}}{j_1! \cdots j_{\tilde{m}}!} \cdot |\boldsymbol{x} -\boldsymbol{x}_{0}|_{2}^{\vartheta} \cdot C \int_{0}^{1} (1-t)^{\tilde{\vartheta}-1} t^{s}  \,{\rm d}t \bigg\} \\
	\le &~  \frac{C}{(\tilde{\vartheta}-1)!} \cdot \tilde{m}^{\tilde{\vartheta}} |\boldsymbol{x} -\boldsymbol{x}_{0}|_{2}^{\vartheta}   \,.
\end{align*}
Hence, we complete the proof of Lemma \ref{lem:taylor}.
$\hfill\Box$

\subsubsection{Proof of Lemma \ref{lem:network-pc}}\label{sec:sub-lem:network-pc}

We subdivide $[-\tilde{a}_n-2\tilde{a}_n/\tilde{M}_n, \tilde{a}_n]^{\tilde{m}}$  into $(\tilde{M}_n+1)^{\tilde{m}}$ cubes of side length $2\tilde{a}_n/\tilde{M}_n$. Let index ${\bf i} = (i_1, \ldots, i_{\tilde{m}}) \in [ \tilde{M}_n +1]^{\tilde{m}}$, and denote the corresponding cube by 
\begin{align*}
	\boldsymbol{C}_{\bf i} =&~\bigg[-\tilde{a}_n+\frac{2(i_1-2)\tilde{a}_n}{\tilde{M}_n}, -a_n + \frac{2(i_1-1)\tilde{a}_n}{\tilde{M}_n}\bigg]\times \cdots \\ &~~\times \bigg[-\tilde{a}_n+\frac{2(i_m-2)\tilde{a}_n}{\tilde{M}_n}, -\tilde{a}_n + \frac{2(i_m-1)\tilde{a}_n}{\tilde{M}_n}\bigg]\,.
\end{align*} 
Moreover, we denote the corners of these cubes by $\boldsymbol{x}_{\bf i}=(x_{{\bf i},1},\ldots, x_{{\bf i}, \tilde{m}})^{\T}$ for ${\bf i} \in [ \tilde{M}_n+2]^{\tilde{m}}$ in the same way, such that for all $\boldsymbol{C}_{\bf i}$, the point $\boldsymbol{x}_{\bf i}$ means the ``bottom left" corner of this cube and the additional indices result from the right border of the whole grid. For any $\boldsymbol{x}=(x_1,\ldots, x_{\tilde{m}})^{\T} \in \boldsymbol{C}_{{\bf i}}$, we have $x_{{\bf i},j} \le x_{j} \le x_{{\bf i+1}, j}$, $j\in[\tilde{m}]$, where ${\bf i+1}$ means that each component of ${\bf i}$ is increased by 1.  This indicates that, with  
$\boldsymbol{v}_{2t-1}=-\boldsymbol{e}_{t}$, $\boldsymbol{v}_{2t} =\boldsymbol{e}_{t}$, $w_{2t-1} = x_{{\bf i},t}$ and $w_{2t} = -x_{{\bf i+1},t}$ for any $t\in[\tilde{m}]$,
\begin{align*}
	\boldsymbol{v}_{k}^{\T}\boldsymbol{x} + w_{k} \le 0 \,,~~~~ \boldsymbol{x} \in \boldsymbol{C}_{{\bf i}}\,,~ k\in[2\tilde{m}]\,, 
\end{align*}
where $\boldsymbol{e}_{v}$ denotes the $v$-th unit vector. Thus,  $\boldsymbol{C}_{{\bf i}}$ is a polytope defined in Lemma \ref{lem:network-cooef} with $H=2\tilde{m}$. 

Let $p_{{\bf i},\tilde{\vartheta}}$ be the Taylor polynomial of  $f$ of order $\tilde{\vartheta}$ around the center of $\boldsymbol{C}_{\bf i}$, which is denoted by $\boldsymbol{x}_{{\bf i},0}=(x_{{\bf i}, 0,1}, \ldots, x_{{\bf i},0, \tilde{m}})^{\T}$, i.e., for any $\boldsymbol{x} \in \mathbb{R}^{\tilde{m}}$,
\begin{align}\label{eq:talot-pi-theta}
	p_{{\bf i},\tilde{\vartheta}}(\boldsymbol{x})  = &~ \sum_{\substack{j_1, \ldots, j_{\tilde{m}} \in\{0\}\cup[\tilde{\vartheta}],\\j_1+\cdots+ j_{\tilde{m}}\le \tilde{\vartheta}}} \bigg\{\frac{1}{j_1! \cdots j_{\tilde{m}}!} \times \frac{\partial^{j_1+\cdots+j_{\tilde{m}}} f}{\partial^{j_1}x_{1}\cdots \partial^{j_{\tilde{m}}}x_{\tilde{m}}}(\boldsymbol{x}_{{\bf i},0}) \notag\\
	&~~~~~~~~~~~~~~~~~~~~~~~~\times(x_{1}-x_{{\bf i},0,1})^{j_1}  \cdots  (x_{m}- x_{{\bf i},0,\tilde{m}})^{j_{\tilde{m}}} \bigg\}\,.
\end{align}
Notice that
\begin{align*}
	p_{{\bf i},\tilde{\vartheta}}(\boldsymbol{x})  
	= &~  \sum_{\substack{j_1, \ldots, j_{\tilde{m}} \in\{0\}\cup[\tilde{\vartheta}],\\j_1+\cdots+ j_{\tilde{m}}\le \tilde{\vartheta}}} \bigg[\frac{1}{j_1! \cdots j_{\tilde{m}}!}  \times \frac{\partial^{j_1+\cdots+j_{\tilde{m}}} f}{\partial^{j_1}x_{1}\cdots \partial^{j_{\tilde{m}}}x_{\tilde{m}}}(\boldsymbol{x}_{{\bf i},0}) \\
	&~~~~~~\times  \bigg\{\sum_{k_1=0}^{j_1} {\rm C}_{j_1}^{k_1}x_1^{k_1} (-x_{{\bf i},0,1})^{j_1-k_1}\bigg\}  \cdots \bigg\{\sum_{k_{\tilde{m}}=0}^{j_{\tilde{m}}} {\rm C}_{j_{\tilde{m}}}^{k_{\tilde{m}}}x_{\tilde{m}}^{k_{\tilde{m}}} (-x_{{\bf i},0,\tilde{m}})^{j_{\tilde{m}}-k_{\tilde{m}}}\bigg\} \bigg]\\
	=&~ \sum_{\substack{j_1, \ldots, j_{\tilde{m}} \in\{0\}\cup[\tilde{\vartheta}],\\j_1+\cdots+ j_{\tilde{m}}\le \tilde{\vartheta}}}  \bigg[\sum_{k_1=0}^{j_1}
	\cdots \sum_{k_{\tilde{m}}=0}^{j_{\tilde{m}}} \bigg\{
	\frac{1}{j_1! \cdots j_{{\tilde{m}}}!} \times \frac{\partial^{j_1+\cdots+j_{\tilde{m}}} f}{\partial^{j_1}x_{1}\cdots \partial^{j_{\tilde{m}}}x_{\tilde{m}}}(\boldsymbol{x}_{{\bf i},0}) \times {\rm C}_{j_1}^{k_1} \cdots {\rm C}_{j_{\tilde{m}}}^{k_{\tilde{m}}} \\
	&~~~~~~~~~~~~~~~~~~~~~~~ \times (-x_{{\bf i},0,1})^{j_1-k_1}  \cdots  (-x_{{\bf i},0,\tilde{m}})^{j_{\tilde{m}}-k_{\tilde{m}}}   x_{1}^{k_1}  \cdots   x_{\tilde{m}}^{k_{\tilde{m}}} \bigg\}\bigg]\\
	=&~ \sum_{\substack{k_1, \ldots, k_{\tilde{m}} \in\{0\}\cup[\tilde{\vartheta}],\\k_1+\cdots+ k_{\tilde{m}}\le \tilde{\vartheta}}}  \bigg[ \sum_{\substack{j_1\ge k_1, \ldots, j_{\tilde{m}} \ge k_{\tilde{m}}  ,\\j_1+\cdots+ j_{\tilde{m}}\le \tilde{\vartheta}}} \bigg\{ \frac{1}{j_1! \cdots j_{\tilde{m}}!}  \times \frac{\partial^{j_1+\cdots+j_{\tilde{m}}} f}{\partial^{j_1}x_{1}\cdots \partial^{j_{\tilde{m}}}x_{\tilde{m}}}(\boldsymbol{x}_{{\bf i},0}) \times {\rm C}_{j_1}^{k_1} \cdots {\rm C}_{j_{\tilde{m}}}^{k_{\tilde{m}}} \\
	&~~~~~~~~~~~~~~~~~~~~~~~ \times (-x_{{\bf i},0,1})^{j_1-k_1} \cdots  (-x_{{\bf i},0,\tilde{m}})^{j_{\tilde{m}}-k_{\tilde{m}}}\bigg\} \bigg] x_{1}^{k_1} \cdots  x_{\tilde{m}}^{k_{\tilde{m}}} \,.
\end{align*}
Given $\delta=\tilde{a}_n\tilde{\eta}_n/(2\tilde{m}\tilde{M}_n)$ and a sufficiently large $\tilde{M}_n$, for any ${\bf i} \in [\tilde{M}_n  +1]^{\tilde{m}}$, by Lemma \ref{lem:network-cooef},   neural networks  $t_{\bf i}(\boldsymbol{x})$ of the type
\begin{align*}
	t_{\bf i}(\boldsymbol{x}) = \sum_{j=1}^{(\tilde{N}+1){\rm C}_{\tilde{m}+\tilde{N}}^{\tilde{m}}} (\mu_{j})_{\bf i}   \sigma\bigg[ \sum_{l=1}^{4\tilde{m}} (\lambda_{j,l})_{\bf i}  \sigma\bigg\{ \sum_{v=1}^{\tilde{m}} (\theta_{l,v})_{\bf i} x_{v} + (\theta_{l,0})_{\bf i}\bigg\} + (\lambda_{j,0})_{\bf i} \bigg] 
\end{align*}
exist, with coefficients bounded as therein, such that
\begin{align}\label{eq:t-polytope-bound}
	&|t_{\bf i}(\boldsymbol{x}) -p_{{\bf i},\tilde{\vartheta}}(\boldsymbol{x})| \le   \frac{\bar{C}_{9} \tilde{m} ({\rm C}_{\tilde{m}+\tilde{N}}^{\tilde{m}})^2 \bar{r}(p_{{\bf i},\tilde{\vartheta}})  a_n^{\tilde{N}+3} }{ (\tilde{M}_n +1)^{\vartheta} }\,, ~~~~ \boldsymbol{x} \in (\boldsymbol{C}_{\bf i})_{\delta}^{\rm o} \cap [-\tilde{a}_n,\tilde{a}_n]^{\tilde{m}} \,, \notag\\
	&~~~~~~~~~~~~ |t_{\bf i}(\boldsymbol{x})| \le \frac{\bar{C}_{10} ({\rm C}_{\tilde{m}+\tilde{N}}^{\tilde{m}})^2 \bar{r}(p_{{\bf i},\tilde{\vartheta}})} { ( \tilde{M}_n +1)^{2\vartheta +\tilde{m}+1} } \,,  ~~~~\boldsymbol{x} \in (\boldsymbol{C}_{\bf i})_{\delta}^{\rm c} \cap  [-\tilde{a}_n,\tilde{a}_n]^{\tilde{m}} \,, \\
	&~~~~~~~~~~~~~~~ |t_{\bf i}(\boldsymbol{x})| \le \bar{C}_{11}  ({\rm C}_{\tilde{m}+\tilde{N}}^{\tilde{m}})^2 \bar{r}(p_{{\bf i},\tilde{\vartheta}}) (\tilde{M}_n +1)^{\tilde{N}\vartheta} \,, ~~~~ \boldsymbol{x} \in \mathbb{R}^{\tilde{m}} \notag \,,
\end{align}
where $\bar{C}_{9}>0$, $\bar{C}_{10}>0$ and $\bar{C}_{11}>0$ are some universal constants only depending on $(\tilde{m}, \tilde{N})$. 
By \eqref{eq:lemm-partial-a} and  the definition of $\bar{r}(p_{{\bf i},\tilde{\vartheta}})$ given in Lemma \ref{lem:polynomials}, we have 
\begin{align}\label{eq:bar-rp}
	\bar{r}(p_{{\bf i},\tilde{\vartheta}})  =&~ \max_{\substack{k_1, \ldots, k_{\tilde{m}} \in\{0\}\cup[\tilde{\vartheta}],\\k_1+\cdots+ k_{\tilde{m}}\le \tilde{\vartheta}}}  \bigg| \sum_{\substack{j_1\ge k_1, \ldots, j_{\tilde{m}} \ge k_{\tilde{m}}  ,\\j_1+\cdots+ j_{\tilde{m}}\le \tilde{\vartheta}}} \bigg\{ \frac{{\rm C}_{j_1}^{k_1} \cdots {\rm C}_{j_{\tilde{m}}}^{k_{\tilde{m}}}}{j_1! \cdots j_{\tilde{m}}!}  \times \frac{\partial^{j_1+\cdots+j_{\tilde{m}}} f}{\partial^{j_1}x_{1}\cdots \partial^{j_{\tilde{m}}}x_{\tilde{m}}}(\boldsymbol{x}_{{\bf i},0})  \notag\\
	&~~~~~~~~~~~~~~~~~~~~~~~~~~~~~~~~~~~~~~~~~ \times (-x_{{\bf i}, 0,1})^{j_1-k_1} \cdots  (-x_{{\bf i},0,\tilde{m}})^{j_{\tilde{m}}-k_{\tilde{m}}} \bigg\} \bigg| \notag\\
	\le&~ \sum_{\substack{j_1, \ldots, j_{\tilde{m}} \in\{0\}\cup[\tilde{\vartheta}],\\j_1+\cdots+ j_{\tilde{m}}\le \tilde{\vartheta}}}  \sum_{k_1=0}^{j_1}
	\cdots \sum_{k_{\tilde{m}}=0}^{j_{\tilde{m}}}  \bigg| \frac{{\rm C}_{j_1}^{k_1} \cdots {\rm C}_{j_{\tilde{m}}}^{k_{\tilde{m}}}}{j_1! \cdots j_{\tilde{m}}!}  \times \frac{\partial^{j_1+\cdots+j_{\tilde{m}}} f}{\partial^{j_1}x_{1}\cdots \partial^{j_{\tilde{m}}}x_{\tilde{m}}}(\boldsymbol{x}_{{\bf i},0})  \notag\\
	&~~~~~~~~~~~~~~~~~~~~~~~~~~~~~~~~~~~~~~~~~ \times (-x_{{\bf i}, 0,1})^{j_1-k_1}   \cdots  (-x_{{\bf i},0,\tilde{m}})^{j_{\tilde{m}}-k_{\tilde{m}}} \bigg|\notag\\ 
	\le  &~ 3^{\tilde{\vartheta}}  B {\rm C}_{\tilde{m}+ \tilde{\vartheta}}^{\tilde{m}}   \cdot \tilde{a}_n ^{\tilde{\vartheta}}\,,~~~~~~~~~ {\bf i}\in [\tilde{M}_n +1]^{\tilde{m}}\,.
\end{align}
Set $t(\boldsymbol{x}) = \sum_{{\bf i} \in [ \tilde{M}_n +1]^{\tilde{m}}} t_{\bf i}(\boldsymbol{x}) $. For any $\boldsymbol{x} \in (\boldsymbol{C}_{\bf i})_{\delta}^{\rm o} \cap [-\tilde{a}_n,\tilde{a}_n]^{\tilde{m}}$, it holds that $\boldsymbol{x} \in (\boldsymbol{C}_{\bf j})_{\delta}^{\rm c}$ for any ${\bf j}\in [\tilde{M}_n +1]^{\tilde{m}}\backslash\{{\bf i}\}$. For $\boldsymbol{x} \in (\boldsymbol{C}_{\bf i})_{\delta}^{\rm o} \cap [-\tilde{a}_n,\tilde{a}_n]^{\tilde{m}}$, by Lemma \ref{lem:taylor}, \eqref{eq:t-polytope-bound}  and \eqref{eq:bar-rp}, we have
\begin{align}\label{eq:tx-fx-bound}
	|t(\boldsymbol{x})-f(\boldsymbol{x})| \le&~ |t_{\bf i}(\boldsymbol{x}) - p_{{\bf i},\tilde{\vartheta}}(\boldsymbol{x})| + |p_{{\bf i},\tilde{\vartheta}}(\boldsymbol{x}) -f(\boldsymbol{x})| + \bigg|\sum_{{\bf j}\in [\tilde{M}_n +1]^{\tilde{m}}\backslash\{{\bf i}\}}t_{\bf j}(\boldsymbol{x}) \bigg| \notag \\
	\le &~  \bar{C}_{9} \tilde{m} ({\rm C}_{\tilde{m}+\tilde{N}}^{\tilde{m}})^2 \bar{r}(p_{{\bf i},\tilde{\vartheta}}) \tilde{a}_n^{\tilde{N}+3} (\tilde{M}_n +1)^{-\vartheta} +  C\tilde{C}_{8} \tilde{m}^{\tilde{\vartheta} + \vartheta/2} \tilde{a}_n^{\vartheta}\tilde{M}_n^{-\vartheta } \notag\\
	&+ \bar{C}_{10} \{(\tilde{M}_n +1)^{\tilde{m}} -1\} ({\rm C}_{\tilde{m}+\tilde{N}}^{\tilde{m}})^2  (\tilde{M}_n +1)^{-2\vartheta -\tilde{m}-1}   \max_{{\bf j}\in [\tilde{M}_n +1]^{\tilde{m}}}\bar{r}(p_{{\bf j},\tilde{\vartheta}}) \notag\\
	\le &~ \bar{C}_{12} \{({\rm C}_{\tilde{m}+\tilde{N}}^{\tilde{m}})^3 + \tilde{m}^{\tilde{\vartheta} + \vartheta/2}\} \cdot  \tilde{a}_n ^{\tilde{N}+3 + \tilde{\vartheta}}  \tilde{M}_n ^{-\vartheta}  \,,
\end{align}
where $\tilde{C}_8>0$ is specified in Lemma \ref{lem:taylor}, and $\bar{C}_{12}>0$ is a universal constant only depending on $(\tilde{m},\tilde{N},B)$.  Recall $\delta=\tilde{a}_n\tilde{\eta}_n/(2\tilde{m}\tilde{M}_n)$. 
Notice that \eqref{eq:tx-fx-bound} holds for all $\boldsymbol{x}\in [-\tilde{a}_n,\tilde{a}_n]^{\tilde{m}} $ which are not contained in
\begin{align}\label{eq:set-measure}
	\bigcup_{j\in[\tilde{m}]} \bigcup_{{\bf i} \in[\tilde{M}_n+2]^{\tilde{m}}} \{\boldsymbol{x}\in \mathbb{R}^{\tilde{m}}: |x_{j}- x_{{\bf i},j}| < \delta\}\,.
\end{align} 
For each fixed $j\in[\tilde{m}]$, by slightly shifting the whole grid of cubes along the $j$-th component (i.e., modifying all $x_{\mathbf{i},j}$ by the same additional summand which is less than $2 \tilde{a}_n\tilde{M}_n^{-1}$),  we can construct more than 
\begin{align*}
	\bigg\lfloor\frac{2 \tilde{a}_n / \tilde{M}_n}{2 \delta}\bigg\rfloor =\bigg\lfloor\frac{2 \tilde{m}}{\tilde{\eta}_n}\bigg\rfloor \geq \frac{\tilde{m}}{ \tilde{\eta}_n}
\end{align*}
different versions of $t(\boldsymbol{x})$ that still satisfy \eqref{eq:tx-fx-bound} for any $\boldsymbol{x} \in [-\tilde{a}_n, \tilde{a}_n]^{\tilde{m}}$ up to corresponding
disjoint versions of $\cup_{{\bf i} \in[\tilde{M}_n+2]^{\tilde{m}}} \{\boldsymbol{x}\in \mathbb{R}^{\tilde{m}}: |x_{j}- x_{{\bf i},j}| < \delta\}$. Because the sum of the $\mu$-measures of these sets is less than or equal to $1$, at least one of them must have $\mu$-measure less than or equal to $\tilde{\eta}_n/\tilde{m}$. Hence,  we can shift the $\boldsymbol{x}_{{\bf i}}$ such that \eqref{eq:set-measure} has $\mu$-measure less than or equal to $\tilde{\eta}_n$,    which implies \eqref{eq:tx-fx-bound} holds for all $\boldsymbol{x} \in [-\tilde{a}_n,\tilde{a}_n]^{\tilde{m}} $ up to a set of $\mu$-measure less than or equal to $\tilde{\eta}_n$. Furthermore,
by Lemma \ref{lem:network-cooef} and \eqref{eq:bar-rp}, the coefficients of  $t_{\bf i}(\boldsymbol{x})$ satisfy
\begin{align*}
	&|(\mu_{j})_{\bf i}|\le  \bar{C}_{13}   ({\rm C}_{\tilde{m}+\tilde{N}}^{\tilde{m}} )^2  \tilde{a}_n^{\tilde{\vartheta}} (\tilde{M}_n+1)^{\tilde{N}\vartheta} \,, \qquad |(\lambda_{j,l})_{\bf i}| \le   \tilde{C}_{7} (\tilde{M}_n+1)^{\tilde{m}+1+\vartheta(\tilde{N}+2)}\,,  \\
	&~~~~~~~~~~~~~~~~~~~~~~~~~~ |(\theta_{l,v})_{\bf i}| \le    4\tilde{\eta}_n^{-1} \tilde{m} (\tilde{M}_n+1)^{\tilde{m}+2+\vartheta(2\tilde{N}+3)}  
\end{align*}
for any ${\bf i} \in [\tilde{M}_n  +1]^{\tilde{m}}$,  $j\in [(\tilde{N}+1)(\tilde{M}_n+1)^{\tilde{m}}{\rm C}_{\tilde{m}+\tilde{N}}^{\tilde{m}}]$, $l\in[4\tilde{m}]\cup\{0\}$ and $v\in[\tilde{m}]\cup\{0\}$. Here $\bar{C}_{13}>0$ is a universal constant only depending on $(\tilde{m},\tilde{N},B)$, and $\tilde{C}_7>0$ is specified in Lemma \ref{lem:network-cooef}. 
Hence, we complete the proof of Lemma \ref{lem:network-pc}.
$\hfill\Box$

\section{Proof of Lemma \ref{lem:nn-epsdts-epdt}}\label{sec:nn-epsdts-epdt}
Recall $\varepsilon_{i,j} = U_{i,j}- f_{j}(\bW_i)$, $\delta_{i,k} = V_{i,k}- g_{k}(\bW_i)$, $ \tilde{\varepsilon}_{i,j} = \hat{U}_{i,j}^{(w)} - \hat{f}_{j}(\hat{\bW}_{i}^{(w)})$ and $ \tilde{\delta}_{i,k} = \hat{V}_{i,k}^{(w)} - \hat{g}_{k}(\hat{\bW}_{i}^{(w)})$. 
\begin{align*}
	&\frac{1}{n_{3}} \sum_{t\in \mathcal{D}_3} ( \tilde{\varepsilon}_{t,j} - \varepsilon_{t,j})( \tilde{\delta}_{t,k}-\delta_{t,k}) \\
	&~~~~=  \frac{1}{n_{3}} \sum_{t\in \mathcal{D}_3} \big[(\hat{U}_{t,j}^{(w)}- U_{t,j}) - \{\hat{f}_{j} (\hat{\bW}_{t}^{(w)}) - f_{j} (\bW_{t})\}\big]  \big[(\hat{V}_{t,k}^{(w)}- V_{t,k}) - \{\hat{g}_{k} (\hat{\bW}_{t}^{(w)}) - g_{k} (\bW_{t})\}\big]\\
	&~~~~=  \underbrace{\frac{1}{n_{3}} \sum_{t\in \mathcal{D}_3} (\hat{U}_{t,j}^{(w)}- U_{t,j})(\hat{V}_{t,k}^{(w)}- V_{t,k})}_{\tilde{\textrm{G}}_1(j,k)} -\underbrace{\frac{1}{n_{3}} \sum_{t\in \mathcal{D}_3} \{\hat{g}_{k} (\hat{\bW}_{t}^{(w)}) - g_{k} (\bW_{t})\} (\hat{U}_{t,j}^{(w)}- U_{t,j})}_{\tilde{\textrm{G}}_2(j,k)}\\
	&~~~~~~~~~ -  \underbrace{\frac{1}{n_{3}} \sum_{t\in \mathcal{D}_3} \{\hat{f}_{j} (\hat{\bW}_{t}^{(w)}) - f_{j} (\bW_{t})\}(\hat{V}_{t,k}^{(w)}- V_{t,k})}_{\tilde{\textrm{G}}_3(j,k)} \\
	&~~~~~~~~~+   \underbrace{\frac{1}{n_{3}} \sum_{t\in \mathcal{D}_3} \{\hat{f}_{j} (\hat{\bW}_{t}^{(w)}) - f_{j} (\bW_{t})\}  \{\hat{g}_{k} (\hat{\bW}_{t}^{(w)}) - g_{k} (\bW_{t})\} }_{\tilde{\textrm{G}}_4(j,k)}\,.
\end{align*}
As we will show in Sections \ref{sec:sub-tilde-g1}--\ref{sec:sub-tilde-g4}, 
\begin{align}
	\max_{j\in[p],\,k\in[q]}|\tilde{\rm G}_{1}(j,k)| =&~ O_{\rm p}\{n^{-\kappa} \log^{2}(\tilde{d}n) \} + O_{\rm p}\{n^{-1/2}(\log n)^{1/2}\log^{1/2}(\tilde{d}n)\} \label{eq:tilde-g1}\,,\\
	\max_{j\in[p],\,k\in[q]} |\tilde{\rm G}_{2}(j,k)| =&~O_{\rm p} \{n^{-\kappa} (\log n)\log^{2}(\tilde{d}n)\} + O_{\rm p}\{n^{-1/2} (\log n)\log (\tilde{d}n)\} \notag\\
	=&~ \max_{j\in[p],\,k\in[q]} |\tilde{\rm G}_{3}(j,k)| \label{eq:tilde-g2} 
\end{align}
provided that $\log(\tilde{d}n) \ll n^{1-\kappa}(\log n)^{-1/2} $, and
\begin{align}
	\max_{j\in[p],\, k\in[q]}|\tilde{{\rm G}}_{4}(j,k)|=&~ O_{\rm p}\{n^{-2\vartheta/(4\vartheta+m_{*})} (m^2\log n)^{(\vartheta+2m_{*}\tilde{\vartheta}+3m_{*} )/(4\vartheta)} (\log n)^{2}\log^{3/2} (\tilde{d}n) \} \notag\\
	& + O_{\rm p}\{n^{-\kappa/2-\vartheta/(4\vartheta+m_{*})} (m^2\log n)^{(\vartheta+2m_{*}\tilde{\vartheta}+3m_{*} )/(8\vartheta)} (\log n)^{2}\log^{7/4} (\tilde{d}n) \} \notag\\
	&      +O_{\rm p}\{n^{-1/2}m (\log n) \log  (\tilde{d}n)\}  +  O_{\rm p}\{n^{-\kappa} m^2   (\log n)^2\log^2 (\tilde{d}n) \} \notag\\
	& + O_{\rm p}\{n^{-\kappa/2-1/4}m^{1/2} (\log n)^{3/2} \log^{3/2} (\tilde{d}n)\}    \label{eq:tilde-g4}
\end{align}
provided that $\log(\tilde{d}n) \ll n^{1-\kappa}(\log n)^{-1/2} $ and $m \lesssim n$.   Hence, we have 
\begin{align*}
	&\max_{j\in[p],\, k\in[q]}\bigg|\frac{1}{n_{3}} \sum_{t\in \mathcal{D}_3} ( \tilde{\varepsilon}_{t,j} - \varepsilon_{t,j})( \tilde{\delta}_{t,k}-\delta_{t,k})\bigg| \\
	&~~~~~~ = O_{\rm p}\{n^{-2\vartheta/(4\vartheta+m_{*})} (m^2\log n)^{(\vartheta+2m_{*}\tilde{\vartheta}+3m_{*} )/(4\vartheta)} (\log n)^{2}\log^{3/2} (\tilde{d}n) \} \notag\\
	&~~~~~~~~~ + O_{\rm p}\{n^{-\kappa/2-\vartheta/(4\vartheta+m_{*})} (m^2\log n)^{(\vartheta+2m_{*}\tilde{\vartheta}+3m_{*} )/(8\vartheta)} (\log n)^{2}\log^{7/4} (\tilde{d}n) \} \notag\\
	&~~~~~~~~~ +O_{\rm p}\{n^{-1/2}m (\log n) \log  (\tilde{d}n)\}  +  O_{\rm p}\{n^{-\kappa} m^2   (\log n)^2\log^2 (\tilde{d}n) \} \notag\\
	&~~~~~~~~~ + O_{\rm p}\{n^{-\kappa/2-1/4}m^{1/2} (\log n)^{3/2} \log^{3/2} (\tilde{d}n)\}
\end{align*}
provided that $\log(\tilde{d}n) \ll n^{1-\kappa}(\log n)^{-1/2} $ and $m \lesssim n$.   We complete the proof of Lemma \ref{lem:nn-epsdts-epdt}.
$\hfill\Box$

\subsection{Proof of \eqref{eq:tilde-g1}}\label{sec:sub-tilde-g1}
Recall $\tilde{d}=p\vee q\vee m$ and $U_{i,j}^{*} =  U_{i,j}I(|U_{i,j}|\le M_1) + M_1 \cdot{\rm sign}(U_{i,j})I(|U_{i,j}|>M_1)$ with $M_1=\sqrt{2\log n_3}$. Analogously, define $V_{i,k}^{*} =  V_{i,k}I(|V_{i,k}|\le M_1) + M_1 \cdot{\rm sign}(V_{i,k})I(|V_{i,k}|>M_1)$. Then, 
\begin{align*} 
	\tilde{{\rm G}}_{1}(j,k)
	=&~
	\underbrace{\frac{1}{n_{3}} \sum_{t\in \mathcal{D}_3} (\hat{U}_{t,j}^{(w)} -U_{t,j}^{*})(\hat{V}_{t,k}^{(w)}-V_{t,k}^{*})}_{\tilde{\textrm{G}}_{11}(j,k)}  + \underbrace{\frac{1}{n_{3}} \sum_{t\in \mathcal{D}_3} \hat{V}_{t,k}^{(w)}(U_{t,j}^{*} - U_{t,j})}_{\tilde{\textrm{G}}_{12}(j,k)} \notag \\
	&+\underbrace{\frac{1}{n_{3}} \sum_{t\in \mathcal{D}_3} \hat{U}_{t,j}^{(w)}(V_{t,k}^{*} - V_{t,k})}_{\tilde{\textrm{G}}_{13}(j,k)} - \underbrace{\frac{1}{n_{3}} \sum_{t\in \mathcal{D}_3} (U_{t,j}^{*}V_{t,k}^{*} -U_{t,j}V_{t,k})}_{\tilde{\textrm{G}}_{14}(j,k)}\,.
\end{align*}
Recall $n_1\asymp n$ and $n_3 \asymp n^{\kappa}$ for some constant $0<\kappa<1$. Using the similar arguments for the proof of the convergence rates of $\max_{j\in[p],\,k\in[q]}|\tilde{\rm H}_{1}(j,k)|$ and $\max_{j\in[p],\,k\in[q]}|\tilde{\rm H}_{2}(j,k)|$ in Sections \ref{sec:sub-g11} and \ref{sec:sub-g12} for the proof of Lemma \ref{lem:uh-u-delta}, it holds that 
\begin{align}\label{eq:uhw-u}
	&\max_{j\in[p]}\frac{1}{n_{3}}\sum_{t\in \mathcal{D}_3} |\hat{U}_{t,j}^{(w)} -U_{t,j}^{*}| \notag\\
	&~~~~~~~ \le \max_{j\in[p]}\frac{1}{n_{3}}\sum_{t\in \mathcal{D}_3} |\hat{U}_{t,j}^{(w)} -U_{t,j}^{*}| I(|U_{t,j}|\le M_1) + \max_{j\in[p]}\frac{1}{n_{3}}\sum_{t\in \mathcal{D}_3} |\hat{U}_{t,j}^{(w)} -U_{t,j}^{*}|I(|U_{t,j}| > M_1) \notag\\
	&~~~~~~~= O_{\rm p} \{n^{-\kappa} (\log n)^{1/2}  \log(\tilde{d}n)\} + O_{\rm p}\{n^{-1/2} \log^{1/2}(\tilde{d}n)\}
\end{align}
provided that $\log(\tilde{d}n) \ll n^{1-\kappa}(\log n)^{-1/2}$.
Recall $\max_{i\in\mathcal{D}_3,\,k\in[q]}|\hat{V}_{i,k}^{(w)}| \le \sqrt{2\log n_1}$. We the have
\begin{align*}
	\max_{j\in[p],\,k\in[q]} |\tilde{\rm G}_{11}(j,k)| \le&~ C_1\sqrt{\log n } \times \max_{j\in[p]} \frac{1}{n_{3}}\sum_{t\in \mathcal{D}_3} |\hat{U}_{t,j}^{(w)} -U_{t,j}^{*}|\\
	=&~O_{\rm p} \{n^{-\kappa}(\log n)  \log(\tilde{d}n)\} + O_{\rm p}\{n^{-1/2}(\log n)^{1/2}\log^{1/2}(\tilde{d}n)\}
\end{align*}
provided that $\log(\tilde{d}n) \ll n^{1-\kappa}(\log n)^{-1/2}$.  
Given $Q>M_1$, it holds that
\begin{align*} 
	\frac{1}{n_{3}} \sum_{t\in \mathcal{D}_3}  |U_{t,j}^{*} - U_{t,j}|  = &~\frac{1}{n_{3}} \sum_{t\in \mathcal{D}_3} \underbrace{\big[|U_{t,j}^{*} - U_{t,j}| I( |U_{t,j}|\le Q)   -\mathbb{E}\{|U_{t,j}^{*} - U_{t,j}|I( |U_{t,j}|\le Q) \}\big] }_{\tilde{\textrm{G}}_{121}(t,j)}\notag\\
	& +\frac{1}{n_{3}} \sum_{t\in \mathcal{D}_3} \underbrace{|U_{t,j}^{*} - U_{t,j}| I(|U_{t,j}|>Q)   }_{\tilde{\textrm{G}}_{122}(t,j)} +\underbrace{\mathbb{E}\{|U_{t,j}^{*} - U_{t,j}| I( |U_{t,j}|\le Q)  \} }_{\tilde{\textrm{G}}_{123}(t,j)}\,.
\end{align*}
Recall $U_{t,j} \sim \mathcal{N}(0,1)$ and $\tilde{d}=p \vee q\vee m$. Since $  |U_{t,j} -U_{t,j}^{*} | \le 2 | U_{t,j} |I(|U_{t,j}|> M_1) $, then
\begin{align*}
	\max_{t\in \mathcal{D}_3,\,j\in[p]}\var\{|U_{t,j}^{*} - U_{t,j}| I( |U_{i,j}|\le Q)\}  \le C_2\max_{t\in \mathcal{D}_3,\, j\in[p] }\mathbb{E}\{U_{t,j}^2 I(|U_{t,j}|>M_1)\} \lesssim M_1 e^{-M_1^2/2}\,.
\end{align*} 
By Bonferroni inequality and Bernstein inequality, it holds that
\begin{align*} 
	\max_{j\in[p]}\bigg|\frac{1}{n_{3}} \sum_{t\in \mathcal{D}_3} \tilde{{\rm G}}_{121}(t,j)\bigg|=O_{\rm p} \{n_3^{-1/2}M_1^{1/2}e^{-M_1^2/4}(\log \tilde{d})^{1/2}\} + O_{\rm p} (n_3^{-1}Q \log \tilde{d})\,.
\end{align*}   
Analogous to the derivation of \eqref{eq:k12}, we have 
\begin{align*} 
	\max_{j\in[p]}\bigg|\frac{1}{n_{3}} \sum_{t\in \mathcal{D}_3} \tilde{{\rm G}}_{122}(t,j)\bigg|=o_{\rm p}(n^{-1})
\end{align*}
provided that $\log(\tilde{d}n) \lesssim Q^2$. Furthermore,  due to $  |U_{t,j} -U_{t,j}^{*} | \le 2 | U_{t,j} |I(|U_{t,j}|> M_1) $, then
\begin{align*}
	\max_{t\in \mathcal{D}_3\,j\in[p]}|\tilde{{\rm G}}_{123}(t,j)| \le  2 \max_{t\in \mathcal{D}_3,\,j\in[p]}\mathbb{E}\{|U_{t,j}| I(|U_{t,j}| >M_1 )\} \lesssim    e^{-M_1^2/2} \,.
\end{align*}    
Recall $n_3 \asymp n^{\kappa}$ for some constant $0< \kappa < 1$. By selecting $Q =C_3 \log^{1/2} (\tilde{d}n)$ for some sufficiently large constant $C_3> 0$,   it holds that
\begin{align}\label{eq:ustar-u}
	\max_{j\in[p]}\frac{1}{n_{3}} \sum_{t\in \mathcal{D}_3} |U_{t,j}^{*} - U_{t,j}| = O_{\rm p}\{n^{-\kappa}\log^{3/2}(\tilde{d}n) \}  \,.
\end{align}  
Recall  
$\max_{i\in\mathcal{D}_3,\,k\in[q]}|\hat{V}_{i,k}^{(w)}| \le \sqrt{2\log n_1}$. Then,
\begin{align*}
	\max_{j\in[p],\,k\in[q]}|\tilde{\rm G}_{12}(j,k)|  \le C_4\sqrt{\log n } \times \max_{j\in[p]}\frac{1}{n_{3}} \sum_{t\in \mathcal{D}_3} |U_{t,j}^{*} - U_{t,j}| = O_{\rm p}\{n^{-\kappa}(\log n)^{1/2} \log^{3/2}(\tilde{d}n) \}  \,.
\end{align*}  
Analogously, we can show such convergence rate also holds for $\max_{j\in[p],\,k\in[q]}|\tilde{\rm G}_{13}(j,k)|$.  Furthermore, using the similar arguments for the proof of Lemma \ref{lem:usvs-uv} with $M_1=\sqrt{2\log n_3}$, we have 
\begin{align*}
	\max_{j\in[p],\,k\in[q]}|\tilde{\rm G}_{14}(j,k)| =  O_{\rm p}\{n^{-\kappa} \log^{2}(\tilde{d}n) \}  \,.
\end{align*}
Hence, it holds that
\begin{align*}
	\max_{j\in[p],\,k\in[q]}|\tilde{\rm G}_{1}(j,k)| \le&~ \max_{j\in[p],\,k\in[q]}|\tilde{\rm G}_{11}(j,k)| + \max_{j\in[p],\,k\in[q]}|\tilde{\rm G}_{12}(j,k)|\\
	&+ \max_{j\in[p],\,k\in[q]}|\tilde{\rm G}_{13}(j,k)| +\max_{j\in[p],\,k\in[q]}|\tilde{\rm G}_{14}(j,k)|\\
	=&~ O_{\rm p}\{n^{-\kappa} \log^{2}(\tilde{d}n) \} + O_{\rm p}\{n^{-1/2}(\log n)^{1/2}\log^{1/2}(\tilde{d}n)\}  
\end{align*}
provided that $\log(\tilde{d}n) \ll n^{1-\kappa}(\log n)^{-1/2}$. Then \eqref{eq:tilde-g1} holds.
$\hfill\Box$

\subsection{Proof of \eqref{eq:tilde-g2}}\label{sec:sub-tilde-g2}

Recall   $\max_{k\in[q]}|g_{k}|_{\infty} \le \tilde{C}$ and   $\max_{t\in\mathcal{D}_3,\,k\in[q]}|\hat{g}_{k}(\hat{\bW}_{t}^{(w)})| \le \tilde{\beta}_{n}$. By \eqref{eq:uhw-u} and \eqref{eq:ustar-u}, it holds that
\begin{align*}
	\max_{j\in[p],\,k\in[q]} |\tilde{\rm G}_{2}(j,k)| \le&~ C_1\tilde{\beta}_{n} \times \bigg\{\max_{j\in[p]}\frac{1}{n_{3}}\sum_{t\in \mathcal{D}_3} |\hat{U}_{t,j}^{(w)} -U_{t,j}^{*}| + \max_{j\in[p]}\frac{1}{n_{3}}\sum_{t\in \mathcal{D}_3} | U_{t,j}^{*} -U_{t,j}|\bigg\}\\
	=&~O_{\rm p} \{n^{-\kappa}\tilde{\beta}_{n}\log^{3/2}(\tilde{d}n)\} + O_{\rm p}\{n^{-1/2}\tilde{\beta}_{n}\log^{1/2}(\tilde{d}n)\}
\end{align*}
provided that $\log(\tilde{d}n) \ll n^{1-\kappa}(\log n)^{-1/2}$. Since $\max_{j\in[p]}|f_{j}|_{\infty} \le \tilde{C}$ and $\max_{t\in\mathcal{D}_3,\,j\in[p]}|\hat{f}_{j}(\hat{\bW}_{t}^{(w)})| \le \tilde{\beta}_{n}$,  using the similar arguments, we can show such result also holds for $\max_{j\in[p],\,k\in[q]} |\tilde{\rm G}_{3}(j,k)|$. 
Due to $\tilde{\beta}_n =(\log n)\log^{1/2}(\tilde{d}n)$, then \eqref{eq:tilde-g2} holds.
$\hfill\Box$

\subsection{Proof of \eqref{eq:tilde-g4}}\label{sec:sub-tilde-g4}
Notice that
\begin{align}\label{eq:tilde-g4-dec}
	\tilde{{\rm G}}_{4}(j,k) 
	=&~\underbrace{\frac{1}{n_{3}}\sum_{t\in \mathcal{D}_3} \{\hat{f}_{j} (\hat{\bW}_{t}^{(w)})  - f_{j} (\hat{\bW}_{t}^{(w)})\}   \{\hat{g}_{k} (\hat{\bW}_{t}^{(w)}) - g_{k} (\hat{\bW}_{t}^{(w)})\} }_{\tilde{\textrm{G}}_{41}(j,k)} \notag\\
	&+\underbrace{\frac{1}{n_{3}}\sum_{t\in \mathcal{D}_3} \{f_{j} (\hat{\bW}_{t}^{(w)})- f_{j} (\bW_{t})\}  \{\hat{g}_{k} (\hat{\bW}_{t}^{(w)}) - g_{k} (\hat{\bW}_{t}^{(w)})\} }_{\tilde{\textrm{G}}_{42}(j,k)}\\
	&+\underbrace{\frac{1}{n_{3}}\sum_{t\in \mathcal{D}_3} \{\hat{f}_{j} (\hat{\bW}_{t}^{(w)})  - f_{j} (\bW_{t})\}  \{g_{k} (\hat{\bW}_{t}^{(w)}) - g_{k} (\bW_{t})\} }_{\tilde{\textrm{G}}_{43}(j,k)}\notag \,.
\end{align}
Recall $\mathcal{W}_{\mathcal{D}_j}=\{(\bX_i,\bY_i,\bZ_i): i\in \mathcal{D}_j\}$ for $j\in[3]$, where $\mathcal{D}_1, \mathcal{D}_2$ and $\mathcal{D}_3$ are three disjoint subsets of $[n]$ with $|\mathcal{D}_1|=n_1\asymp n$, $|\mathcal{D}_2|=n_2\asymp n$ and $|\mathcal{D}_3|=n_3\asymp n^{\kappa}$  for some constant $0<\kappa<1$ and $n_1+n_2+n_3=n$. For any $t\in\mathcal{D}_3$, define
\begin{align*}
	&~~~~~~~\tilde{\mu}_{2,j,k} = \mathbb{E}\big[\{\hat{f}_{j} (\hat{\bW}_{t}^{(w)}) - f_{j} (\hat{\bW}_{t}^{(w)})\}  \{\hat{g}_{k} (\hat{\bW}_{t}^{(w)}) - g_{k} (\hat{\bW}_{t}^{(w)})\} \,|\, \mathcal{W}_{\mathcal{D}_1}, \mathcal{W}_{\mathcal{D}_2}\big] \,, \\
	&\tilde{\sigma}^2_{2,j,k}=  \mathbb{E}  \big\{ \big[\{\hat{f}_{j} (\hat{\bW}_{t}^{(w)}) - f_{j} (\hat{\bW}_{t}^{(w)})\}  \{\hat{g}_{k} (\hat{\bW}_{t}^{(w)}) - g_{k} (\hat{\bW}_{t}^{(w)})\}  -\tilde{\mu}_{2,j,k} \big]^2  \,|\, \mathcal{W}_{\mathcal{D}_1}, \mathcal{W}_{\mathcal{D}_2}\big\}\,.
\end{align*}
Recall 
\begin{align*}
	K(n, m ,\tilde{d}) = \bigg\{ \frac{ \tilde{C}_4(m^2\log n)^{(\vartheta+2m_{*}\tilde{\vartheta}+3m_{*} )/(4\vartheta)}  \tilde{\beta}_{n}^2\log^{1/2} (\tilde{d}n)}{n^{ 2\vartheta /(4\vartheta+m_{*})}}   + \frac{\tilde{C}_4m\tilde{\beta}_{n}\log^{1/2}(\tilde{d}n)}{n^{1/2}}+\frac{\tilde{C}_4m^2 \tilde{\beta}_n }{n^{\kappa}}\bigg\}^{1/2} 
\end{align*}  
with some sufficiently large constant $\tilde{C}_4> 0$ specified in Section \ref{sec:sub-g1} for the proof of Lemma \ref{lem:fhwh-fwh}.
Analogous to the derivation of \eqref{eq:sigma2-z}, it holds that
\begin{align}
	\mathbb{P}\bigg(\max_{k\in[q]}\mathbb{E}\big[\{\hat{g}_{k} (\hat{\bW}_{t}^{(w)}) - g_{k} (\hat{\bW}_{t}^{(w)})\}^2\,|\, \mathcal{W}_{\mathcal{D}_1}, \mathcal{W}_{\mathcal{D}_2}\big]   >  K^2(n, m,\tilde{d})
	\bigg)  \lesssim  n^{-1 }   \label{eq:ghwh-gwh-tail}
\end{align} 
provided that  $ \log (\tilde{d} n ) \ll n^{1-\kappa}(\log n)^{-1/2}$ and $m \lesssim n$.  
By  Cauchy-Schwarz inequality, \eqref{eq:sigma2-z} and \eqref{eq:ghwh-gwh-tail}, it holds that
\begin{align}\label{eq:mu2-bound}
	\max_{j\in[p],\,k\in[q]}|\tilde{\mu}_{2,j,k}| \le&~  \max_{j\in[p]}\big(\mathbb{E} [\{\hat{f}_{j} (\hat{\bW}_{t}^{(w)}) - f_{j} (\hat{\bW}_{t}^{(w)})\}^2 \,|\, \mathcal{W}_{\mathcal{D}_1}, \mathcal{W}_{\mathcal{D}_2}]\big)^{1/2} \notag\\
	&~~\times \max_{k\in[q]}\big(\mathbb{E} [ \{\hat{g}_{k} (\hat{\bW}_{t}^{(w)}) - g_{k} (\hat{\bW}_{t}^{(w)})\}^2 \,|\, \mathcal{W}_{\mathcal{D}_1}, \mathcal{W}_{\mathcal{D}_2}]\big)^{1/2} \notag\\
	\le &~ O_{\rm p}\{n^{-2\vartheta/(4\vartheta+m_{*})} (m^2\log n)^{(\vartheta+2m_{*}\tilde{\vartheta}+3m_{*} )/(4\vartheta)} \tilde{\beta}_n^{2}\log^{1/2} (\tilde{d}n) \} \notag\\
	&~+ O_{\rm p}\{n^{-1/2}m\tilde{\beta}_n \log^{1/2} (\tilde{d}n)\}  +O_{\rm p}(n^{-\kappa}m^2\tilde{\beta}_n ) 
\end{align}
provided that  $ \log (\tilde{d} n ) \ll n^{1-\kappa} (\log n)^{-1/2}$ and $m \lesssim n$.  Due to  $\max_{t\in\mathcal{D}_3,\,j\in[p]}|\hat{f}_{j}(\hat{\bW}_{t}^{(w)})| \le \tilde{\beta}_{n}$ and $\max_{j\in [p]}|f_{j}|_{\infty} \le \tilde{C}$ with $\tilde{\beta}_n =(\log n)\log^{1/2}(\tilde{d}n)$, for any $t\in \mathcal{D}_3$, we have
\begin{align*}
	\tilde{\sigma}^2_{2,j,k}  \le &~ \mathbb{E}\big[\{\hat{f}_{j} (\hat{\bW}_{t}^{(w)}) - f_{j} (\hat{\bW}_{t}^{(w)})\}^2  \{\hat{g}_{k} (\hat{\bW}_{t}^{(w)}) - g_{k} (\hat{\bW}_{t}^{(w)})\}^2\,|\, \mathcal{W}_{\mathcal{D}_1}, \mathcal{W}_{\mathcal{D}_2} \big]\\
	\le &~ C_1\tilde{\beta}_{n}^2   \mathbb{E}\big[\{\hat{g}_{k} (\hat{\bW}_{t}^{(w)}) - g_{k} (\hat{\bW}_{t}^{(w)})\}^2 \,|\, \mathcal{W}_{\mathcal{D}_1}, \mathcal{W}_{\mathcal{D}_2}\big]
\end{align*}
for sufficiently large $n$. By \eqref{eq:ghwh-gwh-tail}, it holds that
\begin{align*}
	\mathbb{P}\bigg\{\max_{j\in[p],\,k\in[q]}\tilde{\sigma}^2_{2,j,k} >      C_1 \tilde{\beta}_n^2 K^2(n, m,\tilde{d})   \bigg\} \lesssim  n^{-1 } 
\end{align*}
provided that  $ \log (\tilde{d} n ) \ll n^{1-\kappa}(\log n)^{-1/2}$ and $m \lesssim n$.   Using the similar arguments for the derivation of \eqref{eq:fhwh-fwh-cov}, we have 
\begin{align*} 
	&\max_{j\in[p],\,k\in[q]}\bigg|\frac{1}{n_{3}} \sum_{t\in \mathcal{D}_3} \big[\{\hat{f}_{j} (\hat{\bW}_{t}^{(w)}) - f_{j} (\hat{\bW}_{t}^{(w)})\}  \{\hat{g}_{k} (\hat{\bW}_{t}^{(w)}) - g_{k} (\hat{\bW}_{t}^{(w)})\} - \tilde{\mu}_{2,j,k} \big]\bigg| \notag\\
	&~~~~~~~~~~~~ = O_{\rm p}\{n^{-\kappa/2-\vartheta/(4\vartheta+m_{*})} (m^2\log n)^{(\vartheta+2m_{*}\tilde{\vartheta}+3m_{*} )/(8\vartheta)} \tilde{\beta}_n^{2}\log^{3/4} (\tilde{d}n) \} \notag\\
	&~~~~~~~~~~~~~~~+ O_{\rm p}\{n^{-\kappa/2-1/4}m^{1/2}\tilde{\beta}_n^{3/2} \log^{3/4} (\tilde{d}n)\}    +  O_{\rm p}\{n^{-\kappa} m \tilde{\beta}_n^{2} \log (\tilde{d}n) \}
\end{align*}
provided that  $ \log (\tilde{d}n) \ll n^{1-\kappa}(\log n)^{-1/2}$ and $m \lesssim n$.  
Together with \eqref{eq:mu2-bound}, we have 
\begin{align}\label{eq:tg-41} 
	\max_{j\in[p],\, k\in[q]}|\tilde{{\rm G}}_{41}(j,k)|=&~ O_{\rm p}\{n^{-2\vartheta/(4\vartheta+m_{*})} (m^2\log n)^{(\vartheta+2m_{*}\tilde{\vartheta}+3m_{*} )/(4\vartheta)} (\log n)^{2}\log^{3/2} (\tilde{d}n) \} \notag\\
	& + O_{\rm p}\{n^{-\kappa/2-\vartheta/(4\vartheta+m_{*})} (m^2\log n)^{(\vartheta+2m_{*}\tilde{\vartheta}+3m_{*} )/(8\vartheta)} (\log n)^{2}\log^{7/4} (\tilde{d}n) \} \notag\\
	&      +O_{\rm p}\{n^{-1/2}m (\log n) \log  (\tilde{d}n)\}  +  O_{\rm p}\{n^{-\kappa} m^2   (\log n)^2\log^2 (\tilde{d}n) \} \notag\\
	& + O_{\rm p}\{n^{-\kappa/2-1/4}m^{1/2} (\log n)^{3/2} \log^{3/2} (\tilde{d}n)\}
\end{align}
provided that $ \log (\tilde{d}n) \ll n^{1-\kappa}(\log n)^{-1/2}$ and $m \lesssim n$.   Applying the similar arguments for the derivation of the convergence rate of $\max_{j\in[p],\,k\in[q]}|{\rm H}_{1}(j,k)|$ in Section \ref{sec:sub-g21} for proof of Lemma \ref{lem:fwh-fw-delta}, it holds that
\begin{align}\label{eq:fwh-fw}
	&\max_{j\in[p]}\frac{1}{n_{3}}\sum_{t\in \mathcal{D}_3} |f_{j} (\hat{\bW}_{t}^{(w)})- f_{j} (\bW_{t})| \notag\\
	&~~~~~~~= O_{\rm p}\{n^{-\kappa}m^{2} \log^{1/2} (\tilde{d}n)\} + O_{\rm p} \{ n^{-1/2}   m\log^{1/2} (\tilde{d}n)\} 
\end{align}
provided that $\log(\tilde{d}n) \ll n^{1-\kappa}(\log n)^{-1/2}$. Since $\max_{k\in [q]}|g_{k}|_{\infty} \le \tilde{C}$ and $\max_{t\in\mathcal{D}_3,\,k\in[q]}|\hat{g}_{k}(\hat{\bW}_{t}^{(w)})| \le \tilde{\beta}_{n}$ with $\tilde{\beta}_n =(\log n)\log^{1/2}(\tilde{d}n)$, then
\begin{align}\label{eq:tg-42}
	\max_{j\in[p],\, k\in[q]}|\tilde{{\rm G}}_{42}(j,k)| \le&~ C_2\tilde{\beta}_{n} \times \max_{j\in[p]}\frac{1}{n_{3}}\sum_{t\in \mathcal{D}_3} |f_{j} (\hat{\bW}_{t}^{(w)})- f_{j} (\bW_{t})| \notag\\
	=&~  O_{\rm p}\{n^{-\kappa} m^{2} (\log n)\log (\tilde{d}n)\}  + O_{\rm p} \{n^{-1/2} m(\log n)\log (\tilde{d}n)\}  
\end{align}
provided that $\log(\tilde{d}n) \ll n^{1-\kappa}(\log n)^{-1/2}$. Using the similar arguments for the derivation of \eqref{eq:tg-42}, we can show such convergence rate also holds for $\max_{j\in[p],\, k\in[q]}|\tilde{{\rm G}}_{43}(j,k)|$. 
Combining \eqref{eq:tg-41} and \eqref{eq:tg-42}, by \eqref{eq:tilde-g4-dec}, we have 
\begin{align*}
	\max_{j\in[p],\, k\in[q]}|\tilde{{\rm G}}_{4}(j,k)| \le&~ \max_{j\in[p],\, k\in[q]}|\tilde{{\rm G}}_{41}(j,k)| + \max_{j\in[p],\, k\in[q]}|\tilde{{\rm G}}_{42}(j,k)|  + \max_{j\in[p],\, k\in[q]}|\tilde{{\rm G}}_{43}(j,k)| \\
	=&~ O_{\rm p}\{n^{-2\vartheta/(4\vartheta+m_{*})} (m^2\log n)^{(\vartheta+2m_{*}\tilde{\vartheta}+3m_{*} )/(4\vartheta)} (\log n)^{2}\log^{3/2} (\tilde{d}n) \} \notag\\
	& + O_{\rm p}\{n^{-\kappa/2-\vartheta/(4\vartheta+m_{*})} (m^2\log n)^{(\vartheta+2m_{*}\tilde{\vartheta}+3m_{*} )/(8\vartheta)} (\log n)^{2}\log^{7/4} (\tilde{d}n) \} \notag\\
	&      +O_{\rm p}\{n^{-1/2}m (\log n) \log  (\tilde{d}n)\}  +  O_{\rm p}\{n^{-\kappa} m^2   (\log n)^2\log^2 (\tilde{d}n) \} \notag\\
	& + O_{\rm p}\{n^{-\kappa/2-1/4}m^{1/2} (\log n)^{3/2} \log^{3/2} (\tilde{d}n)\} 
\end{align*}
provided that $ \log (\tilde{d}n) \ll n^{1-\kappa}(\log n)^{-1/2}$ and $m \lesssim n$. Then \eqref{eq:tilde-g4} holds.
$\hfill\Box$

\section{Proof of Lemma \ref{lem:nn-cov-theta}}\label{sec:nn-cov-theta}
Recall $\bTheta=\mathbb{E}(\bet_i\bet_i^{\T}) - \mathbb{E}(\bet_i)\mathbb{E}(\bet_i^{\T})$ and $\tilde{\bTheta} =n_3^{-1}\sum_{i\in\mathcal{D}_3}\tilde{\bet}_i \tilde{\bet}_{i}^{\T}-(n_3^{-1}\sum_{i\in\mathcal{D}_3}\tilde{\bet}_i )(n_3^{-1}\sum_{i\in\mathcal{D}_3}\tilde{\bet}_i)^{\T}$ with $\bet_i=\boldsymbol{\varepsilon}_{i} \otimes \bdelta_i$ and $\tilde{\bet}_i=\tilde{\boldsymbol{\varepsilon}}_{i} \otimes \tilde{\bdelta}_i$. Then
\begin{align}\label{eq:ttheta-theta-dec}
	|\tilde{\bTheta}-\bTheta|_{\infty} \le&~ \underbrace{\max_{j,k\in[p],\,l,t\in[q]}\bigg|\frac{1}{n_3}\sum_{i\in\mathcal{D}_3}\big\{\ve_{i,j}\ve_{i,k}\delta_{i,l}\delta_{i,t}-\mathbb{E}(\ve_{i,j}\ve_{i,k}\delta_{i,l}\delta_{i,t})\big\}\bigg|}_{\tilde{{\textrm S}}_{1}} \notag\\
	&~+ \underbrace{\max_{j,k\in[p],\,l,t\in[q]}\bigg|\frac{1}{n_3}\sum_{i\in\mathcal{D}_3}\tilde{\ve}_{i,j}\tilde{\ve}_{i,k}\tilde{\delta}_{i,l}\tilde{\delta}_{i,t}  -\frac{1}{n_3}\sum_{i\in\mathcal{D}_3}\ve_{i,j}\ve_{i,k}\delta_{i,l}\delta_{i,t}  \bigg|}_{\tilde{{\textrm S}}_{2}} \notag\\
	&~+\underbrace{\max_{j,k\in[p],\,l,t\in[q]}\bigg| \bigg(\frac{1}{n_3}\sum_{i\in\mathcal{D}_3}\ve_{i,j}\delta_{i,l} \bigg)\bigg(\frac{1}{n_3}\sum_{i\in\mathcal{D}_3}\ve_{i,k}\delta_{i,t}\bigg) -\mathbb{E}(\ve_{i,j}\delta_{i,l})\mathbb{E}(\ve_{i,k}\delta_{i,t})\bigg|}_{\tilde{{\textrm S}}_{3}} \notag\\
	&~+\max_{j,k\in[p],\,l,t\in[q]}\bigg| \bigg(\frac{1}{n_3}\sum_{i\in\mathcal{D}_3}\tilde{\ve}_{i,j}\tilde{\delta}_{i,l} \bigg)\bigg(\frac{1}{n_3}\sum_{i\in\mathcal{D}_3}\tilde{\ve}_{i,k}\tilde{\delta}_{i,t} \bigg) \notag\\
	&~~~~\underbrace{~~~~~~~~~~~~~~~~~~~~~- \bigg(\frac{1}{n_3}\sum_{i\in\mathcal{D}_3}\ve_{i,j}\delta_{i,l}\bigg)\bigg( \frac{1}{n_3}\sum_{i\in\mathcal{D}_3}\ve_{i,k}\delta_{i,t}\bigg)\bigg|}_{\tilde{{\textrm S}}_{4}} \,.
\end{align}
Recall $\tilde{d}=p\vee q\vee m$, $n_3\asymp n^{\kappa}$ for some constant $0<\kappa<1$,   $\mathbb{P}(|\ve_{i,j}|>x) \le C_1 e^{- x^2/4}$ and $\mathbb{P}(|\delta_{i,k}|>x) \le C_1 e^{- x^2/4}$ for any $x>0$. 
Identical to the arguments for deriving the convergence rate of ${\rm R}_2$ in Section \ref{sub:sec-h0-R2} for ${\rm R}_2$  defined in \eqref{eq:SigmaGamma}, we have
\begin{align}\label{eq:ts1-bound}
    \tilde{{\rm S}}_1 
	= O_{\rm p}\{n^{-\kappa/2}(\log \tilde{d})^{1/2}\} + O_{\rm p}\{n^{-\kappa}(\log \tilde{d})\log^2 (\tilde{d}n) \}\,.
\end{align}
Notice that $\max_{k\in[p],\,t\in[q]}|\mathbb{E}(\ve_{i,k}\delta_{i,t})|=O(1)$, $\max_{k\in[p],\,t\in[q]} \var(\ve_{i,j}\delta_{i,k}) \le O(1)$ and
\begin{align*}
	\tilde{\rm S}_{3} \le&~2\max_{j,k\in[p],\,l,t\in[q]}\bigg|\frac{1}{n_3}\sum_{i \in \mathcal{D}_3} \big\{\ve_{i,j}\delta_{i,l}-\mathbb{E}(\ve_{i,j}\delta_{i,l})\big\}\mathbb{E}(\ve_{i,k}\delta_{i,t})\bigg|\\
	&+\max_{j\in[p],\,l\in[q]}\bigg|  \frac{1}{n_3}\sum_{i \in \mathcal{D}_3} \big\{\ve_{i,j}\delta_{i,l}-\mathbb{E}(\ve_{i,j}\delta_{i,l}) \big\} \bigg|^2\,.
\end{align*}
Using the similar arguments for the derivation of \eqref{eq:uv-bound-2}, it holds that   
\begin{align}\label{eq:eps-delta-n3} 
	\max_{j\in[p],\,k\in[q]}\bigg|\frac{1}{n_3}\sum_{i\in \mathcal{D}_3} \big\{\ve_{i,j}\delta_{i,k} - \mathbb{E}(\ve_{i,j}\delta_{i,k}) \big\}\bigg|=O_{\rm p}\big\{n_3^{-1/2}(\log \tilde{d})^{1/2}\big\}
\end{align}
provided that $\log \tilde{d} \lesssim n_3^{1/3}$.  Then
\begin{align}\label{eq:ts3-bound}
	 \tilde{{\rm S}}_3 = O_{\rm p}\{n^{-\kappa/2}(\log \tilde{d})^{1/2}\}  
\end{align}
provided that $\log \tilde{d} \lesssim n^{\kappa/3}$.  
As we will show in Sections \ref{sec:sub-tilde-s2} and \ref{sec:sub-tilde-s4}, 
\begin{align}\label{eq:tilde-s1}
	\tilde{\rm S}_{2}  =&~ O_{\rm p}\{n^{-\vartheta/(4\vartheta+m_{*})} (m^2\log n)^{(\vartheta+2m_{*}\tilde{\vartheta}+3m_{*} )/(8\vartheta)} (\log n)^4\log^{9/4}(\tilde{d}n) \} \notag \\
	& +  O_{\rm p}\{n^{-1/4}  m^{1/2}  (\log n)^{7/2}\log^{2} (\tilde{d}n) \}  + O_{\rm p}\{n^{-\kappa/2}m (\log n)^{7/2}\log^{7/4}  (\tilde{d}n) \}\\
	&+  O_{\rm p}\{n^{-\kappa} m^2 (\log n)^3\log^{2} (\tilde{d}n)\} \notag
\end{align}
provided that $\log(\tilde{d}n) \ll \min\{n^{1-\kappa} (\log n)^{-1/2}, n^{2\kappa/5}(\log n)^{-2/5} \}$ and $m \lesssim n$,  and
\begin{align}\label{eq:tilde-s4}
	\tilde{\rm S}_{4} = &~ O_{\rm p}\{n^{-2\vartheta/(4\vartheta+m_{*})} (m^2\log n)^{(\vartheta+2m_{*}\tilde{\vartheta}+3m_{*} )/(4\vartheta)} (\log n)^{2}\log^{3/2} (\tilde{d}n) \} \notag\\
	& + O_{\rm p}\{n^{-\kappa/2-\vartheta/(4\vartheta+m_{*})} (m^2\log n)^{(\vartheta+2m_{*}\tilde{\vartheta}+3m_{*} )/(8\vartheta)} (\log n)^{2}\log^{7/4} (\tilde{d}n) \} \notag\\
	&  +O_{\rm p}\{n^{-1/2}m (\log n) \log  (\tilde{d}n)\}  +  O_{\rm p}\{n^{-\kappa} m^2   (\log n)^2\log^2 (\tilde{d}n) \} \notag\\
	&  + O_{\rm p}\{n^{-\kappa/2-1/4}m^{1/2} (\log n)^{3/2} \log^{3/2} (\tilde{d}n)\}
\end{align} 
provided that  $m \ll \min [ n^{4\vartheta^2/\{\varrho(4\vartheta+m_{*})\}} (\log n)^{-4\vartheta/\varrho-1/2}\{\log (\tilde{d}n)\}^{-3\vartheta/\varrho}, n^{\kappa/2}(\log n)^{-1}\{\log (\tilde{d}n )\}^{-1} ]$  and $\log(\tilde{d}n) \ll \min\{n^{1-\kappa} (\log n)^{-1/2},n^{\kappa/3}, n^{4\vartheta/(12\vartheta+3m_{*})}(\log n)^{-4/3-\varrho/(6\vartheta)} \}$. Combining \eqref{eq:ts1-bound} and \eqref{eq:ts3-bound}--\eqref{eq:tilde-s4}, by \eqref{eq:ttheta-theta-dec}, we then have
\begin{align*} 
	|\tilde{\bTheta}-\bTheta|_{\infty} =&~ O_{\rm p}\{n^{-\vartheta/(4\vartheta+m_{*})} (m^2\log n)^{(\vartheta+2m_{*}\tilde{\vartheta}+3m_{*})/(8\vartheta)} (\log n)^4\log^{9/4}(\tilde{d}n) \} \notag \\
	& +  O_{\rm p}\{n^{-1/4}  m^{1/2}  (\log n)^{7/2}\log^{2} (\tilde{d}n) \}  + O_{\rm p}\{n^{-\kappa/2}m (\log n)^{7/2}\log^{7/4}  (\tilde{d}n) \}
\end{align*}
provided that  $m \ll \min[n^{4\vartheta^2/\{\varrho(4\vartheta+m_{*})\}} (\log n)^{-4\vartheta/\varrho-1/2}\{\log (\tilde{d}n)\}^{-3\vartheta/\varrho}, n^{\kappa/2}(\log n)^{-1}\{\log (\tilde{d}n )\}^{-1}]$  and $\log(\tilde{d}n) \ll \min\{n^{1-\kappa} (\log n)^{-1/2},n^{\kappa/3}, n^{4\vartheta/(12\vartheta+3m_{*})}(\log n)^{-4/3-\varrho/(6\vartheta)} \}$.   Thus, we complete the proof of Lemma \ref{lem:nn-cov-theta}. 
$\hfill\Box$

\subsection{Proof of \eqref{eq:tilde-s1}}\label{sec:sub-tilde-s2}
Analogous to \eqref{eq:4expend}, $n_3^{-1}\sum_{i\in\mathcal{D}_3}(\tilde{\ve}_{i,j}\tilde{\ve}_{i,k}\tilde{\delta}_{i,l}\tilde{\delta}_{i,t} - \ve_{i,j}\ve_{i,k}\delta_{i,l}\delta_{i,t})$ can be decomposed into 15 terms. To derive the convergence rate of $\tilde{{\rm S}}_2$, by the symmetry, we only consider the convergence rates of the following terms:
\begin{align*}
	\tilde{{\rm S}}_{21}= &\max_{j,k\in[p],\,l,t\in[q]}\bigg|\frac{1}{n_3}\sum_{i\in\mathcal{D}_3}(\tilde{\ve}_{i,j}-\ve_{i,j})\ve_{i,k}\delta_{i,l}\delta_{i,t}\bigg|\,, \\
	\tilde{{\rm S}}_{22}= &\max_{j,k\in[p],\,l,t\in[q]}\bigg|\frac{1}{n_3}\sum_{i\in\mathcal{D}_3}(\tilde{\ve}_{i,j}-\ve_{i,j})(\tilde{\ve}_{i,k} - \ve_{i,k})\delta_{i,l}\delta_{i,t}\bigg|\,,\\  
	\tilde{{\rm S}}_{23}=& \max_{j,k\in[p],\,l,t\in[q]}\bigg|\frac{1}{n_3}\sum_{i\in\mathcal{D}_3}(\tilde{\ve}_{i,j}-\ve_{i,j})(\tilde{\delta}_{i,l}-\delta_{i,l}) \ve_{i,k}\delta_{i,t}\bigg|\,,\\  
	\tilde{{\rm S}}_{24}= &\max_{j,k\in[p],\,l,t\in[q]}\bigg|\frac{1}{n_3}\sum_{i\in\mathcal{D}_3}(\tilde{\ve}_{i,j}-\ve_{i,j})(\tilde{\ve}_{i,k}-\ve_{i,k})(\tilde{\delta}_{i,l}-\delta_{i,l})\delta_{i,t}\bigg|\,, \\
	\tilde{{\rm S}}_{25}= &\max_{j,k\in[p],\,l,t\in[q]}\bigg|\frac{1}{n_3}\sum_{i\in\mathcal{D}_3}(\tilde{\ve}_{i,j}-\ve_{i,j})(\tilde{\ve}_{i,k}-\ve_{i,k})(\tilde{\delta}_{i,l}-\delta_{i,l})(\tilde{\delta}_{i,t}-\delta_{i,t})\bigg|\,.
\end{align*}
Recall $\varepsilon_{i,j} = U_{i,j}- f_{j}(\bW_i)$ and $ \tilde{\varepsilon}_{i,j} = \hat{U}_{i,j}^{(w)} - \hat{f}_{j}(\hat{\bW}_{i}^{(w)})$. We have
\begin{align*}
	\tilde{{\rm S}}_{21} \le&~  \underbrace{\max_{j,k\in[p],\,l,t\in[q]}\bigg|\frac{1}{n_{3}}\sum_{i\in\mathcal{D}_3} \big \{|\hat{U}_{i,j}^{(w)}-U_{i,j}| (|\ve_{i,k}| + |\tilde{\ve}_{i,k}|)   (|\delta_{i,l}|+ |\tilde{\delta}_{i,l} | )  (|\delta_{i,t}| + |\tilde{\delta}_{i,t} |) \big\}\bigg| }_{\tilde{{\textrm S}}_{211}}\\
	&+\underbrace{\max_{j,k\in[p],\,l,t\in[q]}\bigg|\frac{1}{n_{3}}\sum_{i\in\mathcal{D}_3}\big \{ |\hat{f}_{j}(\hat{\bW}_{i}^{(w)}) - f_{j}(\hat{\bW}^{(w)}_i)  | (|\ve_{i,k}| + |\tilde{\ve}_{i,k}|)   (|\delta_{i,l}|+ |\tilde{\delta}_{i,l} | )  (|\delta_{i,t}| + |\tilde{\delta}_{i,t} |) \big\} \bigg| }_{\tilde{{\textrm S}}_{212}}\\
	&+\underbrace{\max_{j,k\in[p],\,l,t\in[q]}\bigg|\frac{1}{n_{3}}\sum_{i\in\mathcal{D}_3}\big\{|f_{j}(\hat{\bW}_{i}^{(w)}) - f_{j}(\bW_i) | (|\ve_{i,k}| + |\tilde{\ve}_{i,k}|)   (|\delta_{i,l}|+ |\tilde{\delta}_{i,l} | )  (|\delta_{i,t}| + |\tilde{\delta}_{i,t} |) \big\}\bigg| }_{\tilde{{\textrm S}}_{213}}\,.
\end{align*}
Write $\aleph(i,k,l,t) = (|\ve_{i,k}| + |\tilde{\ve}_{i,k}|)   (|\delta_{i,l}|+ |\tilde{\delta}_{i,l} | )  (|\delta_{i,t}| + |\tilde{\delta}_{i,t} |)$ for any $i\in[n_3]$, $k\in[p]$ and $l,t\in[q]$. 
Given $Q>0$, it holds that
\begin{align*}
	\tilde{\rm S}_{211} \le&~ \underbrace{\max_{j,k\in[p],\,l,t\in[q]}\bigg|\frac{1}{n_{3}}\sum_{i\in\mathcal{D}_3}|\hat{U}_{i,j}^{(w)}-U_{i,j}|\aleph(i,k,l,t) I(|\ve_{i,k}|,|\delta_{i,l}|,|\delta_{i,t}|\le Q)\bigg|}_{\tilde{{\textrm S}}_{2111}}\\
	&+ \underbrace{\max_{j,k\in[p],\,l,t\in[q]}\bigg|\frac{1}{n_{3}}\sum_{i\in\mathcal{D}_3}|\hat{U}_{i,j}^{(w)}-U_{i,j}|\aleph(i,k,l,t) I(|\ve_{i,k}|,|\delta_{i,l}|\le Q) I(|\delta_{i,t}|> Q)\bigg|}_{\tilde{{\textrm S}}_{2112}}\\
	&+ \underbrace{\max_{j,k\in[p],\,l,t\in[q]}\bigg|\frac{1}{n_{3}}\sum_{i\in\mathcal{D}_3}|\hat{U}_{i,j}^{(w)}-U_{i,j}| \aleph(i,k,l,t) I(|\ve_{i,k}|\le Q)I(|\delta_{i,l}|> Q)  \bigg|}_{\tilde{{\textrm S}}_{2113}}\\
	&+ \underbrace{\max_{j,k\in[p],\,l,t\in[q]}\bigg|\frac{1}{n_{3}}\sum_{i\in\mathcal{D}_3}|\hat{U}_{i,j}^{(w)}-U_{i,j}| \aleph(i,k,l,t) I(|\ve_{i,k}|> Q)  \bigg|}_{\tilde{{\textrm S}}_{2114}} \,.
\end{align*}
Due to $\max_{i\in\mathcal{D}_3,\,j\in[p]}|\hat{f}_{j}(\hat{\bW}_{i}^{(w)})| \le \tilde{\beta}_n$ with $\tilde{\beta}_n=(\log n)\log^{1/2}(\tilde{d}n)$, and $\max_{i\in\mathcal{D}_3,\,j\in[p]}|\hat{U}_{i,j}^{(w)}|\le \sqrt{2 \log n_1}$,  we  have $\max_{i\in\mathcal{D}_3,\, j\in[p]}| \tilde{\varepsilon}_{i,j}| < 2\tilde{\beta}_{n}$. Analogously, we  also have  $\max_{i\in\mathcal{D}_{3},k\in[q]}|\tilde{\delta}_{i,k}| < 2\tilde{\beta}_{n}$.  Recall $U_{i,j}^{*}=U_{i,j}I(|U_{i,j}| \le M_1) + M_{1} \cdot {\rm sign}(U_{i,j})I(|U_{i,j}|>M_1)$ with $M_1=\sqrt{2\log n_3}$.   
By \eqref{eq:uhw-u} and \eqref{eq:ustar-u}, it holds that
\begin{align*}
	\tilde{\rm S}_{2111} \le&~ \max_{j\in[p]}\frac{C_1(Q^3+\tilde{\beta}^{3}_n)}{n_{3}}\sum_{i\in \mathcal{D}_3}|\hat{U}_{i,j}^{(w)}-U_{i,j} | \\
	\le&~ \max_{j\in[p]}\frac{C_1(Q^3 +\tilde{\beta}^{3}_n)}{n_{3}}\sum_{i\in \mathcal{D}_3}|\hat{U}_{i,j}^{(w)}-U_{i,j}^{*}| + \max_{j\in[p]}\frac{C_1(Q^3+ \tilde{\beta}^{3}_n)}{n_{3}}\sum_{i\in \mathcal{D}_3}|U_{i,j}^{*}-U_{i,j}|\\
	=&~O_{\rm p} \{(Q^3+ \tilde{\beta}^{3}_n) n^{-\kappa} \log^{3/2}(\tilde{d}n)\} + O_{\rm p}\{(Q^3+\tilde{\beta}^{3}_n)n^{-1/2} \log^{1/2}(\tilde{d}n)\}
\end{align*}
provided that $\log(\tilde{d}n) \ll n^{1-\kappa}(\log n)^{-1/2}$. Analogous to the derivation of the convergence rate of $\max_{j\in[p],\,k\in[q]}|\tilde{\rm H}_{4}(j,k)|$ in Section \ref{sec:sub-uh-u-delta} for the proof of Lemma \ref{lem:uh-u-delta}, we have $\tilde{\rm S}_{2112} =o_{\rm p}(n^{-1}) =\tilde{\rm S}_{2113}$ and $\tilde{\rm S}_{2114} =o_{\rm p}(n^{-1})$ provided that $\log(\tilde{d}n) \lesssim Q^2$.   By selecting $Q=C_2 \log^{1/2}(\tilde{d}n)$ for some sufficiently large constant $C_2>0$, it holds that
\begin{align}\label{eq:tilde-s211}
	\tilde{\rm S}_{211} = O_{\rm p} \{n^{-\kappa} (\log n)^3 \log^{3}(\tilde{d}n)\} + O_{\rm p}\{ n^{-1/2}(\log n)^3 \log^{2}(\tilde{d}n)\}
\end{align}
provided that $\log(\tilde{d}n) \ll n^{1-\kappa} (\log n)^{-1/2}$. Recall $\mathcal{W}_{\mathcal{D}_j}=\{(\bX_i,\bY_i,\bZ_i): i\in \mathcal{D}_j\}$ for $j\in[3]$, where $\mathcal{D}_1, \mathcal{D}_2$ and $\mathcal{D}_3$ are three disjoint subsets of $[n]$ with $|\mathcal{D}_1|=n_1\asymp n$, $|\mathcal{D}_2|=n_2\asymp n$ and $|\mathcal{D}_3|=n_3\asymp n^{\kappa}$  for some constant $0<\kappa<1$ and $n_1+n_2+n_3=n$.  Using the similar arguments for the derivation of \eqref{eq:fhwh-fwh-cov}, by \eqref{eq:sigma2-z}, we have 
\begin{align*} 
	&\max_{j\in[p]}\bigg|\frac{1}{n_{3}} \sum_{i\in \mathcal{D}_3} \big[|\hat{f}_{j} (\hat{\bW}_{i}^{(w)}) - f_{j} (\hat{\bW}_{i}^{(w)})| - \mathbb{E}\big\{|\hat{f}_{j} (\hat{\bW}_{i}^{(w)}) - f_{j} (\hat{\bW}_{i}^{(w)})| \,\big|\, \mathcal{W}_{\mathcal{D}_1}, \mathcal{W}_{\mathcal{D}_2} \big\} \big]\bigg| \notag\\
	&~~~~~~~~~~ =  O_{\rm p}\{n^{-\kappa/2-\vartheta/(4\vartheta+m_{*})} (m^2\log n)^{(\vartheta+2m_{*}\tilde{\vartheta}+3m_{*} )/(8\vartheta)} (\log n)\log^{5/4}(\tilde{d}n) \}  \\
	&~~~~~~~~~~~~~~ +  O_{\rm p}\{n^{-\kappa/2-1/4}  m^{1/2}  (\log n)^{1/2}\log (\tilde{d}n) \}  + O_{\rm p}\{n^{-\kappa}m (\log n) \log^{3/2}  (\tilde{d}n) \}  
\end{align*}
provided that $\log(\tilde{d}n) \ll n^{1-\kappa} (\log n)^{-1/2}$ and $m \lesssim n$.   Since $\mathbb{E}\{|\hat{f}_{j} (\hat{\bW}_{i}^{(w)}) - f_{j} (\hat{\bW}_{i}^{(w)})| \,|\, \mathcal{W}_{\mathcal{D}_1}, \mathcal{W}_{\mathcal{D}_2}\} \le  [ \mathbb{E}\{|\hat{f}_{j} (\hat{\bW}_{i}^{(w)}) - f_{j} (\hat{\bW}_{i}^{(w)})|^2 \,|\, \mathcal{W}_{\mathcal{D}_1}, \mathcal{W}_{\mathcal{D}_2}\} ]^{1/2}$
	for any $i\in\mathcal{D}_3$, by \eqref{eq:sigma2-z}, it holds that
	\begin{align}\label{eq:fhwh-fwh}
		&\max_{j\in [p]} \frac{1}{n_{3}}\sum_{i\in\mathcal{D}_3}|\hat{f}_{j}(\hat{\bW}_{i}^{(w)}) - f_{j}(\hat{\bW}^{(w)}_i)| \notag  \\
		&~~~~~~~~~~ = O_{\rm p}\{n^{-\vartheta/(4\vartheta+m_{*})} (m^2\log n)^{(\vartheta+2m_{*}\tilde{\vartheta}+3m_{*} )/(8\vartheta)} (\log n)\log^{3/4}(\tilde{d}n) \} \\
		&~~~~~~~~~~~~~~ +  O_{\rm p}\{n^{-1/4}  m^{1/2}  (\log n)^{1/2}\log^{1/2} (\tilde{d}n) \}  + O_{\rm p}\{n^{-\kappa/2}m (\log n)^{1/2}\log^{1/4}  (\tilde{d}n) \}  \notag
	\end{align}
	provided that $\log(\tilde{d}n) \ll \min\{n^{1-\kappa} (\log n)^{-1/2}, n^{2\kappa/5}(\log n)^{-2/5} \}$ and $m \lesssim n$. Applying the similar arguments for the derivation of \eqref{eq:tilde-s211}, by \eqref{eq:fhwh-fwh}, we have
	\begin{align}\label{eq:tilde-s212}
		\tilde{\rm S}_{212}  =&~ O_{\rm p}\{n^{-\vartheta/(4\vartheta+m_{*})} (m^2\log n)^{(\vartheta+2m_{*}\tilde{\vartheta}+3m_{*} )/(8\vartheta)} (\log n)^4\log^{9/4}(\tilde{d}n) \} \notag \\
		& +  O_{\rm p}\{n^{-1/4}  m^{1/2}  (\log n)^{7/2}\log^{2} (\tilde{d}n) \}  + O_{\rm p}\{n^{-\kappa/2}m (\log n)^{7/2}\log^{7/4}  (\tilde{d}n) \}
	\end{align}
	provided that $\log(\tilde{d}n) \ll \min\{n^{1-\kappa} (\log n)^{-1/2}, n^{2\kappa/5}(\log n)^{-2/5} \}$ and $m \lesssim n$. Analogously, by \eqref{eq:fwh-fw}, it holds that
	\begin{align*}
		\tilde{\rm S}_{213} =  O_{\rm p}\{n^{-\kappa} m^2 (\log n)^3\log^{2} (\tilde{d}n)\} 
		+ O_{\rm p} \{ n^{-1/2} m  (\log n)^3 \log^{2} (\tilde{d}n)\} 
	\end{align*}
	provided that $\log(\tilde{d}n) \ll n^{1-\kappa} (\log n)^{-1/2}$.   Together with \eqref{eq:tilde-s211} and \eqref{eq:tilde-s212},  we have
	\begin{align}\label{eq:ts21-bound}
		\tilde{\rm S}_{21} \le&~ \tilde{\rm S}_{211}+ \tilde{\rm S}_{212}+ \tilde{\rm S}_{213} \notag\\ 
		=&~O_{\rm p}\{n^{-\vartheta/(4\vartheta+m_{*})} (m^2\log n)^{(\vartheta+2m_{*}\tilde{\vartheta}+3m_{*} )/(8\vartheta)} (\log n)^4\log^{9/4}(\tilde{d}n) \} \notag \\
		& +  O_{\rm p}\{n^{-1/4}  m^{1/2}  (\log n)^{7/2}\log^{2} (\tilde{d}n) \}  + O_{\rm p}\{n^{-\kappa/2}m (\log n)^{7/2}\log^{7/4}  (\tilde{d}n) \} \notag\\
		& +O_{\rm p}\{n^{-\kappa} m^2 (\log n)^3\log^{2} (\tilde{d}n)\}
	\end{align}
	provided that $\log(\tilde{d}n) \ll \min\{n^{1-\kappa} (\log n)^{-1/2}, n^{2\kappa/5}(\log n)^{-2/5} \}$ and $m \lesssim n $. 
	Since $\tilde{\rm S}_{22}$, $\tilde{\rm S}_{23}$, $\tilde{\rm S}_{24}$ and $\tilde{\rm S}_{25}$ can also be bounded by $\tilde{\rm S}_{211}+ \tilde{\rm S}_{212}+ \tilde{\rm S}_{213}$, we know the  convergence rate specified in \eqref{eq:ts21-bound} also holds for $\tilde{\rm S}_{22}$, $\tilde{\rm S}_{23}$, $\tilde{\rm S}_{24}$ and $\tilde{\rm S}_{25}$. Hence,  \eqref{eq:tilde-s1} holds.
	$\hfill\Box$

	\subsection{Proof of \eqref{eq:tilde-s4}}\label{sec:sub-tilde-s4}
	Notice that
	\begin{align}\label{eq:tilde-s4-dep}
		\tilde{{\rm S}}_{4} \le&~2\max_{j,k\in[p],\,l,t\in[q]}\bigg|\bigg\{\frac{1}{n_{3}}\sum_{i\in \mathcal{D}_3}(\tilde{\ve}_{i,j}\tilde{\delta}_{i,l}-\ve_{i,j}\delta_{i,l})\bigg\}\bigg(\frac{1}{n_{3}}\sum_{i\in \mathcal{D}_3}\ve_{i,k}\delta_{i,t} \bigg)\bigg| \notag \\
		&+ \max_{j\in[p],\,l\in[q]}\bigg|\frac{1}{n_{3}}\sum_{i\in \mathcal{D}_3}(\tilde{\ve}_{i,j}\tilde{\delta}_{i,l}-\ve_{i,j}\delta_{i,l}) \bigg|^2\,.
	\end{align}
	Due to $\tilde{\varepsilon}_{t,j}\tilde{\delta}_{t,k} - \varepsilon_{t,j}\delta_{t,k}= (\tilde{\varepsilon}_{t,j} - \varepsilon_{t,j})\delta_{t,k}  +  (\tilde{\delta}_{t,k} - \delta_{t,k})\varepsilon_{t,j} +(\tilde{\varepsilon}_{t,j} - \varepsilon_{t,j})(\tilde{\delta}_{t,k}-\delta_{t,k})$, by Lemmas \ref{lem:nn-eh-e-delta} and \ref{lem:nn-epsdts-epdt}, we have 
	\begin{align*}
		&\max_{j\in[p],\,l\in[q]}\bigg|\frac{1}{n_{3}}\sum_{i\in \mathcal{D}_{3}}(\tilde{\ve}_{i,j} \tilde{\delta}_{i,l} -\ve_{i,j}\delta_{i,l}) \bigg| \\
		&~~~~~~~~~=  O_{\rm p}\{n^{-2\vartheta/(4\vartheta+m_{*})} (m^2\log n)^{(\vartheta+2m_{*}\tilde{\vartheta}+3m_{*} )/(4\vartheta)} (\log n)^{2}\log^{3/2} (\tilde{d}n) \} \notag\\
		&~~~~~~~~~~~~ + O_{\rm p}\{n^{-\kappa/2-\vartheta/(4\vartheta+m_{*})} (m^2\log n)^{(\vartheta+2m_{*}\tilde{\vartheta}+3m_{*} )/(8\vartheta)} (\log n)^{2}\log^{7/4} (\tilde{d}n) \} \notag\\
		&~~~~~~~~~~~~ +O_{\rm p}\{n^{-1/2}m (\log n) \log  (\tilde{d}n)\}  +  O_{\rm p}\{n^{-\kappa} m^2   (\log n)^2\log^2 (\tilde{d}n) \} \notag\\
		&~~~~~~~~~~~~ + O_{\rm p}\{n^{-\kappa/2-1/4}m^{1/2} (\log n)^{3/2} \log^{3/2} (\tilde{d}n)\}
	\end{align*}
	provided that $\log(\tilde{d}n) \ll n^{1-\kappa} (\log n)^{-1/2}$ and $m \lesssim n$. Recall $n_3\asymp n^{\kappa}$ for some constant $0<\kappa<1$. By $\max_{k\in[p],\,t\in[q]}|\mathbb{E}(\ve_{i,k}\delta_{i,t})|=O(1)$ and \eqref{eq:eps-delta-n3}, it holds that $\max_{k\in[p],\,t\in[q]}|n_{3}^{-1}\sum_{i\in \mathcal{D}_3}\ve_{i,k}\delta_{i,t}|=O_{\rm p}(1)$ provided that $\log \tilde{d} \lesssim n^{\kappa/3}$.  Hence, by \eqref{eq:tilde-s4-dep}, we have 
	\begin{align*}
		\tilde{{\rm S}}_{4}=&~ O_{\rm p}\{n^{-2\vartheta/(4\vartheta+m_{*})} (m^2\log n)^{(\vartheta+2m_{*}\tilde{\vartheta}+3m_{*} )/(4\vartheta)} (\log n)^{2}\log^{3/2} (\tilde{d}n) \} \notag\\
		& + O_{\rm p}\{n^{-\kappa/2-\vartheta/(4\vartheta+m_{*})} (m^2\log n)^{(\vartheta+2m_{*}\tilde{\vartheta}+3m_{*} )/(8\vartheta)} (\log n)^{2}\log^{7/4} (\tilde{d}n) \} \notag\\
		&  +O_{\rm p}\{n^{-1/2}m (\log n) \log  (\tilde{d}n)\}  +  O_{\rm p}\{n^{-\kappa} m^2   (\log n)^2\log^2 (\tilde{d}n) \} \notag\\
		&  + O_{\rm p}\{n^{-\kappa/2-1/4}m^{1/2} (\log n)^{3/2} \log^{3/2} (\tilde{d}n)\}
	\end{align*} 
	provided that  $m \ll \min[n^{4\vartheta^2/\{\varrho(4\vartheta+m_{*})\}} (\log n)^{-4\vartheta/\varrho-1/2}\{\log (\tilde{d}n)\}^{-3\vartheta/\varrho}, n^{\kappa/2}(\log n)^{-1}\{\log (\tilde{d}n )\}^{-1} ]$  and $\log(\tilde{d}n) \ll \min\{n^{1-\kappa} (\log n)^{-1/2},n^{\kappa/3}, n^{4\vartheta/(12\vartheta+3m_{*})}(\log n)^{-4/3-\varrho/(6\vartheta)} \}$ with $\varrho =    \vartheta+2m_{*}\tilde{\vartheta}+3m_{*} $.  Then \eqref{eq:tilde-s4} holds.
	$\hfill\Box$

\section{Proof of Lemma \ref{lem:epsdeth-epsdet} }\label{sec:sub-eps-epsh}  
To prove Lemma \ref{lem:epsdeth-epsdet}, we need Lemmas \ref{lem:w1e}--\ref{lem:coeff}, with their proofs given in Sections \ref{sec:sub-w1e}--\ref{sec:sub-coeff}, respectively. Recall $\tilde{d}=p\vee q\vee m$ and $s=(\max_{j\in[p]}|\balpha_j |_{0}) \vee (\max_{k\in[q]}|\bbeta_k |_{0})$.

\begin{lem}\label{lem:w1e}
Under \eqref{eq:regressionUV} and Condition  {\rm \ref{cn:subgaussian}}{\rm(i)}, if $ \log \tilde{d} \ll n^{1/10} (\log n)^{-1/2}$, there exist universal constants $K_3>0$ and $K_4 >0$ such that  
\begin{align*}
&\mathbb{P}\bigg\{\max_{l \in [m],\,j\in[p]}\bigg|\frac{1}{n}\sum_{i=1}^{n}\hat{W}_{i,l}\ve_{i,j}\bigg|> xn^{-1/2}\log^{1/2} (pm)\bigg\} \le  K_3\exp\{-K_4x^2\log (pm) \} + O\{(\tilde{d}n)^{-2}\} \,,\\
&\mathbb{P}\bigg\{\max_{l \in [m],\,k\in[q]}\bigg|\frac{1}{n}\sum_{i=1}^{n}\hat{W}_{i,l}\delta_{i,k}\bigg|> xn^{-1/2}\log^{1/2} (qm)\bigg\} \le K_3  \exp\{-K_4x^2\log (qm) \} + O\{(\tilde{d}n)^{-2}\}
\end{align*}
for any $x\in[\bar{C},\breve{C}]$ with some sufficiently large constants $\breve{C}>\bar{C}>1$.
\end{lem}

\begin{lem}\label{lem:wh}
It holds that
\begin{align*}
|\hat{\bSigma}_{W}-\bSigma_{W} |_{\infty}=  O_{\rm p}\{n^{-1/2}(\log n)\log (\tilde{d}n)\}
\end{align*}
provided that $\log \tilde{d} \lesssim n^{1/3}$. 
\end{lem}

\begin{lem}\label{lem:coeff}
Assume \eqref{eq:regressionUV} and Condition {\rm \ref{cn:subgaussian}} hold. If  $s \ll   n^{1/2}(\log n)^{-1}\{\log (\tilde{d}n)\}^{-1}$,  it holds that
\begin{align*}
&\max_{j \in [p]}|\hat{\balpha}_j-\balpha_j |_{1} =O_{{\rm p}}\big\{sn^{-1/2} (\log \tilde{d})^{1/2}\big\} =\max_{k \in [q]}|\hat{\bbeta}_k-\bbeta_k |_{1}\,,\\
&~~~~~~~~~~~~~~\max_{j \in [p]}|\hat{\balpha}_j|_{1} =O_{{\rm p}}(\sqrt{s}) =\max_{k \in [q]}|\hat{\bbeta}_k|_{1}
\end{align*}
provided that $ \log \tilde{d} \ll n^{1/10}(\log n)^{-1/2}$.
\end{lem}
Recall
\begin{align*}
\tilde{\delta}_{4,k}(U_{s,j})=&~\mathbb{E} \big[e^{U_{i,j}^2/2}  \big\{I(U_{s,j}\le U_{i,j})-\Phi(U_{i,j})\big\} \delta_{i,k} I\{|U_{i,j}|\le \sqrt{3(\log n)/5}\}\,\big|\,U_{s,j} \big]\,,\\
\tilde{\delta}_{5,j}(V_{s,k})=&~\mathbb{E} \big[e^{V_{i,k}^2/2} \big\{I(V_{s,k}\le V_{i,k})-\Phi(V_{i,k})\big\} \varepsilon_{i,j} I\{|V_{i,k}|\le \sqrt{3(\log n)/5} \}\,\big|\,V_{s,k} \big]
\end{align*}
with  $i\ne s$.
Notice that
\begin{align*}
\frac{1}{n}\sum_{i=1}^{n}\big(\hat{\ve}_{i,j}\hat{\delta}_{i,k}-\ve_{i,j}\delta_{i,k} \big)=&~\underbrace{\frac{1}{n}\sum_{i=1}^{n}(\hat{\ve}_{i,j}-\ve_{i,j})\delta_{i,k}}_{{\textrm T}_{1}(j,k)} + \underbrace{\frac{1}{n}\sum_{i=1}^{n}(\hat{\delta}_{i,k}-\delta_{i,k})\ve_{i,j}}_{{\textrm T}_{2}(j,k)}\\
&+\underbrace{\frac{1}{n}\sum_{i=1}^{n} (\hat{\ve}_{i,j}-\ve_{i,j})(\hat{\delta}_{i,k}-\delta_{i,k})   }_{{\textrm T}_{3}(j,k)} \,.
\end{align*}
As we will show in Sections \ref{sec:sub-t12} and \ref{sec:sub-t3}, 
\begin{align}
{\rm T}_{1}(j,k) =&~  \frac{\sqrt{2\pi}(n-1)}{n(n+1)}\sum_{s=1}^{n} \tilde{\delta}_{4,k}(U_{s,j})    + {\rm Rem}_{11}(j,k) \label{eq:t1-2con}\,,\\
{\rm T}_{2}(j,k) = &~ \frac{\sqrt{2\pi}(n-1)}{n(n+1)}\sum_{s=1}^{n} \tilde{\delta}_{5,j}(V_{s,k})  + {\rm Rem}_{12}(j,k) \label{eq:t2-2con}
\end{align}
with 
\begin{align*}
\max_{j\in[p],\, k\in[q]}|{\rm Rem}_{11}(j,k)| =&~   O_{\rm p} \{s^{1/2}n^{-7/10} \log^{3/2}(\tilde{d}n)\}   +O_{\rm p} \{s^{1/2}n^{-13/20}(\log n)^{-3/4}\log(\tilde{d}n)\}\\
=&~\max_{j\in[p],\, k\in[q]}|{\rm Rem}_{12}(j,k)|
\end{align*}
provided that $s \ll   n^{1/2}(\log n)^{-1}\{\log (\tilde{d}n)\}^{-1}$ and $ \log \tilde{d} \ll n^{1/10} (\log n)^{-1/2}$,   and  
\begin{align}\label{eq:t3}
\max_{j\in[p],\, k\in[q]}|{\rm T}_{3}(j,k)|  =&~    O_{\rm p} \{sn^{-7/10} (\log n) ^{1/2}\}   + O_{\rm p} \{s^{3/2}n^{-1}(\log n) (\log \tilde{d})  \log^{1/2}(\tilde{d}n)\}   \notag\\
&+ O_{\rm p}(s^2n^{-1} \log \tilde{d})   
\end{align}
provided that $s \ll   n^{1/2}(\log n)^{-1}\{\log (\tilde{d}n)\}^{-1}$ and $ \log \tilde{d} \ll n^{1/10}(\log n)^{-1/2}$.  Hence, we have
\begin{align*}
\frac{1}{n}\sum_{i=1}^{n} \hat{\ve}_{i,j}\hat{\delta}_{i,k} - \frac{1}{n}\sum_{i=1}^{n}\ve_{i,j}\delta_{i,k} 
=   \frac{\sqrt{2\pi}(n-1)}{n(n+1)}\sum_{s=1}^{n} \big\{\tilde{\delta}_{4,k}(U_{s,j}) + \tilde{\delta}_{5,j}(V_{s,k})\big\}  + {\rm Rem}_1(j,k)
\end{align*} 
with
\begin{align*}
\max_{j\in[p],\, k\in[q]}|{\rm Rem}_1(j,k)| = &~  O_{\rm p} \{s  n^{-7/10} \log^{3/2}(\tilde{d}n)\} +  O_{\rm p} \{s^{1/2} n^{-13/20}(\log n)^{-3/4} \log(\tilde{d}n)\}
\end{align*}
provided that  $s \lesssim n^{3/10}(\log \tilde{d})^{1/2}$ and $ \log \tilde{d} \ll n^{1/10}(\log n)^{-1/2}$.  We complete the proof of Lemma \ref{lem:epsdeth-epsdet}.
$\hfill\Box$

\subsection{Proof of Lemma \ref{lem:w1e}}\label{sec:sub-w1e} 
Notice that 
\begin{align}\label{eq:lem10dec}
\frac{1}{n}\sum_{i=1}^{n}\hat{W}_{i,l}\delta_{i,k} =\underbrace{\frac{1}{n}\sum_{i=1}^{n}W_{i,l}\delta_{i,k}}_{{\textrm L}_{1}(l,k)}+ \underbrace{\frac{1}{n}\sum_{i=1}^{n}(\hat{W}_{i,l}-W_{i,l})\delta_{i,k}}_{{\textrm L}_2(l,k)}\,.
\end{align}
As we will show in Sections \ref{sec:sub-l1} and \ref{sec:sub-l2},  it holds that
\begin{align}\label{eq:l1-tail}
\mathbb{P}\bigg\{\max_{l \in [m],\,k\in[q]}|{\rm L}_{1} (l,k)|>x\bigg\} \le&~ 2mq \exp(-\bar{C}_1nx^2) \notag\\
&+ 2mq\exp\bigg\{-\frac{\bar{C}_2nx}{\log(\tilde{d}n)}\bigg\} + \bar{C}_3(\tilde{d}n)^{-2}
\end{align} 
for any $x>n^{-1}$ with some universal constants $\bar{C}_1, \bar{C}_2, \bar{C}_3>0$, and if $\log \tilde{d} \ll n^{1/10}(\log n)^{-1/2}$, it holds that
\begin{align}\label{eq:l2-con}
&\mathbb{P}\bigg\{\max_{l\in[m],\,k\in[q]}|{\rm L}_{2}(l,k)| > x\bigg\} \notag\\
&~~~~~~~\le mq \bar{C}_5\bigg[\exp\bigg\{-\frac{\bar{C}_4n^{17/10}(\log n)^{1/2} x^2}{\log(\tilde{d}n)}\bigg\} + \exp\bigg\{-\frac{\bar{C}_4n^{7/10}x}{\log^{1/2}(\tilde{d}n)}\bigg\}\\
&~~~~~~~~~~~~~~~~~~~~~ +\exp\bigg\{-\frac{\bar{C_4}n^{4/5}x^{2/3}}{\log^{1/3}(\tilde{d}n)} \bigg\}
+ \exp\bigg\{-\frac{\bar{C_4}n^{17/20}x^{1/2}}{\log^{1/4}(\tilde{d}n)} \bigg\} \bigg] + \bar{C}_5(\tilde{d}n)^{-2} \notag
\end{align}
for any $x> \bar{C}_6 \{n^{-7/10}\log^{3/2}(\tilde{d}n) + n^{-13/20}(\log n)^{-3/4}\log(\tilde{d}n)\}$ with some universal constants $\bar{C}_4, \bar{C}_5, \bar{C}_6>0$. 
Hence, by \eqref{eq:lem10dec}, for any $ A_1\in[\bar{C}, \breve{C}]$  with  some sufficiently large constants $\breve{C}>\bar{C}>1$, it holds that
\begin{align*}
&\mathbb{P}\bigg\{\max_{l \in [m],\,k\in[q]}\bigg|\frac{1}{n}\sum_{i=1}^{n}\hat{W}_{i,l}\delta_{i,k}\bigg|> A_1 n^{-1/2}\log^{1/2} (qm)\bigg\} \\
&~~~~~~~\le  \mathbb{P}\bigg\{\max_{l \in [m],\,k\in[q]}|{\rm L}_{1} (l,k)|>  \frac{A_1  \log^{1/2} (qm)}{2n^{1/2}}\bigg\} +  \mathbb{P}\bigg\{\max_{l \in [m],\,k\in[q]}|{\rm L}_{2} (l,k)|> \frac{A_1  \log^{1/2} (qm)}{2n^{1/2}}\bigg\}\\
&~~~~~~~\le \bar{C}_8 qm \exp\{-\bar{C}_7A_1^2\log (qm)\}  +\bar{C}_8(\tilde{d}n)^{-2}  \\
&~~~~~~~\le   \bar{C}_8  \exp\{-\bar{C_7}(1-\bar{C}_7^{-1}A_1^{-2})A_1^2\log (qm)\} + \bar{C}_8(\tilde{d}n)^{-2}\\
&~~~~~~~\le  \bar{C}_8  \exp\{-\bar{C}_7(1-\bar{C}_7^{-1}\bar{C}^{-2})A_1^2\log (qm)\} + \bar{C}_8(\tilde{d}n)^{-2}
\end{align*}
with some universal constants $\bar{C}_7,\bar{C}_8 >0$, provided that  $\log \tilde{d} \ll n^{1/10}(\log n)^{-1/2}$. Analogously, we   also have the similar result for $\max_{j\in[p],\,l\in[m]}|n^{-1}\sum_{i=1}^{n}\hat{W}_{i,l}\ve_{i,j}|$.  We complete the proof of Lemma \ref{lem:w1e}. $\hfill\Box$

\subsubsection{Proof of \eqref{eq:l1-tail}}\label{sec:sub-l1}
Recall $W_{i,l}, V_{i,k} \sim \mathcal{N}(0,1)$ and $\mathbb{E}(\delta_{i,k}\,|\, W_{i,l})=0$. By \eqref{eq:regressionUV} and Condition \ref{cn:subgaussian}(i), we have
\begin{align}\label{eq:delta-tail}
\mathbb{P}(|\delta_{i,k}|>x)  = &~\mathbb{P}(|V_{i,k} - \bbeta_k^{\T}\bW_i|>x)  
\leq  \mathbb{P}\bigg(|V_{i,k}| >\frac{x}{2}\bigg) + \mathbb{P}\bigg(|\bbeta_k^{\T}\bW_i|>\frac{x}{2}\bigg) \notag\\
\leq&~ 2e^{-x^2/4} +  c_6 e^{-c_7 x^2/4} \le C_1 e^{-\tilde{c}x^2}
\end{align}
for any $x>0$, $i \in [n]$ and $k \in [q]$, where $\tilde{c}=(1 \wedge c_7)/4$ and $C_1=2+c_6$. Then  $\mathbb{E}(\delta_{i,k}^4) \leq C_2$ and $\var(W_{i,l}\delta_{i,k}) \le \{\mathbb{E}(W_{i,l}^4)\mathbb{E}(\delta_{i,k}^4)\} ^{1/2}\leq \sqrt{3C_2}$.
Since $\mathbb{E}(W_{i,l}\delta_{i,k})=\mathbb{E}\{\mathbb{E}(\delta_{i,k}\,|\,W_{i,l})\}=0$, it holds that
\begin{align*}
{\rm L}_1(l,k) =&~\underbrace{\frac{1}{n}\sum_{i=1}^{n}\big[W_{i,l}\delta_{i,k}I(|W_{i,l}|,\,|\delta_{i,k}|\le Q) -\mathbb{E}\{W_{i,l}\delta_{i,k}I(|W_{i,l}|,\,|\delta_{i,k}|\le Q) \}\big]}_{\textrm{L}_{11}(l,k)} \\
&+\underbrace{\frac{1}{n}\sum_{i=1}^{n} W_{i,l}\delta_{i,k}I(|W_{i,l}|\le Q)I(|\delta_{i,k}|> Q)}_{\textrm{L}_{12}(l,k)} + \underbrace{\frac{1}{n}\sum_{i=1}^{n} W_{i,l}\delta_{i,k}I(|W_{i,l}|> Q)}_{\textrm{L}_{13}(l,k)} \\
&-\underbrace{\big[\mathbb{E}(W_{i,l}\delta_{i,k}) - \mathbb{E}\{W_{i,l}\delta_{i,k}I(|W_{i,l}|,\,|\delta_{i,k}|\le Q)\}\big]}_{\textrm{L}_{14}(j,k)} 
\end{align*}
for any $Q>0$. Analogous to the derivation of \eqref{eq:r31'-tail}, for any $x>0$, we have
\begin{align*}
\mathbb{P}\bigg\{\max_{l \in [m],\,k\in[q]}|{\rm L}_{11} (l,k)|>x\bigg\} \le 2qm\exp\bigg(-\frac{nx^2}{C_3+C_4Q^2x} \bigg)\,.
\end{align*}
By \eqref{eq:delta-tail} and the fact that $W_{i,l} \sim \mathcal{N}(0,1)$, it holds that
\begin{align*}
\mathbb{P}\bigg\{\max_{l \in [m],\,k\in[q]}|{\rm L}_{12} (l,k)|>x\bigg\} \le&~  nq \max_{i\in[n],\,k\in[q]}\mathbb{P}(|\delta_{i,k}|> Q) \le C_1nqe^{-\tilde{c}Q^2}\,,\\
\mathbb{P}\bigg\{\max_{l \in [m],\,k\in[q]}|{\rm L}_{13} (l,k)|>x\bigg\} \le&~ nm\max_{i\in[n],\,l\in[m]}\mathbb{P}(|W_{i,l}|> Q) \le 2nme^{-Q^2/2}
\end{align*}
for any $x>0$. Furthermore, we have
\begin{align*}
\max_{l\in [m],\,k \in[q]}|{\rm L}_{14}(l ,k)| \lesssim &~  \max_{i\in[n],\,l\in[m]} \big[\mathbb{E}\{I(|W_{i,l}|> Q)\}\big]^{1/2} + \max_{i\in[n],\,k\in[q]} \big[\mathbb{E}\{I(|\delta_{i,k}|> Q)\}\big]^{1/2}\\  \lesssim &~  Q^{-1/2}e^{-Q^2/4}  +  e^{-\tilde{c}Q^2/2}\,.
\end{align*}
Recall $\tilde{d}=p\vee q\vee m$. With selecting $Q=C\log^{1/2}(\tilde{d}n)$ for some sufficiently large constant $C>\sqrt{3/\tilde{c}}$, for any $x > n^{-1}$, by
\begin{align*}
\mathbb{P}\bigg\{\max_{l \in [m],\,k\in[q]}|{\rm L}_{1} (l,k)|>x\bigg\} \le &~ \mathbb{P}\bigg\{\max_{l \in [m],\,k\in[q]}|{\rm L}_{11} (l,k)|>\frac{x}{4}\bigg\} + \mathbb{P}\bigg\{\max_{l \in [m],\,k\in[q]}|{\rm L}_{12} (l,k)|>\frac{x}{4}\bigg\} \notag\\
& + \mathbb{P}\bigg\{\max_{l \in [m],\,k\in[q]}|{\rm L}_{13} (l,k)|>\frac{x}{4}\bigg\} + \mathbb{P}\bigg\{\max_{l \in [m],\,k\in[q]}|{\rm L}_{14} (l,k)|>\frac{x}{4}\bigg\}\,,
\end{align*} 
we have \eqref{eq:l1-tail} holds.
$\hfill\Box$

\subsubsection{Proof of \eqref{eq:l2-con}}\label{sec:sub-l2}
Recall $\hat{W}_{i,l}=\Phi^{-1} \{n(n+1)^{-1}\hat{F}_{\bZ,l}(Z_{i,l})\}$ and $W_{i,l}=\Phi^{-1}\{F_{\bZ,l}(Z_{i,l})\}$. Given $M_1=\sqrt{9(\log n)/5}$ and $M_2=\sqrt{3(\log n)/5}$, define $W_{i,l}^{*} =  W_{i,l}I(|W_{i,l}|\le M_1) + M_1 \cdot{\rm sign}(W_{i,l})I(|W_{i,l}|>M_1)$ and
\begin{align*}
\tilde{\delta}_{3,k}(W_{s,l})=\mathbb{E} \big[e^{W_{i,l}^2/2}\big\{I(W_{s,l}\le W_{i,l})-\Phi(W_{i,l})\big\}I(|W_{i,l}|\le M_2)\delta_{i,k} I(|\delta_{i,k}|\le Q)\,\big|\,W_{s,l} \big]
\end{align*}
with  $i \ne s$ and some  $Q>M_2$. Then
\begin{align}\label{eq:l2-dec}
&\frac{1}{n}\sum_{i=1}^{n} (\hat{W}_{i,l} - W_{i,l}) \delta_{i,k} \notag\\
&~~~~~~~= 	\frac{1}{n}\sum_{i=1}^{n} (\hat{W}_{i,l} - W_{i,l}^{*}) \delta_{i,k}I(|W_{i,l}|\le M_1)I(|\delta_{i,k}|\le Q) \notag\\
&~~~~~~~~~~+	\frac{1}{n}\sum_{i=1}^{n} (\hat{W}_{i,l} - W_{i,l}^{*}) \delta_{i,k}I(|W_{i,l}|> M_1)I(|\delta_{i,k}|\le Q) \notag\\
&~~~~~~~~~~+ 	\frac{1}{n}\sum_{i=1}^{n} (W_{i,l}^{*}-W_{i,l})  \delta_{i,k}I(|\delta_{i,k}|\le Q) + 	\frac{1}{n}\sum_{i=1}^{n} (\hat{W}_{i,l} - W_{i,l}) \delta_{i,k}I(|\delta_{i,k}|> Q) \notag\\
&~~~~~~~= \frac{1}{n}\sum_{i=1}^{n}\bigg(
\bigg[\Phi^{-1} \bigg\{\frac{n}{n+1}\hat{F}_{\bZ,l}(Z_{i,l})\bigg\} - \Phi^{-1}\{F_{\bZ,l}(Z_{i,l})\} \bigg] \delta_{i,k}I(|W_{i,l}|\le M_2)I(|\delta_{i,k}|\le Q)\notag \\
&~~~~~~~~~~~ \underbrace{~~~~~~~~~~~~-\frac{\sqrt{2\pi}}{n+1}\sum_{s:\,s\ne i}\tilde{\delta}_{3,k}(W_{s,l})\bigg)~~~~~~~~~~~~~~~~~~~~~~~~~~~~~~~~~~~~~~~~~~~~~~~~~~~~~~~~~~~}_{\textrm{L}_{21}(l,k)} \notag\\
&~~~~~~~~~~+\underbrace{\frac{1}{n}\sum_{i=1}^{n}
	\bigg[\Phi^{-1} \bigg\{\frac{n}{n+1}\hat{F}_{\bZ,l}(Z_{i,l})\bigg\} - \Phi^{-1}\{F_{\bZ,l}(Z_{i,l})\} \bigg] \delta_{i,k}I(M_2<|W_{i,l}|\le M_1)I(|\delta_{i,k}|\le Q) }_{\textrm{L}_{22}(l,k)} \notag\\
&~~~~~~~~~~+\underbrace{\frac{1}{n}\sum_{i=1}^{n} (\hat{W}_{i,l} - W_{i,l}^{*}) \delta_{i,k}I(|W_{i,l}|> M_1)I(|\delta_{i,k}|\le Q)}_{\textrm{L}_{23}(l,k)} +\underbrace{\frac{1}{n}\sum_{i=1}^{n} (W_{i,l}^{*}-W_{i,l})  \delta_{i,k}I(|\delta_{i,k}|\le Q)}_{\textrm{L}_{24}(l,k)} \notag\\
&~~~~~~~~~~ + \underbrace{	\frac{1}{n}\sum_{i=1}^{n} (\hat{W}_{i,l} - W_{i,l}) \delta_{i,k}I(|\delta_{i,k}|> Q) }_{\textrm{L}_{25}(l,k)} +  \underbrace{\frac{\sqrt{2\pi}(n-1)}{n(n+1)}\sum_{s=1}^{n}\tilde{\delta}_{3,k}(W_{s,l})}_{\textrm{L}_{26}(l,k)}\,.
\end{align}
Recall $\hat{F}^{(i)}_{\bZ,l}(Z_{i,l})=(n-1)^{-1}\sum_{s:\,s\ne i}I(Z_{s,l}\le Z_{i,l})$.
Then, for any $i\in [n]$ and $l \in [m]$, we have 
\begin{align*}
\frac{n}{n+1} \hat{F}_{\bZ,l}(Z_{i,l}) -  F_{\bZ,l}(Z_{i,l})= \frac{n-1}{n+1} \big\{\hat{F}^{(i)}_{\bZ,l}(Z_{i,l}) -F_{\bZ,l}(Z_{i,l})\big\} - \frac{2}{n+1}F_{\bZ,l}(Z_{i,l}) + \frac{1}{n+1} \,.\notag
\end{align*}
By the Taylor's expression and   \eqref{eq:Derivative12}, it holds that
\begin{align*}
&~{\rm L}_{21}(l,k)\\
=&~ \frac{\sqrt{2\pi}}{n(n+1)}\sum_{1\leq i_1 \ne i_2 \leq n} \big\{e^{W_{i_1,l}^2/2}\big\{I(W_{i_2,l} \le W_{i_1,l}) -\Phi(W_{i_1,l})\big\}  \delta_{i_1,k} \\
&~~\underbrace{~~~~~~~~~~~~~~~~~~~~~~~~~\times I{(|W_{i_1,l}| \le M_2)}I(|\delta_{i_1,k}|\le Q)-\tilde{\delta}_{3,k}(W_{i_2,l})\big\} }_{\textrm{L}_{211}(l,k)} \\
&+ \underbrace{\frac{\sqrt{2\pi}}{n(n+1)}\sum_{i=1}^{n}e^{W_{i,l}^2/2} \big\{1-2\Phi(W_{i,l})\big\} \delta_{i,k}I{(|W_{i,l}| \le M_2)} I(|\delta_{i,k}|\le Q)}_{\textrm{L}_{212}(l,k)} \\
&+ \underbrace{\sum_{l=2}^{\infty}\frac{1}{n\cdot l!}\sum_{i=1}^{n}(\Phi^{-1})^{(l)}\{F_{\bZ,l}(Z_{i,l})\}\bigg\{\frac{n}{n+1}\hat{F}_{\bZ,l}(Z_{i,l})-  F_{\bZ,l}(Z_{i,l}) \bigg\}^{l}\delta_{i,k}I{(|W_{i,l}| \le M_2)}I(|\delta_{i,k}|\le Q) }_{\textrm{L}_{213}(l,k)} \,.
\end{align*} 
Analogous to the derivation of \eqref{eq:k111-tail}, for any $x>0$, we have
\begin{align}\label{eq:l211-tail}
&\mathbb{P}\bigg\{\max_{l\in[m],\,k\in[q]}|{\rm L}_{211}(l,k)| > x\bigg\} \notag\\
&~~~~~~~  \le C_1mq\exp \bigg\{-\frac{1}{C_1}\min\bigg(\frac{n^2M_2x^2}{Q^2e^{M_2^2/2}}, \frac{nx}{ Qe^{M_2^2/2}}, \frac{nx^{2/3}}{Q^{2/3}e^{M_2^2/3}}, \frac{nx^{1/2}}{Q^{1/2}e^{M_2^2/4}} \bigg)\bigg\} \,. 
\end{align}

Let $\mu_{2}(i,l,k) = \mathbb{E}[e^{W_{i,l}^2/2}  \{1-2\Phi(W_{i,l})\} \delta_{i,k}I{(|W_{i,l}| \le M_2)} I(|\delta_{i,k}|\le Q)]$.
Applying the similar arguments for the derivation of \eqref{eq:i112-tail}, it holds that
\begin{align*} 
&\mathbb{P}\bigg(\max_{l\in[m],\, k\in[q]}\bigg|\frac{1}{n}\sum_{i=1}^{n}\big[e^{W_{i,l}^2/2} \big\{1-2\Phi(W_{i,l})\big\} \delta_{i,k}I{(|W_{i,l}| \le M_2)} I(|\delta_{i,k}|\le Q)  -  \mu_2(i,l,k)\big]\bigg| > x\bigg) \\
&~~~~~~~~~~  \le 2mq \exp\bigg(-\frac{nx^2}{C_2Q^{2}M_2^{-1} e^{M_2^2/2} + C_3Qe^{M_2^2/2}x }\bigg) 
\end{align*}
for any $x>0$. Recall $W_{i,l}\sim \mathcal{N}(0,1)$. We then have $\max_{i\in[n],\,l\in[m],\,k\in[q]}|\mu_2(i,l,k)|\lesssim QM_2$. Hence, for any $x>C_{4}n^{-1}QM_2$ with some sufficiently large constant $C_{4}>0$, we have
\begin{align}\label{eq:l212-tail}
\mathbb{P}\bigg\{\max_{l\in[m],\,k\in[q]}|{\rm L}_{212}(l,k)| > x\bigg\} \le  2mq\exp\bigg(-\frac{n^3x^2}{C_{5}Q^{2}M_2^{-1} e^{M_2^2/2} + C_{6}Qe^{M_2^2/2}nx }\bigg)   \,.
\end{align}

Recall $\tilde{d} =p\vee q\vee m$. Define the event
\begin{align*}
\mathcal{H}_{7} = \bigg\{ \max_{i\in[n],\,l\in[m]}|\hat{F}^{(i)}_{\bZ,l}(Z_{i,l})-  F_{\bZ,l}(Z_{i,l})| \le 2K_2^{-1/2} n^{-1/2}\log^{1/2}(\tilde{d}n) \bigg\}\,,
\end{align*}
where $K_2$ is specified in Lemma \ref{lem:ecdf}. Similar to the derivation of \eqref{eq:i12-bound}, restricted on $\mathcal{H}_{7}$, if $\log(\tilde{d}n) \ll ne^{-M_2^2}M_2^{-2}$, it holds that
\begin{align*}
|{\rm L}_{213}(l,k)| \le&~ \frac{C_{7}QM_2\log (\tilde{d}n)}{n} \times \frac{1}{n}\sum_{i=1}^{n}e^{W_{i,l}^2}I(|W_{i,l}| \le M_2) \\
=&~ \underbrace{\frac{C_{7}QM_2\log (\tilde{d}n)}{n} \times \frac{1}{n}\sum_{i=1}^{n}\big[e^{W_{i,l}^2}I(|W_{i,l}| \le M_2)-\mathbb{E}\{e^{W_{i,l}^2}I(|W_{i,l}| \le M_2)\}\big]}_{\textrm{L}_{2131}(l,k)}\\
&+ \underbrace{\frac{C_{7}QM_2\log (\tilde{d}n)}{n}\times \mathbb{E}\{e^{W_{i,l}^2}I(|W_{i,l}| \le M_2)\}}_{\textrm{L}_{2132}(l,k)}\,. 
\end{align*}
Analogous to \eqref{eq:i12-tail},  for any $x>0$, we have
\begin{align*}
&\mathbb{P}\bigg\{\max_{l\in[m],\,k\in[q]}|{\rm L}_{2131}(l,k)| > x\bigg\} \\
&~~~~~~~~~~~~~\le  2m\exp\bigg\{-\frac{n^3x^2}{C_{8} Q^{2}M_2 e^{3M_2^2/2}\log^{2} (\tilde{d}n) + C_{9}e^{M_2^2}QM_2\log (\tilde{d}n)nx }\bigg\} \,,
\end{align*}
and $\max_{l\in[m],\,k\in[q]}|{\rm L}_{2132}(l,k)| \lesssim n^{-1}Qe^{M_2^2/2}\log(\tilde{d}n)$. Identical to \eqref{eq:h1-c}, we  also have $\mathbb{P}(\mathcal{H}_{7}^{\rm c}) \le K_1(\tilde{d}n)^{-3}$. Hence, for any $x>C_{10}n^{-1}Qe^{M_2^2/2}\log(\tilde{d}n)$ with some sufficiently large constant $C_{10}>0$, it holds that
\begin{align}\label{eq:l213-tail}
&\mathbb{P}\bigg\{\max_{l \in [m],\,k\in[q]}|{\rm L}_{213} (l,k)|>x\bigg\}\notag \\
&~~~~~~~~\le \mathbb{P}\bigg\{\max_{l \in [m],\,k\in[q]}|{\rm L}_{213} (l,k)|>x,\, \mathcal{H}_{7}\bigg\} + \mathbb{P}(\mathcal{H}_{7}^{\rm c})\\
&~~~~~~~~\le  2m\exp\bigg\{-\frac{n^3x^2}{C_{11} Q^{2}M_2 e^{3M_2^2/2}\log^{2} (\tilde{d}n) + C_{12}e^{M_2^2}QM_2\log (\tilde{d}n)nx }\bigg\} + K_1(\tilde{d}n)^{-3} \notag
\end{align}
provided that $\log(\tilde{d}n) \ll  ne^{-M_2^2}M_2^{-2} $.  Notice that 
\begin{align*}
\mathbb{P}\bigg\{\max_{l\in[m],\,k\in[q]}|{\rm L}_{21}(l,k)| > x\bigg\} \le &~ \mathbb{P}\bigg\{\max_{l \in [m],\,k\in[q]}|{\rm L}_{211} (l,k)|>\frac{x}{3}\bigg\} + \mathbb{P}\bigg\{\max_{l \in [m],\,k\in[q]}|{\rm L}_{212} (l,k)|>\frac{x}{3}\bigg\}  \notag\\
&+ \mathbb{P}\bigg\{\max_{l \in [m],\,k\in[q]}|{\rm L}_{213} (l,k)|>\frac{x}{3}\bigg\} 
\end{align*}
for any $x>0$. Combining with \eqref{eq:l211-tail}--\eqref{eq:l213-tail}, for any $x>C_{13}n^{-1}Qe^{M_2^2/2}\log(\tilde{d}n)$ with some sufficiently large constant $C_{13}>0$, we have
\begin{align}\label{eq:l21-tail}
&\mathbb{P}\bigg\{\max_{l\in[m],\,k\in[q]}|{\rm L}_{21}(l,k)| > x\bigg\}   \\
&~~~~~~~~~\le C_{14}mq\exp \bigg\{-\frac{1}{C_{14}}\min\bigg(\frac{n^2M_2x^2}{Q^2e^{M_2^2/2}}, \frac{nx}{ Qe^{M_2^2/2}}, \frac{nx^{2/3}}{Q^{2/3}e^{M_2^2/3}}, \frac{nx^{1/2}}{Q^{1/2}e^{M_2^2/4}} \bigg)\bigg\} + K_1(\tilde{d}n)^{-3} \notag
\end{align}
provided that  $\log (\tilde{d}n) \lesssim  n^{1/2}e^{-M_2^2/2}M_2^{-1} $.

Let $K(W_{i,l}, \tilde{d},n) = 4 n^{-1/2}  [\Phi(W_{i,l}) \{1-\Phi(W_{i,l})\}]^{1/2} \log^{1/2}(\tilde{d}n)  + 7n^{-1} \log(\tilde{d}n)$. Define the event
\begin{align*}
\mathcal{H}_{8} =  \bigcap_{i\in[n],\,l\in[m]}\big\{|\hat{F}_{\bZ,l}^{(i)}(Z_{i,l})-  F_{\bZ,l}(Z_{i,l})| \le K(W_{i,l}, \tilde{d},n)\big\} \,.
\end{align*}
Analogous to the derivation of \eqref{eq:i2-dec}, restricted on $\mathcal{H}_{8}$, if $\log(\tilde{d}n) \ll  ne^{-M_1^2/2}M_1^{-1} $, it holds that
\begin{align*}
&|{\rm L}_{22}(l,k)|\le   \frac{C_{15}Q\log^{1/2} (\tilde{d}n)}{n^{1/2}M_2^{1/2}} \times \frac{1}{n}\sum_{i=1}^{n}  e^{W_{i,l}^2/4}I(M_2<|W_{i,l}| \le M_1) \\
&~~~~~=\underbrace{\frac{C_{15}Q\log^{1/2} (\tilde{d}n)}{n^{1/2}M_2^{1/2}} \times \frac{1}{n}\sum_{i=1}^{n} \big[e^{W_{i,l}^2/4} I(M_2<|W_{i,l}| \le M_1) -\mathbb{E}\big\{e^{W_{i,l}^2/4}I(M_2<|W_{i,l}| \le M_1) \big\}\big]}_{\textrm{L}_{221}(l,k)}\\
&~~~~~~~~+\underbrace{\frac{C_{15}Q\log^{1/2} (\tilde{d}n)}{n^{1/2}M_2^{1/2}} \times  \mathbb{E}\big\{e^{W_{i,l}^2/4} I(M_2<|W_{i,l}| \le M_1) \big\}}_{\textrm{L}_{222}(l,k)} \,.
\end{align*}
Recall $W_{i,l}\sim \mathcal{N}(0,1)$. Since $\max_{i\in[n],\,l\in[m]}\mathbb{E}\{e^{W_{i,l}^2/4}I(M_2< |W_{i,l}| \le M_1)\}\lesssim M_2^{-1} e^{- M_2^2/4}$ and $\max_{i\in[n],\,l\in[m]}\var\{e^{W_{i,l}^2/4}I(M_2< |W_{i,l}| \le M_1)\}\lesssim M_1$, by Bernstein inequality,  for any $x>0$, it holds that
\begin{align*}
&\mathbb{P}\bigg( \bigg|\frac{1}{n}\sum_{i=1}^{n}\big[e^{W_{i,l}^2/4}I(M_2<|W_{i,l}| \le M_1)-\mathbb{E}\big\{e^{W_{i,l}^2/4}I(M_2<|W_{i,l}| \le M_1)\big\}\big]\bigg|>x\bigg)\\
&~~~~~~~~~~~ \le 2 \exp\bigg(-\frac{nx^2}{C_{16} M_1  + C_{17}e^{M_1^2/4} x }\bigg) \,,
\end{align*}
which implies
\begin{align*}
&\mathbb{P}\bigg\{\max_{l\in[m],\,k\in[q]}|{\rm L}_{221}(l,k)| > x\bigg\} \\
&~~~~~~~\le  2mq\exp\bigg\{-\frac{n^2x^2}{C_{18}M_1M_2^{-1}Q^{2}\log(\tilde{d}n)  + C_{19}e^{M_1^2/4} n^{1/2}QM_2^{-1/2}\log^{1/2}(\tilde{d}n)x }\bigg\} 
\end{align*}
for any $x>0$. Notice that $\max_{l\in[m],\,k\in[q]}|{\rm L}_{222}(l,k)| \lesssim n^{-1/2}QM_{2}^{-3/2}e^{-M_2^2/4}\log^{1/2}(\tilde{d}n)$.  Similar to \eqref{eq:h2-c}, we  also have $\mathbb{P}(\mathcal{H}_{8}^{\rm c})\le 4(\tilde{d}n)^{-2}$.  Using the same arguments for the derivation of \eqref{eq:l213-tail}, for any $x> C_{20} n^{-1/2}QM_{2}^{-3/2}e^{-M_2^2/4}\log^{1/2}(\tilde{d}n)$ with some sufficiently large constant $C_{20}>0$,  it holds that
\begin{align}\label{eq:l22-tail}
&\mathbb{P}\bigg\{\max_{l\in[m],\,k\in[q]}|{\rm L}_{22}(l,k)| > x\bigg\} \\
&~~~~~~\le  2mq\exp\bigg\{-\frac{n^2x^2}{C_{21}M_1M_2^{-1}Q^{2}\log(\tilde{d}n)  + C_{22}e^{M_1^2/4} n^{1/2}QM_2^{-1/2}\log^{1/2}(\tilde{d}n)x }\bigg\} + 4(\tilde{d}n)^{-2}\notag
\end{align}
provided that $\log(\tilde{d}n) \ll  ne^{-M_1^2/2}M_1^{-1} $.

Parallel to \eqref{eq:Uhatstar}, we can show $\max_{i\in[n],\,l\in[m]}|\hat{W}_{i,l}- W_{i,l}^{*}| \le 2\sqrt{2\log(n+1)}$. Then
\begin{align*} 
\max_{l\in[m],\, k\in[q]}|{\rm L}_{23}(l,k)| \lesssim&~ Q\sqrt{\log n} \times \max_{  l\in[m]}\bigg|\frac{1}{n}\sum_{i=1}^{n} \big[I(|W_{i,l}|> M_1) - \mathbb{E}\{I(|W_{i,l}|> M_1)\}\big]\bigg|  \\
&+Q\sqrt{\log n} \times \max_{  l\in[m]}|\mathbb{E}\{I(|W_{i,l}|> M_1)\}|\,.
\end{align*}
Due to $W_{i,l}\sim \mathcal{N}(0,1)$, we have  $\mathbb{E}\{I(|W_{i,l}|> M_1)\} \lesssim M_1^{-1}e^{-M_1^2/2}$ and ${\rm Var}\{I(|W_{i,l}|> M_1)\} \lesssim M_1^{-1}e^{-M_1^2/2} $. By  Bonferroni inequality and Bernstein inequality,  for any $x > C_{23}M_1^{-1}e^{-M_1^2/2}Q(\log n)^{1/2}$ with some sufficiently large constant $C_{23}>0$, it holds that
\begin{align}\label{eq:l23-tail}
&\mathbb{P}\bigg\{\max_{l\in[m],\,k\in[q]}|{\rm L}_{23}(l,k)| > x\bigg\} \notag\\
&~~~~~~~\le  2mq\exp\bigg\{-\frac{nx^2}{C_{24} Q^{2}M_1^{-1}e^{-M_1^2/2}\log n  + C_{25} Q(\log n)^{1/2}x}\bigg\}\,. 
\end{align}

Recall  $W_{i,l}^{*} =  W_{i,l}I(|W_{i,l}|\le M_1) + M_1 \cdot{\rm sign}(W_{i,l})I(|W_{i,l}|>M_1)$. Define $\acute{W}_{i,l} =   W_{i,l} -M_1\cdot {\rm sign}(W_{i,l})$. We have $  W_{i,l} -W_{i,l}^{*} = \acute{W}_{i,l}I(|W_{i,l}|>M_1)$ and 
\begin{align*}
{\rm L}_{24}(l,k) =&~-\frac{1}{n}\sum_{i=1}^{n}\bigg[\acute{W}_{i,l}\delta_{i,k}I(M_1 <|W_{i,l}|\le Q)I(|\delta_{i,k}| \le Q)\\ &~~~~~\underbrace{~~~~~~~~~-\mathbb{E}\{\acute{W}_{i,l}\delta_{i,k}I(M_1 <|W_{i,l}|\le Q)I(|\delta_{i,k}| \le Q)\} \bigg]~}_{\textrm{L}_{241}(l,k)}\\
&~-\underbrace{\frac{1}{n}\sum_{i=1}^{n}\acute{W}_{i,l}\delta_{i,k}I( |W_{i,l}|> Q)I(|\delta_{i,k}| \le Q)}_{\textrm{L}_{242}(l,k)} - \underbrace{\mathbb{E}\{\acute{W}_{i,l}\delta_{i,k}I(M_1 <|W_{i,l}|\le Q)I(|\delta_{i,k}| \le Q)\} }_{\textrm{L}_{243}(l,k)} \,.
\end{align*}
Recall $W_{i,l} \sim \mathcal{N}(0,1)$. Since
\begin{align*}
&\max_{i\in[n],\,l\in[m],\, k\in[q]} \var\{\acute{W}_{i,l}\delta_{i,k}I(M_1 <|W_{i,l}|\le Q)I(|\delta_{i,k}| \le Q)\}  \\
&~~~~~~~~~\le    Q^2 \max_{i\in[n],\,l\in[m] }\mathbb{E}\{\acute{W}_{i,l}^2 I(M_1 <|W_{i,l}|\le Q)\}    
\lesssim Q^2 M_1e^{-M_1^{2}/2}\,,
\end{align*}
by  Bonferroni inequality and Bernstein inequality, it holds that
\begin{align*}
\mathbb{P}\bigg\{\max_{l\in[m],\,k\in[q]}|{\rm L}_{241}(l,k)| > x\bigg\} \le&~  2mq\exp\bigg(-\frac{nx^2}{C_{26}Q^{2}M_1e^{-M_1^2/2}   + C_{27} Q^2x }\bigg) 
\end{align*}
for any $x>0$. Due to $W_{i,l} \sim \mathcal{N}(0,1)$, we have
\begin{align*}
\mathbb{P}\bigg\{\max_{l\in[m],\,k\in[q]}|{\rm L}_{242}(l,k)| > x\bigg\} \le&~  nm \max_{i\in[n],\,l\in[m]}\mathbb{P}(|W_{i,l}|>Q)\le  2nmQ^{-1}e^{-Q^2/2}
\end{align*}
for any $x>0$. Since $\mathbb{E}\{\acute{W}_{i,l}\delta_{i,k}I(M_1 <|W_{i,l}|\le Q)\} = \mathbb{E}\{\acute{W}_{i,l}I(M_1 <|W_{i,l}|\le Q) \mathbb{E}(\delta_{i,k}\,|\,W_{i,l})\}=0$, by \eqref{eq:delta-tail}, we have
\begin{align*} 
\max_{l\in[m],\, k\in[q]}|{\rm L}_{243}(l,k)| =&~  \max_{l\in[m],\, k\in[q]}\big|\mathbb{E}\{\acute{W}_{i,l}\delta_{i,k}I(M_1 <|W_{i,l}|\le Q)I(|\delta_{i,k}|> Q)\} \big|\notag\\
\le&~ Q \max_{  k\in[q]}|\mathbb{E}\{ \delta_{i,k} I(|\delta_{i,k}|> Q)\} |\\
\le&~ Q  \max_{  k\in[q]}\bigg\{Q\mathbb{P}(|\delta_{i,k}|>Q) +   \int_{Q}^{\infty}   \mathbb{P}(|\delta_{i,k}|>x) \,{\rm d}x \bigg\} \lesssim   Q^2 e^{-\tilde{c}Q^2}\,.\notag
\end{align*}
Notice that
\begin{align*}
\mathbb{P}\bigg\{\max_{l\in[m],\,k\in[q]}|{\rm L}_{24}(l,k)| > x\bigg\}  
\le&~\mathbb{P}\bigg\{\max_{l \in [m],\,k\in[q]}|{\rm L}_{241} (l,k)|>\frac{x}{3}\bigg\} + \mathbb{P}\bigg\{\max_{l \in [m],\,k\in[q]}|{\rm L}_{242} (l,k)|>\frac{x}{3}\bigg\} \\
&+ \mathbb{P}\bigg\{\max_{l \in [m],\,k\in[q]}|{\rm L}_{243} (l,k)|>\frac{x}{3}\bigg\}
\end{align*}
for any $x>0$. It holds that 
\begin{align}\label{eq:l24-tail}
\mathbb{P}\bigg\{\max_{l\in[m],\,k\in[q]}|{\rm L}_{24}(l,k)| > x\bigg\} \le &~ 2mq\exp\bigg(-\frac{nx^2}{C_{28}Q^{2}M_1e^{-M_1^2/2}   + C_{29} Q^2x }\bigg) \notag\\
&+ 2nmQ^{-1}e^{-Q^2/2} 
\end{align}
for any $x> C_{30}  Q^2 e^{-\tilde{c}Q^2}$ with some sufficiently large constant $C_{30}>0$. By \eqref{eq:delta-tail} again, it holds that
\begin{align}\label{eq:l25-tail}
\mathbb{P}\bigg\{\max_{l\in[m],\, k\in[q]}|{\rm L}_{25}(l,k)|>x\bigg\} \le \max_{i\in[n],\,k\in[q]} nq \mathbb{P}(|\delta_{i,k}|> Q) \lesssim  nqe^{-\tilde{c}Q^2} 
\end{align}
for any $x>0$. 

Since
\begin{align*}
&\mathbb{E} \big[e^{W_{i,l}^2/2} \big\{I(W_{s,l}\le W_{i,l})-\Phi(W_{i,l})\big\}I(|W_{i,l}|\le M_2)\delta_{i,k} \,\big|\,W_{s,l}=a \big]\\
&~~~~~~~~~~~=\mathbb{E} \big[e^{W_{i,l}^2/2}\big\{I(a\le W_{i,l})-\Phi(W_{i,l})\big\}I(|W_{i,l}|\le M_2) \delta_{i,k} \big]\\
&~~~~~~~~~~~=\mathbb{E} \big[e^{W_{i,l}^2/2}\big\{I(a\le W_{i,l})-\Phi(W_{i,l})\big\}I(|W_{i,l}|\le M_2) \mathbb{E}(\delta_{i,k} \,|\, W_{i,l}) \big]=0
\end{align*}
for any $s\in[n]$, $s\ne i $ and $a\in \mathbb{R}$, we have
\begin{align*}
\tilde{\delta}_{3,k}(W_{s,l}) =- \mathbb{E} \big[e^{W_{i,l}^2/2} \big\{I(W_{s,l}\le W_{i,l})-\Phi(W_{i,l})\big\}I(|W_{i,l}|\le M_2)\delta_{i,k}I(|\delta_{i,k}|>Q) \,\big|\, W_{s,l}\big]\,.
\end{align*}
By \eqref{eq:delta-tail}, it holds that
\begin{align*} 
\mathbb{E}\{\delta_{i,k}^2I(|\delta_{i,k}|>Q)\} = Q^2 \mathbb{P}(|\delta_{i,k}|>Q) + 2 \int_{Q}^{\infty} x \mathbb{P}(|\delta_{i,k}|>x) \,{\rm d}x \lesssim    Q^2 e^{-\tilde{c}Q^2}\,.
\end{align*}
Due to $W_{i,l} \sim \mathcal{N}(0,1)$, then
\begin{align}\label{eq:delta-Q}
&\Big|\mathbb{E} \big[e^{W_{i,l}^2/2} \big\{I(W_{s,l}\le W_{i,l})-\Phi(W_{i,l})\big\}I(|W_{i,l}|\le M_2)\delta_{i,k}I(|\delta_{i,k}|>Q) \,\big|\, W_{s,l}\big]\Big|\notag\\
&~~~~~~\le    \mathbb{E} \big\{e^{W_{i,l}^2/2} I(|W_{i,l}|\le M_2)|\delta_{i,k}|I(|\delta_{i,k}|>Q)\big\} \notag\\
&~~~~~~\le   \big[\mathbb{E} \{e^{W_{i,l}^2} I(|W_{i,l}|\le M_2)\}\big]^{1/2} \big[\mathbb{E}\{\delta_{i,k}^2I(|\delta_{i,k}|>Q)\}\big]^{1/2} \notag\\
&~~~~~~\lesssim  M_2^{-1/2} Q e^{M_2^2/4}  e^{-\tilde{c}Q^2/2}\,,
\end{align}
which implies 
\begin{align}\label{eq:l26-con}
\max_{l\in[m],\,k\in[q]}|{\rm L}_{26}(l,k)| =O(  M_2^{-1/2} Q e^{M_2^2/4}  e^{-\tilde{c}Q^2/2} )\,.
\end{align}

Recall $\tilde{d}=p\vee q\vee m$, $M_1=\sqrt{9(\log n)/5}$ and $M_2=\sqrt{3(\log n)/5}$. Combining  with \eqref{eq:l21-tail}--\eqref{eq:l25-tail} and \eqref{eq:l26-con}, with selecting $Q=C\log^{1/2}(\tilde{d}n)$ for some sufficiently large constant $C>\sqrt{3/\tilde{c}}$, by \eqref{eq:l2-dec}, for any $x > C_{31}\{n^{-7/10}\log^{3/2}(\tilde{d}n) + n^{-13/20}(\log n)^{-3/4}\log(\tilde{d}n)\}$ with some sufficiently large constant $C_{31}>0$, we have
\begin{align*}
&\mathbb{P}\bigg\{\max_{l\in[m],\,k\in[q]}|{\rm L}_{2}(l,k)| > x\bigg\} \notag\\
&~~~~~~~~\le mq C_{33}\bigg[\exp\bigg\{- \frac{C_{32}n^{17/10}(\log n)^{1/2}x^2}{\log(\tilde{d}n)}\bigg\} + \exp\bigg\{-\frac{C_{32}n^{7/10}x}{\log^{1/2}(\tilde{d}n)}\bigg\}\notag\\
&~~~~~~~~~~~~~~~~~~~~~~ +\exp\bigg\{-\frac{C_{32}n^{4/5}x^{2/3}}{\log^{1/3}(\tilde{d}n)}\bigg\}
+ \exp\bigg\{-\frac{C_{32}n^{17/20}x^{1/2}}{\log^{1/4}(\tilde{d}n)} \bigg\} \bigg] + C_{33}(\tilde{d}n)^{-2}
\end{align*}
provided that  $\log \tilde{d}\ll n^{1/10}(\log n)^{-1/2}$.  Then \eqref{eq:l2-con} holds. 
$\hfill\Box$

\subsection{Proof of Lemma \ref{lem:wh}}\label{sec:sub-wh} 
Notice that 
\begin{align*}
&\frac{1}{n}\sum_{i=1}^{n}\hat{W}_{i,j}\hat{W}_{i,k}-\mathbb{E}(W_{i,j}W_{i,k})= 	\underbrace{\frac{1}{n}\sum_{i=1}^{n}(\hat{W}_{i,j}\hat{W}_{i,k}-W_{i,j}W_{i,k})}_{{\textrm S}_{1}'(j,k)}+ \underbrace{\frac{1}{n}\sum_{i=1}^{n}\{W_{i,j}W_{i,k}-\mathbb{E}(W_{i,j}W_{i,k})\} }_{{\textrm S}_{2}'(j,k)}\,.
\end{align*}
Recall $\tilde{d}=p\vee q\vee m$,  $\hat{W}_{i,j}=\Phi^{-1} \{n(n+1)^{-1}\hat{F}_{\bZ,j}(Z_{i,j})\}$  and  $W_{i,j}=\Phi^{-1}\{F_{\bZ,j}(Z_{i,j})\}$ for $j\in[m]$. Applying the similar arguments for deriving the convergence rate of ${\rm R}_{4}'$ in Section \ref{sub:sec-h0-R4} for ${\rm R}_{4}'$ defined in \eqref{eq:r4'-dep}, we have
\begin{align}\label{eq:s'1}
\max_{j,k\in[m]}|{\rm S}'_1(j,k)| =    O_{\rm p}\{n^{-1/2}(\log n)(\log \tilde{d})^{1/2}\log^{1/2} (\tilde{d}n)\}
\end{align}
provided that $\log \tilde{d} \lesssim n^{5/12}(\log n)^{-1/2}$. Analogous to the derivation of the convergence rate of $\max_{j\in[p],\, k\in[q]}|{\rm R}'_{3}(j,k)|$ in Section \ref{sub:sec-h0-R3}, 
it holds that
\begin{align}\label{eq:s'2}
\max_{j,k \in[m]} |{\rm  S}_{2}'(j,k)| = O_{\rm p}  \{n^{-1/2}(\log \tilde{d})^{1/2} \}
\end{align}
provided that $\log \tilde{d} \lesssim n^{1/3}$.
Combining \eqref{eq:s'1} and \eqref{eq:s'2}, it holds that 
\begin{align*}
|\hat{\bSigma}_{W}-\bSigma_{W} |_{\infty}   = O_{\rm p}\{n^{-1/2}(\log n)\log (\tilde{d}n)\}
\end{align*}
provided that  $\log \tilde{d} \lesssim n^{1/3}$.  We complete the proof of Lemma \ref{lem:wh}.
$\hfill\Box$

\subsection{Proof of Lemma \ref{lem:coeff}}\label{sec:sub-coeff} 
For each $j \in[p]$, define 
\begin{align*}
\mathcal{F}_j=\bigg\{\max_{l \in [m]}\bigg|\frac{1}{n}\sum_{i=1}^{n}\hat{W}_{i,l}\ve_{i,j}\bigg| \le A_1n^{-1/2}\log^{1/2} (pm)\bigg\}
\end{align*}
for some constant $A_1 \in[\bar{C}, \breve{C}]$, where the constants $\bar{C}$ and $\breve{C}$ are specified in Lemma \ref{lem:w1e}. Write $\balpha_j=(\alpha_{j,1},\ldots,\alpha_{j,m})^\T$ and $S_j=\{l\in[m]:\alpha_{j,l}\neq 0\}$. Then $s_j:=|S_j|\leq s$. Since $s \ll  n^{1/2}(\log n)^{-1}\{\log (\tilde{d}n)\}^{-1}$, there exists $\kappa_n=o(1)$ such that $ n^{-1/2}(\log n)\log (\tilde{d}n) \ll \kappa_n\ll s^{-1}$.  Define 
\begin{align*}
\mathcal{G}=\big\{\big|\hat{\bSigma}_W-\bSigma_W\big|_\infty\leq \kappa_n\big\}.
\end{align*}
It follows from Lemma \ref{lem:wh} that $\mathbb{P}(\mathcal{G})\rightarrow1$ as $n\rightarrow\infty$ provided that  $\log \tilde{d} \lesssim n^{1/3}$. Restricted on $\mathcal{G}$, by Lemma 6.17 of \cite{Buhlmann2011} and Condition \ref{cn:subgaussian}(ii), when $n$ is sufficiently large, we have
\begin{align*}
\balpha^{\T}\hat{\bSigma}_W\balpha \geq&~ \balpha^{\T}\bSigma_W\balpha \cdot \{1-O(s\kappa_n)\} \\
\geq&~\frac{\balpha^{\T}\bSigma_W\balpha}{2}\geq \frac{|\balpha|_1^2}{s_j}\cdot\frac{\lambda_{\min}(\bSigma_W)}{2}
\end{align*}
for any $\balpha$ satisfying 
$
|\balpha_{S_j^{\rm c}}|_1\leq 3|\balpha_{S_j}|_1$. 
Recall $C_1n^{-1/2}\log^{1/2} (pm)\le  \lambda_{\balpha,j}   \le C_2n^{-1/2}\log^{1/2} (pm)$ for any $j\in[p]$ with some sufficiently large constants $C_1>0$ and $C_2>0$. When $ \lambda_{\balpha,j}   \ge 4A_1n^{-1/2}\log^{1/2} (pm)$ for any $j \in [p]$,   Theorem 6.1 of \cite{Buhlmann2011} implies that $ |\hat{\balpha}_j-\balpha_j |_{1} \le C_3s_jn^{-1/2}\log^{1/2} (pm)$ restricted on $\mathcal{F}_j\cap \mathcal{G}$.
We then have 
\begin{align*}
\max_{j \in [p]}|\hat{\balpha}_j-\balpha_j |_{1} \le C_3s n^{-1/2}\log^{1/2}(pm)
\end{align*}
restricted on $\mathcal{G}\cap\mathcal{F}$ with $\mathcal{F}:=\bigcap_{j=1}^{p}\mathcal{F}_j$. Recall $\tilde{d}=p\vee q\vee m$. By Bonferroni inequality and Lemma \ref{lem:w1e}, for some sufficiently large $n$ , it holds that
\begin{align*}
\mathbb{P}(\mathcal{F}^{\rm c}) \le &~\sum_{j=1}^{p}\mathbb{P}\bigg\{\max_{l \in [m]}\bigg|\frac{1}{n}\sum_{i=1}^{n}\hat{W}_{i,l}\ve_{i,j}\bigg| >A_1n^{-1/2}\log^{1/2} (pm)\bigg\}  \\
\le &~  K_3 p \exp\{-K_4 A_1^2\log (pm )\} +o(1)
\end{align*}
provided that $ \log \tilde{d} \ll n^{1/10}(\log n)^{-1/2}$. With selecting a large enough $A_1$, we have $\mathbb{P}(\mathcal{F}^{\rm c}) \to 0$ as $n\to \infty$. 
Thus, 
\begin{align*}
\max_{j \in [p]}|\hat{\balpha}_j-\balpha_j |_{1}=O_{{\rm p}} \{s n^{-1/2}(\log \tilde{d})^{1/2} \}
\end{align*}
provided that $ \log \tilde{d} \ll n^{1/10}(\log n)^{-1/2}$. Analogously, we can also show 
\begin{align*}
\max_{k \in [q]}|\hat{\bbeta}_k-\bbeta_k |_{1} = O_{{\rm p}} \{s n^{-1/2}(\log \tilde{d})^{1/2} \}\,.
\end{align*}

Recall $U_{i,j} \sim \mathcal{N}(0,1)$, $U_{i,j}= \balpha_{j}^{\T}\bW_i +\ve_{i,j}$ and $\bSigma_{W}=\cov(\bW)$. Notice that $\mathbb{E}(\ve_{i,j}^2) \le C_4$ by \eqref{eq:ve_tail} under Condition \ref{cn:subgaussian}(i). Due to $\var(\balpha_{j}^{\T}\bW_i) = \balpha_{j}^{\T} \bSigma_{W} \balpha_{j}$, we have
\begin{align*}
\lambda_{\min}(\bSigma_{W}) |\balpha_{j}|_2^2 \le  \var(\balpha_{j}^{\T}\bW_i)  =\var(U_{i,j}- \ve_{i,j})  \le 2\mathbb{E}(U_{i,j}^2) + 2\mathbb{E}(\ve_{i,j}^2) \le C_5\,,
\end{align*}
where $\lambda_{\min}(\bSigma_{W})$ is the smallest eigenvalues of $\bSigma_{W}$.
By Condition \ref{cn:subgaussian}(ii) and $|\balpha_{j}|_1 \le \sqrt{s}|\balpha_{j}|_2$, we have $\max_{j \in [p]}| \balpha_j |_{1}  \lesssim \sqrt{s}$. Analogously, we also have $ \max_{k \in [q]}| \bbeta_k |_{1} \lesssim \sqrt{s}$. Then
\begin{align*}
\max_{j\in[p]}|\hat{\balpha}_j|_1 \le&~ \max_{j \in [p]}|\hat{\balpha}_j-\balpha_j |_{1} + \max_{j \in [p]}| \balpha_j |_{1} = O_{\rm p}(\sqrt{s})\,,\\
\max_{k\in[q]}|\hat{\bbeta}_k|_1 \le&~ \max_{k \in [q]}|\hat{\bbeta}_k-\bbeta_k |_{1} + \max_{k \in [q]}| \bbeta_k |_{1} = O_{\rm p}(\sqrt{s})
\end{align*}
provided that  $s \ll   n^{1/2}(\log n)^{-1}\{\log (\tilde{d}n)\}^{-1}$ and  $ \log \tilde{d} \ll n^{1/10}(\log n)^{-1/2}$. We complete the proof of Lemma \ref{lem:coeff}.
$\hfill\Box$

\subsection{Proofs of \eqref{eq:t1-2con} and \eqref{eq:t2-2con}}\label{sec:sub-t12}
Recall $\ve_{i,j}= U_{i,j}- \balpha_{j}^{\T}\bW_{i}$ and $\hat{\ve}_{i,j}= \hat{U}_{i,j}- \hat{\balpha}_{j}^{\T}\hat{\bW}_{i}$. Then
\begin{align*}
{\rm T}_{1}(j,k) =&~ \frac{1}{n} \sum_{i=1}^{n}\big(\hat{U}_{i,j} - \hat{\balpha}_{j}^{\T}\hat{\bW}_{i}  - U_{i,j} +  \balpha_{j}^{\T}\bW_{i} \big)\delta_{i,k}\\
=  &~\underbrace{(\balpha_j - \hat{\balpha}_{j})^{\T} \bigg(\frac{1}{n}\sum_{i=1}^{n}\bW_{i}\delta_{i,k}\bigg)}_{{\textrm T}_{11}(j,k)} - \underbrace{\hat{\balpha}_{j}^{\T}\bigg\{\frac{1}{n}\sum_{i=1}^{n}(\hat{\bW}_{i} - \bW_{i}) \delta_{i,k} \bigg\} }_{{\textrm T}_{12}(j,k)}+ \underbrace{ \frac{1}{n}\sum_{i=1}^{n}(\hat{U}_{i,j} - U_{i,j}) \delta_{i,k}}_{{\textrm T}_{13}(j,k)}\,.
\end{align*}  
Recall  $\tilde{d}=p\vee q\vee m$.  By \eqref{eq:lem10dec}--\eqref{eq:l2-con} in Section \ref{sec:sub-w1e} for the proof of Lemma \ref{lem:w1e}, we have 
\begin{align*}
\max_{l \in [m],\,k\in[q]}\bigg|\frac{1}{n}\sum_{i=1}^{n}W_{i,l}\delta_{i,k}\bigg| =O_{\rm p}\{n^{-1/2} (\log \tilde{d})^{1/2}\}
\end{align*}
provided that $\log \tilde{d} \lesssim n^{1/3}$, and 
\begin{align*}
\max_{l \in [m],\,k\in[q]}\bigg|\frac{1}{n}\sum_{i=1}^{n}(\hat{W}_{i,l}-W_{i,l})\delta_{i,k}\bigg| =O_{\rm p}\{n^{-7/10}\log^{3/2}(\tilde{d}n)\} + O_{\rm p}\{n^{-13/20}(\log n)^{-3/4}\log(\tilde{d}n)\} 
\end{align*}
provided that  $\log \tilde{d} \ll n^{1/10}(\log n)^{-1/2}$. 
By Lemma \ref{lem:coeff}, it holds that
\begin{align}
\max_{j\in[p],\, k\in[q]}|{\rm T}_{11}(j,k)|  \le&~ \max_{j\in[p]}|\hat{\balpha}_{j} - \balpha_j|_{1} \max_{l\in[m],\, k\in[q]} \bigg|\frac{1}{n}\sum_{i=1}^{n}W_{i,l}\delta_{i,k}\bigg| = O_{\rm p}\big(sn^{-1}\log \tilde{d}\big) \,,\label{eq:t11-con}\\
\max_{j\in[p],\, k\in[q]}|{\rm T}_{12}(j,k)|  \le &~ \max_{j\in[p]}|\hat{\balpha}_{j}|_1\max_{l\in[m],\,k\in[q]} \bigg|\frac{1}{n}\sum_{i=1}^{n}(\hat{W}_{i,l} - W_{i,l}) \delta_{i,k} \bigg|\notag\\
=&~ O_{\rm p}\{s^{1/2}n^{-7/10}\log^{3/2}(\tilde{d}n)\} + O_{\rm p}\{s^{1/2}n^{-13/20}(\log n)^{-3/4}\log(\tilde{d}n)\}  \label{eq:t12-con}
\end{align}
provided that $s \ll   n^{1/2}(\log n)^{-1}\{\log (\tilde{d}n)\}^{-1}$ and $ \log \tilde{d} \ll n^{1/10} (\log n)^{-1/2} $.
As we will show in Section \ref{sec:sub-t14}, 
\begin{align}\label{eq:t13-con}
{\rm T}_{13}(j,k)  =\frac{\sqrt{2\pi}(n-1)}{n(n+1)} \sum_{s=1}^{n}\tilde{\delta}_{4,k}(U_{s,j}) + {\rm Rem}_{13}(j,k)
\end{align}   
with 
\begin{align*}
\max_{j\in[p],\, k\in[q]}|{\rm Rem}_{13}(j,k)| =    O_{\rm p} \{n^{-7/10} \log^{3/2}(\tilde{d}n)\} +O_{\rm p} \{n^{-13/20}(\log n)^{-3/4} \log(\tilde{d}n)\} 
\end{align*}
provided that $\log \tilde{d} \ll n^{1/10}(\log n)^{-1/2}$. Together with \eqref{eq:t11-con}--\eqref{eq:t13-con}, we have 
\begin{align*}
{\rm T}_{1}(j,k) = \frac{\sqrt{2\pi}(n-1)}{n(n+1)} \sum_{s=1}^{n}\tilde{\delta}_{4,k}(U_{s,j}) + {\rm Rem}_{11}(j,k)
\end{align*}   
with  
\begin{align*}
\max_{j\in[p],\, k\in[q]}|{\rm Rem}_{11}(j,k)| =    O_{\rm p} \{s^{1/2}n^{-7/10} \log^{3/2}(\tilde{d}n)\}   +O_{\rm p} \{s^{1/2}n^{-13/20}(\log n)^{-3/4}\log(\tilde{d}n)\}
\end{align*}
provided that $s \ll   n^{1/2}(\log n)^{-1}\{\log (\tilde{d}n)\}^{-1}$ and $ \log \tilde{d} \ll n^{1/10} (\log n)^{-1/2}$. Then, \eqref{eq:t1-2con} holds.
Analogously, we  also  have
\begin{align*}
{\rm T}_{2}(j,k) =  \frac{\sqrt{2\pi}(n-1)}{n(n+1)}\sum_{s=1}^{n} \tilde{\delta}_{5,j}(V_{s,k}) + {\rm Rem}_{12}(j,k)
\end{align*}
with 
\begin{align*}
\max_{j\in[p],\, k\in[q]}|{\rm Rem}_{12}(j,k)| = O_{\rm p} \{s^{1/2}n^{-7/10} \log^{3/2}(\tilde{d}n)\}  +O_{\rm p} \{s^{1/2}n^{-13/20}(\log n)^{-3/4}\log(\tilde{d}n)\}
\end{align*}
provided that $s \ll   n^{1/2}(\log n)^{-1}\{\log (\tilde{d}n)\}^{-1}$ and $ \log \tilde{d} \ll n^{1/10}(\log n)^{-1/2}$.  Then \eqref{eq:t2-2con} holds.
$\hfill\Box$

\subsubsection{Proof of \eqref{eq:t13-con} }\label{sec:sub-t14}
Recall
\begin{align*}
\tilde{\delta}_{4,k}(U_{s,j})=\mathbb{E} \big[e^{U_{i,j}^2/2}  \big\{I(U_{s,j}\le U_{i,j})-\Phi(U_{i,j})\big\}\delta_{i,k} I\{|U_{i,j}|\le \sqrt{3(\log n)/5}\}\,\big|\,U_{s,j} \big] 
\end{align*}
with $i\ne s$. Given $Q> \sqrt{3(\log n)/5}$, define
\begin{align*}
&\tilde{\delta}_{41,k}(U_{s,j})=\mathbb{E} \big[e^{U_{i,j}^2/2} \big\{I(U_{s,j}\le U_{i,j})-\Phi(U_{i,j})\big\}\delta_{i,k} I\{|U_{i,j}|\le \sqrt{3(\log n)/5}\}I(|\delta_{i,k}|\le Q)\,\big|\,U_{s,j} \big]\,,\\
&\tilde{\delta}_{42,k}(U_{s,j})=\mathbb{E} \big[e^{U_{i,j}^2/2} \big\{I(U_{s,j}\le U_{i,j})-\Phi(U_{i,j})\big\}\delta_{i,k} I\{|U_{i,j}|\le \sqrt{3(\log n)/5}\}I(|\delta_{i,k}|> Q)\,\big|\,U_{s,j} \big]
\end{align*}
with $i \ne s$. Recall $\hat{U}_{i,j}=\Phi^{-1} \{n(n+1)^{-1}\hat{F}_{\bX,j}(X_{i,j})\}$ and $U_{i,j}=\Phi^{-1}\{F_{\bX,j}(X_{i,j})\}$.  Define $U_{i,j}^{*} =  U_{i,j}I(|U_{i,j}|\le M_1) + M_1 \cdot{\rm sign}(U_{i,j})I(|U_{i,j}|>M_1)$ with $M_1=\sqrt{9(\log n)/5}$. Let $M_2=\sqrt{3(\log n)/5}$. Then
\begin{align*}
&~\frac{1}{n}\sum_{i=1}^{n} (\hat{U}_{i,j} - U_{i,j}) \delta_{i,k} \\
= &~	\frac{1}{n}\sum_{i=1}^{n} (\hat{U}_{i,j} - U_{i,j}^{*}) \delta_{i,k}I(|U_{i,j}|\le M_1)I(|\delta_{i,k}|\le Q) \\
&+	\frac{1}{n}\sum_{i=1}^{n} (\hat{U}_{i,j} - U_{i,j}^{*}) \delta_{i,k}I(|U_{i,j}|> M_1)I(|\delta_{i,k}|\le Q) \\
&+ 	\frac{1}{n}\sum_{i=1}^{n} (U_{i,j}^{*}-U_{i,j})  \delta_{i,k}I(|\delta_{i,k}|\le Q) + 	\frac{1}{n}\sum_{i=1}^{n} (\hat{U}_{i,j} - U_{i,j}) \delta_{i,k}I(|\delta_{i,k}|> Q) \\
=&~ \frac{1}{n}\sum_{i=1}^{n}\bigg(
\bigg[\Phi^{-1} \bigg\{\frac{n}{n+1}\hat{F}_{\bX,j}(X_{i,j})\bigg\} - \Phi^{-1}\{F_{\bX,j}(X_{i,j})\} \bigg] \delta_{i,k}I(|U_{i,j}|\le M_2)I(|\delta_{i,k}|\le Q) \\
&  \underbrace{~~~~~~~~~~~~-\frac{\sqrt{2\pi}}{n+1}\sum_{s:\,s\ne i}\tilde{\delta}_{41,k}(U_{s,j})\bigg)~~~~~~~~~~~~~~~~~~~~~~~~~~~~~~~~~~~~~~~~~~~~~~~~~~~~~~~~~~~~~~}_{\textrm{T}_{131}(j,k)} \\
&+\underbrace{\frac{1}{n}\sum_{i=1}^{n}
	\bigg[\Phi^{-1} \bigg\{\frac{n}{n+1}\hat{F}_{\bX,j}(X_{i,j})\bigg\} - \Phi^{-1}\{F_{\bX,j}(X_{i,j})\} \bigg] \delta_{i,k}I( M_2<|U_{i,j}|\le M_1)I(|\delta_{i,k}|\le Q) }_{\textrm{T}_{132}(j,k)} \\
&+\underbrace{\frac{1}{n}\sum_{i=1}^{n} (\hat{U}_{i,j} - U_{i,j}^{*}) \delta_{i,k}I(|U_{i,j}|> M_1)I(|\delta_{i,k}|\le Q)}_{\textrm{T}_{133}(j,k)} +\underbrace{\frac{1}{n}\sum_{i=1}^{n} (U_{i,j}^{*}-U_{i,j})  \delta_{i,k}I(|\delta_{i,k}|\le Q)}_{\textrm{T}_{134}(j,k)}\\
& + \underbrace{	\frac{1}{n}\sum_{i=1}^{n} (\hat{U}_{i,j} - U_{i,j}) \delta_{i,k}I(|\delta_{i,k}|> Q) }_{\textrm{T}_{135}(j,k)} -  \underbrace{\frac{\sqrt{2\pi}(n-1)}{n(n+1)}\sum_{s=1}^{n}\tilde{\delta}_{42,k}(U_{s,j})}_{\textrm{T}_{136}(j,k)}  + \frac{\sqrt{2\pi}(n-1)}{n(n+1)}\sum_{s=1}^{n}\tilde{\delta}_{4,k}(U_{s,j})\,.
\end{align*}
Recall $U_{i,j} \sim \mathcal{N}(0,1)$ and $\tilde{d}=p\vee q\vee m$. Using the similar arguments for the derivations of \eqref{eq:l21-tail}--\eqref{eq:l23-tail} and \eqref{eq:l25-tail} in Section \ref{sec:sub-l2} for the proof of Lemma \ref{lem:w1e}, respectively, it holds that
\begin{align*} 
\max_{j\in[p],\, k\in[q]}|{\rm T}_{131}(j,k)| = O_{\rm p}\big\{n^{-1}Qe^{M_2^2/2}\log (\tilde{d}n)\big\}
\end{align*}
provided that $\log (\tilde{d}n) \lesssim  n^{1/2}e^{-M_2^2/2}M_2^{-1} $,
\begin{align*} 
\max_{j\in[p],\, k\in[q]}|{\rm T}_{132}(j,k)| = O_{\rm p} \{n^{-1/2}QM_2^{-3/2} e^{-M_2^2/4}\log^{1/2} (\tilde{d}n)\} 
\end{align*}
provided that $ \log (\tilde{d}n) \ll n e^{-M_1^2/2} M_1^{-1}$,
\begin{align*}
\max_{j\in[p],\, k\in[q]}|{\rm T}_{133}(j,k)| =  O_{\rm p}\{M_1^{-1} Qe^{-M_1^2/2} (\log n)^{1/2}\} 
\end{align*}
provided that $ \log  \tilde{d} \lesssim n e^{-M_1^2/2} M_1^{-1}$, and 
\begin{align*}
\max_{j\in[p],\, k\in[q]}|{\rm T}_{135}(j,k)|  = o_{\rm p}(n^{-1})
\end{align*}
provided that $\log(\tilde{d}n) \lesssim Q^2$.
Analogous to the derivation of \eqref{eq:k1},  we have
\begin{align*}
&\max_{j\in[p], \,k\in[q]}|{\rm T}_{134}(j,k)| =    O_{\rm p}(n^{-1}Q^2 \log \tilde{d}) + O_{\rm p}(Qe^{-M_1^2/2}) 
\end{align*}
provided that $\log \tilde{d} \lesssim ne^{-M_1^2/2}M_1^{-1}$.
Due to $U_{i,j} \sim \mathcal{N}(0,1)$, using the similar arguments for the derivation of \eqref{eq:delta-Q} in Section \ref{sec:sub-w1e} for the proof of Lemma \ref{lem:w1e}, it holds that
\begin{align}\label{eq:delta_42}
\max_{s\in[n],\,j\in[p],\,k\in[q]}|\tilde{\delta}_{42,k}(U_{s,j})|  \lesssim n^{3/20} (\log n)^{-1/4} Q  e^{-\tilde{c}Q^2/2}\,,
\end{align}
where  $\tilde{c}=(1\wedge c_7)/4$. Hence, we have 
\begin{align*} 
\max_{j\in[p],\,k\in[q]}|{\rm T}_{136}(j,k)| =O\{n^{3/20} (\log n)^{-1/4} Q e^{-\tilde{c}Q^2/2} \} \,. 
\end{align*} 
With selecting $Q = C \log^{1/2}(\tilde{d}n)$ for some sufficiently large constant $C > \sqrt{5/(2\tilde{c})}$, it holds that
\begin{align*}
{\rm T}_{13}(j,k)  =\frac{\sqrt{2\pi}(n-1)}{n(n+1)} \sum_{s=1}^{n}\tilde{\delta}_{4,k}(U_{s,j}) + {\rm Rem}_{13}(j,k)
\end{align*}   
with 
\begin{align*}
\max_{j\in[p],\, k\in[q]}|{\rm Rem}_{13}(j,k)| \le  &\max_{j\in[p],\,k\in[q]}|{\rm T}_{131}(j,k)| + \max_{j\in[p],\,k\in[q]}|{\rm T}_{132}(j,k)|+ \max_{j\in[p],\,k\in[q]}|{\rm T}_{133}(j,k)| \\
&~ +\max_{j\in[p],\,k\in[q]}|{\rm T}_{134}(j,k)| + \max_{j\in[p],\,k\in[q]}|{\rm T}_{135}(j,k)| + \max_{j\in[p],\,k\in[q]}|{\rm T}_{136}(j,k)| \\
= &~ O_{\rm p}\big\{n^{-1}e^{M_2^2/2}\log^{3/2} (\tilde{d}n)\big\} + O_{\rm p} \{n^{-1/2}M_2^{-3/2} e^{-M_2^2/4}\log(\tilde{d}n)\}\\
&~ + O_{\rm p}\{M_1^{-1}  e^{-M_1^2/2} (\log n)^{1/2} \log^{1/2}(\tilde{d}n)\} + O_{\rm p}\{e^{-M_1^2/2}\log^{1/2}(\tilde{d}n)\} \\
&~ + O_{\rm p}\{n^{-1}(\log \tilde{d})\log(\tilde{d} n)\}
\end{align*}
provided that $\log (\tilde{d}n) \ll \max\{ne^{-M_1^2/2}M_1^{-1} , n^{1/2}e^{-M_2^2/2}M_2^{-1} \}$. Recall $M_{1}=\sqrt{9(\log n)/5}$ and $M_2=\sqrt{3(\log n)/5}$. Then \eqref{eq:t13-con} holds. 
$\hfill\Box$

\subsection{Proof of \eqref{eq:t3}}\label{sec:sub-t3}
Recall $\ve_{i,j}= U_{i,j}- \balpha_{j}^{\T}\bW_{i}$, $\delta_{i,k}= V_{i,k}- \bbeta_{k}^{\T}\bW_{i}$,  $\hat{\ve}_{i,j}= \hat{U}_{i,j}- \hat{\balpha}_{j}^{\T}\hat{\bW}_{i}$ and $\hat{\delta}_{i,k}= \hat{V}_{i,k}- \hat{\bbeta}_{k}^{\T}\hat{\bW}_{i}$. Then
\begin{align*}
{\rm T}_3(j,k) 
=&~\underbrace{\frac{1}{n}\sum_{i=1}^{n}(\hat{U}_{i,j} -U_{i,j})(\hat{V}_{i,k} - V_{i,k}) }_{{\textrm T_{31}}(j,k)}  - \underbrace{\frac{1}{n}\sum_{i=1}^{n}   (\hat{U}_{i,j} -U_{i,j})(\hat{\bbeta}_{k}^{\T}\hat{\bW}_{i}  - \bbeta_{k}^{\T}\bW_{i})  }_{{\textrm T_{32}}(j,k)} \\
&-\underbrace{\frac{1}{n}\sum_{i=1}^{n}   (\hat{V}_{i,k} - V_{i,k}) (\hat{\balpha}_{j}^{\T}\hat{\bW}_{i}  - \balpha_{j}^{\T}\bW_{i} ) }_{{\textrm T_{33}}(j,k)} + \underbrace{\frac{1}{n}\sum_{i=1}^{n}  (\hat{\balpha}_{j}^{\T}\hat{\bW}_{i}  - \balpha_{j}^{\T}\bW_{i} )(\hat{\bbeta}_{k}^{\T}\hat{\bW}_{i} -\bbeta_{k}^{\T}\bW_{i})   }_{{\textrm T_{34}}(j,k)}\,.
\end{align*}
As we will show in Sections \ref{sec:sub-t31}--\ref{sec:sub-t34}, 
\begin{align}\label{eq:t31}
\max_{j\in[p],\, k\in[q]} |{\rm T}_{31}(j,k)| =&~    O_{\rm p} \{n^{-7/10} (\log n)^{1/2}\}  + O_{\rm p}\{n^{-1}(\log \tilde{d}) \log (\tilde{d}n)\}  
\end{align}
provided that  $ \log \tilde{d} \lesssim n^{1/8}\log n$,  and 
\begin{align}
\max_{j\in[p],\, k\in[q]}|{\rm T}_{32}(j,k)| =&~    O_{\rm p} \{s^{1/2}n^{-7/10} (\log n)^{1/2}\}  + O_{\rm p} \{sn^{-1}(\log n) (\log \tilde{d})\log^{1/2}(\tilde{d}n)\}   \notag\\
=&~\max_{j\in[p],\, k\in[q]}|{\rm T}_{33}(j,k)| \,,\label{eq:t32} \\
\max_{j\in[p],\, k\in[q]}|{\rm T}_{34}(j,k)|   =&~    O_{\rm p} \{sn^{-7/10} (\log n)^{1/2} \}   + O_{\rm p} \{s^{3/2}n^{-1}(\log n) (\log \tilde{d})  \log^{1/2}(\tilde{d}n)\}  \notag\\
&+ O_{\rm p}(s^2n^{-1} \log \tilde{d}) \label{eq:t34}  
\end{align}   
provided that $s \ll   n^{1/2}(\log n)^{-1}\{\log (\tilde{d}n)\}^{-1}$ and $ \log \tilde{d} \ll n^{1/10}(\log n)^{-1/2}$. Combining \eqref{eq:t31}--\eqref{eq:t34}, we have \eqref{eq:t3} holds.
$\hfill\Box$

\subsubsection{Proof of \eqref{eq:t31}}\label{sec:sub-t31}
Define $U_{i,j}^{*} =  U_{i,j}I(|U_{i,j}|\le M_1) + M_1 \cdot{\rm sign}(U_{i,j})I(|U_{i,j}|>M_1)$ with $M_1=\sqrt{7(\log n)/5}$. Then we have 
\begin{align}\label{eq:t31dec}
{\rm T}_{31}(j,k)
=&~
\underbrace{\frac{1}{n}\sum_{i=1}^{n} (\hat{U}_{i,j} -U_{i,j}^{*})(\hat{V}_{i,k}-V_{i,k}^{*})}_{\textrm{T}_{311}(j,k)}  + \underbrace{\frac{1}{n}\sum_{i=1}^{n}\hat{V}_{i,k}(U_{i,j}^{*} - U_{i,j})}_{\textrm{T}_{312}(j,k)} \notag \\
&+\underbrace{\frac{1}{n}\sum_{i=1}^{n}  \hat{U}_{i,j}(V_{i,k}^{*} - V_{i,k})}_{\textrm{T}_{313}(j,k)} - \underbrace{\frac{1}{n}\sum_{i=1}^{n} (U_{i,j}^{*}V_{i,k}^{*} -U_{i,j}V_{i,k})}_{\textrm{T}_{314}(j,k)}\,.
\end{align}
Recall $\tilde{d}=p\vee q \vee m$. By Lemma \ref{lem:huhv}, we have
\begin{align*}
\max_{j\in[p], \,k\in[q]}|{\rm T}_{311}(j,k)| = O_{\rm p}\{n^{-7/10} (\log n)^{1/2}\} 
\end{align*}
provided that $ \log \tilde{d} \lesssim n^{1/8} \log n$.
By Lemma \ref{lem:usvs-uv},  it holds that
\begin{align*}
\max_{j\in[p], \,k\in[q]}|{\rm T}_{314}(j,k)|=  O_{\rm p} \{n^{-7/10} (\log n)^{1/2}\} + O_{\rm p} \{n^{-1} (\log \tilde{d}) \log (\tilde{d}n) \}  
\end{align*}
provided that $\log \tilde{d} \lesssim n^{3/10}(\log n)^{-1/2}$. 

Note that $U_{i,j}^{*} - U_{i,j} = \{M_1 \cdot{\rm sign}(U_{i,j}) -U_{i,j}\}I(|U_{i,j}|>M_1)$. Given $Q> M_1$,
\begin{align*}
\max_{j\in[p],\, k\in[q]}|{\rm T}_{312}(j,k)| = &~ \max_{j\in[p],\, k\in[q]}\bigg|\frac{1}{n}\sum_{i=1}^{n} \hat{V}_{i,k}\{M_1 \cdot{\rm sign}(U_{i,j}) -U_{i,j}\} I(|U_{i,j}|> M_1)\bigg| \\
\le &~ \underbrace{\max_{i\in[n],\,k\in[q]}|\hat{V}_{i,k}|\cdot\max_{j\in[p] } \frac{1}{n}\sum_{i=1}^{n}|M_1 \cdot{\rm sign}(U_{i,j}) -U_{i,j}| I(M_1< |U_{i,j}|\le Q)  }_{\textrm{T}_{3121} }\\
&+ \underbrace{\max_{j\in[p],\, k\in[q]}\bigg|\frac{1}{n}\sum_{i=1}^{n} \hat{V}_{i,k}\{M_1 \cdot{\rm sign}(U_{i,j}) -U_{i,j}\} I(|U_{i,j}|> Q)\bigg| }_{\textrm{T}_{3122} }\,.
\end{align*}
Due to $U_{i,j} \sim \mathcal{N}(0,1)$, it holds that
\begin{align*}
\max_{i\in[n],\,j\in[p] }\mathbb{E}\big\{|M_1 \cdot{\rm sign}(U_{i,j}) -U_{i,j}| I(M_1< |U_{i,j}|\le Q) \big\} \lesssim &~  e^{-M_1^2/2}\,,\\
\max_{i\in[n],\,j\in[p] }\var\big\{|M_1 \cdot{\rm sign}(U_{i,j}) -U_{i,j}| I(M_1< |U_{i,j}|\le Q) \big\} \lesssim &~  M_1e^{-M_1^2/2}\,.
\end{align*}
Recall  $\tilde{d}=p\vee q\vee m$. By Bernstein inequality, we have
\begin{align*}
\max_{j\in[p] } \frac{1}{n}\sum_{i=1}^{n}|M_1 \cdot{\rm sign}(U_{i,j}) -U_{i,j}| I(M_1< |U_{i,j}|\le Q)  = O_{\rm p}  (e^{-M_1^2/2}) + O_{\rm p}(n^{-1}Q \log \tilde{d})
\end{align*}
provided that $\log \tilde{d} \lesssim ne^{-M_1^2/2}M_{1}^{-1}$. Using the similar arguments for the derivation of \eqref{eq:uh-bound}, it holds that $\max_{i\in[n],\,k\in[q]}|\hat{V}_{i,k}| \le \sqrt{2\log(n+1)}$. Then 
\begin{align*}
{\rm T}_{3121}  = O_{\rm p}  \{e^{-M_1^2/2} (\log n)^{1/2}\} + O_{\rm p}\{n^{-1}Q (\log n)^{1/2}\log \tilde{d}\}
\end{align*}
provided that $\log \tilde{d} \lesssim ne^{-M_1^2/2}M_{1}^{-1}$.
Applying the similar arguments for deriving 
\eqref{eq:k12},  we also have $ {\rm T}_{3122} =o_{\rm p}(n^{-1})$ provided that $\log (\tilde{d}n) \lesssim Q^2$. Recall $M_1=\sqrt{7(\log n)/5}$. With selecting $Q= C\log^{1/2}(\tilde{d}n)$ for some  sufficiently large constant $C> 2$, it holds that
\begin{align*}
\max_{j\in[p],\, k\in[q]}|{\rm T}_{312}(j,k)| =&~O_{\rm p} \big\{n^{-7/10}  (\log n)^{1/2}\big\} + O_{\rm p}\{n^{-1} (\log n)^{1/2}(\log \tilde{d})\log^{1/2} (\tilde{d}n)\}
\end{align*}
provided that $\log \tilde{d} \lesssim n^{3/10}(\log n)^{-1/2}$.  Analogously, we can also show such convergence rate also holds for $\max_{j\in[p],\, k\in[q]}|{\rm T}_{313}(j,k)|$. 
By \eqref{eq:t31dec}, we have \eqref{eq:t31} holds.
$\hfill\Box$

\subsubsection{Proof of \eqref{eq:t32}}\label{sec:sub-t323}
Notice that
\begin{align}\label{eq:t32-dec}
\max_{j\in[p],\, k\in[q]}| {\rm T}_{32}(j,k) | \le &~ \underbrace{\max_{j\in[p],\, k\in[q]}\bigg|\frac{1}{n}\sum_{i=1}^{n} (\hat{U}_{i,j} -U_{i,j})\hat{\bbeta}_{k}^{\T}(\hat{\bW}_{i} - \bW_{i})\bigg|}_{\textrm{T}_{321}}  \notag\\
&~ + \underbrace{\max_{j\in[p],\, k\in[q]}\bigg|\frac{1}{n}\sum_{i=1}^{n} (\hat{U}_{i,j} -U_{i,j})(\hat{\bbeta}_{k}  - \bbeta_{k})^{\T} \bW_{i}\bigg| }_{\textrm{T}_{322}} \,.
\end{align} 
Recall $U_{i,j},W_{i,l}\sim \mathcal{N}(0,1)$. Analogous to the derivation of \eqref{eq:t31}, we can show
\begin{align*}
&\max_{j\in[p],\, l\in[m]}\bigg|\frac{1}{n}\sum_{i=1}^{n}  (\hat{U}_{i,j} -U_{i,j})(\hat{W}_{i,l} - W_{i,l})    \bigg| =    O_{\rm p} \{n^{-7/10}(\log n)^{1/2} \}  + O_{\rm p}\{n^{-1}(\log \tilde{d}) \log (\tilde{d}n)\}  
\end{align*}
provided that  $ \log \tilde{d} \lesssim n^{1/8}(\log n) $.
By Lemma \ref{lem:coeff}, it holds that
\begin{align}\label{eq:t321}
{\rm T}_{321} \le&~ \max_{k\in[q]}|\hat{\bbeta}_{k}|_1 \max_{j\in[p],\, l\in[m]}\bigg|\frac{1}{n}\sum_{i=1}^{n} (\hat{U}_{i,j} -U_{i,j})(\hat{W}_{i,l} - W_{i,l})\bigg| = O_{\rm p} \{s^{1/2}n^{-7/10} (\log n)^{1/2} \}   
\end{align}
provided that $s \ll   n^{1/2}(\log n)^{-1}\{\log (\tilde{d}n)\}^{-1}$ and $ \log \tilde{d} \ll n^{1/10}(\log n)^{-1/2}$.

Define $\hat{U}_{i,j}^{*}  = \hat{U}_{i,j} -U_{i,j}^{*}$ and $\tilde{U}_{i,j} = U_{i,j} - U_{i,j}^{*}$ with $U_{i,j}^{*} =  U_{i,j}I(|U_{i,j}|\le M_1) + M_1 \cdot{\rm sign}(U_{i,j})I(|U_{i,j}|>M_1)$, where $M_1 =\sqrt{7(\log n)/5}$. Then $\hat{U}_{i,j}-U_{i,j} = \hat{U}_{i,j}^{*}-\tilde{U}_{i,j}$ and 
\begin{align*}
\frac{1}{n}\sum_{i=1}^{n} (\hat{U}_{i,j} -U_{i,j})W_{i,l}=    \underbrace{\frac{1}{n}\sum_{i=1}^{n} \hat{U}^{*}_{i,j}W_{i,l}}_{{\textrm T}_{3221}(j,l)} - \underbrace{\frac{1}{n}\sum_{i=1}^{n} \tilde{U}_{i,j}W_{i,l}}_{{\textrm T}_{3222}(j,l)}\,.
\end{align*}
Recall $\tilde{d}=p\vee q\vee m$. Analogous to the derivation of the convergence rates of ${\rm R}_{43}'$ and ${\rm R}_{41}'$ in Section \ref{sub:sec-h0-R4}, we can show 
\begin{align*}
&~~~~~~~\max_{j\in[p],\, l\in[m]}|{\rm T}_{3221}(j,l)| = O_{\rm p} \{n^{-1/2}(\log n) (\log \tilde{d})^{1/2} \log^{1/2}(\tilde{d}n)\}\,,\\
&\max_{j\in[p],\, l\in[m]}|{\rm T}_{3222}(j,l)| =  O_{\rm p}\{ n^{-7/10}\log^{1/2}(\tilde{d}n)\} + O_{\rm p} \{n^{-1} (\log \tilde{d}) \log(\tilde{d}n)\}
\end{align*}
provided that $\log \tilde{d} \lesssim n^{3/10}(\log n)^{-1/2}$. It holds that
\begin{align}\label{eq:uh-uw}
\max_{j\in[p],\, l\in[m]} \bigg|\frac{1}{n}\sum_{i=1}^{n} (\hat{U}_{i,j} -U_{i,j})W_{i,l}\bigg| 
& \leq  \max_{j\in[p],\, l\in[m]}|{\rm T}_{3221}(j,l)|+\max_{j\in[p],\, l\in[m]}|{\rm T}_{3222}(j,l)| \notag\\
&= O_{\rm p} \{n^{-1/2}(\log n) (\log \tilde{d})^{1/2} \log^{1/2}(\tilde{d}n)\}  
\end{align}
provided that  $\log \tilde{d} \lesssim n^{3/10}(\log n)^{-1/2}$. By Lemma \ref{lem:coeff}, we have 
\begin{align*}
{\rm T}_{322} \le  \max_{k\in[q]}|\hat{\bbeta}_{k}  - \bbeta_{k}|_{1} \max_{j\in[p],\, l\in[m]} \bigg|\frac{1}{n}\sum_{i=1}^{n} (\hat{U}_{i,j} -U_{i,j})W_{i,l}\bigg|=  O_{\rm p} \{sn^{-1}(\log n) (\log \tilde{d})\log^{1/2}(\tilde{d}n)\}  
\end{align*}
provided that $s \ll   n^{1/2}(\log n)^{-1}\{\log (\tilde{d}n)\}^{-1}$ and $ \log \tilde{d} \ll n^{1/10}(\log n)^{-1/2}$.
Together with \eqref{eq:t321}, by \eqref{eq:t32-dec}, it holds that
\begin{align*}
\max_{j\in[p],\, k\in[q]}|{\rm T}_{32}(j,k)|  =&~    O_{\rm p} \{s^{1/2}n^{-7/10} (\log n)^{1/2}\}  + O_{\rm p} \{sn^{-1}(\log n) (\log \tilde{d})\log^{1/2}(\tilde{d}n)\}
\end{align*}
provided that $s \ll   n^{1/2}(\log n)^{-1}\{\log (\tilde{d}n)\}^{-1}$ and $ \log \tilde{d} \ll n^{1/10}(\log n)^{1/2}$. Analogously, we can also show such convergence rate  holds for $\max_{j\in[p],\, k\in[q]}|{\rm T}_{33}(j,k)|$. Hence, we have \eqref{eq:t32} holds.
$\hfill\Box$

\subsubsection{Proof of \eqref{eq:t34}}\label{sec:sub-t34}
Notice that
\begin{align}\label{eq:t34-dec}
\max_{j\in[p],\, k\in[q]}|{\rm T}_{34}(j,k)| \le &~ \underbrace{\max_{j\in[p],\, k\in[q]}\bigg|\frac{1}{n}\sum_{i=1}^{n} \hat{\balpha}_{j}^{\T}(\hat{\bW}_{i} - \bW_{i})(\hat{\bW}_{i} - \bW_{i})^{\T}\hat{\bbeta}_{k}\bigg|}_{\textrm{T}_{341}} \notag\\
&~ + \underbrace{\max_{j\in[p],\, k\in[q]}\bigg|\frac{1}{n}\sum_{i=1}^{n} \hat{\balpha}_{j}^{\T}(\hat{\bW}_{i} - \bW_{i}) \bW_{i}^{\T}(\hat{\bbeta}_{k}  - \bbeta_{k})\bigg| }_{\textrm{T}_{342}}\notag\\
&~ + \underbrace{\max_{j\in[p],\, k\in[q]}\bigg|\frac{1}{n}\sum_{i=1}^{n} (\hat{\balpha}_{j}-\balpha_j)^{\T} \bW_{i}(\hat{\bW}_{i} - \bW_{i})^{\T}\hat{\bbeta}_{k}\bigg|}_{\textrm{T}_{343}} \notag\\
&~ + \underbrace{\max_{j\in[p],\, k\in[q]}\bigg|\frac{1}{n}\sum_{i=1}^{n} (\hat{\balpha}_{j}-\balpha_j)^{\T} \bW_{i}\bW_{i}^{\T}(\hat{\bbeta}_{k}  - \bbeta_{k})\bigg| }_{\textrm{T}_{344}} \,.
\end{align}  
Recall $ W_{i,l},W_{i,t}\sim \mathcal{N}(0,1)$. Applying the similar arguments for deriving  \eqref{eq:t31} and \eqref{eq:uh-uw}, respectively, we have
\begin{align*}
&\max_{l,t\in[m]} \bigg|\frac{1}{n}\sum_{i=1}^{n}  (\hat{W}_{i,l} - W_{i,l}) (\hat{W}_{i,t} - W_{i,t})  \bigg| = O_{\rm p} \{n^{-7/10} (\log n)^{1/2}\}  + O_{\rm p}\{n^{-1}(\log \tilde{d}) \log (\tilde{d}n)\} \notag
\end{align*}
provided that  $ \log \tilde{d} \lesssim n^{1/8}\log n$, and 
\begin{align*}
\max_{l,t\in[m]} \bigg|\frac{1}{n}\sum_{i=1}^{n} (\hat{W}_{i,l} -W_{i,l})W_{i,t}\bigg| =  O_{\rm p} \{n^{-1/2}(\log n) (\log \tilde{d})^{1/2} \log^{1/2}(\tilde{d}n)\} 
\end{align*}
provided that  $\log \tilde{d} \lesssim n^{3/10}(\log n)^{-1/2}$.  By Lemma \ref{lem:coeff}, it holds that
\begin{align*}
{\rm T}_{341} \le&~  \max_{j\in[p],\, k\in[q]}|\hat{\balpha}_j|_1 |\hat{\bbeta}_{k}|_1  \cdot \max_{l,t\in[m]} \bigg|\frac{1}{n}\sum_{i=1}^{n} (\hat{W}_{i,l} - W_{i,l}) (\hat{W}_{i,t} - W_{i,t})  \bigg| \\
=&~ O_{\rm p} \{sn^{-7/10} (\log n)^{1/2}\}\,,\\
{\rm T}_{342}  \le&~ \max_{j\in[p],\, k\in[q]}|\hat{\balpha}_j|_1 |\hat{\bbeta}_k -\bbeta_{k}|_1 \cdot \max_{l,t\in[m]} \bigg|\frac{1}{n}\sum_{i=1}^{n}(\hat{W}_{i,l} - W_{i,l})	W_{i,t}\bigg|\\
=&~ O_{\rm p} \{s^{3/2}n^{-1}(\log n) (\log \tilde{d})  \log^{1/2}(\tilde{d}n)\}   
\end{align*}
provided that $s \ll   n^{1/2}(\log n)^{-1}\{\log (\tilde{d}n)\}^{-1}$ and $ \log \tilde{d} \ll n^{1/10}(\log n)^{-1/2}$.  Analogously, we can also show the convergence rate of ${\rm T}_{343}$ is identical to ${\rm T}_{342}$. 

By \eqref{eq:s'2} in Section \ref{sec:sub-wh} for the proof of Lemma \ref{lem:wh}, due to $W_{i,l}, W_{i,t} \sim \mathcal{N}(0,1)$, if $\log \tilde{d} \lesssim n^{1/3}$, we have 
\begin{align*}
\max_{l,t\in[m]}\bigg|\frac{1}{n}\sum_{i=1}^{n} W_{i,l} W_{i,t}\bigg| & \le \max_{l,t\in[m]}\bigg|\frac{1}{n}\sum_{i=1}^{n} \{W_{i,l} W_{i,t} - \mathbb{E}(W_{i,l} W_{i,t})\}\bigg| + \max_{i\in[n],\,l,t\in[m]}|\mathbb{E}(W_{i,l} W_{i,t})|\\  
&= O_{\rm p} \big\{n^{-1/2} (\log \tilde{d})^{1/2}\big\} + O(1) = O_{\rm p} (1)\,.
\end{align*}
By Lemma \ref{lem:coeff} again, it holds that
\begin{align*}
{\rm T}_{344} \le \max_{j\in[p],\, k\in[q]}|\hat{\balpha}_j-\balpha_j|_1 |\hat{\bbeta}_k -\bbeta_{k}|_1 \cdot \max_{l,t\in[m]} \bigg|\frac{1}{n}\sum_{i=1}^{n}  W_{i,l} W_{i,t}\bigg| = O_{\rm p}(s^2n^{-1} \log \tilde{d}) 
\end{align*}
provided that $s \ll   n^{1/2}(\log n)^{-1}\{\log (\tilde{d}n)\}^{-1}$ and $ \log \tilde{d} \ll n^{1/10}(\log n)^{-1/2}$. By \eqref{eq:t34-dec}, we have \eqref{eq:t34} holds. 
$\hfill\Box$

\section{Proof of Lemma \ref{lem:delta-t-4}}\label{sec:sub-delta-t-4} 
Recall
\begin{align*}
&~~~~~~~\tilde{\delta}_{4,k}(U_{s,j})=  \mathbb{E} \big[e^{U_{i,j}^2/2}  \big\{I(U_{s,j}\le U_{i,j})-\Phi(U_{i,j})\big\}\delta_{i,k} I\{|U_{i,j}|\le \sqrt{3(\log n)/5}\}\,\big|\,U_{s,j} \big] \,,\\
&\tilde{\delta}_{42,k}(U_{s,j}) =  \mathbb{E} \big[e^{U_{i,j}^2/2} \big\{I(U_{s,j}\le U_{i,j})-\Phi(U_{i,j})\big\} \delta_{i,k}  I\{|U_{i,j}|\le \sqrt{3(\log n)/5}\}I(|\delta_{i,k}|> Q) \,\big|\,U_{s,j} \big]
\end{align*}
with $i\ne s$. Given $Q> \tilde{M}$ with $\tilde{M}= \sqrt{9(\log n)/(10\tilde{c})}$ for $\tilde{c} =(1\wedge c_7)/4$, let 
\begin{align*}
&~~~~~~~~~~~~~~~~~\tilde{\delta}_{43,k}(U_{s,j}) =  \mathbb{E} \big[e^{U_{i,j}^2/2}  \big\{I(U_{s,j}\le U_{i,j})-\Phi(U_{i,j})\big\}\delta_{i,k} \\
&~~~~~~~~~~~~~~~~~~~~~~~~~~~~~~~~~~~~~\times I\{|U_{i,j}|\le \sqrt{3(\log n)/5}\} I(\tilde{M} <|\delta_{i,k}|\le Q) \,\big|\,U_{s,j} \big] \,,\\
&\tilde{\delta}_{44,k}(U_{s,j}) =  \mathbb{E} \big[e^{U_{i,j}^2/2}  \big\{I(U_{s,j}\le U_{i,j})-\Phi(U_{i,j})\big\}\delta_{i,k}  I\{|U_{i,j}|\le \sqrt{3(\log n)/5}\}I(|\delta_{i,k}|\le \tilde{M}) \,\big|\,U_{s,j} \big] 
\end{align*}
with  $i\ne s$.
Notice that
\begin{align*}
\frac{1}{n}\sum_{s=1}^{n} \tilde{\delta}_{4,k}(U_{s,j})  =  \underbrace{\frac{1}{n}\sum_{s=1}^{n} \tilde{\delta}_{42,k}(U_{s,j})}_{{\textrm I}'_{1}(j,k)} + \underbrace{\frac{1}{n}\sum_{s=1}^{n} \tilde{\delta}_{43,k}(U_{s,j}) }_{{\textrm I}'_{2}(j,k)} + \frac{1}{n}\sum_{s=1}^{n} \tilde{\delta}_{44,k}(U_{s,j})\,.
\end{align*}
By \eqref{eq:delta_42}, we have
\begin{align}\label{eq:i1'}
\max_{j\in[p],\, k\in[q]}|{\rm I}'_{1}(j,k)| = O\{n^{3/20}(\log n)^{-1/4}Qe^{-\tilde{c}Q^2/2}\}\,.
\end{align}
Due to $(U_{i,j}, \delta_{i,k})$ and  $(U_{s,j}, \delta_{s,k})$ are independent for any $s\ne i$, and $U_{s,j} \sim \mathcal{N}(0,1)$, then
\begin{align*}
\mathbb{E}\{\tilde{\delta}_{43,k}(U_{s,j})\} =&~ \mathbb{E} \big[e^{U_{i,j}^2/2}I\{|U_{i,j}|\le \sqrt{3(\log n)/5}\} \delta_{i,k}I(\tilde{M}<|\delta_{i,k}|\le Q)\\
&~~~~\times \mathbb{E}\big\{I(U_{s,j}\le U_{i,j})-\Phi(U_{i,j})\,\big|\, U_{i,j},\delta_{i,k}\big\}\big] =0\,.
\end{align*}
Analogous to the derivation of \eqref{eq:delta-Q} in Section \ref{sec:sub-w1e} for the proof of Lemma \ref{lem:w1e}, we have 
$$\max_{s\in[n],\,j\in[p],\,k\in[q]}|\tilde{\delta}_{43,k}(U_{s,j})| 
\lesssim   n^{3/20} (\log n)^{-1/4} \tilde{M}e^{ -\tilde{c}\tilde{M}^2/2} \lesssim n^{-3/10}(\log n)^{1/4}\,,$$ 
which implies
\begin{align*}
\max_{s\in[n],\,j\in[p],\,k\in[q]}\var\{\tilde{\delta}_{43,k}(U_{s,j})\} \le &~\max_{s\in[n],\,j\in[p],\,k\in[q]}\mathbb{E}\{\tilde{\delta}^2_{43,k}(U_{s,j})\} 
\lesssim   n^{-3/5}(\log n)^{1/2}  \,.
\end{align*}
Recall $d=pq$ and $\tilde{d}=p\vee q\vee m$. By Bonferroni inequality and Bernstein inequality, 
\begin{align*}
\mathbb{P} \bigg\{\max_{j\in[p],\,k\in[q]} |{\rm I}'_{2}(j,k)| >x\bigg\} \le 2d\exp\bigg\{ -\frac{nx^2}{C_{1}n^{-3/5}(\log n)^{1/2} + C_{2}n^{-3/10}(\log n)^{1/4}x}\bigg\}
\end{align*}
for any $x>0$, which implies
\begin{align*} 
\max_{j\in[p],\, k\in[q]} |{\rm I}'_{2}(j,k)| =O_{\rm p}  \{n^{-4/5} (\log n)^{1/4}(\log \tilde{d})^{1/2} \}  + O_{\rm p}  \{n^{-13/10} (\log n)^{1/4}\log \tilde{d} \} \,.
\end{align*} 
Together with \eqref{eq:i1'}, by selecting 
$Q =\bar{C} \log^{1/2}(\tilde{d}n)$ for some sufficiently large constant $\bar{C} > \sqrt{5/(2\tilde{c})}$, we have
\begin{align*}
\frac{1}{n}\sum_{s=1}^{n} \tilde{\delta}_{4,k}(U_{s,j})  =    \frac{1}{n}\sum_{s=1}^{n} \tilde{\delta}_{44,k}(U_{s,j}) + {\rm Rem}_{21}(j,k)
\end{align*}
with $\max_{j\in[p],\,k\in[q]}|{\rm Rem}_{21}(j,k)| =O_{\rm p} \{n^{-4/5} (\log n)^{1/4} (\log \tilde{d})^{1/2} \}  $ provided that $\log \tilde{d} \lesssim n$.
Analogously, we can also show 
\begin{align*}
\frac{1}{n}\sum_{s=1}^{n} \tilde{\delta}_{5,j}(V_{s,k})  =    \frac{1}{n}\sum_{s=1}^{n} \tilde{\delta}_{54,j}(V_{s,k}) + {\rm Rem}_{22}(j,k)
\end{align*}
with $\max_{j\in[p],\,k\in[q]}|{\rm Rem}_{22}(j,k)| =O_{\rm p} \{n^{-4/5} (\log n)^{1/4}(\log \tilde{d})^{1/2} \} $ provided that $\log \tilde{d} \lesssim n$.
Hence, it holds that
\begin{align*}
\frac{1}{n}\sum_{s=1}^{n}\big\{\tilde{\delta}_{4,k}(U_{s,j}) + \tilde{\delta}_{5,j}(V_{s,k})\big\} =  \frac{1}{n}\sum_{s=1}^{n}\big\{\tilde{\delta}_{44,k}(U_{s,j}) + \tilde{\delta}_{54,j}(V_{s,k})\big\} + {\rm Rem}_{2}(j,k)
\end{align*}
with $\max_{j\in[p],\,k\in[q]}|{\rm Rem}_{2}(j,k)| =O_{\rm p} \{n^{-4/5} (\log n)^{1/4}(\log \tilde{d})^{1/2} \} $ provided that $\log \tilde{d} \lesssim n$. We complete the proof of Lemma \ref{lem:delta-t-4}.
$\hfill\Box$

\section{Proof of Lemma \ref{lem:cov}}\label{subsec:G-thetah}
Recall $\bTheta=\mathbb{E}(\bet_i\bet_i^{\T}) - \mathbb{E}(\bet_i)\mathbb{E}(\bet_i^{\T})$ and $\hat{\bTheta}=n^{-1}\sum_{i=1}^{n}\hat{\bet}_i\hat{\bet}_{i}^{\T}-(n^{-1}\sum_{i=1}^{n}\hat{\bet}_i)(n^{-1}\sum_{i=1}^{n}\hat{\bet}_i)^{\T}$ with $\bet_i=\boldsymbol{\varepsilon}_{i} \otimes \bdelta_i$ and $\hat{\bet}_i=\hat{\boldsymbol{\varepsilon}}_{i} \otimes \hat{\bdelta}_i$. Then
\begin{align*}
	|\hat{\bTheta}-\bTheta|_{\infty} \le&~ \underbrace{\max_{j,k\in[p],\,l,t\in[q]}\bigg|\frac{1}{n}\sum_{i=1}^{n}\big\{\ve_{i,j}\ve_{i,k}\delta_{i,l}\delta_{i,t}-\mathbb{E}(\ve_{i,j}\ve_{i,k}\delta_{i,l}\delta_{i,t})\big\}\bigg|}_{{\textrm S}_{1}} \\
	&~+ \underbrace{\max_{j,k\in[p],\,l,t\in[q]}\bigg|\frac{1}{n}\sum_{i=1}^{n}\hat{\ve}_{i,j}\hat{\ve}_{i,k}\hat{\delta}_{i,l}\hat{\delta}_{i,t} -\frac{1}{n}\sum_{i=1}^{n}\ve_{i,j}\ve_{i,k}\delta_{i,l}\delta_{i,t}  \bigg|}_{{\textrm S}_{2}}\\
	&~+\underbrace{\max_{j,k\in[p],\,l,t\in[q]}\bigg| \bigg(\frac{1}{n}\sum_{i=1}^{n}\ve_{i,j}\delta_{i,l} \bigg)\bigg(\frac{1}{n}\sum_{i=1}^{n}\ve_{i,k}\delta_{i,t}\bigg) -\mathbb{E}(\ve_{i,j}\delta_{i,l})\mathbb{E}(\ve_{i,k}\delta_{i,t})\bigg|}_{{\textrm S}_{3}} \\
	&~+\underbrace{\max_{j,k\in[p],\,l,t\in[q]}\bigg| \bigg(\frac{1}{n}\sum_{i=1}^{n}\hat{\ve}_{i,j}\hat{\delta}_{i,l} \bigg)\bigg(\frac{1}{n}\sum_{i=1}^{n}\hat{\ve}_{i,k}\hat{\delta}_{i,t} \bigg)- \bigg(\frac{1}{n}\sum_{i=1}^{n}\ve_{i,j}\delta_{i,l}\bigg)\bigg( \frac{1}{n}\sum_{i=1}^{n}\ve_{i,k}\delta_{i,t}\bigg)\bigg|}_{{\textrm S}_{4}} \,.
\end{align*}
Recall $\tilde{d}=p\vee q\vee m$. Identical to the arguments for deriving the convergence rate of ${\rm R}_2$ in Section \ref{sub:sec-h0-R2} for ${\rm R}_2$  defined in \eqref{eq:SigmaGamma}, it holds that
$${\rm S}_1=O_{\rm p}\{n^{-1/2}(\log \tilde{d})^{1/2}\} + O_{\rm p}\{n^{-1}(\log \tilde{d})\log^2 (\tilde{d}n) \}\,.$$ As we will show in Sections \ref{sec:sub-s2}--\ref{sec:sub-s4}, 
\begin{align}\label{eq:s2}
	{\rm S}_{2}= O_{\rm p}\{s^{2}n^{-1/2} (\log n) (\log \tilde{d})^{1/2}\log^{3/2}(\tilde{d}n)\} 
\end{align}  
provided that $s \ll   n^{1/2}(\log n)^{-1}\{\log (\tilde{d}n)\}^{-1}$ and $ \log \tilde{d} \ll n^{1/10}(\log n)^{-1/2}$,
\begin{align}\label{eq:s3}
	{\rm S}_{3} = O_{{\rm p}}\big\{n^{-1/2} (\log \tilde{d})^{1/2}\big\}
\end{align}
provided that $\log \tilde{d} \lesssim n^{1/3}$, and 
\begin{align}\label{eq:s4}
	{\rm S}_4 =&~   O_{\rm p} \{s  n^{-7/10} \log^{3/2}(\tilde{d}n)\} +  O_{\rm p} \{s^{1/2} n^{-13/20}(\log n)^{-3/4} \log(\tilde{d}n)\} \notag \\
	&+  O_{\rm p} \{n^{-1/2}(\log n)(\log \tilde{d})^{1/2}\}
\end{align}
provided that $s \lesssim n^{3/10}(\log \tilde{d})^{1/2}$ and $ \log \tilde{d} \ll n^{1/10}(\log n)^{-1/2}$. Hence, we have
\begin{align*}
	|\hat{\bTheta}-\bTheta|_{\infty} =   O_{\rm p}\{s^{2}n^{-1/2} (\log n) (\log \tilde{d})^{1/2}\log^{3/2}(\tilde{d}n)\}  
\end{align*} 
provided that $s \lesssim n^{3/10}(\log \tilde{d})^{1/2}$ and $ \log \tilde{d} \ll n^{1/10}(\log n)^{-1/2}$. We complete the proof of Lemma \ref{lem:cov}.
$\hfill\Box$

\subsection{Convergence rate of ${\rm S}_2$}\label{sec:sub-s2}
Analogous to \eqref{eq:4expend}, $n^{-1}\sum_{i=1}^{n}(\hat{\ve}_{i,j}\hat{\ve}_{i,k}\hat{\delta}_{i,l}\hat{\delta}_{i,t}-\ve_{i,j}\ve_{i,k}\delta_{i,l}\delta_{i,t})$ can be decomposed into 15 terms. To derive the convergence rate of ${\rm S}_2$, by the symmetry, we only consider the convergence rates of the following terms:
\begin{align*}
	{\rm S}_{21}=&\max_{j,k\in[p],\,l,t\in[q]}\bigg|\frac{1}{n}\sum_{i=1}^{n}(\hat{\ve}_{i,j}-\ve_{i,j})\ve_{i,k}\delta_{i,l}\delta_{i,t}\bigg|\,, \\
	{\rm S}_{22}=&\max_{j,k\in[p],\,l,t\in[q]}\bigg|\frac{1}{n}\sum_{i=1}^{n}(\hat{\ve}_{i,j}-\ve_{i,j})(\hat{\ve}_{i,k}-\ve_{i,k})\delta_{i,l}\delta_{i,t}\bigg|\,,\\  
	{\rm S}_{23}=&\max_{j,k\in[p],\,l,t\in[q]}\bigg|\frac{1}{n}\sum_{i=1}^{n}(\hat{\ve}_{i,j}-\ve_{i,j})(\hat{\delta}_{i,l}-\delta_{i,l}) \ve_{i,k}\delta_{i,t}\bigg|\,,\\  
	{\rm S}_{24}=&\max_{j,k\in[p],\,l,t\in[q]}\bigg|\frac{1}{n}\sum_{i=1}^{n}(\hat{\ve}_{i,j}-\ve_{i,j})(\hat{\ve}_{i,k}-\ve_{i,k})(\hat{\delta}_{i,l}-\delta_{i,l})\delta_{i,t}\bigg|\,, \\
	{\rm S}_{25}=&\max_{j,k\in[p],\,l,t\in[q]}\bigg|\frac{1}{n}\sum_{i=1}^{n}(\hat{\ve}_{i,j}-\ve_{i,j})(\hat{\ve}_{i,k}-\ve_{i,k})(\hat{\delta}_{i,l}-\delta_{i,l})(\hat{\delta}_{i,t}-\delta_{i,t})\bigg|\,.
\end{align*}
As we will show in Sections \ref{sec:sub-s21}--\ref{sec:sub-s25}, 
\begin{align}
	&{\rm S}_{21} =  O_{\rm p}\{s^{1/2}n^{-1/2} (\log n) (\log \tilde{d})^{1/2}\log^{3/2}(\tilde{d}n)\}  +  O_{\rm p}\{sn^{-1/2}(\log \tilde{d})^{1/2}\}\,, \label{eq:s21} \\ 
	&~~~~~~~~~~~~{\rm S}_{22} = O_{\rm p}\{s n^{-1/2} (\log n) (\log \tilde{d})^{1/2}\log^{3/2}(\tilde{d}n)\}   = {\rm S}_{23}\,, \label{eq:s22}  \\
	&~~~~~~~~~~~~~~~{\rm S}_{24}= O_{\rm p}\{ s^{3/2}n^{-1/2} (\log n) (\log \tilde{d})^{1/2}\log^{3/2}(\tilde{d}n)\} \,,\label{eq:s24}\\
	&~~~~~~~~~~~~~~~~{\rm S}_{25} =  O_{\rm p}\{s^{2}n^{-1/2} (\log n) (\log \tilde{d})^{1/2}\log^{3/2}(\tilde{d}n)\}  \label{eq:s25}
\end{align}   
provided that $s \ll   n^{1/2}(\log n)^{-1}\{\log (\tilde{d}n)\}^{-1}$ and $ \log \tilde{d} \ll n^{1/10}(\log n)^{-1/2}$. Hence,  we have \eqref{eq:s2} holds.
$\hfill\Box$

\subsubsection{Convergence rate of ${\rm S}_{21}$}\label{sec:sub-s21}
Recall $\ve_{i,j}= U_{i,j}- \balpha_{j}^{\T}\bW_{i}$ and $\hat{\ve}_{i,j}= \hat{U}_{i,j}- \hat{\balpha}_{j}^{\T}\hat{\bW}_{i}$. We then have
\begin{align}\label{eq:s21-exp}
	{\rm S}_{21}
	\le&~ \underbrace{\max_{j,k\in[p],\,l,t\in[q]}\bigg|\frac{1}{n}\sum_{i=1}^{n}(\hat{U}_{i,j}-U_{i,j})\ve_{i,k}\delta_{i,l}\delta_{i,t}\bigg|}_{{\textrm S}_{211}} \notag \\
	&+ \underbrace{ \max_{j\in[p]}|\hat{\balpha}_j|_{1} \cdot \max_{ v\in[m],\,k\in[p],\,l,t\in[q]} \bigg|\frac{1}{n}\sum_{i=1}^{n}(\hat{W}_{i,v}-W_{i,v})\ve_{i,k}\delta_{i,l}\delta_{i,t}\bigg|}_{{\textrm S}_{212}}\\
	&+ \underbrace{\max_{j\in[p]}|\hat{\balpha}_j-\balpha_j|_{1} \cdot\max_{v\in[m],\,k\in[p],\,l,t\in[q]}\bigg|\frac{1}{n}\sum_{i=1}^{n}W_{i,v}\ve_{i,k}\delta_{i,l}\delta_{i,t}\bigg|}_{{\textrm S}_{213}}\notag\,.
\end{align}
Recall $\tilde{d}=p\vee q\vee m$. Using the similar arguments for deriving the convergence rate of ${\rm R}_{11}$ in Section \ref{sec:sub-r11} for ${\rm R}_{11}$  defined in \eqref{eq:R1terms}, it holds that
\begin{align}\label{eq:s211}
	{\rm S}_{211} = O_{\rm p}\{n^{-1/2} (\log n) (\log \tilde{d})^{1/2}\log^{3/2}(\tilde{d}n)\}
\end{align}
provided that $\log \tilde{d} \lesssim n^{5/12} (\log n)^{-1/2}$.  Analogously, we can also show such convergence rate holds for $\max_{ v\in[m],\,k\in[p],\,l,t\in[q]}  |n^{-1}\sum_{i=1}^{n}(\hat{W}_{i,v}-W_{i,v})\ve_{i,k}\delta_{i,l}\delta_{i,t} |$.
By Lemma \ref{lem:coeff}, it holds that
\begin{align}\label{eq:s212}
	{\rm S}_{212} =  O_{\rm p}\{s^{1/2}n^{-1/2} (\log n) (\log \tilde{d})^{1/2}\log^{3/2}(\tilde{d}n)\}
\end{align}
provided that $s \ll   n^{1/2}(\log n)^{-1}\{\log (\tilde{d}n)\}^{-1}$ and $ \log \tilde{d} \ll n^{1/10}(\log n)^{-1/2}$.
Recall $W_{i,v}\sim \mathcal{N}(0,1)$. By \eqref{eq:ve_tail}  and \eqref{eq:delta-tail}, it holds that $\mathbb{E}(\ve_{i,j}^4) \leq C$ and $\mathbb{E}(\delta_{i,k}^4) \leq C$. By  Cauchy-Schwarz inequality, 
\begin{align*}
	\mathbb{E}(W_{i,v}\ve_{i,k}\delta_{i,l}\delta_{i,t}) \leq &~ \big\{\mathbb{E}(W_{i,v}^4)\big\}^{1/4}\big\{\mathbb{E}(\ve_{i,k}^4)\big\}^{1/4}\big\{\mathbb{E}(\delta_{i,l}^4)\big\}^{1/4}\big\{\mathbb{E}(\delta_{i,t}^4)\big\}^{1/4}\leq C_1\,.
\end{align*}
Using the same arguments for deriving the convergence rate of ${\rm R}_2$ in Section \ref{sub:sec-h0-R2} for ${\rm R}_2$ defined in \eqref{eq:SigmaGamma}, it holds that
\begin{align*}
	&\max_{v\in[m],\,k\in[p],\,l,t\in[q]}\bigg|\frac{1}{n}\sum_{i=1}^{n}W_{i,v}\ve_{i,k}\delta_{i,l}\delta_{i,t} - \mathbb{E}( W_{i,v}\ve_{i,k}\delta_{i,l}\delta_{i,t})\bigg|\\
	&~~~~~~~~~~~~~~~ = O_{\rm p}\{n^{-1/2}(\log \tilde{d})^{1/2}\} + O_{\rm p}\{n^{-1}\log^2 (\tilde{d}n) \log \tilde{d}\}\,,
\end{align*}
which implies $\max_{v\in[m],\,k\in[p],\,l,t\in[q]} |n^{-1}\sum_{i=1}^{n}W_{i,v}\ve_{i,k}\delta_{i,l}\delta_{i,t} |=  O_{\rm p}(1)$ provided that $\log \tilde{d} \lesssim n^{1/3}$.  By Lemma \ref{lem:coeff} again, we have
\begin{align*}
	{\rm S}_{213} = O_{\rm p}\{sn^{-1/2}(\log \tilde{d})^{1/2}\}
\end{align*}
provided that $s \ll   n^{1/2}(\log n)^{-1}\{\log (\tilde{d}n)\}^{-1}$ and $ \log \tilde{d} \ll n^{1/10}(\log n)^{-1/2}$. Hence, together with \eqref{eq:s211} and \eqref{eq:s212}, by \eqref{eq:s21-exp}, we have  \eqref{eq:s21} holds.
$\hfill\Box$

\subsubsection{Convergence rates of ${\rm S}_{22}$ and ${\rm S}_{23}$}\label{sec:sub-s22}
Recall $\hat{\ve}_{i,j}=\hat{U}_{i,j}-\hat{\balpha}_{j}^{\T}\hat{\bW}_{i}$. By direct calculation, we have
\begin{align}\label{eq:s22-exp}
	{\rm S}_{22} 
	\le &~ \underbrace {\max_{j,k\in[p],\,l,t\in[q]}\bigg|\frac{1}{n}\sum_{i=1}^{n}(\hat{U}_{i,j} - U_{i,j})(\hat{U}_{i,k} - U_{i,k})\delta_{i,l}\delta_{i,t}\bigg| }_{{\rm S}_{221}}\notag\\
	&+ 2 \underbrace{\max_{j\in[p]}|\hat{\balpha}_{j}|_{1} \cdot \max_{j\in[p],\,l,t\in[q],\,k\in[m]}\bigg|\frac{1}{n}\sum_{i=1}^{n}(\hat{U}_{i,j} - U_{i,j})(\hat{W}_{i,k} - W_{i,k})\delta_{i,l}\delta_{i,t}\bigg|}_{{\rm S}_{222}} \notag\\
	&+2\underbrace{ \max_{j\in[p]}|\hat{\balpha}_j -\balpha_j|_{1} \cdot \max_{j\in[p],\,l,t\in[q],\,k\in[m]}\bigg|\frac{1}{n}\sum_{i=1}^{n}(\hat{U}_{i,j} - U_{i,j})W_{i,k}\delta_{i,l}\delta_{i,t}\bigg|}_{{\rm S}_{223}} \\
	&+\underbrace{\max_{j\in[p]} |\hat{\balpha}_j|_{1}^2  \cdot \max_{j,k\in[m],\,l,t\in[q]} \bigg|\frac{1}{n}\sum_{i=1}^{n}(\hat{W}_{i,j}-W_{i,j})(\hat{W}_{i,k}-W_{i,k})\delta_{i,l}\delta_{i,t}\bigg| }_{{\rm S}_{224}} \notag\\
	&+2\underbrace{ \max_{j\in[p]} |\hat{\balpha}_j|_{1} \cdot \max_{j\in[p]}|\hat{\balpha}_j - \balpha_j|_{1} \cdot \max_{j,k\in[m],\,l,t\in[q]}\bigg|\frac{1}{n}\sum_{i=1}^{n}(\hat{W}_{i,j}-W_{i,j})W_{i,k}\delta_{i,l}\delta_{i,t}\bigg|}_{{\rm S}_{225}}\notag\\
	&+\underbrace{\max_{j\in[p]}|\hat{\balpha}_j - \balpha_j|_{1}^2 \cdot \max_{j,k\in[m],\,l,t\in[q]}\bigg|\frac{1}{n}\sum_{i=1}^{n}W_{i,j}W_{i,k}\delta_{i,l}\delta_{i,t}\bigg| }_{{\rm S}_{226}} \,.\notag
\end{align}
Recall $\tilde{d}=p\vee q\vee m$. Using the same arguments for deriving the convergence rate of ${\rm R}_{12}$ in Section \ref{sec:sub-r12}  for ${\rm R}_{12}$ defined in \eqref{eq:R1terms},  we have
\begin{align}\label{eq:s221}
	{\rm S}_{221} = O_{\rm p}\{n^{-1/2} (\log n) (\log \tilde{d})^{1/2}\log^{3/2}(\tilde{d}n)\} 
\end{align}
provided that $\log \tilde{d} \lesssim n^{5/12} (\log n)^{-1/2}$. 
Analogously, we can also show such convergence rate holds for $\max_{j\in[p],\,l,t\in[q],\,k\in[m]} |n^{-1}\sum_{i=1}^{n}(\hat{U}_{i,j} - U_{i,j})(\hat{W}_{i,k} - W_{i,k})\delta_{i,l}\delta_{i,t} |$ and $\max_{j,k\in[m],\,l,t\in[q]} |n^{-1}\sum_{i=1}^{n}$ $(\hat{W}_{i,j}-W_{i,j})(\hat{W}_{i,k}-W_{i,k})\delta_{i,l}\delta_{i,t}|$.
By Lemma \ref{lem:coeff}, it holds that
\begin{equation}\label{eq:s224}
	\begin{split}
		{\rm S}_{222} =&~ O_{\rm p}\{s^{1/2}n^{-1/2} (\log n) (\log \tilde{d})^{1/2}\log^{3/2}(\tilde{d}n)\} \,,  \\
		{\rm S}_{224} =&~   O_{\rm p}\{sn^{-1/2} (\log n) (\log \tilde{d})^{1/2}\log^{3/2}(\tilde{d}n)\}
	\end{split}
\end{equation}
provided that $s \ll   n^{1/2}(\log n)^{-1}\{\log (\tilde{d}n)\}^{-1}$ and $ \log \tilde{d} \ll n^{1/10}(\log n)^{-1/2}$. Analogous to the derivation of \eqref{eq:s211}, we have
\begin{align*}
	\max_{j\in[p],\,l,t\in[q],\,k\in[m]}\bigg|\frac{1}{n}\sum_{i=1}^{n}(\hat{U}_{i,j} - U_{i,j})W_{i,k}\delta_{i,l}\delta_{i,t}\bigg|=&~ O_{\rm p}\{n^{-1/2} (\log n) (\log \tilde{d})^{1/2}\log^{3/2}(\tilde{d}n)\}\\
	=&~\max_{j,k\in[m],\,l,t\in[q]}\bigg|\frac{1}{n}\sum_{i=1}^{n}(\hat{W}_{i,j}-W_{i,j})W_{i,k}\delta_{i,l}\delta_{i,t}\bigg|
\end{align*}
provided that $\log \tilde{d} \lesssim n^{5/12} (\log n)^{-1/2}$. By Lemma \ref{lem:coeff} again, 
\begin{equation}\label{eq:s223}
	\begin{split}
		&{\rm S}_{223} =  O_{\rm p}\{sn^{-1} (\log n) (\log \tilde{d}) \log^{3/2}(\tilde{d}n)\} \,,\\
		&{\rm S}_{225} =  O_{\rm p}\{s^{3/2}n^{-1} (\log n) (\log \tilde{d}) \log^{3/2}(\tilde{d}n)\} 
	\end{split}
\end{equation}
provided that $s \ll   n^{1/2}(\log n)^{-1}\{\log (\tilde{d}n)\}^{-1}$ and $ \log \tilde{d} \ll n^{1/10}(\log n)^{-1/2}$.  Using the similar arguments for deriving the convergence rate of ${\rm S}_{213}$, it holds that
\begin{align*}
	{\rm S}_{226}= O_{\rm p} (s^2n^{-1}\log \tilde{d})
\end{align*}
provided that $s \ll   n^{1/2}(\log n)^{-1}\{\log (\tilde{d}n)\}^{-1} $ and $ \log \tilde{d} \ll n^{1/10}(\log n)^{-1/2}$. Together with \eqref{eq:s221}--\eqref{eq:s223}, by \eqref{eq:s22-exp},  we have 
\begin{align*}
	{\rm S}_{22} =  O_{\rm p}\{s n^{-1/2} (\log n) (\log \tilde{d})^{1/2}\log^{3/2}(\tilde{d}n)\} 
\end{align*}
provided that $s \ll   n^{1/2}(\log n)^{-1}\{\log (\tilde{d}n)\}^{-1}$ and $ \log \tilde{d} \ll n^{1/10}(\log n)^{-1/2}$. Using the similar arguments, we can also show such convergence rate holds for ${\rm S}_{23}$. Hence, \eqref{eq:s22} holds.
$\hfill\Box$

\subsubsection{Convergence rate of ${\rm S}_{24}$}\label{sec:sub-S24}
Recall $\ve_{i,j}= U_{i,j}- \balpha_{j}^{\T}\bW_{i}$, $\hat{\ve}_{i,j}= \hat{U}_{i,j}- \hat{\balpha}_{j}^{\T}\hat{\bW}_{i}$, $\delta_{i,l}= V_{i,l}- \bbeta_{l}^{\T}\bW_{i}$ and $\hat{\delta}_{i,l}= \hat{V}_{i,l}- \hat{\bbeta}_{l}^{\T}\hat{\bW}_{i}$. We then have
\begin{align*}
	{\rm S}_{24}
	\le&~ \underbrace{ \max_{j,k\in[p],\,l,t\in[q]}\bigg|\frac{1}{n} \sum_{i=1}^{n} (\hat{U}_{i,j}-U_{i,j}) (\hat{U}_{i,k}-U_{i,k}) (\hat{V}_{i,l}-V_{i,l}) \delta_{i,t} \bigg|}_{{\textrm S}_{241}}\\
	&+ \underbrace{\max_{j,k\in[p],\,l,t\in[q]}\bigg|\frac{1}{n} \sum_{i=1}^{n} (\hat{U}_{i,j}-U_{i,j}) (\hat{U}_{i,k}-U_{i,k}) (\hat{\bbeta}_{l}^{\T}\hat{\bW}_{i} - \bbeta_{l}^{\T}\bW_{i}) \delta_{i,t} \bigg|}_{{\textrm S}_{242}}\\
	&+2 \underbrace{\max_{j,k\in[p],\,l,t\in[q]}\bigg|\frac{1}{n} \sum_{i=1}^{n} (\hat{U}_{i,j}-U_{i,j}) (\hat{\balpha}_{k}^{\T}\hat{\bW}_{i} -\balpha_{k}^{\T}\bW_{i} ) (\hat{V}_{i,l}-V_{i,l}) \delta_{i,t} \bigg|}_{{\textrm S}_{243}}\\
	&+2 \underbrace{\max_{j,k\in[p],\,l,t\in[q]}\bigg|\frac{1}{n} \sum_{i=1}^{n} (\hat{U}_{i,j}-U_{i,j}) (\hat{\balpha}_{k}^{\T}\hat{\bW}_{i} -\balpha_{k}^{\T}\bW_{i} ) (\hat{\bbeta}_{l}^{\T}\hat{\bW}_{i} - \bbeta_{l}^{\T}\bW_{i}) \delta_{i,t} \bigg|}_{{\textrm S}_{244} }\\
	&+ \underbrace{\max_{j,k\in[p],\,l,t\in[q]}\bigg|\frac{1}{n} \sum_{i=1}^{n} (\hat{\balpha}_{j}^{\T}\hat{\bW}_{i} -\balpha_{j}^{\T}\bW_{i} ) (\hat{\balpha}_{k}^{\T}\hat{\bW}_{i} -\balpha_{k}^{\T}\bW_{i} ) (\hat{V}_{i,l}-V_{i,l}) \delta_{i,t} \bigg|}_{{\textrm S}_{245}}\\
	&+ \underbrace{\max_{j,k\in[p],\,l,t\in[q]}\bigg|\frac{1}{n} \sum_{i=1}^{n} (\hat{\balpha}_{j}^{\T}\hat{\bW}_{i} -\balpha_{j}^{\T}\bW_{i} ) (\hat{\balpha}_{k}^{\T}\hat{\bW}_{i} -\balpha_{k}^{\T}\bW_{i} ) (\hat{\bbeta}_{l}^{\T}\hat{\bW}_{i} - \bbeta_{l}^{\T}\bW_{i}) \delta_{i,t}\bigg| }_{{\textrm S}_{246}}\,.
\end{align*}
Recall $\tilde{d}=p\vee q\vee m$. Using the same arguments for deriving the convergence rate of ${\rm R}_{14}$ in Section \ref{sec:sub-r14}  for ${\rm R}_{14}$ defined in \eqref{eq:R1terms},  we have
\begin{align}\label{eq:s241}
	{\rm S}_{241} =   O_{\rm p}\{n^{-1/2} (\log n) (\log \tilde{d})^{1/2}\log^{3/2}(\tilde{d}n)\}
\end{align}
provided that $\log \tilde{d} \lesssim n^{5/12} (\log n)^{-1/2}$. As we will show later,
\begin{align}
	&{\rm S}_{242} =  O_{\rm p}\{s^{1/2}n^{-1/2} (\log n) (\log \tilde{d})^{1/2}\log^{3/2}(\tilde{d}n)\} = {\rm S}_{243}\,, \label{eq:s242}\\
	&~~{\rm S}_{244}=    O_{\rm p}\{sn^{-1/2} (\log n) (\log \tilde{d})^{1/2}\log^{3/2}(\tilde{d}n)\} ={\rm S}_{245} \label{eq:s244}\,,\\
	&~~~~~~{\rm S}_{246}=     O_{\rm p}\{ s^{3/2}n^{-1/2} (\log n) (\log \tilde{d})^{1/2}\log^{3/2}(\tilde{d}n)\} \label{eq:s248}
\end{align}
provided that $s \ll   n^{1/2}(\log n)^{-1}\{\log (\tilde{d}n)\}^{-1}$ and $ \log \tilde{d} \ll n^{1/10}(\log n)^{-1/2}$. Combining \eqref{eq:s241}--\eqref{eq:s248}, we have \eqref{eq:s24} holds.

\underline{{\it Convergence rates of ${\rm S}_{242}$ and ${\rm S}_{243}$}.} 
Notice that
\begin{align}\label{eq:s242-dec}
	{\rm S}_{242}\le&~ \max_{k\in[q]}|\hat{\bbeta}_{k}|_1 \cdot \max_{j,k\in[p],\, l\in[m],\, t\in[q]}\bigg|\frac{1}{n} \sum_{i=1}^{n} (\hat{U}_{i,j}-U_{i,j}) (\hat{U}_{i,k}-U_{i,k}) (\hat{W}_{i,l}-W_{i,l}) \delta_{i,t}\bigg|\notag\\
	&+ \max_{k\in[q]}|\hat{\bbeta}_{k} -\bbeta_{k}|_1 \cdot \max_{j,k\in[p],\, l\in[m],\, t\in[q]}\bigg|\frac{1}{n} \sum_{i=1}^{n} (\hat{U}_{i,j}-U_{i,j}) (\hat{U}_{i,k}-U_{i,k})  W_{i,l} \delta_{i,t}\bigg| \,.
\end{align}
Analogous to the derivations of \eqref{eq:s241} and \eqref{eq:s221}, respectively,  we have
\begin{align}\label{eq:uuw-d}
	&\max_{j,k\in[p],\, l\in[m],\, t\in[q]}\bigg|\frac{1}{n} \sum_{i=1}^{n} (\hat{U}_{i,j}-U_{i,j}) (\hat{U}_{i,k}-U_{i,k}) (\hat{W}_{i,l}-W_{i,l}) \delta_{i,t}\bigg| \notag\\
	&~~~~~~~~~~~~~~~~~~~~~~~~~~~=   O_{\rm p}\{n^{-1/2} (\log n) (\log \tilde{d})^{1/2}\log^{3/2}(\tilde{d}n)\} \\
	&~~~~~~~~~~~~~~~~~~~~~~~~~~~~~~~~~~~~~=\max_{j,k\in[p],\, l\in[m],\, t\in[q]}\bigg|\frac{1}{n} \sum_{i=1}^{n} (\hat{U}_{i,j}-U_{i,j}) (\hat{U}_{i,k}-U_{i,k})  W_{i,l} \delta_{i,t}\bigg|\notag
\end{align}
provided that $\log \tilde{d} \lesssim n^{5/12} (\log n)^{-1/2}$. By \eqref{eq:s242-dec} and Lemma \ref{lem:coeff}, it holds that
\begin{align*}
	{\rm S}_{242} =   O_{\rm p}\{s^{1/2}n^{-1/2} (\log n) (\log \tilde{d})^{1/2}\log^{3/2}(\tilde{d}n)\} 
\end{align*}
provided that $s \ll   n^{1/2}(\log n)^{-1}\{\log (\tilde{d}n)\}^{-1}$ and $ \log \tilde{d} \ll n^{1/10}(\log n)^{-1/2}$. Analogously, we can also show such convergence rate holds for ${\rm S}_{243}$. Hence, 
we have  \eqref{eq:s242} holds.

\underline{{\it Convergence rates of $ {\rm S}_{244} $ and $ {\rm S}_{245} $}.} 
Notice that
\begin{align*}
	{\rm S}_{244}\le &~
	\underbrace{\max_{j,k\in[p],\, l,t\in[q]}\bigg|\frac{1}{n} \sum_{i=1}^{n} (\hat{U}_{i,j}-U_{i,j})   \hat{\balpha}_{k}^{\T}(\hat{\bW}_{i} - \bW_{i})(\hat{\bW}_{i} - \bW_{i})^{\T} \hat{\bbeta}_{l} \delta_{i,t} \bigg|}_{{\textrm S}_{2441}}\\
	&+ \underbrace{\max_{j,k\in[p],\, l,t\in[q]}\bigg|\frac{1}{n} \sum_{i=1}^{n} (\hat{U}_{i,j}-U_{i,j})   \hat{\balpha}_{k}^{\T}(\hat{\bW}_{i} - \bW_{i})\bW_{i}^{\T} (\hat{\bbeta}_{l} -\bbeta_{l} ) \delta_{i,t}\bigg| }_{{\textrm S}_{2442}}\\
	&+\underbrace{\max_{j,k\in[p],\, l,t\in[q]}\bigg|\frac{1}{n} \sum_{i=1}^{n} (\hat{U}_{i,j}-U_{i,j})   (\hat{\balpha}_{k} -\balpha_{k}) ^{\T}\bW_{i}(\hat{\bW}_{i} - \bW_{i})^{\T} \hat{\bbeta}_{l} \delta_{i,t} \bigg|}_{{\textrm S}_{2443} }\\
	&+\underbrace{\max_{j,k\in[p],\, l,t\in[q]}\bigg|\frac{1}{n} \sum_{i=1}^{n} (\hat{U}_{i,j}-U_{i,j})   (\hat{\balpha}_{k} -\balpha_{k})^{\T} \bW_{i}  \bW_{i}^{\T} (\hat{\bbeta}_{l} -\bbeta_{l}) \delta_{i,t}\bigg| }_{{\textrm S}_{2444}}
\end{align*}
Parallel to  \eqref{eq:uuw-d}, by Lemma \ref{lem:coeff}, it holds that
\begin{align*}
	{\rm S}_{2441}  \le&~   \max_{j \in[p]}|\hat{\balpha}_{j} |_1 \cdot \max_{k\in[q]}|\hat{\bbeta}_{k} |_1 \notag\\
	&~~~~~\times \max_{j\in[p],\, k,l\in[m],\, t\in[q]}\bigg|\frac{1}{n} \sum_{i=1}^{n} (\hat{U}_{i,j}-U_{i,j}) (\hat{W}_{i,k}-W_{i,k}) (\hat{W}_{i,l}-W_{i,l}) \delta_{i,t}\bigg| \notag\\
	= &~   O_{\rm p}\{sn^{-1/2} (\log n) (\log \tilde{d})^{1/2}\log^{3/2}(\tilde{d}n)\} \,, \notag\\
	{\rm S}_{2442} \le&~  \max_{j \in[p]}|\hat{\balpha}_{j} - \hat{\balpha}_{j} |_1 \cdot \max_{k\in[q]}|\hat{\bbeta}_{k} |_1 \notag\\
	&~~~~~\times \max_{j\in[p],\, l,k\in[m],\, t\in[q]}\bigg|\frac{1}{n} \sum_{i=1}^{n} (\hat{U}_{i,j}-U_{i,j}) (\hat{W}_{i,k}-W_{i,k})  W_{i,l}  \delta_{i,t}\bigg| \notag\\
	= &~   O_{\rm p}\{s^{3/2}n^{-1} (\log n) (\log \tilde{d})\log^{3/2}(\tilde{d}n)\}
\end{align*}
provided that $s \ll   n^{1/2}(\log n)^{-1}\{\log (\tilde{d}n)\}^{-1}$ and $ \log \tilde{d} \ll n^{1/10}(\log n)^{-1/2}$.  Similarly, we can also show the convergence rate of 
$ {\rm S}_{2443} $ is identical to $ {\rm S}_{2442} $. 
Parallel to    \eqref{eq:s211}, 
\begin{align}\label{eq:uh-h-1}
	\max_{j\in[p],\, k,l\in[m],\, t\in[q]}\bigg|\frac{1}{n} \sum_{i=1}^{n} (\hat{U}_{i,j}-U_{i,j})  W_{i,k}   W_{i,l} \delta_{i,t}\bigg|=  O_{\rm p}\{n^{-1/2} (\log n) (\log \tilde{d})^{1/2}\log^{3/2}(\tilde{d}n)\} 
\end{align}
provided that $\log \tilde{d} \lesssim n^{5/12} (\log n)^{-1/2}$. By Lemma \ref{lem:coeff} again, it holds that
\begin{align*}
	{\rm S}_{2444}   \le&~   \max_{j \in[p]}|\hat{\balpha}_{j} - \hat{\balpha}_{j} |_1 \cdot \max_{k \in[q]}|\hat{\bbeta}_{k} - \hat{\bbeta}_{k} |_1  \cdot \max_{j\in[p],\, l,k\in[m],\, t\in[q]}\bigg|\frac{1}{n} \sum_{i=1}^{n} (\hat{U}_{i,j}-U_{i,j})  W_{i,k}   W_{i,l}  \delta_{i,t}\bigg|  \\
	= &~    O_{\rm p}\{s^{2}n^{-3/2} (\log n) (\log \tilde{d})^{3/2}\log^{3/2}(\tilde{d}n)\}
\end{align*}
provided that $s \ll   n^{1/2}(\log n)^{-1}\{\log (\tilde{d}n)\}^{-1}$ and $ \log \tilde{d} \ll n^{1/10}(\log n)^{-1/2}$.  Hence, we have  
\begin{align*}
	{\rm S}_{244}=   O_{\rm p}\{sn^{-1/2} (\log n) (\log \tilde{d})^{1/2}\log^{3/2}(\tilde{d}n)\} 
\end{align*}
provided that $s \ll   n^{1/2}(\log n)^{-1}\{\log (\tilde{d}n)\}^{-1}$ and $ \log \tilde{d} \ll n^{1/10}(\log n)^{-1/2}$. Analogously, we can also show such convergence rate holds for ${\rm S}_{245}$. Hence, we have
\eqref{eq:s244} holds.

\underline{{\it Convergence rate of $ {\rm S}_{246} $}.}
Notice that
\begin{align}\label{eq:s248-dec}
	{\rm S}_{246}\le &~ \underbrace{\max_{j,k\in[p],\, l,t\in[q]}\bigg|\frac{1}{n}\sum_{i=1}^{n} \hat{\balpha}_{j}^{\T}(\hat{\bW}_i-\bW_i) \hat{\balpha}_{k}^{\T}(\hat{\bW}_i-\bW_i) \hat{\bbeta}_{l}^{\T}(\hat{\bW}_i-\bW_i)\delta_{i,t}\bigg| }_{{\textrm S}_{2461} } \notag\\
	&+\underbrace{\max_{j,k\in[p],\, l,t\in[q]}\bigg|\frac{1}{n}\sum_{i=1}^{n} \hat{\balpha}_{j}^{\T}(\hat{\bW}_i-\bW_i) \hat{\balpha}_{k}^{\T}(\hat{\bW}_i-\bW_i) (\hat{\bbeta}_{l} -\bbeta_{l})^{\T}\bW_i\delta_{i,t} \bigg|}_{{\textrm S}_{2462}} \notag\\
	&+2\underbrace{\max_{j,k\in[p],\, l,t\in[q]}\bigg|\frac{1}{n}\sum_{i=1}^{n} \hat{\balpha}_{j}^{\T}(\hat{\bW}_i-\bW_i) (\hat{\balpha}_{k} -\balpha_{k})^{\T}\bW_i \hat{\bbeta}_{l}^{\T}(\hat{\bW}_i-\bW_i)\delta_{i,t}\bigg|}_{{\textrm S}_{2463}} \notag\\
	&+2\underbrace{\max_{j,k\in[p],\, l,t\in[q]}\bigg|\frac{1}{n}\sum_{i=1}^{n} \hat{\balpha}_{j}^{\T}(\hat{\bW}_i-\bW_i) (\hat{\balpha}_{k} -\balpha_{k})^{\T}\bW_i (\hat{\bbeta}_{l} -\bbeta_{l})^{\T}\bW_i\delta_{i,t} \bigg|}_{{\textrm S}_{2464}}\\
	&+\underbrace{\max_{j,k\in[p],\, l,t\in[q]}\bigg|\frac{1}{n}\sum_{i=1}^{n} (\hat{\balpha}_{j}  -\balpha_{j})^{\T} \bW_i  (\hat{\balpha}_{k} -\balpha_{k})^{\T}\bW_i \hat{\bbeta}_{l}^{\T}(\hat{\bW}_i-\bW_i)\delta_{i,t}\bigg| }_{{\textrm S}_{2465}}\notag\\
	&+ \underbrace{\max_{j,k\in[p],\, l,t\in[q]}\bigg|\frac{1}{n}\sum_{i=1}^{n} (\hat{\balpha}_{j}  -\balpha_{j})^{\T} \bW_i  (\hat{\balpha}_{k} -\balpha_{k})^{\T}\bW_i (\hat{\bbeta}_{l} -\bbeta_{l})^{\T}\bW_i\delta_{i,t} \bigg|}_{{\textrm S}_{2466}}\,.\notag
\end{align}
Parallel to \eqref{eq:uuw-d}, by Lemma \ref{lem:coeff}, it holds that
\begin{align*}
	{\rm S}_{2461}  
	\le&~ \max_{j\in[p]}|\hat{\balpha}_{j}|_1^2 \cdot \max_{k\in[q]}|\hat{\bbeta}_{k}|_1  \notag\\
	&~~~\times\max_{r_1,r_2,r_3 \in[m],\,t\in[q]}\bigg|\frac{1}{n}\sum_{i=1}^{n} (\hat{W}_{i,r_1}- W_{i,r_1}) (\hat{W}_{i,r_2}- W_{i,r_2}) (\hat{W}_{i,r_3}- W_{i,r_3}) \delta_{i,t} \bigg| \notag\\
	=&~   O_{\rm p}\{ s^{3/2}n^{-1/2} (\log n) (\log \tilde{d})^{1/2}\log^{3/2}(\tilde{d}n)\}\,,\notag\\
	{\rm S}_{2462}   \le&~ \max_{j\in[p]}|\hat{\balpha}_{j}|_1^2  \cdot \max_{k\in[q]}|\hat{\bbeta}_{k} -\bbeta_{k}|_1 \notag\\
	&~~~~\times \max_{r_1,r_2,r_3 \in[m],\,t\in[q]}\bigg|\frac{1}{n}\sum_{i=1}^{n} (\hat{W}_{i,r_1}- W_{i,r_1}) (\hat{W}_{i,r_2}- W_{i,r_2}) W_{i,r_3} \delta_{i,t} \bigg|  \notag \\
	=&~  O_{\rm p}\{s^2n^{-1} (\log n) (\log \tilde{d}) \log^{3/2}(\tilde{d}n)\} 
\end{align*}   
provided that $s \ll   n^{1/2}(\log n)^{-1}\{\log (\tilde{d}n)\}^{-1}$ and $ \log \tilde{d} \ll n^{1/10}(\log n)^{-1/2}$. Analogously, we can  show such derived convergence rate  of $ {\rm S}_{2462} $ also holds for $ {\rm S}_{2463} $.
Parallel to \eqref{eq:uh-h-1},
by Lemma \ref{lem:coeff} again, 
\begin{align*}
	{\rm S}_{2464}  \le&~ \max_{j\in[p]}|\hat{\balpha}_{j}|_1 \cdot \max_{j\in[p]}|\hat{\balpha}_{j} - \balpha_{j}|_1  \cdot  \max_{k\in[q]}|\hat{\bbeta}_{k} -\bbeta_{k}|_1 \notag\\
	&~~~~\times \max_{r_1,r_2,r_3 \in[m],\,t\in[q]}\bigg|\frac{1}{n}\sum_{i=1}^{n} (\hat{W}_{i,r_1}- W_{i,r_1})  W_{i,r_2}  W_{i,r_3} \delta_{i,t} \bigg|  \notag \\
	=&~   O_{\rm p}\{s^{5/2} n^{-3/2} (\log n) (\log \tilde{d})^{3/2}\log^{3/2}(\tilde{d}n)\}  
\end{align*}
provided that $s \ll   n^{1/2}(\log n)^{-1}\{\log (\tilde{d}n)\}^{-1}$ and $ \log \tilde{d} \ll n^{1/10}(\log n)^{-1/2}$.  Analogously, we can  show such    convergence rate also holds for   $ {\rm S}_{2465} $.  
Using the similar arguments for deriving the convergence rate of ${\rm S}_{213}$ in Section \ref{sec:sub-s21}, it holds that
\begin{align*}
	{\rm S}_{2466} \le &~  \max_{j\in[p]}|\hat{\balpha}_{j} - \balpha_{j}|_1^2 \cdot  \max_{k\in[q]}|\hat{\bbeta}_{k} -\bbeta_{k}|_1 \cdot  \max_{r_1,r_2,r_3 \in[m],\,t\in[q]}\bigg|\frac{1}{n}\sum_{i=1}^{n}   W_{i,r_1}   W_{i,r_2}  W_{i,r_3} \delta_{i,t} \bigg|\\
	=&~O_{\rm p}\{s^{3}n^{-3/2}(\log \tilde{d})^{3/2}\}
\end{align*}
provided that $s \ll   n^{1/2}(\log n)^{-1}\{\log (\tilde{d}n)\}^{-1}$ and $ \log \tilde{d} \ll n^{1/10}(\log n)^{-1/2}$.   By \eqref{eq:s248-dec}, we have \eqref{eq:s248} holds.
$\hfill\Box$

\subsubsection{Convergence rate of ${\rm S}_{25}$}\label{sec:sub-s25}
Recall $\ve_{i,j}= U_{i,j}- \balpha_{j}^{\T}\bW_{i}$, $\hat{\ve}_{i,j}= \hat{U}_{i,j}- \hat{\balpha}_{j}^{\T}\hat{\bW}_{i}$, $\delta_{i,l}= V_{i,l}- \bbeta_{l}^{\T}\bW_{i}$ and $\hat{\delta}_{i,l}= \hat{V}_{i,l}- \hat{\bbeta}_{l}^{\T}\hat{\bW}_{i}$. Notice that
\begin{align*}
	{\rm S}_{25}
	\le &~\underbrace{\max_{j,k\in[p],\,l,t\in[q]}\bigg|\frac{1}{n} \sum_{i=1}^{n} (\hat{U}_{i,j}-U_{i,j}) (\hat{U}_{i,k}-U_{i,k}) (\hat{V}_{i,l}-V_{i,l}) (\hat{V}_{i,t}-V_{i,t}) \bigg|}_{{\textrm S}_{251}}\\
	&+ 2\underbrace{\max_{j,k\in[p],\,l,t\in[q]}\bigg|\frac{1}{n} \sum_{i=1}^{n} (\hat{U}_{i,j}-U_{i,j}) (\hat{U}_{i,k}-U_{i,k}) (\hat{V}_{i,l}-V_{i,l}) (\hat{\bbeta}_{t}^{\T}\hat{\bW}_{i} - \bbeta_{t}^{\T}\bW_{i}) \bigg|}_{{\textrm S}_{252}}\\
	&+ \underbrace{\max_{j,k\in[p],\,l,t\in[q]}\bigg|\frac{1}{n} \sum_{i=1}^{n} (\hat{U}_{i,j}-U_{i,j}) (\hat{U}_{i,k}-U_{i,k}) (\hat{\bbeta}_{l}^{\T}\hat{\bW}_{i} - \bbeta_{l}^{\T}\bW_{i}) (\hat{\bbeta}_{t}^{\T}\hat{\bW}_{i} - \bbeta_{t}^{\T}\bW_{i}) \bigg| }_{{\textrm S}_{253}}\\
	&+ 2\underbrace{\max_{j,k\in[p],\,l,t\in[q]}\bigg|\frac{1}{n} \sum_{i=1}^{n} (\hat{U}_{i,j}-U_{i,j}) (\hat{\balpha}_{k}^{\T}\hat{\bW}_{i} -\balpha_{k}^{\T}\bW_{i} ) (\hat{V}_{i,l}-V_{i,l}) (\hat{V}_{i,t}-V_{i,t})\bigg| }_{{\textrm S}_{254}}\\
	&+ 4\underbrace{\max_{j,k\in[p],\,l,t\in[q]}\bigg|\frac{1}{n} \sum_{i=1}^{n} (\hat{U}_{i,j}-U_{i,j}) (\hat{\balpha}_{k}^{\T}\hat{\bW}_{i} -\balpha_{k}^{\T}\bW_{i} ) (\hat{V}_{i,l}-V_{i,l})(\hat{\bbeta}_{t}^{\T}\hat{\bW}_{i} - \bbeta_{t}^{\T}\bW_{i}) \bigg| }_{{\textrm S}_{255} }\\
	&+ 2\underbrace{\max_{j,k\in[p],\,l,t\in[q]}\bigg|\frac{1}{n} \sum_{i=1}^{n} (\hat{U}_{i,j}-U_{i,j}) (\hat{\balpha}_{k}^{\T}\hat{\bW}_{i} -\balpha_{k}^{\T}\bW_{i} ) (\hat{\bbeta}_{l}^{\T}\hat{\bW}_{i} - \bbeta_{l}^{\T}\bW_{i})(\hat{\bbeta}_{t}^{\T}\hat{\bW}_{i} - \bbeta_{t}^{\T}\bW_{i}) \bigg| }_{{\textrm S}_{256} }\\
	&+ \underbrace{\max_{j,k\in[p],\,l,t\in[q]}\bigg|\frac{1}{n} \sum_{i=1}^{n} (\hat{\balpha}_{j}^{\T}\hat{\bW}_{i} -\balpha_{j}^{\T}\bW_{i} ) (\hat{\balpha}_{k}^{\T}\hat{\bW}_{i} -\balpha_{k}^{\T}\bW_{i} ) (\hat{V}_{i,l}-V_{i,l}) (\hat{V}_{i,t}-V_{i,t}) \bigg|}_{{\textrm S}_{257} }\\
	&+ 2 \underbrace{\max_{j,k\in[p],\,l,t\in[q]}\bigg|\frac{1}{n} \sum_{i=1}^{n} (\hat{\balpha}_{j}^{\T}\hat{\bW}_{i} -\balpha_{j}^{\T}\bW_{i} ) (\hat{\balpha}_{k}^{\T}\hat{\bW}_{i} -\balpha_{k}^{\T}\bW_{i} ) (\hat{V}_{i,l}-V_{i,l})(\hat{\bbeta}_{t}^{\T}\hat{\bW}_{i} - \bbeta_{t}^{\T}\bW_{i}) \bigg|}_{{\textrm S}_{258} }\\
	&+ \max_{j,k\in[p],\,l,t\in[q]}\bigg|\frac{1}{n} \sum_{i=1}^{n} (\hat{\balpha}_{j}^{\T}\hat{\bW}_{i} -\balpha_{j}^{\T}\bW_{i} ) (\hat{\balpha}_{k}^{\T}\hat{\bW}_{i} -\balpha_{k}^{\T}\bW_{i} ) \\
	&~~~\underbrace{~~~~~~~~~~~~~~~~~~~~~~~~~~~~~~\times (\hat{\bbeta}_{l}^{\T}\hat{\bW}_{i} - \bbeta_{l}^{\T}\bW_{i})(\hat{\bbeta}_{t}^{\T}\hat{\bW}_{i} - \bbeta_{t}^{\T}\bW_{i}) \bigg| }_{{\textrm S}_{259} }\,.
\end{align*}
Recall $\tilde{d}=p\vee q\vee m$. Notice that ${\rm S}_{251}={\rm R}_{15}$ for ${\rm R}_{15}$ defined in \eqref{eq:R1terms}. By \eqref{eq:r14},   we have
\begin{align}\label{eq:s251}
	{\rm S}_{251}  =O_{\rm p}\{n^{-1/2} (\log n) (\log \tilde{d})^{1/2} \log^{3/2} (\tilde{d}n)\}
\end{align}
provided that $\log \tilde{d} \lesssim n^{5/12} (\log n)^{-1/2}$. As we will show later,
\begin{align}
	&{\rm S}_{252}  =   O_{\rm p}\{s^{1/2}n^{-1/2} (\log n) (\log \tilde{d})^{1/2} \log^{3/2} (\tilde{d}n)\} = {\rm S}_{254}   \label{eq:s252}\,,\\
	&~~{\rm S}_{253}  =  O_{\rm p}\{sn^{-1/2} (\log n) (\log \tilde{d})^{1/2}\log^{3/2}(\tilde{d}n)\} =  {\rm S}_{255}  \label{eq:s254}\,, \\
	&{\rm S}_{256}  =   O_{\rm p}\{s^{3/2}n^{-1/2} (\log n) (\log \tilde{d})^{1/2}\log^{3/2}(\tilde{d}n)\}   = {\rm S}_{258} \label{eq:s258}\,,\\
	&~~~~~~{\rm S}_{257} =  O_{\rm p}\{sn^{-1/2} (\log n) (\log \tilde{d})^{1/2}\log^{3/2}(\tilde{d}n)\} \,,\label{eq:s257}\\
	&~~~~~~{\rm S}_{259}  =  O_{\rm p}\{s^{2}n^{-1/2} (\log n) (\log \tilde{d})^{1/2}\log^{3/2}(\tilde{d}n)\}   \label{eq:s2516}
\end{align}
provided that $s \ll   n^{1/2}(\log n)^{-1}\{\log (\tilde{d}n)\}^{-1}$ and $ \log \tilde{d} \ll n^{1/10}(\log n)^{-1/2}$. Hence, we have \eqref{eq:s25} holds.

\underline{{\it Convergence rates of $ {\rm S}_{252} $ and $ {\rm S}_{254} $}.}
Notice that
\begin{align*}
	{\rm S}_{252}
	\le&~  \underbrace{\max_{j,k\in[p],\,l,t\in[q]}\bigg|\frac{1}{n} \sum_{i=1}^{n} (\hat{U}_{i,j}-U_{i,j}) (\hat{U}_{i,k}-U_{i,k}) (\hat{V}_{i,l}-V_{i,l}) \hat{\bbeta}_{t}^{\T}(\hat{\bW}_{i} - \bW_{i})\bigg|}_{{\textrm S}_{2521}}\\
	&+ \underbrace{\max_{j,k\in[p],\,l,t\in[q]}\bigg|\frac{1}{n} \sum_{i=1}^{n} (\hat{U}_{i,j}-U_{i,j}) (\hat{U}_{i,k}-U_{i,k}) (\hat{V}_{i,l}-V_{i,l}) (\hat{\bbeta}_{t} - \bbeta_{t})^{\T}\bW_{i}\bigg| }_{{\textrm S}_{2522}}\,.
\end{align*}   
Parallel to the   \eqref{eq:s251} and \eqref{eq:uuw-d}, 
by Lemma \ref{lem:coeff}, it then holds that
\begin{align*}
	{\rm S}_{2521} \le &~ \max_{k\in[q]} |\hat{\bbeta}_{k}|_1  \cdot
	\max_{j,k\in[p],\, l\in[q],\, t\in[m]}\bigg|\frac{1}{n} \sum_{i=1}^{n} (\hat{U}_{i,j}-U_{i,j}) (\hat{U}_{i,k}-U_{i,k}) (\hat{V}_{i,l}-V_{i,l}) (\hat{W}_{i,t}-W_{i,t})\bigg|\\
	=&~  O_{\rm p}\{s^{1/2}n^{-1/2} (\log n) (\log \tilde{d})^{1/2} \log^{3/2} (\tilde{d}n)\}\,,\\
	{\rm S}_{2522} \le &~ \max_{k\in[q]} |\hat{\bbeta}_{k} -\bbeta_{k}|_1  \cdot
	\max_{j,k\in[p],\, l\in[q],\, t\in[m]}\bigg|\frac{1}{n} \sum_{i=1}^{n} (\hat{U}_{i,j}-U_{i,j}) (\hat{U}_{i,k}-U_{i,k}) (\hat{V}_{i,l}-V_{i,l}) W_{i,t}\bigg|\\
	=&~ O_{\rm p}\{sn^{-1} (\log n) (\log \tilde{d}) \log^{3/2}(\tilde{d}n)\}
\end{align*}
provided that $s \ll   n^{1/2}(\log n)^{-1}\{\log (\tilde{d}n)\}^{-1}$ and $ \log \tilde{d} \ll n^{-1/10}(\log n)^{-1/2}$, which implies
\begin{align*}
	{\rm S}_{252}  =  O_{\rm p}\{s^{1/2}n^{-1/2} (\log n) (\log \tilde{d})^{1/2} \log^{3/2} (\tilde{d}n)\}
\end{align*}
provided that $s \ll   n^{1/2}(\log n)^{-1}\{\log (\tilde{d}n)\}^{-1}$ and $ \log \tilde{d} \ll n^{1/10}(\log n)^{-1/2}$. Analogously, we can also show such convergence rate holds for $ {\rm S}_{254}$. Hence, we have \eqref{eq:s252}  holds.

\underline{{\it Convergence rates of  $ {\rm S}_{253} $, ${\rm S}_{255} $ and $ {\rm S}_{257} $}.}
Notice that
\begin{align*}
	{\rm S}_{253}\le &~ \underbrace{\max_{j,k\in[p],\,l,t\in[q]}\bigg|\frac{1}{n} \sum_{i=1}^{n} (\hat{U}_{i,j}-U_{i,j}) (\hat{U}_{i,k}-U_{i,k}) \hat{\bbeta}_{l}^{\T}(\hat{\bW}_{i} - \bW_{i}) (\hat{\bW}_{i} -\bW_{i})^{\T}\hat{\bbeta}_{t} \bigg|}_{{\textrm S}_{2531}}\\
	&+ \underbrace{\max_{j,k\in[p],\,l,t\in[q]}\bigg|\frac{1}{n} \sum_{i=1}^{n} (\hat{U}_{i,j}-U_{i,j}) (\hat{U}_{i,k}-U_{i,k}) \hat{\bbeta}_{l}^{\T}(\hat{\bW}_{i} - \bW_{i}) \bW_{i}^{\T} (\hat{\bbeta}_{t} -\bbeta_{t}) \bigg|}_{{\textrm S}_{2532}}\\
	&+ \underbrace{\max_{j,k\in[p],\,l,t\in[q]}\bigg|\frac{1}{n} \sum_{i=1}^{n} (\hat{U}_{i,j}-U_{i,j}) (\hat{U}_{i,k}-U_{i,k}) (\hat{\bbeta}_{l}  -\bbeta_{l})^{\T} \bW_{i}(\hat{\bW}_{i} - \bW_{i})^{\T} \hat{\bbeta}_{t} \bigg|}_{{\textrm S}_{2533} }\\
	&+ \underbrace{\max_{j,k\in[p],\,l,t\in[q]}\bigg|\frac{1}{n} \sum_{i=1}^{n} (\hat{U}_{i,j}-U_{i,j}) (\hat{U}_{i,k}-U_{i,k}) (\hat{\bbeta}_{l}  -\bbeta_{l})^{\T} \bW_{i} \bW_{i}^{\T} (\hat{\bbeta}_{t} -\bbeta_{t}) \bigg|}_{{\textrm S}_{2534}}\,.
\end{align*}
Parallel to the   \eqref{eq:s251} and \eqref{eq:uuw-d},  by Lemma \ref{lem:coeff}, it holds that
\begin{align*}
	{\rm S}_{2531}   \le&~\max_{k\in[q]}|\hat{\bbeta}_k|_1^2 \cdot \max_{j,k\in[p], \,l,t\in[m]}\bigg|\frac{1}{n} \sum_{i=1}^{n} (\hat{U}_{i,j}-U_{i,j}) (\hat{U}_{i,k}-U_{i,k}) (\hat{W}_{i,l}-W_{i,l}) (\hat{W}_{i,t}-W_{i,t})\bigg|\\
	=&~   O_{\rm p}\{sn^{-1/2} (\log n) (\log \tilde{d})^{1/2}\log^{3/2}(\tilde{d}n)\}\,,\\
	{\rm S}_{2532}  \le &~ \max_{k\in[q]}|\hat{\bbeta}_k|_1 \cdot \max_{k\in[q]}|\hat{\bbeta}_k-\bbeta_{k}|_1 \\
	&~~~~~\times \max_{j,k\in[p], \,l,t\in[m]}\bigg|\frac{1}{n} \sum_{i=1}^{n} (\hat{U}_{i,j}-U_{i,j}) (\hat{U}_{i,k}-U_{i,k}) (\hat{W}_{i,l}-W_{i,l})W_{i,t}\bigg|\\
	=&~   O_{\rm p}\{s^{3/2}n^{-1} (\log n) (\log \tilde{d}) \log^{3/2}(\tilde{d}n)\}  \,,\\
	{\rm S}_{2534}  \le &~ \max_{k\in[q]}|\hat{\bbeta}_k-\bbeta_{k}|_1^2 \cdot \max_{j,k\in[p],\, l,t\in[m]}\bigg|\frac{1}{n} \sum_{i=1}^{n} (\hat{U}_{i,j}-U_{i,j}) (\hat{U}_{i,k}-U_{i,k})  W_{i,l}W_{i,t}\bigg|\\
	=&~  O_{\rm p}\{s^{2}n^{-3/2} (\log n) (\log \tilde{d})^{3/2} \log^{3/2}(\tilde{d}n)\}
\end{align*}
provided that $s \ll   n^{1/2}(\log n)^{-1}\{\log (\tilde{d}n)\}^{-1}$ and $ \log \tilde{d} \ll n^{1/10}(\log n)^{-1/2}$.  Analogously, we can  show such derived convergence rate of ${\rm S}_{2532}$ also holds for ${\rm S}_{2533}$. Hence, we have
\begin{align*}
	{\rm S}_{253}  = O_{\rm p}\{sn^{-1/2} (\log n) (\log \tilde{d})^{1/2}\log^{3/2}(\tilde{d}n)\}
\end{align*}
provided that $s \ll   n^{1/2}(\log n)^{-1}\{\log (\tilde{d}n)\}^{-1}$ and $ \log \tilde{d} \ll n^{1/10}(\log n)^{-1/2}$. Analogously, we can also show such convergence rate holds for $ {\rm S}_{255} $ and $ {\rm S}_{257} $. Hence, we have \eqref{eq:s254} and \eqref{eq:s257}  hold.

\underline{{\it Convergence rates of  $ {\rm S}_{256} $ and ${\rm S}_{258}$}.}
Notice that
\begin{align}\label{eq:s256-dec}
	{\rm S}_{256}\le&~ \underbrace{\max_{j,k\in[p],\,l,t\in[q]}\bigg|\frac{1}{n}\sum_{i=1}^{n} \hat{\balpha}_{k}^{\T}(\hat{\bW}_i-\bW_i) \hat{\bbeta}_{l}^{\T}(\hat{\bW}_i-\bW_i) \hat{\bbeta}_{t}^{\T}(\hat{\bW}_i-\bW_i)(\hat{U}_{i,j}-U_{i,j})\bigg|}_{{\textrm S}_{2561}} \notag\\
	&+2 \underbrace{\max_{j,k\in[p],\,l,t\in[q]}\bigg|\frac{1}{n}\sum_{i=1}^{n} \hat{\balpha}_{k}^{\T}(\hat{\bW}_i-\bW_i) \hat{\bbeta}_{l}^{\T}(\hat{\bW}_i-\bW_i) (\hat{\bbeta}_{t}-\bbeta_{t})^{\T}\bW_i(\hat{U}_{i,j}-U_{i,j})\bigg| }_{{\textrm S}_{2562} } \notag\\
	&+\underbrace{\max_{j,k\in[p],\,l,t\in[q]}\bigg|\frac{1}{n}\sum_{i=1}^{n} \hat{\balpha}_{k}^{\T}(\hat{\bW}_i-\bW_i) (\hat{\bbeta}_{l}-\bbeta_{l})^{\T}\bW_i (\hat{\bbeta}_{t}-\bbeta_{t})^{\T}\bW_i(\hat{U}_{i,j}-U_{i,j})\bigg| }_{{\textrm S}_{2563}}\\
	&+\underbrace{\max_{j,k\in[p],\,l,t\in[q]}\bigg|\frac{1}{n}\sum_{i=1}^{n} (\hat{\balpha}_{k}  -\balpha_{k})^{\T} \bW_i  \hat{\bbeta}_{l}^{\T}(\hat{\bW}_i-\bW_i) \hat{\bbeta}_{t}^{\T}(\hat{\bW}_i-\bW_i)(\hat{U}_{i,j}-U_{i,j})\bigg|}_{{\textrm S}_{2564}} \notag\\
	&+2 \underbrace{\max_{j,k\in[p],\,l,t\in[q]}\bigg|\frac{1}{n}\sum_{i=1}^{n} (\hat{\balpha}_{k}  -\balpha_{k})^{\T} \bW_i \hat{\bbeta}_{l}^{\T}(\hat{\bW}_i-\bW_i) (\hat{\bbeta}_{t}-\bbeta_{t})^{\T}\bW_i(\hat{U}_{i,j}-U_{i,j}) \bigg|}_{{\textrm S}_{2565}} \notag\\
	&+\underbrace{\max_{j,k\in[p],\,l,t\in[q]}\bigg|\frac{1}{n}\sum_{i=1}^{n} (\hat{\balpha}_{k}  -\balpha_{k})^{\T} \bW_i (\hat{\bbeta}_{l}-\bbeta_{l})^{\T}\bW_i (\hat{\bbeta}_{t}-\bbeta_{t})^{\T}\bW_i(\hat{U}_{i,j}-U_{i,j}) \bigg|}_{{\textrm S}_{2566}}\,.\notag
\end{align}
Parallel to the   \eqref{eq:s251} and \eqref{eq:uuw-d},
by Lemma \ref{lem:coeff}, we have
\begin{align*}
	{\rm S}_{2561}   \le&~\max_{j\in[p]}|\hat{\balpha}_{j}|_1 \cdot  \max_{k\in[q]}|\hat{\bbeta}_{k}|_1^2 \notag\\
	&~~~\times \max_{j\in[p],\,r_1,r_2,r_3 \in[m]}\bigg|\frac{1}{n}\sum_{i=1}^{n} (\hat{W}_{i,r_1}- W_{i,r_1}) (\hat{W}_{i,r_2}- W_{i,r_2}) (\hat{W}_{i,r_3}- W_{i,r_3}) (\hat{U}_{i,j}-U_{i,j})\bigg| \notag\\
	=&~ O_{\rm p}\{s^{3/2}n^{-1/2} (\log n) (\log \tilde{d})^{1/2}\log^{3/2}(\tilde{d}n)\}\,, \label{eq:s2581}\\
	{\rm S}_{2562}  \le &~ \max_{j\in[p]}|\hat{\balpha}_{j}|_1 \cdot   \max_{k\in[q]}|\hat{\bbeta}_{k}|_1 \cdot \max_{k\in[q]}|\hat{\bbeta}_{k} - \bbeta_{k}|_1  \notag\\
	&~~~~~\times \max_{j\in[p],\,r_1,r_2,r_3 \in[m]}\bigg|\frac{1}{n}\sum_{i=1}^{n}  W_{i,r_1} (\hat{W}_{i,r_2}- W_{i,r_2}) (\hat{W}_{i,r_3}- W_{i,r_3}) (\hat{U}_{i,j}-U_{i,j})\bigg|  \notag\\
	=&~  O_{\rm p}\{s^2n^{-1} (\log n) (\log \tilde{d}) \log^{3/2}(\tilde{d}n)\}\notag\,,\\
	{\rm S}_{2563}  \le&~ \max_{j\in[p]}|\hat{\balpha}_{j}|_1 \cdot  \max_{k\in[q]}|\hat{\bbeta}_{k} - \bbeta_{k}|_1^2  \\
	&~~~~~\times \max_{j\in[p],\,r_1,r_2,r_3 \in[m]}\bigg|\frac{1}{n}\sum_{i=1}^{n}  W_{i,r_1}  W_{i,r_2}  (\hat{W}_{i,r_3}- W_{i,r_3}) (\hat{U}_{i,j}-U_{i,j})\bigg|  \notag\\
	=&~   O_{\rm p}\{s^{5/2}n^{-3/2} (\log n) (\log \tilde{d})^{3/2} \log^{3/2}(\tilde{d}n)\}
\end{align*}
provided that $s \ll   n^{1/2}(\log n)^{-1}\{\log (\tilde{d}n)\}^{-1}$ and $ \log \tilde{d} \ll n^{1/10}(\log n)^{-1/2}$. Analogously, we can also show such derived convergence rates of $ {\rm S}_{2562} $ and ${\rm S}_{2563}$  hold for $ {\rm S}_{2564} $ and $ {\rm S}_{2565}$, respectively.  Parallel to \eqref{eq:uh-h-1}, by Lemma \ref{lem:coeff} again, we have
\begin{align*}
	{\rm S}_{2566} \le &~\max_{j\in[p]}|\hat{\balpha}_{j}-\balpha_{k}|_1  \cdot \max_{k\in[q]}|\hat{\bbeta}_{k} - \bbeta_{k}|_1^2 \cdot  \max_{j\in[p],\,r_1,r_2,r_3 \in[m]}\bigg|\frac{1}{n}\sum_{i=1}^{n}  W_{i,r_1}  W_{i,r_2}  W_{i,r_3}  (\hat{U}_{i,j}-U_{i,j})\bigg|  \notag\\
	=&~ O_{\rm p}\{s^{3}n^{-2} (\log n) (\log \tilde{d})^{2} \log^{3/2}(\tilde{d}n)\}
\end{align*}
provided that $s \ll   n^{1/2}(\log n)^{-1}\{\log (\tilde{d}n)\}^{-1}$ and $ \log \tilde{d} \ll n^{1/10}(\log n)^{1/2}$.  Hence, by \eqref{eq:s256-dec}, we have 
\begin{align*}
	{\rm S}_{256}  =   O_{\rm p}\{s^{3/2}n^{-1/2} (\log n) (\log \tilde{d})^{1/2}\log^{3/2}(\tilde{d}n)\}
\end{align*}
provided that $s \ll   n^{1/2}(\log n)^{-1}\{\log (\tilde{d}n)\}^{-1}$ and $ \log \tilde{d} \ll n^{1/10}(\log n)^{-1/2}$. Analogously, we can also show such convergence rate holds for $ {\rm S}_{258}$. Then \eqref{eq:s258}  holds.

\underline{{\it Convergence rate of  $ {\rm S}_{259} $}.}
Notice that
\begin{align*}
	{\rm S}_{259}\le &~ \underbrace{\max_{j,k\in[p],\,l,t\in[q]}\bigg|\frac{1}{n}\sum_{i=1}^{n}  \hat{\balpha}_{j}^{\T}(\hat{\bW}_{i} -\bW_{i} ) \hat{\balpha}_{k}^{\T}(\hat{\bW}_{i} -\bW_{i} )  \hat{\bbeta}_{l}^{\T}(\hat{\bW}_{i} - \bW_{i}) \hat{\bbeta}_{t}^{\T}(\hat{\bW}_{i} - \bW_{i})\bigg| }_{{\textrm S}_{2591}}\\
	&+ 2 \underbrace{\max_{j,k\in[p],\,l,t\in[q]}\bigg|\frac{1}{n}\sum_{i=1}^{n}  \hat{\balpha}_{j}^{\T}(\hat{\bW}_{i} -\bW_{i} ) \hat{\balpha}_{k}^{\T}(\hat{\bW}_{i} -\bW_{i} )  \hat{\bbeta}_{l}^{\T}(\hat{\bW}_{i} - \bW_{i}) (\hat{\bbeta}_{t} -\bbeta_{t})^{\T} \bW_{i}\bigg|}_{{\textrm S}_{2592}} \\
	&+ \underbrace{\max_{j,k\in[p],\,l,t\in[q]}\bigg|\frac{1}{n}\sum_{i=1}^{n}  \hat{\balpha}_{j}^{\T}(\hat{\bW}_{i} -\bW_{i} ) \hat{\balpha}_{k}^{\T}(\hat{\bW}_{i} -\bW_{i} )  (\hat{\bbeta}_{l} -\bbeta_{l})^{\T} \bW_{i}(\hat{\bbeta}_{t} -\bbeta_{t})^{\T} \bW_{i}\bigg|}_{{\textrm S}_{2593}}\\
	&+2 \underbrace{\max_{j,k\in[p],\,l,t\in[q]}\bigg|\frac{1}{n}\sum_{i=1}^{n}  \hat{\balpha}_{j}^{\T}(\hat{\bW}_{i} -\bW_{i} ) (\hat{\balpha}_{k}-\balpha_{k})^{\T} \bW_{i}   \hat{\bbeta}_{l}^{\T}(\hat{\bW}_{i} - \bW_{i}) \hat{\bbeta}_{t}^{\T}(\hat{\bW}_{i} - \bW_{i})\bigg|}_{{\textrm S}_{2594}}\\
	&+ 4 \underbrace{\max_{j,k\in[p],\,l,t\in[q]}\bigg|\frac{1}{n}\sum_{i=1}^{n}  \hat{\balpha}_{j}^{\T}(\hat{\bW}_{i} -\bW_{i} ) (\hat{\balpha}_{k}-\balpha_{k})^{\T} \bW_{i}  \hat{\bbeta}_{l}^{\T}(\hat{\bW}_{i} - \bW_{i}) (\hat{\bbeta}_{t} -\bbeta_{t})^{\T} \bW_{i}\bigg|}_{{\textrm S}_{2595}} \\
	&+ 2 \underbrace{\max_{j,k\in[p],\,l,t\in[q]}\bigg|\frac{1}{n}\sum_{i=1}^{n}  \hat{\balpha}_{j}^{\T}(\hat{\bW}_{i} -\bW_{i} ) (\hat{\balpha}_{k}-\balpha_{k})^{\T} \bW_{i}  (\hat{\bbeta}_{l} -\bbeta_{l})^{\T} \bW_{i}(\hat{\bbeta}_{t} -\bbeta_{t})^{\T} \bW_{i}\bigg|}_{{\textrm S}_{2596}}\\
	&+\underbrace{\max_{j,k\in[p],\,l,t\in[q]}\bigg|\frac{1}{n}\sum_{i=1}^{n}  (\hat{\balpha}_{j} -\balpha_{j})^{\T} \bW_{i}(\hat{\balpha}_{k}-\balpha_{k})^{\T} \bW_{i}   \hat{\bbeta}_{l}^{\T}(\hat{\bW}_{i} - \bW_{i}) \hat{\bbeta}_{t}^{\T}(\hat{\bW}_{i} - \bW_{i})\bigg|}_{{\textrm S}_{2597}}\\
	&+ 2 \underbrace{\max_{j,k\in[p],\,l,t\in[q]}\bigg|\frac{1}{n}\sum_{i=1}^{n}  (\hat{\balpha}_{j} -\balpha_{j})^{\T} \bW_{i}(\hat{\balpha}_{k}-\balpha_{k})^{\T} \bW_{i}  \hat{\bbeta}_{l}^{\T}(\hat{\bW}_{i} - \bW_{i}) (\hat{\bbeta}_{t} -\bbeta_{t} )^{\T} \bW_{i}\bigg|}_{{\textrm S}_{2598}} \\
	&+ \underbrace{\max_{j,k\in[p],\,l,t\in[q]}\bigg|\frac{1}{n}\sum_{i=1}^{n}  (\hat{\balpha}_{j} -\balpha_{j})^{\T} \bW_{i} (\hat{\balpha}_{k}-\balpha_{k})^{\T} \bW_{i}  (\hat{\bbeta}_{l} -\bbeta_{l} )^{\T} \bW_{i}(\hat{\bbeta}_{t} -\bbeta_{t} )^{\T} \bW_{i}\bigg|}_{{\textrm S}_{2599}}
\end{align*}
Applying the similar arguments for deriving the convergence rates of $ {\rm S}_{2561} $ and $ {\rm S}_{2562} $, we have
\begin{equation*}
	\begin{split}
		&~~~{\rm S}_{2591}  =  O_{\rm p}\{s^{2}n^{-1/2} (\log n) (\log \tilde{d})^{1/2}\log^{3/2}(\tilde{d}n)\}\,,  \\
		&{\rm S}_{2592}   =   O_{\rm p}\{s^{5/2}n^{-1} (\log n) (\log \tilde{d}) \log^{3/2}(\tilde{d}n)\} ={\rm S}_{2594} 
	\end{split}
\end{equation*}
provided that $s \ll   n^{1/2}(\log n)^{-1}\{\log (\tilde{d}n)\}^{-1}$ and $ \log \tilde{d} \ll n^{1/10}(\log n)^{-1/2}$.  
Using the similar arguments for deriving the convergence rates of $ {\rm S}_{2563}$ and $ {\rm S}_{2566}$, it holds that
\begin{equation*}
	\begin{split}
		&{\rm S}_{2593} =    O_{\rm p}\{s^{3}n^{-3/2} (\log n) (\log \tilde{d})^{3/2} \log^{3/2}(\tilde{d}n)\}={\rm S}_{2595}\,,\\
		&~~~~~~{\rm S}_{2597} =    O_{\rm p}\{s^{3}n^{-3/2} (\log n) (\log \tilde{d})^{3/2} \log^{3/2}(\tilde{d}n)\}\,,\\
		&~~{\rm S}_{2596} =  O_{\rm p}\{s^{7/2}n^{-2} (\log n) (\log \tilde{d})^{2} \log^{3/2}(\tilde{d}n)\} ={\rm S}_{2598} 
	\end{split}
\end{equation*}
provided that $s \ll   n^{1/2}(\log n)^{-1}\{\log (\tilde{d}n)\}^{-1}$ and $ \log \tilde{d} \ll n^{1/10}(\log n)^{-1/2}$.  
Using the similar arguments for deriving the convergence rate of ${\rm S}_{213}$ in Section \ref{sec:sub-s21}, we have
\begin{align*} 
	{\rm S}_{2599}  \le&~  \max_{j\in[p]}|\hat{\balpha}_{j} - \balpha_{j}|_1^2 \cdot \max_{k\in[q]}|\hat{\bbeta}_{k} -\bbeta_{k}|_1^2  \cdot \max_{r_1,r_2,r_3,r_4 \in[m] }\bigg|\frac{1}{n}\sum_{i=1}^{n}   W_{i,r_1}   W_{i,r_2}  W_{i,r_3} W_{i,r_4} \bigg|  \notag \\
	=&~  O_{\rm p} \{s^{4} n^{-2} (\log \tilde{d})^{2} \}
\end{align*}
provided that $s \ll   n^{1/2}(\log n)^{-1}\{\log (\tilde{d}n)\}^{-1}$ and $ \log \tilde{d} \ll n^{1/10}(\log n)^{-1/2}$.  Hence, we have \eqref{eq:s2516} holds.
$\hfill\Box$

\subsection{Convergence rate of ${\rm S}_3$}\label{sec:sub-s3}
Notice that
\begin{align*}
	{\rm S}_{3} \le&~2\max_{j,k\in[p],\,l,t\in[q]}\bigg|\frac{1}{n}\sum_{i=1}^{n}\big\{\ve_{i,j}\delta_{i,l}-\mathbb{E}(\ve_{i,j}\delta_{i,l})\big\}\mathbb{E}(\ve_{i,k}\delta_{i,t})\bigg|\\
	&+\max_{j\in[p],\,l\in[q]}\bigg| \frac{1}{n}\sum_{i=1}^{n} \big\{\ve_{i,j}\delta_{i,l}-\mathbb{E}(\ve_{i,j}\delta_{i,l}) \big\} \bigg|^2\,.
\end{align*}
By \eqref{eq:eps-delta-tail}, we have $\max_{k\in[p],\,t\in[q]}|\mathbb{E}(\ve_{i,k}\delta_{i,t})|=O(1)$ and $ \var(\ve_{i,j}\delta_{i,k}) \le C$. Recall $\tilde{d}=p\vee q\vee m$. Using the similar arguments for the derivation of \eqref{eq:uv-bound-2}, it holds that   
\begin{align}\label{eq:ve-delta}
	\max_{j\in[p],\,k\in[q]}\bigg|\frac{1}{n}\sum_{i=1}^{n} \big\{\ve_{i,j}\delta_{i,k} - \mathbb{E}(\ve_{i,j}\delta_{i,k}) \big\}\bigg|=O_{\rm p}\big\{n^{-1/2}(\log \tilde{d})^{1/2}\big\}
\end{align}
provided that $\log \tilde{d} \lesssim n^{1/3}$.  Then \eqref{eq:s3} holds.
$\hfill\Box$

\subsection{Convergence rate of ${\rm S}_4$}\label{sec:sub-s4}
Notice that
\begin{align}\label{eq:s4-dep}
	{\rm S}_{4} \le&~2\max_{j,k\in[p],\,l,t\in[q]}\bigg|\bigg\{\frac{1}{n}\sum_{i=1}^{n}(\hat{\ve}_{i,j}\hat{\delta}_{i,l}-\ve_{i,j}\delta_{i,l})\bigg\}\bigg(\frac{1}{n}\sum_{i=1}^{n}\ve_{i,k}\delta_{i,t} \bigg)\bigg| \notag \\
	&+ \max_{j\in[p],\,l\in[q]}\bigg|\frac{1}{n}\sum_{i=1}^{n}(\hat{\ve}_{i,j}\hat{\delta}_{i,l}-\ve_{i,j}\delta_{i,l}) \bigg|^2\,.
\end{align}
By Lemmas \ref{lem:epsdeth-epsdet} and \ref{lem:delta-t-4}, we have 
\begin{align*}
	\frac{1}{n}\sum_{i=1}^{n}\big(\hat{\ve}_{i,j}\hat{\delta}_{i,k} - \ve_{i,j}\delta_{i,k}\big) =&~\frac{\sqrt{2\pi}(n-1)}{n(n+1)}\sum_{s=1}^{n}\big\{\tilde{\delta}_{44,k}(U_{s,j}) + \tilde{\delta}_{54,j}(V_{s,k})\big\} \\
	&+ {\rm Rem}_{1}(j,k) + {\rm Rem}_{2}(j,k)\,,
\end{align*}  
with \begin{align*}
	&\max_{j\in[p],\, k\in[q]}|{\rm Rem}_1(j,k)| =  O_{\rm p} \{s  n^{-7/10} \log^{3/2}(\tilde{d}n)\} + O_{\rm p} \{s^{1/2} n^{-13/20}(\log n)^{-3/4} \log(\tilde{d}n)\}\,,\\
	&~~~~~~~~~~~~~~~~~~~\max_{j\in[p],\,k\in[q]}|{\rm Rem}_{2}(j,k)| = O_{\rm p} \{n^{-4/5} (\log n)^{1/4}(\log \tilde{d})^{1/2} \}
\end{align*} 
provided that  $s \lesssim n^{3/10}(\log \tilde{d})^{1/2}$ and $ \log  \tilde{d}  \ll n^{1/10}(\log n)^{-1/2}$, where
\begin{align*}
	&\tilde{\delta}_{44,k}(U_{s,j}) = \mathbb{E} \big[e^{U_{i,j}^2/2} \big\{I(U_{s,j}\le U_{i,j})-\Phi(U_{i,j})\big\}\delta_{i,k}I\{|U_{i,j}|\le \sqrt{3(\log n)/5}\} I(|\delta_{i,k}|\le \tilde{M}) \,\big|\,U_{s,j} \big] \,,\\
	&\tilde{\delta}_{54,j}(V_{s,k}) = \mathbb{E} \big[e^{V_{i,k}^2/2} \big\{I(V_{s,k}\le V_{i,k})-\Phi(V_{i,k})\big\} \varepsilon_{i,j} 
	I\{|V_{i,k}|\le \sqrt{3(\log n)/5}\}I(|\varepsilon_{i,j}|\le \tilde{M}) \,\big|\,V_{s,k} \big] 
\end{align*}
with  $i\ne s$ and $\tilde{M}=\sqrt{9(\log n)/(10\tilde{c})}$ for $\tilde{c} =(1\wedge c_7)/4$. 

Recall $U_{i,j} \sim \mathcal{N}(0,1)$. Since $(U_{i,j}, \delta_{i,k})$ and $(U_{s,j}, \delta_{s,k})$ are independent for any $s\ne i$,
\begin{align*}
	&~\mathbb{E}\{\tilde{\delta}_{44,k}(U_{s,j})\} \notag\\
	=&~ \mathbb{E} \big[e^{U_{i,j}^2/2}  \big\{I(U_{s,j}\le U_{i,j})-\Phi(U_{i,j})\big\}\delta_{i,k}I\{|U_{i,j}|\le \sqrt{3(\log n)/5}\}I(|\delta_{i,k}|\le  \tilde{M}) \big]\\
	=&~ \mathbb{E} \big[e^{U_{i,j}^2/2} I\{|U_{i,j}|\le \sqrt{3(\log n)/5}\}\delta_{i,k}I(|\delta_{i,k}|\le \tilde{M})  \mathbb{E}\big\{I(U_{s,j}\le U_{i,j})-\Phi(U_{i,j}) \,\big|\, U_{i,j},\delta_{i,k}\big\}\big] =0\,. \notag
\end{align*}
Analogously, $\mathbb{E}\{\tilde{\delta}_{54,j}(V_{s,k})\}=0$.
Notice that $\max_{s\in[n],\,j\in[p],\, k\in[q]}|\tilde{\delta}_{44,k}(U_{s,j})| \le 3\sqrt{3}(\log n)/(5\sqrt{\tilde{c}\pi})$ and $\max_{s\in[n],\,j\in[p],\, k\in[q]}|\tilde{\delta}_{54,j}(V_{s,k})| \le 3\sqrt{3}(\log n)/(5\sqrt{\tilde{c}\pi})$. By Bonferroni inequality and  Hoeffding's inequality,  it holds that
\begin{align}\label{eq:tail-delta-4}
	&\mathbb{P}\bigg[\max_{j\in[p],\,k\in[q]}\bigg|\frac{\sqrt{2\pi}(n-1)}{n(n+1)}\sum_{s=1}^{n}\big\{\tilde{\delta}_{44,k}(U_{s,j})+\tilde{\delta}_{54,j}(V_{s,k})\big\} \bigg| > x\bigg] \notag \\
	&~~~~~~~~~~~~~~\le 2pq \exp\bigg\{-\frac{25\tilde{c} nx^2}{432(\log n)^2}\bigg\}
\end{align}
for any $x>0$. 
Recall $\tilde{d}=p\vee q\vee m$. We have
\begin{align*}
	\max_{j\in[p],\,k\in[q]}\bigg|\frac{\sqrt{2\pi}(n-1)}{n(n+1)}\sum_{s=1}^{n}\big\{\tilde{\delta}_{44,k}(U_{s,j}) + \tilde{\delta}_{54,j}(V_{s,k}) \big\}\bigg| = O_{\rm p} \{n^{-1/2}(\log n)(\log \tilde{d})^{1/2}\} \,.
\end{align*}
Hence, it holds that
\begin{align*} 
	\max_{j\in[p],\,l\in[q]}\bigg|\frac{1}{n}\sum_{i=1}^{n}(\hat{\ve}_{i,j}\hat{\delta}_{i,l}-\ve_{i,j}\delta_{i,l}) \bigg| =&~ O_{\rm p} \{s^{1/2} n^{-13/20}(\log n)^{-3/4} \log(\tilde{d}n)\} \\
	&  +  O_{\rm p} \{s  n^{-7/10} \log^{3/2}(\tilde{d}n)\} +  O_{\rm p} \{n^{-1/2}(\log n)(\log \tilde{d})^{1/2}\} 
\end{align*}
provided that  $s \lesssim n^{3/10}(\log \tilde{d})^{1/2}$ and $ \log  \tilde{d}  \ll n^{1/10}(\log n)^{-1/2}$. Due to  $\max_{k\in[p],\,t\in[q]}|\mathbb{E}(\ve_{i,k}\delta_{i,t})| =O(1)$, by \eqref{eq:ve-delta}, it holds that $\max_{k\in[p],\,t\in[q]}|n^{-1}\sum_{i=1}^{n}\ve_{i,k}\delta_{i,t}|=O_{\rm p}(1)$ provided that $\log \tilde{d}\lesssim n^{1/3}$. Hence, by \eqref{eq:s4-dep},  we have \eqref{eq:s4} holds.
$\hfill\Box$

\setcounter{table}{0}
\renewcommand{\thetable}{S\arabic{table}}
\setcounter{figure}{0}
\renewcommand{\thefigure}{S\arabic{figure}}


\section{Real Data Analysis}\label{sec:real-data}

In this section, we use the proposed testing procedures to analyze the dependence and conditional dependence structures in the S\&P 500 stocks. The dataset is downloaded from the Wharton Research Data Services  (WRDS) database on the website {\url{https://wrds-www.wharton.upenn.edu/}},  which consists of the daily closed prices of stocks.  We consider two periods in our analysis: (i) from 1 January 2016 to 31 December 2018 (754 trading days, before COVID-19 period), and (ii) from 1 January 2020 to 31 December 2022 (756 trading days,   during/after COVID-19 period). We select 485 stocks that do not have missing values in these two periods. Based on the Global Industry Classification Standard, these stocks can be classified into 11 sectors: Communication Services (21 stocks), Consumer Discretionary (63 stocks), Consumer Staples (30 stocks), Energy (23 stocks), Financials (63 stocks), Health Care (58 stocks), Industrials (73 stocks),  Information Technology (73 stocks), Materials (23 stocks),  Real Estate (29 stocks), and Utilities (29 stocks).  See Table \ref{tab:name-sec} for detailed information on these sectors.  We are interested in comparing the dependence structures among all the sectors in the above-mentioned two periods, which can be used to understand the impact of the COVID-19 pandemic on the financial market. 

\begin{figure}[htbp] 
\centering 
\subfigure[Before COVID-9 period]{
\label{Fig.sub.1}
\includegraphics[width=0.48\textwidth]{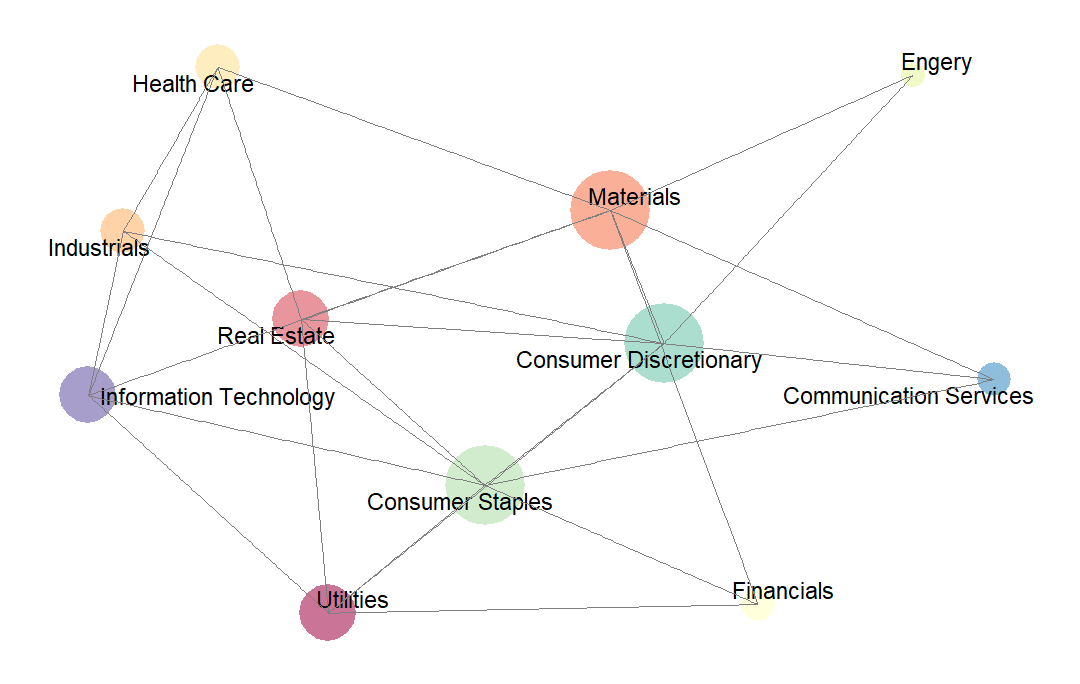}}
\subfigure[During/after  COVID-9 period]{
\label{Fig.sub.2}
\includegraphics[width=0.48\textwidth]{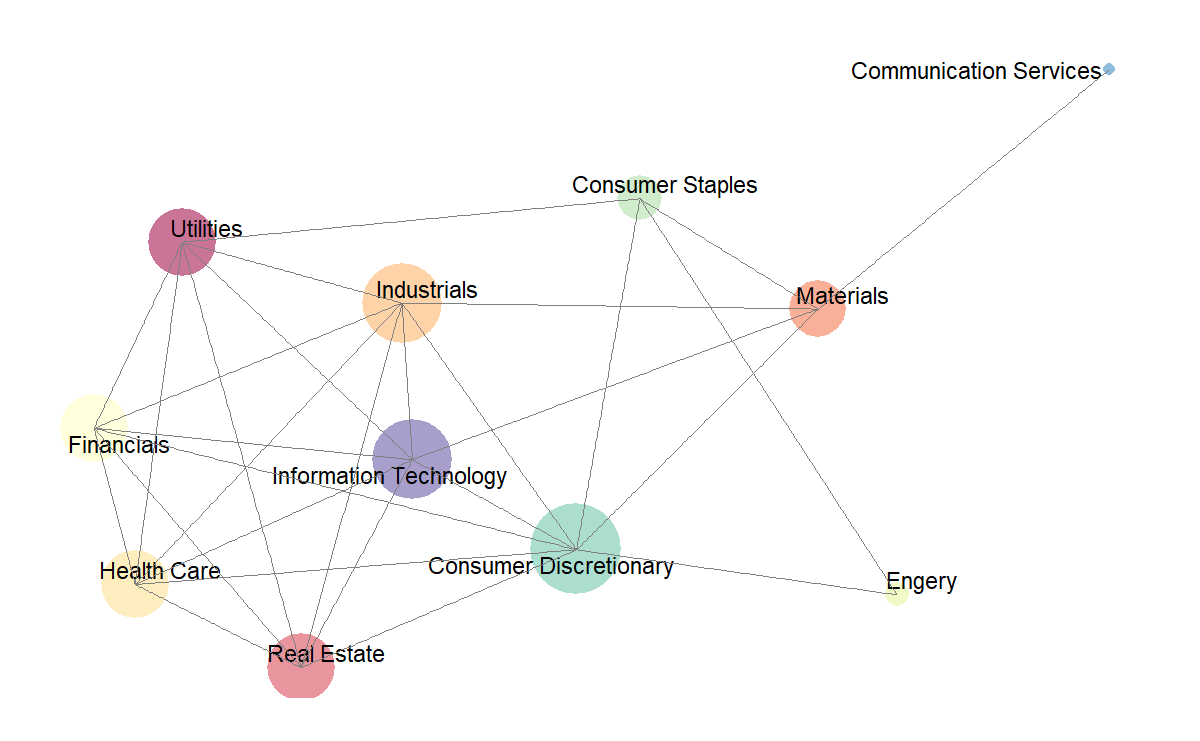}}
\caption{Conditional dependence network of the 11 sectors (denoted by the nodes)  obtained by using the CI-FNN test with Rademacher multiplier. There exists  an edge between two nodes if the conditional independence test between them is significant. The sizes of the nodes are proportional to their degrees.}
\label{Fig.cin-graph}
\end{figure}

Denote by $h=1$ and $h=2$, respectively, the before COVID-19 period and the during/after COVID-19 period. Let ${\bf R}^{(h)}_{s}$ be the $p_s$-dimensional daily stock return vector of the $s$-th sector.  The daily stock return is defined as the log difference of the daily closing prices. For each given $h\in\{1,2\}$ and $(s_1,s_2)$ with $s_1<s_2$, we first test the independence between $\mathbf{R}^{(h)}_{s_1}$ and $\mathbf{R}^{(h)}_{s_2}$. 
Given the observed closed prices of the stocks, we can obtain 11 sequences $\{{\bf R}^{(h)}_{1,t}\},\ldots, \{{\bf R}^{(h)}_{11,t}\}$ for each $h\in \{1,2\}$. Based on the standard financial theory that the stock prices follow geometric Brownian Motions, each $\{{\bf R}^{(h)}_{s,t}\}$ is an i.i.d. sequence. 
For each given $h\in\{1,2\}$, we apply the proposed independence test with Rademacher multiplier to these $55$ hypothesis testing problems and find that all the associated 55 p-values are smaller than $0.0001$.  Applying the BH procedure \citep{Benjamini1995} to the 55 p-values with controlling the false discovery rate (FDR) at the level 0.01  for $h = 1$ and $h=2$, respectively, we know that all 11 sectors are pairwise dependent. Recall that the FDR is defined as the expected ratio of the number of false discoveries to the total number of rejections of the null. Controlling the FDR at the level 0.01 here ensures the expected number of false discoveries does not exceed $55\times 0.01 =0.55 < 1$ asymptotically. This motivates us to further investigate the conditional independence structure among the 11 sectors. More specifically, we can use a network with 11 nodes to characterize such conditional independence structure, where each node represents a sector and there exists an edge between the nodes $s_1$ and $s_2$ if the null hypothesis of the following hypothesis testing problem is rejected: 
\begin{align*}
	\mathbb{H}^{(h)}_{0,(s_1, s_2)}: \mathbf{R}^{(h)}_{s_1}  \Vbar \mathbf{R}^{(h)}_{s_2} \,|\, \mathbf{R}^{(h)}_{-(s_1,s_2)} ~~~\mbox{ versus} ~~~ \mathbb{H}^{(h)}_{1,(s_1, s_2)}:  \mathbf{R}^{(h)}_{s_1}  \not\Vbar \mathbf{R}^{(h)}_{s_2} \,|\, \mathbf{R}^{(h)}_{-(s_1,s_2)} 
\end{align*}  
with $\mathbf{R}^{(h)}_{-(s_1,s_2)} = (\mathbf{R}_{1}^{(h),\T}, \ldots,\mathbf{R}_{s_1-1}^{(h),\T},\mathbf{R}_{s_1+1}^{(h),\T},\ldots,\mathbf{R}_{s_2-1}^{(h),\T},\mathbf{R}_{s_2+1}^{(h),\T},\ldots, \mathbf{R}_{11}^{(h),\T})^{\T}$.
For each given $h\in\{1,2\}$, we apply the CI-FNN test with Rademacher multiplier to these $55$ hypothesis testing problems. The associated 55 p-values are reported in Table \ref{tab:pval-bf}. 
Applying the BH procedure to these p-values with controlling the FDR at the level $0.02$, we reject, respectively, 26 and 29  null hypotheses, in the before COVID-19 period (see panel (a) of Figure \ref{Fig.cin-graph}) and the during/after COVID-19 period (see panel (b) of Figure \ref{Fig.cin-graph}). Controlling the FDR at the level 0.02 here makes the expected numbers of false discoveries in the before COVID-19 period and the during/after COVID-19 period, respectively, do not exceed  $26\times 0.02 =0.52 < 1$ and $29\times 0.02 =0.58 < 1$  asymptotically. Such results indicate that the COVID-19 pandemic has led to significant changes in financial network structure by altering the conditional dependence structure among the 11 sectors.

From Figure \ref{Fig.cin-graph}, it can be observed that the Consumer Discretionary is the most influential sector in the before COVID-19 period, with the most connections with other sectors. As we know, consumption plays a central role in the economy and the Consumer Discretionary sector can act as an indicator for overall economic prosperity. When people are willing to spend more on non-essential goods and services, it indicates economic recovery and growth. After the COVID-19 pandemic, as the economy gradually recovers, the Consumer Discretionary  remains the most influential sector, continuing to interact with and impact other sectors. On the other hand,  during and after the COVID-19 pandemic, some parts of the Consumer Discretionary sector, such as e-commerce and travel, have experienced significant changes. These changes may have strengthened connections with the Information Technology (e-commerce platforms and online services), Health Care (healthcare products), and Financials (payment services and credit cards) sectors. 
Additionally, the Information Technology and  Industrials sectors have more connections to other sectors in the during/after COVID-19 period. With the rapid development of remote work, online education, e-commerce and other related areas, the Information Technology sector has played a crucial role during and after the COVID-19 pandemic. The Industrials sector becomes more closely connected with other industries through global economic recovery and technological advancements. Furthermore, the influence of Health Care sector in the financial network rises in the during/after COVID-19  period, since it plays an important role in the pandemic. 

\begin{figure}[htbp] 
\centering 
\subfigure[Before COVID-9 period]{
\label{Fig.sub.3}
\includegraphics[width=0.48\textwidth]{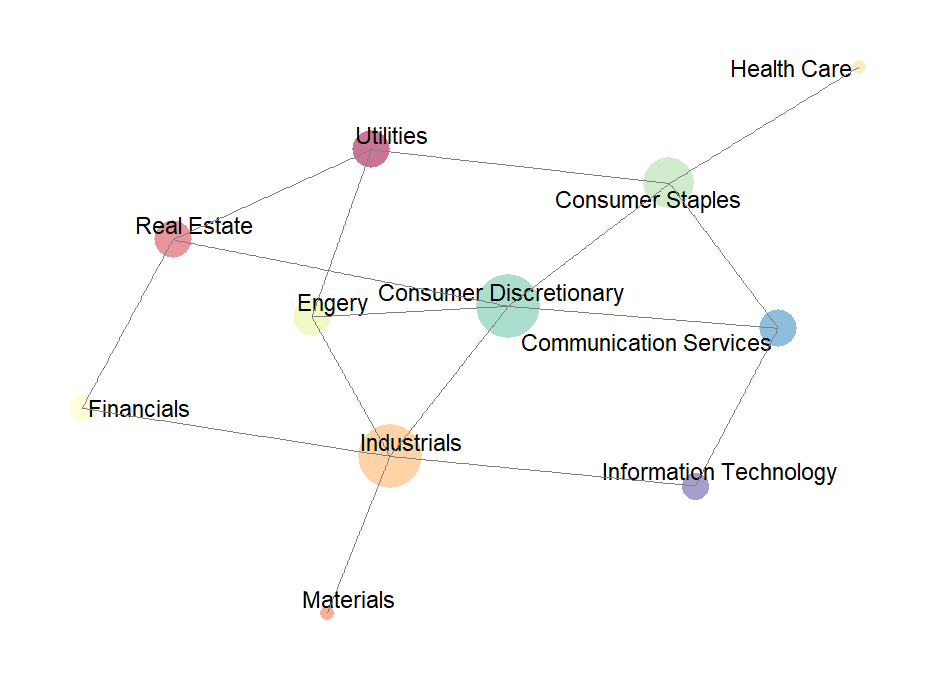}}
\subfigure[During/after  COVID-9 period]{
\label{Fig.sub.4}
\includegraphics[width=0.48\textwidth]{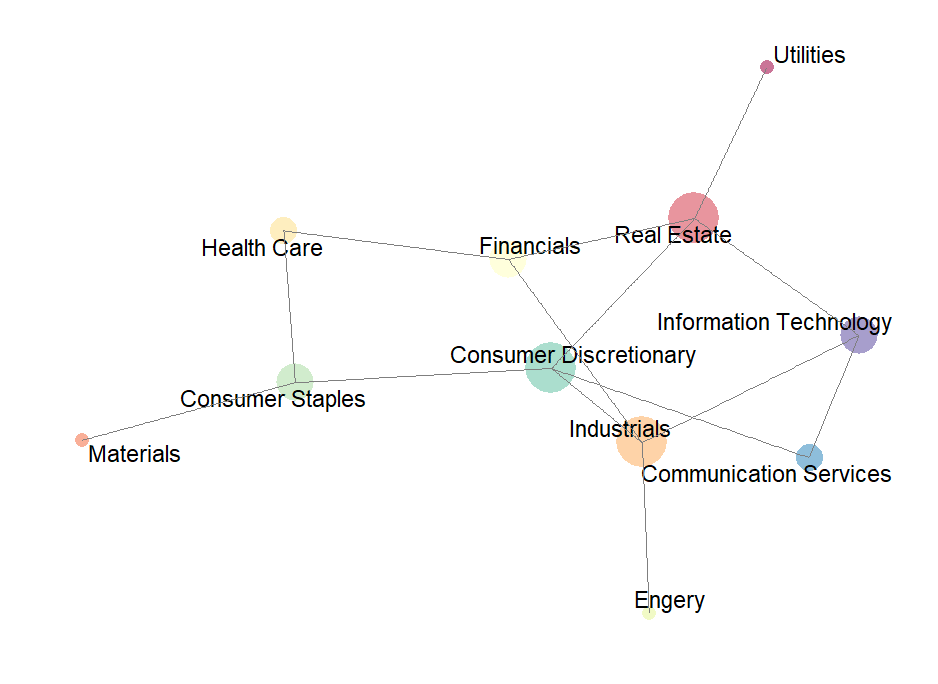}}
\caption{Conditional dependence network of the 11 sectors (denoted by the nodes)  obtained by using the CI-Lasso test with Rademacher multiplier. There exists an edge between two nodes if the conditional independence test between them is significant. The sizes of the nodes are proportional to their degrees.}
\label{Fig.cin-graph-linear}
\end{figure}

The CI-Lasso test with Rademacher multiplier shows similar findings to those of the CI-FNN test with Rademacher multiplier. See Figure \ref{Fig.cin-graph-linear}  for the conditional dependence network of the 11 sectors based on the associated 55 p-values summarized in Table \ref{tab:pval-bf-linear}.  
We have also repeated the above-mentioned analysis for investigating the conditional independence structure among the 11 sectors based on the existing five conditional independence tests mentioned in Section \ref{sc:condindtestsimu}. Table \ref{tab:degree}  summarizes the associated results. Since the PCD test returns invalid results in this real data analysis, its results are omitted. The cdCov test does not reject all null hypotheses in the 55 hypothesis testing problems with all p-values equal to 0.01 in either of the two periods, which cannot provide helpful information for understanding the conditional independence structure of the network. Hence, the results of the cdCov test are also omitted.  The GCM, RCIT, and RCoT tests yield findings consistent with those of our proposed methods.

\begin{table}[htbp]
\scriptsize
\caption{The degrees of nodes associated with the 11 sectors in the networks constructed based on the proposed conditional independence tests with Rademacher multiplier and the three competing methods (GCM, RCIT, RCoT), respectively. } 
\label{tab:degree} 
\begin{spacing}{1.3}
\resizebox{\textwidth}{!}{
	\begin{threeparttable}
		\begin{tabular}{lccccc|ccccc}
			\hline
			\hline	 
			&    \multicolumn{5}{c|}{Before COVID-19 period} & \multicolumn{5}{c}{During/after COVID-19 period}   \\				
			\hline 
			& \multicolumn{2}{c}{Proposed Methods}  & \multirow{2}{*} {GCM} &  \multirow{2}{*} {RCIT} & \multirow{2}{*} {RCoT}  &  \multicolumn{2}{c}{Proposed Methods} &\multirow{2}{*} {GCM} &  \multirow{2}{*} {RCIT} & \multirow{2}{*} {RCoT}\\
			& CI-FNN & CI-Lasso &   &   &  & CI-FNN &  CI-Lasso &   &   &  \\
			\hline
			Communication Services & 3  &3   & 3     & 2     & 2     & 1  &2   & 3     & 3     & 4 \\
			Consumer Discretionary & 7  &5   & 5     & 5     & 5     & 8   &4  & 4     & 4     & 4 \\
			Consumer Staples  & 7   &4  & 4     & 4     & 4     & 4  &3   & 3     & 3     & 3 \\
			Energy & 2   &3  & 3     & 3     & 3     & 2  &1   & 1     & 2     & 1 \\
			Financials & 3   &2  & 2     & 2     & 2     & 6  &3   & 4     & 3     & 3 \\
			Health Care & 4  &1   & 1     & 2     & 2     & 6  &2   & 2     & 3     & 3 \\
			Industrials & 4  &5   & 5     & 7     & 7   & 7  &4   & 5     & 5     & 5 \\
			Information Technology & 5  &2   & 2     & 1     & 1     & 7   &3  & 3     & 4     & 3 \\
			Materials & 7    &1 & 2     & 2     & 1     & 5   &1  & 2     & 2     & 1 \\
			Real Estate & 5  &3   & 3     & 3     & 3     & 6    &4  & 4     & 3     & 3 \\
			Utilities & 5  &3   & 4     & 5     & 4     & 6   &1  & 1     & 0     & 0 \\
			\hline
			\hline
		\end{tabular}%
	\end{threeparttable}
}
\end{spacing}
\end{table}%

\begin{table}[htbp]
\centering
\caption{\footnotesize{ The p-values of the 55 hypothesis testing problems, which are associated with pairs of different sectors, based on the CI-FNN test with Rademacher multiplier. }}
\label{tab:pval-bf}%
\begin{spacing}{1.3}
\resizebox{145mm}{!}{
	\begin{tabular}{ccccccccccccc}
		\hline
		\hline
		&       & \multicolumn{1}{c}{CS} & \multicolumn{1}{c}{CD} & \multicolumn{1}{c}{CSt} & \multicolumn{1}{c}{Eng} & \multicolumn{1}{c}{Fin} & \multicolumn{1}{c}{HC} & \multicolumn{1}{c}{Ind} & \multicolumn{1}{c}{IT} & \multicolumn{1}{c}{Mat} & \multicolumn{1}{c}{RE} & \multicolumn{1}{c}{Uti} \\
		\hline
		\multirow{11}[0]{*}{Before COVID-19 period} &CS    &       &       &       &       &       &       &       &       &       &       &  \\
		&CD    & 0.0082 &       &       &       &       &       &       &       &       &  \\
		&CSt   & 0.0004  & $<$0.0001  &       &       &       &       &       &       &       &  \\
		&Eng   & 0.0102  & $<$0.0001  & 0.0328  &       &       &       &       &       &       &  \\
		&Fin   & 0.2918  & 0.0994  & $<$0.0001  & 0.7074  &       &       &       &       &       &  \\
		&HC    & 0.3802  & 0.0230  & 0.0546  & 0.0134  & 0.1178  &       &       &       &       &  \\
		&Ind   & 0.2180  & $<$0.0001  & $<$0.0001  &  0.4668  & 0.0382  & $<$0.0001  &       &       &       &  \\
		& IT    & 0.2240  & 0.0992  & $<$0.0001  & 0.0160  & 0.0228  & $<$0.0001  & 0.0002  &       &       &  \\
		& Mat   & 0.0008  & $<$0.0001  & 0.6232  & 0.0006  & 0.0004  & $<$0.0001  &  0.0100  & 0.0010  &       &  \\
		&RE    & 0.0198  & 0.0002  & $<$0.0001  & 0.3658  &  0.0242  & 0.0020  & 0.1188  & 0.0326  & 0.0026  &  \\
		&Uti   & 0.0646  & $<$0.0001  & $<$0.0001  & 0.7696  & 0.0004  & 0.0160  & 0.1592  & 0.0038  & 0.8152  & $<$0.0001  \\
		\hline
		\multirow{11}[0]{*}{During/after COVID-19 period} & CS    &       &       &       &       &       &       &       &       &       &  \\
		&CD    & 0.0140 &       &       &       &       &       &       &       &       &  \\
		&CSt   &  0.2440   & 0.0024  &       &       &       &       &       &       &       &  \\
		&Eng   &  0.2064  & 0.0032  & 0.0002  &       &       &       &       &       &       &  \\
		&Fin   & 0.3356  & 0.0008  & 0.2114  & 0.0336  &       &       &       &       &       &  \\
		&HC    &  0.5602  & $<$0.0001  & 0.0182  & 0.0636  &  0.0006  &       &       &       &       &  \\
		&Ind   &  0.0164  & 0.0002  & 0.0206  & 0.2056  &  0.0002 & 0.0002  &       &       &       &  \\
		&IT    & 0.0400 & $<$0.0001  & 0.0560  & 0.0304  &  $<$0.0001  & 0.0004  & $<$0.0001  &       &       &  \\
		&Mat   &  0.0086 & $<$0.0001  & 0.0018  & 0.0646  &  0.0152  & 0.1458  & 0.0054  & $<$0.0001  &       &  \\
		&RE    &  0.2646  & 0.0102  & 0.3072  & 0.0272  &  $<$0.0001  & 0.0026  & $<$0.0001  & $<$0.0001  & 0.0452  &  \\
		&Uti   & 0.3884 & 0.0346  & 0.0014  & 0.6220  & $<$0.0001 & 0.0004  & $<$0.0001  & $<$0.0001  & 0.0140  & 0.0038  \\
		\hline
		\hline
	\end{tabular}%
}
\end{spacing}
\end{table}%

\begin{table}[htbp]
\centering
\caption{\footnotesize{ The p-values of the 55 hypothesis testing problems, which are associated with pairs of different sectors, based on the CI-Lasso test with Rademacher multiplier. }}
\label{tab:pval-bf-linear}%
\begin{spacing}{1.3}
\resizebox{145mm}{!}{
	\begin{tabular}{ccccccccccccc}
		\hline
		\hline
		&       & \multicolumn{1}{l}{CS} & \multicolumn{1}{l}{CD} & \multicolumn{1}{l}{CSt} & \multicolumn{1}{l}{Eng} & \multicolumn{1}{l}{Fin} & \multicolumn{1}{l}{HC} & \multicolumn{1}{l}{Ind} & \multicolumn{1}{l}{IT} & \multicolumn{1}{l}{Mat} & \multicolumn{1}{l}{RE} & \multicolumn{1}{l}{Uti} \\
		\hline
		\multirow{11}[0]{*}{Before COVID-19 period} & CS    &       &       &       &       &       &       &       &       &       &  \\
          & CD    & 0.0014 &       &       &       &       &       &       &       &       &  \\
          & CSt   & 0.0030 & $<$0.0001 &       &       &       &       &       &       &       &  \\
          & Eng   & 0.8792 & 0.0020 & 0.4536 &       &       &       &       &       &       &  \\
          & Fin   & 0.0160 & 0.0548 & 0.2964 & 0.1748 &       &       &       &       &       &  \\
          & HC    & 0.0480 & 0.4024 & $<$0.0001 & 0.6448 & 0.2726 &       &       &       &       &  \\
          & Ind   & 0.1684 & $<$0.0001 & 0.1102 & $<$0.0001 & $<$0.0001 & 0.0080 &       &       &       &  \\
          & IT    & 0.0038 & 0.0238 & 0.3496 & 0.2070 & 0.2012 & 0.0492 & $<$0.0001 &       &       &  \\
          & Mat   & 0.1900 & 0.3808 & 0.0196 & 0.0238 & 0.0898 & 0.0854 & $<$0.0001 & 0.6032 &       &  \\
          & RE    & 0.2066 & $<$0.0001 & 0.0152 & 0.2116 & $<$0.0001 & 0.9724 & 0.4682 & 0.4702 & 0.0816 &  \\
          & Uti   & 0.0198 & 0.5056 & 0.0010 & 0.0014 & 0.1612 & 0.0300 & 0.0116 & 0.2546 & 0.0116 & $<$0.0001 \\
		\hline
		\multirow{11}[0]{*}{During/after COVID-19 period} & CS    &       &       &       &       &       &       &       &       &       &  \\
          & CD    & $<$0.0001 &       &       &       &       &       &       &       &       &  \\
          & CSt   & 0.0124 & $<$0.0001 &       &       &       &       &       &       &       &  \\
          & Eng   & 0.3706 & 0.4966 & 0.7504 &       &       &       &       &       &       &  \\
          & Fin   & 0.2412 & 0.2040 & 0.0962 & 0.5760 &       &       &       &       &       &  \\
          & HC    & 0.0104 & 0.2186 & $<$0.0001 & 0.4768 & 0.0026 &       &       &       &       &  \\
          & Ind   & 0.0162 & $<$0.0001 & 0.2082 & $<$0.0001 & $<$0.0001 & 0.2382 &       &       &       &  \\
          & IT    & $<$0.0001 & 0.0090 & 0.3518 & 0.4000 & 0.2940 & 0.0138 & $<$0.0001 &       &       &  \\
          & Mat   & 0.2180 & 0.1362 & 0.0016 & 0.0096 & 0.0086 & 0.0414 & 0.0218 & 0.0460 &       &  \\
          & RE    & 0.2934 & 0.0004 & 0.0850 & 0.0832 & $<$0.0001 & 0.0226 & 0.1286 & 0.0036 & 0.4964 &  \\
          & Uti   & 0.1500 & 0.5338 & 0.0152 & 0.0728 & 0.3352 & 0.1546 & 0.0228 & 0.3012 & 0.0698 & 0.0006 \\
		\hline
		\hline
	\end{tabular}%
}
\end{spacing}
\end{table}%

\begin{table}[h!]
\centering
\caption{\footnotesize{ The  11 sectors and 74 industries included in the Global Industry Classification Standard (GICS) structure. The abbreviations of the sector names are presented in the column named `Abbr.'. }}
\label{tab:name-sec} 
\begin{spacing}{1.3}
\resizebox{\textwidth}{!}{
	\begin{tabular}{c|ll||c|ll}
		\hline
		\hline
		Abbr. & Sector & Industry &   Abbr.    & Sector & Industry \\
		\hline
		\multirow{6}[1]{*}{CSt} & \multirow{6}[1]{*}{Consumer Staples} & Consumer Staples Distribution \& Retail &    \multirow{6}[1]{*}{HC}    & \multirow{6}[1]{*}{Health Care} & Health Care Equipment \& Supplies \\
		&       & Beverages &       &       & Health Care Providers \& Services \\
		&       & Food Products &  &       & Health Care Technology \\
		&       & Tobacco &       &       & Biotechnology \\
		&       & Household Products &       &       & Pharmaceuticals \\
		&       & Personal Care Products &       &       & Life Sciences Tools \& Services \\
		\hline
		\multirow{5}[2]{*}{Mat} & \multirow{5}[2]{*}{Materials} & Chemicals & \multirow{5}[2]{*}{CS} & \multirow{5}[2]{*}{Communication Services} & Diversified Telecommunication Services \\
		&       & Construction Materials &       &       & Wireless Telecommunication Services \\
		&       & Containers \& Packaging &       &       & Media \\
		&       & Metals \& Mining &       &       & Entertainment \\
		&       & Paper \& Forest Products &       &       & Interactive Media \& Services \\
		\hline
		\multirow{6}[2]{*}{Fin} & \multirow{6}[2]{*}{Financials} & Banks & \multirow{6}[2]{*}{Uti} & \multirow{6}[2]{*}{Utilities} & Electric Utilities \\
		&       & Financial Services &       &       & Gas Utilities \\
		&       & Consumer Finance &       &       & Multi-Utilities \\
		&       & Capital Markets &       &       & Water Utilities \\
		&       & Mortgage Real Estate Investment Trusts (REITs) &       &       & Independent Power and Renewable Electricity Producers \\
		&       & Insurance &       &       &  \\
		\hline
		\multirow{9}[2]{*}{IT} & \multirow{9}[2]{*}{Information Technology} & IT Services  & \multirow{9}[2]{*}{RE} & \multirow{9}[2]{*}{Real Estate} & Diversified REITs \\
		&       & Software &       &       & Industrial REITs \\
		&       & Communications Equipment &       &       & Hotel \& Resort REITs \\
		&       & Technology Hardware, Storage \& Peripherals &       &       & Office REITs \\
		&       & Electronic Equipment, Instruments \& Components &       &       & Health Care REITs \\
		&       & Semiconductors \& Semiconductor Equipment &       &       & Residential REITs \\
		&       &  &       &       & Retail REITs \\
		&       &       &       &       & Specialized REITs \\
		&       &       &       &       & Real Estate Management \& Development \\
		\hline
		\multirow{14}[4]{*}{Ind} & \multirow{14}[4]{*}{Industrials} & Aerospace \& Defense & \multirow{12}[2]{*}{CD} & \multirow{12}[2]{*}{Consumer Discretionary} & Automobile Components \\
		&       & Building Products &       &       & Automobiles \\
		&       & Construction \& Engineering &       &       & Household Durables \\
		&       & Electrical Equipment &       &       & Leisure Products \\
		&       & Industrial Conglomerates &       &       & Textiles, Apparel \& Luxury Goods \\
		&       & Machinery &       &       & Hotels, Restaurants \& Leisure \\
		&       & Trading Companies \& Distributors &       &       & Diversified Consumer Services \\
		&       & Commercial Services \& Supplies &       &       & Distributors \\
		&       & Professional Services &       &       & Broadline Retail \\
		&       & Air Freight \& Logistics &       &       & Specialty Retail \\
		&       & Passenger Airlines &       &       &  \\
		&       & Marine Transportation &       &       &  \\
		\cline{4-6}        &       & Ground Transportation & \multirow{2}[2]{*}{Eng} & \multirow{2}[2]{*}{Energy} & Energy Equipment \& Services \\
		&       & Transportation Infrastructure &       &       & Oil, Gas \& Consumable Fuels \\
		\hline
		\hline
	\end{tabular}%
}
\end{spacing}
\end{table}%

\section{Additional Numerical Studies}\label{sec:add-numerical}

\subsection{Performance of the Extension Method in  Independence Testing}\label{sec:other methods-test}
 
As discussed in Remark \ref{rek:sufficient}, our proposed independence test introduced in Section \ref{sec:indtest-m}  actually uses $|\mathbb{E}(\bU\bV^{\T})|_{\infty}$  as a measure for $\bU \Vbar \bV$, which essentially tests a necessary condition for the independence. Notice that $|\mathbb{E}(\bU\bV^{\T})|_{\infty}$ may be equal to zero when $\bU \not\Vbar \bV$. In this case, we may consider replace the measure $|\mathbb{E}(\bU\bV^{\T})|_{\infty}$ by some other measures: 
(i) the projection correlation (Pcor) in \cite{Zhu2017}, (ii) the  ranks of distances (rdCov) in \cite{Heller2013}, (iii) the distance correlation (dCor) in \cite{Szekely2013}, (iv) the $k$-variate HSIC  (dHSIC) in \cite{Pfister2018}, (v) the rank-based dependence matrix (JdCov\_R) in \cite{Chakraborty2018}, (vi) the generalized distance covariance (GdCov) in \cite{Jin2018}, (vii) the center-outward ranks and signs (Hallin) in \cite{Shi2020}, and (viii) the multivariate ranks  (mrdCov) in \cite{Deb2021}. 
For each of these measures, $\bU \Vbar \bV$ if and only if this measure between $\bU$ and $\bV$ equals to zero. In this section, we give an example where $\bU \not \Vbar \bV$ but $|\mathbb{E}(\bU\bV^{\T})|_{\infty}=0$, and evaluate the associated performance of the independence tests based on $\{(\hat{\bU}_i,\hat{\bV}_i)\}_{i=1}^n$ and the alternative measures. 

Let $S\in\{-1,1\}$ be a Rademacher variable with $\mathbb P(S=1)=\mathbb P(S=-1)=0.5$.
Conditional on $S$, draw $(X,Y)^{\T} \,|\, S \sim \mathcal{N} (\bzero, \bSigma_{S} )$, where   
\begin{align*}
    \bSigma_S:=\begin{pmatrix}1 & 0.8S\\ 0.8 S  & 1
    \end{pmatrix}  \,.
\end{align*}
 Notice that $X$ and $Y$ are not independent, and $\mathbb{E}(XY) =\mathbb{E}\{\mathbb{E}(XY\,|\,S)\} = \mathbb{E}(0.8S) =0$.
Write $\bX=(X_1,\ldots,X_p)^{\T}$ and $\bY=(Y_1,\ldots, Y_p)^{\T}$, where  $\{(X_j,Y_j)\}_{j=1}^{p}$ are i.i.d. copies of $(X,Y)$.   Recall $\bU=(U_1, \ldots, U_p)^{\T}$ and $\bV=(V_1, \ldots, V_p)^{\T}$ with $U_{j}=\Phi^{-1}\{F_{\bX,j}(X_j)\}$ and $V_{j} =\Phi^{-1}\{F_{\bY,j}(Y_j)\}$. Then $U_j=X_j$ and $V_j=Y_j$ for any $j\in[p]$. Hence, $|\mathbb{E}(\bU\bV^{\T})|_{\infty}=0$ but $\bU \not\Vbar \bV$. Let $\{(\bX_i^{\T}, \bY_i^{\T})^{\T}\}_{i=1}^{n}$ be the observations which are i.i.d. copies of $(\bX^{\T},\bY^{\T})^{\T}$.  We set $p\in\{10,100\}$ and $n\in\{50,100\}$. Table \ref{tab:other methods} reports the  empirical powers of the 
independence tests  based on $\{(\hat{\bU}_i,\hat{\bV}_i)\}_{i=1}^n$ and different  measures. For our proposed independence test with  Rademacher multiplier,  since  $|\mathbb{E}(\bU\bV^{\T})|_{\infty}=0$ in this example, it fails to detect the dependence between $\bU$ and $\bV$. Among all the other considered measures, rdCov, dHSIC, and JdCov\_R show reasonable power performance. Although Pcor, dCor, JdCov\_R, GdCov, Hallin, and mrdCov test the necessary and sufficient conditions of $\bU \Vbar \bV$, they are still powerless in this example.

\begin{table}[htbp]
\centering
\caption{\footnotesize{ Empirical  powers of the  independence test (with  Rademacher multiplier)  based on $\{(\hat{\bU}_i,\hat{\bV}_i)\}_{i=1}^n$ and different  measures.  }}
\label{tab:other methods} 
\begin{spacing}{1.3}
\resizebox{\textwidth}{!}{
    \begin{tabular}{ccccccccccc}
    \hline\hline
         $n$ & $p$ & Proposed  Method &   {Pcor} &  {rdCov} &  {dCor}  &  {dHSIC} &  {JdCov\_R} &  {GdCov} &  {Hallin} &  {mrdCov}  \\
          \hline
    {$50$}  &  10  &6.2 & 8.2  & 84.2  & 7.2  & 19.0  & 9.1  & 9.4  & 5.7   & 5.3 \\
          &  100 & 7.0 & 7.9  & 83.5  & 5.6  & 16.3  & 6.0  & 7.6   & 4.5   & 4.7 \\
          \hline
    {$100$}  &  10   &6.3 & 10.2  & 100.0 & 8.8  & 31.2  & 15.0   & 14.1  & 8.1  & 5.1 \\
          &  100 &   4.7    &  7.3    & 99.9      &    5.3  &    19.1  & 7.0 &  8.4     &   5.2    & 5.3 \\
          \hline\hline
 	\end{tabular}%
}
\end{spacing}
\end{table}%

\subsection{The Effect of Coordinatewise  Gaussianization in Independence Testing}\label{sec:test-gaussianization}

The numerical studies in Section \ref{sc:indtestsimu} indicate that most competing methods for independence testing suffer from low power performance in high-dimensional settings and some of them even yield invalid results for heavy-tailed designs. In this section, we investigate whether applying coordinatewise Gaussianization as a preliminary step can  improve the performance of these competing methods. The competing methods used here are, respectively, Pcor \citep{Zhu2017}, rdCov \citep{Heller2013},  dCor \citep{Szekely2013},  dHSIC \citep{Pfister2018},  JdCov\_R \citep{Chakraborty2018}, GdCov \citep{Jin2018},  Hallin \citep{Shi2020} and mrdCov \citep{Deb2021}.   Table  \ref{tab:ind-test-gaussian-n100} reports  the empirical sizes and powers of these competing methods  with coordinatewise Gaussianization.   

Among Examples 1–5, $\bX$ and $\bY$ in Examples 1 and 2 are heavy-tailed. As shown in  Table \ref{tab:ind},  Pcor, rdCov, and JdCov\_R all work for the original data of Examples 1 and 2. With coordinatewise Gaussianization, the powers of both Pcor and rdCov for Examples 1 and 2 can be improved substantially, whereas JdCov\_R exhibits similar performance regardless of whether coordinatewise  Gaussianization is applied or not. This is because JdCov\_R is invariant to monotonic component-wise transformations (See Section 2.4 of \cite{Chakraborty2018}). 
On the other hand,  dCor, dHSIC, GdCov, Hallin and mrdCov all return invalid results for the original data of Example 1 as reported in Table  \ref{tab:ind}. However, all of them with coordinatewise Gaussianization yield valid results. For Example 2, with coordinatewise Gaussianization, dCor, dHSIC, GdCov, and mrdCov exhibit substantial power improvements, but Hallin  is powerless in all settings. For Example 3, all the competing methods  without coordinatewise Gaussianization fail to detect the nonlinear dependence between $\bX$ and $\bY$. However, with coordinatewise Gaussianization, the power performance of Pcor,  rdCov, dCor, dHSIC, GdCov and mrdCov can improve substantially, whereas JdCov\_R and Hallin show little change.  In Gaussian or light-tailed settings with sparse dependence (Examples 4 and 5),  the performance of these methods behaves similarly regardless of whether coordinatewise  Gaussianization is applied or not.

	\begin{table}[htbp]
		\scriptsize
		\caption{Empirical sizes (the rows with $K=0$ in Examples 1--3 and `{\rm null}' in Examples 4--5) and powers (the rows with $K=p/20$ and $p/10$ in Examples 1--3 and `{\rm alternative}' in Examples 4--5) of the independence test with coordinatewise Gaussianization for the competing methods in Examples 1--5 with $n=100$. All numbers reported below are multiplied by 100.  }
		\renewcommand\tabcolsep{4.0pt}
		\label{tab:ind-test-gaussian-n100}
		\begin{spacing}{1.1}
			\resizebox{\textwidth}{!}{
				\begin{threeparttable}
					\begin{tabular}{ccc|cccccccc}
						\hline
						\hline
						&  $p$  & Setting &   {Pcor} &  {rdCov} &  {dCor}  &  {dHSIC} &  {JdCov\_R} &  {GdCov} &  {Hallin} &  {mrdCov}  \\
						\hline 
						
    \multirow{9}{*}{ Example 1} & \multirow{3}{*}{100 } & $K=0$  & 5.5   & 4.8   & 5.1   & 5.3   & 5.9   & 5.1   & 4.6   & 5.2  \\
          &       & $K=p/20$  & 96.5  & 32.4  & 96.3  & 96.4  & 12.4  & 31.2  & 5.3   & 40.3  \\
          &       & $K=p/10$  & 100.0  & 83.5  & 100.0  & 100.0  & 23.6  & 73.6  & 7.1   & 63.2  \\
\cline{2-11}          & \multirow{3}{*}{400 } & $K=0$   & 5.0   & 5.4   & 5.0   & 5.5   & 6.2   & 5.3   & 4.4   & 3.7  \\
          &       & $K=p/20$  & 96.6  & 28.8  & 96.5  & 96.4  & 8.7   & 19.5  & 6.0   & 84.2  \\
          &       & $K=p/10$  & 100.0  & 81.2  & 100.0  & 100.0  & 12.8  & 42.9  & 5.4   & 98.7  \\
\cline{2-11}          & \multirow{3}{*}{1600 } & $K=0$ & 5.3	& 3.9	& 5.2	& 5.0	& 6.0	& 5.4	& 4.6	& 4.1 \\
          &       & $K=p/20$ 	& 97.4	& 27.7	& 97.4	& 97.5	& 9.3	& 9.8	& 5.2	& 100.0 \\
          &       & $K=p/10$ 	& 100.0	& 80.4	& 100.0	& 100.0	& 12.8	& 17.7	& 4.8	& 100.0  \\
    \hline
    \multirow{9}{*}{ Example 2} & \multirow{3}{*}{100 } & $K=0$   & 6.5   & 5.6   & 7.1   & 6.4   & 5.4   & 5.0   & 4.7   & 4.7  \\
          &       & $K=p/20$  & 99.8  & 78.3  & 99.8  & 99.7  & 59.0  & 51.8  & 4.5   & 5.7  \\
          &       & $K=p/10$ & 100.0  & 100.0  & 100.0  & 100.0  & 100.0  & 99.0  & 6.1   & 8.9  \\
\cline{2-11}          & \multirow{3}{*}{400 } & $K=0$  & 4.1   & 4.4   & 4.0   & 4.3   & 4.2   & 4.7   & 4.6   & 6.2  \\
          &       & $K=p/20$   & 100.0  & 100.0  & 100.0  & 100.0  & 99.6  & 94.9  & 3.9   & 89.5  \\
          &       & $K=p/10$ & 100.0  & 100.0  & 100.0  & 100.0  & 100.0  & 100.0  & 5.4   & 100.0  \\
\cline{2-11}          & \multirow{3}{*}{1600 } & $K=0$ & 6.0	& 5.7	& 5.7	& 5.4	& 5.6	& 4.7	& 5.3	& 4.6 \\
          &       & $K=p/20$ & 100.0	& 100.0	& 100.0	& 100.0	& 100.0	& 98.8	& 5.6	& 100.0 \\
          &       & $K=p/10$ & 100.0	& 100.0	& 100.0	& 100.0	& 100.0	& 100.0	& 5.7	& 100.0  \\
    \hline
    \multirow{9}{*}{ Example 3} & \multirow{3}{*}{100 } & $K=0$  & 4.7   & 4.0   & 4.6   & 4.6   & 5.6   & 5.8   & 4.5   & 4.2  \\
          &       & $K=p/20$ & 97.6  & 29.2  & 97.5  & 97.1  & 10.4  & 32.9  & 5.3   & 5.1  \\
          &       & $K=p/10$  & 100.0  & 84.3  & 100.0  & 100.0  & 22.7  & 72.4  & 6.8   & 6.4  \\
\cline{2-11}          & \multirow{3}{*}{400 } & $K=0$ & 5.0	& 5.6	& 4.9	& 4.7	& 5.7	& 5.1	& 4.8	& 4.7 \\
          &       & $K=p/20$  & 96.7  & 27.4  & 96.5  & 96.9  & 8.4   & 18.5  & 5.6   & 7.9  \\
          &       & $K=p/10$ & 100.0	& 78.9	& 100.0	& 100.0	& 12.4	& 40.7	& 6.6	& 50.7 \\
\cline{2-11}          & \multirow{3}{*}{1600 } & $K=0$	& 4.7	& 5.4	& 5.0	& 4.9	& 5.8	& 5.9	& 4.1	& 4.7  \\
          &       & $K=p/20$ & 96.3	& 28.7	& 96.1	& 96.5	& 9.1	& 11.2	& 5.4	& 42.0  \\
          &       & $K=p/10$ & 100.0	& 80.1	& 100.0	& 100.0	& 13.0	& 19.6	& 4.7	& 71.3 \\
    \hline
    \multirow{6}{*}{Example 4} & \multirow{2}{*}{100 } & null  & 5.4   & 4.9   & 5.3   & 4.7   & 4.8   & 5.4   & 5.4   & 5.8  \\
          &       & alternative & 26.5  & 5.1   & 25.9  & 25.8  & 6.8   & 9.0   & 5.3   & 5.0  \\
\cline{2-11}          & \multirow{2}{*}{400 } & null  & 4.8   & 5.5   & 4.9   & 5.1   & 4.4   & 4.6   & 5.2   & 4.8  \\
          &       & alternative  & 8.6   & 5.8   & 8.8   & 7.9   & 4.7   & 5.5   & 5.3   & 4.9  \\
\cline{2-11}          & \multirow{2}{*}{1600 } & null  & 6.2	& 5.0	& 5.8	& 5.7	& 5.2	& 4.6	& 5.0	& 5.7 \\
          &       & alternative & 6.5	& 4.5	& 6.5	& 6.1	& 5.0	& 5.0	& 4.5	& 6.4  \\
    \hline
    \multirow{6}{*}{Example 5} & \multirow{2}{*}{100 } & null     & 5.4   & 4.9   & 5.3   & 4.7   & 4.8   & 5.4   & 5.4   & 5.8  \\
          &       & alternative  & 26.5  & 5.1   & 25.9  & 25.8  & 6.8   & 9.0   & 5.3   & 5.1  \\
\cline{2-11}          & \multirow{2}{*}{400 } & null   & 4.8   & 5.5   & 4.9   & 5.1   & 4.4   & 4.6   & 5.2   & 4.8  \\
          &       & alternative & 8.6	& 5.8	& 8.8	& 7.9	& 4.7	& 5.5	& 5.3	& 5.4  \\
\cline{2-11}          & \multirow{2}{*}{1600 } & null & 6.2	& 5.0	& 5.8	& 5.7	& 5.2	& 4.6	& 5.0	& 5.7 \\
          &       & alternative 	& 6.5	& 4.5	& 6.5	& 6.1	& 5.0	& 5.0	& 4.5	& 5.7 \\ 
	\hline
	\hline
					\end{tabular}%
				\end{threeparttable}
			}
		\end{spacing}
	\end{table}

\subsection{Comparison with Covariance-Based Methods under Gaussian Data}\label{sec:gaussian-test}

In this section, we compare the proposed tests with five covariance-matrix–based tests on Gaussian data: (i) the test based on the so-called Schott type statistic (Schott) in \cite{BaoHuPanZhou2017EJS}, (ii) the classical likelihood tests (LRT) in \cite{JiangYang2013AOS}, (iii) the test based on the Wilks’ statistic (TW) in \cite{bodnar2019testing}, (iv) the test based on Lawley-Hotelling’s trace criterion (TLH) in \cite{bodnar2019testing}, and (v) the test based on Bartlett–Nanda–Pillai’s trace criterion (TBNP) in \cite{bodnar2019testing}.  These five tests are all designed for testing whether $\cov(\bX, \bY) =\bzero$ or not for Gaussian data, which is equivalent to testing $\bX \Vbar \bY$. More specifically, let $(\bX^{\T}, \bY^{\T})^{\T} \sim \mathcal{N}(\bzero, \tilde{\bSigma})$ with $\bX \in \mathbb{R}^{p}$, $\bY \in \mathbb{R}^{p}$ and 
\begin{align*}
    \tilde{\bSigma}=\begin{pmatrix}
\tilde{\bSigma}_{\bX} & \tilde{\bSigma}_{\bX\bY}\\
\tilde{\bSigma}_{\bX\bY}^{\T} & \tilde{\bSigma}_{\bY}
\end{pmatrix} \,,
\end{align*}
where $\tilde{\bSigma}_{\bX} \in \mathbb{R}^{p\times p}$, $\tilde{\bSigma}_{\bY} \in \mathbb{R}^{p\times p}$ and $\tilde{\bSigma}_{\bX\bY} \in \mathbb{R}^{p\times p}$. 
We consider following settings for $\tilde{\bSigma}$:
\begin{description}
    \item[S1.] Set $\tilde{\bSigma}_{\bX} =\mathbf{I}_{p}$, $\tilde{\bSigma}_{\bY} = \mathbf{I}_{p}$ and $\tilde{\bSigma}_{\bX\bY} =\bzero$. In this setting, $\bX \Vbar \bY$. 
    \item[S2.] Let $\mathbf{D}_1 = (D_{1,i,j})_{p\times p}$ with $D_{1,i,j} = 0.5^{|i-j|}$ for $i,j\in[p]$. Set $\tilde{\bSigma}_{\bX} =\mathbf{D}_1$, $\tilde{\bSigma}_{\bY} = \mathbf{D}_1$ and $\tilde{\bSigma}_{\bX\bY} =\bzero$. In this setting, $\bX \Vbar \bY$.
    \item[S3.]  Set $\tilde{\bSigma}_{\bX} =\mathbf{I}_{p}$, $\tilde{\bSigma}_{\bY} = \mathbf{I}_{p}$ and $\tilde{\bSigma}_{\bX\bY} =\mathbf{D}_2$ with $\mathbf{D}_2 =0.5 \mathbf{I}_{p}$. In this setting, $\bX \not\Vbar \bY$. 
    \item[S4.]  Set $\tilde{\bSigma}_{\bX} =\mathbf{D}_1$, $\tilde{\bSigma}_{\bY} = \mathbf{D}_1$ and $\tilde{\bSigma}_{\bX\bY} =\mathbf{D}_2$, where $\mathbf{D}_1$ and $\mathbf{D}_2$ are specified in Settings S1 and S2, respectively. In this setting, $\bX \not\Vbar \bY$.
    \item[S5.]  Set $\tilde{\bSigma} = 0.75 \mathbf{I}_{2p} + 0.25 \mathbf{J}_{2p}$, where $\mathbf{J}_{2p} \in \mathbb{R}^{(2p) \times (2p)}$ with all entries being  equal to 1.  In this setting, $\bX \not\Vbar \bY$.
\end{description}
Table \ref{tab:cov_test} reports the empirical sizes and powers of our proposed independence test (with Rademacher multiplier) and these five competing methods. 
As shown in Table \ref{tab:cov_test}, our proposed independence test can consistently control the empirical sizes around the nominal significance level  0.05 across all settings and  also has good power performance. However, the performance of these five competing methods designed for Gaussian data depends crucially on the relationship between $p$ and $n$. The Schott test only returns valid results when $p<n$, which is undersized in general, and does not have power in some settings. The LRT and TW tests only return valid results when $p<n/2$. Although the LRT test can control size around the nominal significance level 0.05, it is powerless across all settings. The TW test shows good performance  when 
$p< n/2$. Furthermore,  the TLH and TBNP tests cannot control size when $p$ is large. These results demonstrate that, even under Gaussian distributions, our method surpasses these five Gaussian-specific competitors, exhibiting superior robustness to the dimension–sample size relationship.

\begin{table}[htbp]
\centering
\caption{\footnotesize{ Empirical sizes (the rows with S1–S2) and powers (the rows with S3–S5) of the proposed independence test (with  Rademacher multiplier) and covariance-matrix-based competitors under Gaussian data. All numbers reported below are multiplied by 100. The results reported by `NA' indicate that the associated tests return invalid results. }}
\label{tab:cov_test} 
\begin{spacing}{1.2}
\resizebox{\textwidth}{!}{
    \begin{tabular}{cc|cccccc|cccccc}
		\hline
		\hline
           &    & \multicolumn{6}{c|}{$n=50$}    & \multicolumn{6}{c}{$n=100$} \\
     {$p$} & {Setting}& {Proposed Method} & {Schott} & {LRT} & {TW} & {TLH} & {TBNP} & {Proposed Method} & {Schott} & {LRT} & {TW} & {TLH} & {TBNP} \\
    \hline
     \multirow{5}{*}{20}  & S1    & 6.9   & 1.4   & 4.0   & 6.1   & 10.0  & 0.0   & 5.5   & 3.4   & 4.9   & 4.8   & 5.9   & 0.0 \\
          & S2    & 5.9   & 1.4   & 5.4   & 5.3   & 9.7   & 0.0   & 6.1   & 3.5   & 4.2   & 5.4   & 6.1   & 0.0 \\
          & S3    & 100.0 & 100.0 & 0.0   & 99.9  & 94.0  & 0.0   & 100.0 & 100.0 & 0.0   & 100.0 & 100.0 & 98.4 \\
          & S4    & 68.0  & 100.0 & 0.0   & 100.0 & 100.0 & 0.0   & 99.6  & 100.0 & 0.0   & 100.0 & 100.0 & 100.0 \\
          & S5    & 95.2  & 8.0   & 0.2   & 34.6  & 51.6  & 0.0   & 100.0 & 60.3  & 0.0   & 85.1  & 100.0 & 0.0 \\
          \hline
           \multirow{5}{*}{40} & S1    & 6.3   & 0.0   & NA    & NA    & 50.5  & 99.3  & 6.1   & 1.9   & 4.5   & 6.3   & 9.2   & 0.0 \\
          & S2    & 5.9   & 0.0   & NA    & NA    & 49.0  & 99.4  & 6.2   & 1.3   & 5.8   & 5.1   & 7.9   & 0.0 \\
          & S3    & 100.0 & 2.1   & NA    & NA    & 49.9  & 99.8  & 100.0 & 100.0 & 0.0   & 100.0 & 100.0 & 0.0 \\
          & S4    & 70.0  & 33.5  & NA    & NA    & 49.4  & 99.5  & 99.9  & 100.0 & 0.0   & 100.0 & 100.0 & 0.0 \\
          & S5    & 98.2  & 0.0   & NA    & NA    & 49.6  & 99.4  & 100.0 & 10.3  & 0.0   & 54.1  & 89.8  & 0.0 \\
          \hline
           \multirow{5}{*}{80}  & S1    & 6.5   & NA    & NA    & NA    & 81.6  & 96.9  & 5.4   & 0.0   & NA    & NA    & 48.6  & 98.7 \\
          & S2    & 7.6   & NA    & NA    & NA    & 81.3  & 96.5  & 5.4   & 0.0   & NA    & NA    & 48.8  & 99.2 \\
          & S3    & 100.0 & NA    & NA    & NA    & 83.2  & 97.6  & 100.0 & 47.6  & NA    & NA    & 51.5  & 99.4 \\
          & S4    & 63.7  & NA    & NA    & NA    & 81.4  & 97.2  & 100.0 & 99.9  & NA    & NA    & 50.8  & 99.2 \\
          & S5    & 99.2  & NA    & NA    & NA    & 87.6  & 96.9  & 100.0 & 0.0   & NA    & NA    & 48.9  & 99.4 \\
          \hline
           \multirow{5}{*}{200}  & S1    & 6.3   & NA    & NA    & NA    & 86.9  & 99.3  & 7.4   & NA    & NA    & NA    & 84.0  & 99.0 \\
          & S2    & 6.7   & NA    & NA    & NA    & 87.9  & 99.3  & 5.6   & NA    & NA    & NA    & 85.3  & 98.8 \\
          & S3    & 100.0 & NA    & NA    & NA    & 88.4  & 99.4  & 100.0 & NA    & NA    & NA    & 84.3  & 99.2 \\
          & S4    & 58.6  & NA    & NA    & NA    & 87.3  & 99.3  & 100.0 & NA    & NA    & NA    & 85.0  & 99.1 \\
          & S5    & 99.8  & NA    & NA    & NA    & 93.1  & 99.6  & 100.0 & NA    & NA    & NA    & 93.4  & 99.0 \\
    \hline\hline
    \end{tabular}%
}
\end{spacing}
\end{table}%

For the conditional independence test, let  $(\bX^{\T}, \bY^{\T}, \bZ^{\T})^{\T} \sim \mathcal{N}(\bzero, \bar{\bSigma}) $ with $\bX \in \mathbb{R}^{p}$, $\bY \in \mathbb{R}^{p}$, $\bZ \in \mathbb{R}^{m}$ and 
\begin{align*}
    \bar{\bSigma}=\begin{pmatrix}
\bar{\bSigma}_{\bX\bX} & \bar{\bSigma}_{\bX\bY} & \bar{\bSigma}_{\bX\bZ}\\
\bar{\bSigma}_{\bX\bY}^{\T} & \bar{\bSigma}_{\bY\bY}  & \bar{\bSigma}_{\bY\bZ} \\
\bar{\bSigma}_{\bX\bZ}^{\T} & \bar{\bSigma}_{\bY\bZ}^{\T}  & \bar{\bSigma}_{\bZ\bZ}
\end{pmatrix} \,,
\end{align*}
where $\bar{\bSigma}_{\bX\bX} \in \mathbb{R}^{p\times p}$, $\bar{\bSigma}_{\bY\bY} \in \mathbb{R}^{p\times p}$ and $\bar{\bSigma}_{\bZ\bZ} \in \mathbb{R}^{m\times m}$.  The conditional covariance matrix of $(\bX^{\T}, \bY^{\T})^{\T}$ given $\bZ$ is 
\begin{align*}
    \begin{pmatrix}
\bar{\bSigma}_{\bX\bX} & \bar{\bSigma}_{\bX\bY}  \\
\bar{\bSigma}_{\bX\bY}^{\T} & \bar{\bSigma}_{\bY\bY}  
\end{pmatrix} 
-  \begin{pmatrix}
        \bar{\bSigma}_{\bX\bZ}\\
        \bar{\bSigma}_{\bY\bZ} 
\end{pmatrix}  \bar{\bSigma}_{\bZ\bZ}^{-1}  \begin{pmatrix}
        \bar{\bSigma}_{\bX\bZ}^{\T} &
        \bar{\bSigma}_{\bY\bZ}^{\T} 
\end{pmatrix}  \,. 
\end{align*}
Hence, when $(\bX^{\T}, \bY^{\T}, \bZ^{\T})^{\T}$ is jointly Gaussian, we have  $\bX \Vbar \bY \,|\, \bZ$ if and only if $\bar{\bSigma}_{\bX\bY} - \bar{\bSigma}_{\bX\bZ}\bar{\bSigma}_{\bZ\bZ}^{-1} \bar{\bSigma}_{\bY\bZ}^{\T} =\bzero$.  Consider the regression model
\begin{align}\label{eq:linear-model}
    \begin{pmatrix}
        \bX\\
        \bY 
\end{pmatrix}  = \mathbf{H} \bZ + \mathbf{E} \,,
\end{align}
where $\mathbf{H} \in \mathbb{R}^{(2p) \times m}$, $\mathbf{E} \in \mathbb{R}^{2p}$ and  $\mathbb{E}(\bZ \mathbf{E}^{\T}) =\bzero$. With direct calculation, it holds that
\begin{align*}
    \cov(\mathbf{E} ) =  \begin{pmatrix}
\bar{\bSigma}_{\bX\bX} & \bar{\bSigma}_{\bX\bY}  \\
\bar{\bSigma}_{\bX\bY}^{\T} & \bar{\bSigma}_{\bY\bY}  
\end{pmatrix} 
-  \begin{pmatrix}
        \bar{\bSigma}_{\bX\bZ}\\
        \bar{\bSigma}_{\bY\bZ} 
\end{pmatrix}  \bar{\bSigma}_{\bZ\bZ}^{-1}  \begin{pmatrix}
        \bar{\bSigma}_{\bX\bZ}^{\T} &
        \bar{\bSigma}_{\bY\bZ}^{\T} 
\end{pmatrix} \,.
\end{align*}
Write $\mathbf{E} =(\mathbf{E}_{1}^{\T}, \mathbf{E}_{2}^{\T})^{\T}$ with $\mathbf{E}_{1} \in \mathbb{R}^{p}$ and $\mathbf{E}_{2} \in \mathbb{R}^{p}$.  Since $\mathbf{E}$ is jointly Gaussian, we have $\bX \Vbar \bY \,|\, \bZ$ if and only if $\cov(\mathbf{E}_{1}, \mathbf{E}_{2}) =\bzero$. Thus, the conditional independence test can be reformulated as a covariance matrix test. Different from the traditional covariance matrix test, we cannot obtain the observations of the random vectors $\mathbf{E}_1$ and $\mathbf{E}_2$ here. In practice, for given observations $\{(\bX_i^{\T},\bY_i^{\T}, \bZ_i^{\T})^{\T}\}_{i=1}^{n}$, we can first obtain the estimator of the loading matrix $\mathbf{H}$, denoted by  $\hat{\mathbf{H}}$, in \eqref{eq:linear-model}, and then compute 
\begin{align*}
    \begin{pmatrix}
        \hat{\mathbf{E}}_{i,1}\\
        \hat{\mathbf{E}}_{i,2} 
\end{pmatrix}  =  \begin{pmatrix}
        \bX_i\\
        \bY_i 
\end{pmatrix}  - \hat{\mathbf{H}} \bZ_i \,,~~~~~~i=1,\ldots,n\,,
\end{align*}
where $(\hat{\mathbf{E}}_{i,1}^{\T},\hat{\mathbf{E}}_{i,2}^{\T})^{\T}$ provides an approximation of the error $\mathbf{E}_{i}=(\mathbf{X}^{\T}_i,\mathbf{Y}_i^{\T})^{\T}-\mathbf{H}\mathbf{Z}_i$. Based on the obtained $\{(\hat{\mathbf{E}}_{i,1}^{\T},\hat{\mathbf{E}}_{i,2}^{\T})^{\T}\}_{i=1}^n$, we can use the five tests focusing on the covariance matrix mentioned above to test whether $\cov(\mathbf{E}_{1}, \mathbf{E}_{2}) =\bzero$ or not. To evaluate the numerical performance of such idea, we let $\bZ\sim\mathcal N(\mathbf 0,\mathbf{I}_m)$, the entries of $\mathbf{H}\in\mathbb R^{(2p)\times m}$  be i.i.d.  $U(0,1)$, and $\mathbf{E}\sim \mathcal{N}(\bzero, \tilde{\bSigma})$ with $\tilde{\bSigma}$ specified in Settings S1--S5, where $\bZ$, $\mathbf{E} $ and $\mathbf{H}$ are mutually independent.  Therefore, $\bX\Vbar \bY \,|\, \bZ$ with $\tilde{\bSigma}$ specified in Settings S1 and S2, and $\bX \not\Vbar \bY \,|\, \bZ$ with $\tilde{\bSigma}$ specified in Settings S3--S5.  

Table \ref{tab:cov_test-cin} reports the empirical sizes and powers of our proposed conditional independence test  based on linear regressions (with
Rademacher multiplier) and  five competing covariance-matrix-based methods. As shown in Table \ref{tab:cov_test-cin}, our proposed method can consistently control the empirical sizes around the nominal significance level  0.05 across all settings and also has good power performance. However,  the five competing methods exhibit size distortions and their performance depends  crucially
on the relationship between $p$ and $n$.   The Schott, TW and TLH tests are
oversized, whereas the LRT and TBNP tests are undersized,  with the LRT test being  powerless and the TBNP test having power only in Settings S3 and S4 when $p$ is small. A possible explanation for the bad performance of the five competing covariance-matrix-based methods is that the asymptotic distributions of the corresponding test statistics may be changed due to the fact $\{(\mathbf{E}_{i,1}^{\T},\mathbf{E}_{i,2}^{\T})^{\T}\}_{i=1}^n$ is replaced by $\{(\hat{\mathbf{E}}_{i,1}^{\T},\hat{\mathbf{E}}_{i,2}^{\T})^{\T}\}_{i=1}^n$. Overall, these results also indicate that, even under Gaussian distributions, our method outperforms the five Gaussian-specific competitors, displaying greater robustness to the relationship between dimensionality and sample size. 
 
\begin{table}[htbp]
\centering
\caption{\footnotesize{ Empirical sizes (the rows with S1–S2) and powers (the rows with S3–S5) of the proposed conditional independence test based on linear regressions using the Rademacher multiplier and covariance-matrix-based competitors under Gaussian data with $m=5$. All numbers reported below are multiplied by 100. The results reported by `NA' indicate that the associated tests return invalid results. }}
\label{tab:cov_test-cin} 
\begin{spacing}{1.3}
\resizebox{\textwidth}{!}{
    \begin{tabular}{cc|cccccc|cccccc}
		\hline
		\hline
          &    & \multicolumn{6}{c|}{$n=100$}    & \multicolumn{6}{c}{$n=200$} \\
     {$p$} & {Setting} & {Proposed Method} & {Schott} & {LRT} & {TW} & {TLH} & {TBNP} & {Proposed Method} & {Schott} & {LRT} & {TW} & {TLH} & {TBNP} \\
    \hline
    		\multirow{5}{*}{20}& S1    & 7.3   & 19.2  & 0.7   & 26.2  & 28.1  & 0.0   & 5.5   & 9.3   & 2.2   & 10.6  & 11.3  & 0.0 \\
          & S2    & 8.9   & 19.7  & 0.5   & 26.0  & 29.3  & 0.0   & 5.8   & 10.9  & 2.1   & 12.1  & 13.5  & 0.0 \\
          & S3    & 100.0 & 100.0 & 0.0   & 100.0 & 100.0 & 99.1  & 100.0 & 100.0 & 0.0   & 100.0 & 100.0 & 100.0 \\
          & S4    & 95.7  & 100.0 & 0.0   & 100.0 & 100.0 & 100.0 & 100.0 & 100.0 & 0.0   & 100.0 & 100.0 & 100.0 \\
          & S5    & 99.7  & 89.2  & 0.0   & 97.3  & 100.0 & 0.0   & 100.0 & 100.0 & 0.0   & 99.9  & 100.0 & 0.0 \\
          \hline
          \multirow{5}{*}{40}& S1    & 7.6   & 63.9  & 0.0   & 89.7  & 85.9  & 0.0   & 6.3   & 17.4  & 0.3   & 23.2  & 24.9  & 0.0 \\
          & S2    & 7.8   & 64.8  & 0.0   & 90.1  & 85.5  & 0.0   & 5.7   & 19.4  & 0.8   & 24.2  & 25.5  & 0.0 \\
          & S3    & 100.0 & 100.0 & 0.0   & 100.0 & 100.0 & 0.0   & 100.0 & 100.0 & 0.0   & 100.0 & 100.0 & 100.0 \\
          & S4    & 93.5  & 100.0 & 0.0   & 100.0 & 100.0 & 0.0   & 100.0 & 100.0 & 0.0   & 100.0 & 100.0 & 100.0 \\
          & S5    & 99.9  & 88.8  & 0.0   & 99.9  & 100.0 & 0.0   & 100.0 & 93.2  & 0.0   & 99.8  & 100.0 & 0.0 \\
          \hline
          \multirow{5}{*}{80}& S1    & 7.3   & 100.0 & NA    & NA    & 51.9  & 99.7  & 6.2   & 61.1  & NA    & 88.0  & 83.5  & 0.0 \\
          & S2    & 6.8   & 100.0 & NA    & NA    & 49.7  & 99.1  & 5.4   & 61.4  & NA    & 87.8  & 82.2  & 0.0 \\
          & S3    & 100.0 & 100.0 & NA    & NA    & 50.9  & 99.3  & 100.0 & 100.0 & NA    & 100.0 & 100.0 & 0.0 \\
          & S4    & 91.5  & 100.0 & NA    & NA    & 51.5  & 99.2  & 100.0 & 100.0 & NA    & 100.0 & 100.0 & 0.0 \\
          & S5    & 100.0 & 100.0 & NA    & NA    & 48.0  & 99.5  & 100.0 & 89.6  & NA    & 100.0 & 100.0 & 0.0 \\
          \hline
          \multirow{5}{*}{200}& S1    & 7.9   & NA    & NA    & NA    & 83.6  & 99.3  & 6.3   & NA    & NA    & 88.0  & 83.5  & 0.0 \\
          & S2    & 8.2   & NA    & NA    & NA    & 83.7  & 99.0  & 6.8   & NA    & NA    & 87.8  & 82.2  & 0.0 \\
          & S3    & 100.0 & NA    & NA    & NA    & 83.9  & 99.4  & 100.0 & NA    & NA    & 100.0 & 100.0 & 0.0 \\
          & S4    & 88.8  & NA    & NA    & NA    & 82.9  & 99.2  & 100.0 & NA    & NA    & 100.0 & 100.0 & 0.0 \\
          & S5    & 100.0 & NA    & NA    & NA    & 93.4  & 98.8  & 100.0 & NA    & NA    & 100.0 & 100.0 & 0.0 \\
    \hline\hline
    \end{tabular}%
}
\end{spacing}
\end{table}%

\subsection{The Effect of Data Splitting in the Proposed Conditional Independence Test}\label{sec:data-splitting test}

The test statistic $\tilde{G}_n$ given in Section \ref{sec:step-1} for the conditional independence testing based on nonparametric regressions is obtained based on a single  split of the data. In this section, we study whether the randomness in the splitting procedure will affect the  performance of our proposed test.  More specifically, for a given sample size $n$, we first consider 10 different permutations of $\{1,\ldots,n\}$. For each $b=1,\ldots,10$, let $\pi(b) :=\{t_1^{(b)},\ldots, t_n^{(b)}\}$ be the associated permutation of $\{1, \ldots, n\}$, and write $\mathcal{W}_{\mathcal{D}_j}^{(b)}=\{(\bX_i,\bY_i,\bZ_i): i\in \mathcal{D}_j^{(b)}\}$ for $j=1,2,3$, where 
$\mathcal{D}_1^{(b)} =\{t_1^{(b)}, \ldots, t_{\lfloor n/3 \rfloor}^{(b)}\}$, $\mathcal{D}_2^{(b)} =\{ t_{\lfloor n/3 \rfloor +1}^{(b)}, \ldots, t_{\lfloor n/3 \rfloor + \lfloor n/2 \rfloor}^{(b)}\}$, and  $\mathcal{D}_3^{(b)}$ is selected by Algorithm 1 from $\tilde{\mathcal{D}}_3^{(b)} =\{  t_{\lfloor n/3 \rfloor + \lfloor n/2 \rfloor +1}^{(b)},\ldots, t_n^{(b)}\}$. For a given dataset $\{(\bX_i,\bY_i,\bZ_i)\}_{i=1}^n$, we can construct the proposed test statistic $\tilde{G}_n^{(b)}$ and the associated testing result $\Psi^{(b)}$ based on $\mathcal{W}_{\mathcal{D}_1}^{(b)}$, $\mathcal{W}_{\mathcal{D}_2}^{(b)}$ and $ \mathcal{W}_{\tilde{\mathcal{D}}_3}^{(b)}$, where $\Psi^{(b)}=1$ if rejecting the null hypothesis, and $\Psi^{(b)}=0$ otherwise. For a given dataset $\{(\bX_i,\bY_i,\bZ_i)\}_{i=1}^n$, if the randomness in the splitting procedure does not affect the performance of the proposed conditional independence test, we should have $\Psi^{(1)} = \cdots = \Psi^{(10)}$.  Let ${\rm mod}$ be the mode of $\{\Psi^{(1)}, \cdots \Psi^{(10)}\}$, and define
\begin{align*}
    D= \sum_{b=1}^{10} I\{\Psi^{(b)} \neq {\rm mod}\}\,. 
\end{align*}
Such defined $D$
provides a measure of how sensitive our proposed conditional independence test is to the randomness in the splitting procedure. A larger  $D$ indicates greater sensitivity. Table \ref{tab:n3-sens} reports the empirical frequencies of $D\in \{0,1,2,3,4,5\}$ in different settings based on 1000 repetitions, which shows that:
\begin{itemize}
\item When the sample size is large $(n=200)$, the procedure is insensitive to splitting randomness. 

 \item When the sample size is small $(n=100)$, the empirical size is generally insensitive ($K=0$  or $\rho =0$), whereas the power performance is sensitive to the splitting randomness. The power performance will be less sensitive  as the dependence becomes stronger (larger $K$ or larger $\rho$). 
This also verifies the common knowledge that sample-splitting will lead to power loss when the sample size is small.
\end{itemize}

When the sample size is small, 
we may consider using the full sample at each step for constructing the test statistic. 
Table \ref{tab:full-spli} compares  the empirical performance of the proposed conditional independence test based on nonparametric regressions with sample splitting and without sample splitting. 
When the sample size $n<100$, sample splitting leads to severe size inflation.
    A possible reason is that the effective sample size in each subsample becomes too small after sample splitting, and the asymptotic distribution does not provide a reliable approximation to the true null distribution. When $n = 100$, sample splitting provides better size control but leads to a loss of power in comparison to that using the full sample. However,  as the sample size further increases, the power under sample splitting improves significantly and becomes comparable to the full-sample performance. In practice, when the sample size is small $(n \le 100)$, we recommend using our proposed test without sample splitting.


\begin{table}[htbp]
\centering
\caption{\footnotesize{ Empirical frequencies of $D \in \{0,1,2,3,4,5\}$ based on 1000 repetitions with $p=q=100$ and $m=5$. 
Empirical sizes correspond to the rows with $K=0$ in Examples 6 and 8--10 and $\rho=0$ in Example 7, 
while powers correspond to the rows with $K\in \{p/10, p/5\}$ in Examples 6 and 8--10 and $\rho \in \{0.7, 0.8\}$ in Example 7.}}
\label{tab:n3-sens}
\begin{spacing}{1.2}
\resizebox{\textwidth}{!}{
    \begin{tabular}{cc|cccccc|cccccc}
    \hline\hline
          &       & \multicolumn{6}{c|}{$n=100$}&   \multicolumn{6}{c}{$n=200$}  \\
          &   Setting    & $D=0$    & $D=1$    & $D=2$    & $D = 3$  & $D = 4$     & $D=5$   & $D=0$    & $D=1$    & $D=2$    & $D = 3$  & $D = 4$     & $D=5$ \\
          \hline
    \multirow{3}[0]{*}{Example 6} & $K=0$ & 42.2 &39.1& 14.9 &3.6 &0.2 &   0.0  & 48.8 & 35.9&  13.0&  2.0&  0.3&     0.0 \\
          & $K=p/10$ & 0.9 &3.7 &14.5 &25.5 &37.2 &18.2 & 41.1& 40.9  &14.2 &3.5& 0.3  &  0.0\\
          & $K=p/5$ &12.5 &28.3& 31.2& 18.8 &7.4 &1.8 & 80.5 &17.5 &1.7& 0.3 &   0.0   & 0.0\\
          \hline
          \multirow{3}[0]{*}{Example 7} & $\rho=0$   &   44.8    &   36.5    &    14.8   &    3.3   &    0.5   &     0.1  &  53.1 &36.8 & 8.3  &1.6 & 0.2&    0.0 \\
          & $\rho=0.7$ &  76.7     &  20.0     &    3.1   &    0.2   &   0.0    &    0.0   &     99.8&  0.2 &   0.0 &   0.0 &   0.0 &   0.0  \\
          & $\rho=0.8$ &  90.6     &   9.1    &  0.3     &  0.0    &   0.0    &   0.0    & 100.0  &  0.0   &  0.0    &  0.0   &  0.0  &  0.0\\
           \hline
           \multirow{3}[0]{*}{Example 8} & $K=0$   &  43.8     &  38.9     &   14.0    &   2.9    &    0.8   &    0.1   &    46.2& 37.1& 13.5 & 2.4 & 0.6&  0.2\\
          & $K=p/10$ &   1.4    &   2.6    &   9.4    &     24.7   &   39.9    &  22.0     &  33.1 &40.1 &20.8  &  5.0  &  1.0  &  0.0  \\
          & $K=p/5$ & 13.1 & 29.2 &  30.0   &18.0&  8.4  &1.3  & 79.2 &18.7 & 1.9 &  0.2  &  0.0  &  0.0\\
           \hline
           \multirow{3}[0]{*}{Example 9} & $K=0$  & 40.2 & 38.6 &16.7  &3.9 & 0.6  &  0.0 & 48.7 &  36.0 &11.8 & 3.1&  0.4  &  0.0 \\
          & $K=p/10$ &1.0 & 1.2 & 7.4& 23.4 &42.3 &24.7 &36.9 &36.8& 18.5 & 6.5  &1.1  &0.2 \\
          & $K=p/5$ &2.4 &10.7&   24.0& 27.6 &26.3  & 9.0 & 63.2& 30.2  &  6.0 & 0.6 &   0.0   & 0.0\\
           \hline
           \multirow{3}[0]{*}{Example 10} & $K=0$ &   43.9 &37.1 &14.2  &4.3 & 0.4 & 0.1  &  49.9 &36.2& 11.6 & 2.2 & 0.1 &   0.0 \\
          & $K=p/10$ &0.3 & 1.8 & 6.9 &23.5 &42.9 &24.6  & 33.3  &  39.4  &    20.0   &  6.1  &   0.9  &   0.3\\
          & $K=p/5$ &2.6 &14.7 &22.6 &31.1 &19.8 & 9.2 & 67.0 &27.9 & 4.5&  0.6   & 0.0   & 0.0 \\
           \hline\hline
    \end{tabular}%
}
\end{spacing}
\end{table}%

\begin{landscape}  
\begin{table}[htbp]
\centering
\caption{\footnotesize{Empirical sizes (the rows with $K=0$ in Examples 6 and 8--10, and $\rho=0$ in Example 7) and powers (the rows with $K=p/10$ and $p/5$ in Examples 6 and 8--10, $\rho=0.7$ and 0.8 in Example 7) of the proposed conditional independence test based on nonparametric regressions using the Rademacher multiplier, with and without sample-splitting, where $p=q=100$ and $m=5$. All numbers reported below are multiplied by 100.   }}
\label{tab:full-spli}
\begin{spacing}{1.2}
\resizebox{23cm}{!}{
    \begin{tabular}{cc|cc|cc|cc|cc|cc}
    \hline\hline
          &     & \multicolumn{2}{c|}{$n=25$}&   \multicolumn{2}{c|}{$n=50$}   & \multicolumn{2}{c|}{$n=100$}&   \multicolumn{2}{c|}{$n=200$} &   \multicolumn{2}{c}{$n=400$} \\
          &    Setting   & full sample    & splitting sample   & full sample    & splitting sample     & full sample    & splitting sample & full sample    & splitting sample     & full sample    & splitting sample  \\
          \hline
    \multirow{3}[0]{*}{Example 6} & $K=0$ &10.1 & 74.6&  4.9& 15.1 & 6.0 &  7.3     &    6.3   &   7.0    &  6.9     &    5.6      \\
          & $K=p/10$ &100.0 & 76.7&  100.0 & 23.2  &  100.0  &     57.9  &    100.0   &  92.6     &   100.0    &    99.9    \\
          & $K=p/5$ & 100.0& 77.2& 100.0& 33.2 &   100.0     &   79.9    &   100.0    &   97.6    &   100.0    &    100.0    \\
          \hline
          \multirow{3}[0]{*}{Example 7} & $\rho=0$  & 9.2& 77.8& 8.8& 12.7 &    7.2    &    7.2   & 6.4      &  5.1     &   5.6    &    5.1    \\
          & $\rho=0.7$ &100.0 & 81.0& 100.0& 54.4 &    100.0    &  97.9     &   100.0     &   99.9     &     100.0  &  100.0   \\
          & $\rho=0.8$ &100.0 & 85.5 & 100.0& 68.9 &     100.0   &   99.3    &    100.0   &   100.0    &  100.0      &      100.0  \\
           \hline
           \multirow{3}[0]{*}{Example 8} & $K=0$   & 9.9&76.1 &9.7 & 11.9 &    9.5    &    7.8   &    7.8   &   5.0    & 6.8      & 4.9       \\
          & $K=p/10$ &100.0 & 79.3& 100.0& 19.0 &    100.0    &    52.2   &    100.0   &     89.0 &     100.0  &    99.6   \\
          & $K=p/5$ & 100.0& 78.1&100.0 & 31.7 &    100.0    &    80.1   &   100.0    &   97.2    &    100.0   &  100.0    \\
           \hline
           \multirow{3}[0]{*}{Example 9} & $K=0$   &10.7 & 77.8& 10.9 & 12.7 &     10.7   &  7.1     &     8.6  &  5.1     &   7.9    &   5.8     \\
          & $K=p/10$ &100.0& 78.8& 100.0& 18.5 &    100.0    &   44.7    &   100.0    &   89.6    &  100.0     &    99.6    \\
          & $K=p/5$ & 100.0& 78.7& 100.0 & 26.0 &    100.0    &   65.3    &     100.0  &   95.6    &    100.0   &  100.0    \\
           \hline
           \multirow{3}[0]{*}{Example 10} & $K=0$    & 10.3&76.4 &7.9 & 11.8 &   9.4     &   7.8 & 7.1     &  5.9     &    7.1   &   6.7      \\
          & $K=p/10$ &100.0 & 79.2& 100.0& 21.2 &    100.0    &   48.0    &   100.0    &   88.9    & 100.0      &   99.9   \\
          & $K=p/5$ & 100.0& 80.0&100.0 & 24.7 &   100.0    &   67.3    &  100.0     &    96.2   &  100.0     &  100.0      \\
           \hline\hline
    \end{tabular}%
}
\end{spacing}
\end{table}%
\end{landscape}

	\end{document}